\documentclass[11pt,a4paper]{report}
\pdfoutput=1
\usepackage{blindtext}

\usepackage{enumerate}
\usepackage{ifpdf}
\usepackage{sty/packages}
\usepackage{sty/tocbibind}
\usepackage{sty/thesis_layout}
\usepackage{bm}
\usepackage{subfigure}
\renewcommand{\baselinestretch}{1.8}
\usepackage{tikz}
\usepackage{bigints}
\usepackage{multirow}	
\usepackage{graphicx}													

\begin{document}

\begin{titlepage}
\begin{center}

\vspace{2mm}
\mbox{}\vfill {\Large \sc{Sensorless Physiological Control of Implantable Rotary Blood Pumps for Heart Failure Patients Using Modern Control Techniques}}

\vspace{20mm}

{\large {\bf Mohsen A. Bakouri}}

\vfill

\end{center}

\end{titlepage}
\pagenumbering{roman}
\centerline{Abstract}

The most common type of heart disease is left ventricular heart failure (HF). It usually occurs due to excessive load on the left ventricle (LV) as a result of the systemic circulatory insufficiencies and deprivation of oxygen rich blood to the cardiac muscle by narrowed coronary arteries. Sufferers of this disease have a life expectancy of one year and heart  transplantation is usually the only guarantee of survival beyond this period. The number of donor hearts available currently is less than 3,000 per annum worldwide and this number is continually decreasing. This may be due to the fact that HF disease has emerged globally as the largest killer in comparison with other diseases. Apart from the relatively fortunate people who receive donor hearts for transplant, the only alternative for people with HF is the implantation of rotary blood pump (IRBP) type left ventricular assist device (LVAD). It has been shown that, in some cases where recipients have had such devices implanted, recovery of the cardiac muscle has been possible allowing the LVAD to be explanted (bridge to recovery). In fact, an LVAD with its continuous operation requires a more complex controller to achieve basic physiological requirements. The essential control requirement of an LVAD needs to mimic the way that the heart pumps as much blood to the arterial circulation as it receives from the venous circulation. In this report, we focus particularly IRBPs and aim to design, develop and implement novel control strategies combining sensorless and non-invasive data measurements to provide an adaptive and fairly robust preload sensitive controller for IRBPs subjected to varying patient conditions, model uncertainties and external disturbances. We apply concepts and ideas from modern control and estimation theory to achieve the goals. The research consists of the following steps.

A pump flow is important physiological hemodynamic variable to study the CVS. However, this variable has not widely studied for control an LVAD. In this report, a sensorless estimator has been developed to estimate the average pulsatile flow  using two stable dynamical auto-regressive models with linear time variant systems based on the data from animal experiments. We have utilised a pulse width modulation signal as input to the estimator model. The cardiovascular states are simulated using different values of hemodynamic parameters of total blood volume $(V_{total})$, systemic vascular resistance $(R_{sa})$ and those parameters representing the left and right ventricular contractility $(E_{lv})$ and $(E_{rl})$. The estimator estimates the average pulsatile flow very efficiently and the results obtained from linear regression analysis are clinically accepted (i.e., a high correlation between estimated and measured flows $(R^{2}=0.99)$ resulting in minimum mean absolute error $(e=0.22)$ L/min). The merits of the proposed estimator include stability and use of sensorless measurements in comparison with existing methods of estimation. The developed estimator is really important as it will be used subsequently in the design of different robust physiological controllers for IRBPs.

A sensorless control strategies for an LVAD based on average pulsatile flow estimation have been presented, which refers to the crucial issue of controlling the operation of the pump rotational speed to cater for cardiovascular system perturbations and changing metabolic demand for HF patients. A novel tracking control algorithm based on a robust model reference sliding mode control (SMC) technique is developed to track the input reference pump flow signal. The constant and sinusoidal reference signals are used as inputs to the cardiovascular system (CVS) model. Simulations prove that the proposed control algorithm is capable of tracking the reference inputs with minimum mean absolute error $(e=0.46)$ L/min. The controller ensures the safe operating conditions in presence of model uncertainties.  

Furthermore, a robust physiological controller has been designed and developed for LVADs to regulate the estimated average pulsatile flow with a reference signal. This reference signal is dynamically updated based on non-linear function representing actual body's physiological needs and demands. The proposed physiological controller is improved further using feedforward - sliding mode control approach to maintain a motivated perfusion of LVADs. Simulation results depict that the controller responds to sudden perturbation in the CVS very quickly and adjusts the pump flow accordingly to avoid any suction or overperfusion.  

Finally, we have developed a novel physiological control algorithm that mimics the Frank-Starling law of the heart coupled with a robust pole placement sliding mode control. In this mechanism, the stroke volume of the heart increases in response to an increase in the volume of blood filling the left ventricular at the end of diastole. We exploit the linear relation between estimated average pulsatile flow and pump flow pulsatility in our proposed control algorithm. In comparison to other control methods with traditional controller like PI, PD, PID and fuzzy logic, the efficiency of the proposed controller is very high in terms of adjusting the average pulsatile flow accurately using pump flow pulsatility as the feedback parameter. 

The performance of the developed control algorithms is assessed using a lumped parameter model of the CVS that was previously developed using actual data from healthy pigs over a wide range of operating conditions. Immediate responses of the controllers to short-term circulatory changes as well as adaptive characteristics of the controllers in response to long-term changes are examined in a parameter-optimised model of CVS - IRBP interactions. Simulation results prove that the proposed controllers are fairly robust against model uncertainties, parameter variations and external disturbances. The controllers have shown a reasonably good tracking performance with a minimum mean absolute error. It has also been observed that our proposed control strategies are capable of restoring abnormal hemodynamic variables of LVADs back to normal physiological range.

\tableofcontents
\listoffigures
\listoftables
\include{empty} 
\pagenumbering{arabic}

\chapter{Introduction}

\section{Modern Control Techniques for Regulation of Biomedical Systems}

In recent years the availability of powerful low-cost microprocessors has made the implementation of complex nonlinear control strategies very efficient. In particular, the control of uncertain nonlinear systems has become an important subject of research which is in fact motivated by a large amount of important practical biomedical applications. As a result, considerable progresses in nonlinear robust control techniques such as $H^{\infty}$ control \cite{savkin1996robust, petersen2000robust}, sliding mode control \cite{bartoszewicz2007sliding, utkin1992sliding}, back stepping \cite{kwan2000robust}, geometric approach based control \cite{spong2006robot}, nonlinear adaptive control \cite{krstic1995nonlinear} and others, that explicitly account for an imprecise description of the model of the controlled plant, guaranteeing the attainment of the relevant control objectives in the face of modeling error and/or parameter uncertainties, have been attained.

The need for modern controller arises in many clinical situations requiring more accurate, reliable and autonomous regulation of physiological variables \cite{javed2012recent}. In general, modern control theory plays an important role in biomedical systems. For example, Chee et al. \cite{chee2005optimal} applied $H^{\infty}$ optimal control theory to insulin injection for the regulation of blood glucose in diabetic patients. An expert PID control system has also been used by the author to regulate the blood glucose in diabetic patients \cite{chee2003expert}. In addition, Su et al. \cite{su2007identification, su2007oxygen} developed a novel integrated approach for heart rate regulation during treadmill exercise using an $H^{\infty}$ control. In a similar fashion Cheng et al. \cite{cheng2008nonlinear} used a nonlinear controller consisting of feedforward and feedback for the regulation of heart rate during treadmill exercises. Furthermore, model predictive control methodology has been used to regulate blood volume and heart rate during hemodialysis \cite{javed2010model, javed2011identification}.

To date, the continuous operation of ventricular assist devices (VADs) to treat end stage heart failure patients is still need a more complex controller to achieve basic physiological requirement. The traditional control strategies shows a different limitation to adapt the cardiac with the physiological demand. In this field, different control strategies have been designed and implemented by various research groups. For instance,  Giridharan et al. \cite{giridharan2002nonlinear} selected PI controller to control the differential pressure accrued between aorta and the left ventricle. Similarly, Wu et al.\cite{wu2007modeling} proposed optimal PI controller to construct the control algorithm based on aortic pressure rather than pump differential pressure. Compared to pump differential pressure, Smith et al.[10] proved that pump flow is a more pertinent physiological parameter for the control of VADs. The main drawback related to these control strategies is the regulation of differential pressure and pump flow at fixed speed, which non physiological operation condition of these devices as stated by clinicians.

\section{Research Motivation}

Heart failure (HF) or congestive heart failure (CHF) is the final stage of heart disease and a major cause of mortality worldwide. This disease may arise due to  the factors internal to the heart like valvular and coronary heart disease, and/or  external factors which place excessive demands on the heart muscle such as hypertension and excessive volume load \cite{guytontextbook}. Currently there are an estimated 11.2 million sufferers of CHF worldwide and from amongst them, an estimated one million have a life expectancy of less than one year \cite{alomari2013developments}. Most importantly, this figure is increasing by 10\% per year due to poor diet, low exercise and increasing stress levels. Sufferers of end-stage HF have a life expectancy of one year and transplantation is usually the only guarantee of survival beyond this period for them. Currently there are about 3000 donor hearts available per annum worldwide. The ratio of the number of donor hearts available to the number of potential recipients is decreasing. This may be due to the fact that heart disease has emerged globally as the largest killer in comparison with other diseases \cite{fauci2008harrison}. The most common type of heart disease is left-side heart failure. This usually occurs due to excessive load on the LV as a result of the systemic circulatory insufficiencies and deprivation of oxygen rich blood to the cardiac muscle by the narrowed coronary arteries. Consequently it causes  the destruction of the cardiac muscle.

The traditional treatment to end-stage HF is heart transplantation. However, some patients are not eligible for a transplant because of age or health constraints. Even if the patients are eligible for a transplant, the severely limited supply of donor heart can play a significant role by offering approximately 10\% of patients an annual transplantation \cite{wu2004design}. In recent years, blood pumps have emerged as valuable solution to treat HF disease. These pumps have shown several relative advantages over the traditional treatments such as  low cost, no limits and may also provide treatment for the patients ineligible for heart transplants. Since a large percent of HF is attributed to LV failure,  an implantable rotary blood pump (IRBP) type of left ventricular assist device (LVAD) appears to be a promising alternative to cardiac transplantation and support method to extend the survival of HF patients in the last decade. It can revise the systematic abnormalities in advanced HF patients by improving the systemic end-organ perfusion. Currently, sensorless control of rotary blood pump has become one of the most important goals in providing long-term alternative treatment for HF patients. However, the implantation of additional sensors is not desirable which may result in thrombus formation, reduce system reliability, increase the cost and require regular calibration due to measurement drifts.

The LVADs are small size devices that are aimed to be implanted in HF patients. One of the main goals required to improve the clinical application of LVADs technology includes the development of a control strategy which automatically adjusts the pump rotational speed to cater for cardiovascular system (CVS) perturbations and the changing metabolic demand. In a healthy individual, the frank-starling mechanism ensures that the stroke volume of left ventricle (LV) is adjusted appropriately to compensate for changes in LV end-diastolic pressure such that the LV ejects entire volume of blood  received from the right ventricle \cite{guytontextbook}. Salamonsen et al. \cite{salamonsen2011response} found that the responses of an IRBP, when maintained at a fixed speed, to changes in preload and after load are very different from the natural heart. Also, they have insufficient preload sensitivity to inherently sense the amount of blood and are affected by variations in left ventricular afterload as well. Therefore, it is imperative that a pump control strategy must maintain a safe operating range where pump outflow matches the right heart output.

A significant number of control systems have been designed previously to provide physiological control of IRBP output to regulate the control variable at the given set point which often become inappropriate with the change in circulatory circumstances \cite{yi2007physiological}. The goal of an  efficient controller should be to achieve a balance between the venous return from the pulmonary circulation and pump output in the presence of varying conditions. In this research, we aim to  design a physiological control strategy to drive a pump rotational speed in accordance with the body metabolic demand. We use a sensorless linear time variant (LTV) dynamical model of average pulsatile flow to describe its behavior with bounded disturbances in the system dynamics \cite{bakouri2013feasible, bakouri2013sliding}.  Sliding mode control (SMC) approach has been chosen carefully to implement an efficient physiological control  system. SMC is a special discontinuous control technique applicable to various practical systems \cite{young1996control}.  A few of the the key advantages of using a SMC  include good transient response,  simple implementation , disturbance rejection, insensitivity to parameter variations and particularly robustness with respect to system uncertainties and external disturbances. Therefore, SMC has been proved to be applicable to a wide range of problems in many highly nonlinear uncertain systems \cite{slotine1991applied}.

\section{Aims and Objectives}

A physiological control algorithm using long-term reliable signals is essential for the permanent IRBPs. Clinically, the desired goal of physiological controller is to improve the intersection between IRBP and the CVS. Therefore, this controller should capable to restore the Frank-Starling law of the heart to prevent suction or over perfusion.

\emph{The broad aim of this research is to refine  pump estimation and state identification algorithms (i.e., to derive an algorithm for non-invasive estimation of pump flow). Then design, develop and implement different physiological control algorithms for ventricular assist devices (VADs) based on modern SMC approaches using suitable sensorless measurements of pump motor.}

\section{Report Contributions}

The major contributions of this report are:
\begin{itemize}
	\item \textbf{A State Space Average Pulsatile Flow Estimator Model}

In this report state space average pulsatile flow estimator model based on sensorless measurements of pulse-width modulation (PWM) has been developed using dog's data collected in a real time experiment. In this model, the cardiovascular states are simulated by using different values of hemodynamic parameters including total blood volume, afterload and parameters representing the left and right contractility. In addition, severity of HF is represented by the diversity of the parameters. Linear regression analysis between estimated and actual flow resulted in significant correlation with minimum mean absolute error ($e$). This estimation model will play a crucial role in developing robust physiological control algorithms.
	
	\item \textbf{A Novel Physiological Controller using Model Reference SMC}

This report presents a novel control methodology based on model reference  SMC to design a physiological controller for IRBPs by regulating average pulsatile flow using constant as well as  sinusoidal reference inputs. In this method, for the first time a model reference SMC approach to control IRBPs has been used. The controller response  is evaluated using a parameter-optimized model of the CVS - RBP where different levels of preload, afterload, left ventricular contractility and right ventricular contractility have been simulated.
	
	\item \textbf{A Novel Physiological Control Algorithm using Feedforward - SMC Approach}

To the author's knowledge, a novel physiological control algorithm using feedforward - SMC approach has been developed for the first time in this report to control IRBPs. The developed controller regulates the average pulsatile flow based on updated dynamical reference signal to prevent suction or over perfusion. A lumped parameter model of CVS, that was previously developed using data from animal experiments of healthy pigs, has been used to validate the control algorithm. Simulation results show that the average pulsatile flow estimation error is small, and that the abnormal hemodynamic variables of HF patient are restored back to normal physiological range.
	
	\item \textbf{A Novel Physiological Control Stretagy based on Frank-Starling Law}

This report presents a novel physiological controller that mimics the Frank-Starling law of the heart. In this method we used a pole placement  SMC approach to construct our control algorithm. The controller adjusts average pulsatile flow using pump flow pulsatility as a feedback parameter. The immediate response of the controller in presence of different conditions has been evaluated using a lumped parameter model of the CVS - RBP,. e.g., the changes in rest and changes from rest to exercise which in fact impose very different challenging operating conditions for the controller. \end{itemize}

\section{Report Organisation}

This report consists of six chapters which document the assessment of novel physiological control algorithms for IRBPs of HF patients. The structure of this report is as follows:
\begin{itemize}
		
		\item $\boldsymbol{Chapter}$ $\boldsymbol{2}$ develops a sensorless stable dynamical model using two ARX models with LTV system to estimate the pulsatile flow. The first ARX model uses PWM signal that was acquired sensorless from the pump controller to estimate the pulsatility index of rotational speed. The second ARX has been used to model the pulsatility index of rotational speed in order to estimate pulsatile flow.
		
		\item $\boldsymbol{Chapter}$ $\boldsymbol{3}$ presents a novel sensorless control strategy for an IRBP based on pulsatile flow estimation. Model reference SMC approach is used to develop a physiological control algorithm to regulate the estimated pulsatile flow with a desired reference signal considered as pump flow. A sinusoidal and constant reference signals are proposed to drive this strategy. The system uncertainties are assumed to be replaced with the upper and lower bounded values. The controller is  evaluated using a lumped-parameter model of the CVS - RBP.
		
		\item $\boldsymbol{Chapter}$ $\boldsymbol{4}$ presents the design of a feedforward - SMC physiological controller to drive IRBPs in the presence of model uncertainty and with varying degrees of HFs. The control algorithm is developed to regulate the estimated average pulsatile flow with a reference signal that was dynamically updated based on a non-linear function. The proposed control strategy and associated  advantages have been evaluated using  numerical simulations.
		
		\item $\boldsymbol{Chapter}$ $\boldsymbol{5}$ presents a novel physiological controller which emulates the Frank-Starling law of the heart using a novel robust pole placement SMC technique. The controller has been designed to adjust average pulsatile flow using pump flow pulsatility as a feedback parameter. The immediate response of the controller to changes in rest and changes from rest to exercise which impose very different operating conditions for the controller have been evaluated using a lumped parameter model of the CVS.
		
		\item $\boldsymbol{Chapter}$ $\boldsymbol{6}$ briefly summarises the report contributions and present some future recommendations which can be pursued to further enhance the implementation and validation of physiological controller to IRBPs for HF patients.
\end{itemize}


\chapter{Mathematical Model of Average Pulsatile Flow Estimator} \label{ch2}

\section{Overview}

The sensorless device operation has become quite challenging nowadays especially in the case of IRBP flow and pressure estimation.  Most recently, this field has received a lot of attention worldwide \cite{alomari2011non, alomari2009non, ayre2000sensorless, ayre2003non, bertram2005measurement, funakubo2002flow, karantonis2007noninvasive, karantonis2010noninvasive, malagutti2007noninvasive, nakata2000estimation, tanaka2003vivo, tsukiya1997use, tsukiya2001application, wakisaka1998development, wakisaka1997noninvasive, wakisaka1997establishment, yoshizawa2002sensorless}.  As the use of additional sensors is not desirable in LVADs due to some inherent drawback like thrombus formation, reduced system reliability and need for periodic recalibration \cite{lim2008noninvasive}. Therefore, the main design goal is to make IRBPs estimate the pump flow and head pressure without sensors with a reasonable accuracy and reliability.

In this chapter, we have proposed two auto-regressive (ARX) models to estimate the average pulsatile flow using linear time variant (LTV) systems \cite{bakouri2013feasible, bakouri2013sliding}. The first ARX model uses pulse-width modulation signal that is acquired sensorless from the pump controller to estimate the pulsatility index of rotational speed. The second ARX is used to model the pulsatility index of rotational speed to  estimate  pulsatile average pump flow. The data collected from dogs' experiments have been analysed carefully under steady flow condition for a wide range of pump operations and used for the estimation of model. Finally linear regression analysis between estimated and actual flow proves a highly significant correlation and minimum of mean absolute error.

\section{Methods}

\subsection{Software Simulation Environment} \label{model}
\begin{figure*}[!ht]
\centering
\includegraphics[width=5.2in]{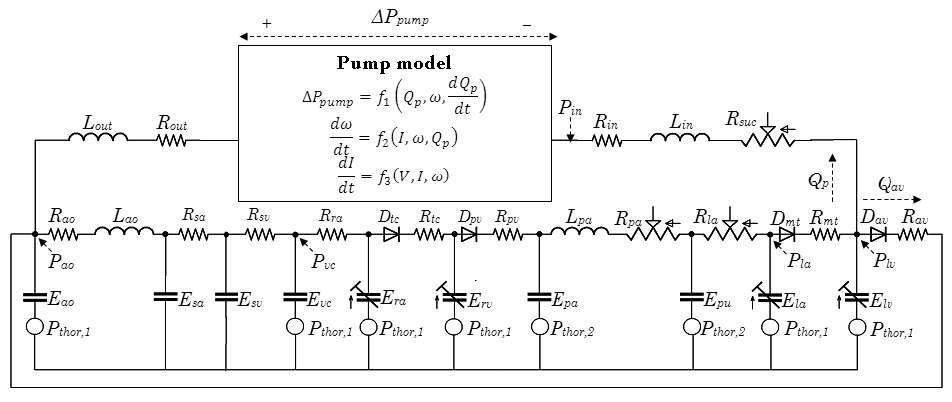}
\caption{Electrical equivalent circuit analogue of CVS - LVAD interaction. $R_{in}$: inlet cannulae resistances; $R_{out}$: outlet cannulae resistances; $L_{in}$: inlet cannulae inertances; $L_{out}$: outlet cannulae inertances; $R_{suc}$: suction resistance; $P_{thor,1}$ \& $P_{thor,2}$: intrathoracic pressures \cite{lim2010parameter}.}
\label{3fig:1}
\end{figure*}

A schematic diagram of the CVS - RBP system model that is used to develop this estimator is shown in Fig. \ref{3fig:1}. The model of the CVS consists of an arbitrary number of lumped parameter blocks; each characterised by its own resistances $(R)$, elastances $(E)$, inertances $(L)$, diodes $(D)$ and pressure $(P)$. In its simplest configuration, the CVS has ten compartments including the right and left sides of the heart as well as the pulmonary and systemic circulations. Each compartment in the CVS model is formulated based on  well-established experimental observations \cite{guytontextbook, epstein1967characterization}. The CVS model parameters have been tuned accordingly to reproduce pressure, flow and volume distributions in a healthy subject \cite{guytontextbook}. As reported in  \cite{lim2010parameter}, the model has been carefully validated using published data from literature as well as using actual data from healthy pigs implanted with an LVAD. 

\subsection{The Experiments}

This model has potentially studied by our research group \cite{alomari2011non} and developed in this report. In this study, the cardiovascular states are simulated using different values of hemodynamic parameters including; total blood volume $(V_{total})$, systemic vascular resistance $(R_{sa})$ and those parameters representing the left and right ventricular contractility $(E_{lv})$and $(E_{rl})$. Since the implementation of LVADs  is internal to the patients with abnormal hearts, the diversity of such parameters may represent varying degrees of heart failure severity. For  example, total blood volume can indicate the preload variation during rest condition, systemic vascular resistance represents the afterload variation during exercise condition and the left and right contractilities represent a change of weak heart conditions of the patients.

During the experimental, pump rotational speed has been varied from 1400 to 3200 rpm in stepwise increments of 100 rpm with each step of 20 seconds duration. The experimental data are recorded using a sampling rate of 150Hz. However, the data will be down-sampled to 50Hz in any future studies. In each experiment, pulse width modulation signal $(\overline {PWM})$, voltage of pump motor $(V)$, motor current $(I)$, pump flow $(Q_{p})$ and pump rotational speed $(\omega)$ are continuously monitored and recorded. Furthermore, for each cardiac cycle, the pulsatility indices, i.e., the amplitude of the signal, are extracted for pump rotational speed $(PI_{\omega})$ and  pump flow $(PI_{Q_{p}})$.

\subsection{Dynamic Modelling}

The design and development of the model is based on two dynamical time variant single-input single-output auto-regressive models with exogenous input (ARX) as described below: We used two ARX models as given by \ref{3eq:1} and \ref{3eq:2}. The first proposed ARX model represents relation between average pulsatility index of pump rotational speed ($\overline{PI}_{\omega}$) and average pulse-width modulation signal ($\overline{PWM}$) as:
\begin{equation} \begin{array}{rl}
\hat{q}_{1}(k+1)+\sum_{i=1}^{n}a_{i}(k-i+1)\hat{q}_{1}(k-i+1) \\
       =  \sum_{j=1}^{m}b_{j}(k-j+1)u(k-j-l+1)+e_{1}(k), \label{3eq:1}
\end{array} \end{equation}
where $\hat{q}_{1}(k)$ is the $\overline{PI}_{\omega}$, $u(k)$ is the $\overline{PWM}$, $a(k)$, $b(k)$ are the output and input time-varying system parameters respectively, $e_{1}(k)$ is the model noise, $l$ is the delay value, $k$ is the sampling time, $n$ and $m$ represent the model output and input orders respectively.

Similarly, the second proposed ARX model represents average pulsatile flow ($\overline {Q}_{p}$) and the estimated of $\overline{PI}_{\omega}$ as:
\begin{equation} \begin{array}{rl}
\hat{q}_{2}(k+1)+\sum_{i=1}^{n}c_{i}(k-i+1)\hat{q}_{2}(k-i+1) \\
       =  \sum_{j=1}^{m}d_{j}(k-j+1)\hat{q}_{1}(k-j-q+1)+e_{2}(k), \label{3eq:2}
\end{array} \end{equation}
where $\hat{q}_{2}(k)$ is the estimated $\overline{Q}_{p}$, $\hat{q}_{1}(k)$ is the estimated $\overline {PI}_{\omega}$ obtained from (\ref{3eq:1}), $c(k)$, $d(k)$ are the output and input time-varying system parameters of the model, $q$ is the system delay, $k$ is the sampling time, $e_{2}(k)$ is the model noise, $n$ and $m$ represent the model output and input orders respectively.

\begin{figure}[htbp]
\centering
\includegraphics[scale=0.5]{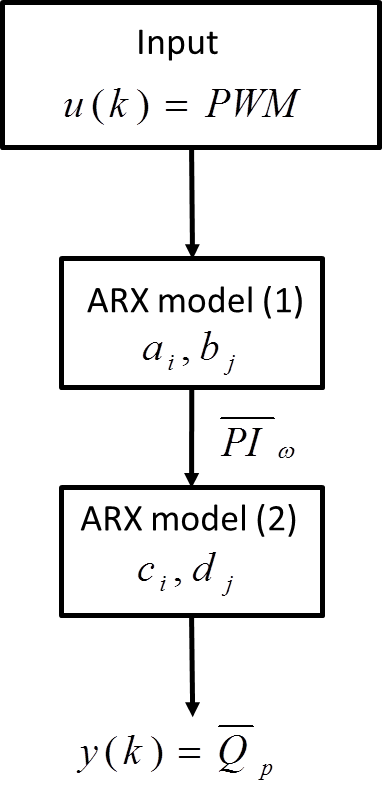}
\caption{The block diagram of ARX models used to estimate $\overline {PI}_{\omega}$ and $\overline {Q}_{p}$.}
\label{3fig:2}
\end{figure}

Fig. \ref{3fig:2} illustrates the ARX models with output $\overline{PI}_{\omega}$ in model \ref{3eq:1} and $\overline{Q}_{p}$ in model \ref{3eq:2} respectively. The $PWM$ signals from the experiments, $\omega$, $Q_{p}$ and  $PI_{\omega}$ are averaged over every 10 seconds. The model identification is accomplished using data samples from first 5 seconds and remaining data are used for model validation.   The system's transient response has been identified by including the data changes in the average pump rotational speed resulting from the variations in the pump target speed. The input and output model orders $n$ and $m$  are set from 1 to 10. The cross-correlation analysis between the input and the output signals are used to determine the delay values by estimating the impulse response of the system.

In both equations (\ref{3eq:1}) and (\ref{3eq:2}), recursive least square method was used to estimate the system parameters. These parameters were tracked on-line using a constant forgetting factor method \cite{ljung1999system}. This factor accommodates the variations of parameters and the change of preload, afterload and heart contractilities. In the case of sudden changes of venous return due to body posture when the body is straining or coughing, an updated forgetting factor is needed to allow fast track of the system parameters and given as:

\begin{equation}
	H(k+1)=\frac{1}{\chi}\left\{H(k)-\frac{H(k)\Omega (k+1)\Omega ^{T}(k+1)H(k)}{\chi +\Omega ^{T}(k+1)H(k)\Omega (k+1)}\right\},  \label{3eq:4}
\end{equation}
where $\Omega(k+1)$: is the estimated parameters at time $(k+1)$, $y(k+1)$: is the observed output at time $(k+1)$, $\hat{y}(k+1)$: is the prediction of $y(k+1)$ up to time $(k)$ and $\chi$: is the forgetting factor $(0<\chi\leq 1)$.

The estimated parameters were minimised at each time step $k$ using the following least squares cost function:

\begin{equation}
	V(y,\Omega)=\frac{1}{2}\sum ^{k} _{i=1} \chi^{k-i} \left (y(i)-\Omega(i) \right )^{2},  \label{3eq:8}
\end{equation}

During the system identification, the correlation coefficient $(R^{2})$ and the mean absolute error $(e)$ between measured and estimated values of $\overline {PI}_{\omega}$ and $\overline {Q}_{p}$ were used to evaluate the accuracy of the system model performance for different values of total circulatory volume ($V_{total}$), systemic vascular resistance ($R_{sa}$) and left ventricular contractility ($E_{lv}$) as:

\begin{equation}
	R^{2}=\frac{\sum ^{N} _{i=1} (y_{meas}(k)-\overline {y}_{meas}(k))(y_{est}(k)-\overline {y}_{est}(k))}{\left (\sum ^{N} _{i=1} (y_{meas}(k)-\overline {y}_{meas}(k))^{2}\sum ^{N} _{i=1} (y_{est}(k)-\overline {y}_{est}(k))^{2} \right )^{1/2}},    \label{3eq:9}
\end{equation}

\begin{equation}
	e=\frac{1}{N} \sum ^{N} _{i=1} (y_{meas}(k)-y_{est}(k))^{2},    \label{3eq:10}
\end{equation}
where $N$ is the length of data, $\overline {y}_{meas}$ and $\overline {y}_{est}$ are the average values of the measured and estimated pump flow.

Remark: In both models the best results have been obtained when the system orders are set to $n=m=1$ and the delay values as $l=q=1$.

The resulting system models are described by the following difference equations:
\begin{equation}
\hat{q}_{1}(k+1)+a(k)\hat{q}_{1}(k)=b(k)u(k)+e_{1}(k), \label{3eq:11}
\end{equation}
\begin{equation}
\hat{q}_{2}(k+1)+c(k)\hat{q}_{2}(k)=d(k)\hat{q}_{1}(k)+e_{2}(k), \label{3eq:12}
\end{equation}
where

$\hat{q}_{1}(k)$: is the estimated $\overline {PI}_{\omega}$,

$\hat{q}_{2}(k)$:  is the estimated $\overline {Q}_{p}$,

$u(k)$: is the $\overline {PWM}$,

$a(k)$, $b(k)$, $c(k)$ and $d(k)$: The time-varying of system parameters,

$e_{1}(k)$ and $e_{2}(k)$: The model noise.

Practically the difference equations (\ref{3eq:11}) and (\ref{3eq:12}) can be  converted to a state space representation as follows:
\begin{equation} \begin{array}{rcl}
q(k+1) & = & A(k)q(k)+B(k)u(k)+\zeta(k) \\
    y(k) &=& C(k)q(k),     \label{3eq:13}
\end{array} \end{equation}
where
$A(k)=$
$ \begin{bmatrix}
 -a(k)  &  0 \\
d(k) & -c(k) \\
\end{bmatrix}$,
$B(k)=$
$\begin{bmatrix}
b(k) \\
0\\
\end{bmatrix}$,
$C(k)=$
$\begin{bmatrix}
0 & 1\\
\end{bmatrix}$

here, $q(k)=$
$\begin{bmatrix}
q_{1}(k) & q_{2}(k) \\
\end{bmatrix}^{T}$
Where $q_{1}(k)$ is the $\overline{PI}_{\omega}$, $q_{2}(k)$is the $\overline{Q}_{p}$, $u(k)$ is the $\overline{PWM}$ which is the real pump control input, $\delta A$ is system parameter variations, $\zeta(k)$ is the system noise, $y(k)$ is the system output, $A , B$ and $C$ are the compatibly dimensioned matrices.

\section{Model Results}
\begin{figure}[!ht]
\centering
\includegraphics[scale=0.4]{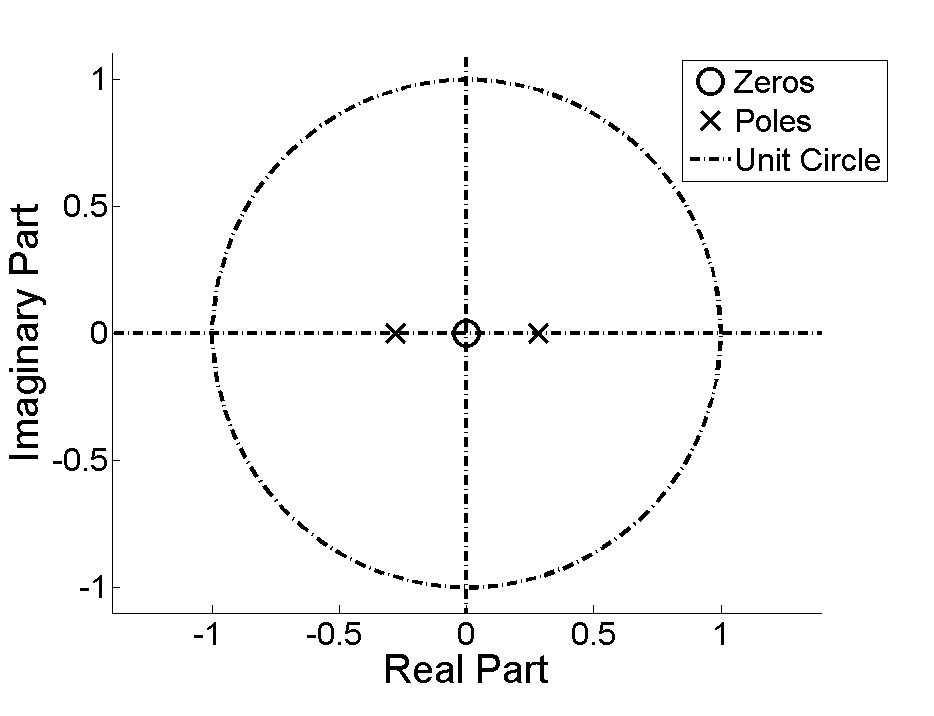}
\caption{Poles-zeros of the model estimator.}
\label{3fig:3}
\end{figure}

The correlation coefficient $(R)$ and mean absolute error $(e)$ are used to assess the performance of model as described in \ref{3eq:13}. It has been observed that the system orders $n=m=1$ and the delay values were $l=q=1$ give the best results in terms of minimum error and highest correlation between the estimated and measured values of $\overline{PI}_{\omega}$ and $\overline{Q}_{p}$.  Fig. \ref{3fig:3} shows that all poles and  zeros  lie inside the  unit circle satisfying the stability criteria of the system.
Consider the model in (\ref{3eq:13}), it has been verified from the experimental results that the variations of $a(k)$, $c(k)$ and $d(k)$ are bounded. Also, the parameter $b(k)$ is close to a constant. Therefore, the system model (\ref{3eq:13}) can be re-written as:

\begin{equation} \begin{array}{rcl}
q(k+1) & = & Aq(k)+\delta Aq(k)+Bu(k)+\zeta(k) \\
    y(k) &=& Cq(k),     \label{3eq:14}
\end{array} \end{equation}
where $\delta A$ is the system parameter variation and $\zeta (k)$ is the system disturbance.

\begin{figure}[!ht]
\centering
\includegraphics[width=5in]{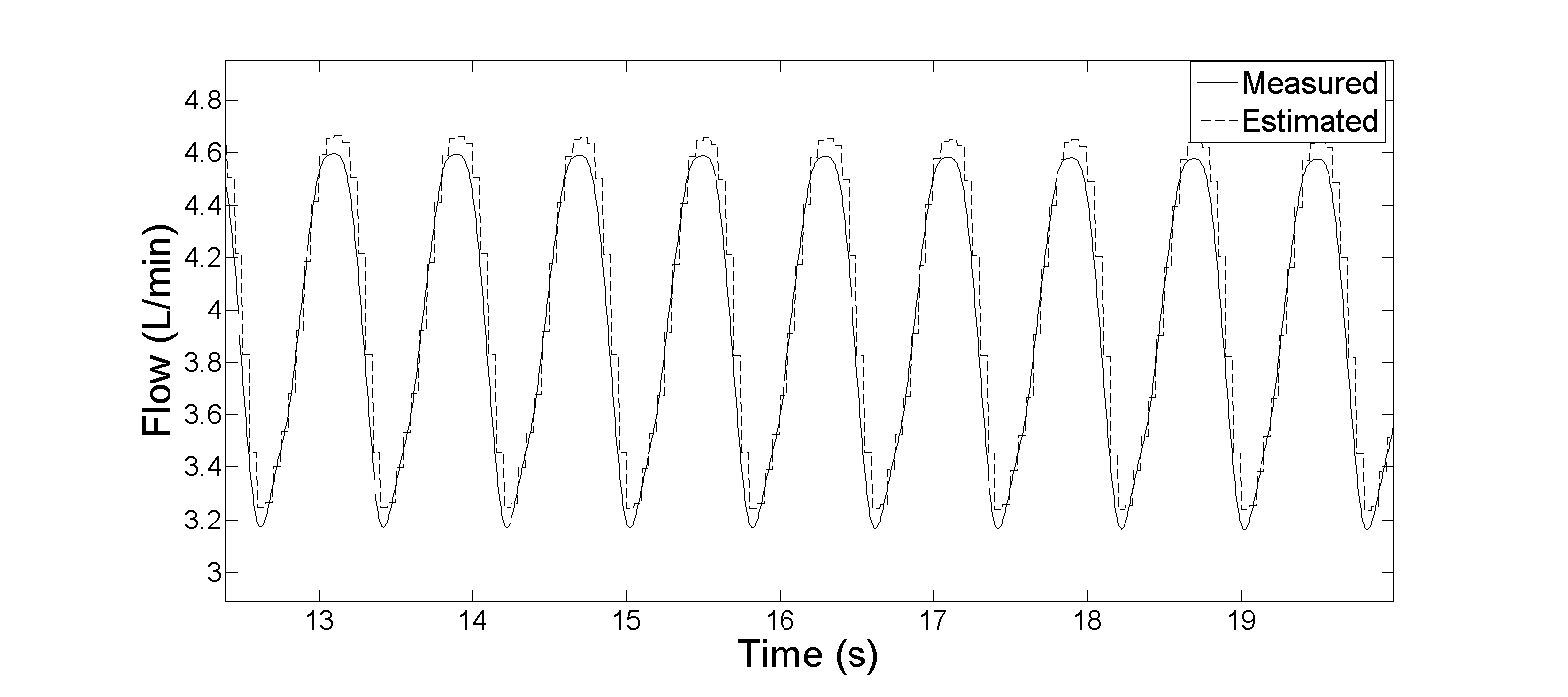}
\caption{Estimated average pulsatile flow $\overline{Q}_{p}$ compared with the measured flow $(Q_{meas})$ in one animal experiment at $\omega$ = 2900 rpm.}
\label{3fig:4}
\end{figure}

\begin{figure}[!ht]
\centering
\includegraphics[width=5in]{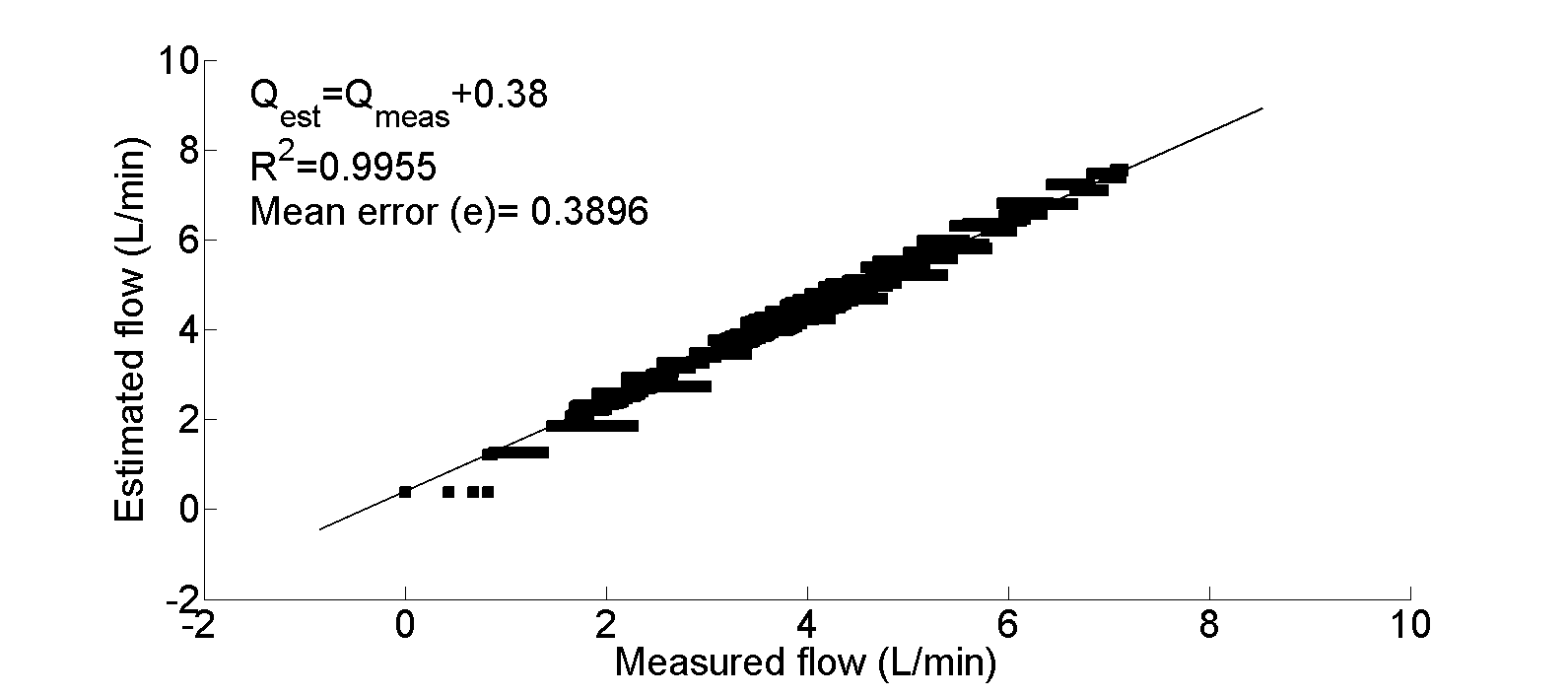}
\caption{Linear regression plot between estimated steady state flow $(Q_{est})$ against measured steady state flow $(Q_{meas})$ at $\omega$ = 2900 rpm.}
\label{3fig:5}
\end{figure}

\begin{figure}[!ht]
\centering
\includegraphics[width=5in]{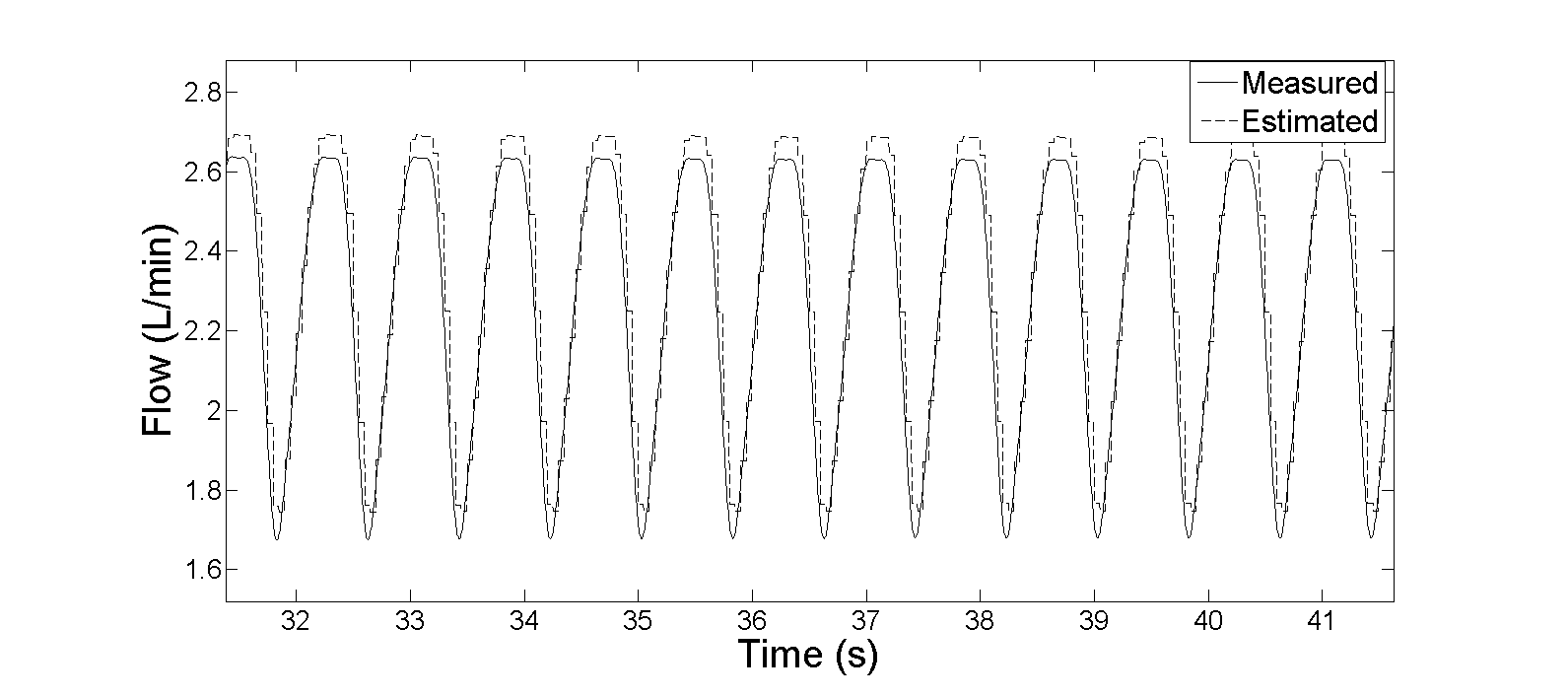}
\caption{Estimated average pulsatile flow $\overline{Q}_{p}$ compared with the measured flow $(Q_{meas})$ in one animal experiment at $\omega$ = 2100 rpm.}
\label{3fig:6}
\end{figure}

\begin{figure}[!ht]
\centering
\includegraphics[width=5in]{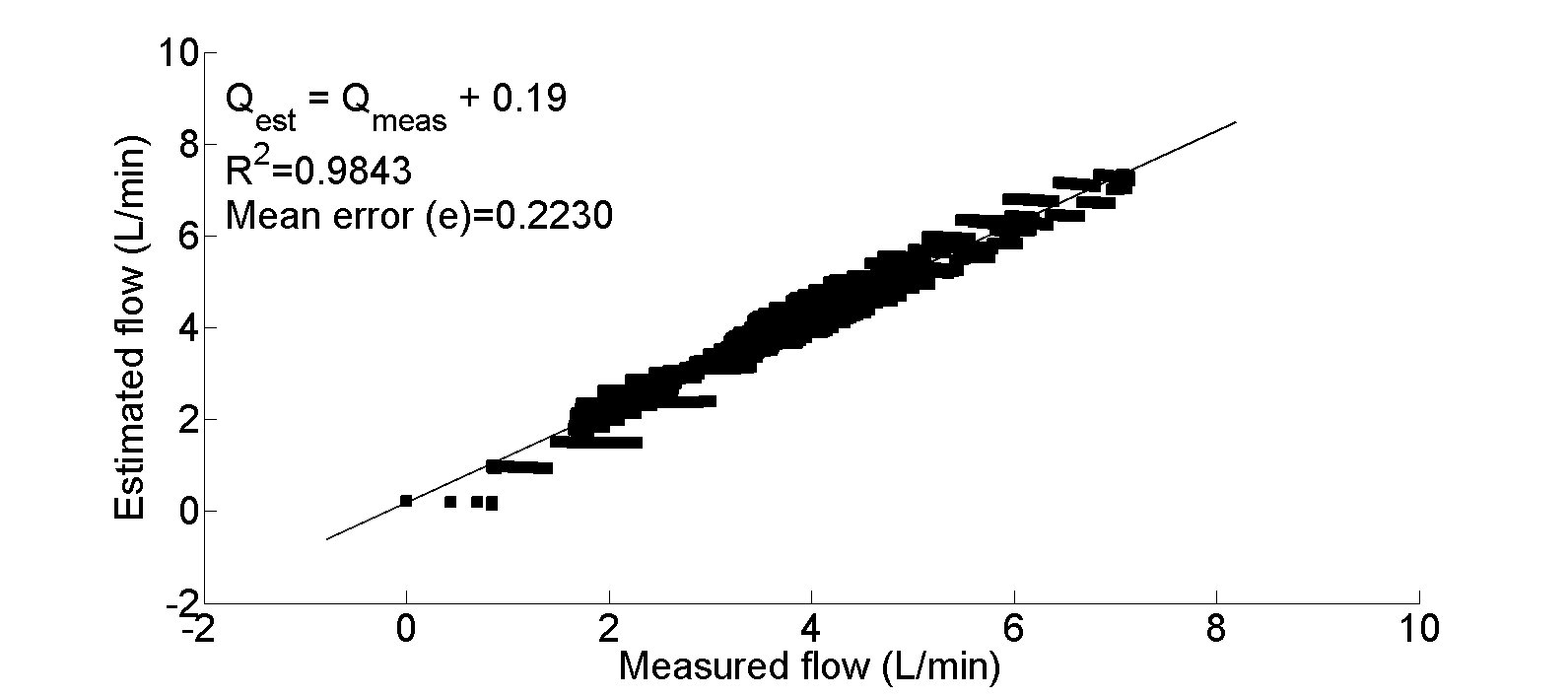}
\caption{Linear regression plot between estimated steady state flow $(Q_{est})$ against measured steady state flow $(Q_{meas})$ at $\omega$ = 2100 rpm.}
\label{3fig:7}
\end{figure}

Figures $\ref{3fig:4}-\ref{3fig:7}$  show the estimated steady state flow $(Q_{est})$ corresponding to a range of measured $(Q_{meas})$  at two different  speeds. At $\omega = 2900$ rpm, the correlation between estimated and measured  flow is highly significant, i.e., $(R^{2}=0.9955)$ with a small mean absolute error ($e=0.3896$ L/min), and the slope of linear regression line  is unity. Similarly at $\omega = 2100$ rpm, similar results have been obtained as shown in Fig. $\ref{3fig:7}$ with a highly significant correlation coefficient $(R^{2}=0.9843)$ and small mean absolute error ($e=0.2230$ L/min). Again the mean slope of the linear regression line is also unity.

\section{Discussions}
The feasibility of sensorless signal approach based on  accelerometery has been clearly demonstrated and justified in this chapter.  The reliability of this signal in  representing cardiac preload requires integration with other physiological signals such as heart rate and respiratory rate. However, heart failure patients have been observed to often have limited changes in these signals. So we believe that integration of sensorless pressure and/or flow may be required to effectively monitor and control cardiac preload so as to give the pump system a natural Frank-Starling behaviour.

Recently, non-invasive estimation of pulsatile flow has been carefully investigated due to its relevance to physiological design.  Most researches have used measurable quantities such as power and speed to map the flow. For  example, the first successful method was developed in \cite{wakisaka1997noninvasive, wakisaka1997establishment} where  the algorithm was derived using centrifugal pump and steady flow mock loop with whole goat blood for a full range of HCT ($21.5 - 42\%$). The curves of power and speed were   analysed to estimate the output flow. The method was reported with an average error of 0.5 L/min over a range of 2.3 - 8.1 L/min and with a high significant correlation average between real and estimated flow ($R^{2}=0.988$). However, the main drawback of this method is that the viscosity effects were investigated at only a single target speed (2800 rpm) with  only one animal ($n=1$).

Similarly, Funakubo et al. \cite{funakubo2002flow} developed a pump system  to estimate flow rate and pressure head non-invasively.  They analysed motor current and speed to estimate flow rate ($I^{'}=0.3062Q^{'}+17.445$) using  a Kyocera C1E3 centrifugal pump in a steady flow mock loop; where $I^{'}$ current consumption  and $Q^{'}$ flow are both functions of speed. While the pressure head was estimated as the function of speed only ($P^{'}=0.00007\omega^{2}-0.00076\omega+20.568$; where $\omega$ is angular velocity). They found that the average error of flow rate between estimated and measured is (0.65 L/min), while the average difference between the estimated pressure head and measured pressure head is (30.7 mmHg).

Although the previous models were implemented practically in different  scenarios, Kitamura et al. \cite{kitamura2000physical} attempted to estimate blood pressure and flow using centrifugal pump. Non-invasive measurements such as motor current and rotational  speed were used in this method. The model was evaluated in both in-vitro using Capiox pump (Terumo Corp., Tokoyo, Japan) and in-vivo using a 45 Kg sheep. In case of in-vitro, results showed a significant linear correlation between actual and estimated differential pressure ($R^{2}=0.994$) and pump flow ($R^{2}=0.992$), while in-vivo's results proved that the proposed algorithm needs robustness for the convergence of viscosity estimates.

\section{Conclusion}

The development of two stable dynamical  models using sensorless measurement of pulse-width modulation has been presented in this chapter. The models will be used to estimate the average pulsatile flow using LTV systems and are based on the data from animal experiments. We have used correlation coefficient and mean absolute error as performance measures to assess the overall performance of the proposed models. The results from linear regression analysis show that there exists a high correlation between estimated and measured flows resulting in minimum mean absolute error. The developed model is of high importance as it will be used to  develop  robust control systems  in the subsequent chapters especially the design of robust controller for pump flow particularly to cope with the changing physiological demands.


\chapter{Sensorless Physiological Control Algorithm of Implantable Rotary Blood  Pumps Using Model Reference Sliding Mode Control}

\section{Overview}

This Chapter presents a sensorless control strategy for a left ventricular assist device (LVAD) based on average pulsatile flow estimation which is referred as  the crucial issue in the usefulness of LVADs, i.e., how to control the operation of the pump rotational speed to cater for cardiovascular system perturbations and changing metabolic demands for heart failure patients \cite{bakouri2013feasible}. In order to assess left ventricular (LV) support in heart failure patients, a lumped parameter model of a rotary blood pump and the cardiovascular system has been used to investigate different control strategies. A tracking control algorithm based on model reference sliding mode (MRSMC) technique is developed to track the error difference between the reference pump flow and estimated average pulsatile flow. A lumped parameter model of cardiovascular system in combination with the stable dynamical model of average pulsatile flow estimation is used to evaluate the controller. The control algorithm has been tested using constant and sinusoidal reference pump flow inputs under healthy and heart failure conditions. Tracking control is evaluated  in the presence of modelling uncertainty and disturbance. Simulation results demonstrate that the control algorithm fairly tracks the reference input with minimal error in the presence of model uncertainty.

\section{Control Strategy} \label{4sub:vad}
 A novel sliding mode control (SMC) approach based on model reference is proposed to overcome the limited degree of adaptability to cardiac demand and clinical conditions of the heart that have plagued traditional control strategies \cite{bullister2002physiologic}. The desired closed-loop behaviour is achieved utilising a model reference in order to control the pump flow by tracking the error between the reference input and the estimated average pulsatile flow of the LVAD. The main goal of the proposed control system would be to force the system dynamics to achieve the characeteristics of an ideal model.  Obviously the controller's objective is to force the error to zero as time tends to infinity which means that the plant's output follows the model output faithfully \cite{edwards1998sliding}.

It is important to define a few terms which will be used in the coming sections. First is the reaching condition: a discrete system is said to satisfy a reaching condition if the resulting system possesses the following conditions:
\begin{equation} \begin{array}{lr}
\boldsymbol{\eta}(k) > \nu \rightarrow - \nu \leq \boldsymbol{\eta}(k+1) < \boldsymbol{\eta}(k) \\
\boldsymbol{\eta}(k) > - \nu \rightarrow \boldsymbol{\eta}(k) < \boldsymbol{\eta}(k+1) \leq \boldsymbol{\eta}(k) \\
\boldsymbol{\eta}(k)\leq \nu \rightarrow \left|\boldsymbol{\eta}(k+1)\right| \leq \nu,     \label{eq:1}
\end{array} \end{equation}
where $\boldsymbol{\eta}(k)$ is the switching function and $\nu$ is a positive constant.

Second is the matched uncertainty: any uncertainty which lies within the range space of the input disturbance matrix is described as matched uncertainty. Similarly, any uncertainty which does not lies within the range space of the input disturbance matrix is described as unmatched uncertainty \cite{edwards1998sliding}.

\begin{figure}[htbp]
\centering
\includegraphics[scale =0.5]{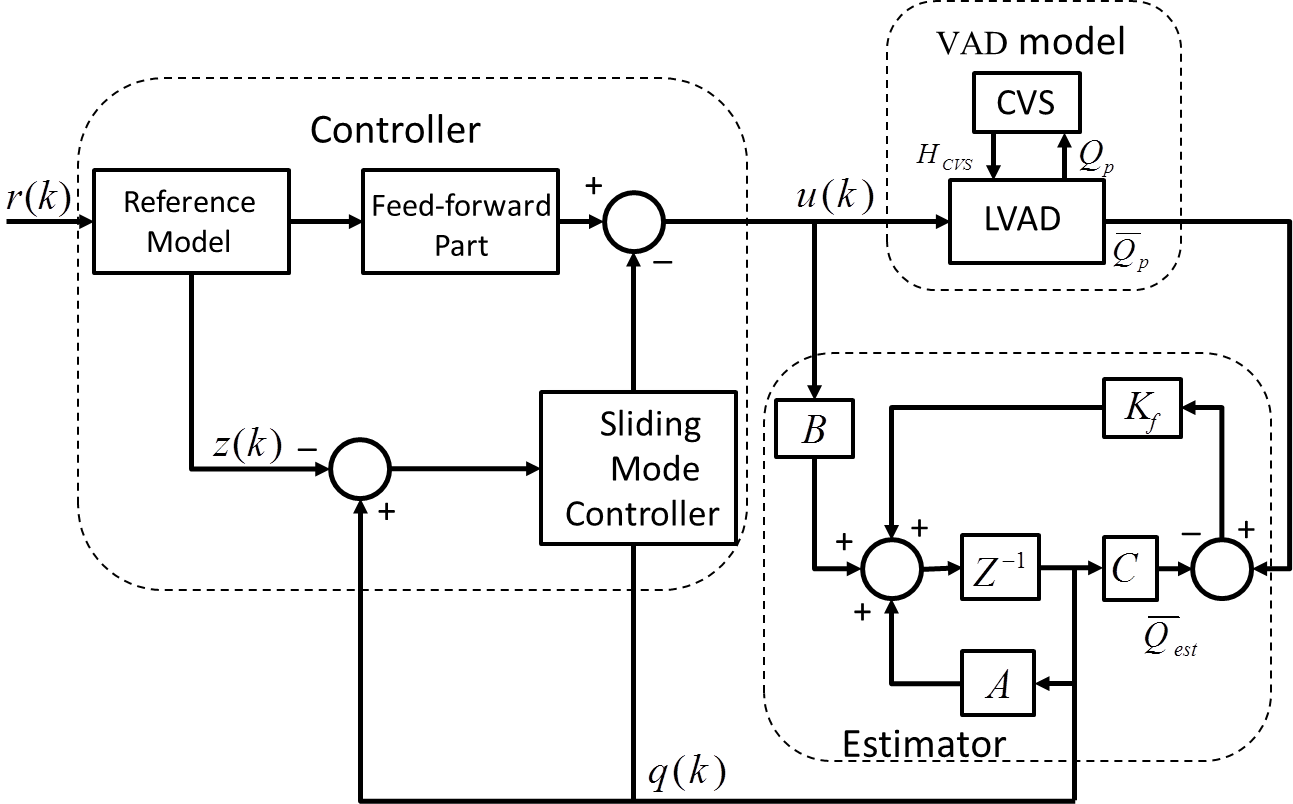}
\caption{Block diagram of control system.}
\label{4fig:1}
\end{figure}

Consider the block diagram of the control system shown in Fig. \ref{4fig:1}. In this control strategy, the controller  aims to achieve $e\rightarrow 0$ as $t\rightarrow \infty$. A reference model is also designed as a part to drive the control system. The states of the model estimator are used to calculate the states of the reference model. This reference model is devoted to the replication of the desired closed-loop behaviour and  combined with SMC, the controller has been designed to control the reference pump flow. In this strategy, two reference inputs are simulated: constant reference input, i.e.,  $r(k)\equiv a$, where $a=constant >0$ and sinusoidal reference input, i.e., $r(k)=a+b \text{sin}(2\pi t/T+\phi)$ where $a$, $b$ and $\phi$ = constant, $a>b$, while  $T$ is the heart period. The advantage of the SMC is the ability to drive the system error to zero within a minimum possible sampling period \cite{hung1993variable}.

\section{Materials and Methods} \label{4sec:mat}

\subsection{Modelling of VAD Assisted Circulatory System}
The proposed controller has been evaluated with a software model that incorporates a model of the human circulatory system (CVS) with a stable dynamical model of a left ventricular assist device (LVAD). As reported in \cite{lim2010parameter}, the model of the CVS consists of an arbitrary number of lumped parameter blocks, each  characterised by its own resistance, compliance, pressure and volume of blood. In its simplest configuration, the CVS has ten compartments including the right and left sides of the heart as well as the pulmonary and systemic circulations. 

\subsection{Simulation Study Protocols} \label{4sub:simpro}
In this protocol, constant and sinusoidal signals of varying mean, amplitude and phase shift has been applied to the reference pump flow input to study the hemodynamic response of the CVS under various changing conditions. In all simulations, frequencies of the sinusoidal signals are chosen to be equal to the heart rate.  The design parameters of the switching function in (\ref{4eq:30}) are $\boldsymbol{\Gamma}=[0.9413 ~~-0.0805]$ and those of the control law in the same equation are $\tau T=0.05$, $\epsilon T=0.025$ and $\chi=0.5$. The resulting values of $K$ and $G$ are $[0.9413 ~~-0.0805]$  and  $(-131.5705)$ respectively.

The transition simulation has been carried out from normal to rest in order to evaluate the tracking performance of the control algorithm in the presence of variations in the model parameters. The HF condition is taken as the baseline state (Table \ref{4tab:t1}) and the model parameters have been changed linearly at the middle of each period time for 30s, 60s, 90s and 120s respectively. These changes include the linearly decrease  in  the reference pump flow from $5+2.3 \text{sin}(2\pi t/T+\phi)$  to  $3.2+2.3 \text{sin}(2\pi t/T+\phi)$ and a linear decrease of total blood volume ($V_{total}$) by 500 mL.

On the other hand, the transition simulation from normal to exercise has been carried out to determine the controller's ability in combination with the CVS to provide the hemodynamic support required during normal daily activities. The model parameters have been changed linearly at the middle of over a period time for 30s, 60s, 90s and 120s respectively. These changes include a linear decrease in the reference pump flow from $5+2.3 \text{sin}(2\pi t/T+\phi)$  to  $5.5+2.3 \text{sin}(2\pi t/T+\phi)$, the systemic peripheral resistance ($R_{sa}$) (decreased linearly by $20\%$), total blood volume ($V_{total}$) (increased linearly by 500 mL), left and right ventricular contractility ($E_{lv}$, $E_{rv}$)  (increased linearly by 15\%) and heart rate (increased linearly by 30 bpm). We refer to Table \ref{4tab:t1} for other details HF condition with parameter changes.

\begin{table}[htbp]
 \caption{Changes in important model parameters to simulate the HF condition.}
  \small\addtolength{\tabcolsep}{5pt}
  \label{4tab:t1}
\begin{center}
\scalebox{0.85}{
		\begin{tabular}{ c  c  c c c}
			\hline
	Variable & Symbol & Unit & Healthy & Heart failure\\ \hline	
	Left ventricular contractility  & $E_{lv}$ & $ mmHg.mL^{-1}$      &    3.54    & 0.71   \\
	Right ventricular contractility & $E_{rv}$ & $ mmHg.mL^{-1}$      &    1.75    & 0.53   \\
	Systemic peripheral resistance  & $R_{sa}$ &  $mmHg.s.mL^{-1}$    &    0.74    & 1.11   \\
	Total blood volume              & $V_{total}$ & $(mL)$            &    5300    & 5800    \\
			\hline
		\end{tabular}}
		\end{center}
\end{table}

\subsection{Implementation of the Control Algorithm} \label{4sec:prob}

Consider the model in (\ref{3eq:14}) without uncertainty and disturbance as:
\begin{equation} \begin{array}{rcl}
q(k+1) & = & Aq(k)+Bu(k) \\
    y(k) &=& Cq(k),     \label{4eq:2}
\end{array} \end{equation}
and assume the corresponding reference model as:

\begin{equation} \begin{array}{rcl}
z(k+1) & = & A_{t}w(k)+B_{t}r(k), \\
      \label{4eq:3}
\end{array} \end{equation}
where $q\in \mathbb{R}^{n}$ and $z\in \mathbb{R}^{n}$ are the state vectors of the estimator and reference model respectively, $u\in \mathbb{R}^{m}$ is the control vector, $r\in \mathbb{R}^{r}$ is an input vector, $y(k)$ is the system estimator output and $A, B, A_{t}, B_{t}$ and $C$ are compatibly dimensioned matrices.  We assume that the pairs $(A,B)$ and $(A_{t},B_{t})$ are controllable and that the reference model is stable i.e., the eigenvalues of $A_{t}$ have negative real parts.

Consider the model reference in  (\ref{4eq:3}), it is required to determine $(A_{t},B_{t})$ to be tractable by the system in  (\ref{4eq:2}). This determination is clesarly illustrated by considering the following non-proofing theorem in linear algebra \cite{strang1997linear}.

\begin{theorem}
Let $A$ be an $n\times m$ matrix and $c$ be a vector in $\mathbb{R}^{m}$. Then the system of equations $Aq=c$ has a solution if and only if $rank[A]=rank[A_{c}]$;
where $A_{c}$ is $(m\times(n+1))$ augmented matrix.
\end{theorem}

If the conditions in (\ref{4eq:4}) and (\ref{4eq:5})  can be satisfied:

\begin{equation}
rank [B    ~~~A_{t}-A]= rank [B],
\label{4eq:4}
\end{equation}

\begin{equation}
rank [B    ~~~B_{t}]= rank [B],
\label{4eq:5}
\end{equation}
then one important implication of the previous theorem is that there exists compatibly dimensioned matrices $K$ and $G$ such that:

\begin{equation}
A_{t}=A-BK,
\label{4eq:6}
\end{equation}
and,
\begin{equation}
B_{t}=BG,
\label{4eq:7}
\end{equation}
where $K$ and $G$ are state feedback and feed-forward matrices respectively. The state feedback gain $K= [K_{1}  ~K_{2} \cdot \cdot \cdot K_{n}]$ can be designed to make the closed loop dynamic systems asymptotically stable using pole placement method or the Ackerman's formula \cite{khalil1996nonlinear,goodwin2001control}. Theoretically pole placement is to synthesise a controller such that the desired closed-loop poles are in  predefined locations and the closed-loop system achieves the desired response.

The feed-forward steady state gain can be calculated as the inverse of  triple components $(C,(A+BF),B)$ as described in \cite{goodwin2001control}.

Consider a new state feedback law of the form: $u(k)=Kq(k)+Gr(k)$; the resulting closed-loop state equation is:
\begin{equation} \begin{array}{rcl}
q(k+1) & = & (A+BK)x(k)+BGr(k) \\
    y(k) &=& Cq(k),     \label{eq:13}
\end{array} \end{equation}
 where $G$ is the feed-forward steady state gain. The reference input $r(k)$ is now multiplied by a gain $G$  such  that for a reference input $r(k)=R$, $t\geq 0$, the steady state of the output is $R$ such that:
  \begin{equation} \begin{array}{rcl}
    y_{ss}\equiv \lim \limits_{t=k\to\ \infty} y(k)=R,
      \label{eq:14}
\end{array} \end{equation}
To obtain an expression for $G$, we proceed as follows:
\begin{itemize}
	\item For the constant reference input $r(k)=R$, $\tau\geq 0$, steady state corresponds to an equilibrium condition for the closed-loop state equation involving an equilibrium state denoted by $q_{ss}$. Hence the state equation satisfies:
	\begin{equation} \begin{array}{rcl}
    q(k+1)=(A+BK)q_{ss}+BGR=0,
      \label{eq:15}
\end{array} \end{equation}

	\item The steady-state output is obtained from:
	\begin{equation} \begin{array}{rcl}
    y_{ss}=Cq_{ss}=C(A+BK)^{-1}BGR,
      \label{eq:16}
\end{array} \end{equation}

	\item By referring to the stated limit condition:
	\begin{equation} \begin{array}{rcl}
    R=C(A+BK)^{-1}BGR,
      \label{eq:17}
\end{array} \end{equation}

Solving the above equation for $G$, yields the feed-forward steady state gain represented as:
\begin{equation} \begin{array}{rcl}
    G=[C(A+BK)^{-1}]^{-1}.
      \label{4eq:8}
\end{array} \end{equation}
\end{itemize}

In this  study, we consider error trajectory tracking problem where  tracking reference trajectories are predefined. We define the state tracking error as:
\begin{equation}
e_{q}(k)=q(k)-z(k), \label{4eq:9}
\end{equation}

The error needs to tend to zero asymptotically by forcing the closed loop system to track the desired output (i.e., estimated flow). The state error dynamics can be obtained  from (\ref{4eq:2}) and (\ref{4eq:3}) as follows:

\begin{equation}
e_{q}(k+1)=q(k+1)-z(k+1), \label{4eq:10}
\end{equation}

Substituting (\ref{4eq:2}) and (\ref{4eq:3}) into (\ref{4eq:10}) gives:

\begin{equation}
e_{q}(k+1)=Aq(k)+Bu(k)-(A_{t}z(k)+B_{t}r(k)), \label{4eq:11}
\end{equation}

Re-arranging  (\ref{4eq:11}) yields:

\begin{equation}
e_{q}(k+1)=Aq(k)+Bu(k)-A_{t}z(k)-B_{t}r(k), \label{4eq:12}
\end{equation}

Adding and subtracting  a term $A_{t}q(k)$ to  (\ref{4eq:12}) yields:

\begin{equation} \begin{array}{rcl}
e_{q}(k+1)  = & Aq(k)-A_{t}q(k)+A_{t}q(k) \\
          & -A_{t}z(k)+Bu(k)-B_{t}r(k), \label{4eq:13} \\
\end{array} \end{equation}

Re-arranging  (\ref{4eq:13}) we get:

\begin{equation}
e_{q}(k+1)=A_{t}e_{q}(k)+(A-A_{t})q(k)+Bu(k)-B_{t}r(k). \label{4eq:14}
\end{equation}

The main design objective is to construct a control law $u(k)$ which guarantees a minimum robust tracking error  ensuring the stability of the system.
The controller includes the flow rate estimator, a reference model and a SMC to determine the required change in pump rotational speed.
Generally, discrete sliding mode controllers  are normally developed mainly using state-space models \cite{furuta1990sliding,sarpturk1987stability, chan1994robust}.
In the case of state space model, the design of the SMC is divided into two steps: 1) the choice of an appropriate switching function and 2) determination of a control law.

Let us define the switching function for this model as:

\begin{equation}
\boldsymbol{\eta}(k)=\boldsymbol{\Gamma }e_{q}(k), \label{4eq:15}
\end{equation}

From  (\ref{4eq:15}), we can write:

\begin{equation}
\boldsymbol{\eta}(k+1)=\boldsymbol{\Gamma} e_{q}(k+1), \label{4eq:16}
\end{equation}
where $\boldsymbol{\Gamma}$ is a constant vector designed based on quadratic minimisation method. In this method, switching function is established based on minimising a cost function \cite{utkin1978methods, edwards1998sliding}.

In SMC, different control techniques have been introduced. One of these techniques is known as the reaching law method, which was firstly proposed by Gao et al \cite{gao1995discrete}. This reaching law describes the necessary conditions to guarantee the ideal sliding motion as:

\begin{equation}
\boldsymbol{\eta}(k)*(\boldsymbol{\eta}(k+1)-\boldsymbol{\eta}(k))\leq 0. \label{4eq:17}
\end{equation}

The equivalent discrete-time reaching law that satisfies our model is:

\begin{equation}
\boldsymbol{\eta}(k+1)=(1-\tau T)\boldsymbol{\eta}(k)-\epsilon T \text{sign}(\boldsymbol{\eta}(k)), \label{4eq:18}
\end{equation}

In the equation above the following conditions must be satisfied:

\begin{equation} \begin{array}{rl}
0<1-\tau T<1 \\   0<\epsilon T<1, \label{4eq:19}
\end{array} \end{equation}
where $T>0$ is the sampling period, $\epsilon >0$ is the reaching velocity and $\tau>0$ is the converging exponential.

Substituting  (\ref{4eq:14}) into  (\ref{4eq:16})  yields:

\begin{equation}
\boldsymbol{\eta}(k+1)=\boldsymbol{\Gamma}(A_{t}e_{q}(k)+(A-A_{t})q(k)+Bu(k)-B_{t}r(k)), \label{4eq:116}
\end{equation}

Equating (\ref{4eq:116}) and  (\ref{4eq:18}) gives:

\begin{equation} \begin{array}{lr}
\boldsymbol{\Gamma}(A_{t}e_{q}(k)+(A-A_{t})q(k)+Bu(k)-B_{t}r(k))=\\
 \qquad(1-\tau T)(\boldsymbol{\eta}(k)-\epsilon T \text{sign}(\boldsymbol{\eta}(k)),
    \label{4eq:20}
\end{array} \end{equation}

Solving the above equation for the command signal $u(k)$ yields:

\begin{align}
    \begin{aligned}
 u(k) &= -(\boldsymbol{\Gamma} B)^{-1}(\boldsymbol{\Gamma} A_{t}e_{q}(k)+\boldsymbol{\Gamma}(A-A_{t})q(k)\\
  &\qquad -\boldsymbol{\Gamma} B_{t}r(k)-(1-\tau T)\boldsymbol{\eta}(k)+\epsilon T \text{sign}(\boldsymbol{\eta}(k))).
  \end{aligned}
\end{align}

\begin{figure}[!ht]
\centering
\includegraphics[scale=0.35]{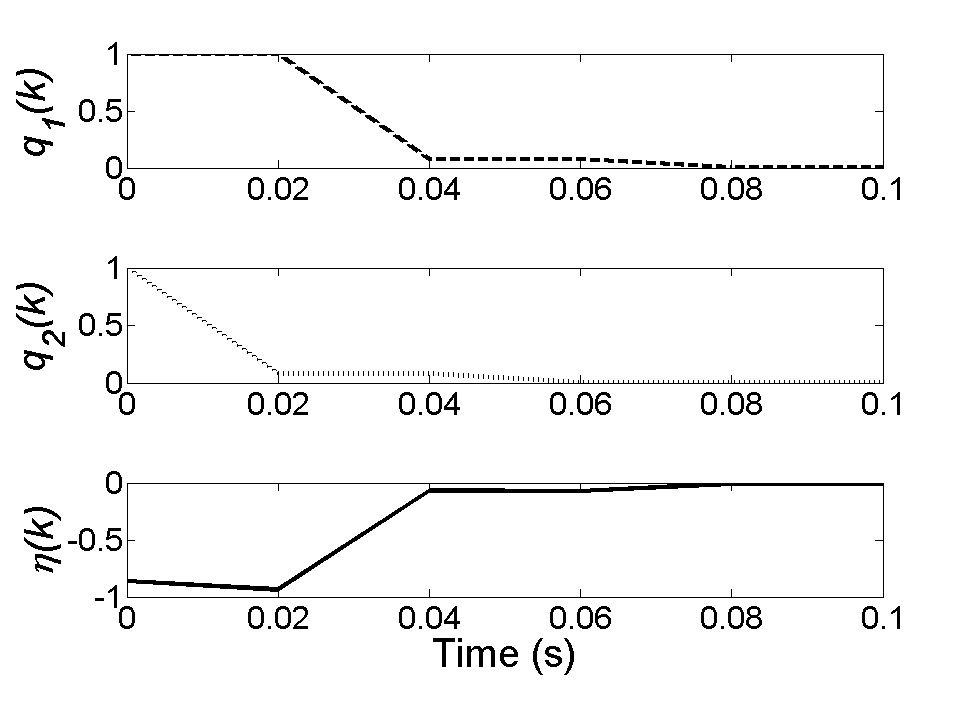}
\caption{Responses of state variables to the switching plane.}
\label{5f:2}
\end{figure}

During design the switching function $\boldsymbol{\Gamma}$, the matrix pair $(A_{t},B)$ is used and a corresponding hyperplane technique is considered as in \cite{spurgeon1992hyperplane}, where  $\Gamma$  is chosen such that $(\boldsymbol{\Gamma} B)^{-1}$ is non-singular. Fig. \ref{5f:2} illustrates the system response of the state variables $q_{1}(k)$ and $q_{2}(k)$ with convergence to the switching plane. 

So far, the control law has been designed for the nominal plant model that has been defined by  (\ref{4eq:2}). However, in presence of disturbance or system parameters uncertainties, we invoke the model in  (\ref{3eq:14}) as follows:

\begin{equation} \begin{array}{rcl}
q(k+1) & = & Aq(k)+\delta Aq(k)+Bu(k)+\zeta(k) \\
    y(k) &=& Cq(k)+\psi(k),     \label{4eq:22}
\end{array} \end{equation}
where $\delta A$ is system parameter variation, $\zeta (k)\in \mathbb{R}^{n}$ and $\psi (k)\in \mathbb{R}^{r}$ are unmatched system disturbance and measurement noise respectively.

From  (\ref{4eq:10}) and (\ref{4eq:22}), the state tracking error can be expressed as:

\begin{equation}\begin{array}{rl}
e_{q}(k+1)= & Aq(k)+\delta Aq(k)+Bu(k)+\zeta(k)\\
            & -(A_{t}z(k)+B_{t}r(k)), \label{4eq:23}\\
\end{array}\end{equation}

Re-arranging the above equation we get:

\begin{equation}\begin{array}{rl}
e_{q}(k+1)=& Aq(k)+\delta Aq(k)+Bu(k)+\zeta(k)\\
           & -A_{t}z(k)-B_{t}r(k), \label{4eq:24}\\
\end{array}\end{equation}

Adding and subtracting the term ($A_{t}q(k)$) to  (\ref{4eq:24}) yields:

\begin{equation} \begin{array}{rl}
e_{q}(k+1)  = & Aq(k)+\delta Aq(k)-A_{t}q(k)+A_{t}q(k) \\
          & -A_{t}z(k)+Bu(k)-B_{t}r(k)+\zeta(k), \label{4eq:25} \\
\end{array} \end{equation}

Re-arranging  (\ref{4eq:25}) we obtain:

\begin{equation}\begin{array}{cl}
e_{q}(k+1)= & A_{t}e_{q}(k)+(A-A_{t})q(k)+\delta Aq(k)\\
            & +Bu(k)-B_{t}r(k) + \zeta(k). \label{4eq:26}
\end{array} \end{equation}

Based on  (\ref{4eq:16}), we can write   (\ref{4eq:26}) as:

\begin{equation} \begin{array}{cl}
\boldsymbol{\eta}(k+1)=  &= \boldsymbol{\Gamma}(A_{t}e_{q}(k)+(A-A_{t})q(k)+\delta Aq(k)\\
						 & +Bu(k)-B_{t}r(k)+\zeta(k)), \label{4eq:27}\\
\end{array} \end{equation}

Re-calling  (\ref{4eq:18}) and equalising with  (\ref{4eq:27}) we get:

\begin{equation} \begin{array}{cl}
(1-\tau T)\boldsymbol{\eta}(k)-\epsilon T \text{sign}(\boldsymbol{\eta}(k))  &= \boldsymbol{\Gamma}(A_{t}e_{q}(k)+(A-A_{t})q(k)+\delta Aq(k)\\
						 & +Bu(k)-B_{t}r(k)+\zeta(k)), \label{4eq:127}\\
\end{array} \end{equation}

Solving the above equation for the command signal $u(k)$ we can write:

\begin{equation} \begin{array}{rl}
u(k) = & -(\boldsymbol{\Gamma} B)^{-1}(\boldsymbol{\Gamma} A_{r}e_{x}(k)+\boldsymbol{\Gamma}(A-A_{t})x(k)-\boldsymbol{\Gamma} B_{t}r(k) \\
       & -(1-\tau T)\boldsymbol{\eta}(k)+\epsilon T \text{sign}(\boldsymbol{\eta}(k))) \\
       & -(\boldsymbol{\Gamma} B)^{-1}(\boldsymbol{\Gamma} \delta Aq(k)+\boldsymbol{\Gamma} \zeta(k)). \label{4eq:28}  \\
\end{array} \end{equation}

Assumption: As $\delta A$ and $\zeta(k)$ are unknown, so the control law cannot be implemented unless we assume that the upper and lower bounds of the value $(\Gamma \delta Aq(k)+\Gamma \omega(k))$ are known and given as:

\begin{equation}
-\chi<(\boldsymbol{\Gamma} \delta Aq(k)+\boldsymbol{\Gamma} \zeta(k))<\chi,
\label{4eq:29}
\end{equation}

Finally, the control law can be re-written as:

\begin{equation} \begin{array}{rl}
u(k) = & -(\boldsymbol{\Gamma} B)^{-1}(\boldsymbol{\Gamma} A_{t}e_{q}(k)+\boldsymbol{\Gamma}(A-A_{t})q(k)-\boldsymbol{\Gamma} B_{t}r(K) \\
       & -(1-\tau T)\boldsymbol{\Gamma} e_{q}(k)+\epsilon T \text{sign}(\boldsymbol{\Gamma} e_{q}(k))) \\
       & -(\boldsymbol{\Gamma} B)^{-1}\chi \text{sign}(\boldsymbol{\Gamma} e_{q}(k)).\label{4eq:30}  \\
\end{array} \end{equation}

Since the whole state $q(k)$ is not available to our controller, so we need to estimate $q(k)$ based on the measured output $y(k)$. We use the steady-state Kalman estimator \cite{welch1995introduction, simon2006optimal} for this purpose as:

\begin{equation} \begin{array}{rcl}
\hat {q}(k+1) & = & A\hat {q}(k)+Bu(k)+K_{f}(y(k)-C\hat{q}(k)) \\
    \hat{y}_(k) &=& C\hat{q}(k),     \label{4eq:32}
\end{array} \end{equation}
where $\hat{q}(k)$ is the estimated of the state $q(k)$ and $K_{f}$ is the  "optimal Kalman" gain given as:

\begin{equation}
K_{f}=PC^{T}R^{-1},
\label{4eq:33}
\end{equation}
and $P$ is the solution of the following algebraic Riccati equation which is given by:

 \begin{equation} \begin{array}{rcl}
AP+PA^{T}-PC^{T}R^{-1}CP+Q=0, \\
       \label{4eq:34}
\end{array} \end{equation}

The effectiveness of the presented algorithm in combination with LQE is  justified by numerical simulations using MATLAB-Simulink (The Math-Works Inc., Natick, MA, USA).

\section{Simulation Results} \label{4sec:res}

\subsection{Results in Rest Condition}

\begin{figure*}[htbp]
\centering
\subfigure[LV volume versus LV pressure before and after Parameter Change.]{
   \includegraphics[scale =0.16752] {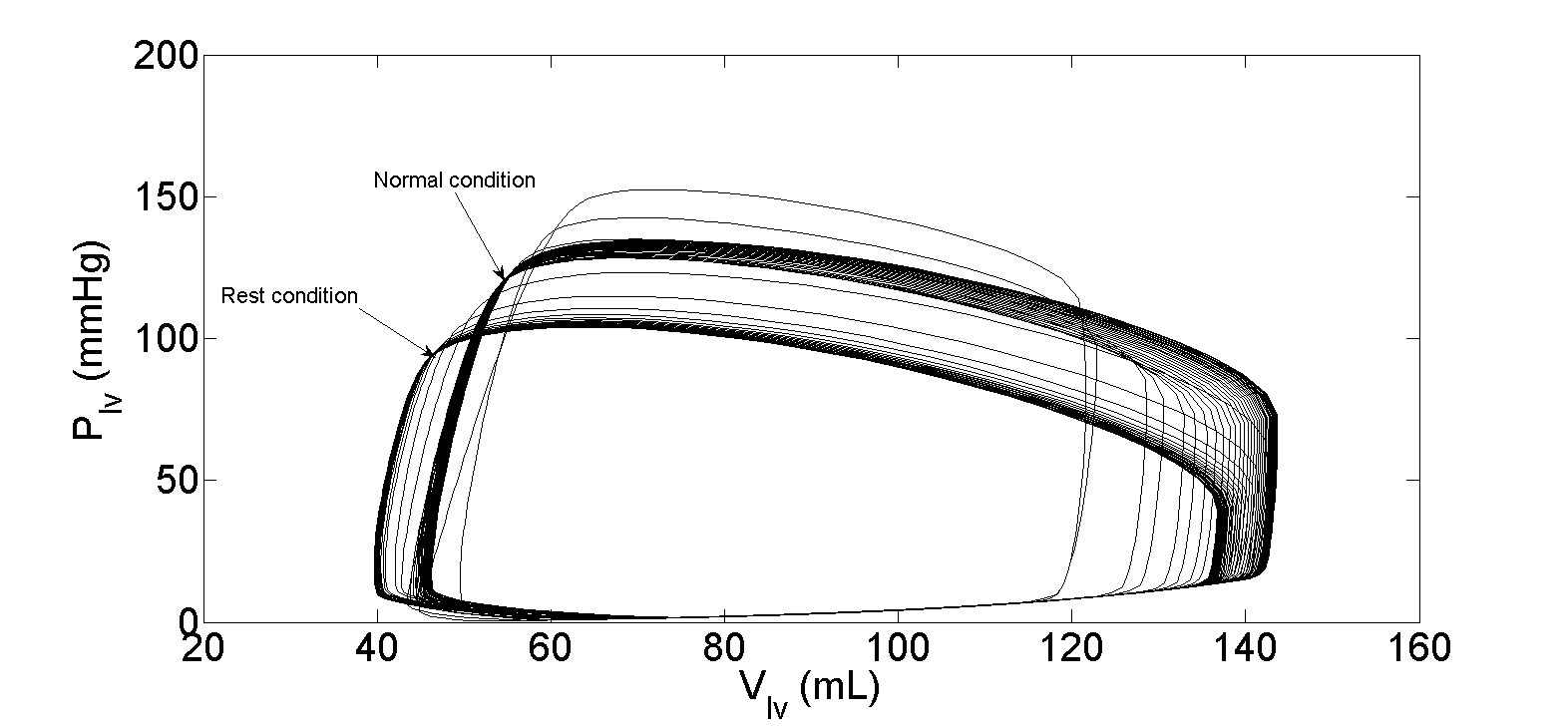}
   \label{43a}
 }
\subfigure[RV volume versus RV pressure before and after Parameter Change.]{
   \includegraphics[scale =0.16752] {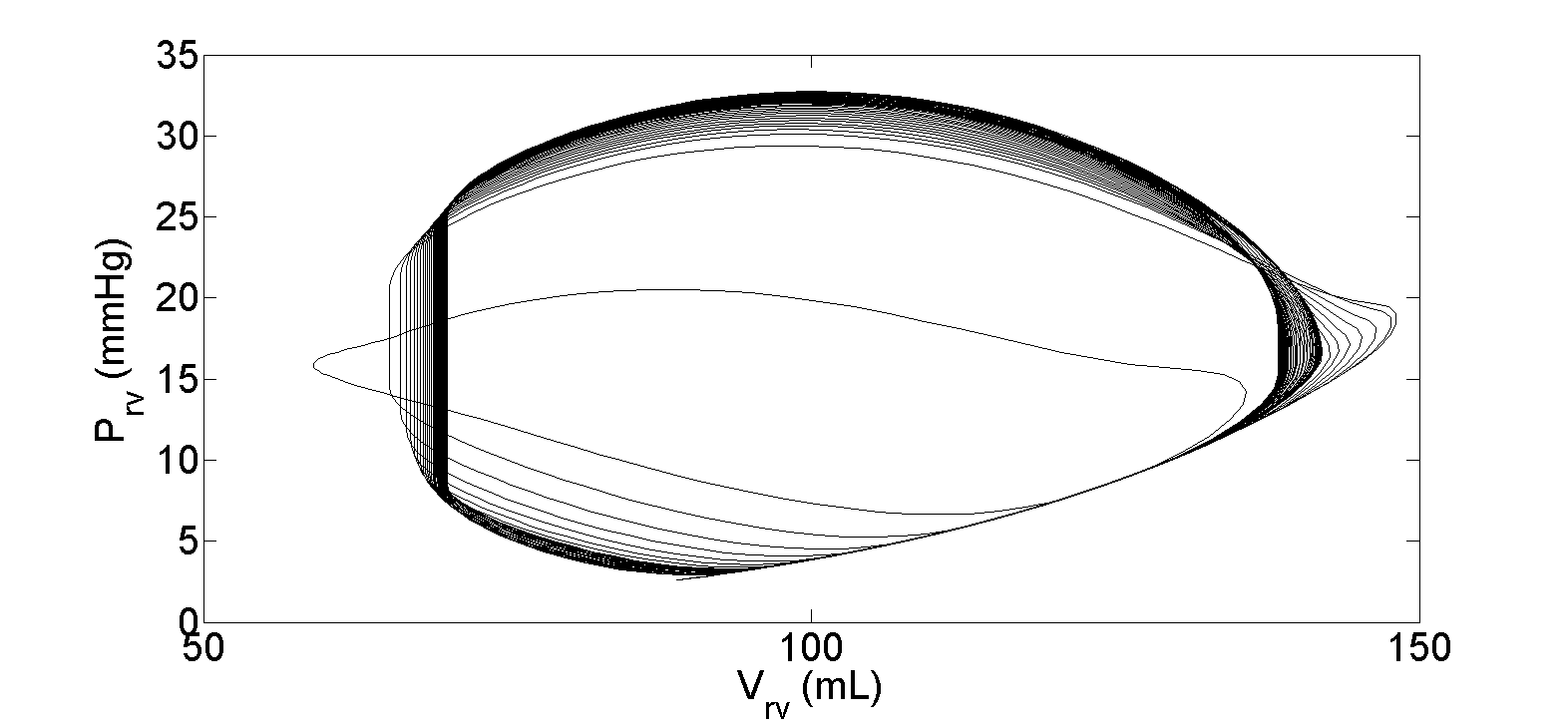}
   \label{43b}
 }

 \subfigure[Aortic pressure.]{
   \includegraphics[scale =0.16752] {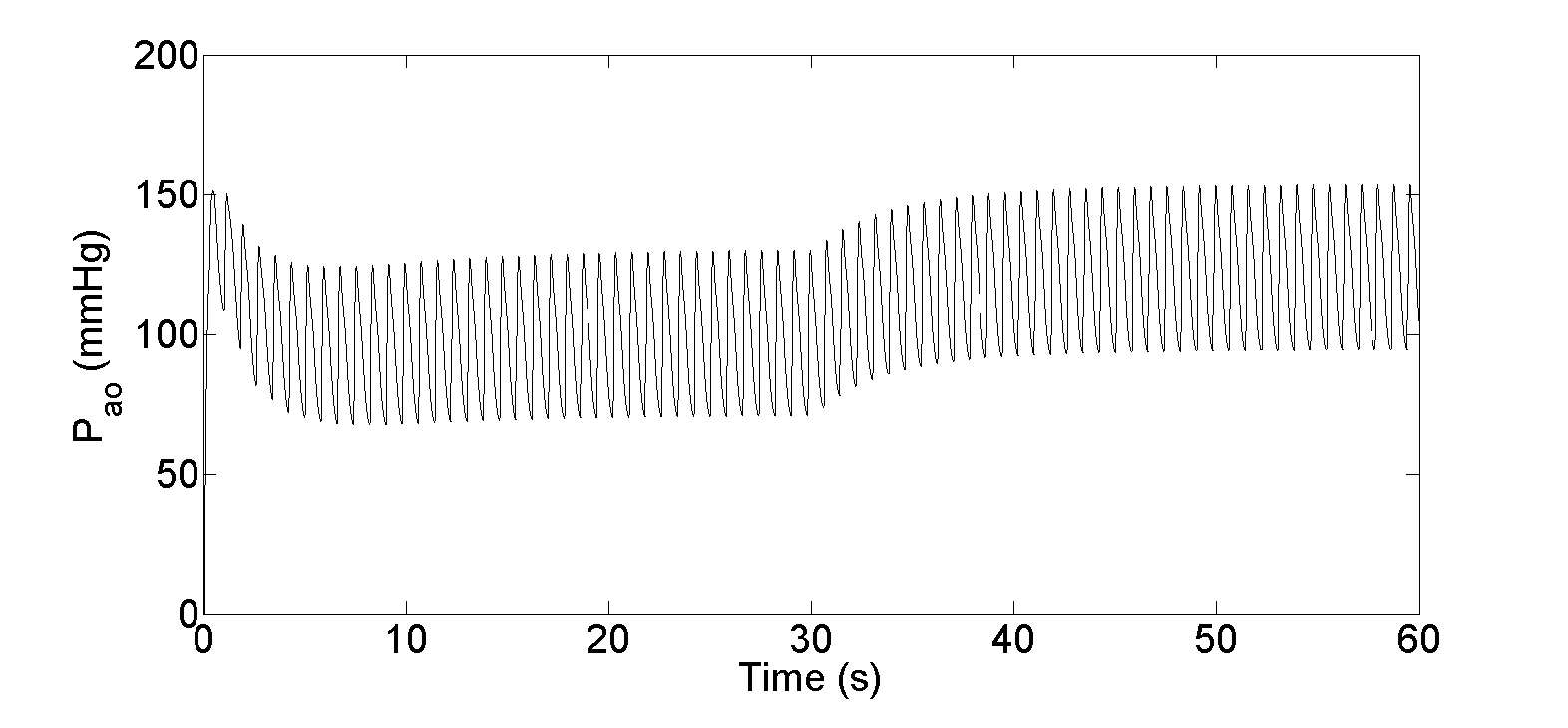}
   \label{43c}
 }
  \subfigure[Left atrial pressure.]{
   \includegraphics[scale =0.16752] {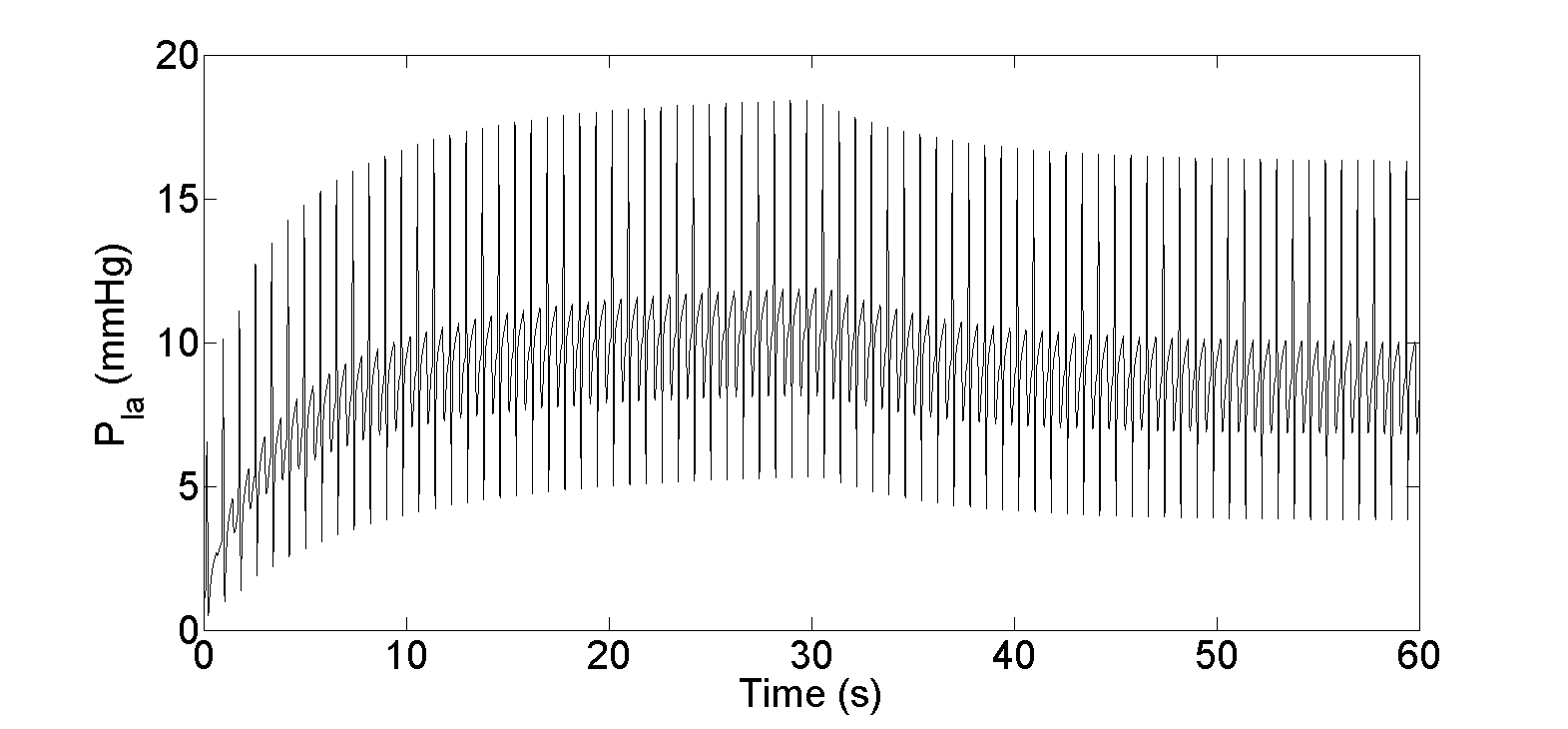}
   \label{43d}
 }

\subfigure[Right atrial pressure.]{
   \includegraphics[scale =0.16752] {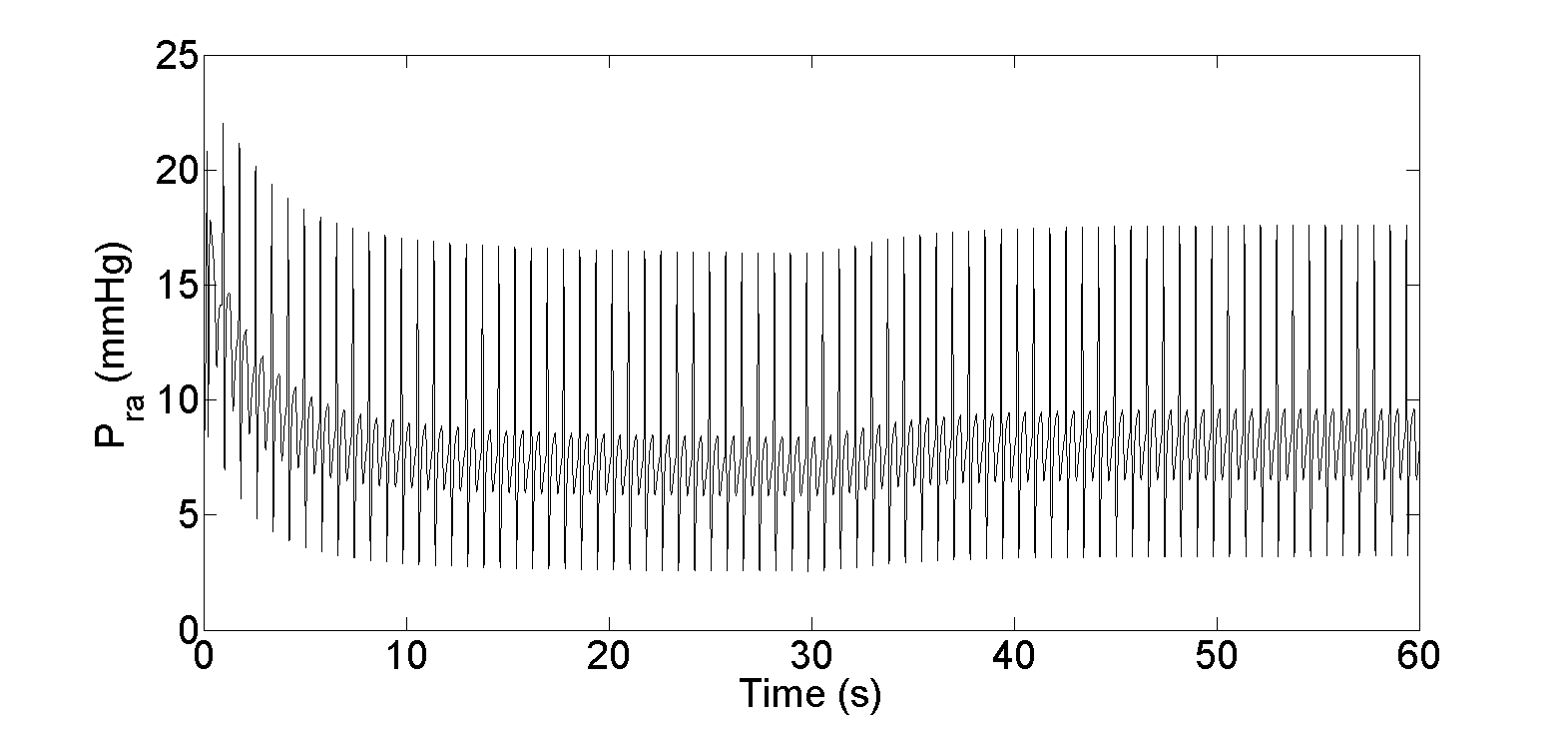}
   \label{43e}
 }
\caption{Hemodynamic variables results in rest condition when the system induced at 30s.}
\label{4:30a}
\end{figure*}

\begin{figure*}[htbp]
\centering
\subfigure[LV volume versus LV pressure before and after Parameter Change.]{
   \includegraphics[scale =0.16752] {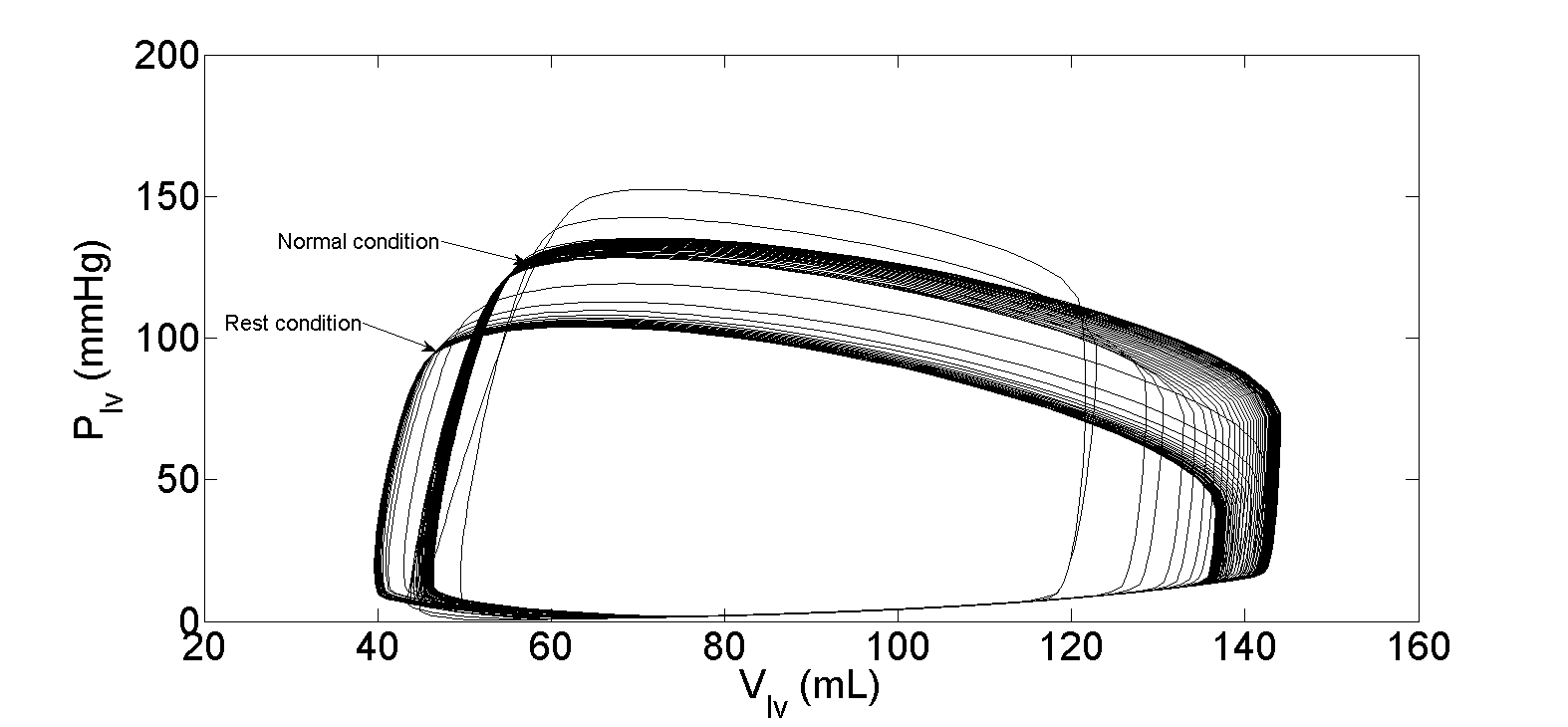}
   \label{46a}
 }
\subfigure[RV volume versus RV pressure before and after Parameter Change.]{
   \includegraphics[scale =0.16752] {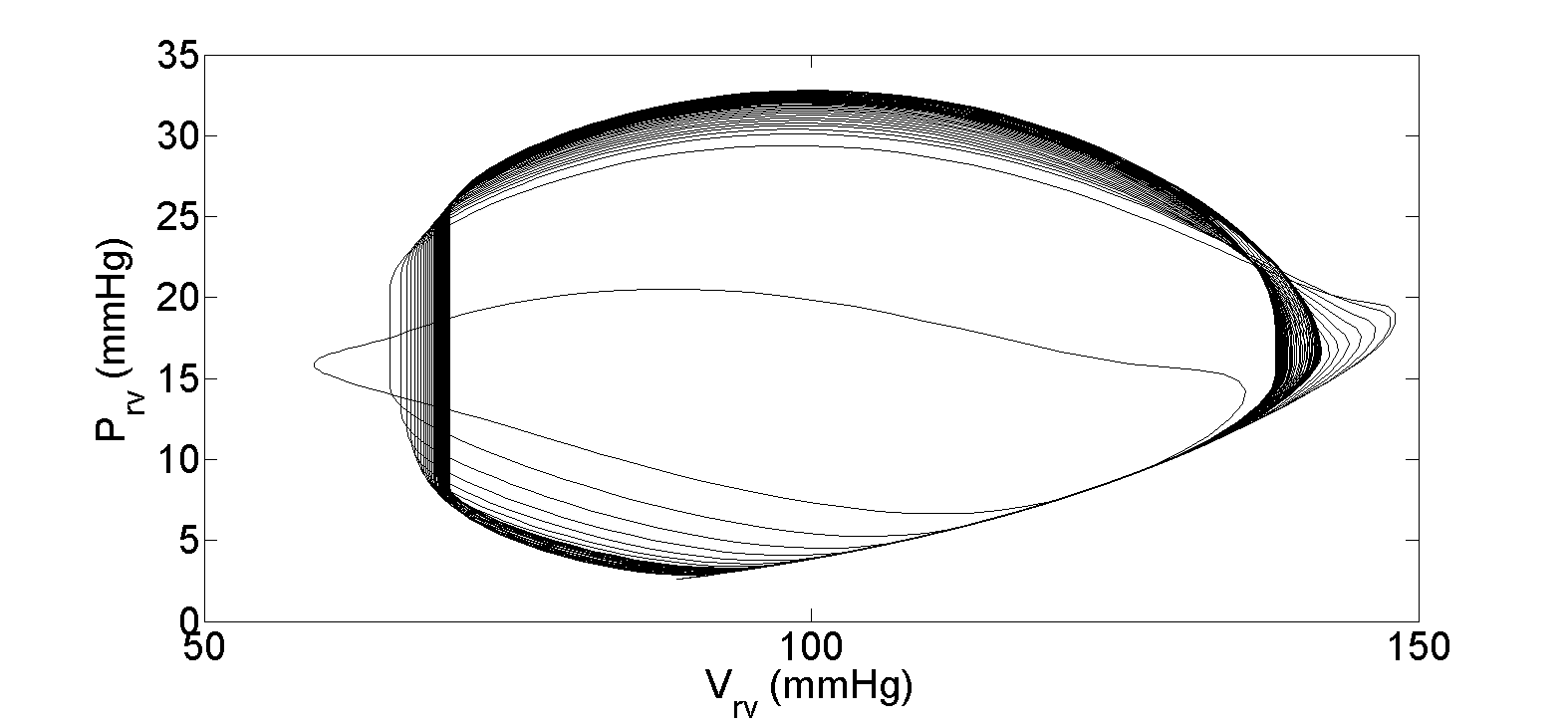}
   \label{46b}
 }

 \subfigure[Aortic pressure.]{
   \includegraphics[scale =0.16752] {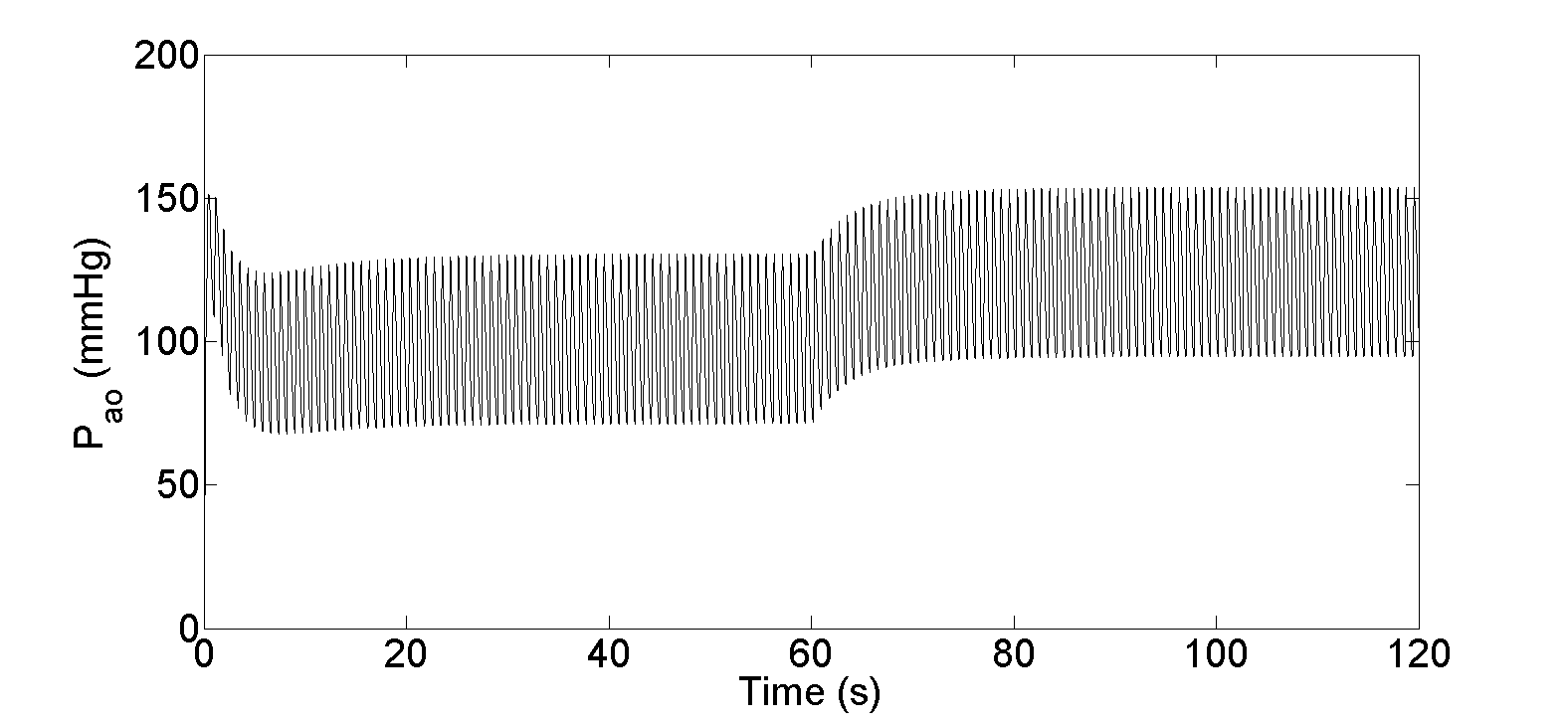}
   \label{46c}
 }
  \subfigure[Left atrial pressure.]{
   \includegraphics[scale =0.16752] {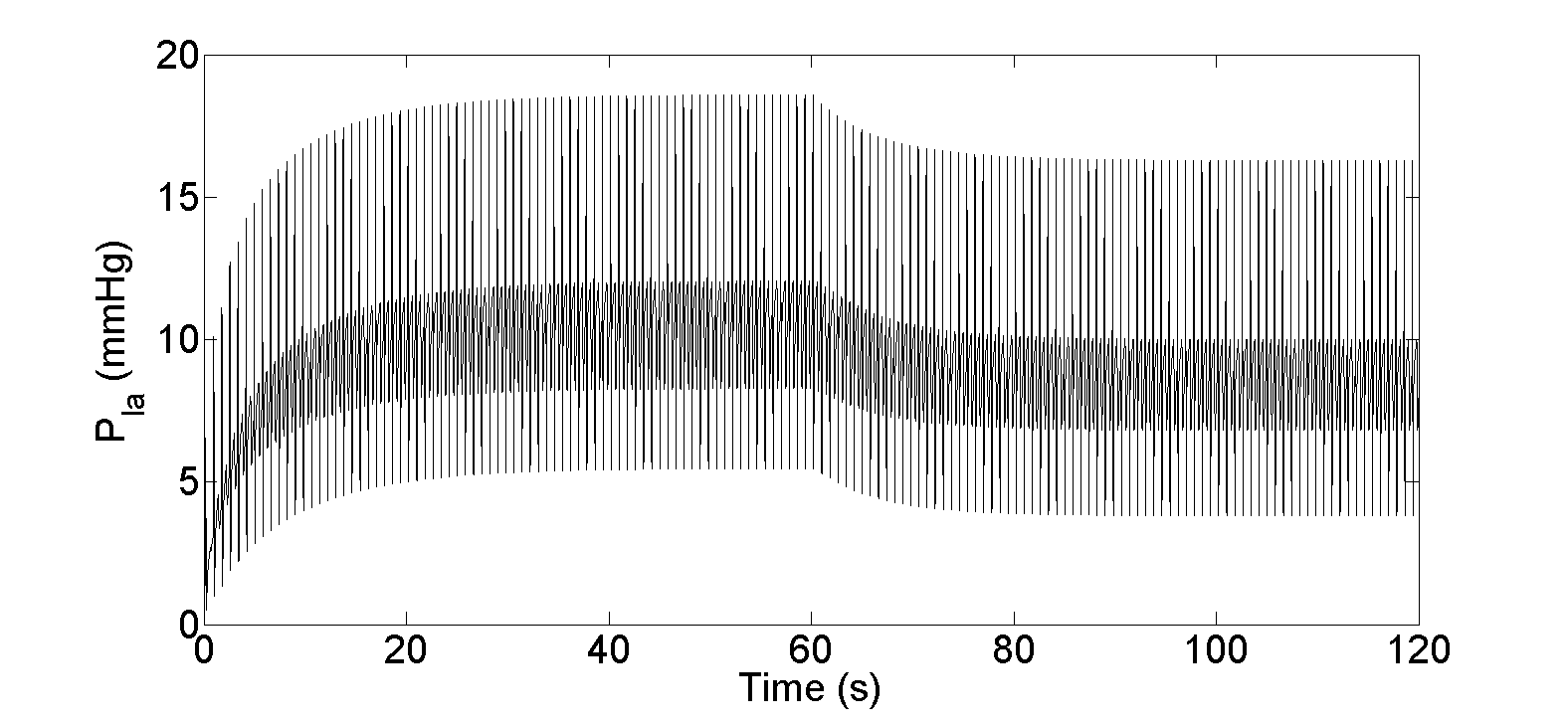}
   \label{46d}
 }

\subfigure[Right atrial pressure.]{
   \includegraphics[scale =0.16752] {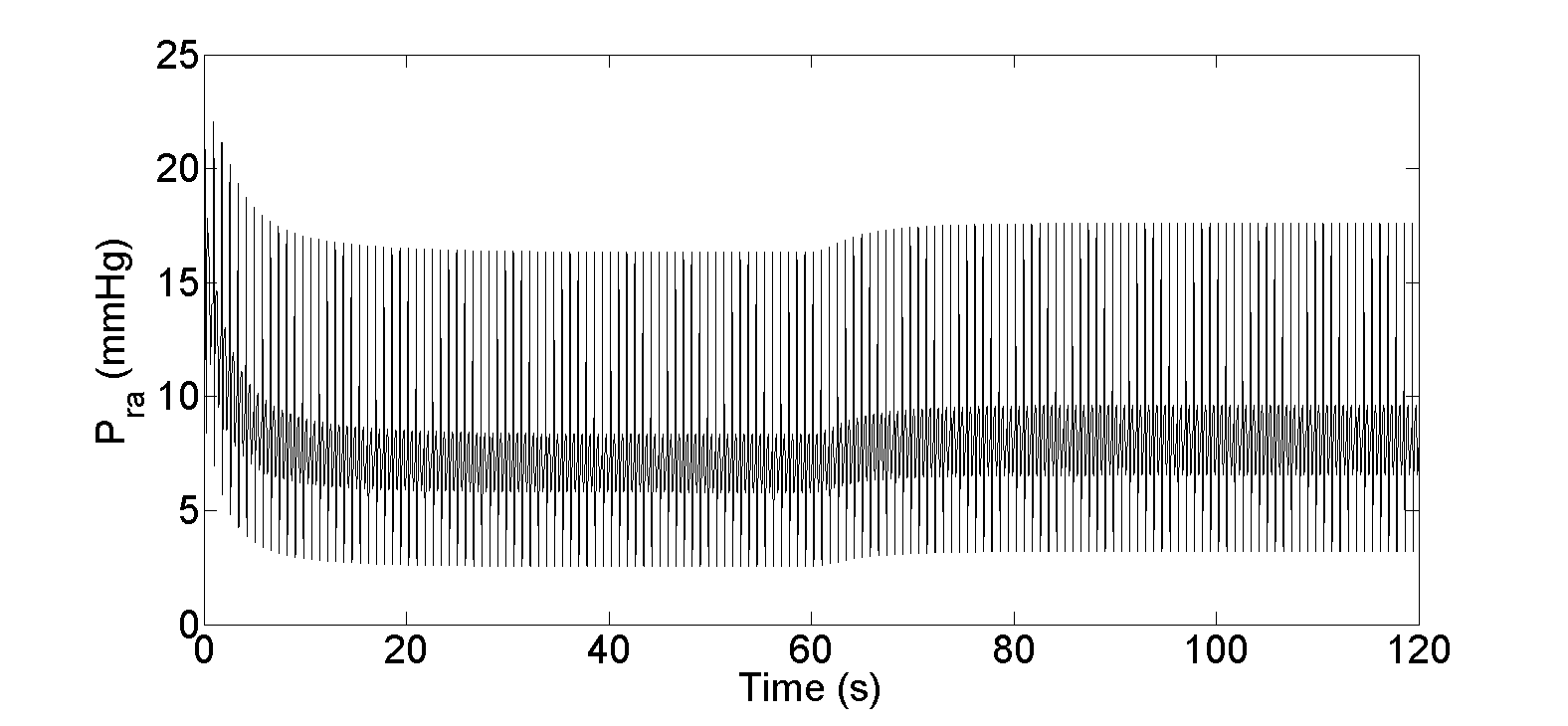}
   \label{46e}
 }
\caption{Hemodynamic variables results in rest condition when the system induced at 60s.}
\label{4:60a}
\end{figure*}

\begin{figure*}[htbp]
\centering
\subfigure[LV volume versus LV pressure before and after Parameter Change.]{
   \includegraphics[scale =0.16752] {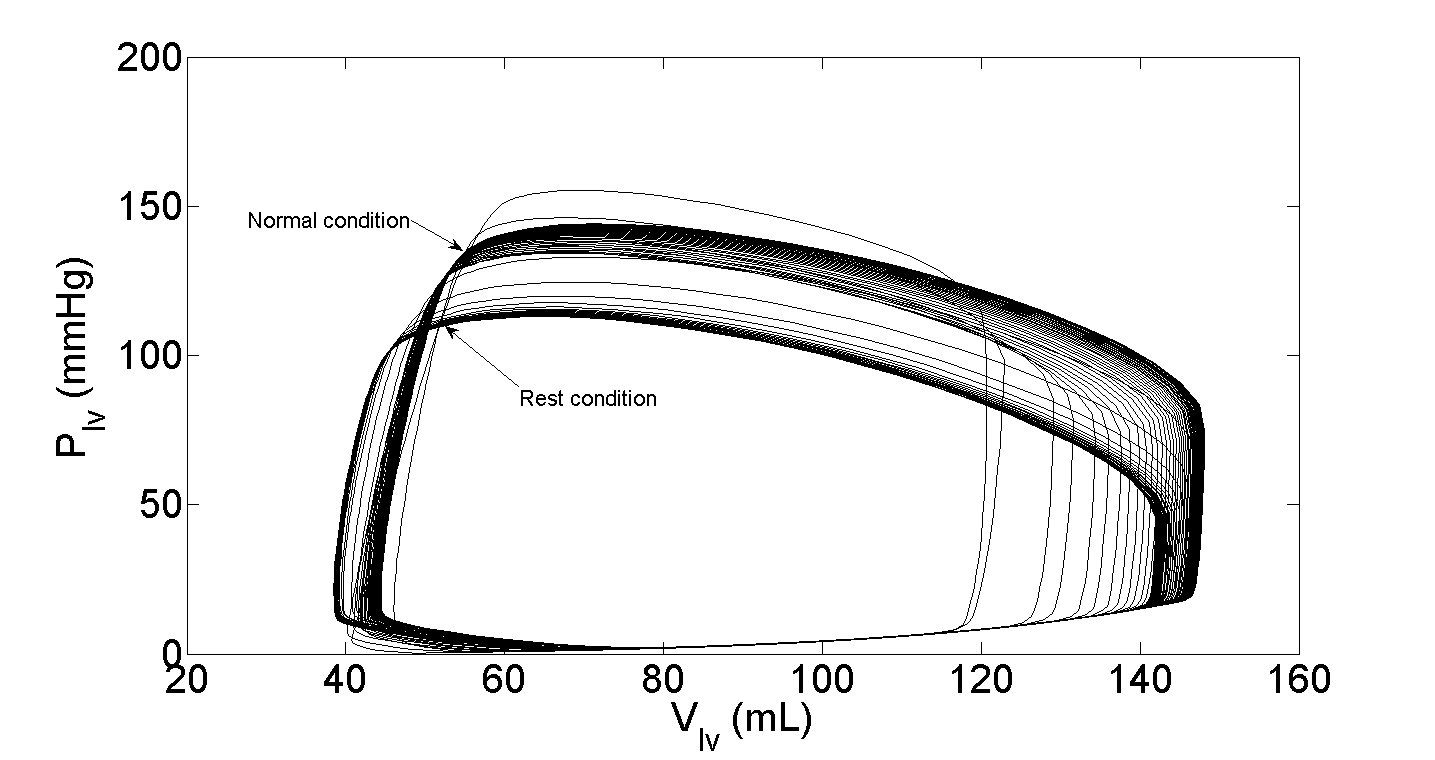}
   \label{49a}
 }
\subfigure[RV volume versus RV pressure before and after Parameter Change.]{
   \includegraphics[scale =0.16752] {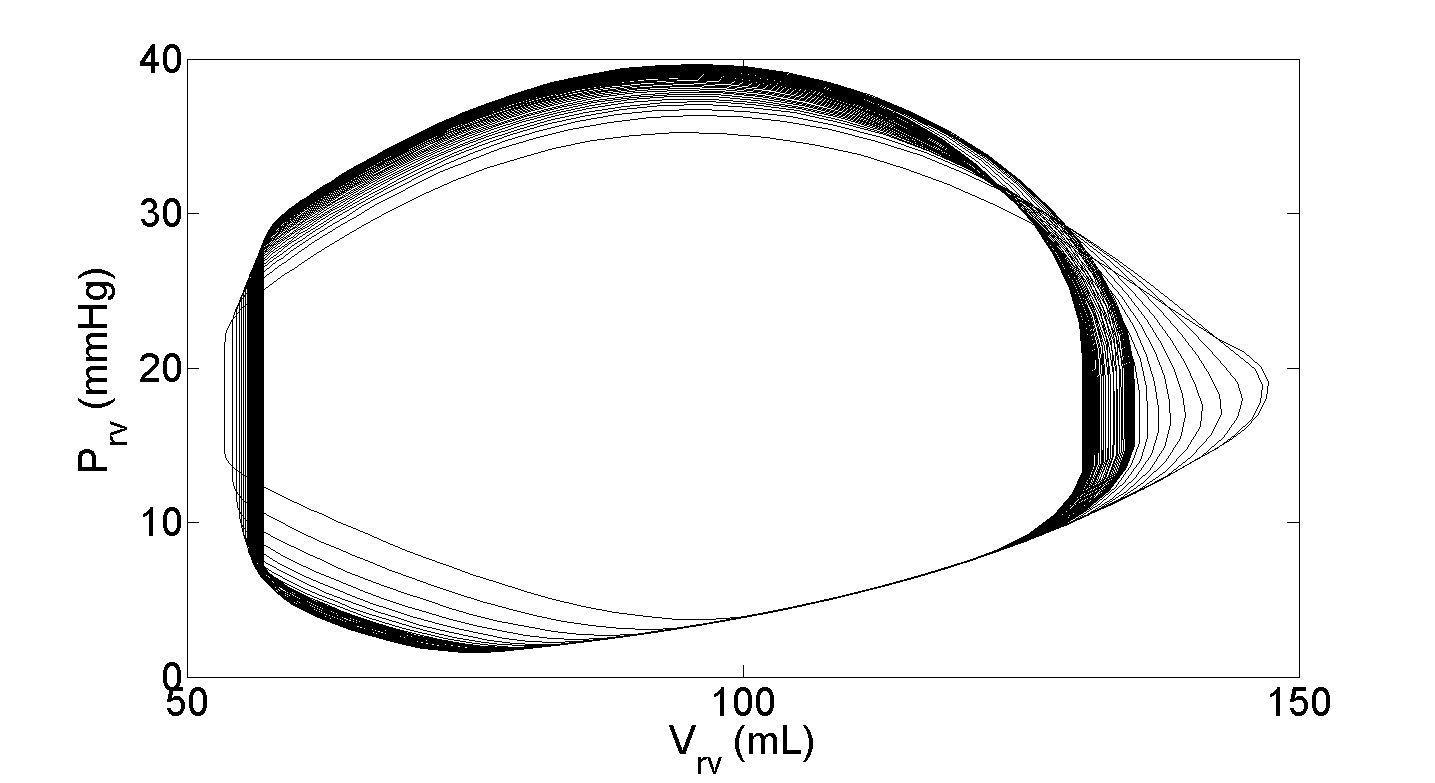}
   \label{49b}
 }

 \subfigure[Aortic pressure.]{
   \includegraphics[scale =0.16752] {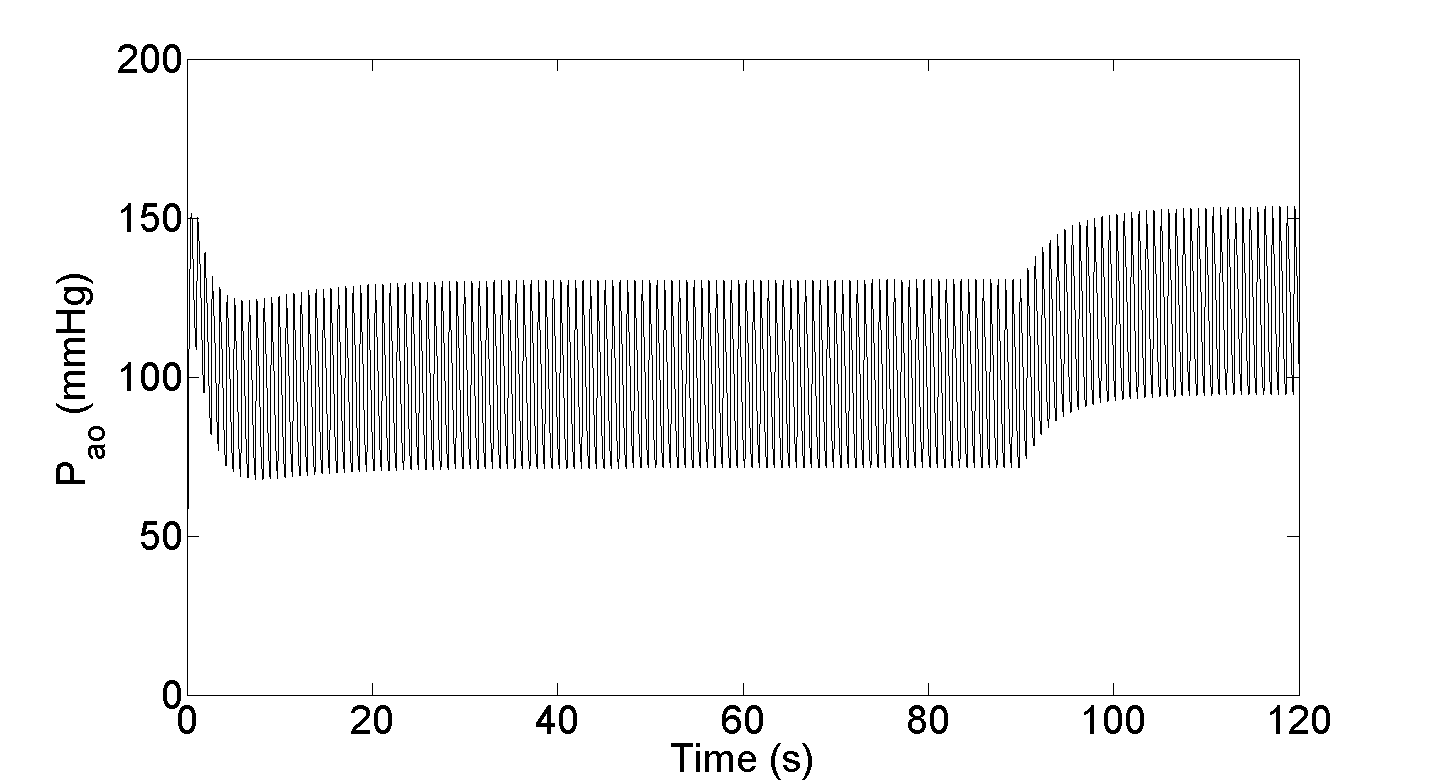}
   \label{49c}
 }
  \subfigure[Left atrial pressure.]{
   \includegraphics[scale =0.16752] {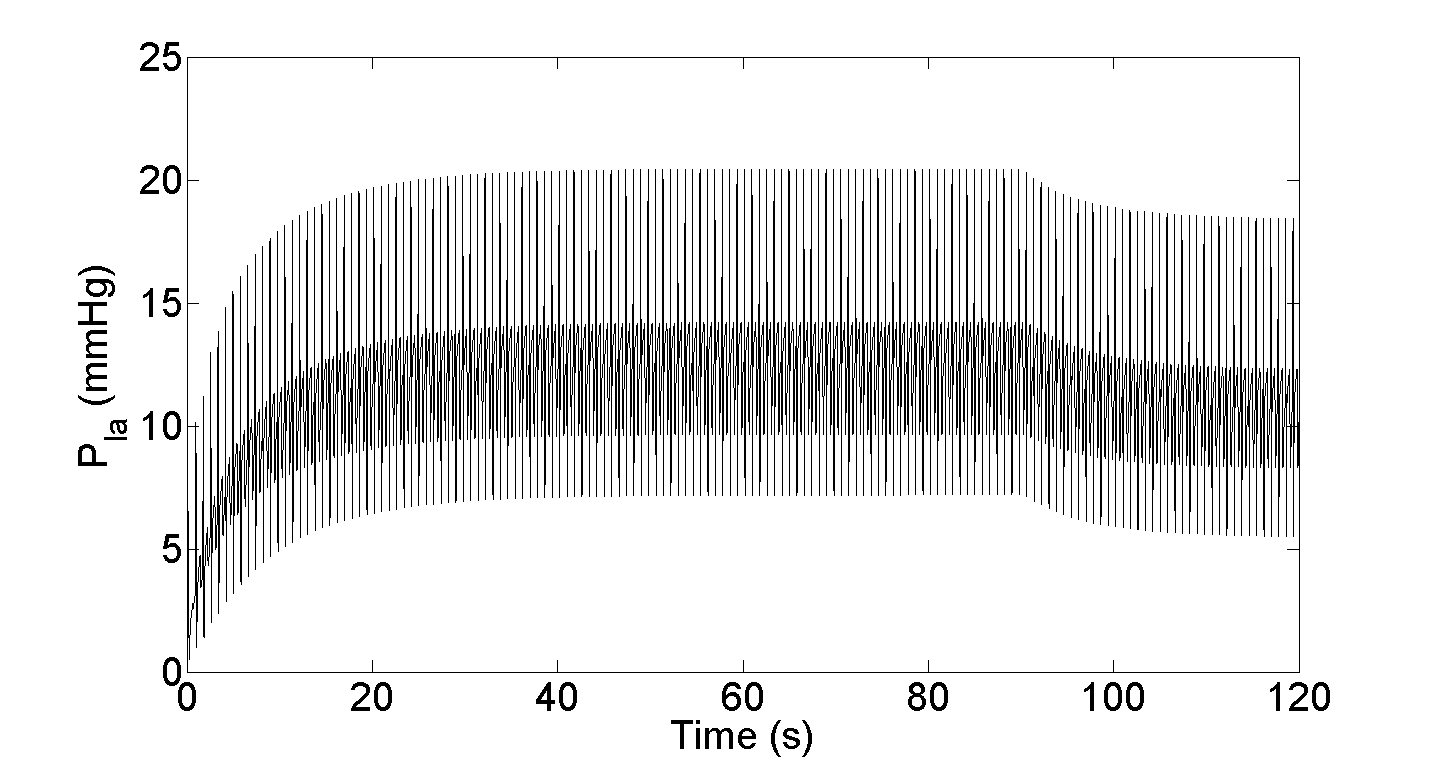}
   \label{49d}
 }

\subfigure[Right atrial pressure.]{
   \includegraphics[scale =0.16752] {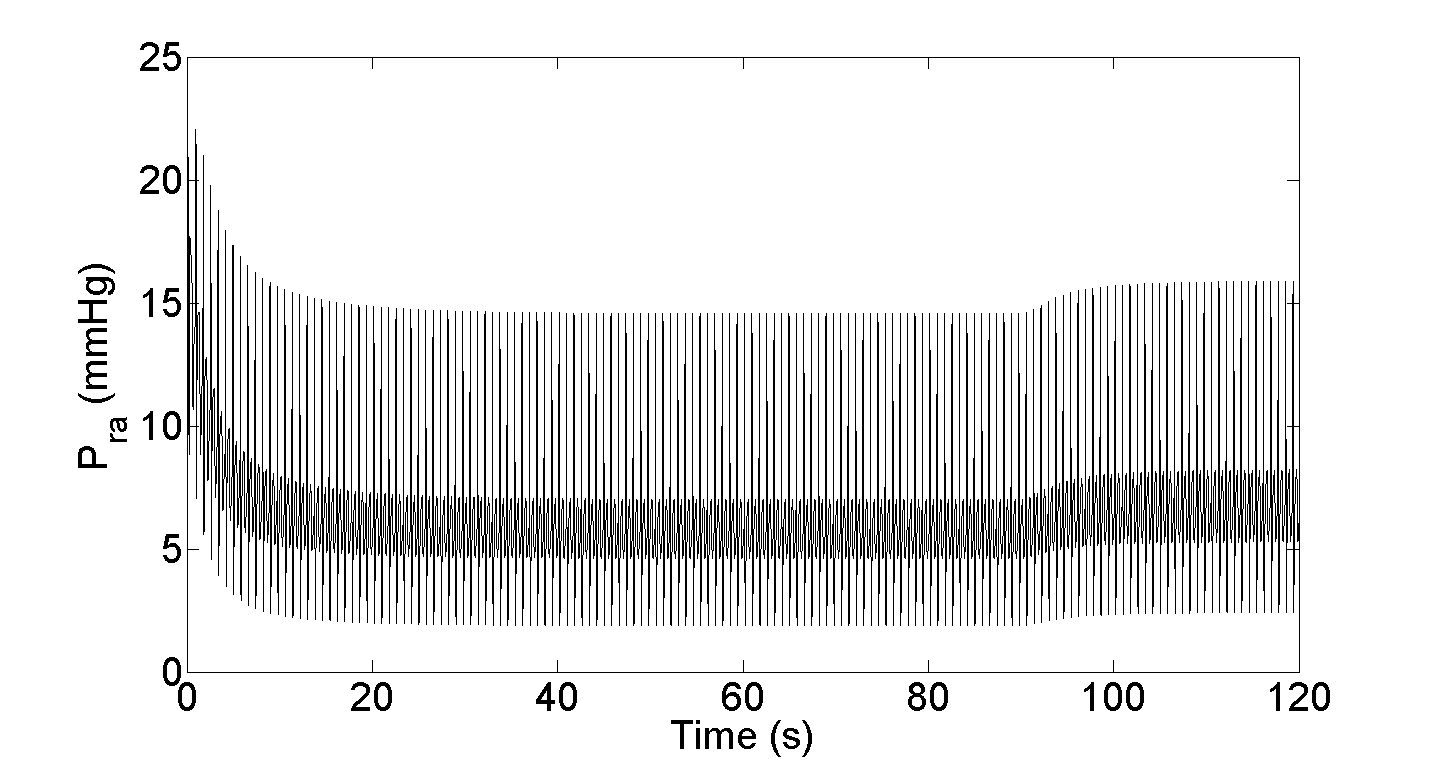}
   \label{49e}
 }
\caption{Hemodynamic variables results at rest condition when the system induced at 90s.}
\label{4:90a}
\end{figure*}

\begin{figure*}[htbp]
\centering
\subfigure[LV volume versus LV pressure before and after Parameter Change.]{
   \includegraphics[scale =0.16752] {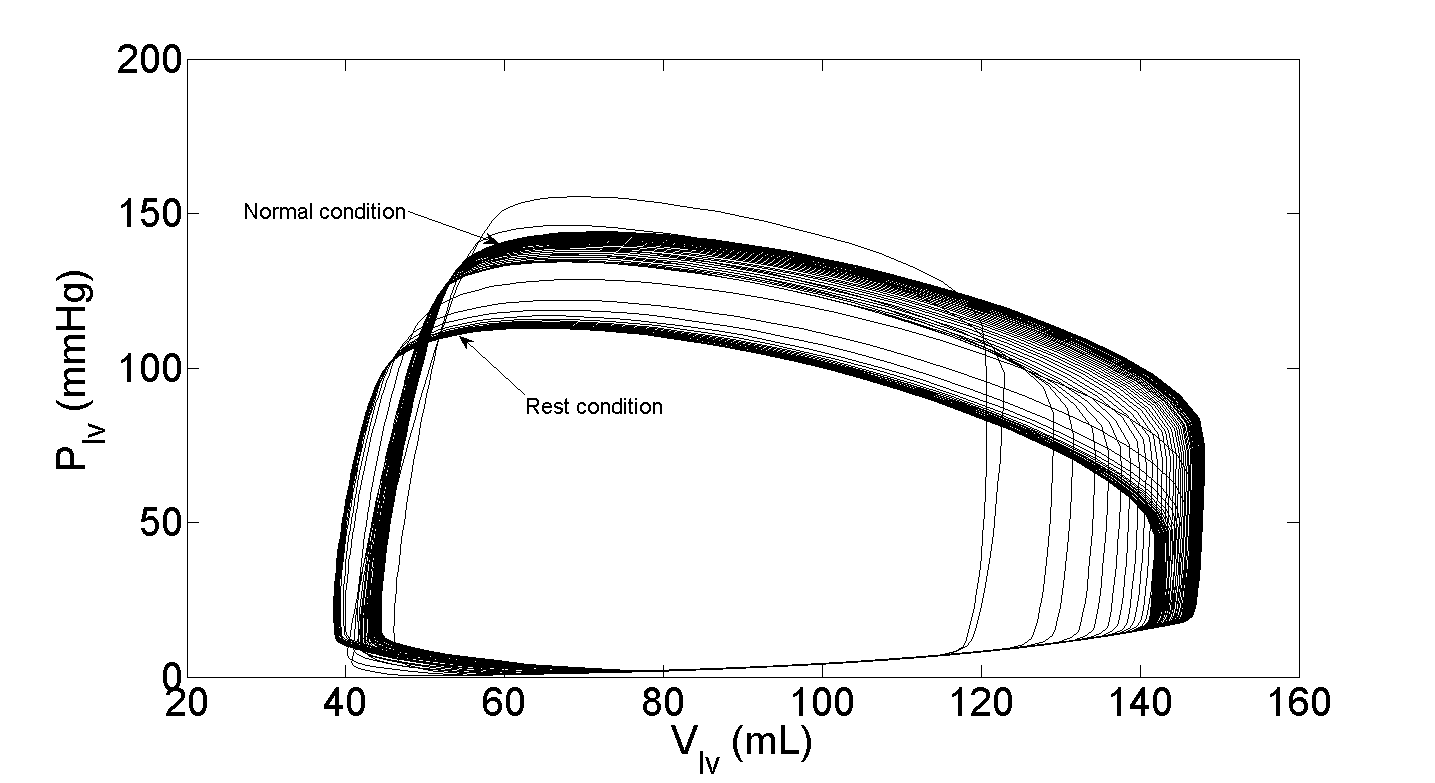}
   \label{42a}
 }
\subfigure[RV volume versus RV pressure before and after Parameter Change.]{
   \includegraphics[scale =0.16752] {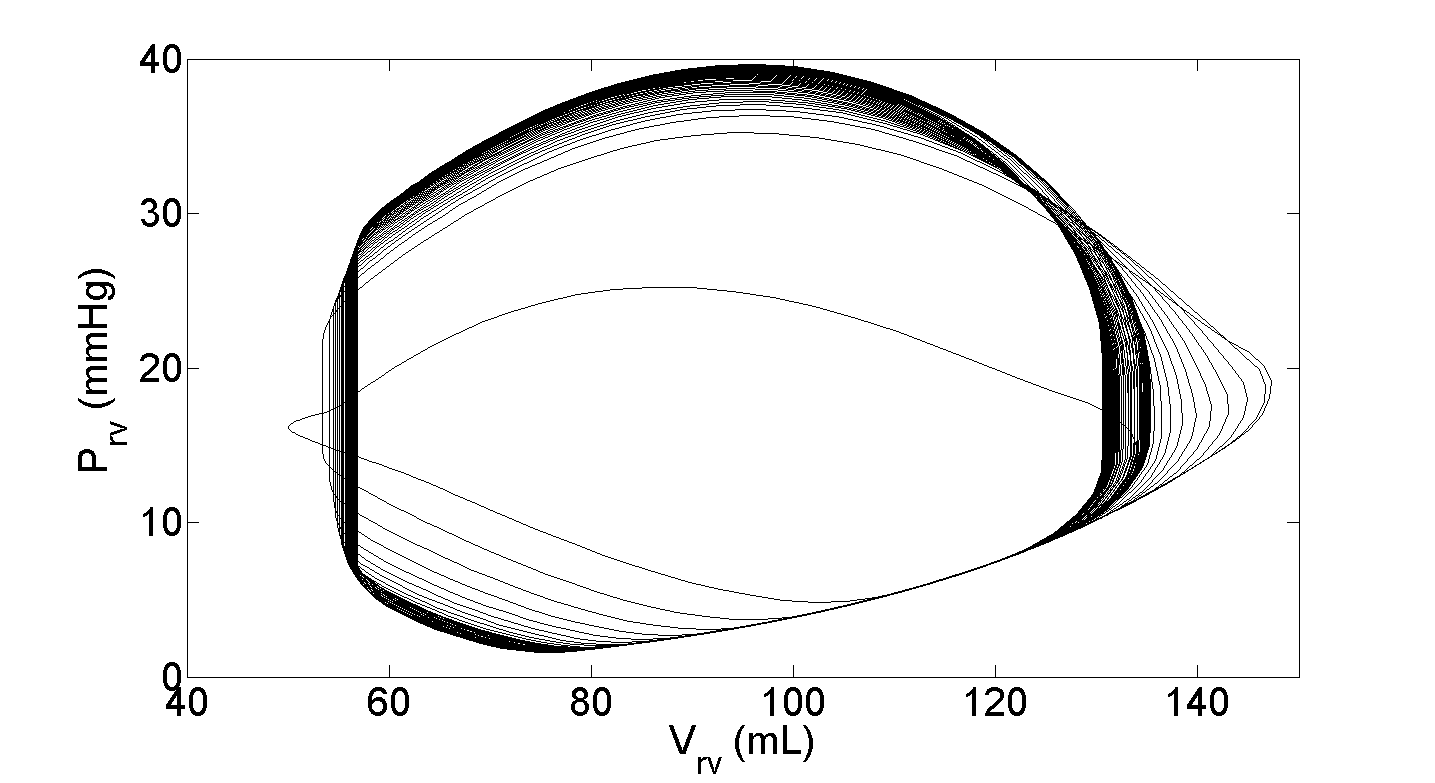}
   \label{42b}
 }

 \subfigure[Aortic pressure.]{
   \includegraphics[scale =0.16752] {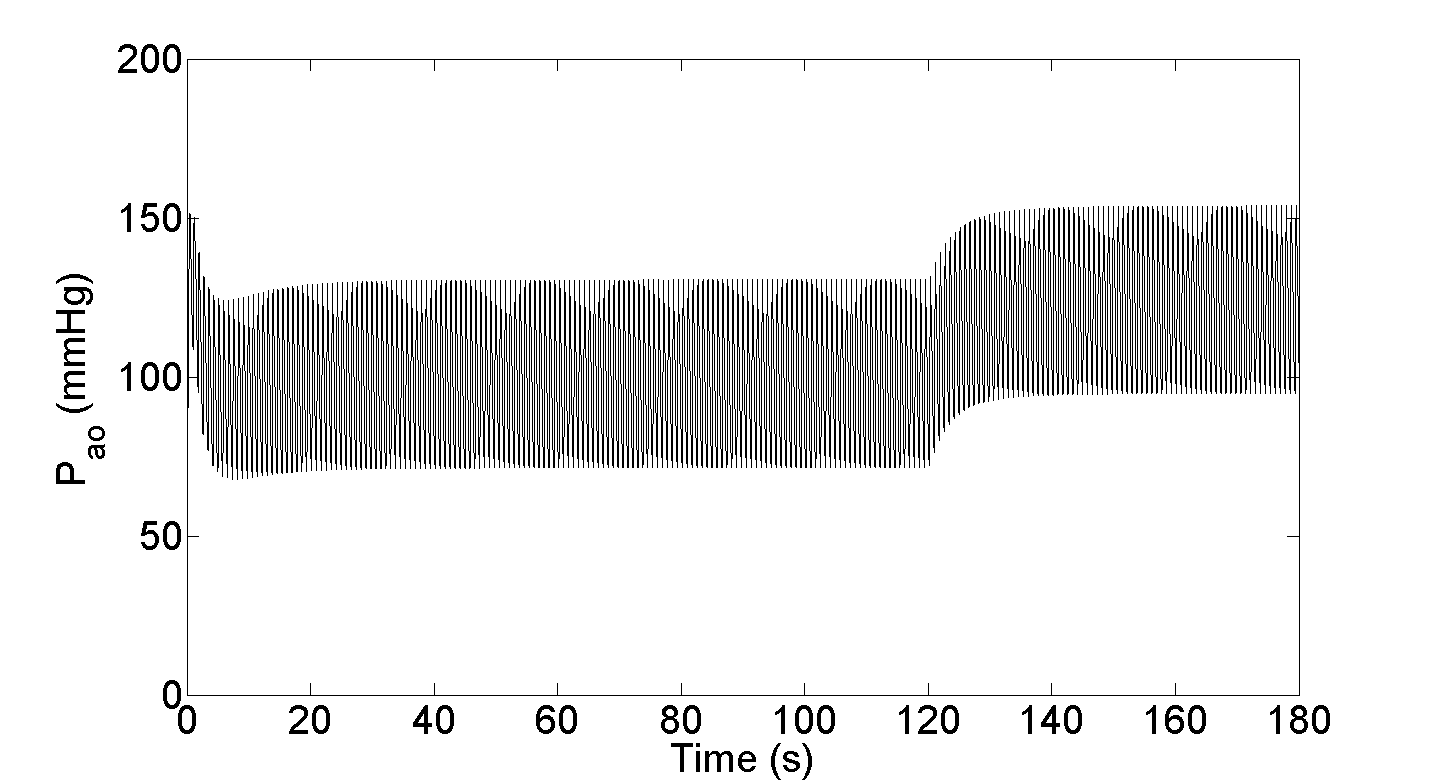}
   \label{42c}
 }
  \subfigure[Left atrial pressure.]{
   \includegraphics[scale =0.16752] {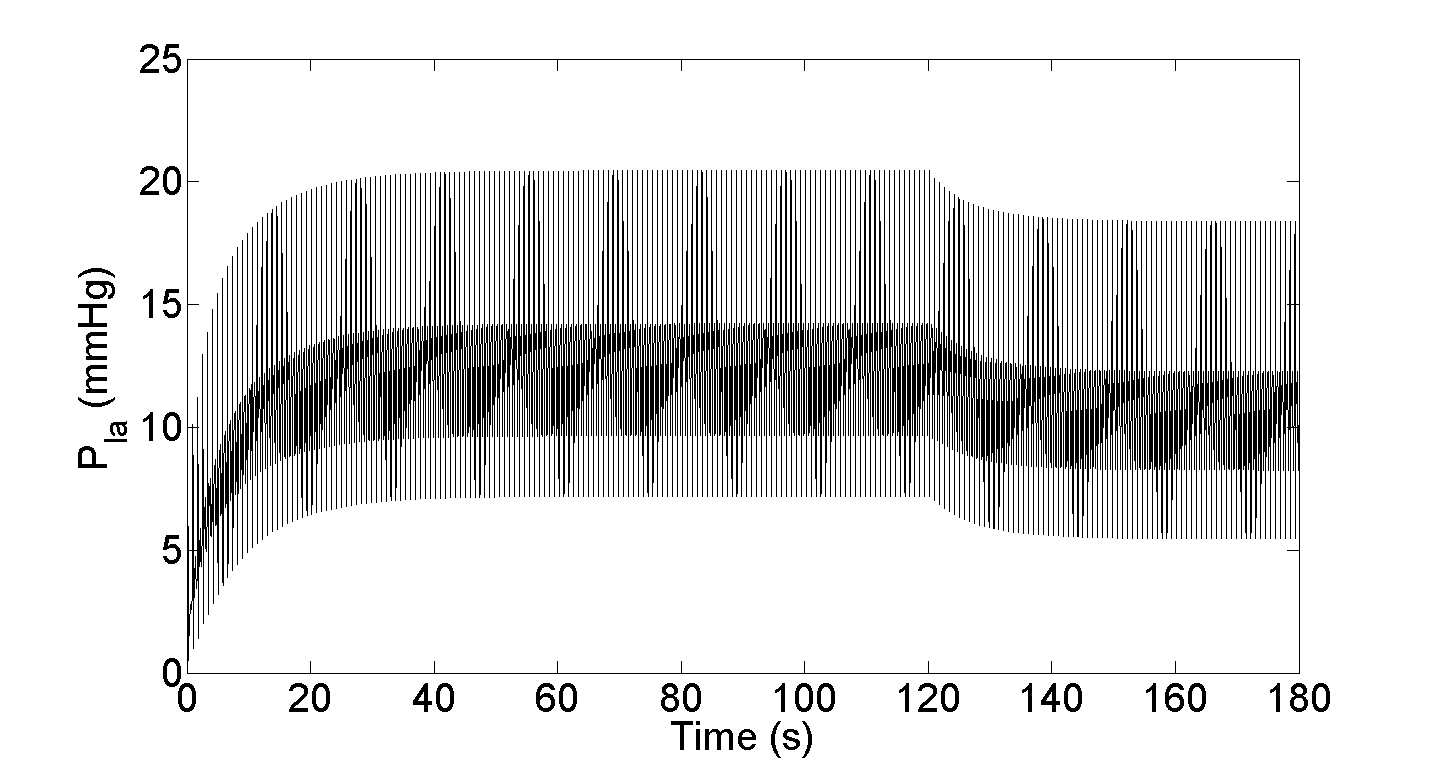}
   \label{42d}
 }

\subfigure[Right atrial pressure.]{
   \includegraphics[scale =0.16752] {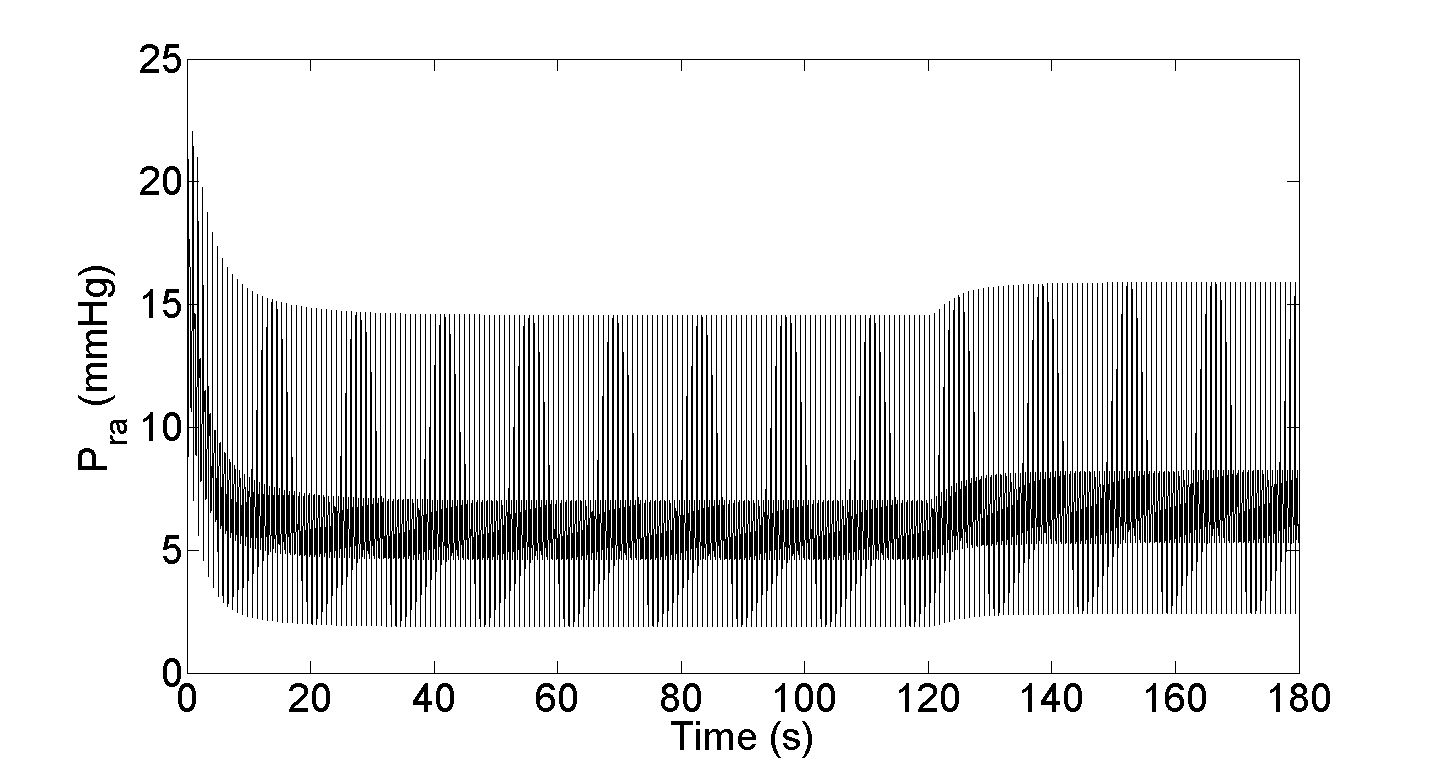}
   \label{42e}
 }
\caption{Hemodynamic variables results at rest condition when the system induced at 120s.}
\label{4:20a}
\end{figure*}

This scenario shows an immediate response of the controller corresponding to the blood loss. The reduction in blood volume has caused a reduction in stroke volume (SV) of the right ventricular (RV). This is in fact associated with a shift to the left of the left ventricular (LV) pressure-volume loop, causing a reduction in LV end-diastolic pressure ($P_{lved}$), LV end-diastolic volume ($V_{lved}$) and LV end-systolic volumes ($V_{lves}$) while a slight shift can be seen in the right of the RV pressure-volume loops. As a result, the LVAD successfully increases the aortic pressure $P_{ao}$ and decreases  the left atrial pressure $P_{la}$ and keeps the right atrial pressure $P_{ra}$ within the safe operating range. These results can be observed from hemodynamic waveforms (see, e.g., Figures (\ref{4:30a}, \ref{4:60a}, \ref{4:90a} and \ref{4:20a})).

\begin{figure*}[htbp]
\centering
\subfigure[Average pump speed.]{
   \includegraphics[scale =0.16752] {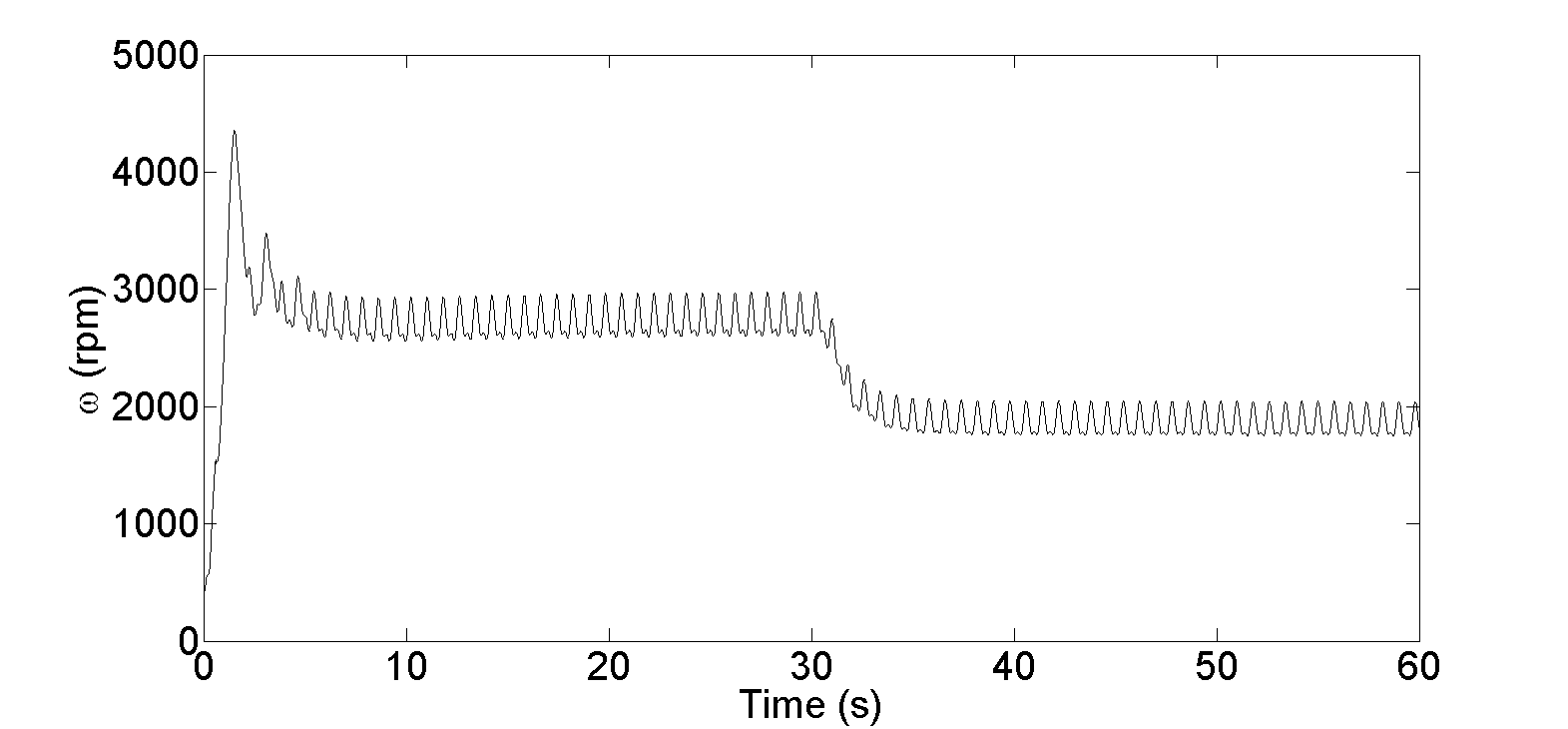}
   \label{43f}
 }

  \subfigure[Pump flow compared with desired reference flow.]{
   \includegraphics[scale =0.25] {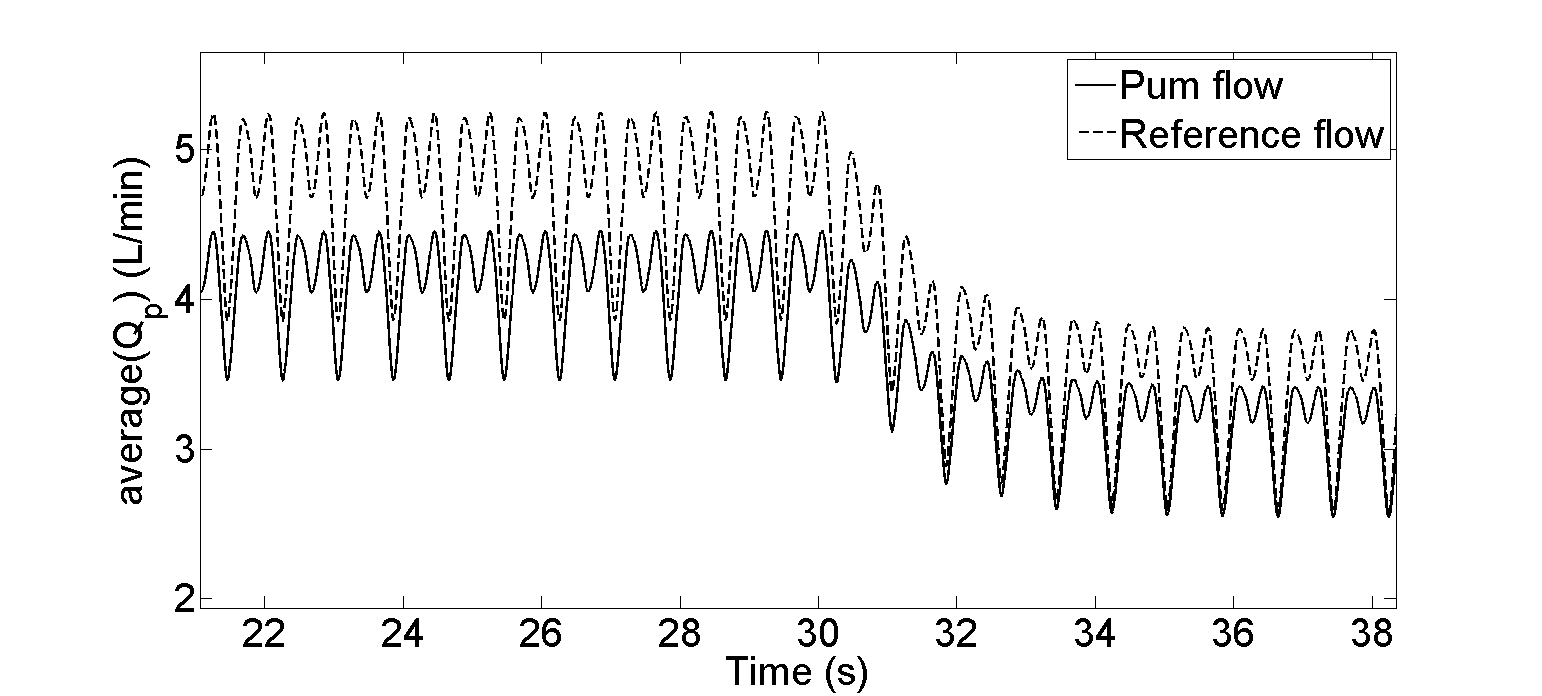}
   \label{43h}
 }

\subfigure[Measured steady state pump flow against estimated pump flow.]{
   \includegraphics[scale =0.16752] {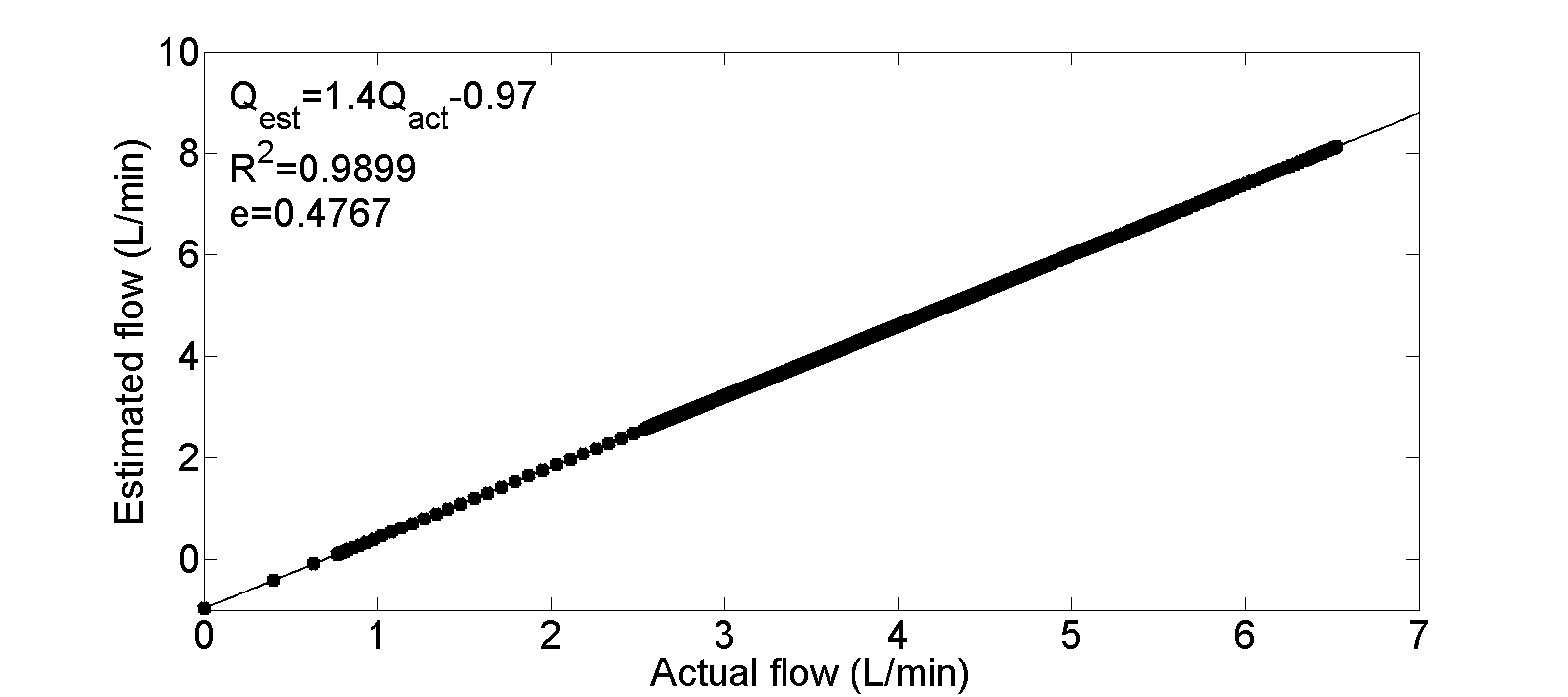}
   \label{43i}
 }

\caption{Pump variable results in rest condition when the system induced at 30s.}
\label{4:30b}
\end{figure*}

\begin{figure}[htbp]
\centering
\subfigure[Average pump speed.]{
   \includegraphics[scale =0.16752] {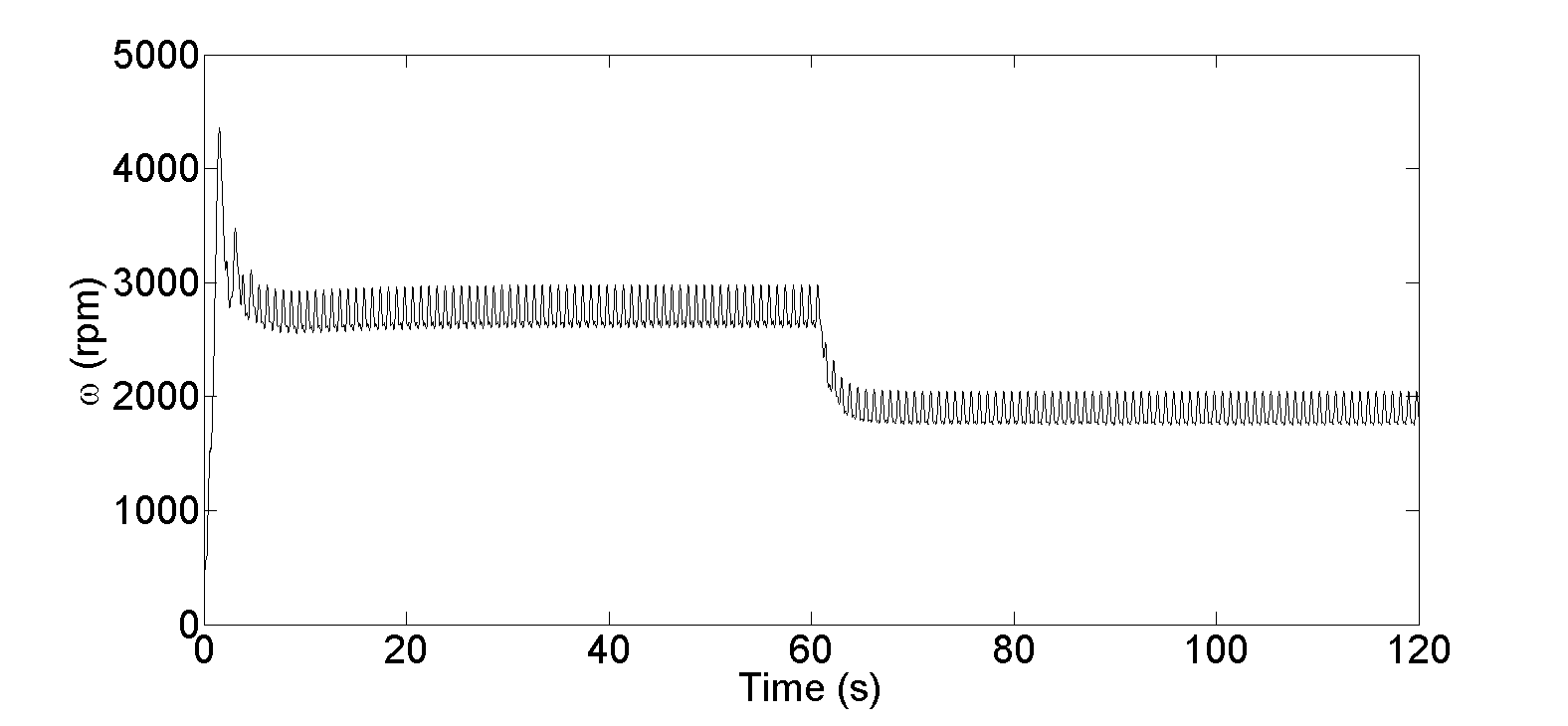}
   \label{46f}
 }

  \subfigure[Pump flow compared with desired reference flow.]{
   \includegraphics[scale =0.16752] {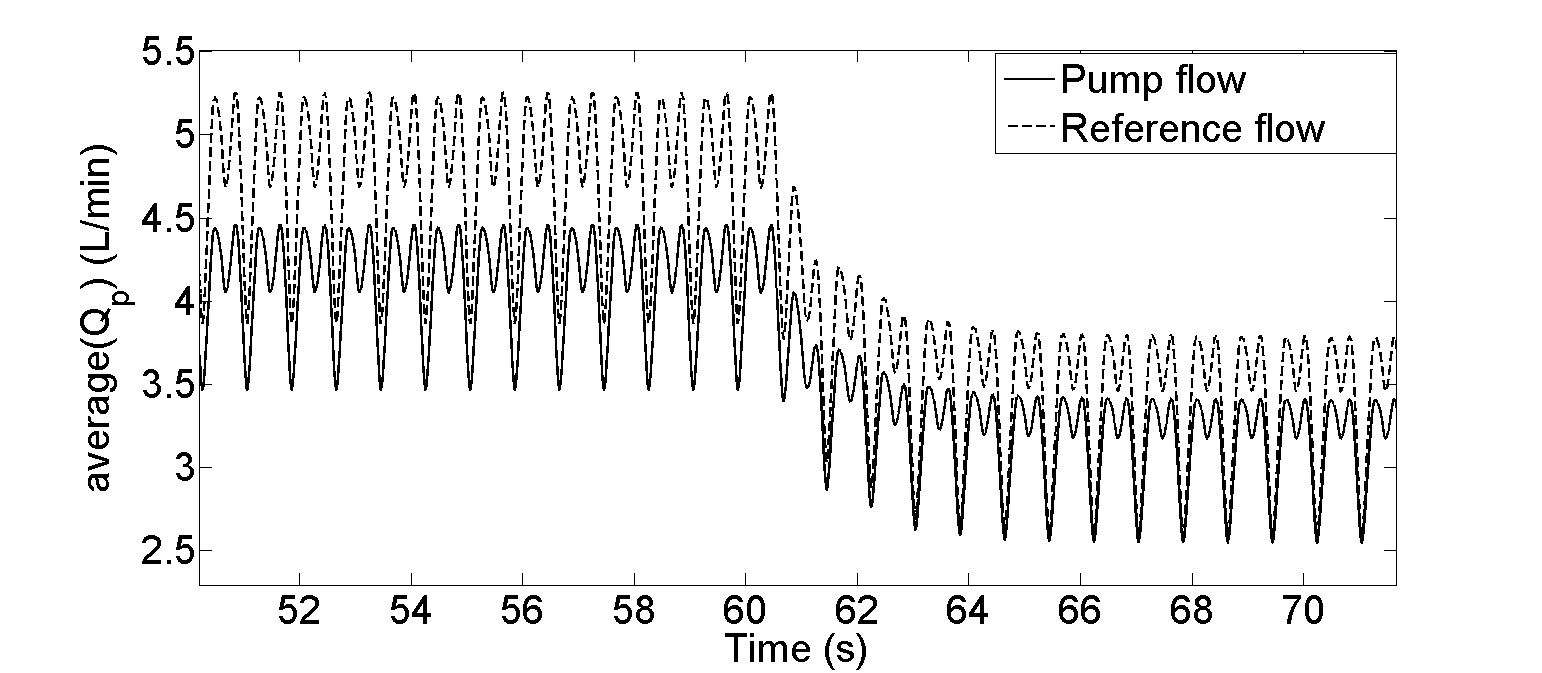}
   \label{46h}
 }

\subfigure[Measured steady state pump flow against estimated pump flow.]{
   \includegraphics[scale =0.16752] {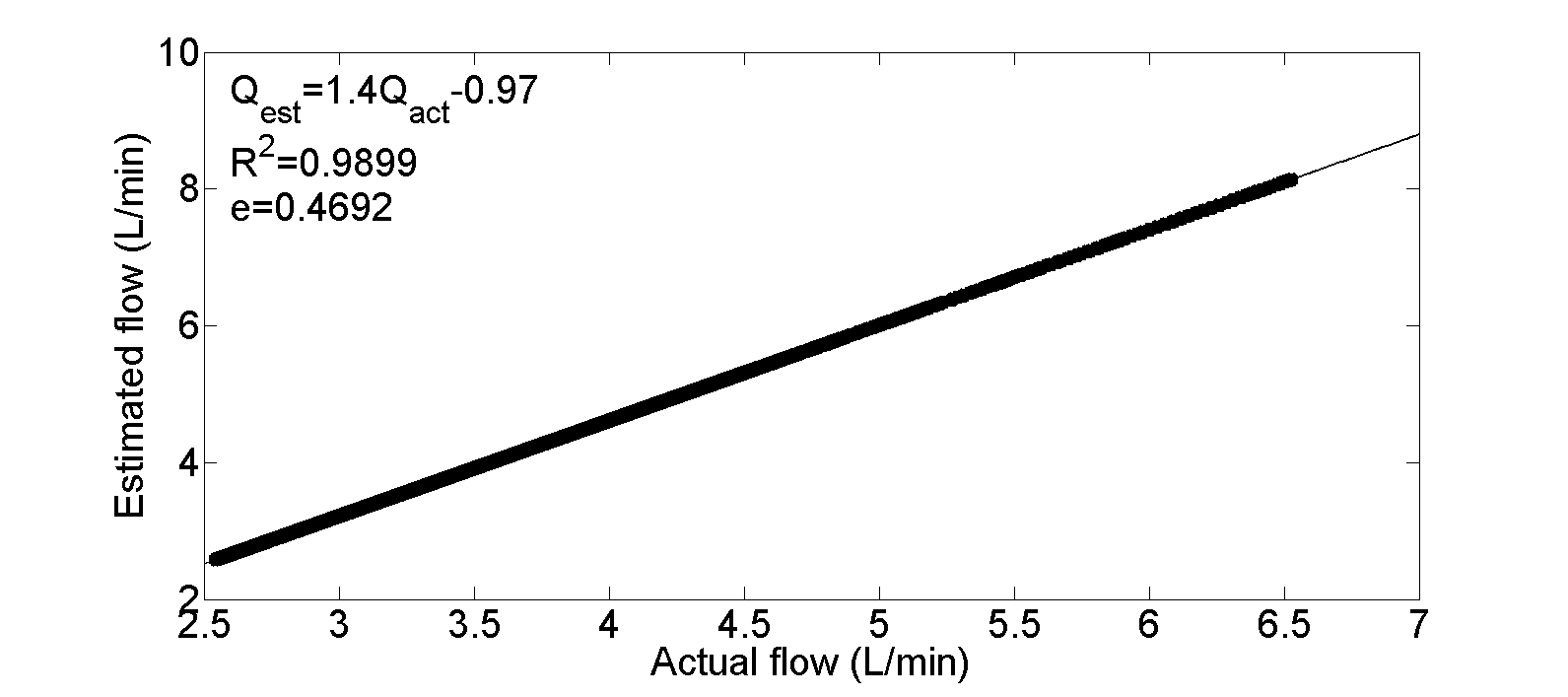}
   \label{46i}
 }

\caption{Pump variable results in rest condition when the system induced at 60s.}
\label{4:60b}
\end{figure}

\begin{figure}[htbp]
\centering
\subfigure[Average pump speed.]{
   \includegraphics[scale =0.1752] {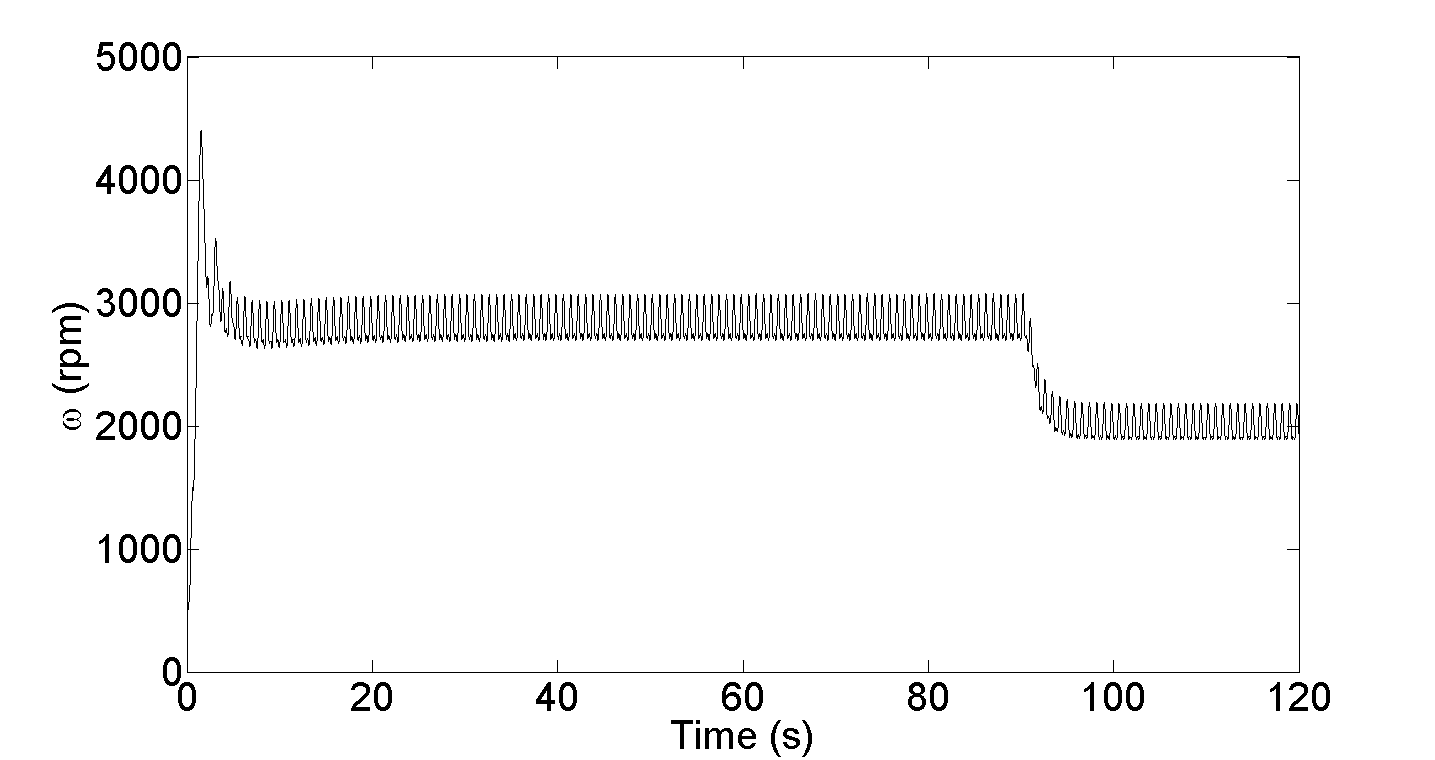}
   \label{49f}
 }

  \subfigure[Pump flow compared with desired reference flow.]{
   \includegraphics[scale =0.1752] {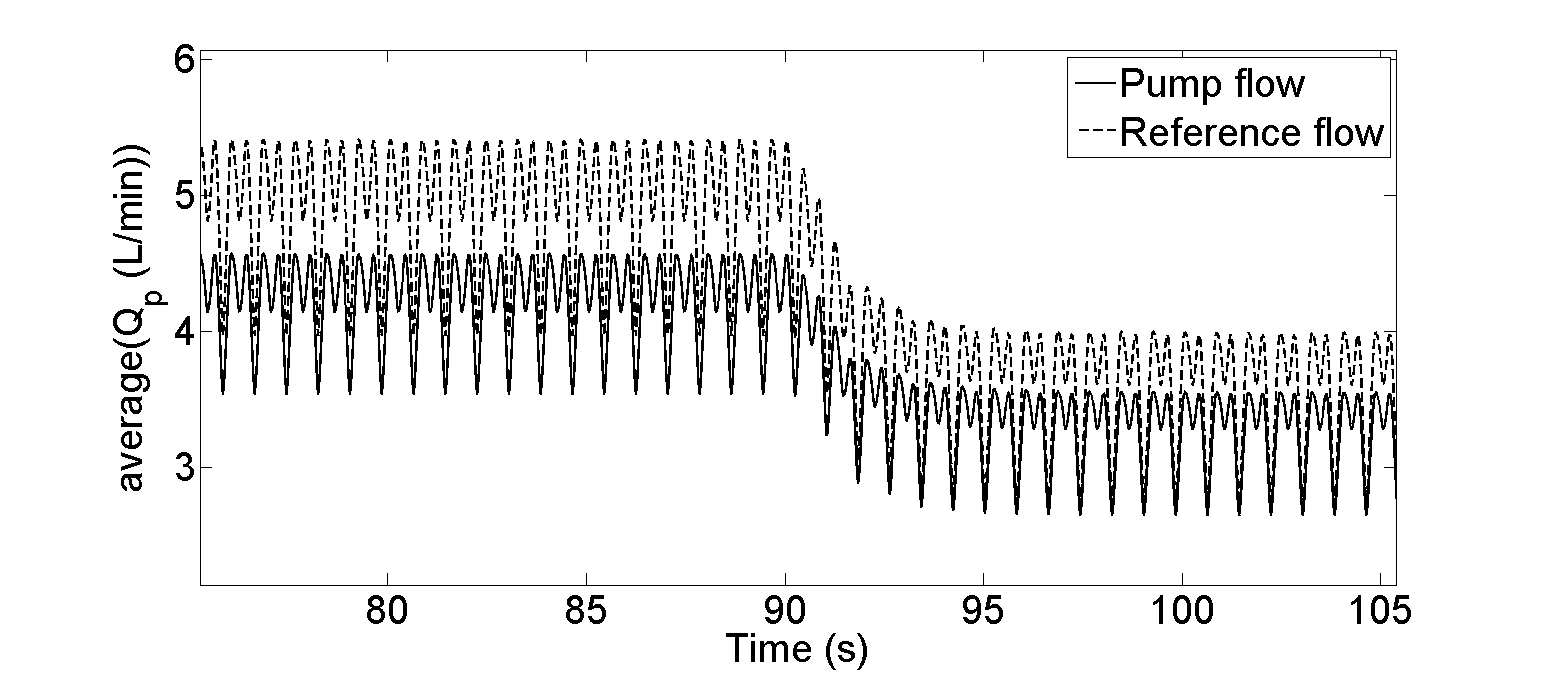}
   \label{49h}
 }

\subfigure[Measured steady state pump flow against estimated pump flow.]{
   \includegraphics[scale =0.1752] {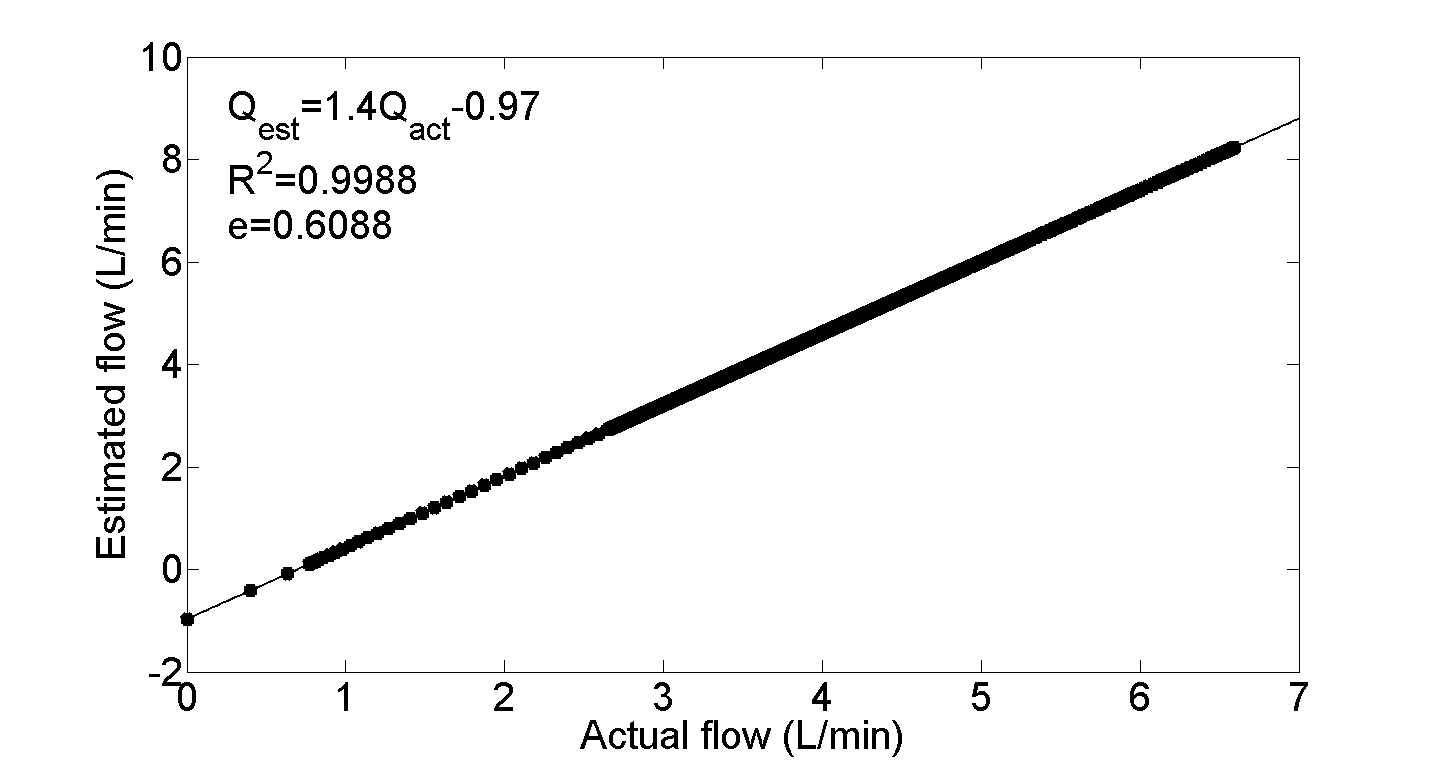}
   \label{49i}
 }

\caption{Pump variable results at rest condition when the system induced at 90s.}
\label{4:90b}
\end{figure}

\begin{figure}[htbp]
\centering
\subfigure[Average pump speed.]{
   \includegraphics[scale =0.1752] {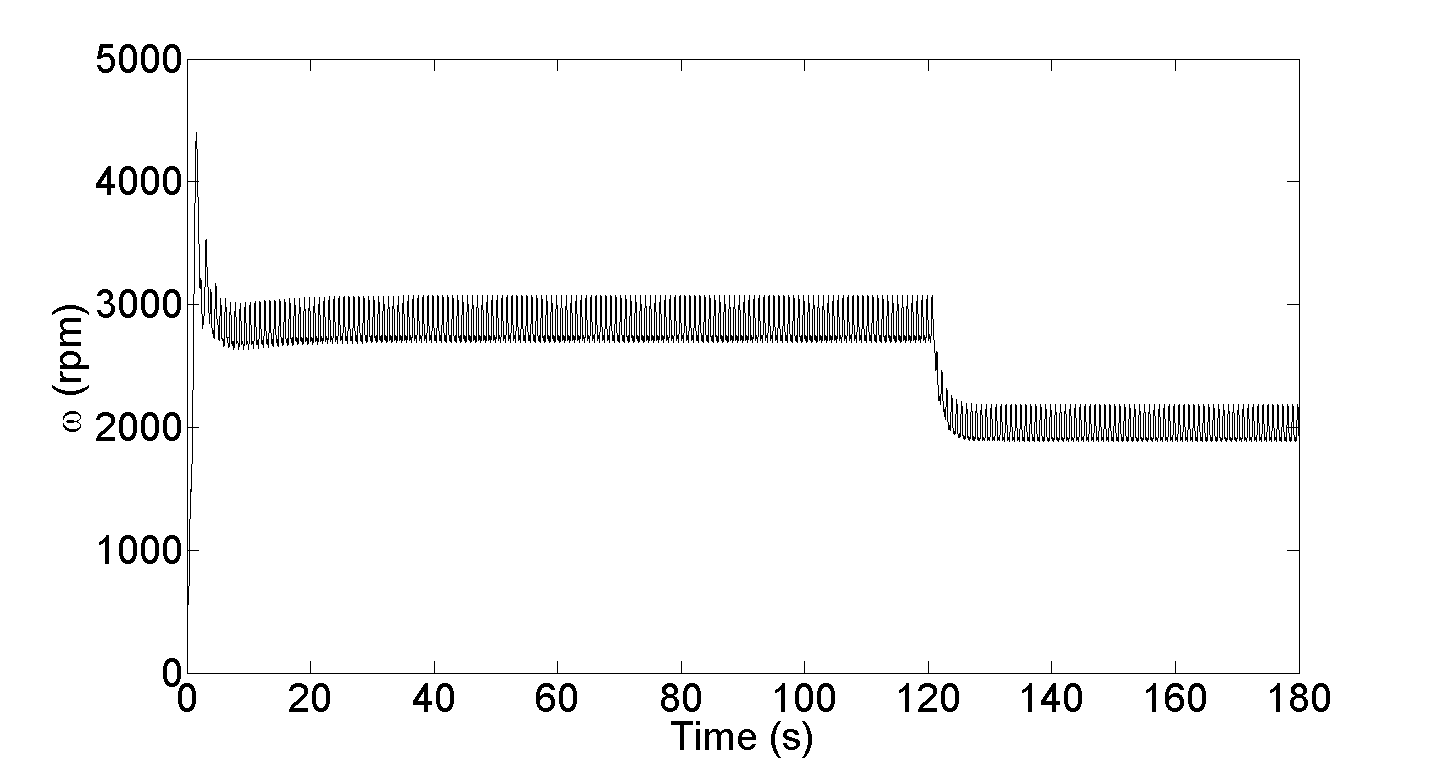}
   \label{42f}
 }

   \subfigure[Pump flow compared with desired reference flow.]{
   \includegraphics[scale =0.1752] {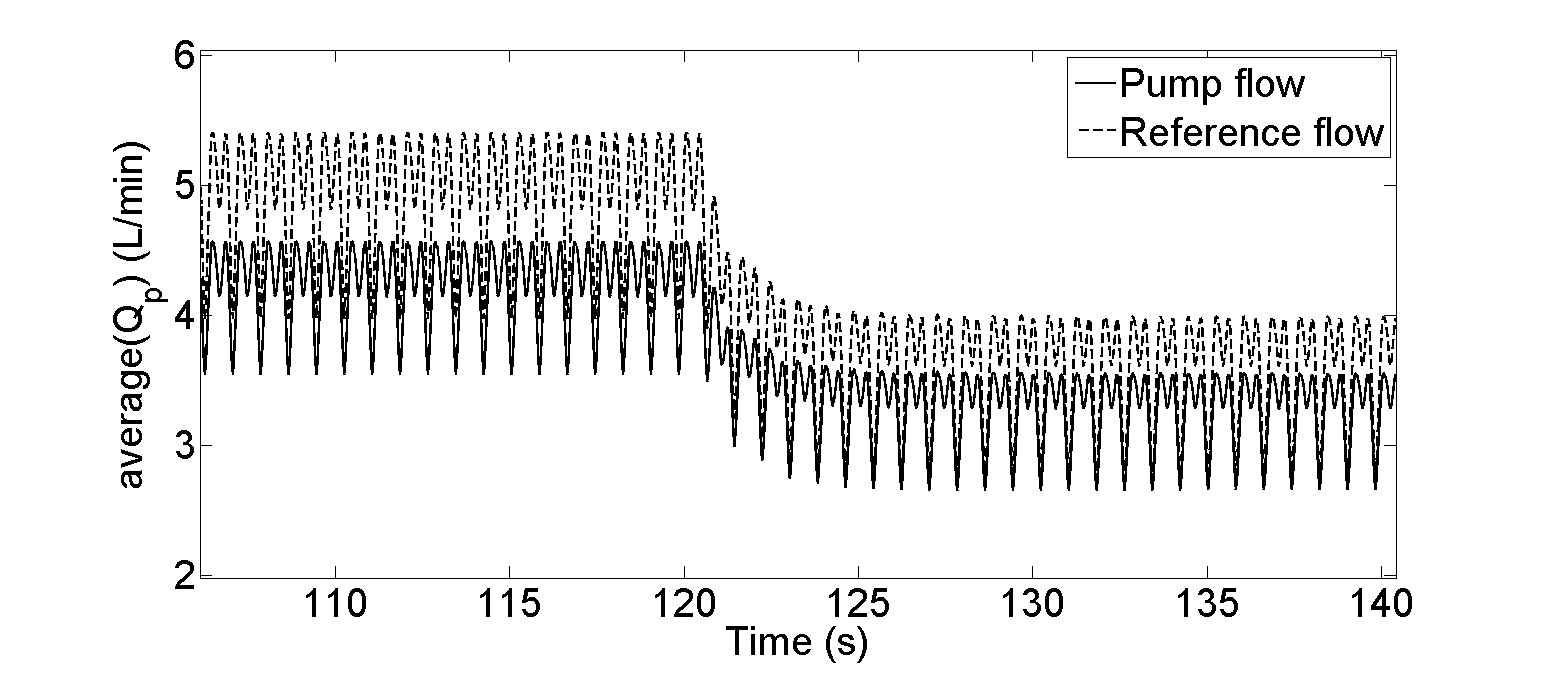}
   \label{42h}
 }

\subfigure[Measured steady state pump flow against estimated pump flow.]{
   \includegraphics[scale =0.1752] {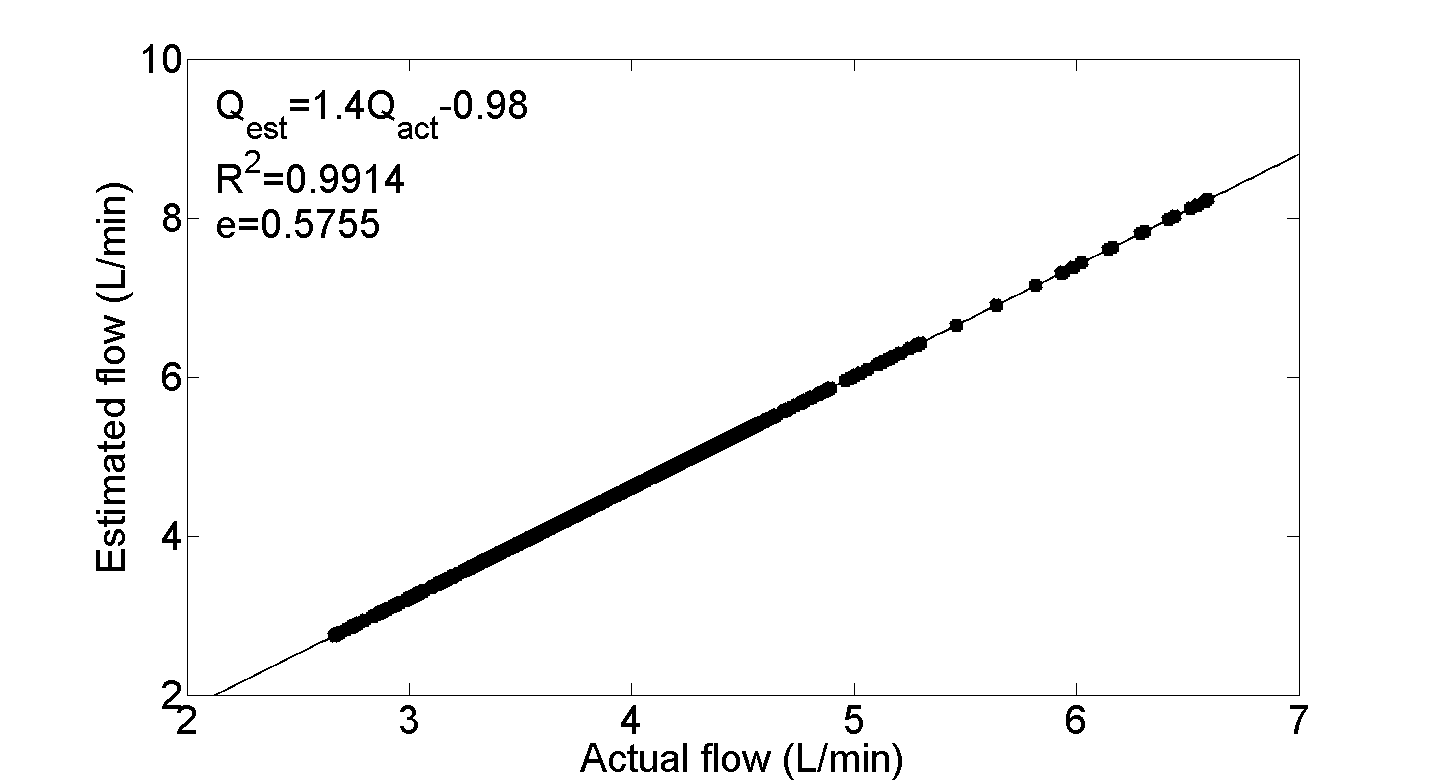}
   \label{42i}
 }

\caption{Pump variable results at rest condition when the system induced at 120s.}
\label{4:20b}
\end{figure}

Figures (\ref{4:30b}, \ref{4:60b}, \ref{4:90b} and \ref{4:20b}) illustrate the Pump variable results during parameter variations. The controller responds to the decrease in LV preload by decreasing average pump rotational speed from 2950 rpm to 2040 rpm and actual average pulsatile flow from 4.5 L/min to 3.4 L/min. These changes are substantially completed within four heartbeats. More importantly  it can be seen from Figures (\ref{43h}, \ref{46h}, \ref{49h} and \ref{42h}) that the simulated pump flow tracks the desired reference flow accurately within an error of $\pm$ 0.7 L/min. In addition, Figures (\ref{43i}, \ref{46i}, \ref{49i} and \ref{42i}) show that actual and estimated flows are highly correlated.


\subsection{Results in Exercise Condition}

\begin{figure*}[htbp]
\centering
\subfigure[LV volume versus LV pressure before and after Parameter Change.]{
   \includegraphics[scale =0.16752] {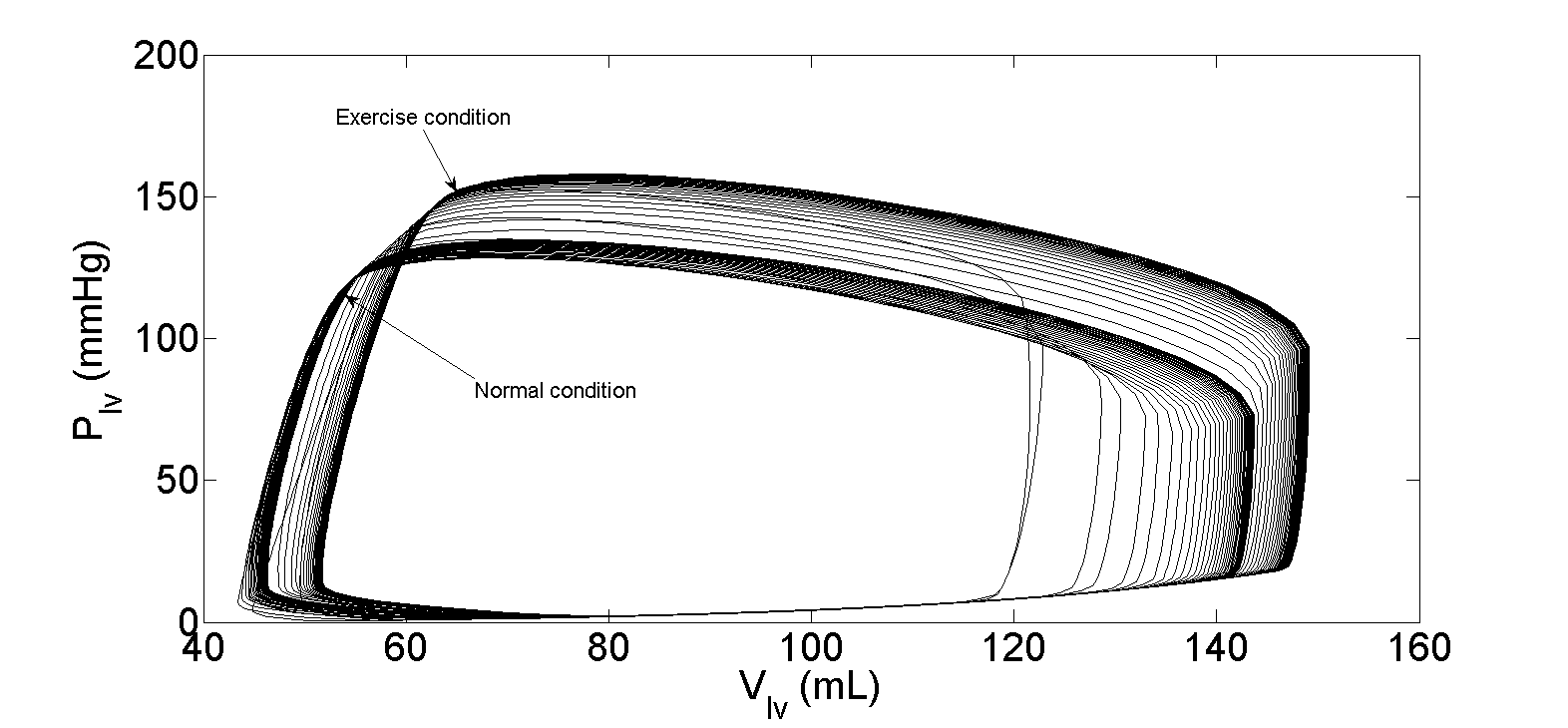}
   \label{43ea}
 }
\subfigure[RV volume versus RV pressure before and after Parameter Change.]{
   \includegraphics[scale =0.16752] {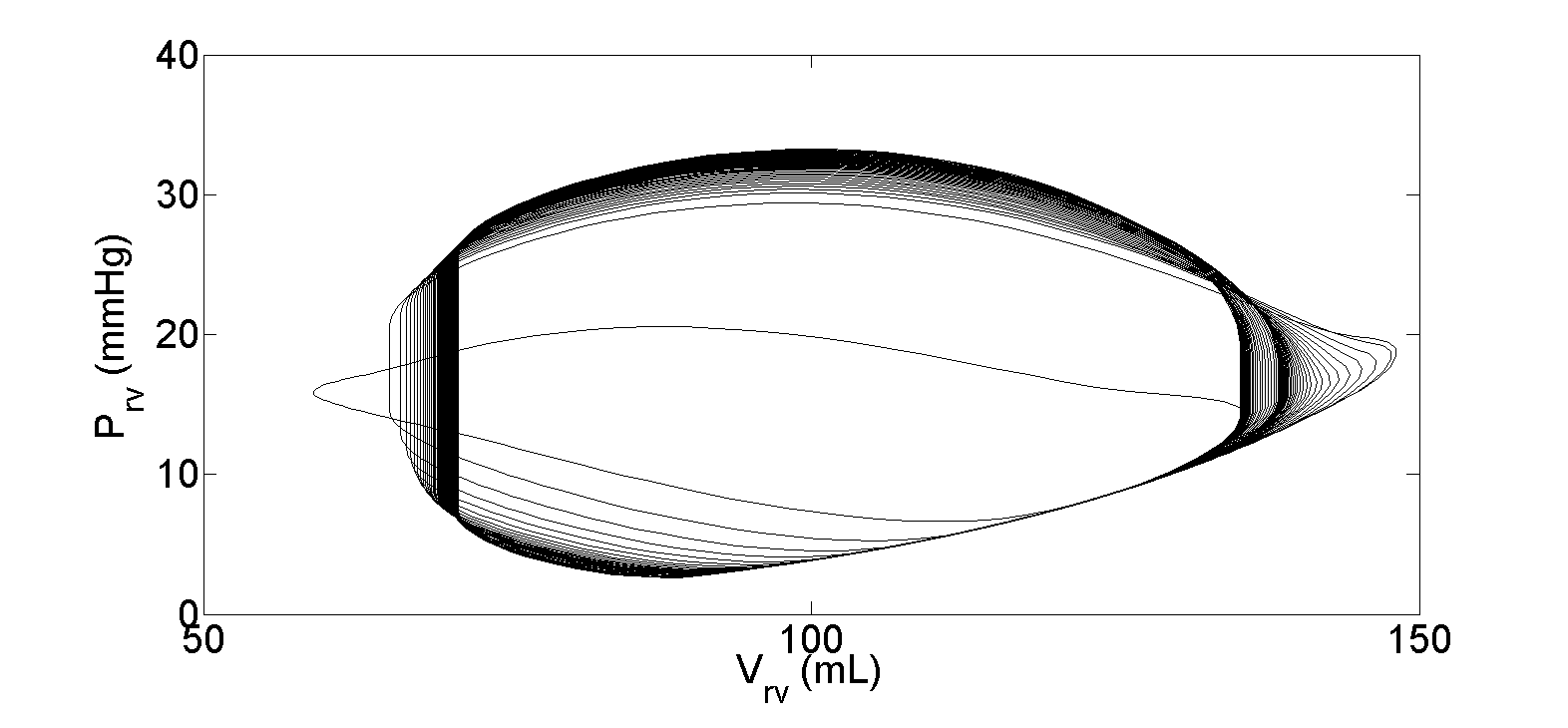}
   \label{43eb}
 }

 \subfigure[Aortic pressure.]{
   \includegraphics[scale =0.16752] {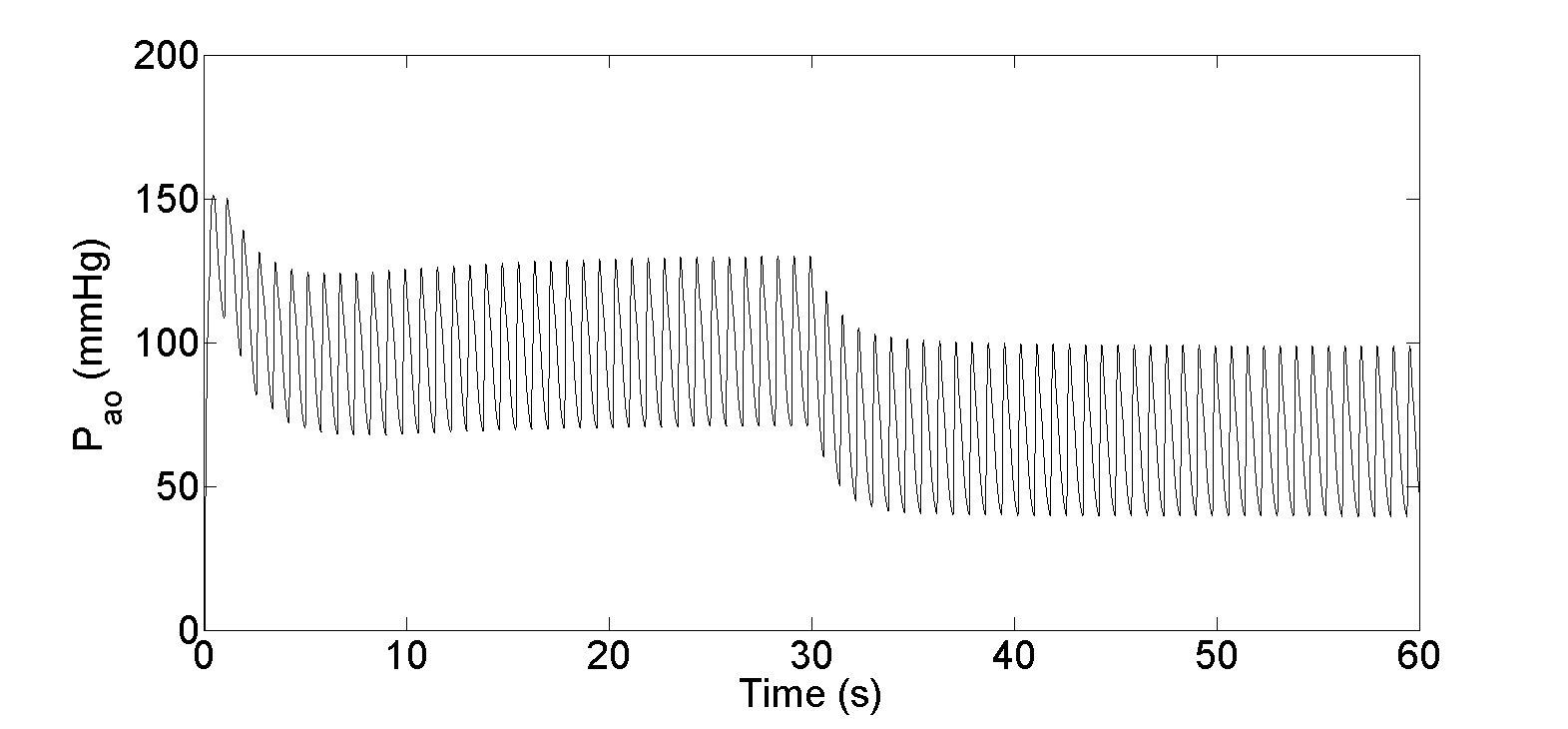}
   \label{43ec}
 }
  \subfigure[Left atrial pressure.]{
   \includegraphics[scale =0.16752] {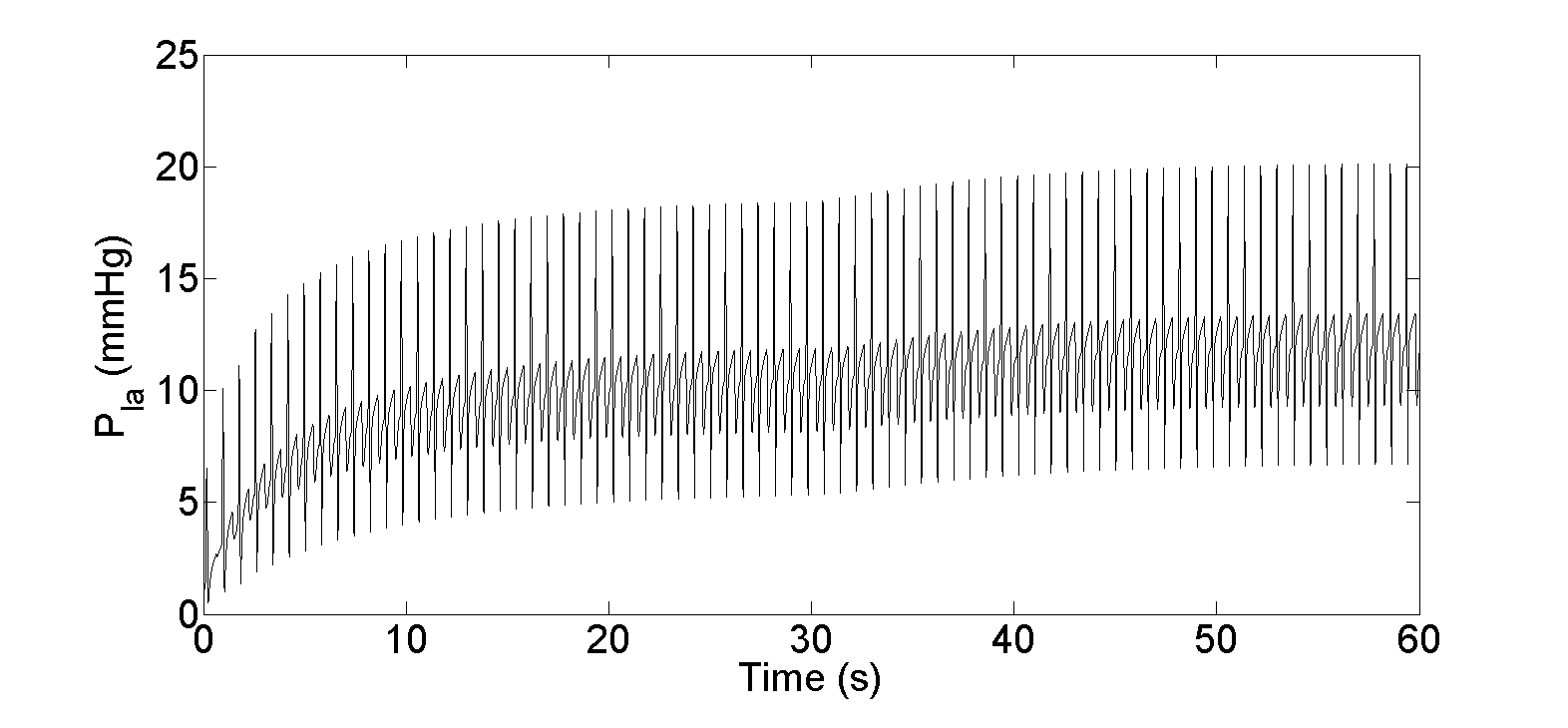}
   \label{43ed}
 }

\subfigure[Right atrial pressure.]{
   \includegraphics[scale =0.16752] {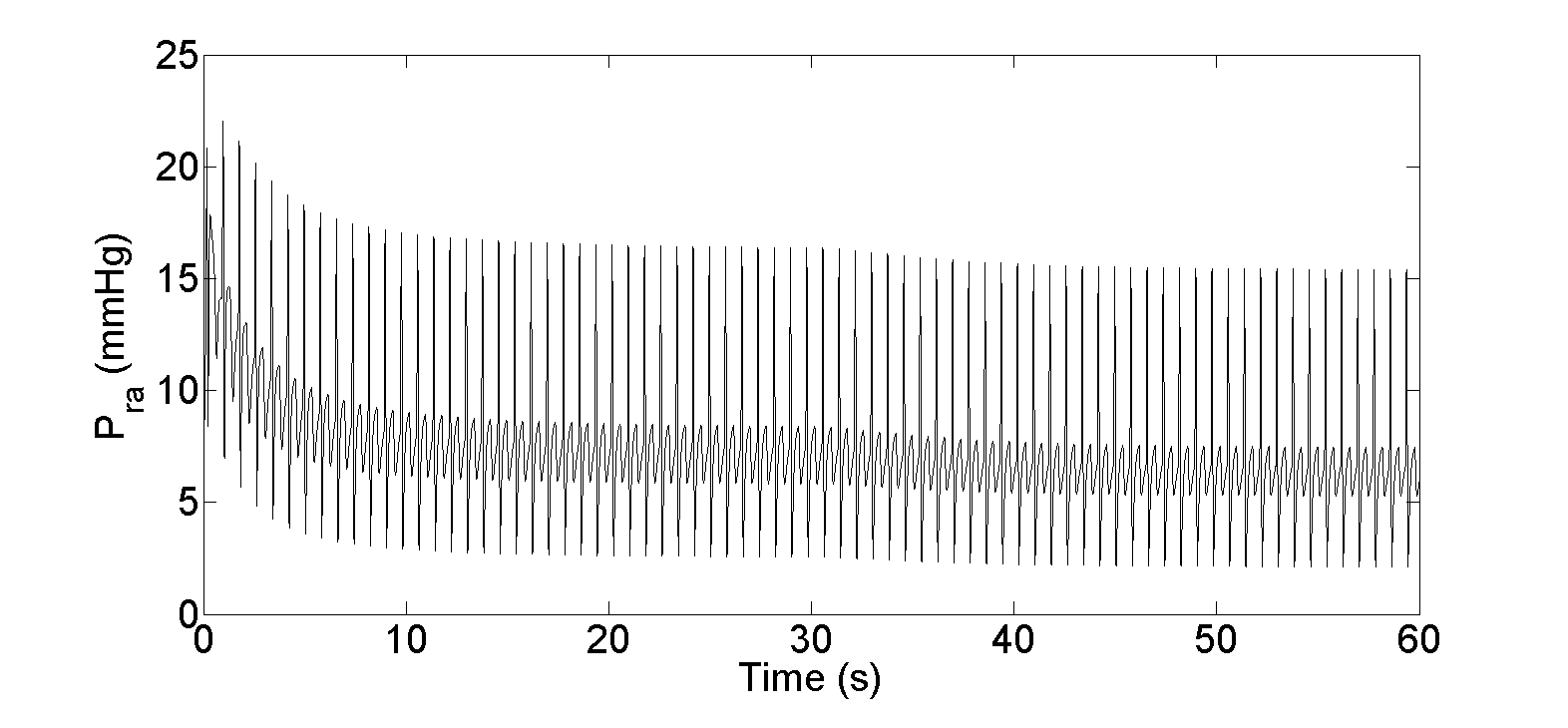}
   \label{43ee}
 }
\caption{Hemodynamic variables results at exercise condition when the system induced at 30s.}
\label{4:30ea}
\end{figure*}

\begin{figure*}[htbp]
\centering
\subfigure[LV volume versus LV pressure before and after Parameter Change.]{
   \includegraphics[scale =0.16752] {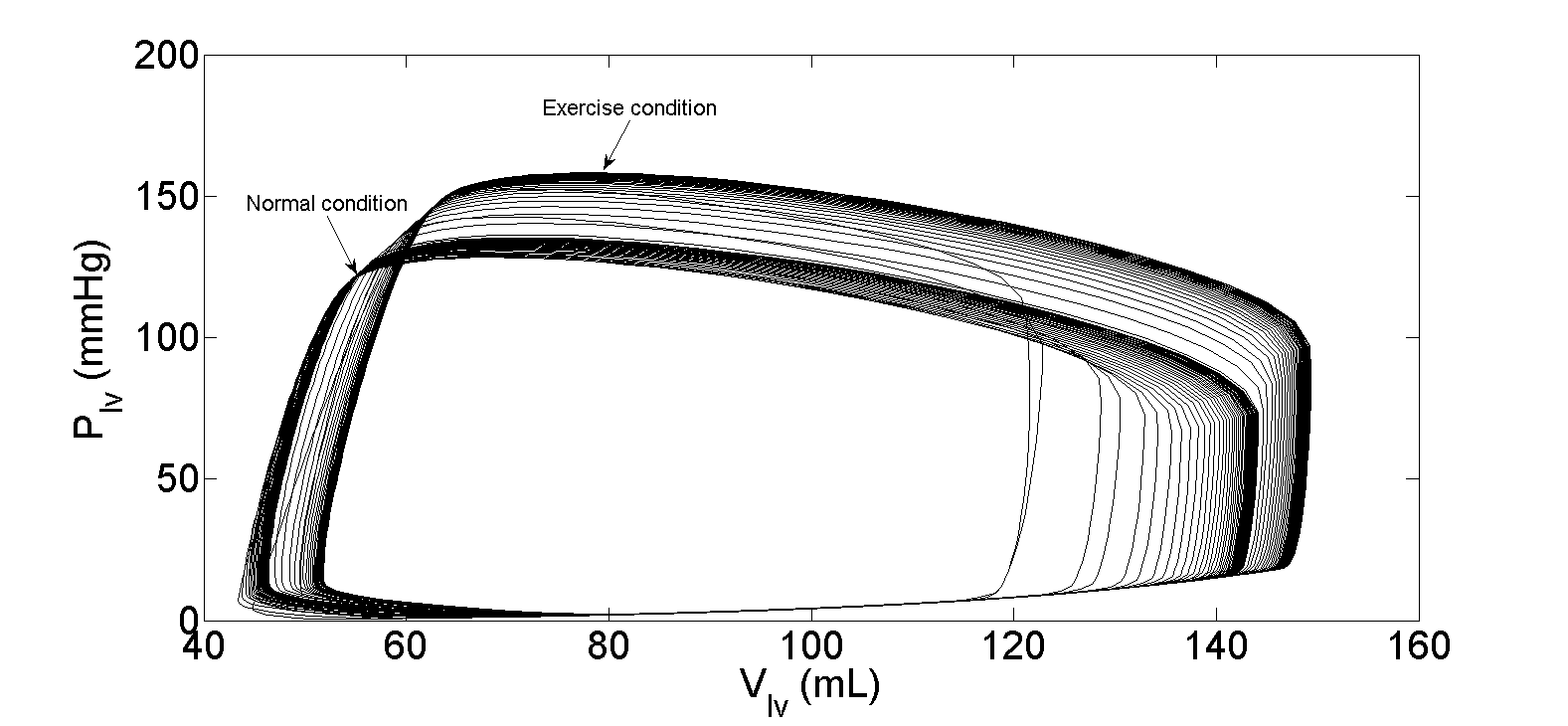}
   \label{46ea}
 }
\subfigure[RV volume versus RV pressure before and after Parameter Change.]{
   \includegraphics[scale =0.16752] {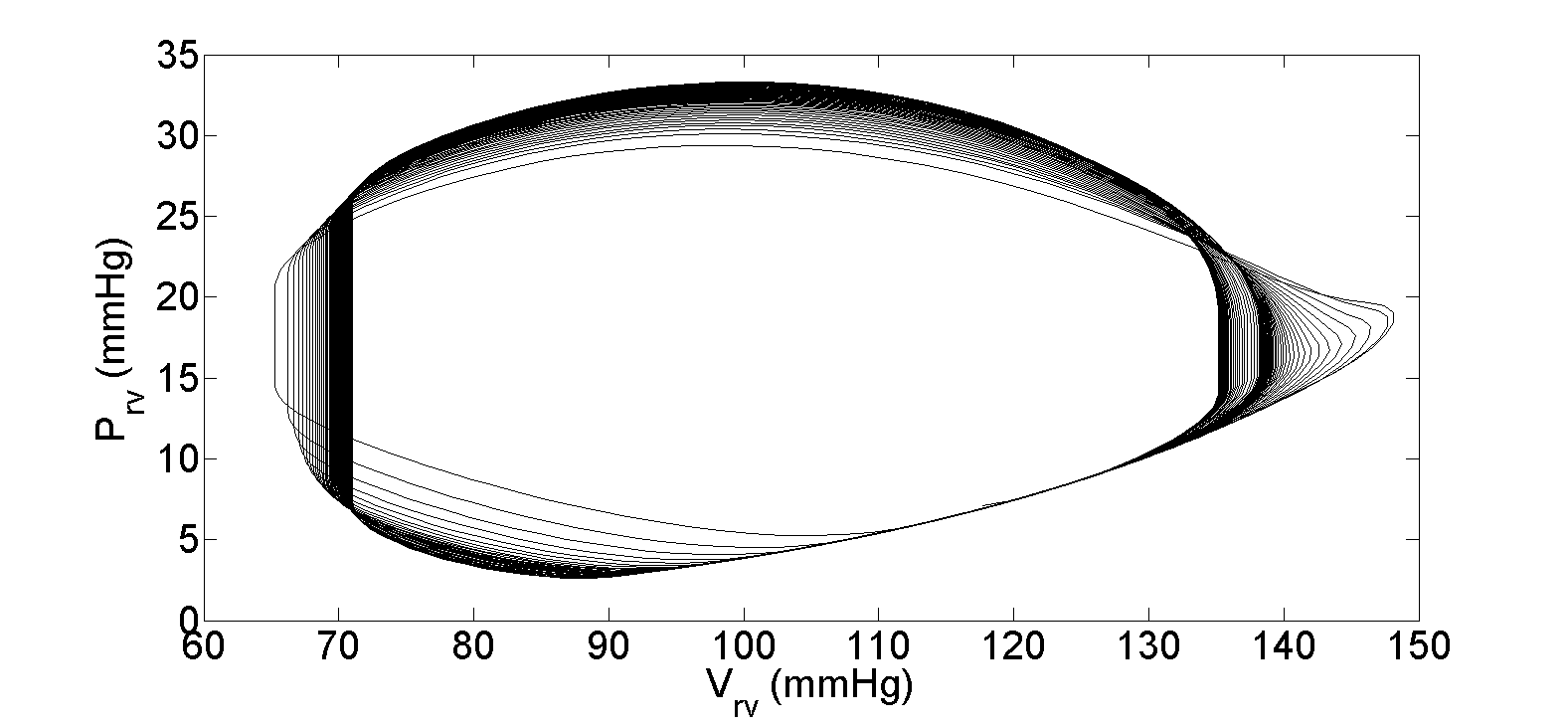}
   \label{46eb}
 }

 \subfigure[Aortic pressure.]{
   \includegraphics[scale =0.16752] {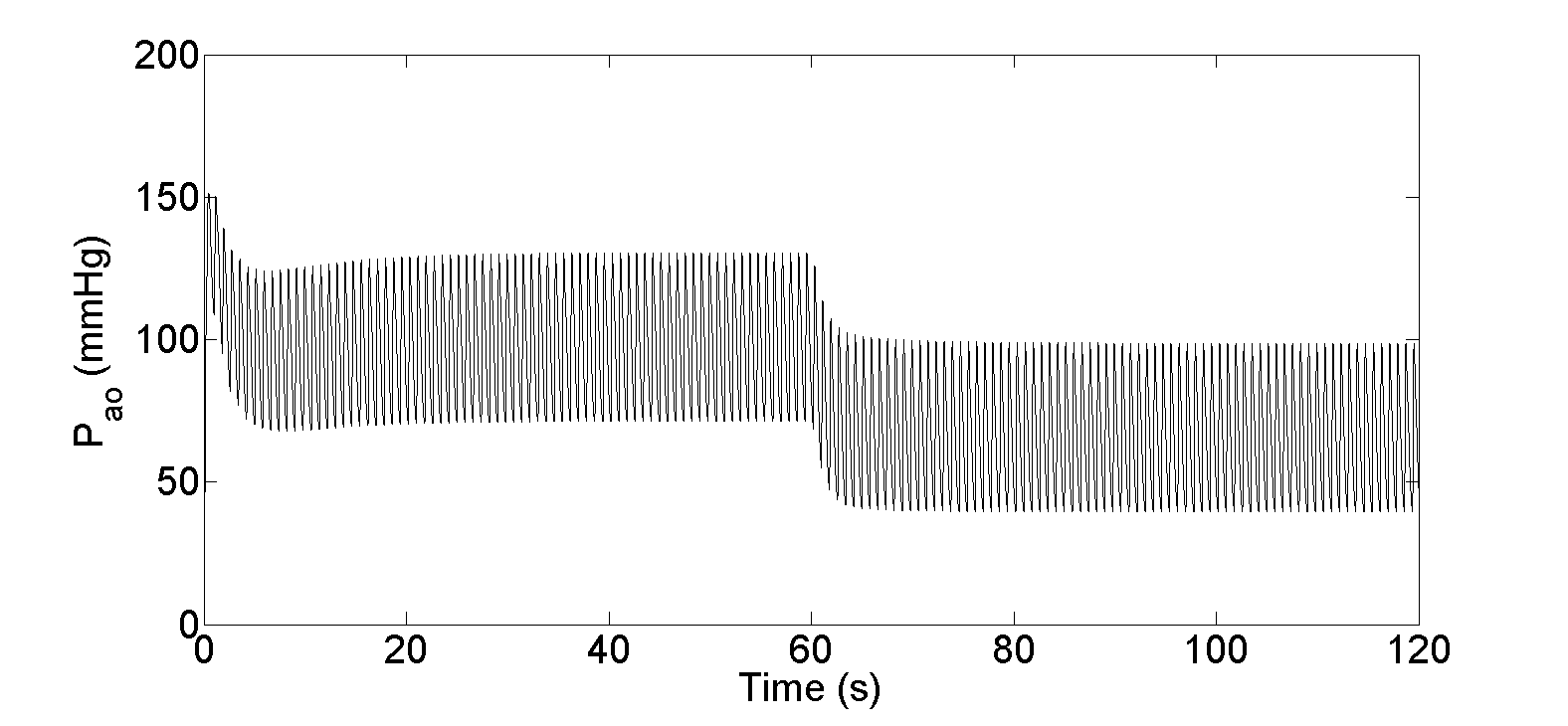}
   \label{46ec}
 }
  \subfigure[Left atrial pressure.]{
   \includegraphics[scale =0.16752] {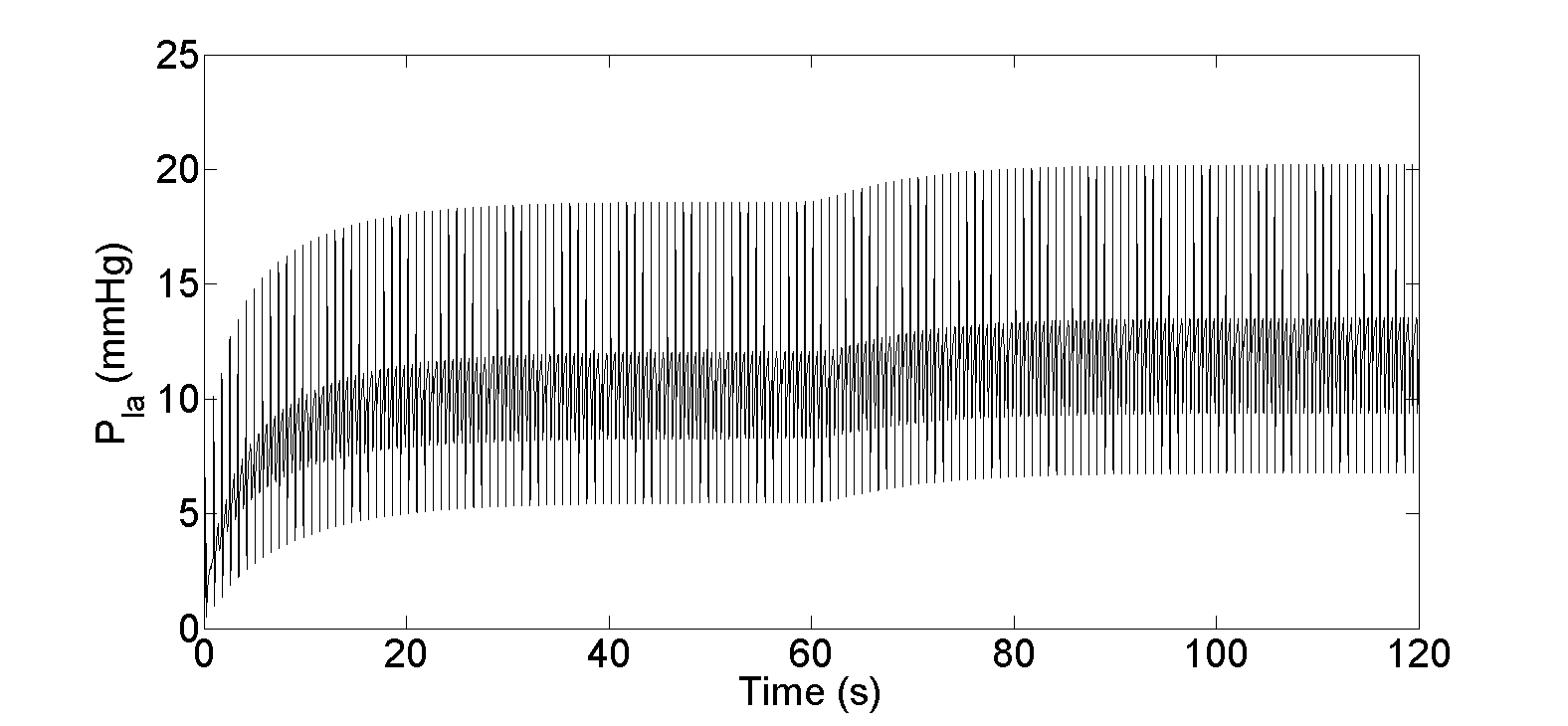}
   \label{46ed}
 }

\subfigure[Right atrial pressure.]{
   \includegraphics[scale =0.16752] {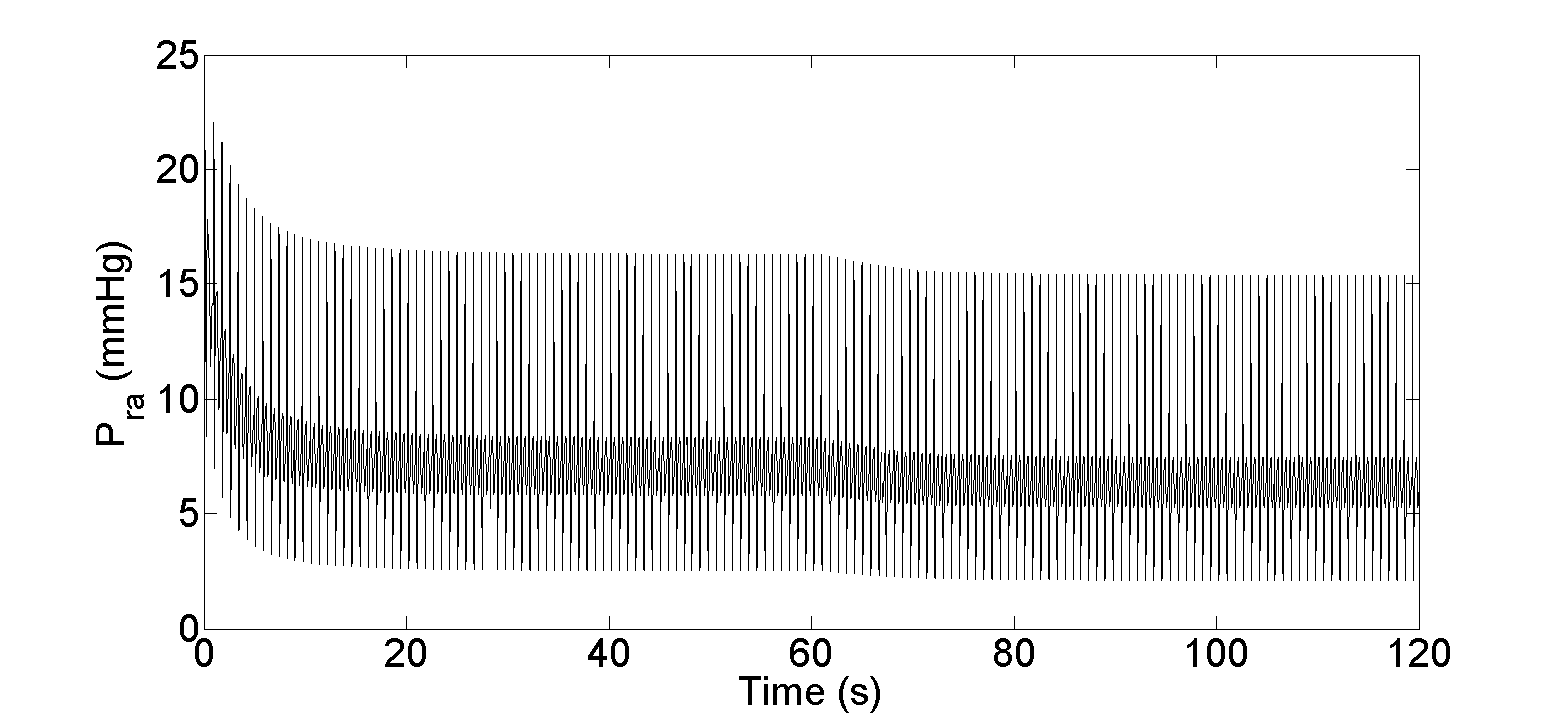}
   \label{46ee}
 }
\caption{Hemodynamic variables results at exercise condition when the system induced at 60s.}
\label{4:60ea}
\end{figure*}

\begin{figure*}[htbp]
\centering
\subfigure[LV volume versus LV pressure before and after Parameter Change.]{
   \includegraphics[scale =0.16752] {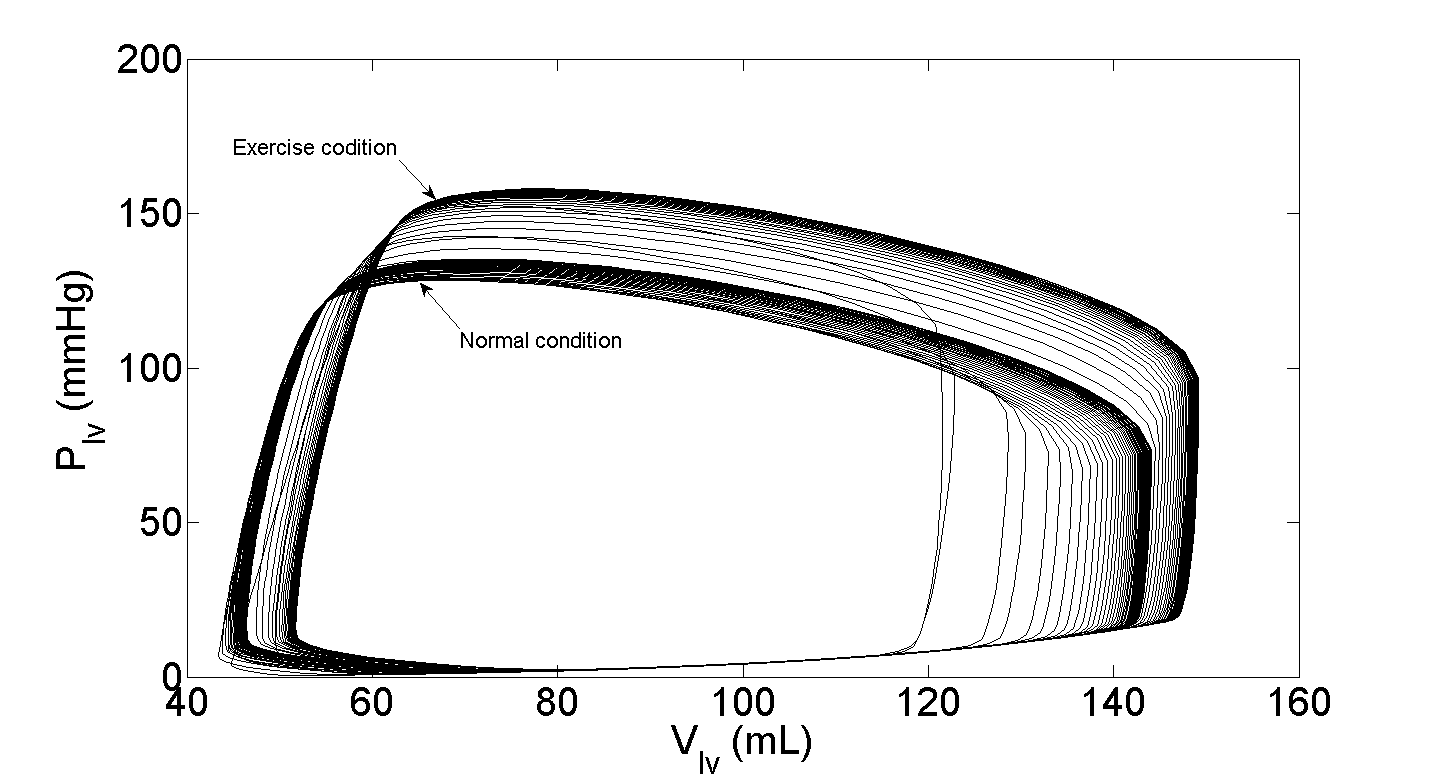}
   \label{49ea}
 }
\subfigure[RV volume versus RV pressure before and after Parameter Change.]{
   \includegraphics[scale =0.16752] {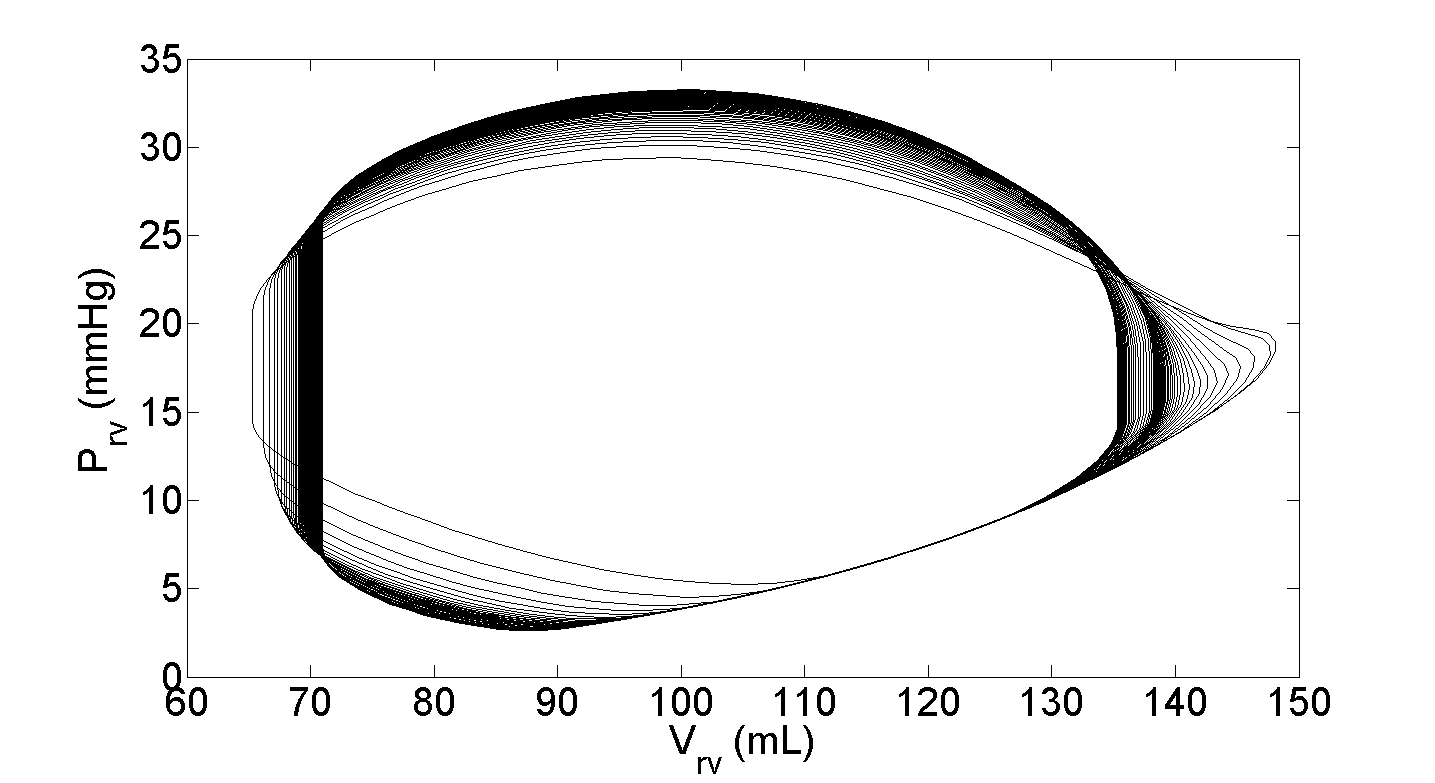}
   \label{49eb}
 }

 \subfigure[Aortic pressure.]{
   \includegraphics[scale =0.16752] {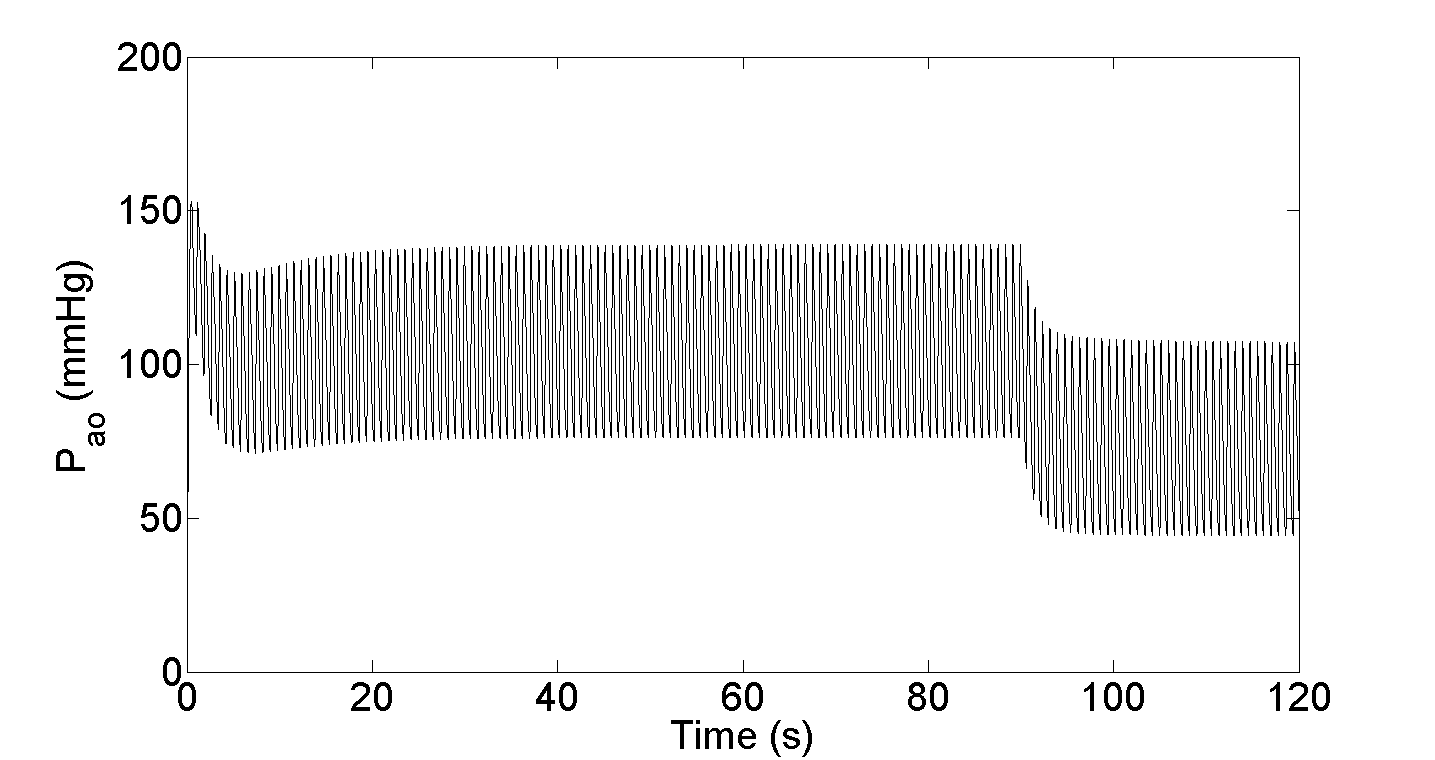}
   \label{49ec}
 }
  \subfigure[Left atrial pressure.]{
   \includegraphics[scale =0.16752] {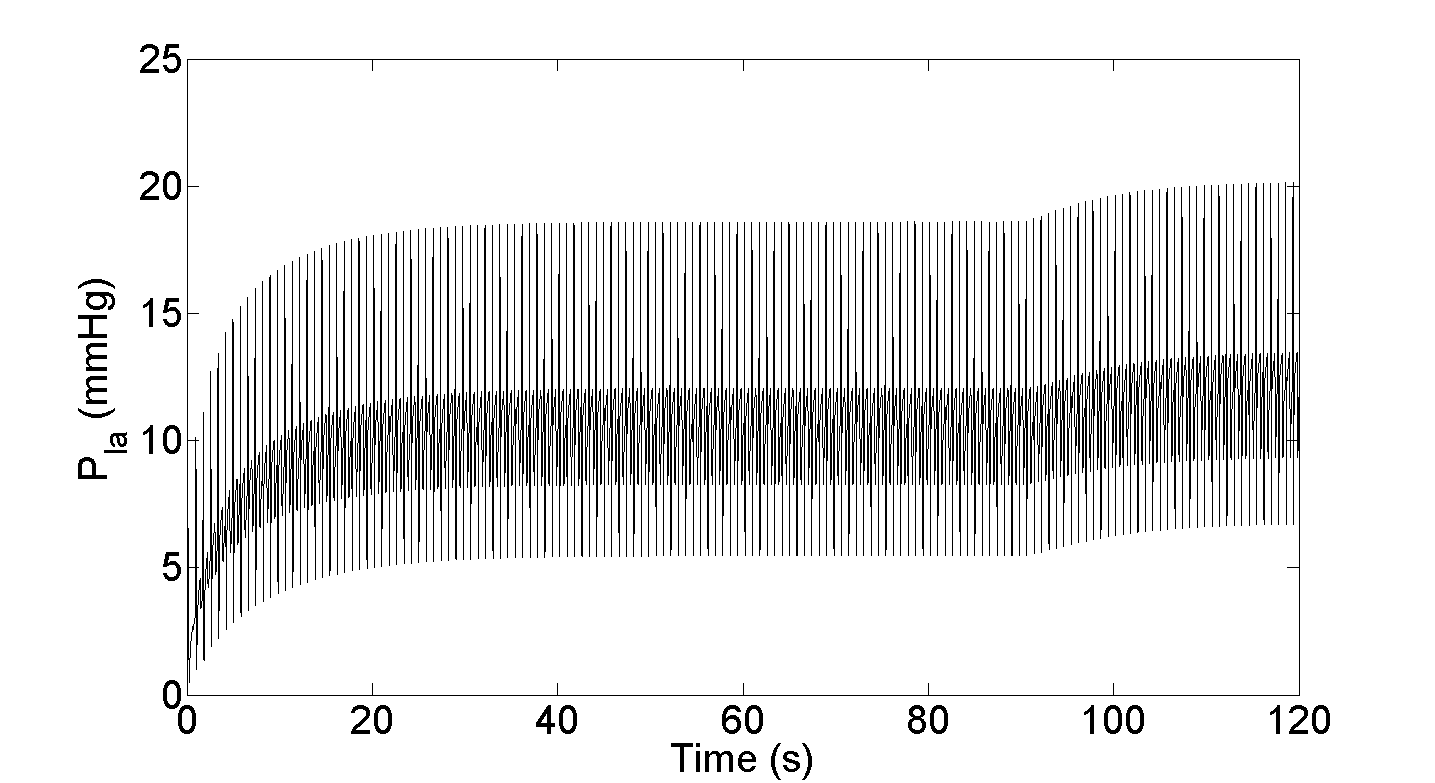}
   \label{49ed}
 }

\subfigure[Right atrial pressure.]{
   \includegraphics[scale =0.16752] {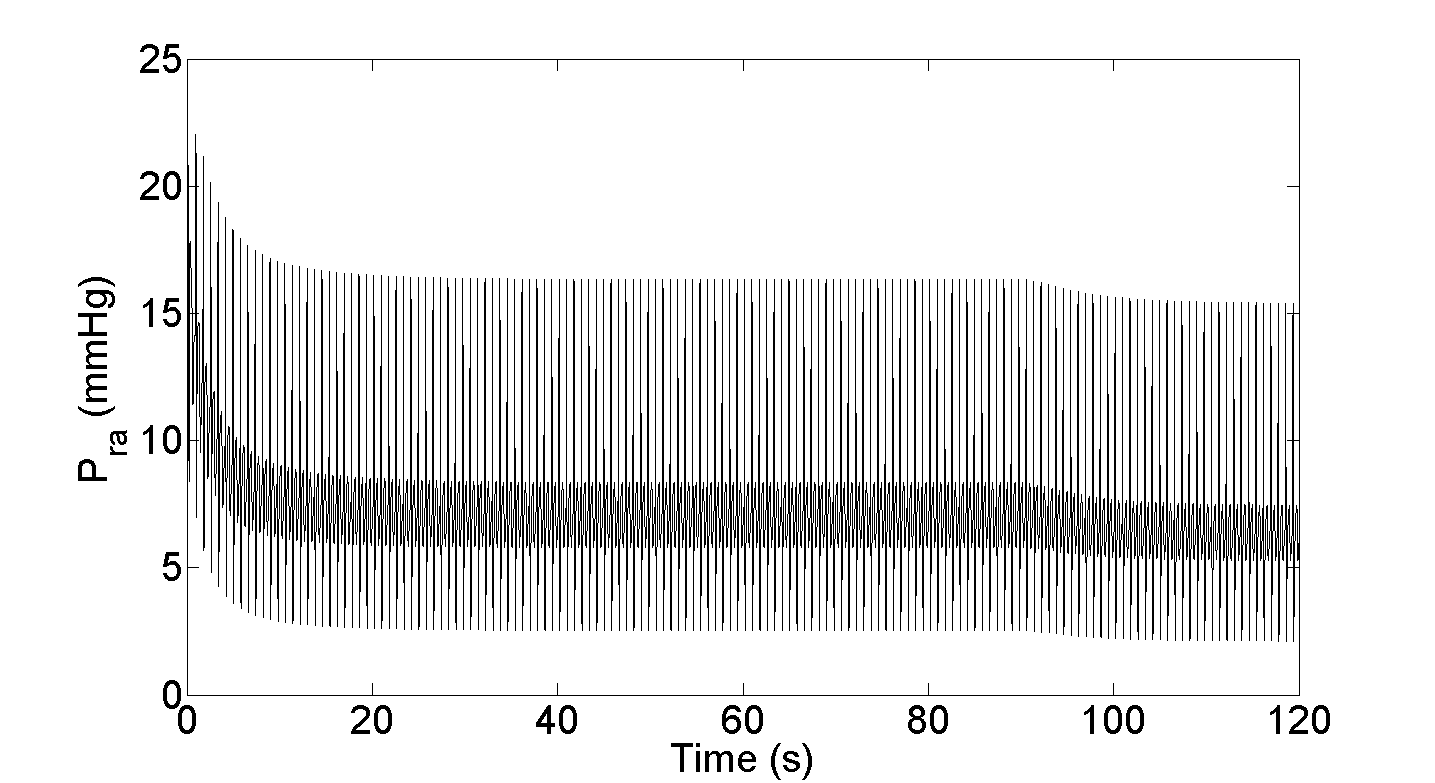}
   \label{49ee}
 }
\caption{Hemodynamic variables results at exercise condition when the system induced at 90s.}
\label{4:90ea}
\end{figure*}

\begin{figure*}[htbp]
\centering
\subfigure[LV volume versus LV pressure before and after Parameter Change.]{
   \includegraphics[scale =0.16752] {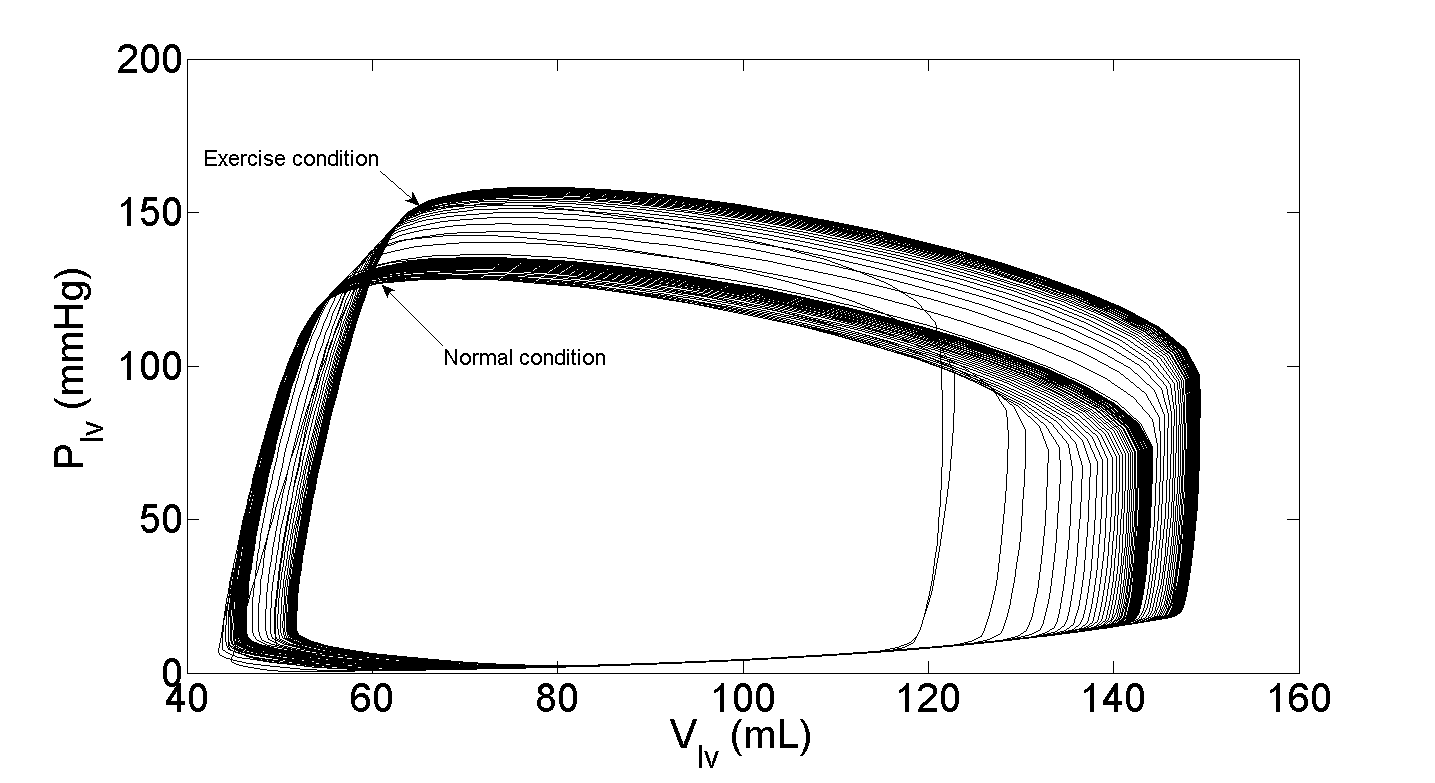}
   \label{42ea}
 }
\subfigure[RV volume versus RV pressure before and after Parameter Change.]{
   \includegraphics[scale =0.16752] {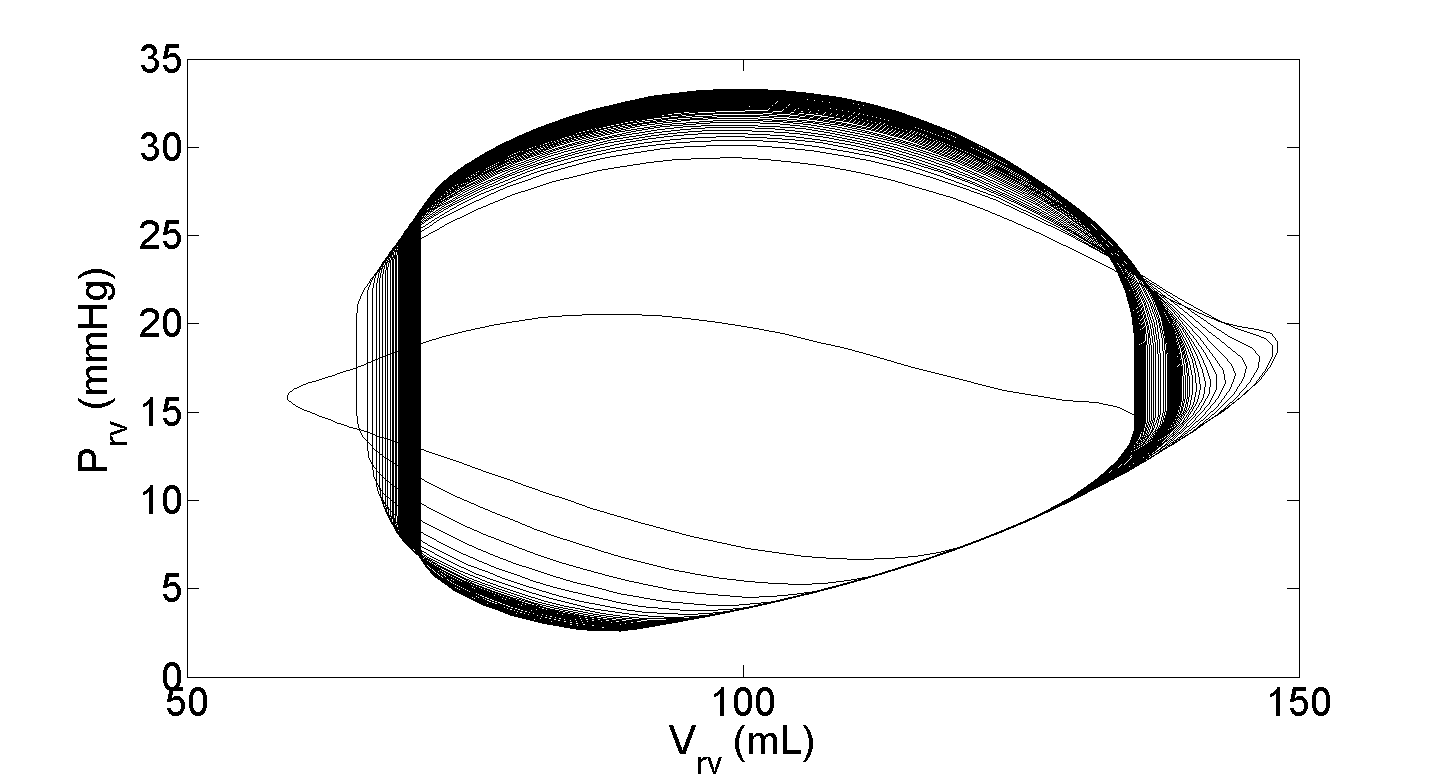}
   \label{42eb}
 }

 \subfigure[Aortic pressure.]{
   \includegraphics[scale =0.16752] {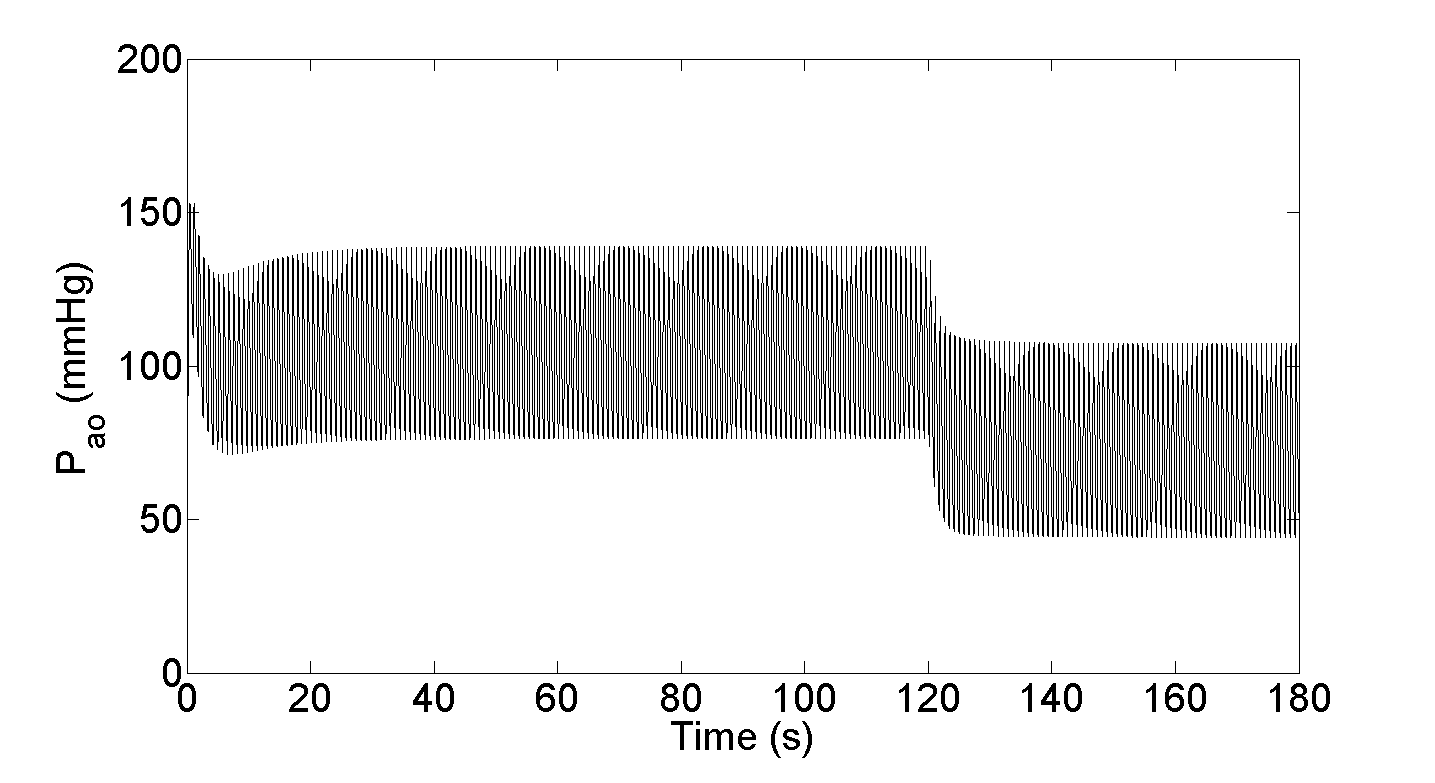}
   \label{42ec}
 }
  \subfigure[Left atrial pressure.]{
   \includegraphics[scale =0.16752] {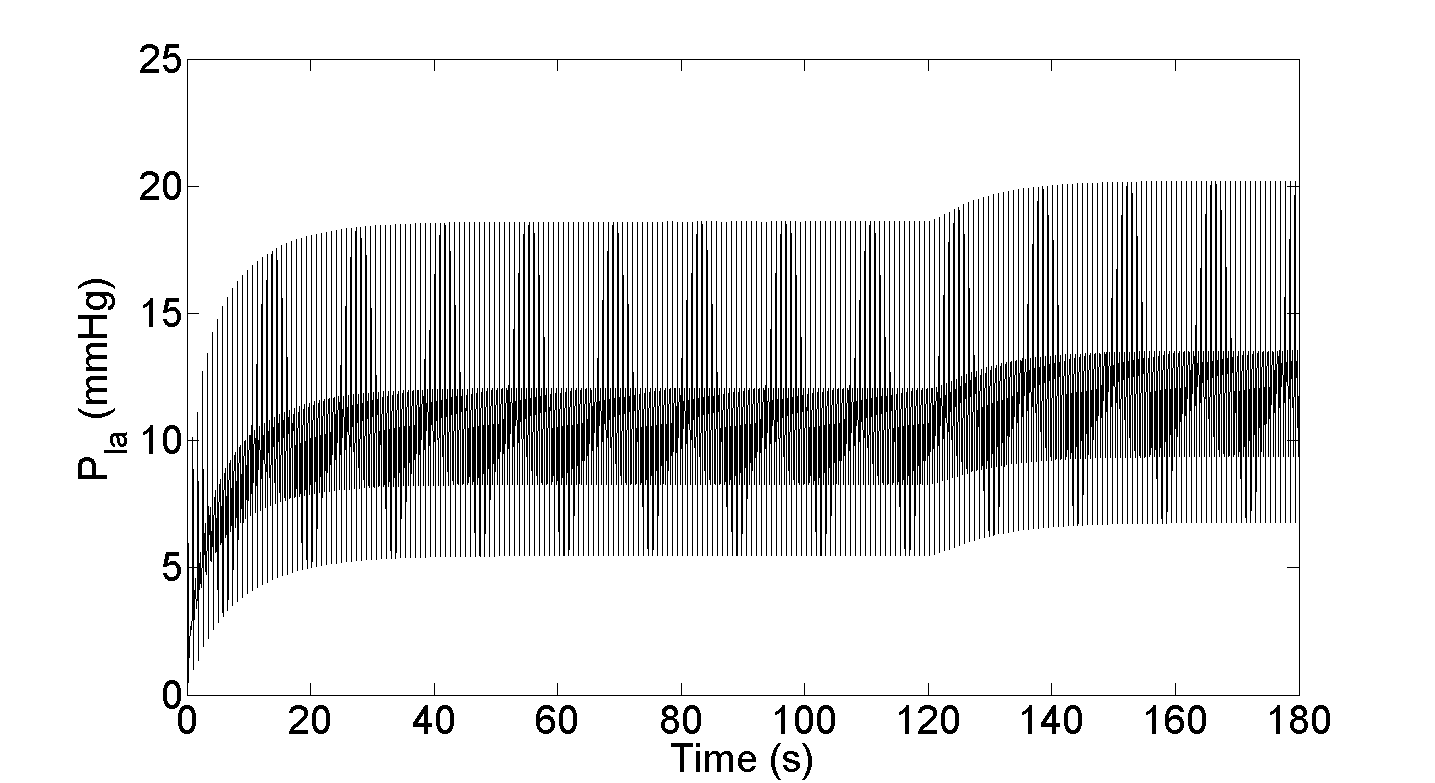}
   \label{42ed}
 }

\subfigure[Right atrial pressure.]{
   \includegraphics[scale =0.16752] {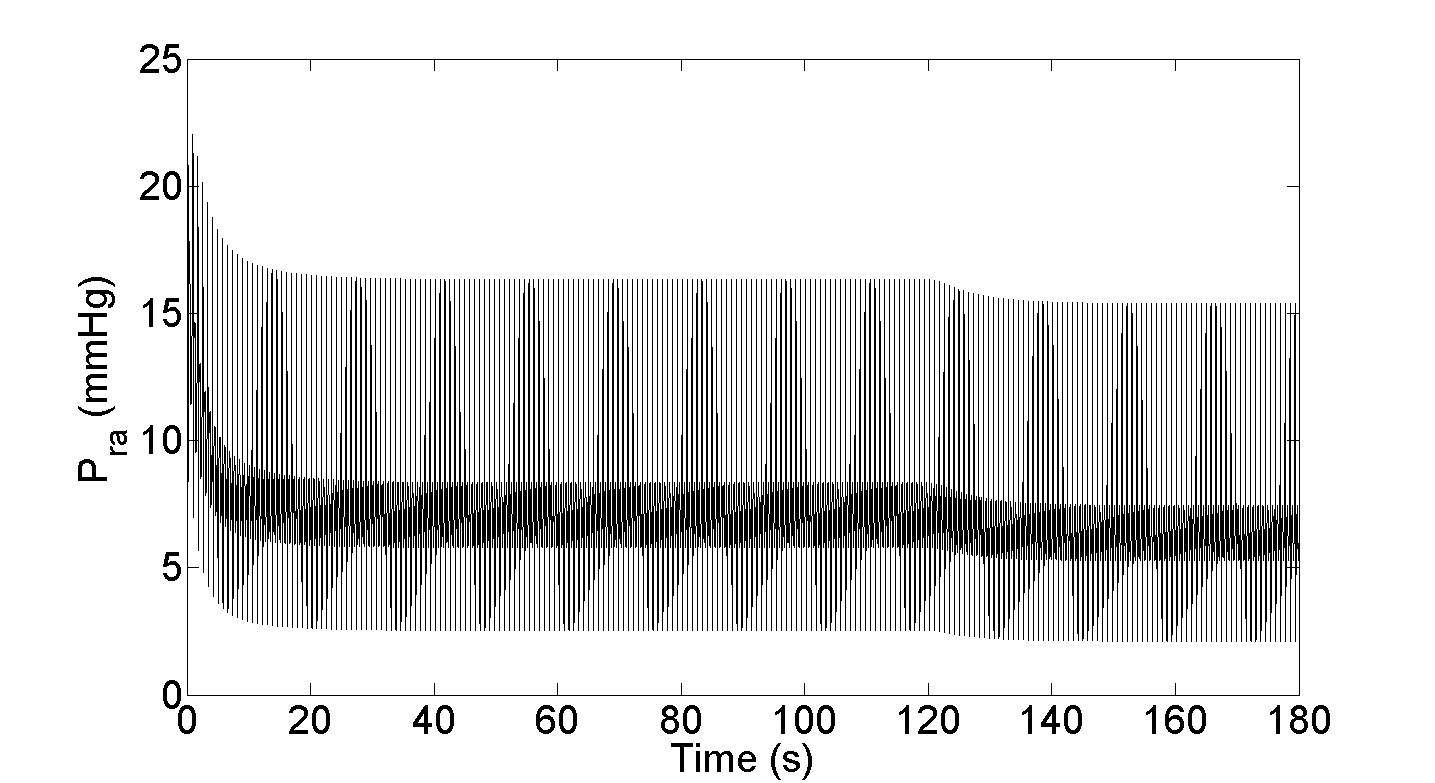}
   \label{42ee}
 }
\caption{Hemodynamic variables results at exercise condition when the system induced at 120s.}
\label{4:20ea}
\end{figure*}

During an exercise, the model parameters have  been changed to simulate the transition from normal to exercise. Figures (\ref{4:30ea}, \ref{4:60ea}, \ref{4:90ea} and \ref{4:20ea}) illustrate the hemodynamic variables corresponding to immediate response of the controller. Changing the parameters of the system at the middle of each period of time produces a rightward shift of LV pressure volume loops combined with a major increase in LV stroke volume, and similar increase in LV end-systolic pressure. This is associated with a shift to the left of the RV pressure-volume loop causing a reduction in LV end-diastolic and end-systolic volumes and pressure. As a result, the LVAD successfully decreases the aortic pressure $P_{ao}$ and increases the left atrial pressure $P_{la}$ and keeps the right atrial pressure $P_{ra}$ within safe operating  range.

\begin{figure*}[htbp]
\centering
\subfigure[Average pump speed.]{
   \includegraphics[scale =0.1752] {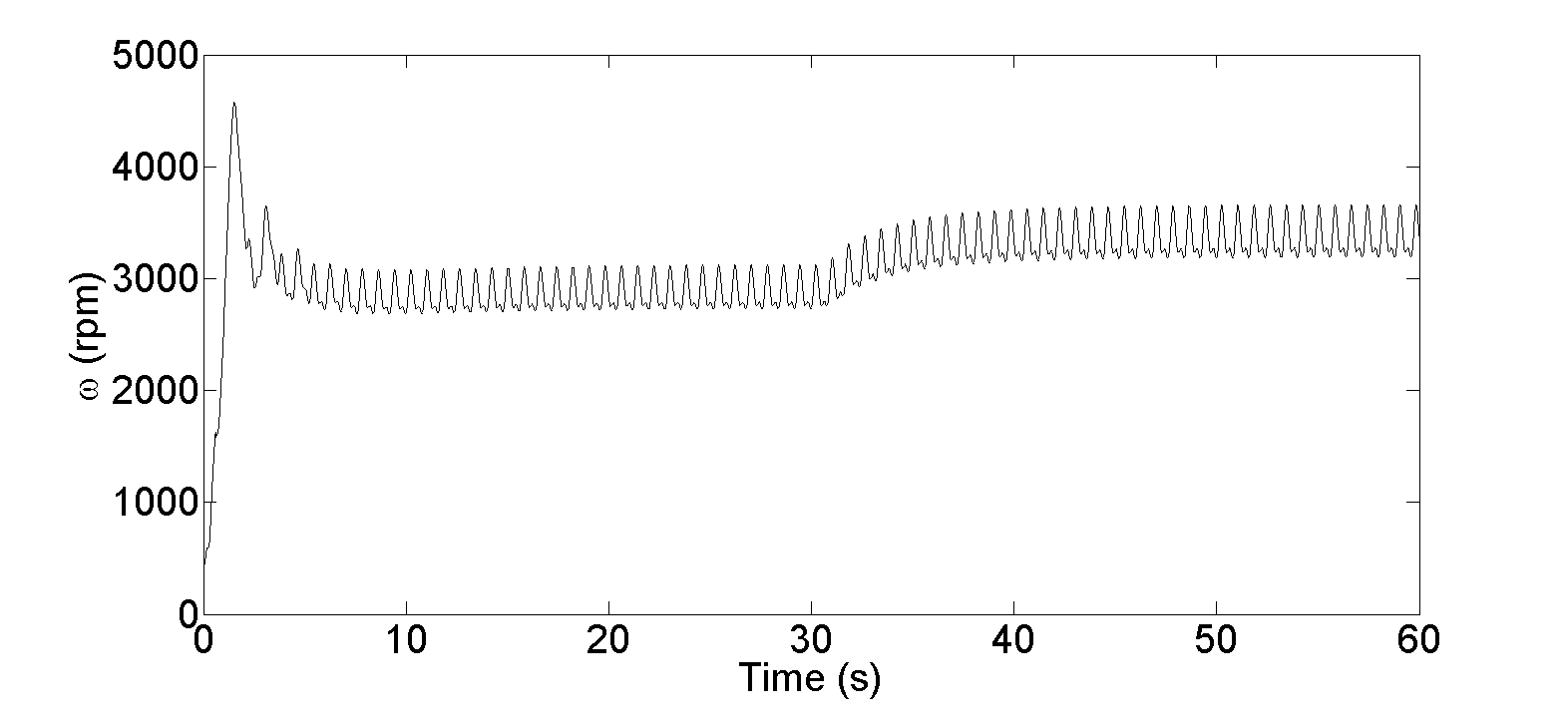}
   \label{43ef}
 }

  \subfigure[Pump flow compared with desired reference flow.]{
   \includegraphics[scale =0.1752] {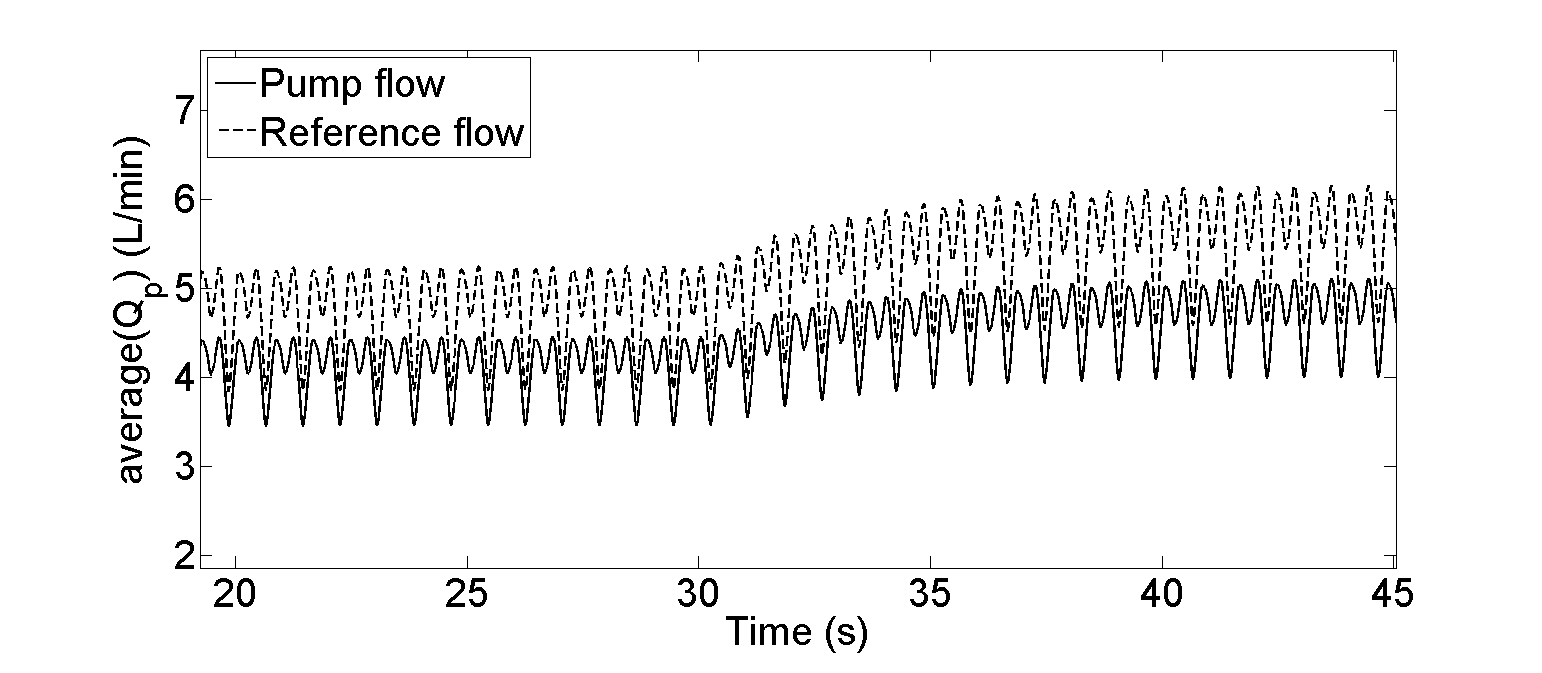}
   \label{43eh}
 }

\subfigure[Measured steady state pump flow against estimated pump flow.]{
   \includegraphics[scale =0.1752] {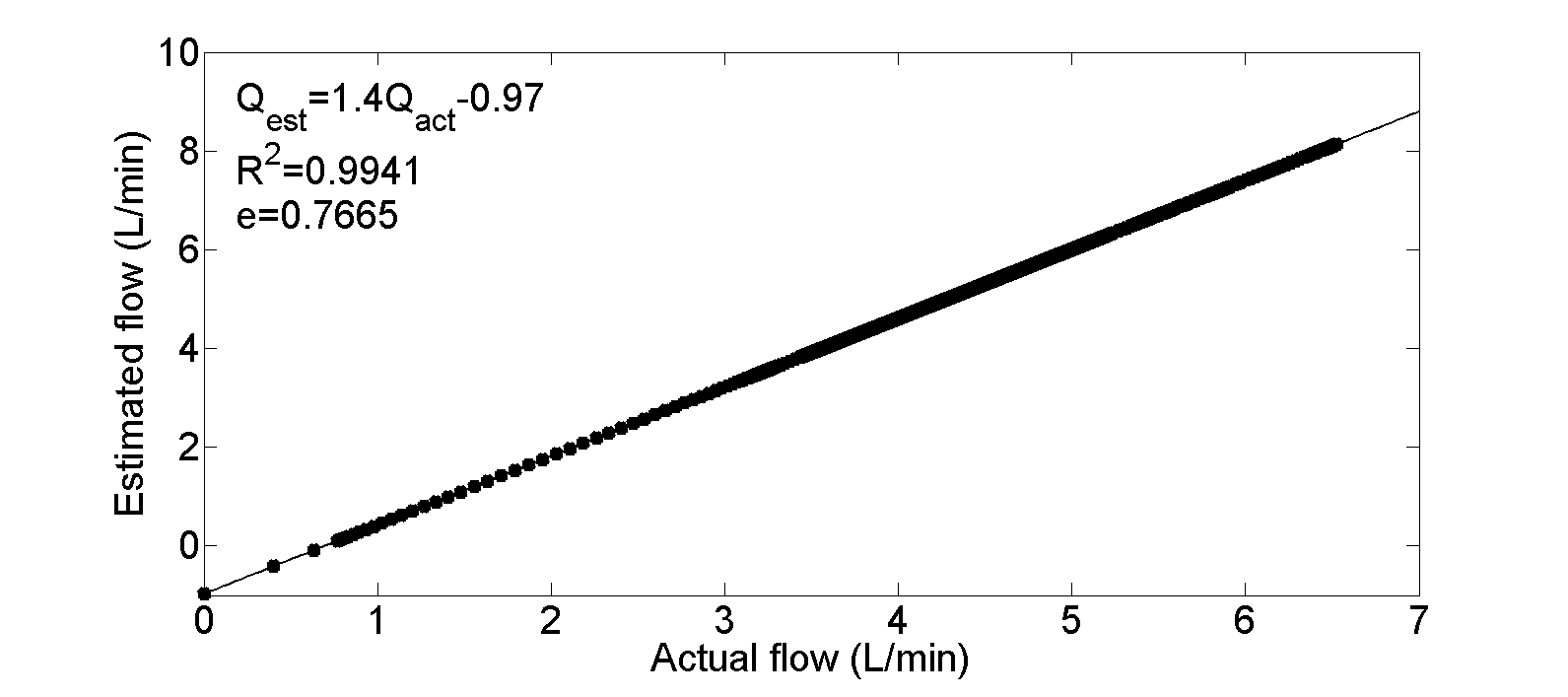}
   \label{43ei}
 }

\caption{Pump variable results at exercise condition when the system induced at 30s.}
\label{4:30eb}
\end{figure*}

\begin{figure}[htbp]
\centering
\subfigure[Average pump speed.]{
   \includegraphics[scale =0.16752] {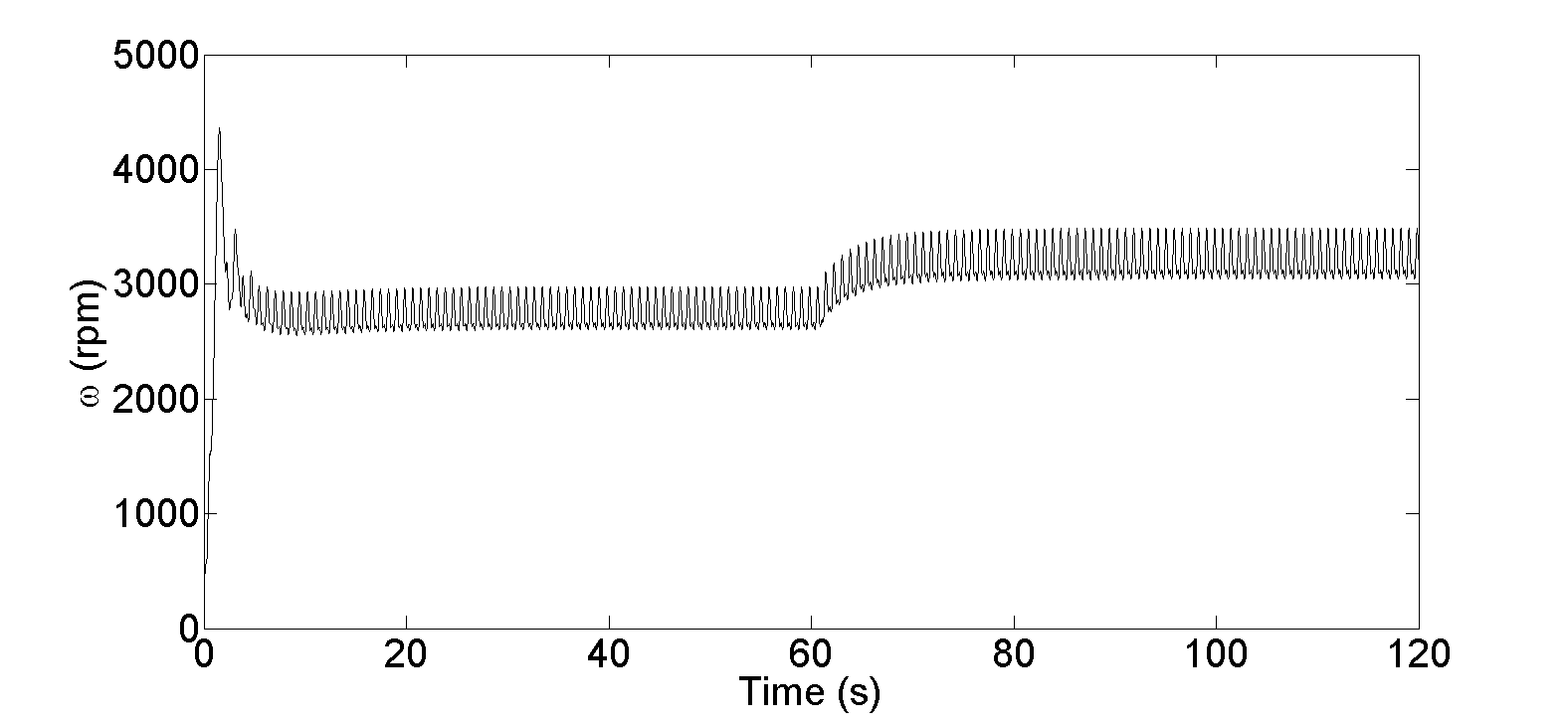}
   \label{46ef}
 }

  \subfigure[Pump flow compared with desired reference flow.]{
   \includegraphics[scale =0.16752] {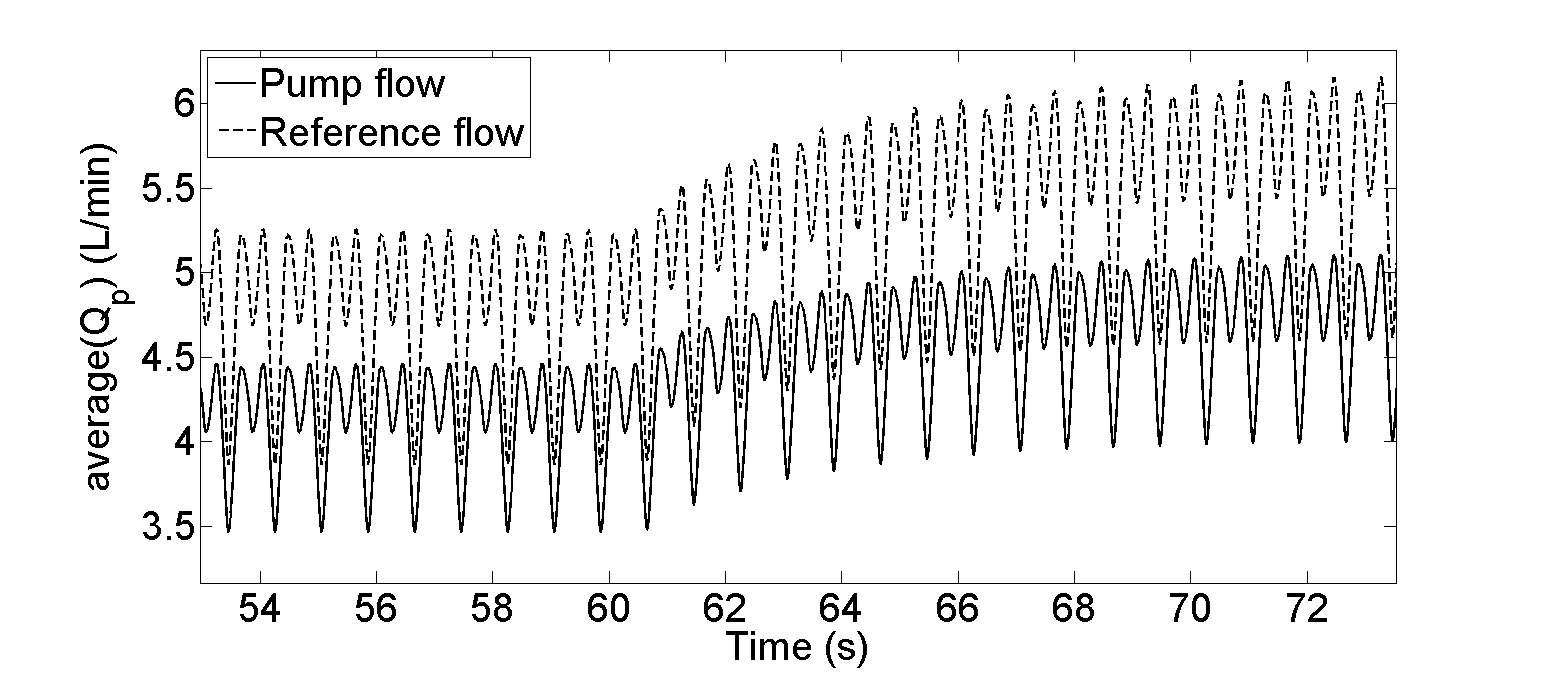}
   \label{46eh}
 }

\subfigure[Measured steady state pump flow against estimated pump flow.]{
   \includegraphics[scale =0.16752] {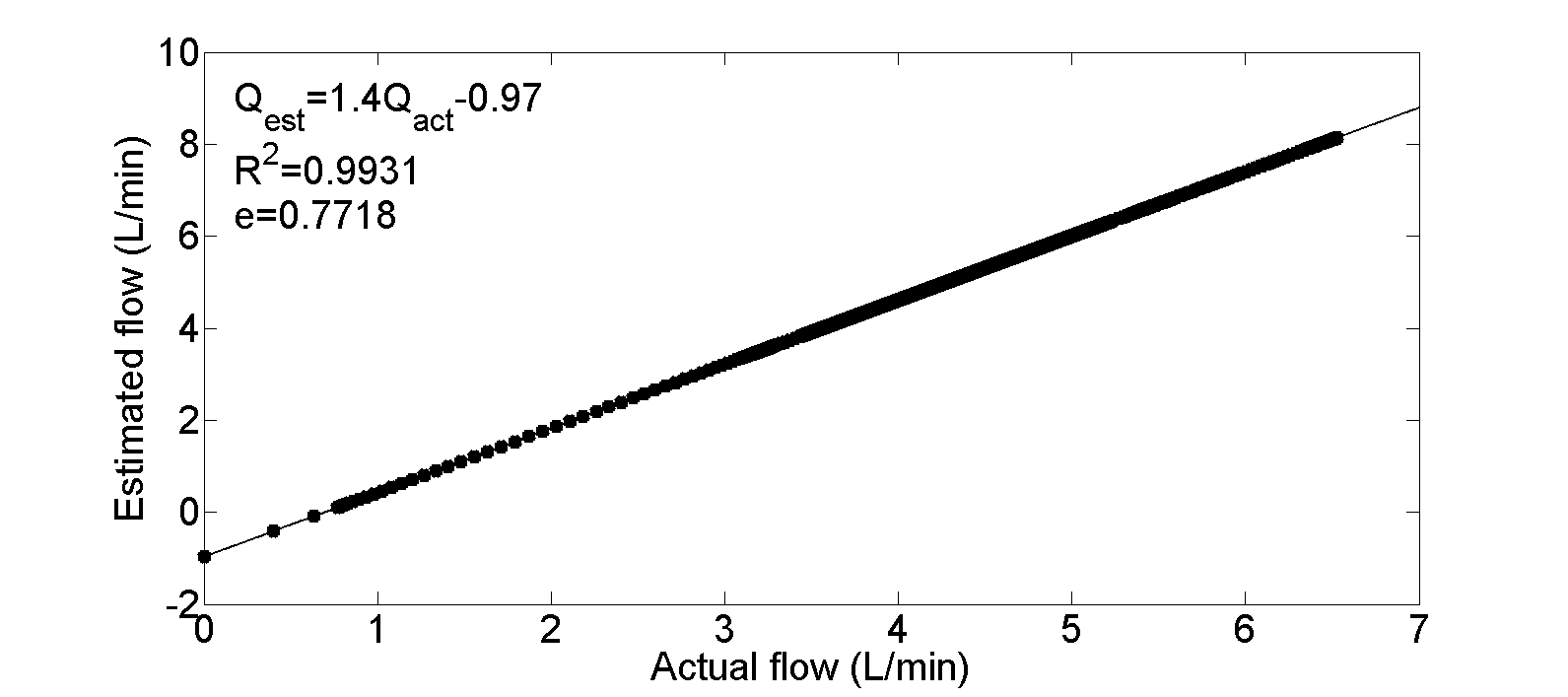}
   \label{46ei}
 }

\caption{Pump variable results at exercise condition when the  system induced at 60s.}
\label{4:60eb}
\end{figure}

\begin{figure}[htbp]
\centering
\subfigure[Average pump speed.]{
   \includegraphics[scale =0.1752] {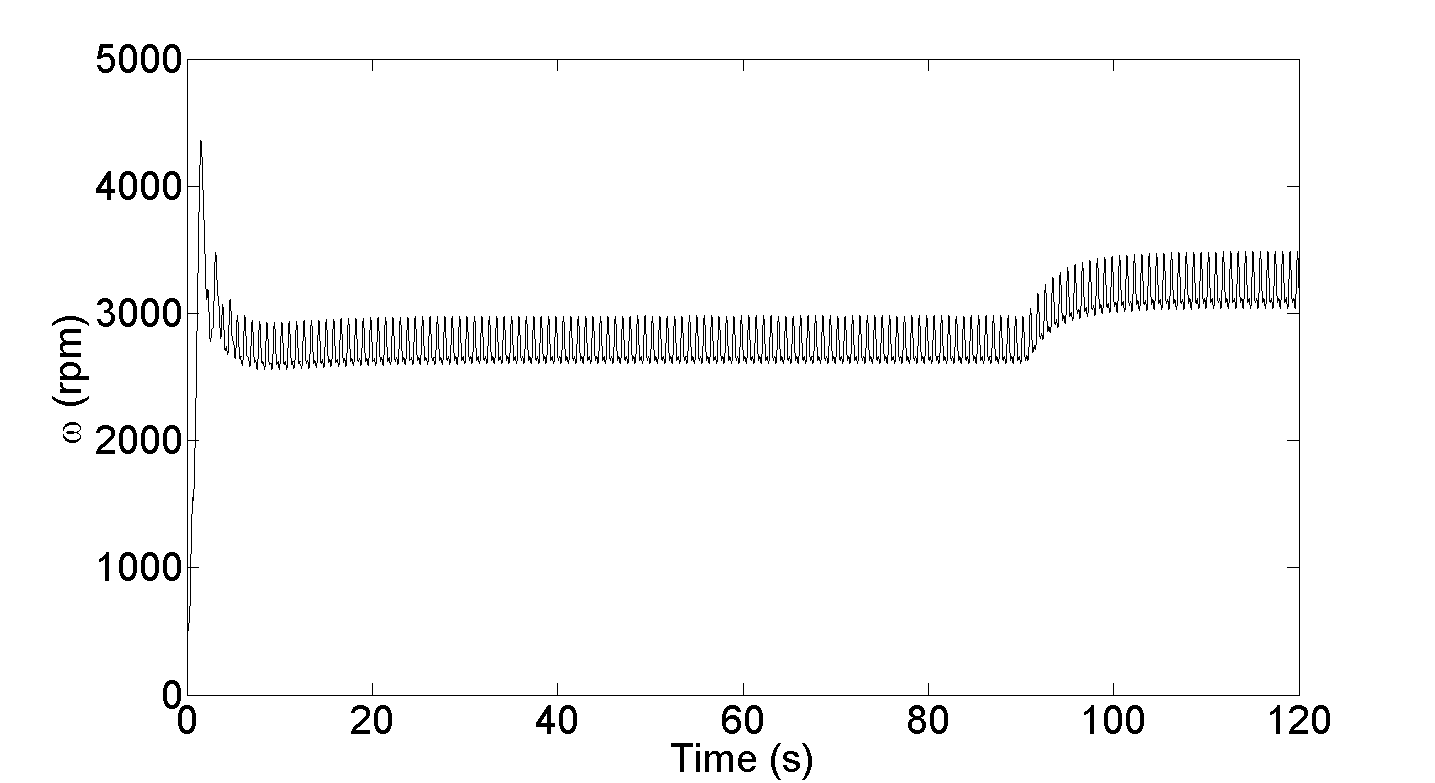}
   \label{49ef}
 }

  \subfigure[Pump flow compared with desired reference flow.]{
   \includegraphics[scale =0.1752] {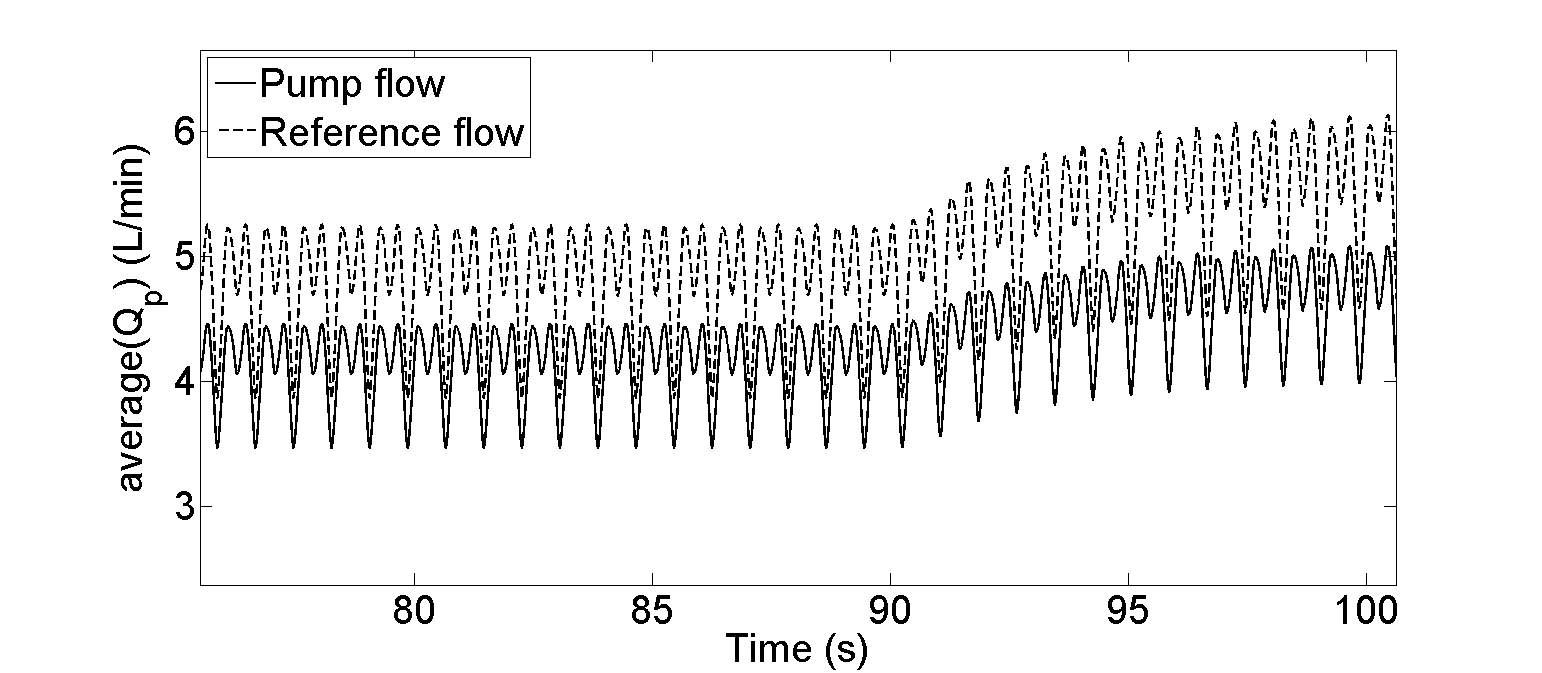}
   \label{49eh}
 }

\subfigure[Measured steady state pump flow against estimated pump flow.]{
   \includegraphics[scale =0.1752] {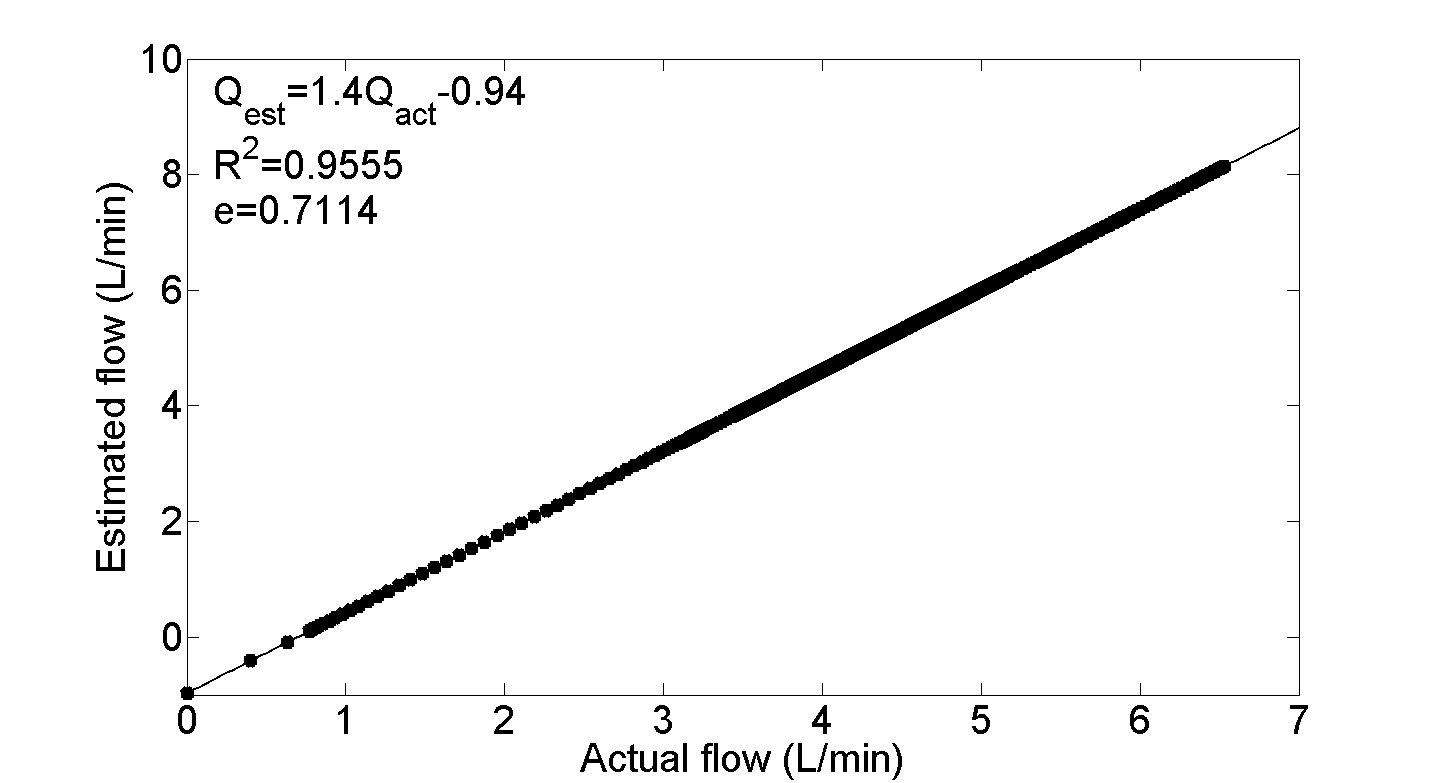}
   \label{49ei}
 }

\caption{Pump variable results at exercise condition when the system induced at 90s.}
\label{4:90eb}
\end{figure}

\begin{figure}[htbp]
\centering
\subfigure[Average pump speed.]{
   \includegraphics[scale =0.1752] {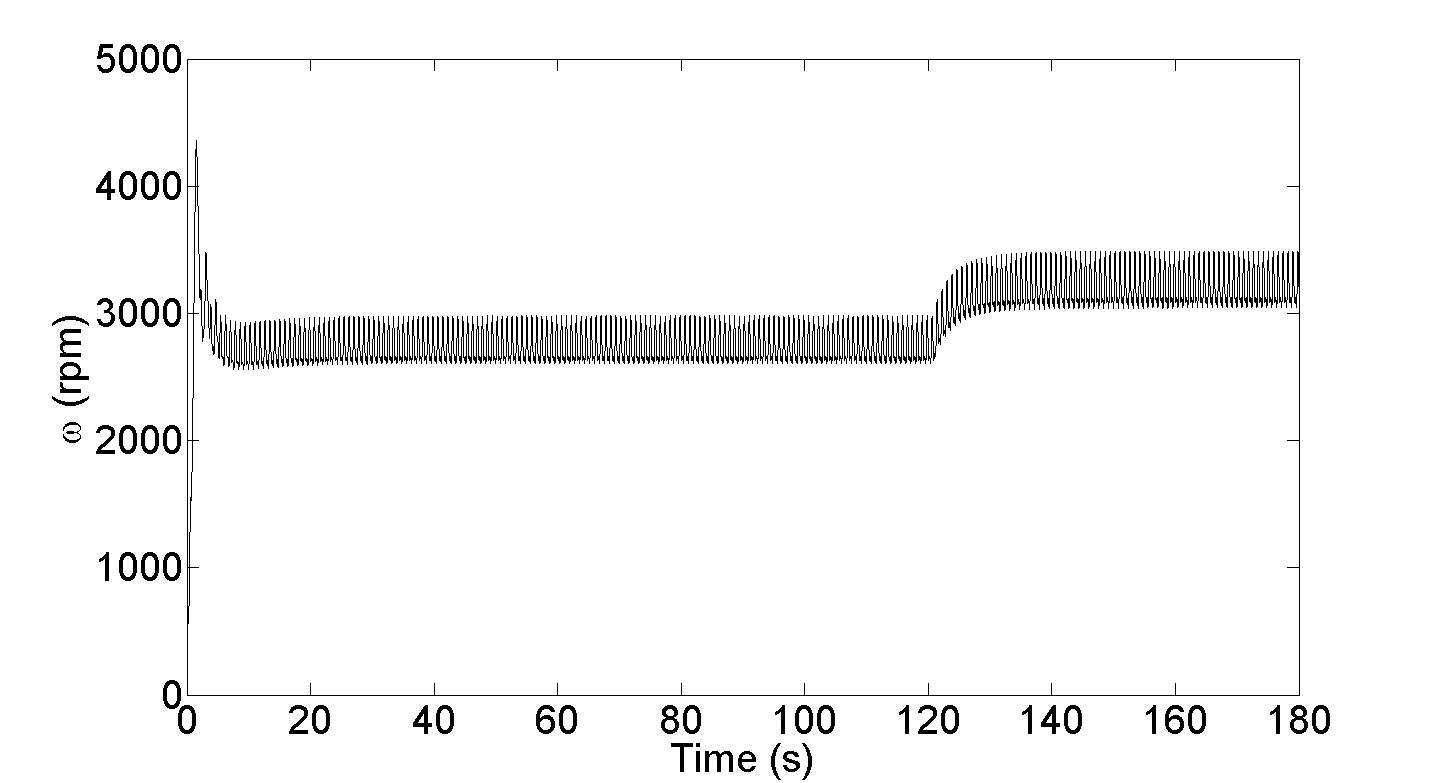}
   \label{42ef}
 }

  \subfigure[Pump flow compared with desired reference flow.]{
   \includegraphics[scale =0.1752] {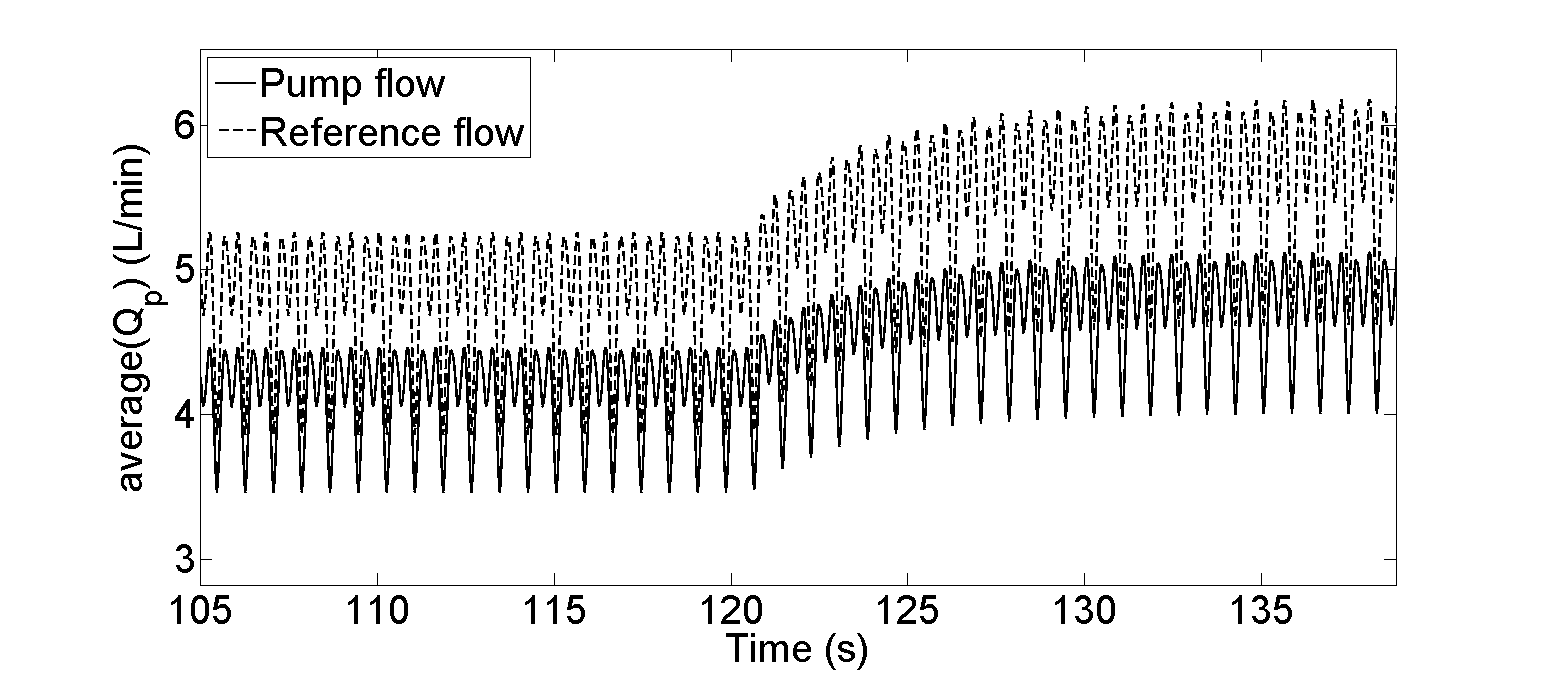}
   \label{42eh}
 }

\subfigure[Measured steady state pump flow against estimated pump flow.]{
   \includegraphics[scale =0.1752] {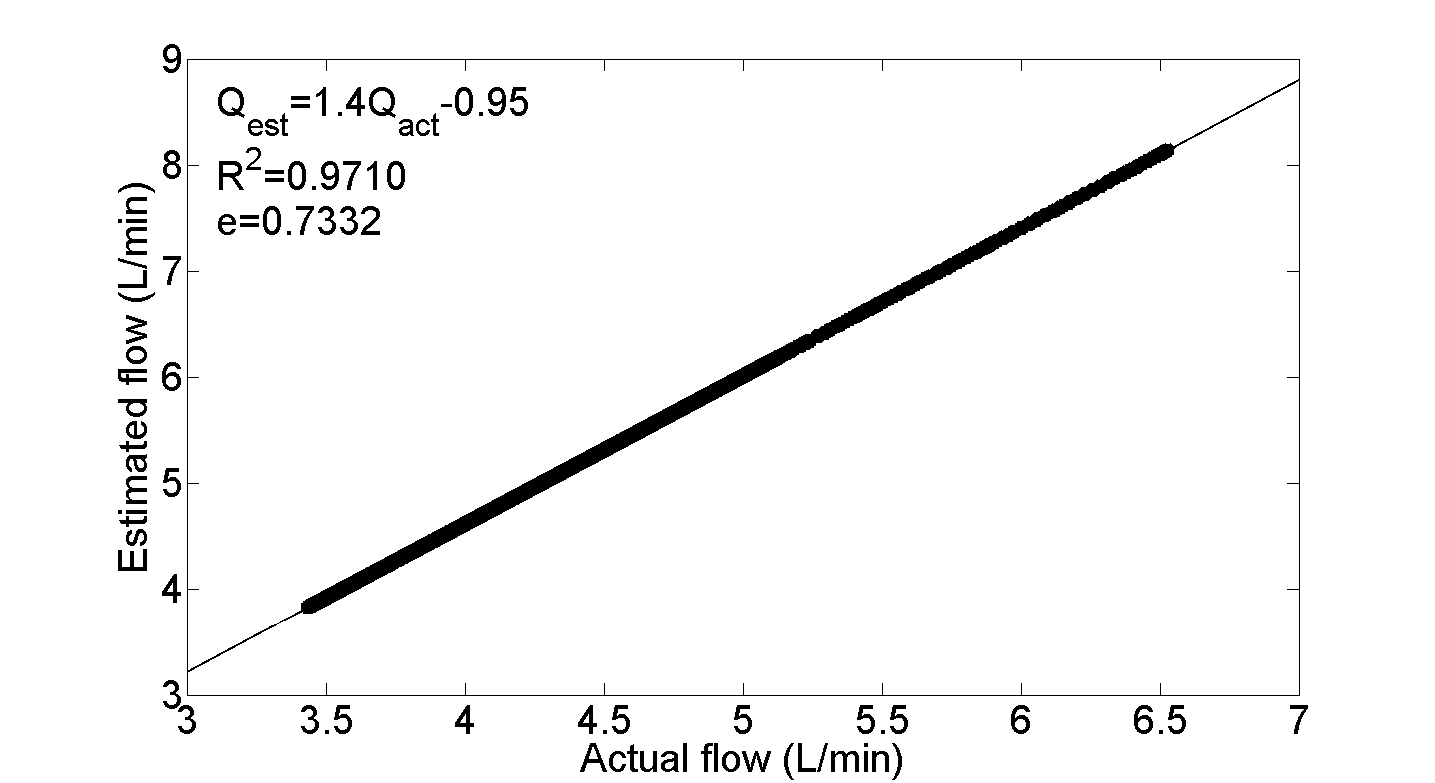}
   \label{42ei}
 }

\caption{Pump variable results at exercise condition when the system induced at 120s.}
\label{4:20eb}
\end{figure}

The waveforms of pump variables are shown in Figures (\ref{4:30eb}, \ref{4:60eb}, \ref{4:90eb} and \ref{4:20eb}). The controller responds to the increase in LV preload by increasing average pump rotational speed from 2950 rpm to 3200 rpm and actual average pulsatile flow from 4.5 L/min to 5.4 L/min. These changes have been substantially completed within four heartbeats. It can also be seen from Figures (\ref{43eh}, \ref{46eh}, \ref{49eh} and \ref{42eh}) that the simulated pump flow is accurately tracking the desired reference flow within an error of $\pm$ 0.7 L/min.  A significantly high correlation between actual and estimated flows can been seen from Figures (\ref{43ei}, \ref{46ei}, \ref{49ei} and \ref{42ei}).

Table \ref{4tab:t2}  summarises the salient hemodynamic variables, specific for the heart failure condition before and after perturbations of blood loss and exercise. While Table \ref{4t2} shows the comparison between the values of the model correlation $R^{2}$, slope $S$ and mean absolute error $e$ for each period of time.


\begin{table}[htbp]
 \caption{Associated hemodynamic variables at rest and exercise conditions.}
 \small\addtolength{\tabcolsep}{5pt}
  \label{4tab:t2}
	\begin{center}
		\begin{tabular}{ c  c  c  c  c c c}
			\hline
	     \multirow{2}{*}{Variables} & \multirow{2}{*}{Unit}       &  \multicolumn{3}{c }{Heart failure  plus LVAD} \\\cline{3-5} 	
	                            &       &   Normal     &  Rest (Blood loss)   &   Exercise  \\  \hline
	     $P_{lved}$         & mmHg      &  10.00    &  7.07       &  19.65     \\
	     $P_{rved}$         & mmHg      &   7.00    &  4.60       &  6.78      \\
	     $V_{lves}$         & L/min     &  38.00    &  47.70     &  62.36       \\
			 $V_{lved}$	        & L/min     &  140.00   &  138.10    &  149.00       \\
			 $SV$               & mL        &  100.00   &   98.70     &  96.21      \\
			 $\overline P_{ao}$ & mmHg      &  80.00    &   152.09    &  97.00     \\
			 $\overline P_{la}$ & mmHg      &   15.00   &   10.00     &  13.35       \\
			 $\overline P_{ra}$ & mmHg      &   6.00    &   9.60      &  7.35       \\
			 $\overline Q_{act}$& L/min     &   4.50    &   3.80     &   5.07        \\
			 $\overline Q_{est}$& L/min     &   4.70   &   3.50      &   6.01        \\
			\hline
		\end{tabular}
	\end{center}
\end{table}

\begin{table}[htbp]
 \caption{The model correlation (R), slope (S) and mean absolute error (e) at different instants of time.}
 \small\addtolength{\tabcolsep}{5pt}
  \label{4t2}
	\begin{center}
		\begin{tabular}{ c  c  c  c c c c}
			\hline
			\multirow {3} {*} {Time (s)} & & &\multicolumn{2}{c}{Heart failure  + LVAD} \\ \cline{3-6}	
	                  &   \multicolumn{3}{c }{Blood loss}    &    \multicolumn{3}{c }{Exercise}  \\ \cline{2-7}
										&    $R^{2}$        &  S        &   $e$ (L/min)          &   $R^{2}$ & S       & $e$ (L/min)   \\   \hline
	     30           &    0.9899         &  1.4      &  0.4767      &    0.9941      &  1.4    &  0.7665  \\
	     60           &    0.9899         &  1.4      &  0.4692      &    0.9931      &  1.4    &  0.7718   \\
	     90           &    0.9988         &  1.4      &  0.6088      &    0.9555      &  1.4    &  0.7114   \\
			 120	        &    0.9914         &  1.4      &  0.5755      &    0.9710      &  1.4    &  0.7332    \\
			 \hline
		\end{tabular}
	\end{center}
\end{table}


\section{Discussions} \label{4sec:dis}

Pump flow is an important parameter which determines total blood flow for the body. This flow can be derived with limited accuracy from sensorless indicators of metabolic demand such as heart rate and acceleration \cite{schima2006first}. However, current sensor technology is not reliable enough for long-term use. Consequently, several investigators are trying to develop controllers  free from implanted sensors. This chapter has presented a sensorless pump  flow control strategy based on a stable dynamical model of average pulsatile flow estimation \cite{alomari2011non}. In this strategy, we design a novel control algorithm based on model reference sliding mode control approach to track error between estimated and reference  flows. The primary advantage of this strategy is its quick response to sudden perturbation in the CVS and adjusting pump flow accordingly to avoid suction or regurgitation, as well as meeting the body's metabolic demand. The simulation results prove that the controller is efficient enough to track the changes in the states of the system during different physiological conditions while keeping the control input within the given constraints.

Generally, pump flow is considered as  more relevant physiological parameter \cite{smith1999system}. Furthermore, a rotary blood pump flow normally contributes completely to total cardiac output in a severe heart failure patient, therefore pump flow is important parameter which determines total blood flow to the body. Pump flow control of an IRBP has not been widely studied yet, and despite potential concerns regarding their non-physiological hemodynamics as well as the new possibilities it may offer in the field of control. Lim et al. \cite{lim2009noninvasive} has proposed and tested a deadbeat controller to control the pump flow. The results show that the controller tracks the reference input with minimal error in the presence of model uncertainty, and by using non-invasive measurements.

In view of this, a number of studies \cite{vandenberghe2003unloading, nishida2008clinical, vandenberghe2005hemodynamic} have attempted to generate pulsatile pump flow in IRBPs or during cardiopulmonary bypass settings. The most commonly used method is by varying the voltage on the motor to switch between a low and high rotor speed after achieving a desired average flow rate \cite{nishida2008clinical}. A few important artificial pulse parameters have been varied including beat rate, the minimum and maximum pump rotational speeds, the sharpness of speed changes and the systolic interval \cite{bourque2006vivo}. The main drawback of the pulsatile speed control strategy is that it offers a limited degree of adaptability to cardiac demand and pathological state of the heart. Since a centrifugal pump is highly afterload dependent, the resultant pump flow does not only depend on pump rotational speed but varies substantially with changing cardiovascular parameters.

The design of an automatic, robust and efficient control system to control the pump flow effectively according to the body�s physiological needs and perturbations and adapting for preload is still under preliminary phases. Mostly the available IRBPs are currently operating at a fixed target speed determined by the physician based on the activity level of the patient. However, fixed speed is not physiological since it effectively decouples the pump from the cardiovascular system. A number of control strategies for IRBPs are available in the literature  \cite{choi2007hemodynamic, wu2009adaptive} including both modern techniques and traditional control strategies such as PID and fuzzy control. In general, LVAD devices require precise control and traditional control algorithms demonstrate a limited degree of adaptability to cardiac demand and clinical conditions of the heart  \cite{smith1999system}\cite{giridharan2003control}. The modern control strategies such as SMC  have proven to be a robust approach in various applications since its first development in the 1950's \cite{utkin1977variable}. The SMC has the potential to solve the difficult problems applying physiological control to LVADs. More  recently, SMC methods for discrete-time systems have received a lot of attention \cite{furuta1990sliding, sarpturk1987stability,spurgeon1992hyperplane} because the SMC retains the robustness/insensitivity of the control system to bounded model uncertainties, parameter variations and external disturbances that has led intense interest in biological systems.

Although the proposed control algorithm was shown to be able to respond quickly enough to sudden perturbations in the cardiovascular system accordingly without inducing suction, but still there are some important issues should be highlighted. Regulating of constant reference flow, sometimes is not provided physiological demand. Therefore, adapting the reference flow according to metabolic demand is needed (will discussed in next chapter). In addition, transient overshoot was observed at first two seconds in each scenario on Figs.\ref{43f}, \ref{46f}, \ref{49f} and \ref{42f} at blood loss scenario and Figs. \ref{43ef}, \ref{46ef}, \ref{49ef} and \ref{42ef} at exercise scenario. Despite, the overshoot is increases the average pulsatile flow up to 7.2 L/min, overperfusion is not occurred with this level.

\section{Conclusion} \label{4sec:con}

A robust control strategy based on average pulsatile flow estimation has been presented for LVADs using the sensorless measurements. A robust  SMC approach is proposed to drive this strategy effectively. The uncertainties are made bounded within the given  upper and lower  limits. The proposed robust controller has been evaluated using a lumped-parameter model of the cardiovascular system. It  has been proved that the controller ensures the system states are driven towards the proposed sliding surface and remain on it thereafter ensuring the stability of closed-loop system. Simulation results  demonstrate that the control input drives the tracking error close to zero in the presence of the bounded disturbance, and a stable transient response has been achieved.

\chapter{Sensorless Physiological Control Algorithm of Implantable Rotary Blood  Pumps Using Feedforward Sliding Mode Control }

\section{Overview}

This Chapter presents the design of a physiological controller using feedforward - sliding mode control (FFSMC) methods for LVADs to maintain a motivated perfusion. The purpose of the proposed control strategy is to adjust the pump speed automatically to cater for changes in metabolic demand of LVADs. We propose an innovative approach combining a feedforward and a dynamic sliding mode control to determine the physiologically optimum pump flow and calculate the appropriate pump speed to achieve it. Our computer simulations are based on two physiological conditions ranging from rest to exercise. The performance of the proposed control strategy is assessed using a previously developed lumped parameter model of the cardiovascular system based on the experimental data of healthy pigs. It is shown that the developed adaptive controller properly tracks the reference signal in presence of model uncertainties and external disturbances for LVADs. It is also shown that the abnormal hemodynamic variables of LVADs are restored back to normal physiological range.

\section{Control Strategy}
\begin{figure}[!ht]
\centering
\includegraphics[scale =0.5]{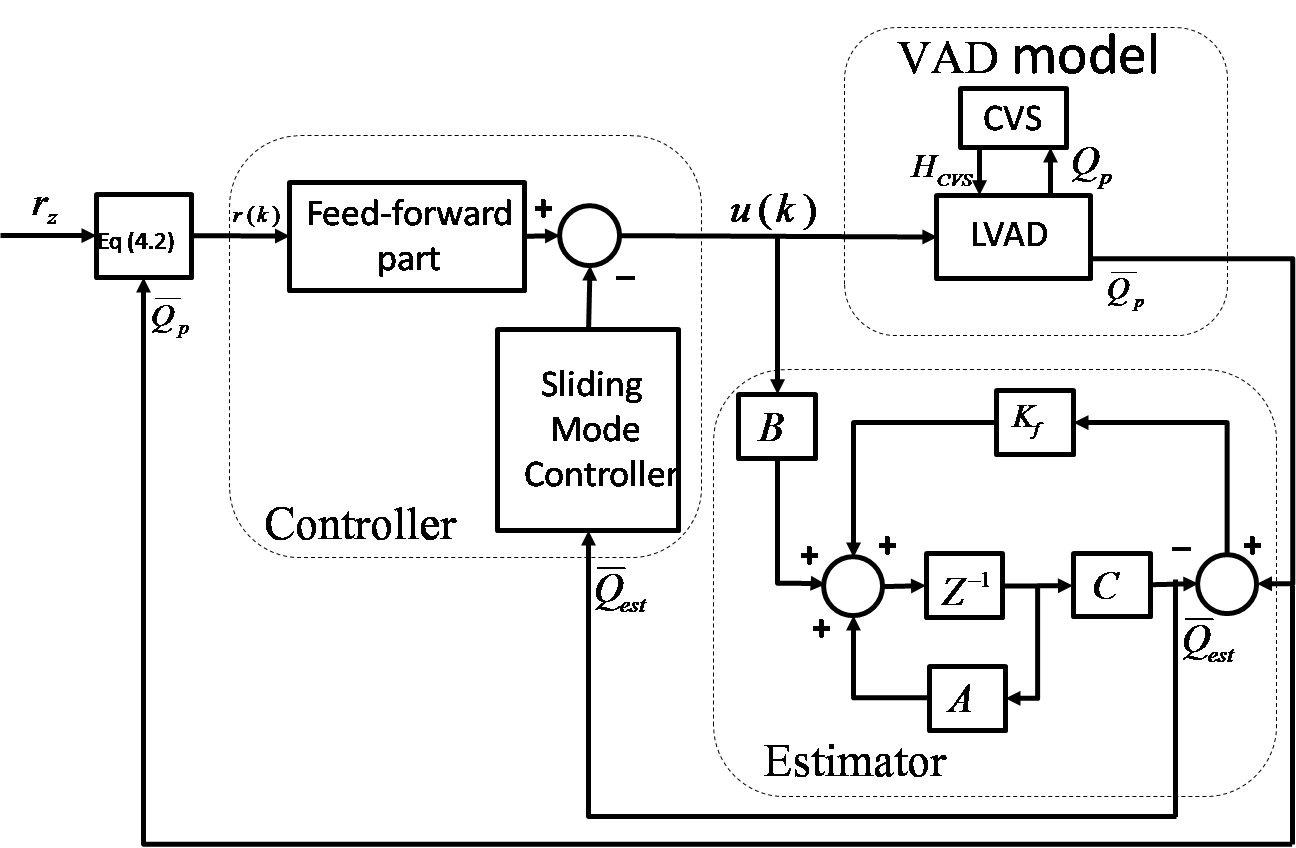}
\caption{Block diagram of the control system.}
\label{5fig:1}
\end{figure}

Consider the block diagram of the control system as shown in Fig. \ref{5fig:1}. In this method, it is assumed that while the aortic valve is closed, the total cardiac output can be approximated as pump flow. Hence the output of an LVAD is used as a part to design a non-linear function that is generating the dynamical reference signal to adequate the motivated perfusion of the body. Therefore, the designed controller is used to regulate estimated average pulsatile flow ($\overline Q_{est}$) and dynamical reference input signal ($r(k)$) considered as pump flow. The input reference signal to the system has been selected arbitrarily to be $r_{z}\equiv 5 L/min$ at $t=0$ in order to  initiate the LVAD controller. The system is derived with feedforward - SMC controller, where the SMC is implemented as feedback part. In  general, the only implementation of SMC as feedback controller in order to track the  reference signal may introduce some phase lag, which is detrimental to the system states. The design method of feedforward part, as proposed here, is an appropriate technique to compensate this phase lag in a tracking control system with minimum error within a minimum possible sampling period. The input to the feedforward part is the reference signal and the input to the SMC is the estimated pump flow, while the output of the controller is PWM voltage signal to the rotary pump, represented as $u(k)$ in the system.

\section{Reference Signal}

An ideal IRBP should mimic the Frank-Starling mechanism of the heart where stroke volume is adjusted according to the ventricular end diastolic volume such that the heart ejects all the blood volume it receives from the circulation \cite{guytontextbook}. The parameter of the LV end diastolic volume is known as preload and considered as an indicator for the physiological need of a human body. The value of this parameter should be maintained within physiological range for a healthy person. A few essential criteria have been suggested for safe operation of an IRBP \cite{boston1998intelligent,wu2007modeling} including:

\begin{itemize}
	\item sufficient cardiac output to meet metabolic requirements,
	\item maintaining systolic arterial pressure within a physiological range to maintain sufficient liver or kidney perfusion and to avoid over perfusion or under perfusion,
	\item maintaining left atrial pressure within a normal physiological range to avoid pulmonary congestion and suction.
\end{itemize}
In addition, it is advisable to maintain positive  LV outflow for a portion of the cardiac cycle (during systole) and to avoid regurgitation blood back from the aorta to the left ventricle in diastole. Due to this reason, we assume that the aortic valve is totally closed and the cardiac output represented as pump flow. Therefore, if the pump flow is lower than the body's physiological need, the reference signal should be increased to maintain the pump flow. However, if the blood flow is higher than the body's physiological need, the reference signal needs to be decreased to lower the pump flow. Therefore, any type of control algorithm can be implemented to modify the reference signal according to the above principle  where required.

In this method, it has been proposed that the reference signal $r(k)$ should be regulated according to the following equation:
\begin{equation}
r(k)=r_{z}+f(\Delta Q),
\label{5eq:1}
\end{equation}
where $r_{z}$ is a preset constant equal to the normal value of pump flow, i.e., 5 L/min at $t=0$ and $f(\Delta Q)$ is a function defined as:

\begin{equation}
	f(\Delta Q) = \left\{ \begin{array}{cl}
	0 &  \textrm{$a\leq 0, b\leq 0$}\\
	\lambda(\overline Q_{p})-(a+b)   &  \textrm{$a>0, b>0$},
	\end{array}\right.
\label{5eq:2}\end{equation}
where $\overline Q_{p}$ is the average pulsatile flow, $\lambda >0$, $ a > 0$ and $b > 0$ are three designed parameters.

Simply, the function $f$ in (\ref{5eq:1}) is a typical dead zone function to prevent any suction and over perfusion.

\section{Methods}

\subsection{Software Model of Cardiovascular System}
The proposed controller has been evaluated with a software model  incorporating a model of the human circulatory  system with a stable dynamical model of an LVAD. As reported in \cite{lim2010parameter}, the model of the CVS consists of an arbitrary number of lumped parameter blocks; each characterised by its own resistance, compliance, pressure and volume of blood. In its simplest configuration, the CVS has ten  compartments including the right and left sides of the heart as well as the pulmonary and systemic circulations. 

\subsection{Simulation Protocols}

The responses of the controller as described above have been studied through a series of computer simulations. The computer simulation test is implemented in MATLAB/ Simulink  using its inbuilt Ordinary Differential Equation (ODE) solver suite. In this study, model parameters as reported in \cite{lim2010parameter} corresponding to health condition are used as baseline values for a normal subject. In order to evaluate the performance of the controller to track the reference signal, we have simulated two different scenarios by varying the values of matrix linearly for systemic veins unstressed volume (preload) ($V_{0,sv}$), systemic peripheral resistance afterload ($R_{sa}$), left and RV contractility ($E_{es,lv}$, $E_{es,rv}$).

Firstly, we decreased the simulated ($V_{0,sv}$ at the rest condition (blood loss) linearly by 500 mL and $r_{z}$ linearly from 5 to 2.5.) to  check  if the LVAD can provide an essential support to the HF patient under a minimal condition. In the second scenario, the system parameters have been  varied to simulate the activities from normal to exercise. These variations  include: ($R_{sa}$ decreased by $50\%$), ($E_{es,lv}$ , $E_{es,rv}$ increased by $20\%$), ($V_{0,sv}$ increased by 500 mL and the $r_{z}$ was linearly increased from 5 to 5.5). In fact this test has been conducted to validate the designed physiological controller with the LVAD. It has been observed that the designed controller is capable  of providing the fundamental support during normal daily life. In all simulations, the sinusoidal signal frequencies  are  selected equal to the heart rate.

It is important to note that the rest condition test is conducted to evaluate if LVAD can provide an essential support to the HF patient under minimal condition. And the controller is tested under varying conditions like from rest to exercise in order to verify that the designed physiological controller provides fundamental support to LVAD accommodating CVS during a normal daily life. In all simulations, the design parameters of the sliding surface in   (\ref{5eq:17})  are  $\boldsymbol{\Gamma}=[0.9413 ~~-0.0805]$ and those of the control law in the same equation are $\tau T=0.03$ and $\sigma T=0.05$. The model parameters have been changed linearly over four different period of time, which are 30s, 60s, 90s and 120s respectively. These periods of times have been chosen to verify the system stability at different  time instants. For each period, the system has been induced at the middle and the simulations are continued for other half of the same period to allow the system to reach to a steady state corresponding to new parameter values.

\subsection{Implementation of Control Algorithm}
An empirical sensorless stable dynamical model of an LVAD to estimate $\overline Q_{est}$ has been developed and validated based on actual animal experimental data (see Chapter \ref{ch2}). This model can be described by following state space equation as:

\begin{equation} \begin{array}{rcl}
q(k+1) & = & Aq(k)+\delta Aq(k)+Bu(k)+\zeta(k) \\
    y(k) &=& Cq(k),     \label{5eq:3}
\end{array} \end{equation}
where, $q(k)=$
$\begin{bmatrix}
q_{1}(k) & q_{2}(k) \\
\end{bmatrix}^{T}$
are the system states, $q_{1}(k)$ is the $\overline {PI} _{\omega}$, $q_{2}(k)$ is the $\overline Q_{est}$, $u(k)$ is the $\overline {PWM} $, $\delta A$ is system parameter variations, $\zeta(k)$ is the system noise, $y(k)$ is the system output, $A , B$ and $C$ are the compatibly dimensioned matrices.

The design methodology of the control algorithm consists of the following parts:

\subsubsection{Part 1: Sliding Model Controller Design}

In general, different control techniques with multiple reaching laws have been introduced in SMC \cite{bartolini1995adaptive, chan1997discrete, chen1998robust, bartoszewicz1998discrete, perruquetti2002sliding, bartoszewicz2007sliding, xia2010robust}. One of them is known as the Gao reaching law method firstly proposed by Gao et al. \cite{gao1995discrete}. The reaching law that describes the ideal conditions to guarantee strong reachability and ideal sliding motion is used to implement the control algorithm.

The Gao's reaching law is given as:
\begin{equation}
\boldsymbol{\eta}(k+1)=(1-\tau T)\boldsymbol{\eta}(k)-\sigma T \text{sign}(\boldsymbol{\eta}(k)),   \\
\label{5eq:5} \end{equation}
where, $T>0$ is the sampling period,
$\sigma>0$,
$\tau \geq0$
such that $0<(1-\tau T)<1$.

\begin{figure}[htbp]
\centering
\includegraphics[width=4in]{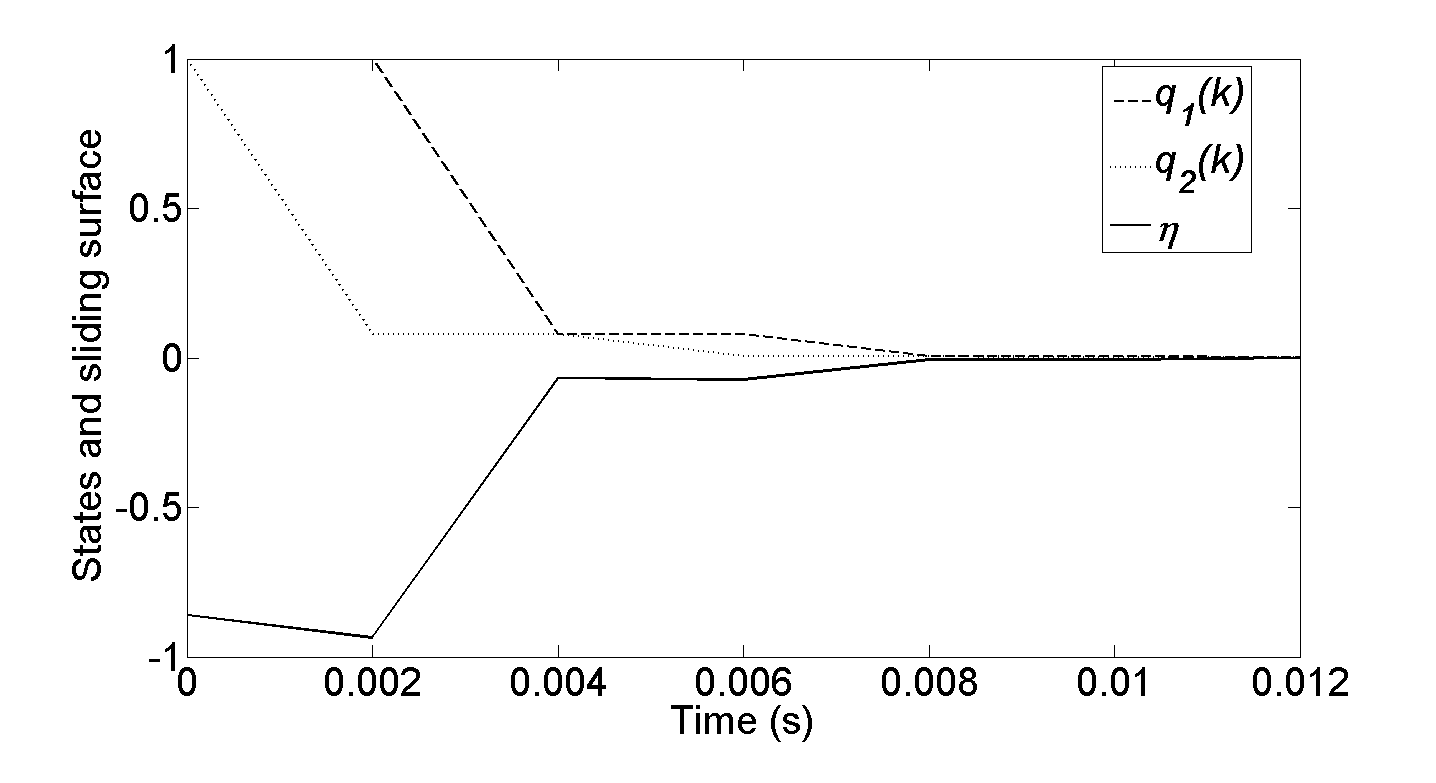}
\caption{System response.}
\label{fig:ss1}
\end{figure}

A common sliding surface is chosen as:
\begin{equation}
\boldsymbol{\eta}(k)=\boldsymbol{\Gamma} q(k)  \label{5eq:6} \\
\end{equation}
where $\boldsymbol{\Gamma}$ is a constant vector designed based on robust pole placement method to guarantee that $q(k)$ is asymptotically stable on $\eta(k)=0$  \cite{spurgeon1992hyperplane, edwards1998sliding}. Fig. \ref{fig:ss1} shows the performance of the state variables $q_{1}(k)$ and $q_{2}(k)$ with convergence to the sliding surface. 

To satisfy the reaching law in   (\ref{5eq:5}), we  have:

\begin{equation}
\boldsymbol{\eta}(k+1)=\boldsymbol{\Gamma} q(k+1),   \label{5eq:7}  \\
\end{equation}
from  (\ref{5eq:3}) and  (\ref{5eq:7}) we get:

\begin{equation}
\boldsymbol{\eta}(k+1)=\boldsymbol{\Gamma} q(k+1)=\boldsymbol{\Gamma} (Aq(k)+\delta Aq(k)+Bu(k)+\zeta(k)) \label{5eq:8}.  \\
\end{equation}
The equalisation of (\ref{5eq:5}) and (\ref{5eq:8}) gives:

\begin{equation} \begin{array}{cl}
&\boldsymbol{\Gamma} Aq(k)+\boldsymbol{\Gamma}\delta Aq(k)+\boldsymbol{\Gamma} Bu(k)+\boldsymbol{\Gamma}\zeta(k) \\
                  =&(1-\tau T)\boldsymbol{\eta}(k)-\sigma T\text{sign}(\boldsymbol{\eta}(k)).  \label{5eq:10}  \\
\end{array} \end{equation}
Solving the above equation for the command signal $u(k)$ yields:
\begin{equation} \begin{array}{rl}
u(k) = & -(\boldsymbol{\Gamma} B)^{-1}(\boldsymbol{\Gamma} Aq(k)-(\tau T-1)\boldsymbol{\Gamma} q(k) \\
       &+\sigma T\text{sign}(\eta(k)))-(\boldsymbol{\Gamma} B)^{-1}(\boldsymbol{\Gamma} \delta Aq(k)+\boldsymbol{\Gamma} \zeta(k)) \\
        \label{5eq:11}  \\
\end{array} \end{equation}

As $\delta A$ and $\zeta(k)$ are unknown, the control algorithm cannot be implemented unless we assume the upper and lower bounds of $(\boldsymbol{\Gamma} \delta Aq(k)+\boldsymbol{\Gamma} \zeta(k))$  as:
\begin{equation}
-\beta<(\boldsymbol{\Gamma} \delta Aq(k)+\boldsymbol{\Gamma} \zeta(k))<\beta. \label{5eq:111}  \\
\end{equation}
Hence the final control algorithm can be re-written as:
\begin{equation} \begin{array}{rl}
u(k) = & -(\boldsymbol{\Gamma} B)^{-1}(\boldsymbol{\Gamma} Aq(k)-(\tau T-1)\boldsymbol{\Gamma} q(k) \\
       &+(\sigma T+\beta)\text{sign}(\boldsymbol{\Gamma} q(k))), \\
        \label{5eq:17}  \\
\end{array}
\end{equation}
where
\begin{equation}
(-\beta<(\boldsymbol{\Gamma} \delta Aq(k)+\boldsymbol{\Gamma} \zeta(k))<\beta)\equiv (\beta \text{sign}(\boldsymbol{\Gamma} q(k))), \label{5eq:112}  \\
\end{equation}

\subsubsection{Part 2: Feedforward Design}
In order to achieve robust output tracking, it is necessary to design a feedforward part based on closed-loop system. Wang et al. \cite{wang2003robust} stated that SMC in the closed-loop structure is independent of the system  dynamics when the system is on the sliding surface. Therefore, the feedforward part can be implemented based on the inverse of a closed loop transfer function of the system as in \cite{tomizuka1987zero}.

The closed-loop system can be obtained by substituting (\ref{5eq:17}) into (\ref{5eq:3}) as:
  \begin{equation} \begin{array}{rl}
  q(k+1)= & Aq(k)+B(-(\boldsymbol{\Gamma} B)^{-1}(\boldsymbol{\Gamma} Aq(k)\\
          & -(1- \tau T)\boldsymbol{\Gamma} x(k)+(\sigma T +\beta)\text{sign}(\boldsymbol{\Gamma} q(k)))),  \label{5eq:18}  \\
  \end{array} \end{equation}
 by simplification the above equation we can get:
\begin{equation} \begin{array}{rl}
  q(k+1)= & (A-B(\boldsymbol{\Gamma} B)^{-1}\boldsymbol{\Gamma} A-B(\boldsymbol{\Gamma} B)^{-1}(\tau T-1)\boldsymbol{\Gamma})q(k)\\
          & -B(\boldsymbol{\Gamma} B)^{-1}(\sigma T+\beta)\text{sign}(\boldsymbol{\Gamma} q(k))),  \label{5eq:19}  \\
  \end{array} \end{equation}
from  (\ref{5eq:19}), if we define $\Omega$ as:

\begin{equation} \begin{array}{rl}
  \Omega= A-B(\boldsymbol{\Gamma} B)^{-1}\boldsymbol{\Gamma} A-B(\boldsymbol{\Gamma} B)^{-1}(1- \tau T)\boldsymbol{\Gamma},
\end{array} \end{equation} \label{5eq:20}
It can be noticed that the closed loop system in (\ref{5eq:18}) is an autonomous non-linear system and the state dynamics of the system are calculated using system matrix $\Omega$.
In order to ensure the stability of the system, all eigenvalues of $\Omega$ must lie inside the unit circle and these eigenvalues are calculated by the choice of the switching function $\boldsymbol{\Gamma}$ in (\ref{5eq:7}) and the time factor in the reaching law (\ref{5eq:5}). The feedforward part is implemented based on the inverse of a closed loop transfer function of the system \cite{tomizuka1987zero}. The feedforward path is then used as a pre-filter for the reference input signal to compensate for the phase lag of the feedback system and consequently to achieve a good tracking \cite{van1996accurate}. To achieve this goal the reference signal  should be known in advance for at least a few sampling intervals. It means that the inverse of the model is non-causal. Therefore, the design method can be completed by converting the system model in (\ref{5eq:3}) into controllable canonical form in state space. After that the discontinuous part of the plant states will change its sign at each successive step in QSM band such that  the average control action during QSM band is a simple state feedback controller which is in fact an  average closed loop transfer function.

From (\ref{5eq:19}) the closed loop transfer function can be written as:

\begin{equation}
\Lambda(z)=C(zI-\Omega)^{-1}B, \label{5eq:21}
\end{equation}
The direct inverse of  (\ref{5eq:21}) will result in a non-causal feedforward part $\Lambda(z)^{-1}$. By assuming that the reference signal $r(k)$ is known in advance, the implementation of a non-causal is then possible.

\subsubsection{Part 3: Kalman Filter Design}

In our problem the whole state $q(k)$ is not available to our controller, so we need to estimate $q(k)$ based on the measured output $y(k)$. For this purpose, we use the steady-state Kalman estimator \cite{welch1995introduction, simon2006optimal} as follows:

\begin{equation} \begin{array}{rcl}
\hat {q}(k+1) & = & A\hat {q}(k)+Bu(k)+K_{f}(y(k)-C\hat{q}(k)) \\
    \hat{y}(k) &=& C\hat{q}(k),     \label{5eq:23}
\end{array} \end{equation}
where $\hat{q}(k)$ is the estimated of the state $q(k)$ and $K_{f}$ is the  "optimal Kalman gain" given as:

\begin{equation}
K_{f}=PC^{T}R^{-1}
\label{5eq:24}
\end{equation}
and $P$ is the solution of the following algebraic Riccati equation given as:

 \begin{equation} \begin{array}{rcl}
AP+PA^{T}-PC^{T}R^{-1}CP+Q=0, \\
       \label{5eq:25}
\end{array} \end{equation}

\section{Simulation Results}

\subsection{Results in Rest Condition (Blood loss)}
\begin{figure*}[htbp]
\centering
\subfigure[LV volume versus LV pressure before and after Parameter Change.]{
   \includegraphics[scale =0.16752] {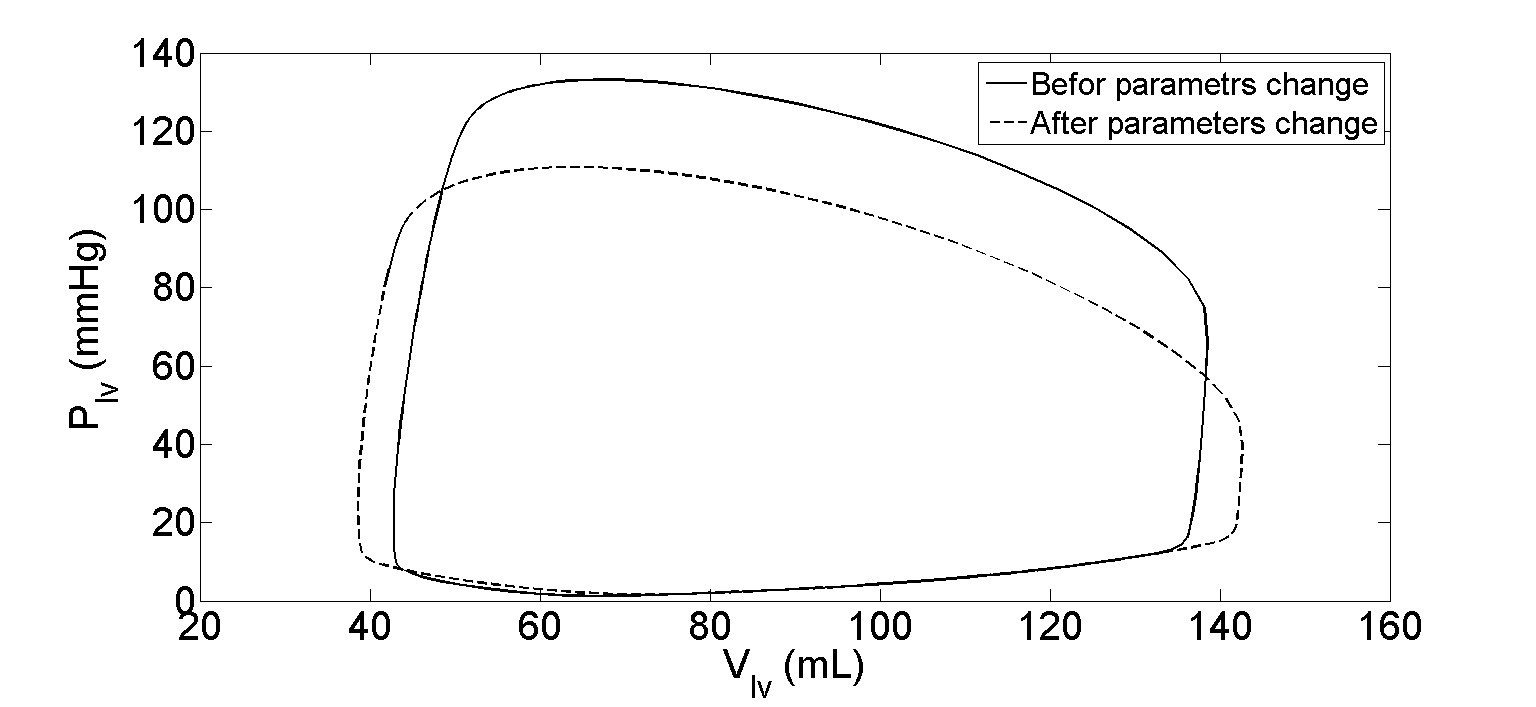}
   \label{53a}
 }
\subfigure[RV volume versus RV pressure before and after Parameter Change.]{
   \includegraphics[scale =0.16752] {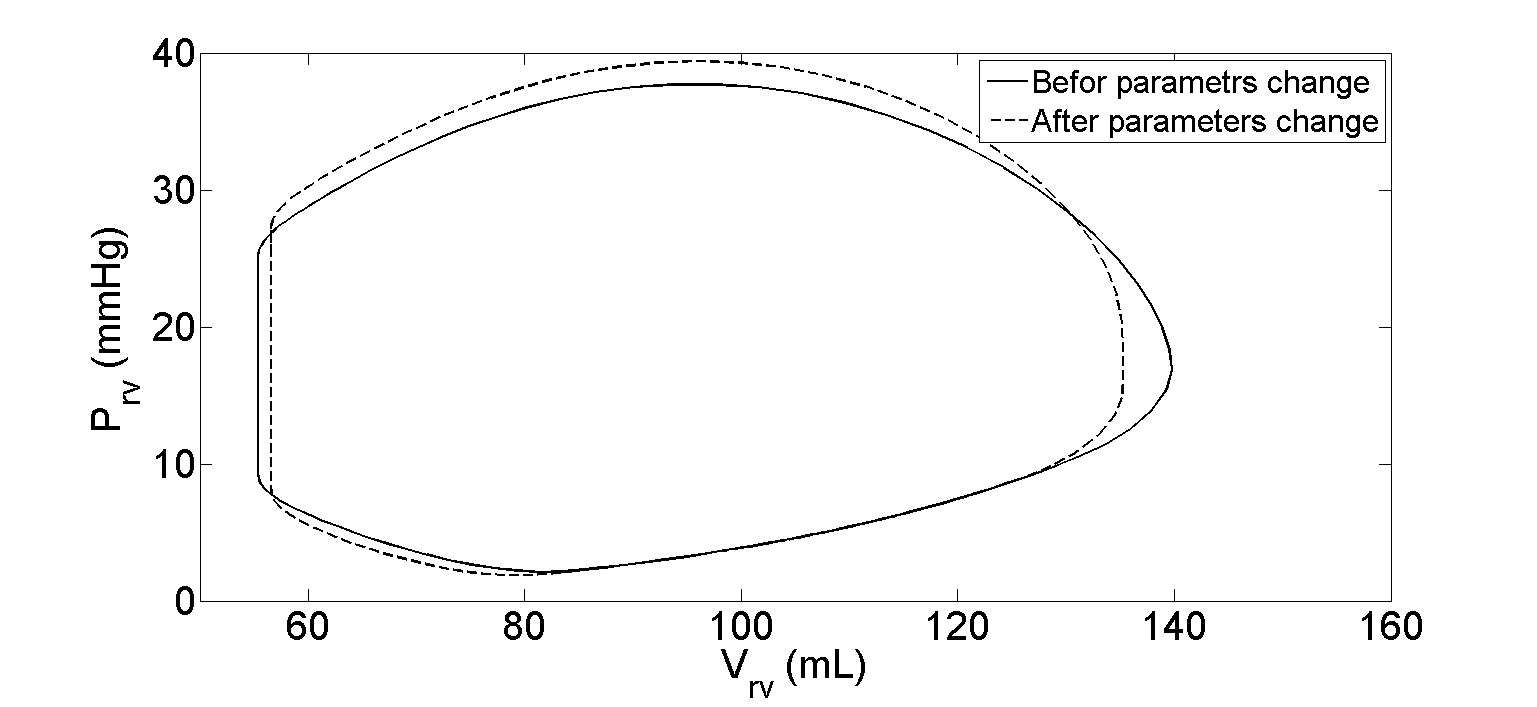}
   \label{53b}
 }

 \subfigure[Aortic pressure.]{
   \includegraphics[scale =0.16752] {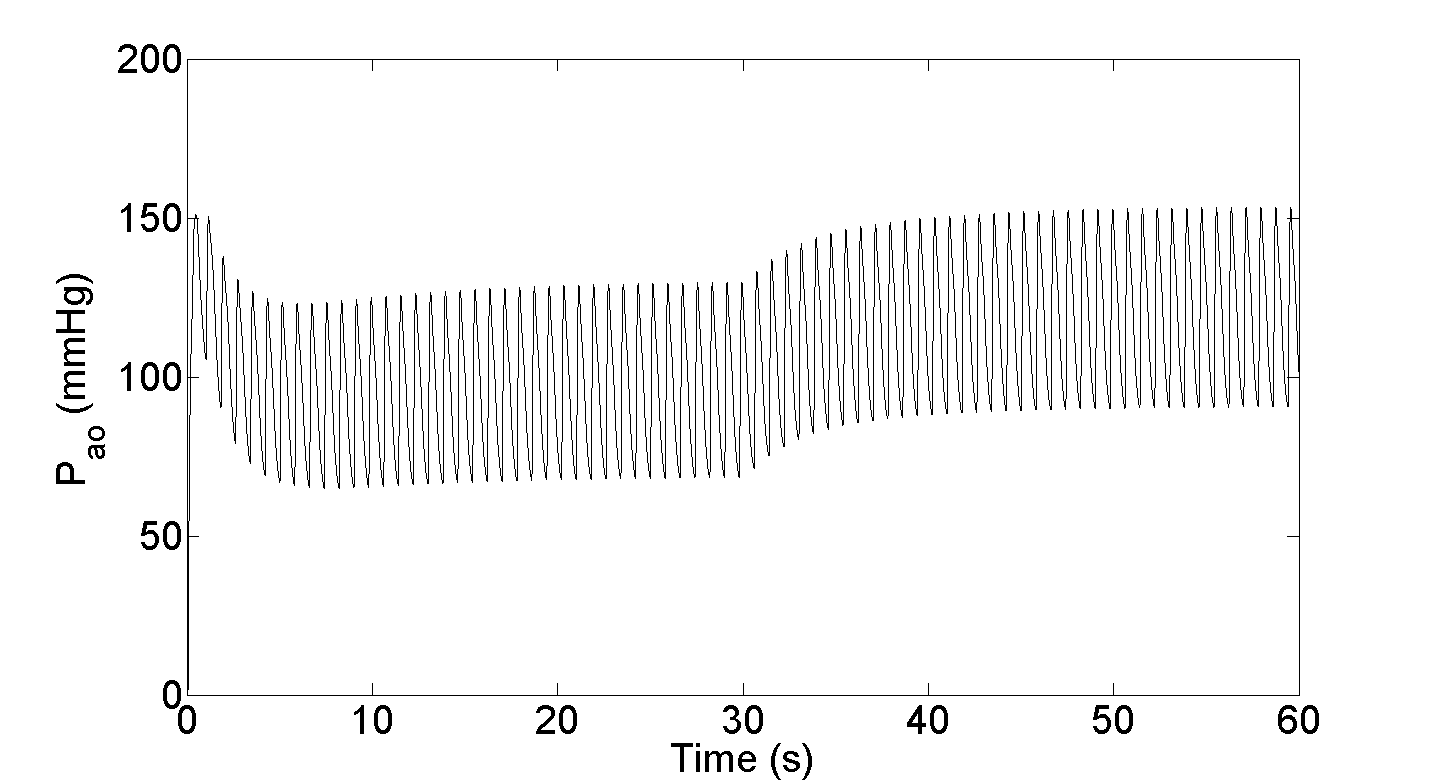}
   \label{53c}
 }
  \subfigure[Left atrial pressure.]{
   \includegraphics[scale =0.16752] {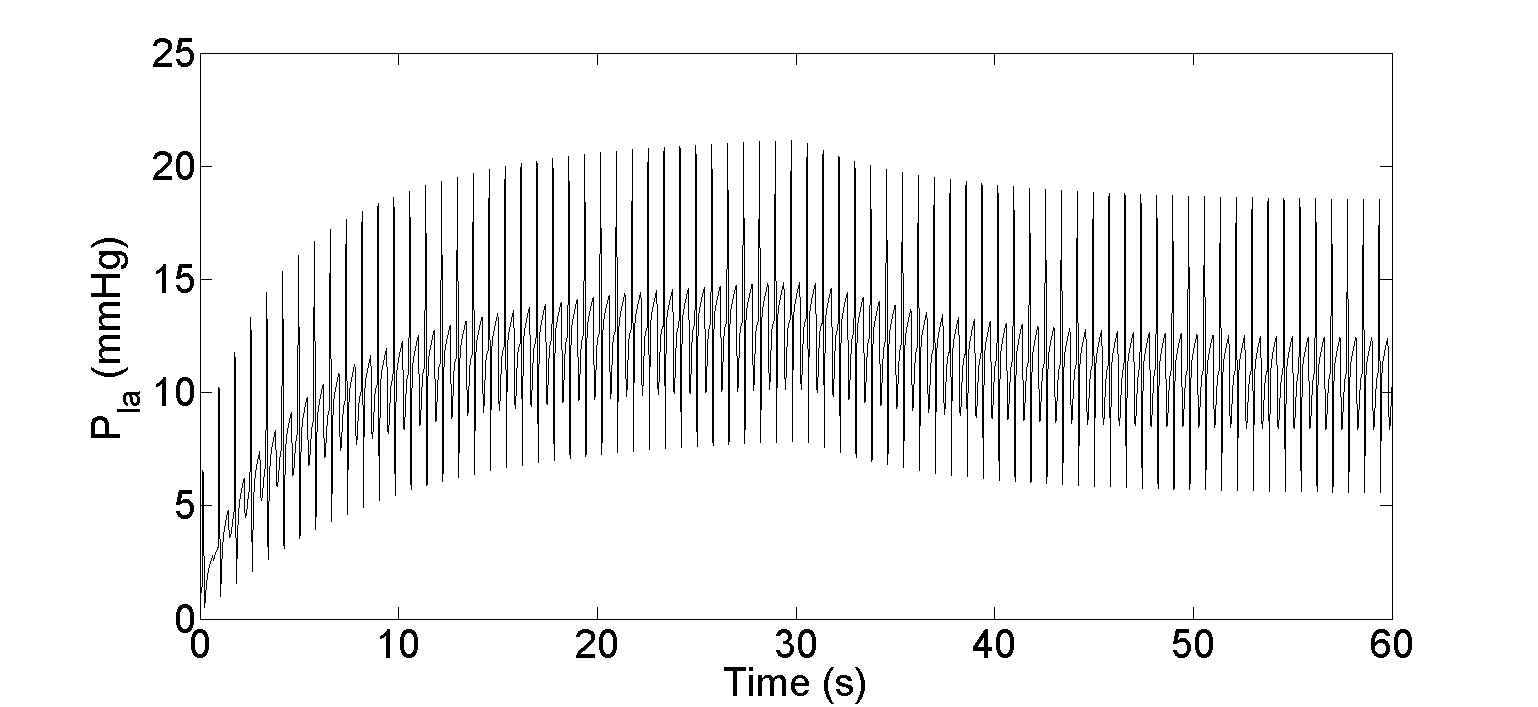}
   \label{53d}
 }

\subfigure[Right atrial pressure.]{
   \includegraphics[scale =0.16752] {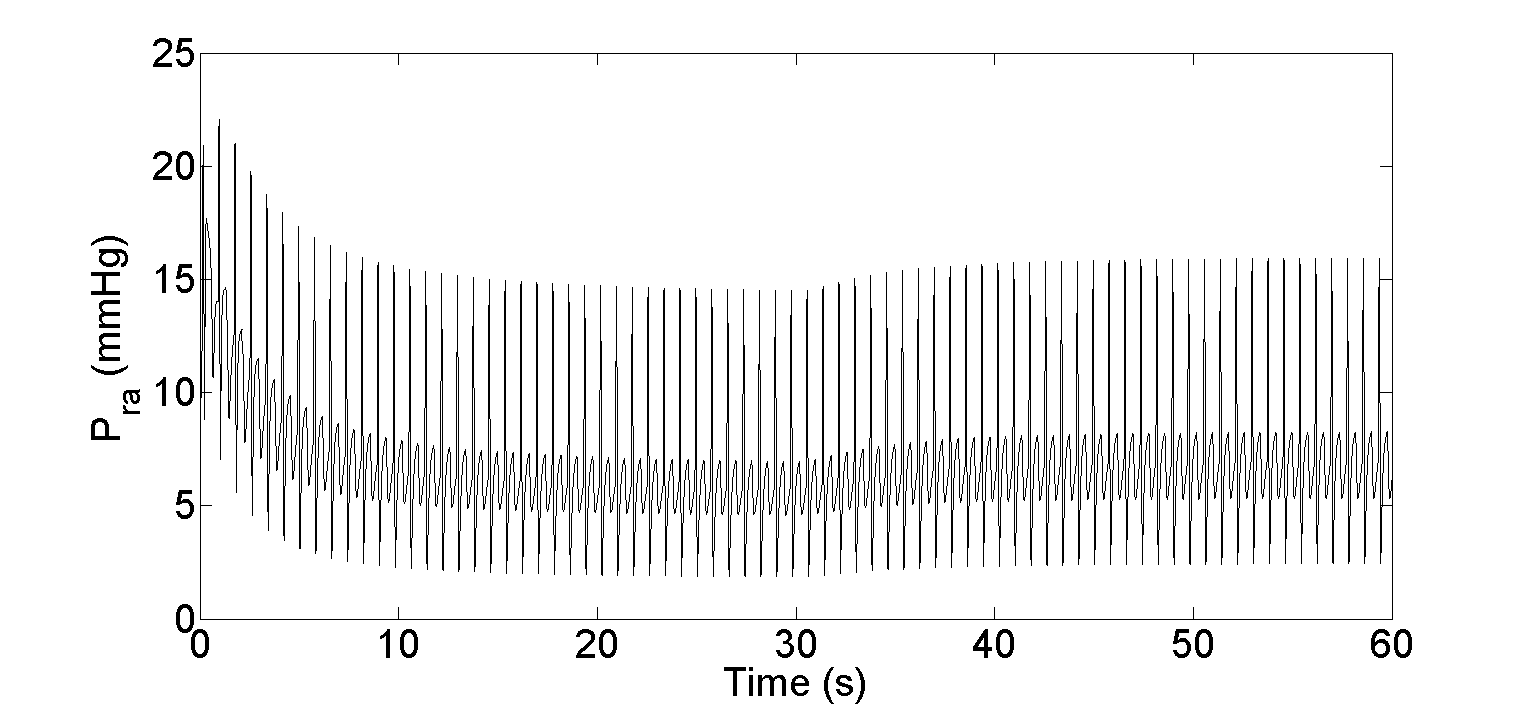}
   \label{53e}
 }
\caption{Hemodynamic variables results in rest condition when the system induced at 30s.}
\label{5:30a}
\end{figure*}

\begin{figure*}[htbp]
\centering
\subfigure[Average pump speed.]{
   \includegraphics[scale =0.16752] {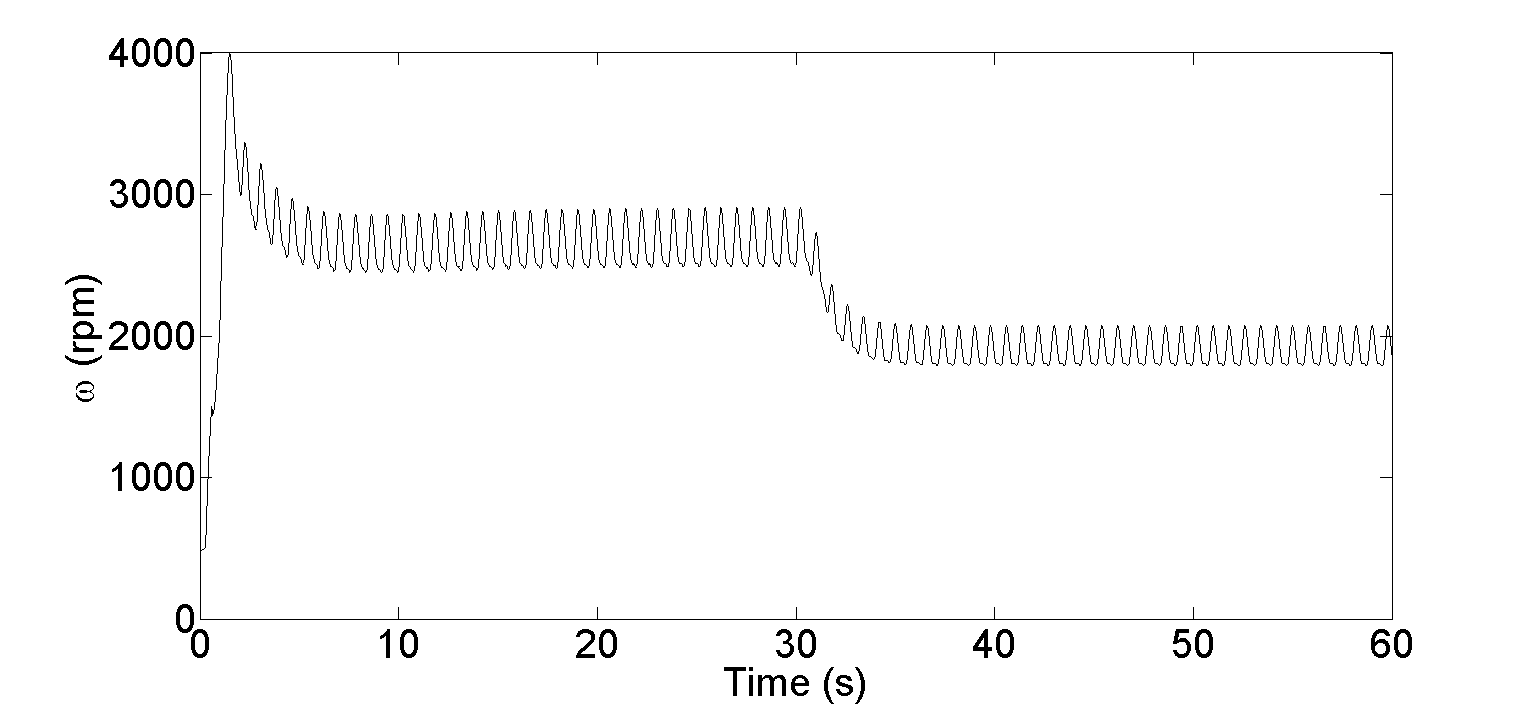}
   \label{53f}
 }

  \subfigure[Pump flow compared with reference signal.]{
   \includegraphics[scale =0.16752] {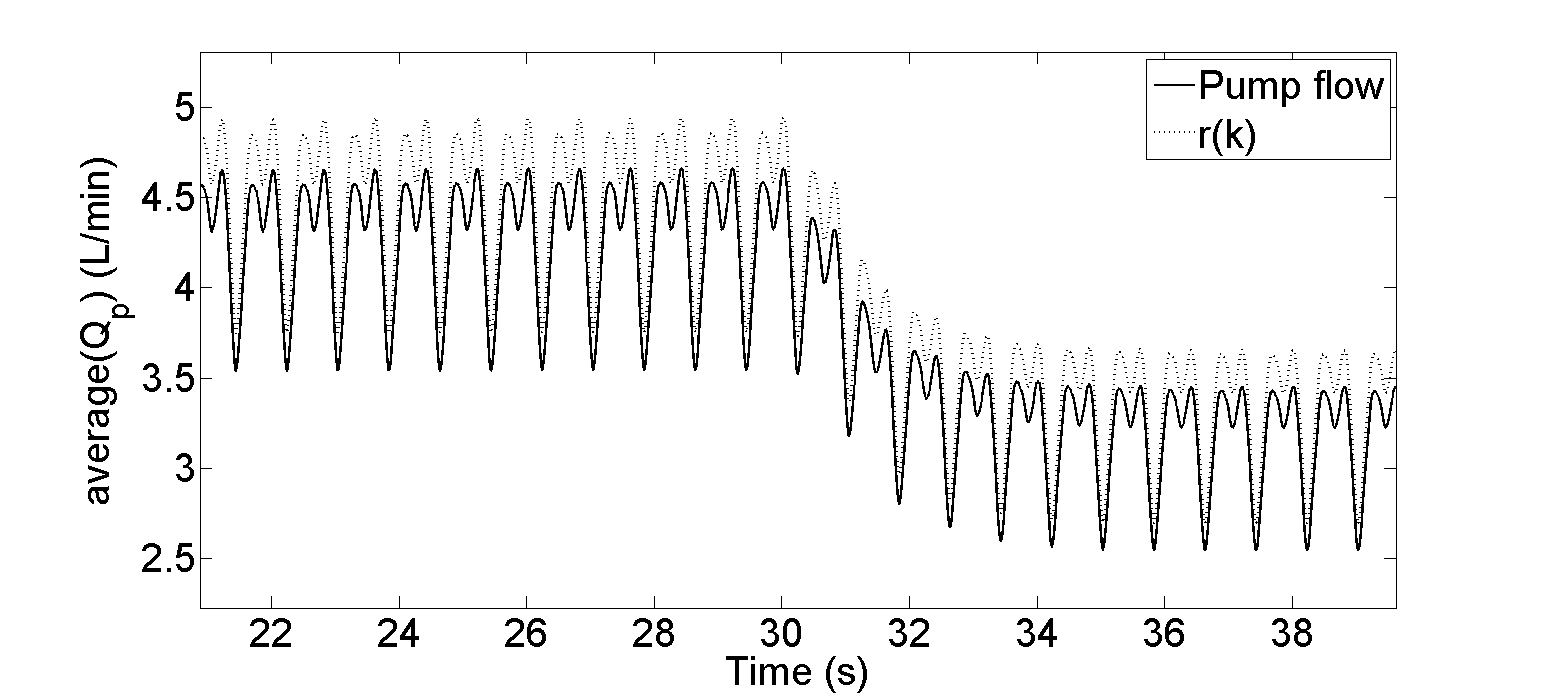}
   \label{53h}
 }

\subfigure[Measured steady state pump flow against estimated pump flow.]{
   \includegraphics[scale =0.16752] {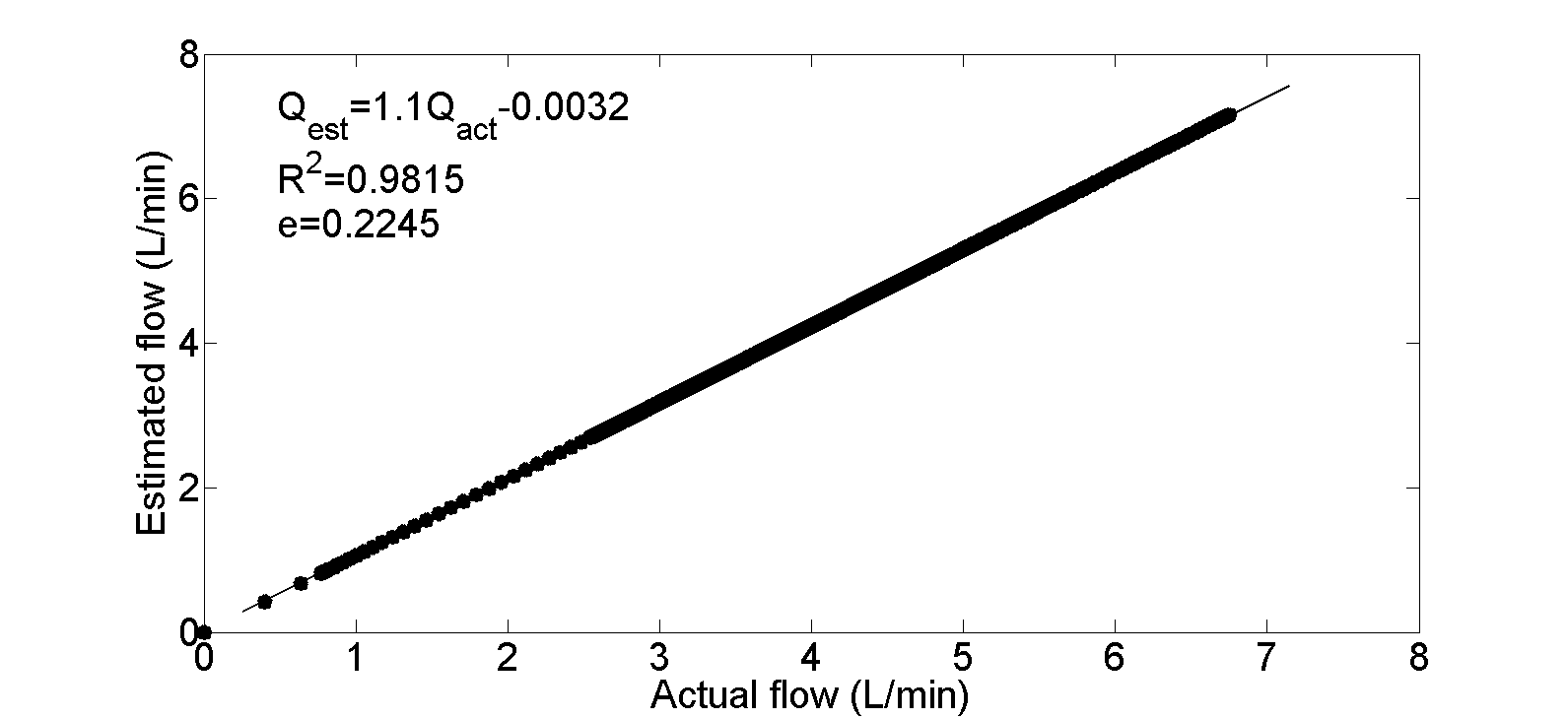}
   \label{53i}
 }

\caption{Pump variable results in rest condition when the system induced at 30s.}
\label{5:30b}
\end{figure*}

\begin{figure*}[htbp]
\centering
\subfigure[LV volume versus LV pressure before and after Parameter Change.]{
   \includegraphics[scale =0.16752] {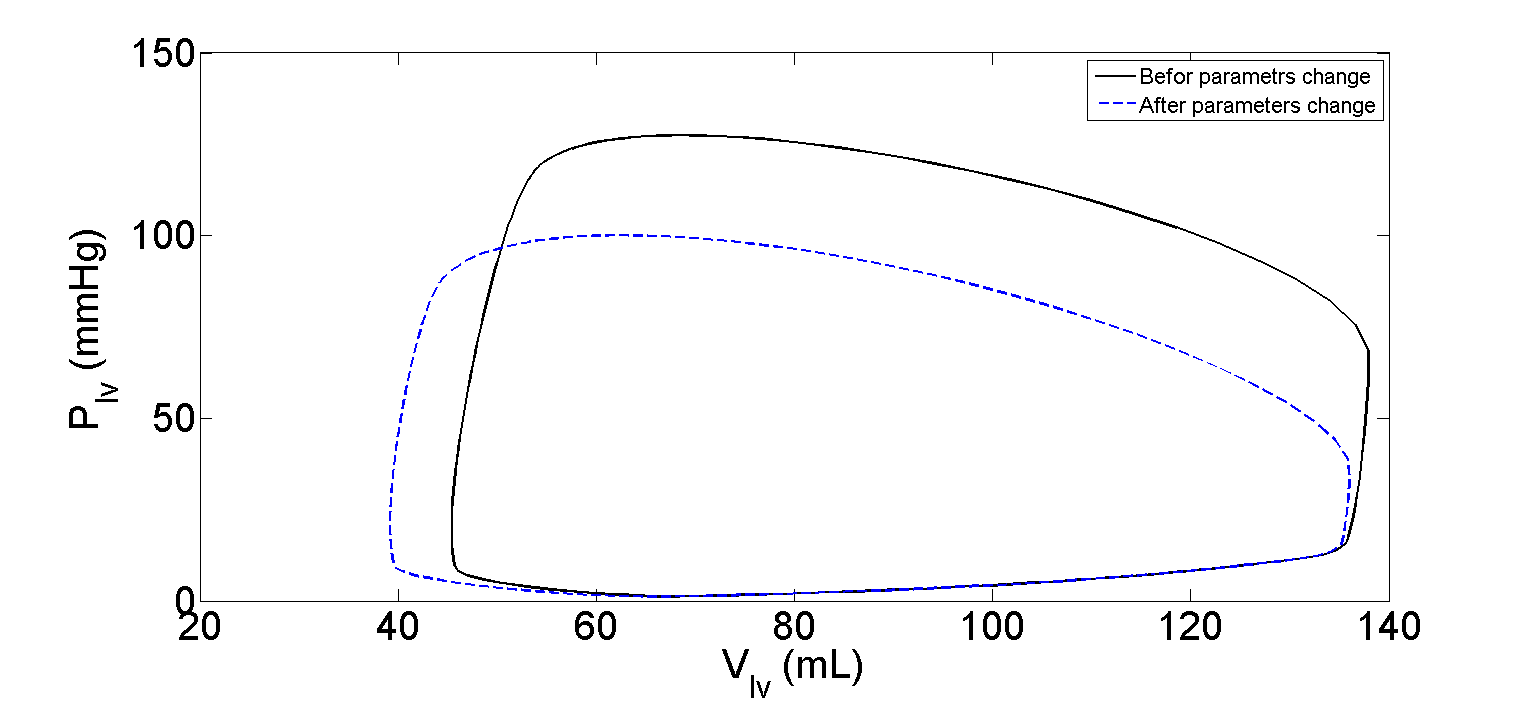}
   \label{56a}
 }
\subfigure[RV volume versus RV pressure before and after Parameter Change.]{
   \includegraphics[scale =0.16752] {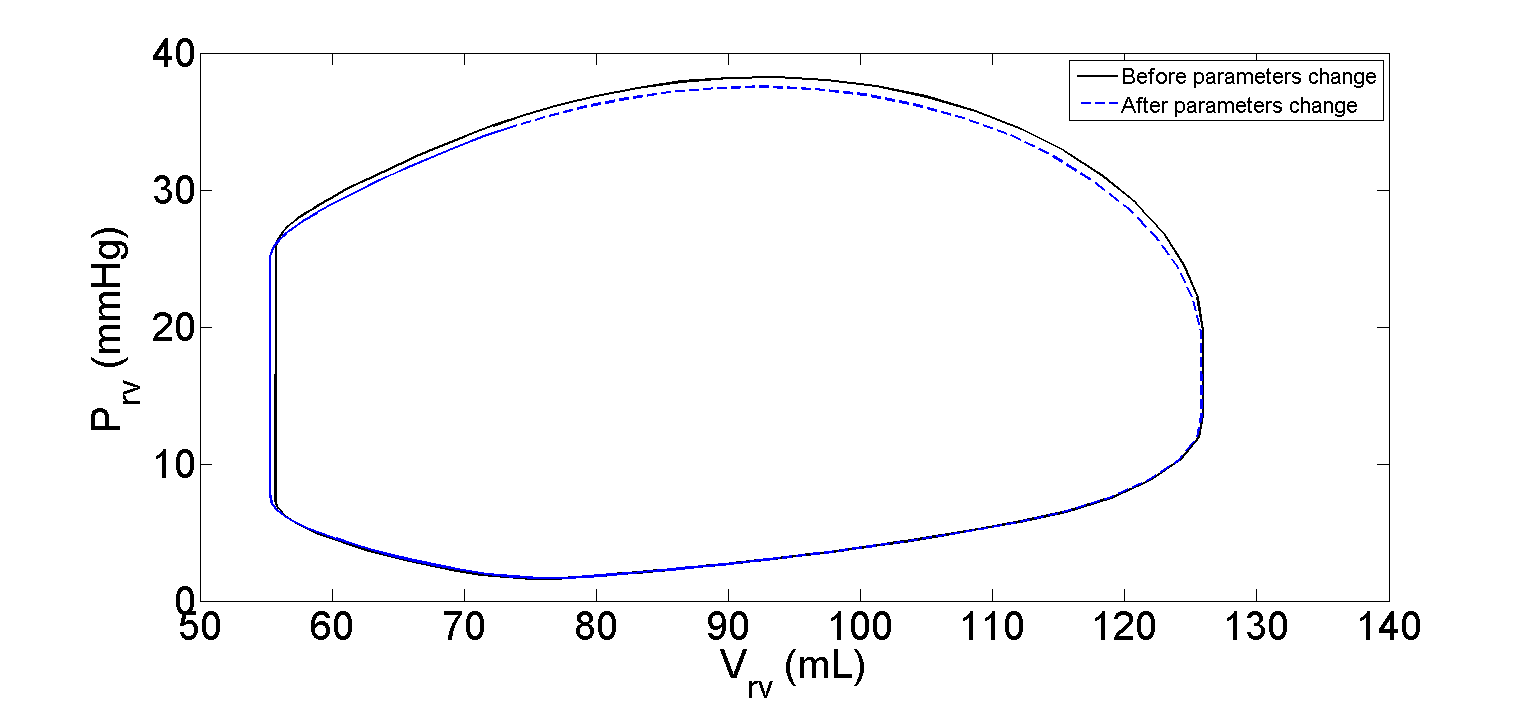}
   \label{56b}
 }

 \subfigure[Aortic pressure.]{
   \includegraphics[scale =0.16752] {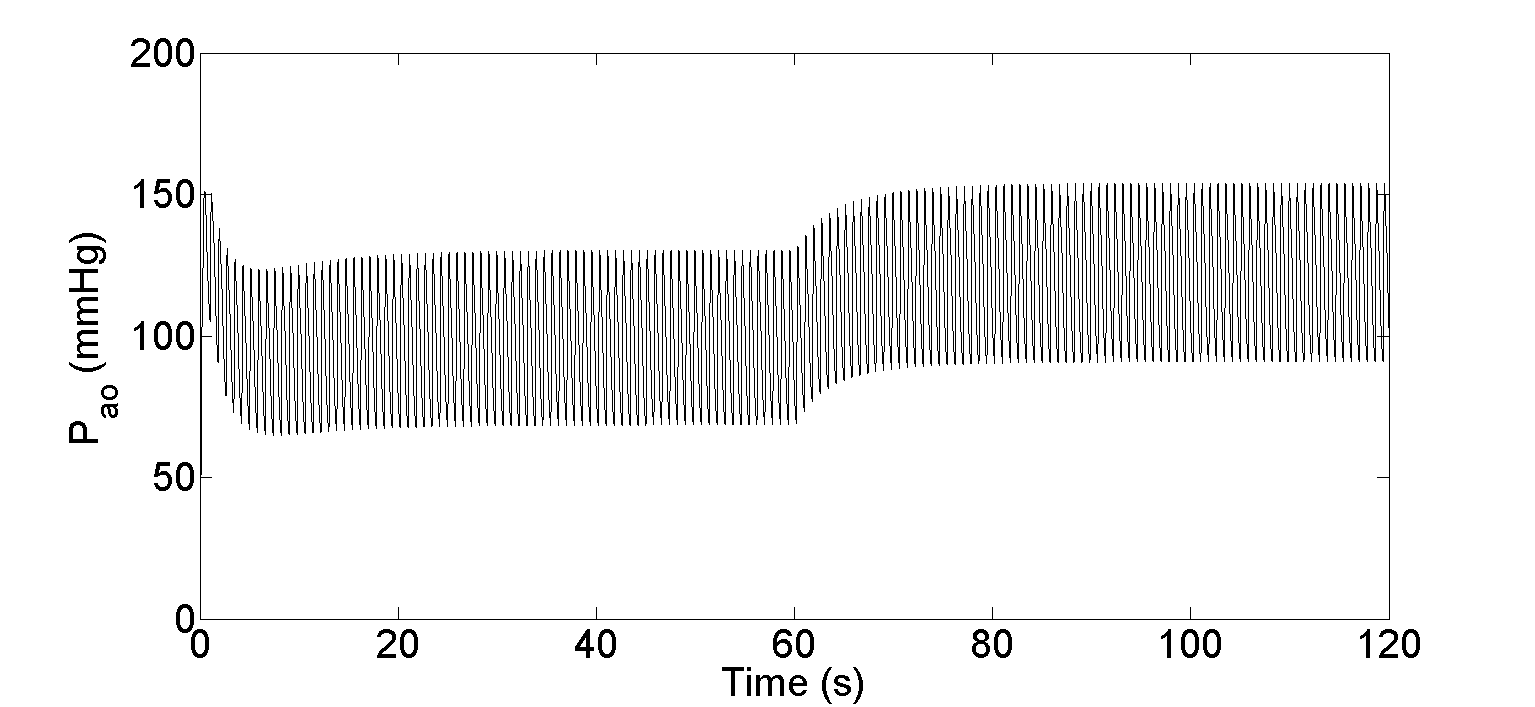}
   \label{56c}
 }
  \subfigure[Left atrial pressure.]{
   \includegraphics[scale =0.16752] {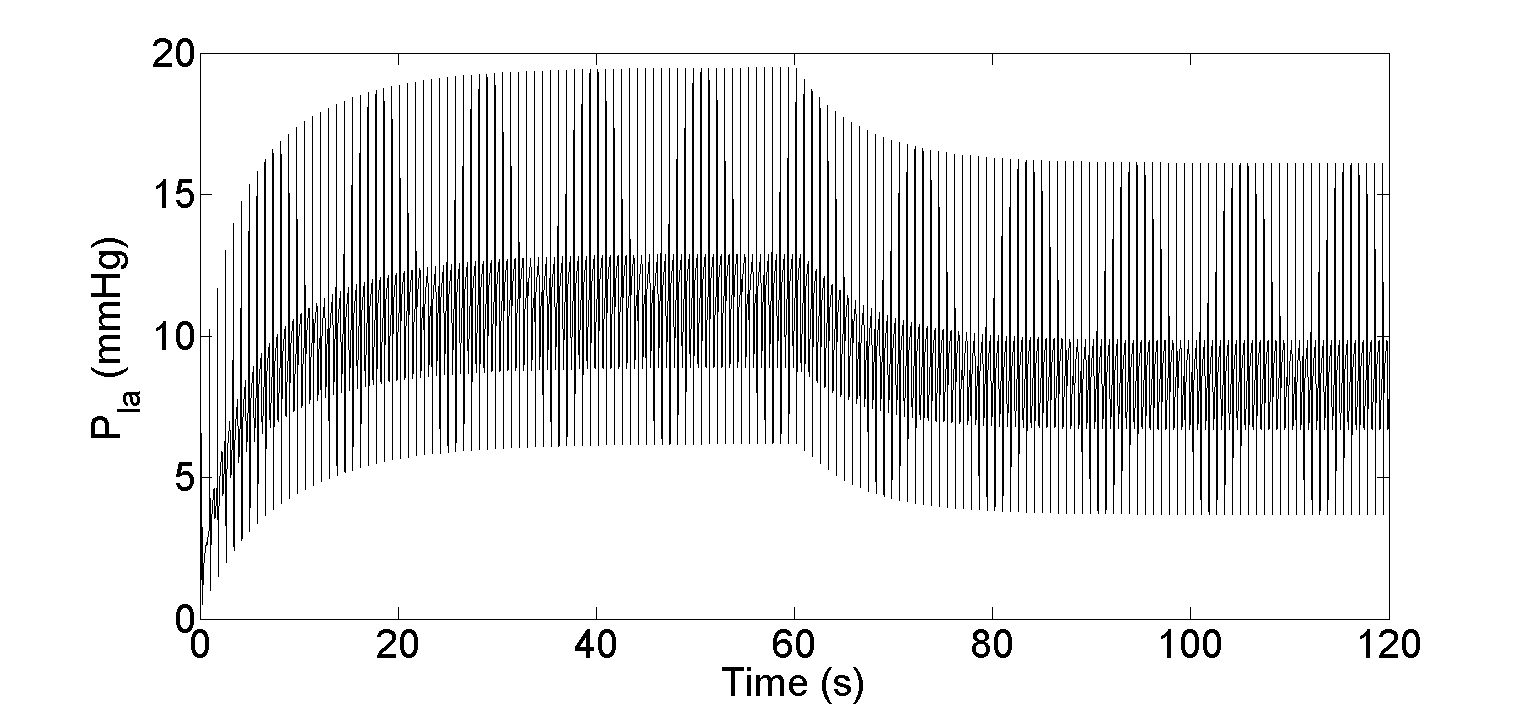}
   \label{56d}
 }

\subfigure[Right atrial pressure.]{
   \includegraphics[scale =0.16752] {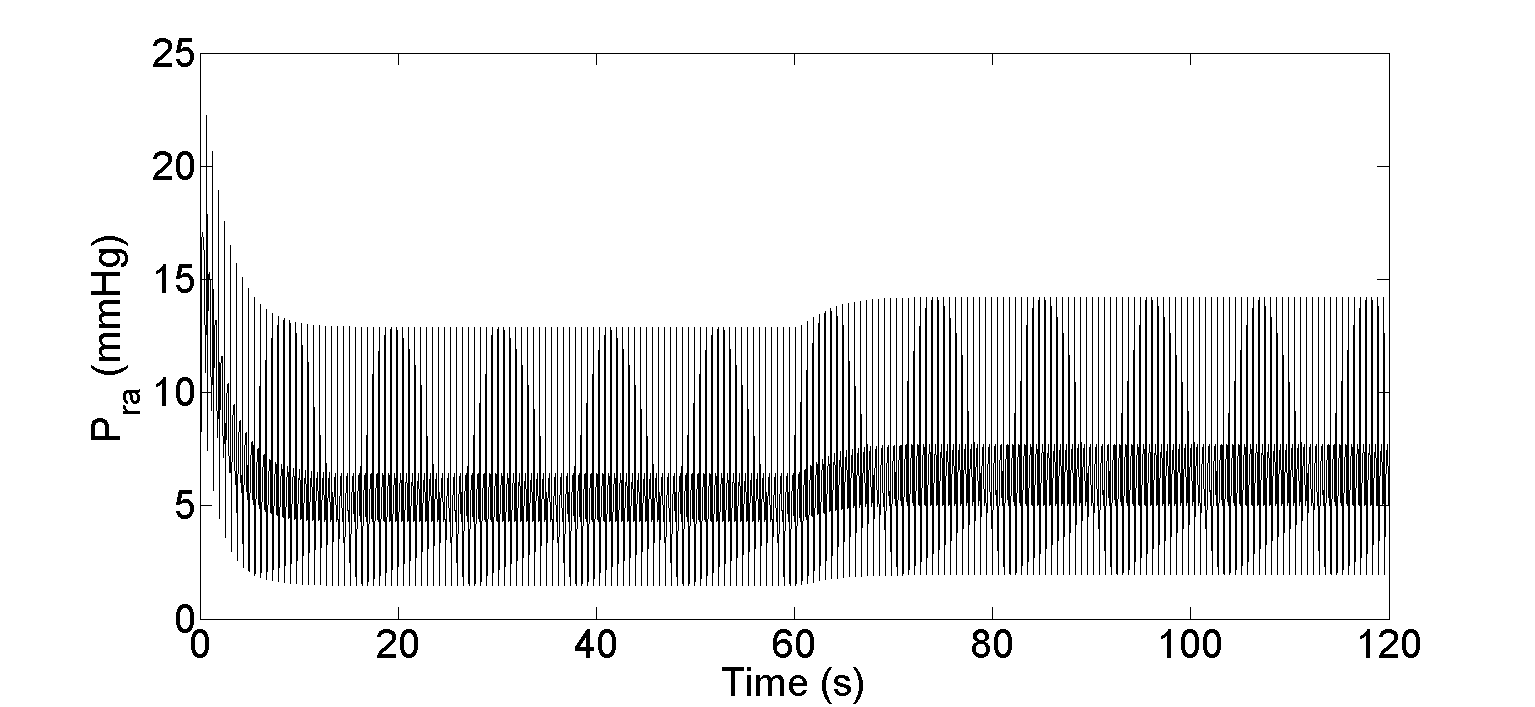}
   \label{56e}
 }
\caption{Hemodynamic variables results in rest condition when the system induced at 60s.}
\label{5:60a}
\end{figure*}

\begin{figure}[htbp]
\centering
\subfigure[Average pump speed.]{
   \includegraphics[scale =0.16752] {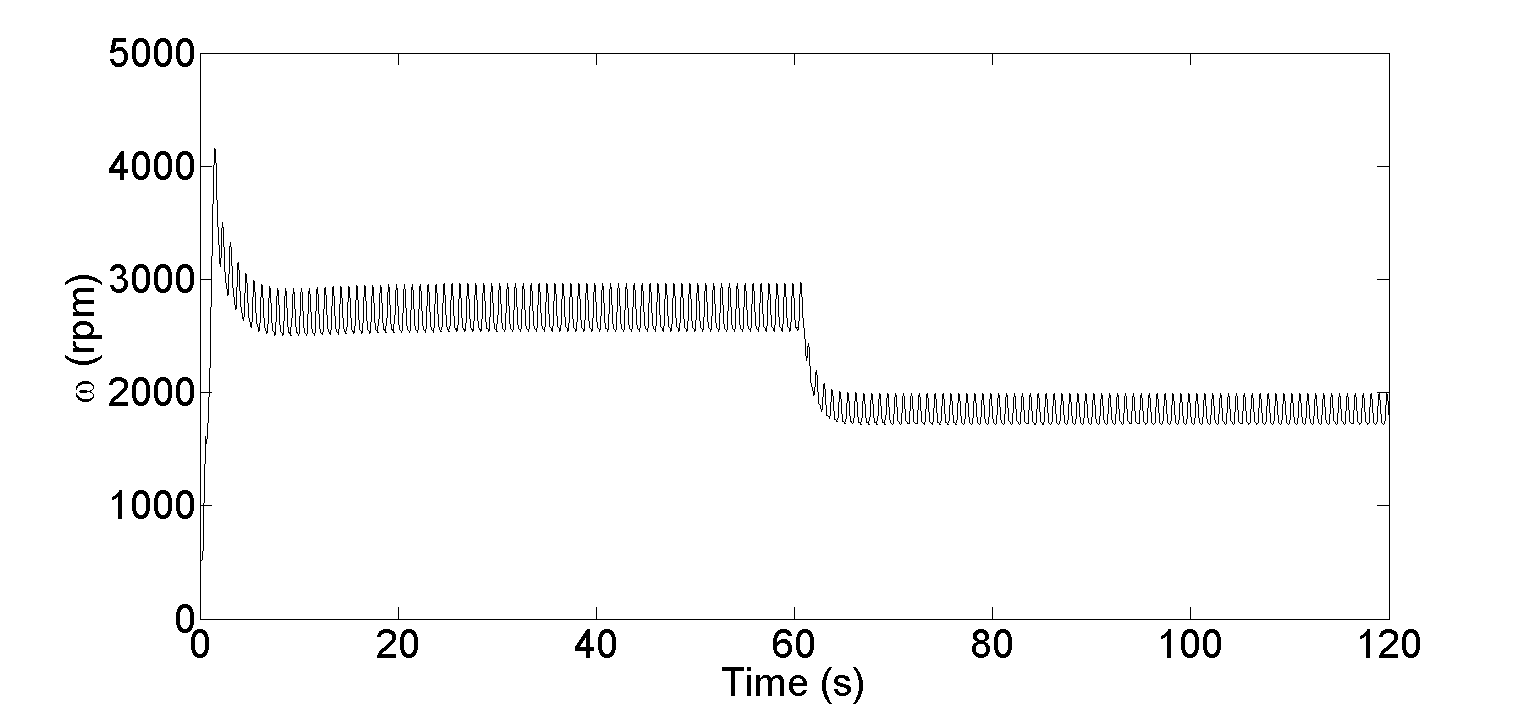}
   \label{56f}
 }

  \subfigure[Pump flow compared with  reference signal.]{
   \includegraphics[scale =0.16752] {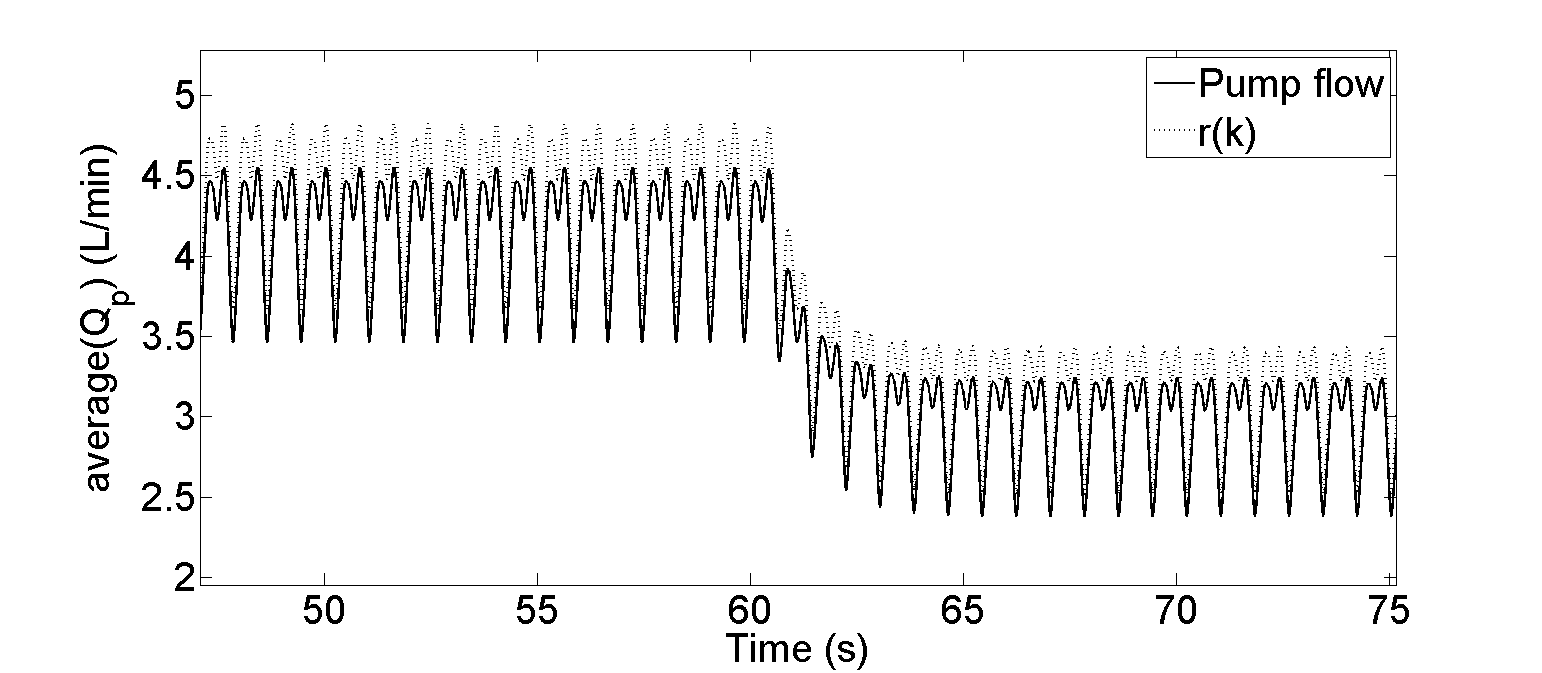}
   \label{56h}
 }

\subfigure[Measured steady state pump flow against estimated pump flow.]{
   \includegraphics[scale =0.16752] {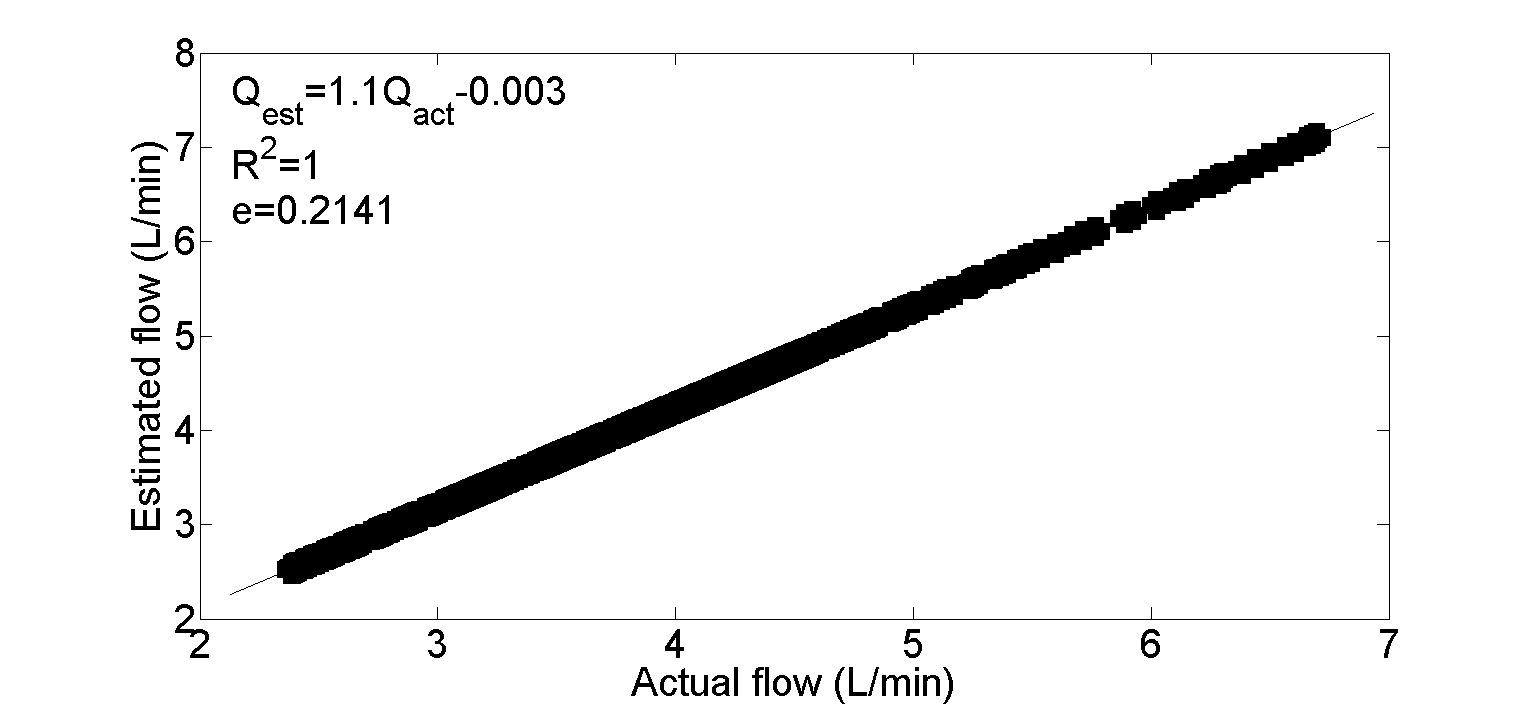}
   \label{56i}
 }

\caption{Pump variable results in rest condition when the system induced at 60s.}
\label{5:60b}
\end{figure}

\begin{figure*}[htbp]
\centering
\subfigure[LV volume versus LV pressure before and after Parameter Change.]{
   \includegraphics[scale =0.16752] {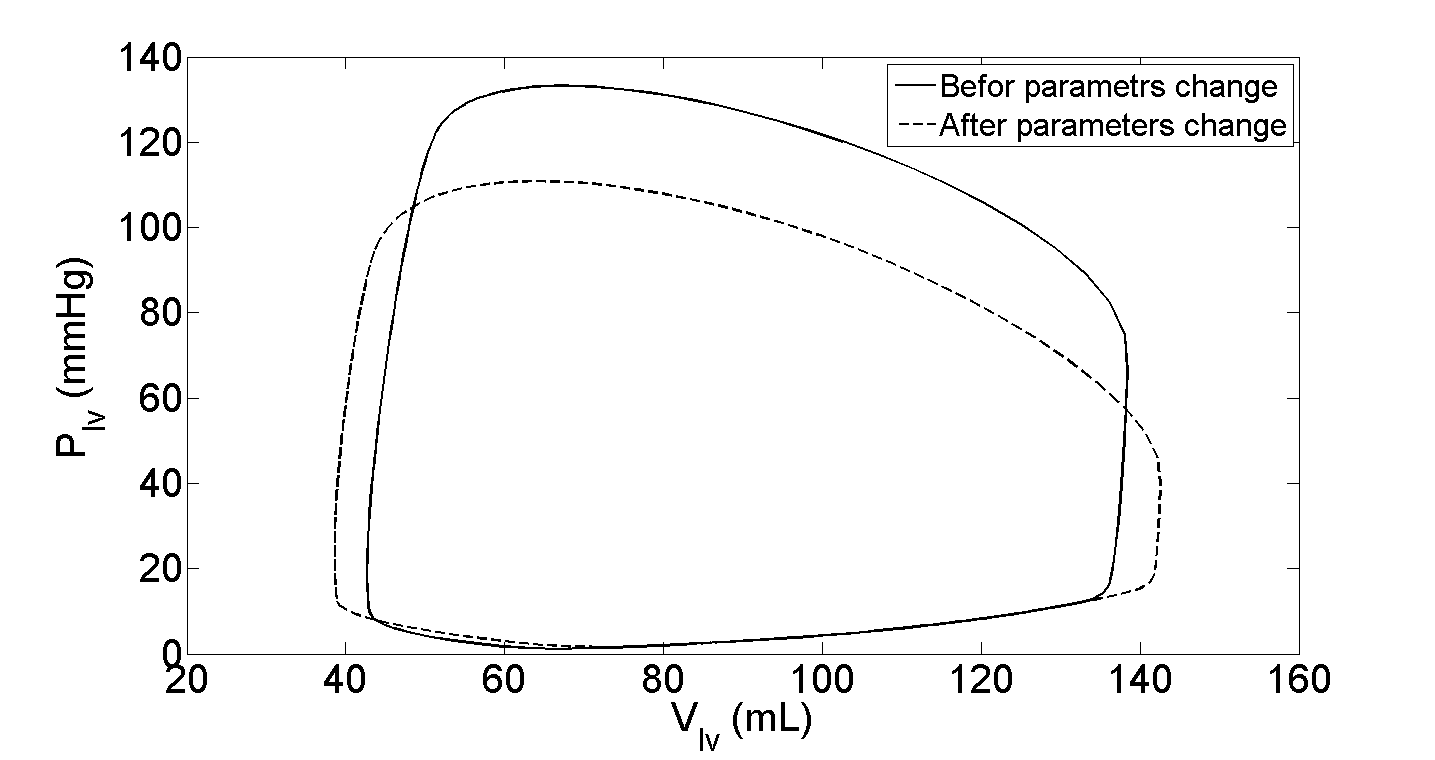}
   \label{59a}
 }
\subfigure[RV volume versus RV pressure before and after Parameter Change.]{
   \includegraphics[scale =0.16752] {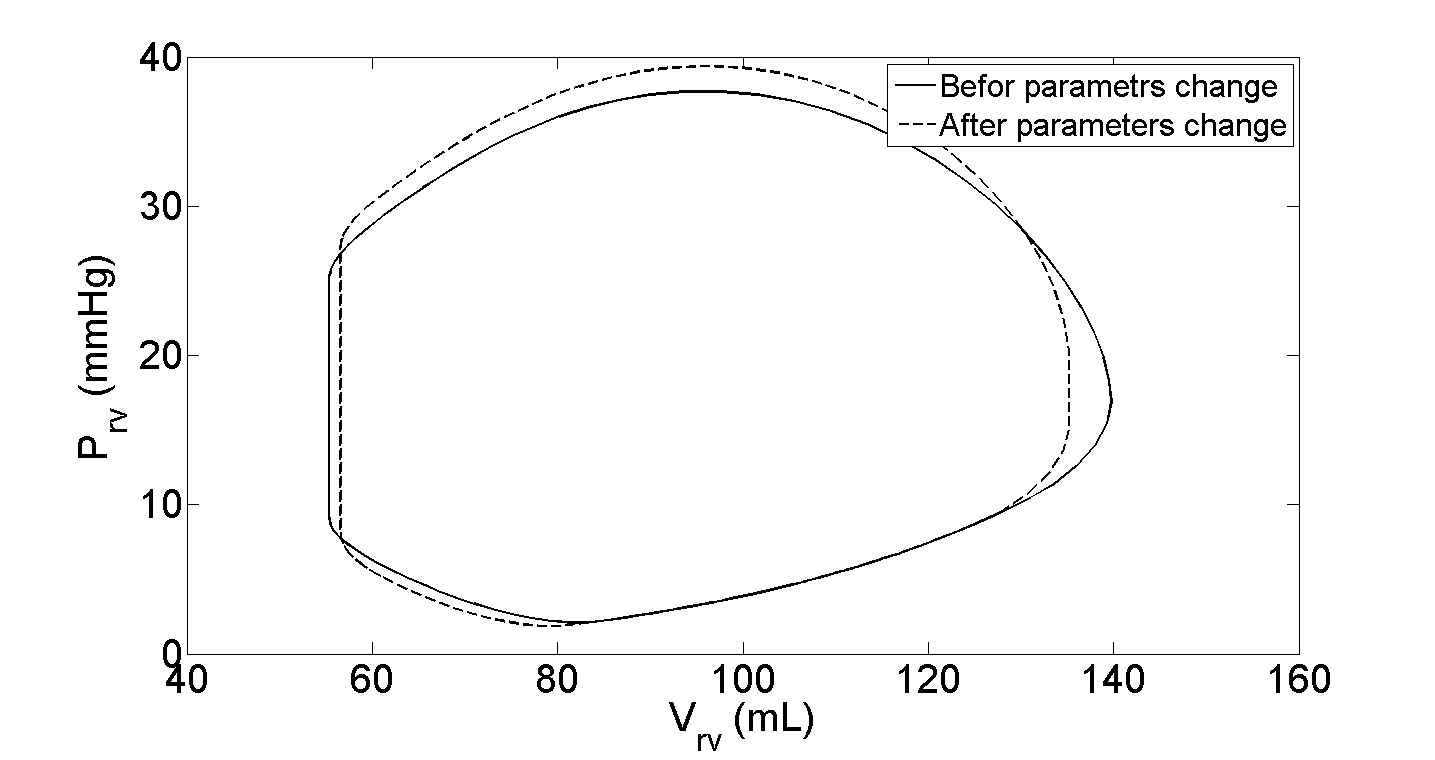}
   \label{59b}
 }

 \subfigure[Aortic pressure.]{
   \includegraphics[scale =0.16752] {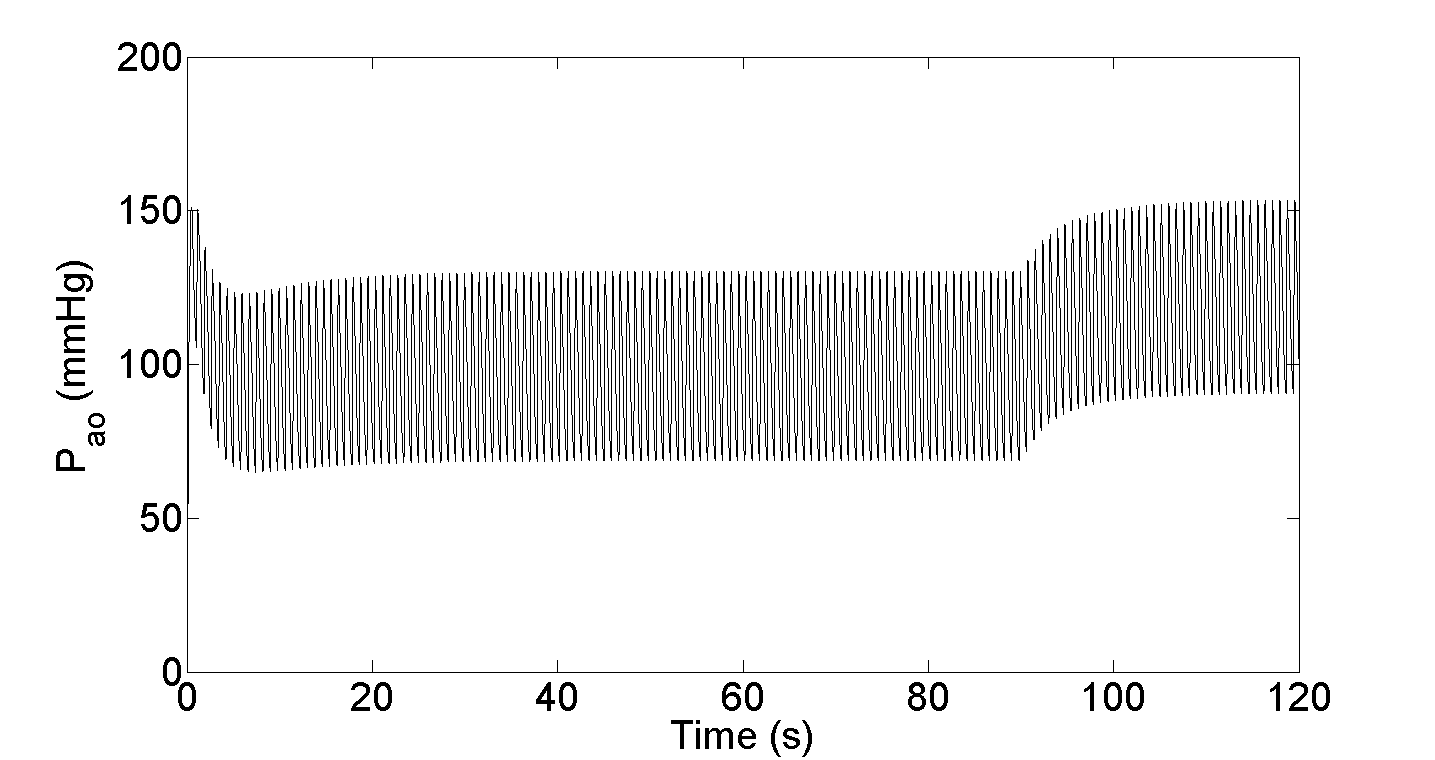}
   \label{59c}
 }
  \subfigure[Left atrial pressure.]{
   \includegraphics[scale =0.16752] {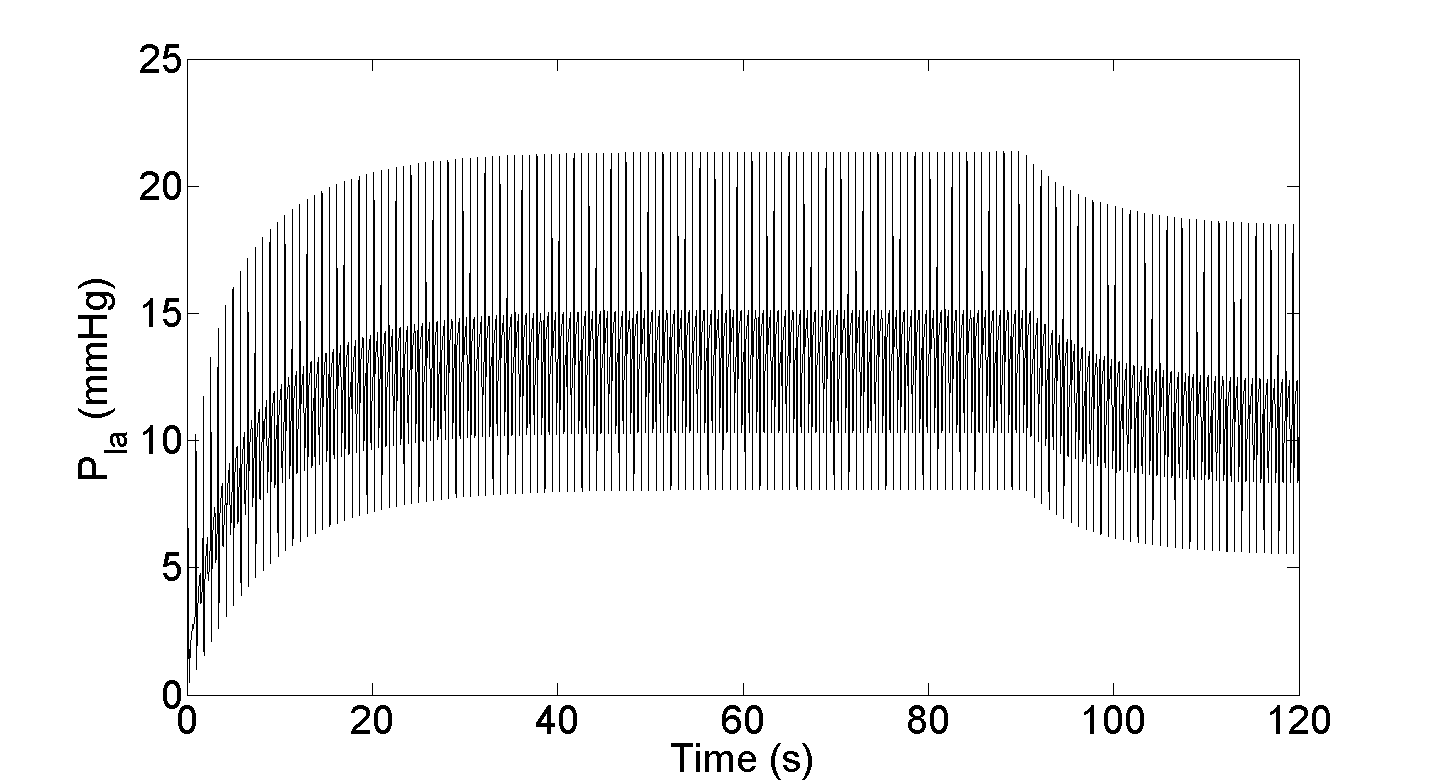}
   \label{59d}
 }

\subfigure[Right atrial pressure.]{
   \includegraphics[scale =0.16752] {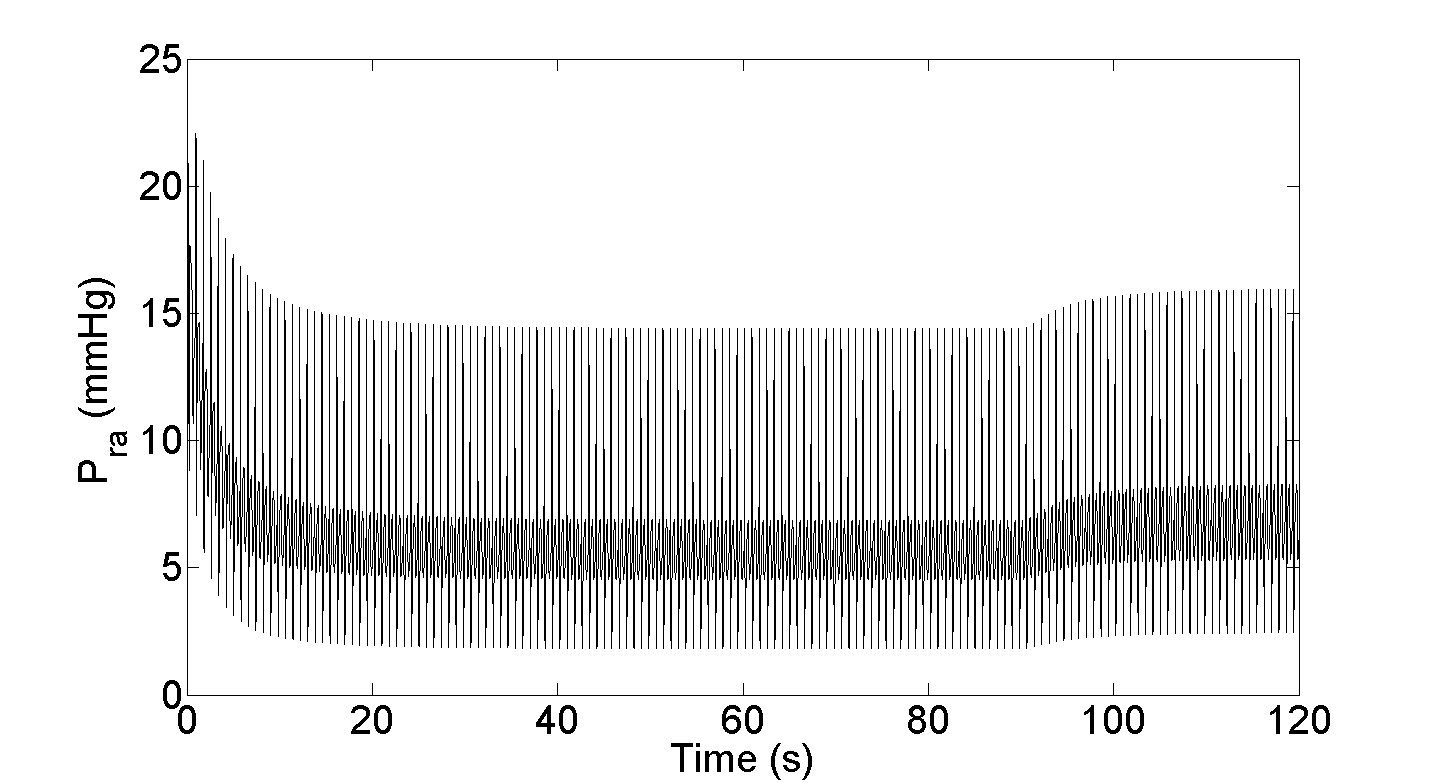}
   \label{59e}
 }
\caption{Hemodynamic variables results in rest condition when the system induced at 90s.}
\label{5:90a}
\end{figure*}

\begin{figure}[htbp]
\centering
\subfigure[Average pump speed.]{
   \includegraphics[scale =0.1752] {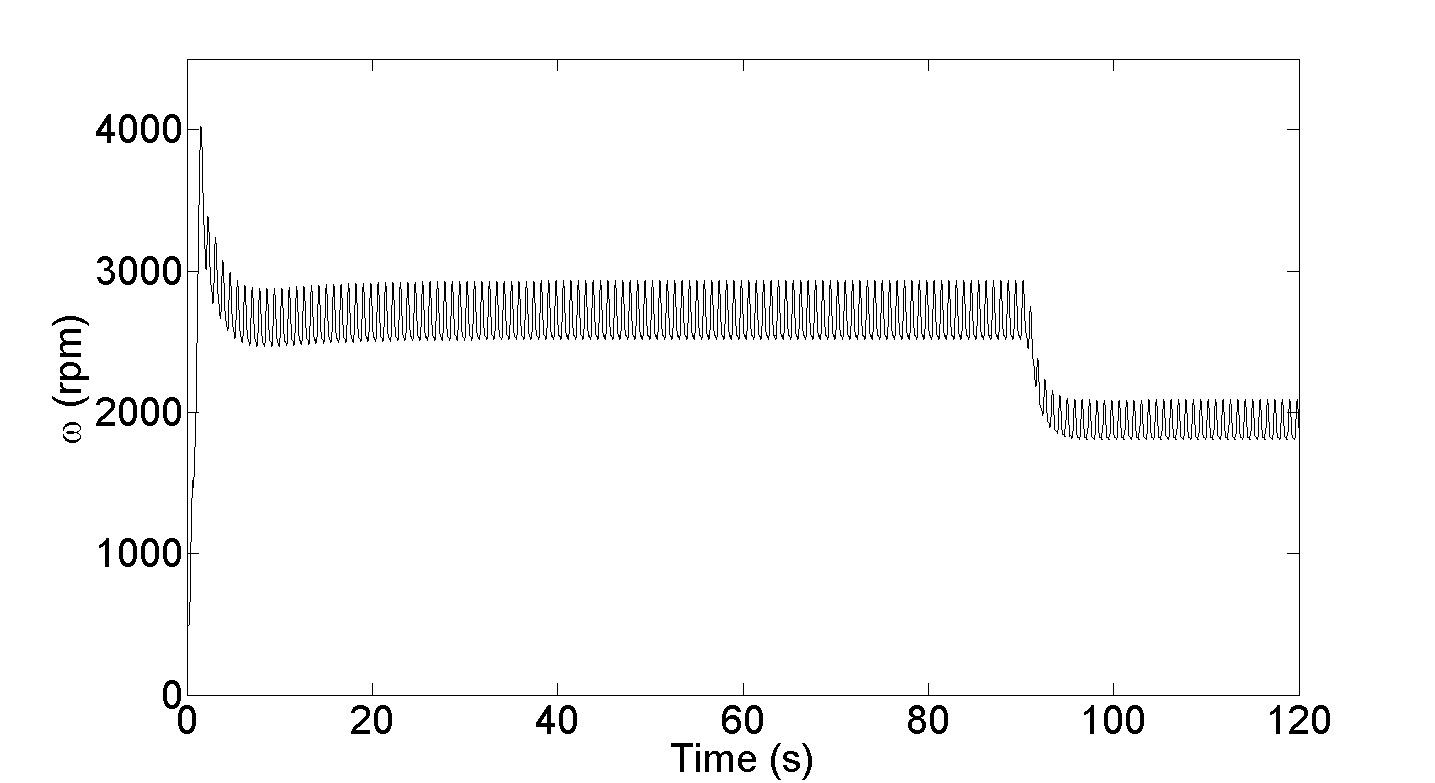}
   \label{59f}
 }

  \subfigure[Pump flow compared with reference signal.]{
   \includegraphics[scale =0.1752] {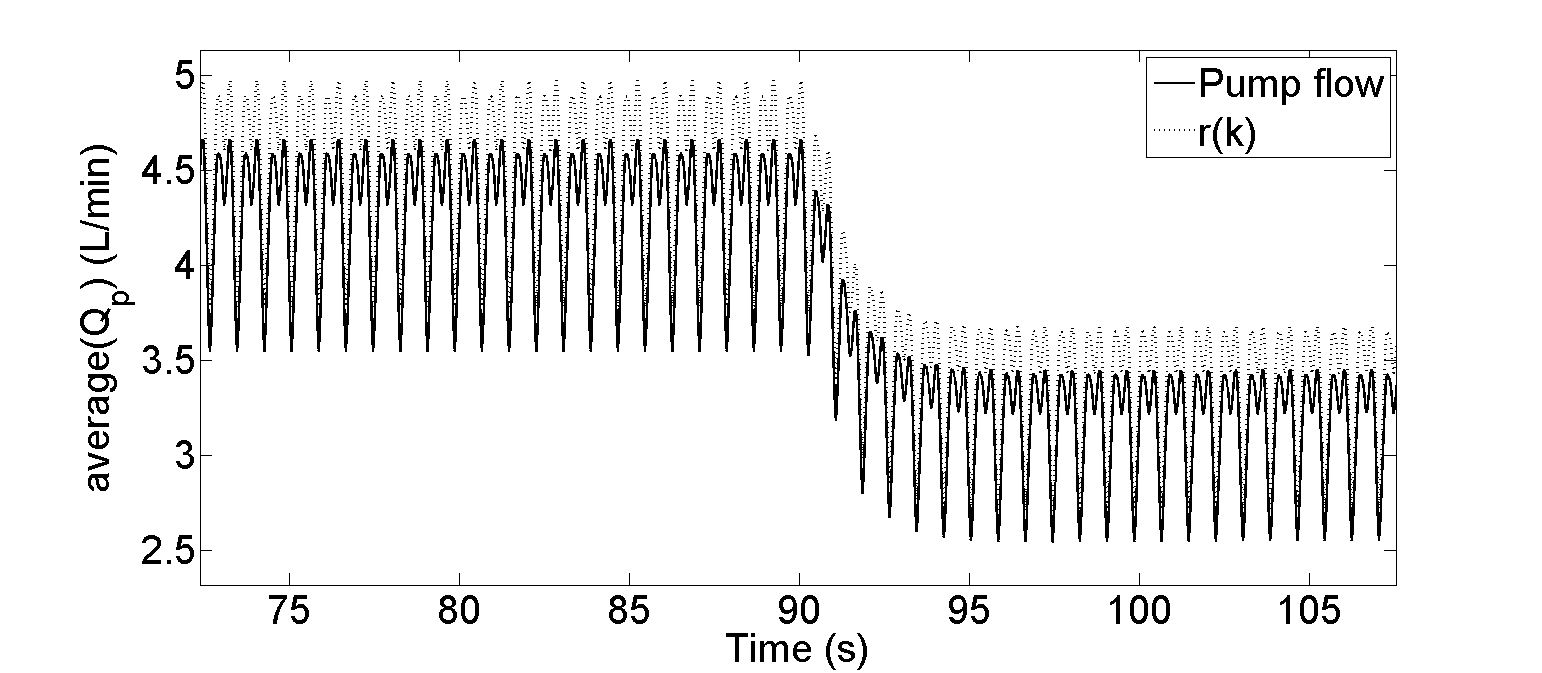}
   \label{59h}
 }

\subfigure[Measured steady state pump flow against estimated pump flow.]{
   \includegraphics[scale =0.1752] {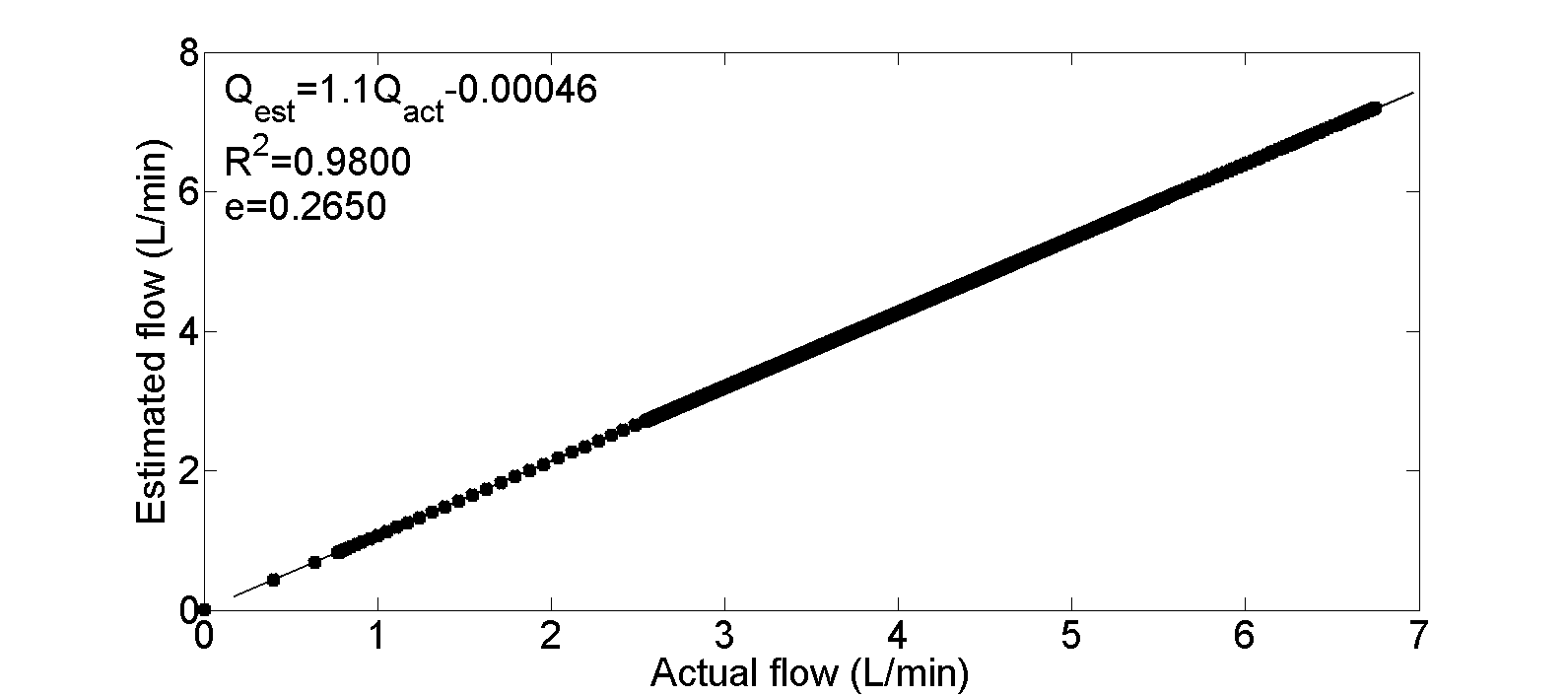}
   \label{59i}
 }

\caption{Pump variable results in rest condition when the system induced at 90s.}
\label{5:90b}
\end{figure}

\begin{figure*}[htbp]
\centering
\subfigure[LV volume versus LV pressure before and after Parameter Change.]{
   \includegraphics[scale =0.16752] {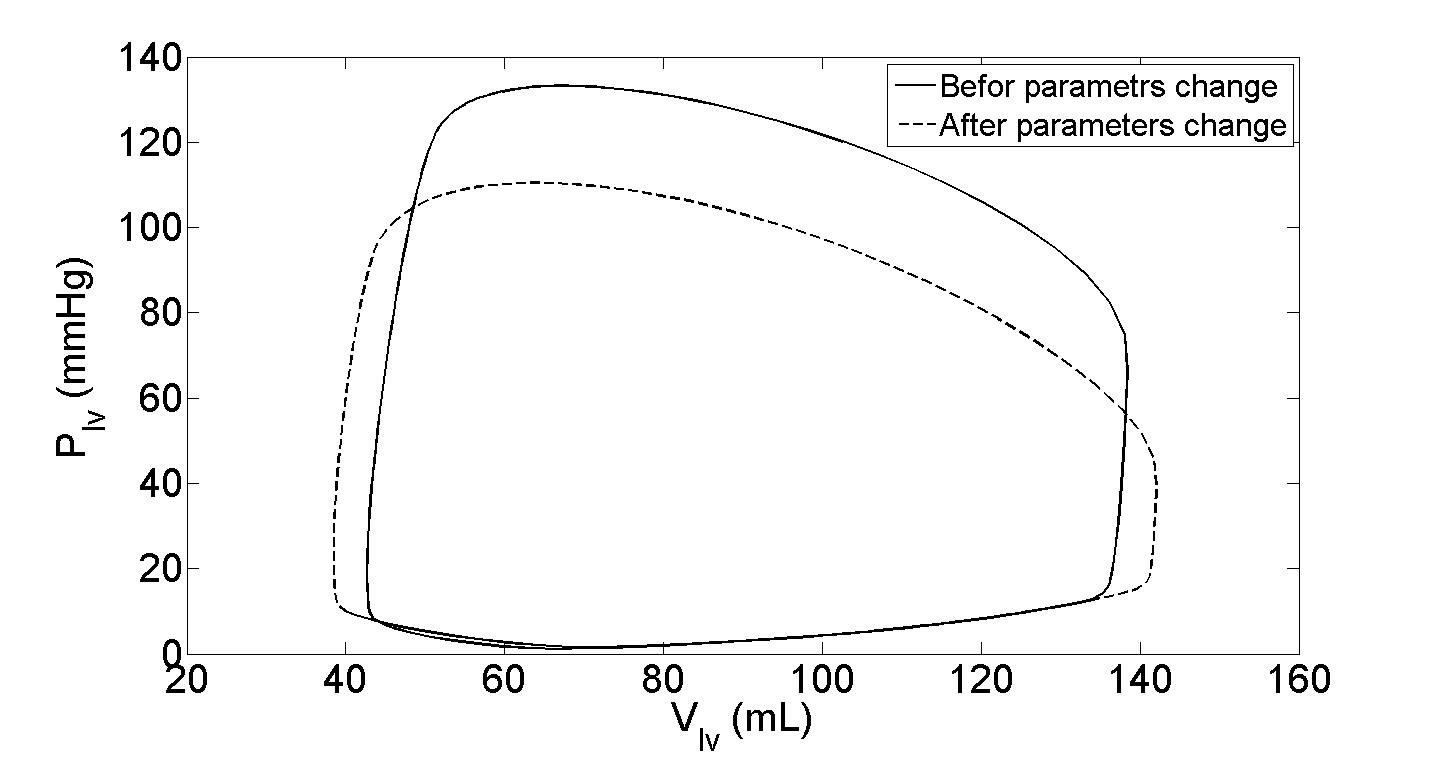}
   \label{52a}
 }
\subfigure[RV volume versus RV pressure before and after Parameter Change.]{
   \includegraphics[scale =0.16752] {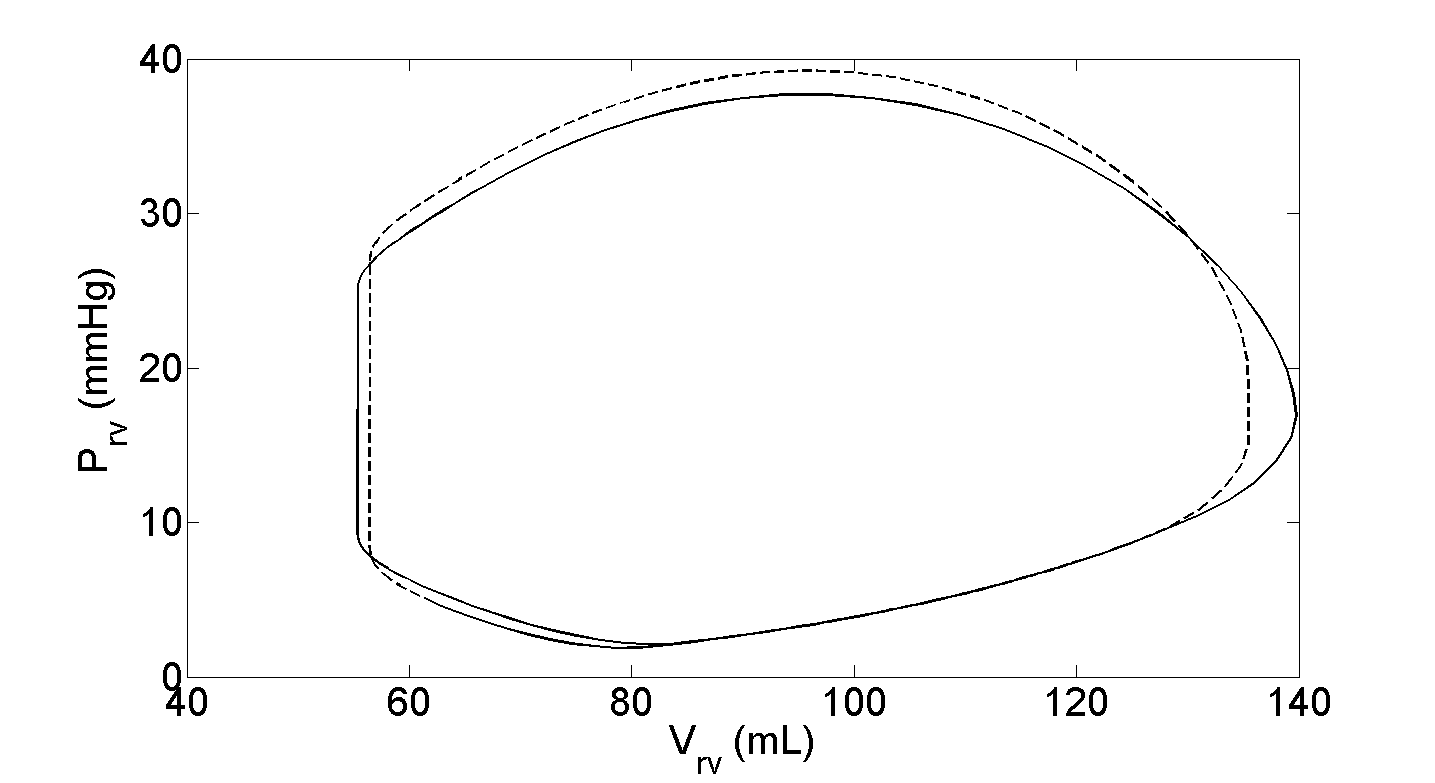}
   \label{52b}
 }

 \subfigure[Aortic pressure.]{
   \includegraphics[scale =0.16752] {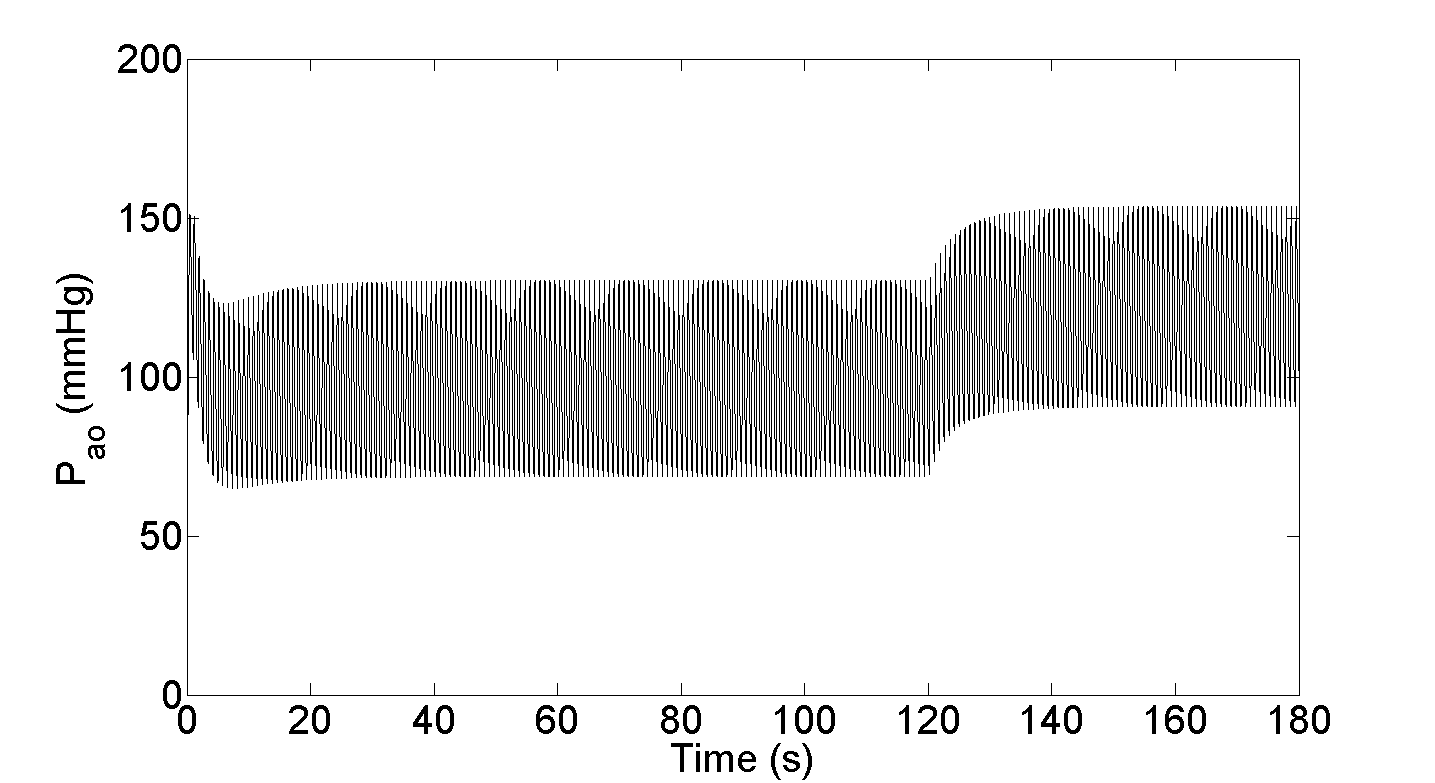}
   \label{52c}
 }
  \subfigure[Left atrial pressure.]{
   \includegraphics[scale =0.16752] {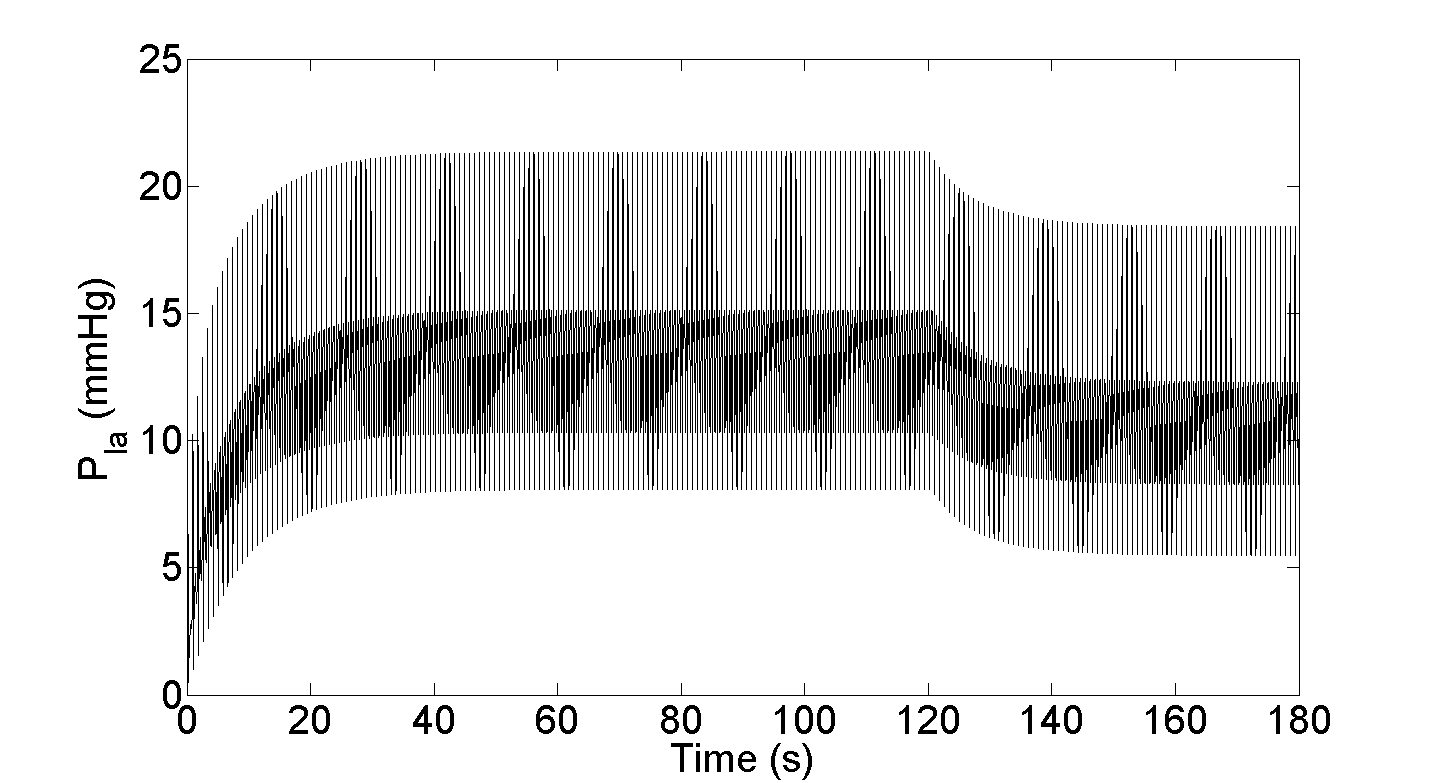}
   \label{52d}
 }

\subfigure[Right atrial pressure.]{
   \includegraphics[scale =0.16752] {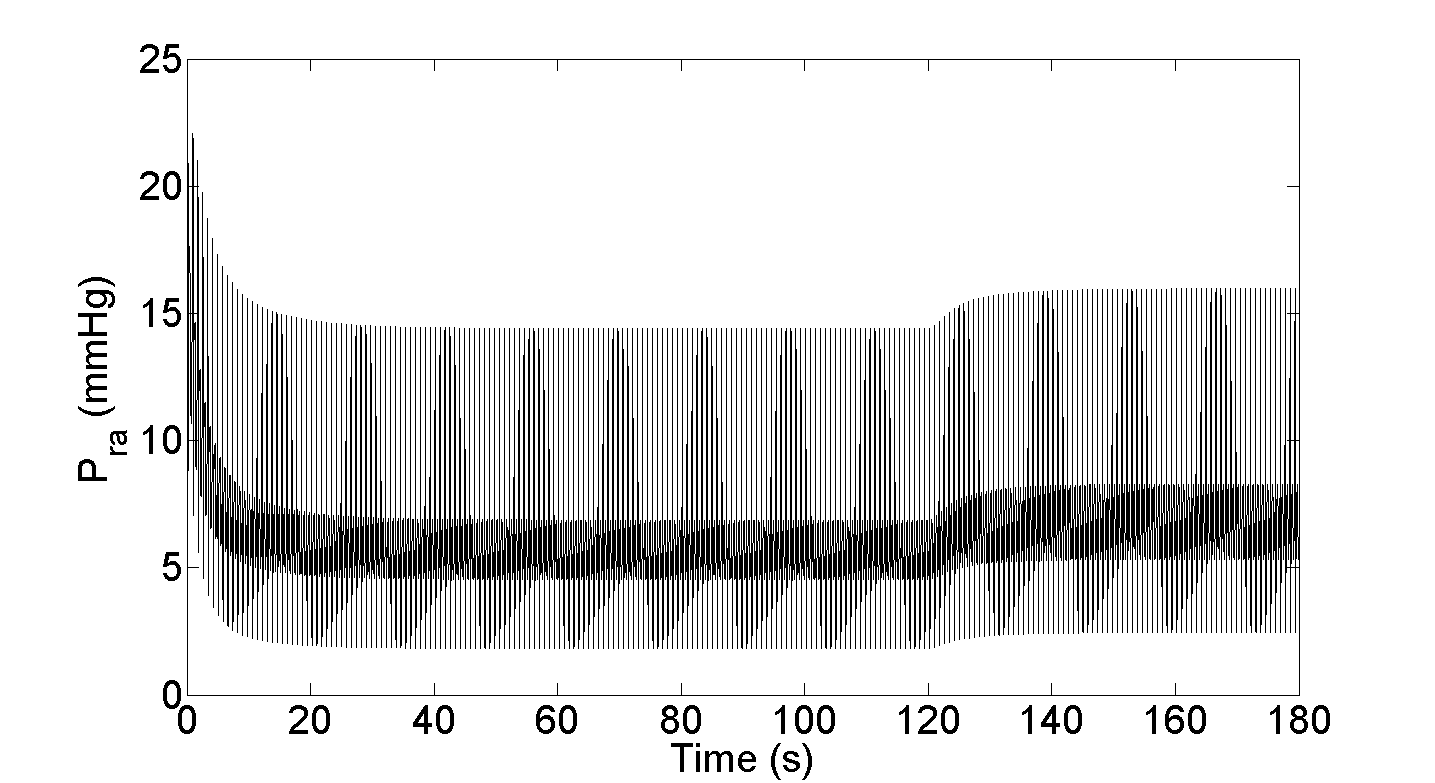}
   \label{52e}
 }
\caption{Hemodynamic variables results in rest condition when the system induced at 120s.}
\label{5:20a}
\end{figure*}

\begin{figure}[htbp]
\centering
\subfigure[Average pump speed.]{
   \includegraphics[scale =0.1752] {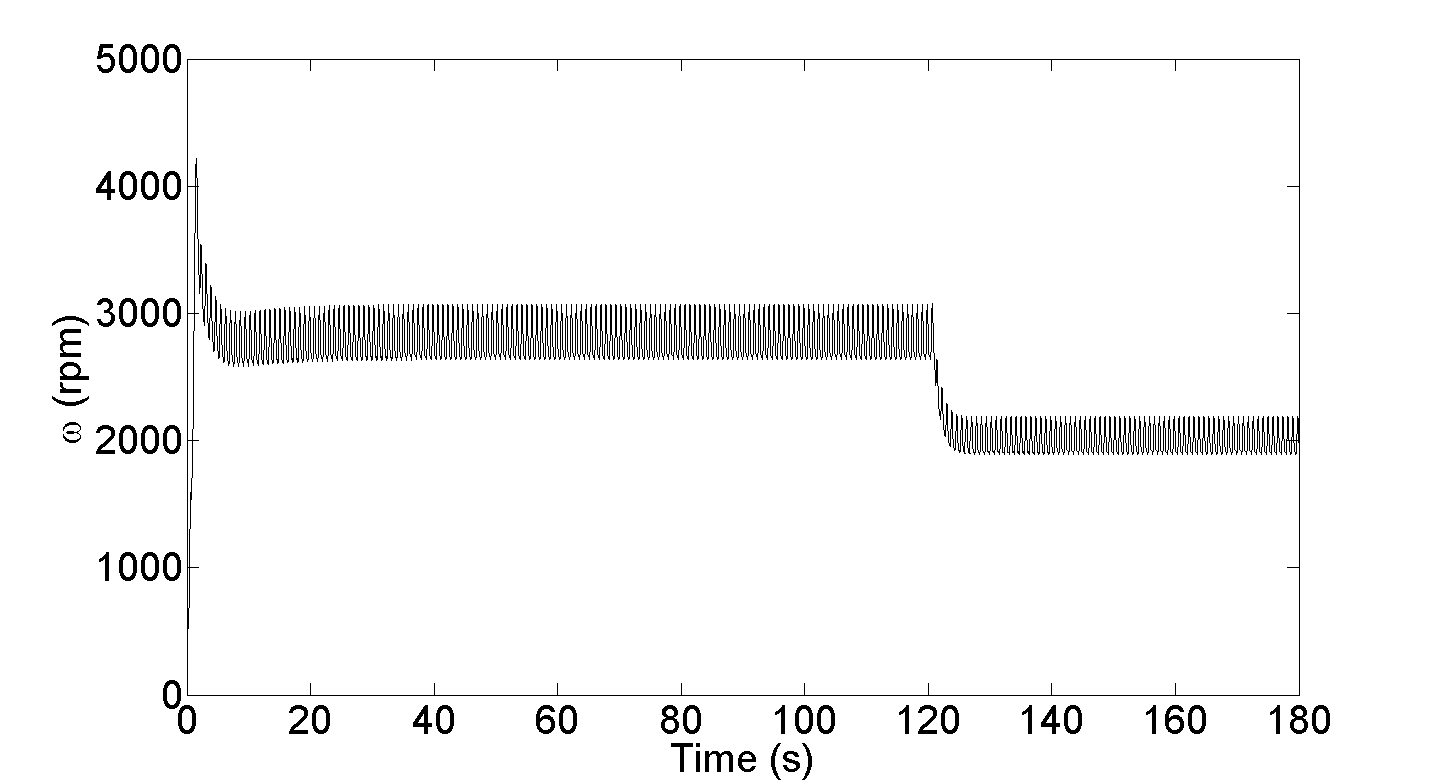}
   \label{52f}
 }

  \subfigure[Pump flow compared with reference signal.]{
   \includegraphics[scale =0.1752] {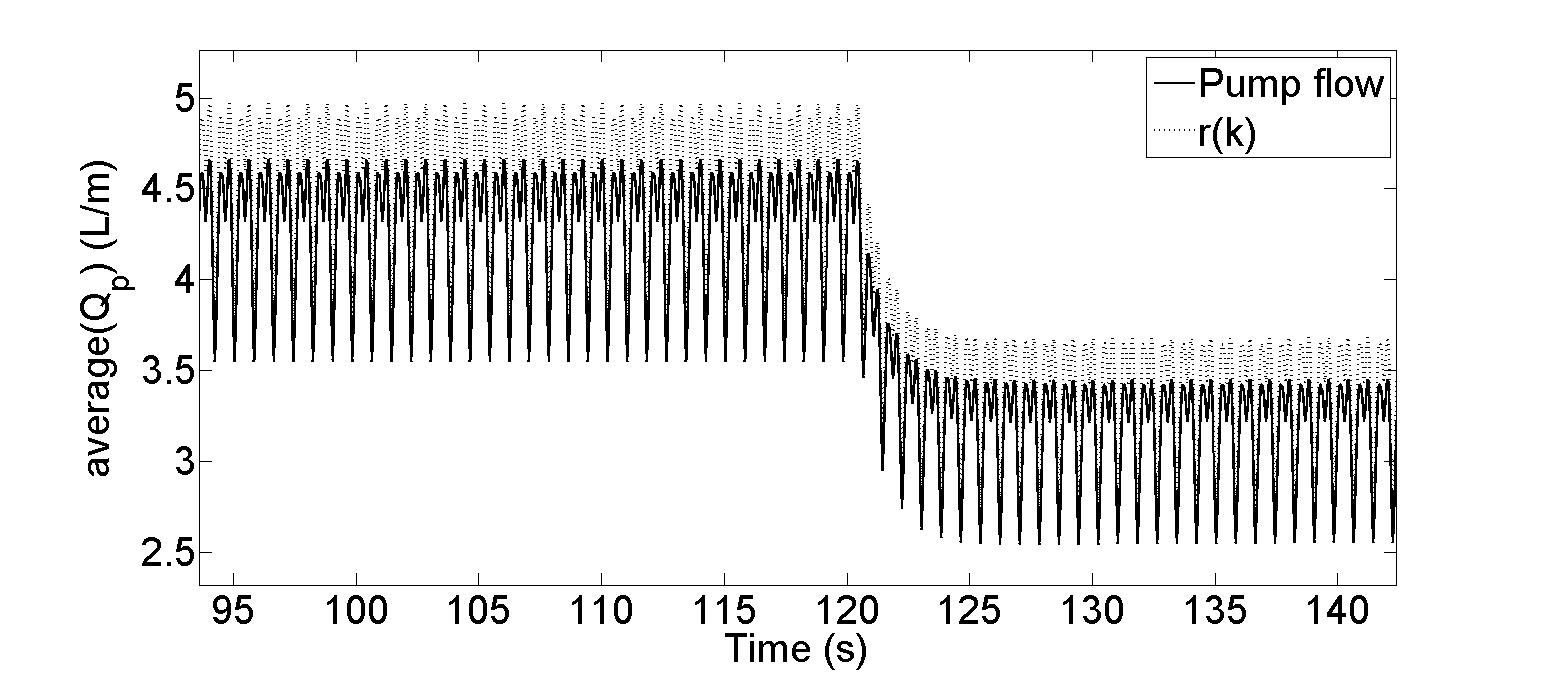}
   \label{52h}
 }

\subfigure[Measured steady state pump flow against estimated pump flow.]{
   \includegraphics[scale =0.1752] {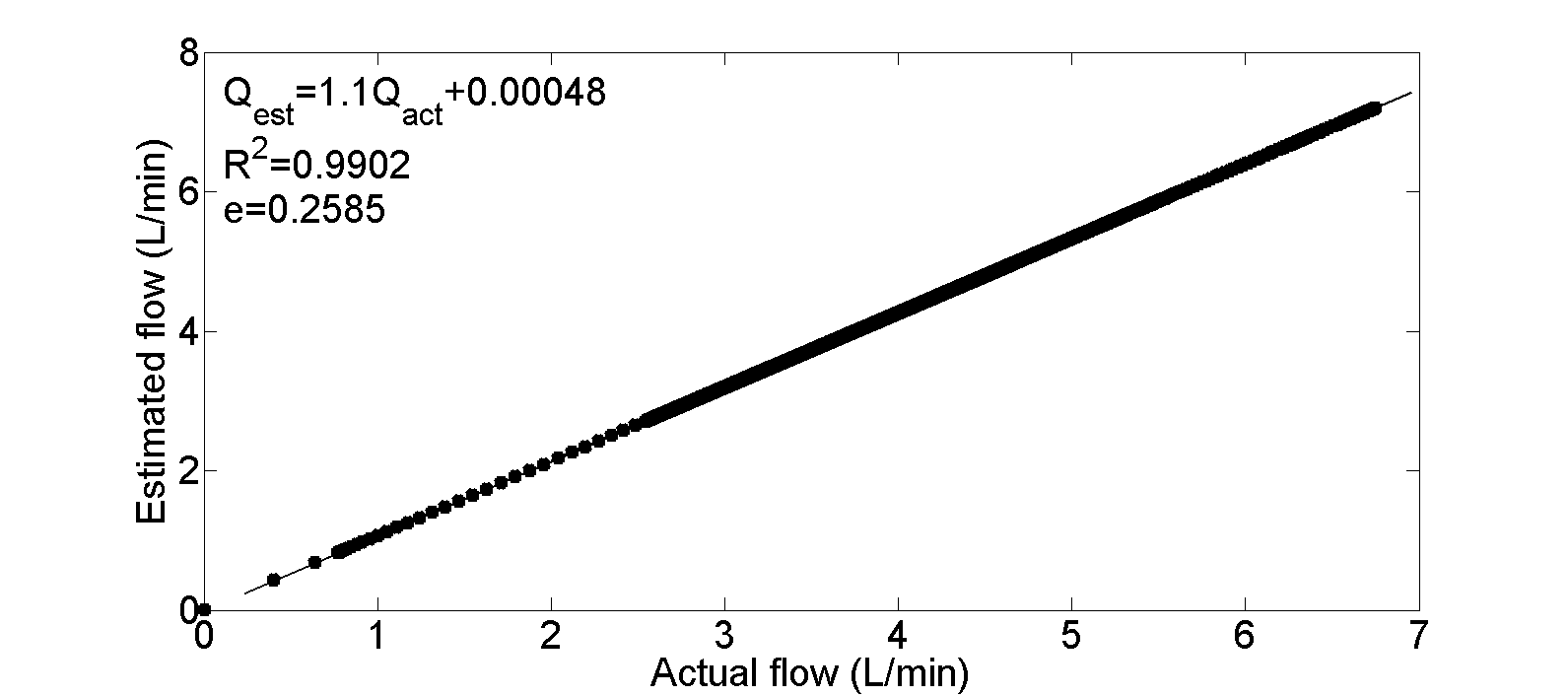}
   \label{52i}
 }

\caption{Pump variable results in rest condition when the system induced at 120s.}
\label{5:20b}
\end{figure}

Figures \ref{5:30a} - \ref{5:20b} illustrate the simulation results of the immediate response of the controller corresponding to the blood loss (rest condition) for a HF patients with an LVAD. It can be seen that the reduction in blood volume caused a reduction in stroke volume of the right ventricle. This is associated with a shift to the left of the LV pressure-volume loop, causing a reduction in LV end-diastolic and end-systolic volumes and pressure, while a  slight  shift to the right of the RV pressure-volume loops. As a result, the LVAD successfully increases the aortic pressure $P_{ao}$ and  decreases the left atrial pressure $P_{la}$ and keeps the right atrial pressure $P_{ra}$ within safe operating mode as can be seen in Figures \ref{5:30a}, \ref{5:60a}, \ref{5:90a} and \ref{5:20a}.

The Pump variable results are illustrated in Figures (\ref{5:30b}, \ref{5:60b}, \ref{5:90b} and \ref{5:20b}). It can be seen that the simulated pump flow accurately tracks the reference signal within an error of $\pm$ 0.34 L/min. During parameters variations, the controller responds to the  decrease average pump rotational speed from 2900 rpm to 2000 rpm and average actual pump flow from 4.6 L/min to 3.2 L/min. And consequently, estimated average pulsatile flow has been decreased from 4.8 L/min to 3.4 L/min. These changes are substantially completed within four heartbeats. Furthermore, both actual and estimated pump flows indicate that there  is extremely close correlation between the two flows. Also, the correlation between actual and estimated flow is highly significant, and the slope is 1.1 for the linear regression (see Table \ref{5t2}).


\subsection{Results in Exercise Condition}

\begin{figure*}[htbp]
\centering
\subfigure[LV volume versus LV pressure before and after Parameter Change.]{
   \includegraphics[scale =0.16752] {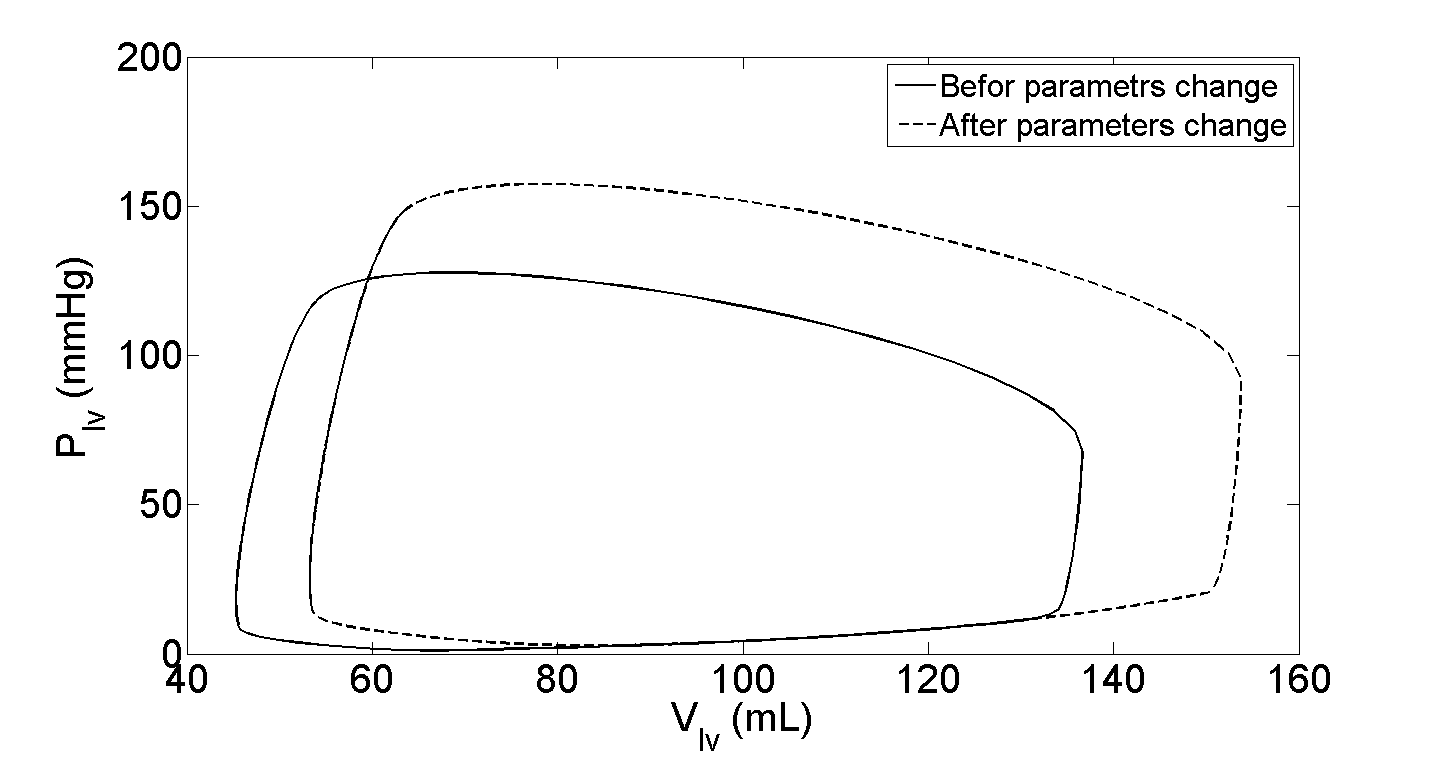}
   \label{53ea}
 }
\subfigure[RV volume versus RV pressure before and after Parameter Change.]{
   \includegraphics[scale =0.16752] {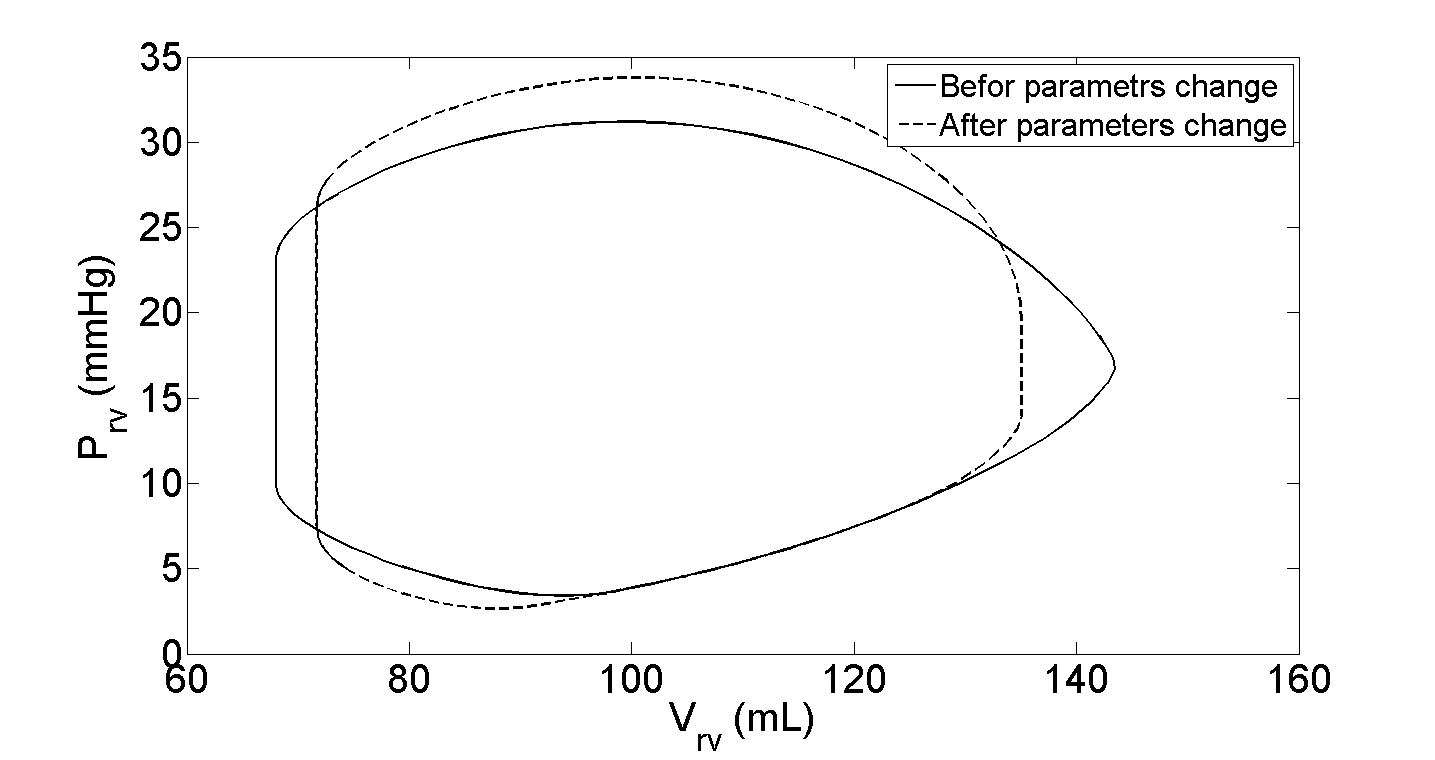}
   \label{53eb}
 }

 \subfigure[Aortic pressure.]{
   \includegraphics[scale =0.16752] {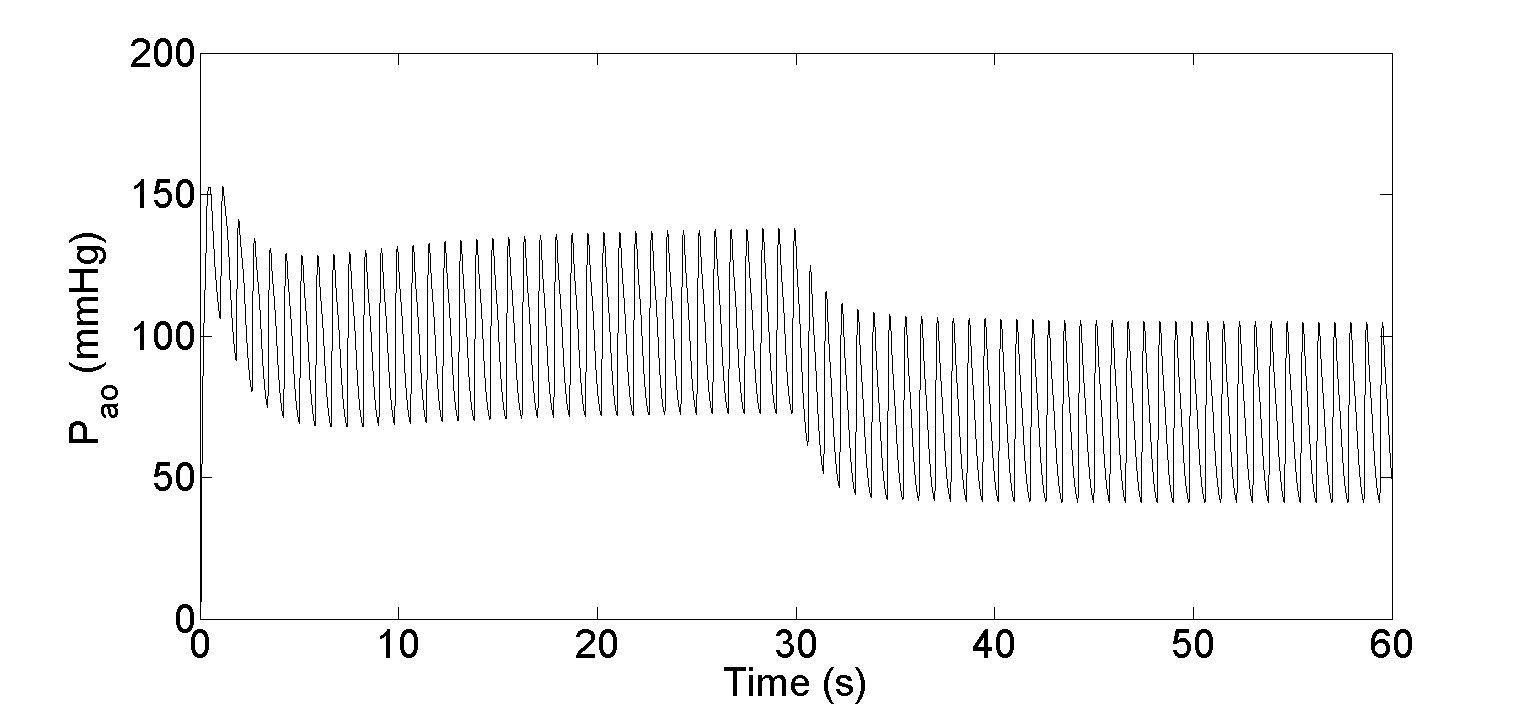}
   \label{53ec}
 }
  \subfigure[Left atrial pressure.]{
   \includegraphics[scale =0.16752] {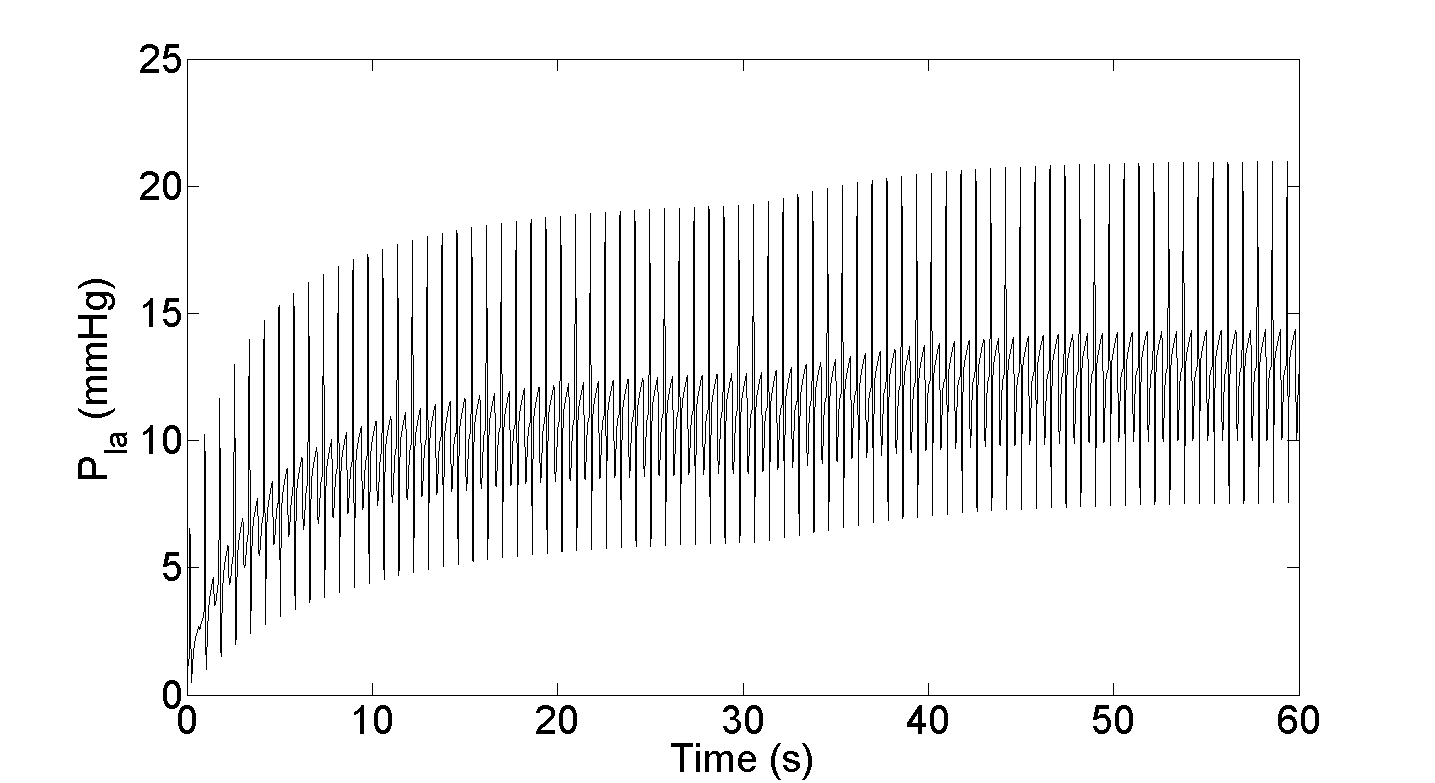}
   \label{53ed}
 }

\subfigure[Right atrial pressure.]{
   \includegraphics[scale =0.16752] {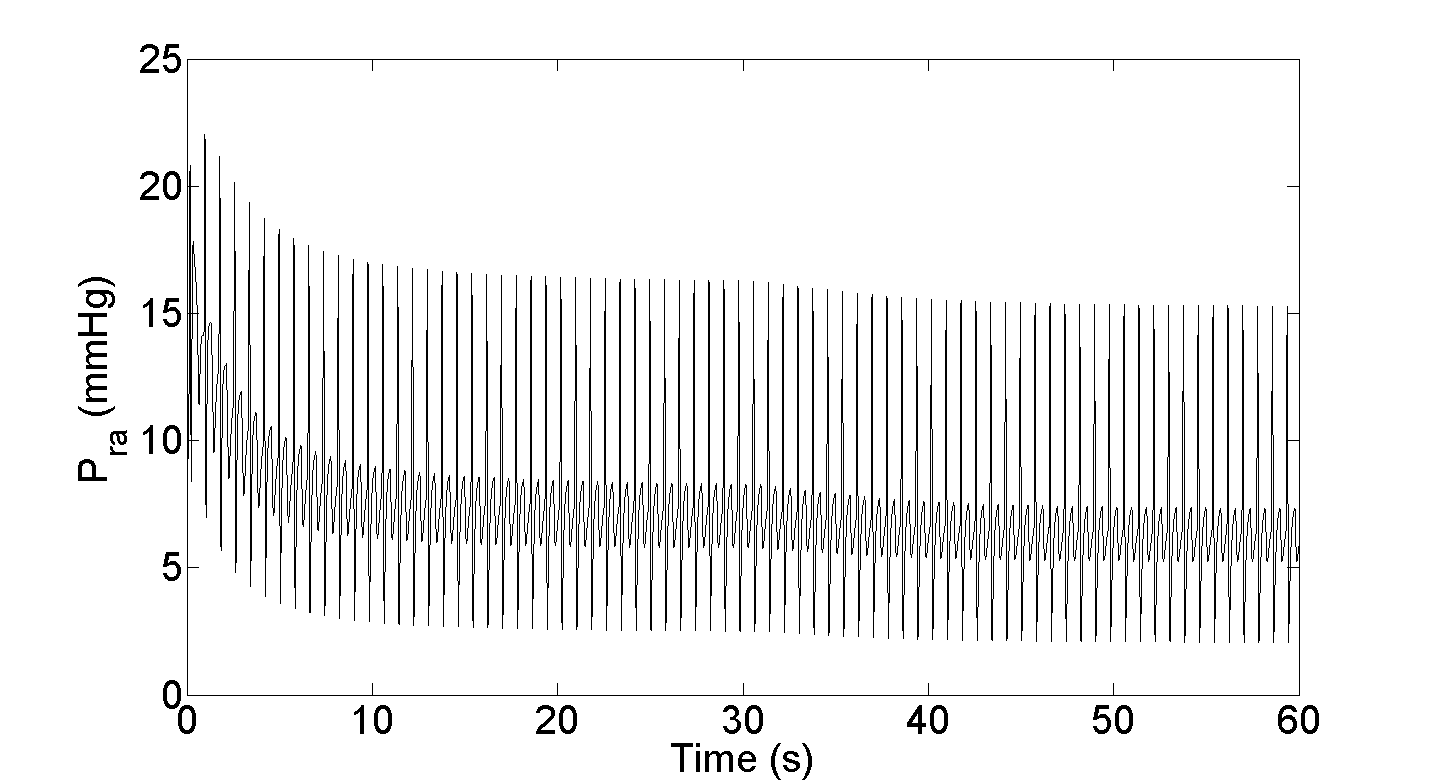}
   \label{53ee}
 }
\caption{Hemodynamic variables results in exercise condition when the system induced at 30s.}
\label{5:30ea}
\end{figure*}

\begin{figure*}[htbp]
\centering
\subfigure[Average pump speed.]{
   \includegraphics[scale =0.1752] {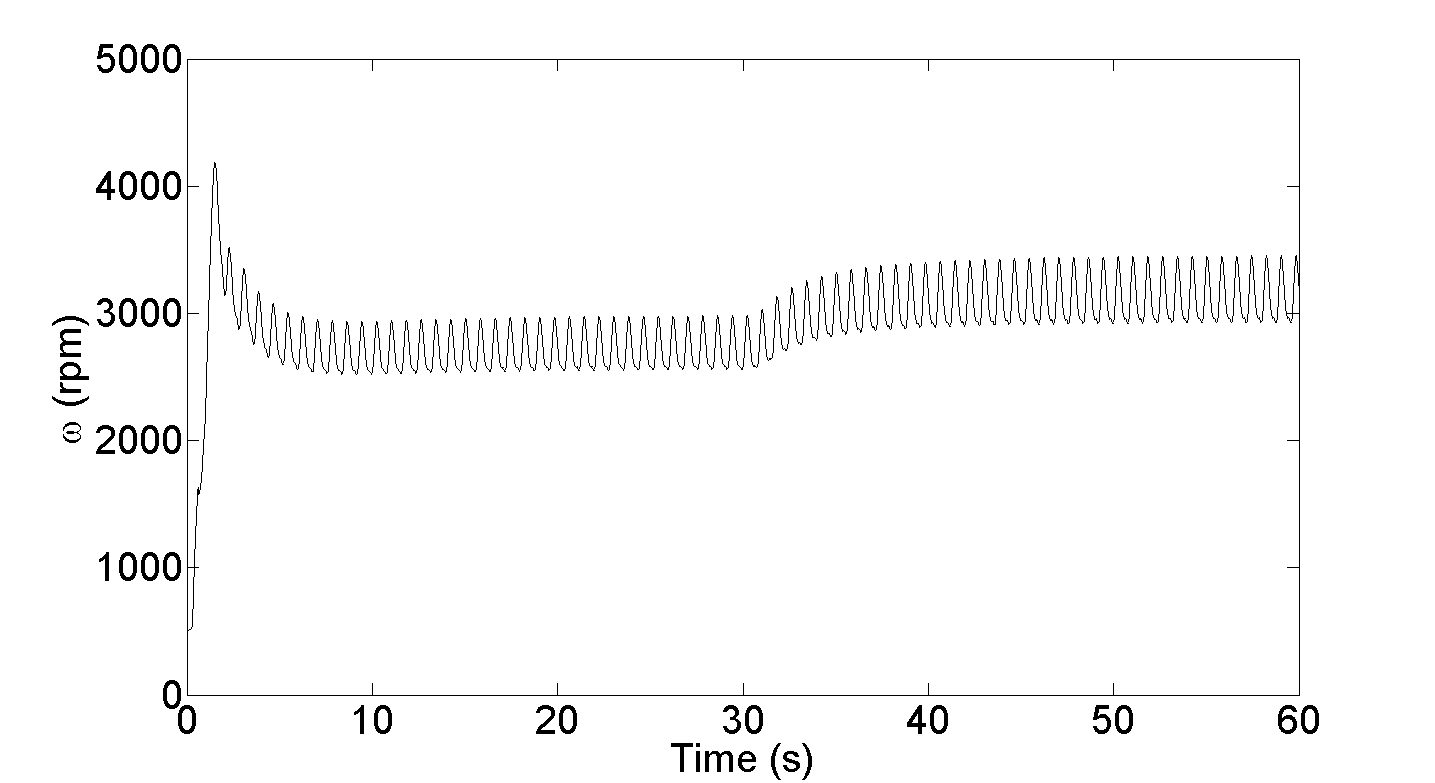}
   \label{53ef}
 }

  \subfigure[Pump flow compared with reference signal.]{
   \includegraphics[scale =0.1752] {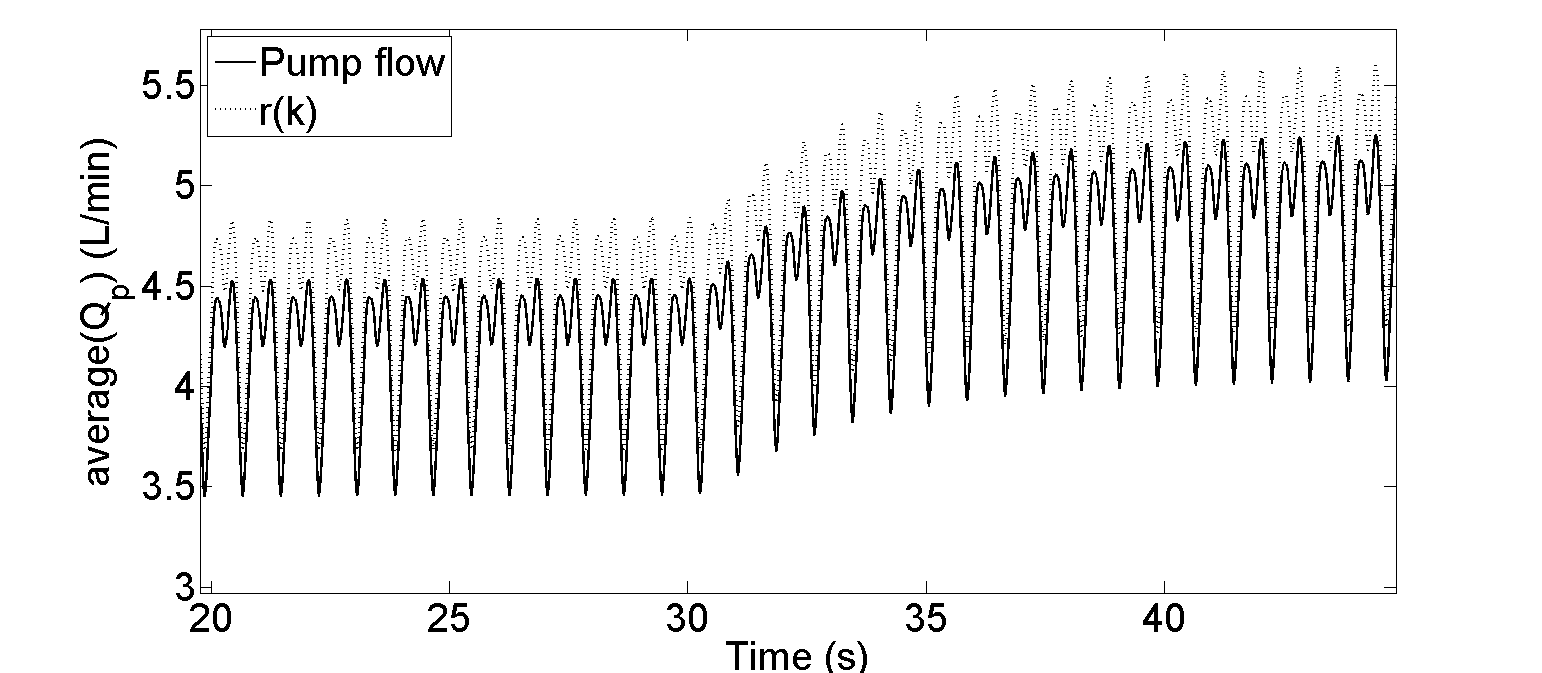}
   \label{53eh}
 }

\subfigure[Measured steady state pump flow against estimated pump flow.]{
   \includegraphics[scale =0.1752] {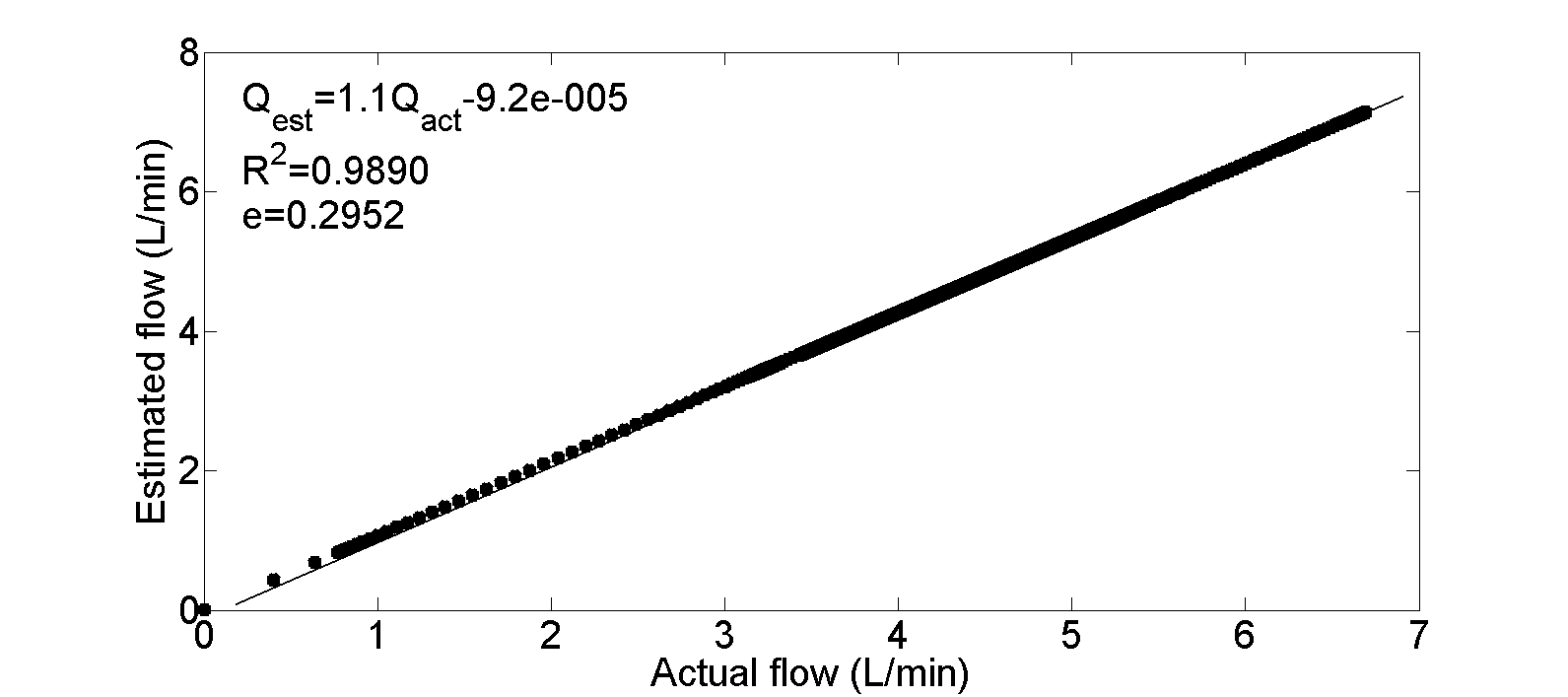}
   \label{53ei}
 }

\caption{Pump variable results in exercise condition when the system induced at 30s.}
\label{5:30eb}
\end{figure*}

\begin{figure*}[htbp]
\centering
\subfigure[LV volume versus LV pressure before and after Parameter Change.]{
   \includegraphics[scale =0.16752] {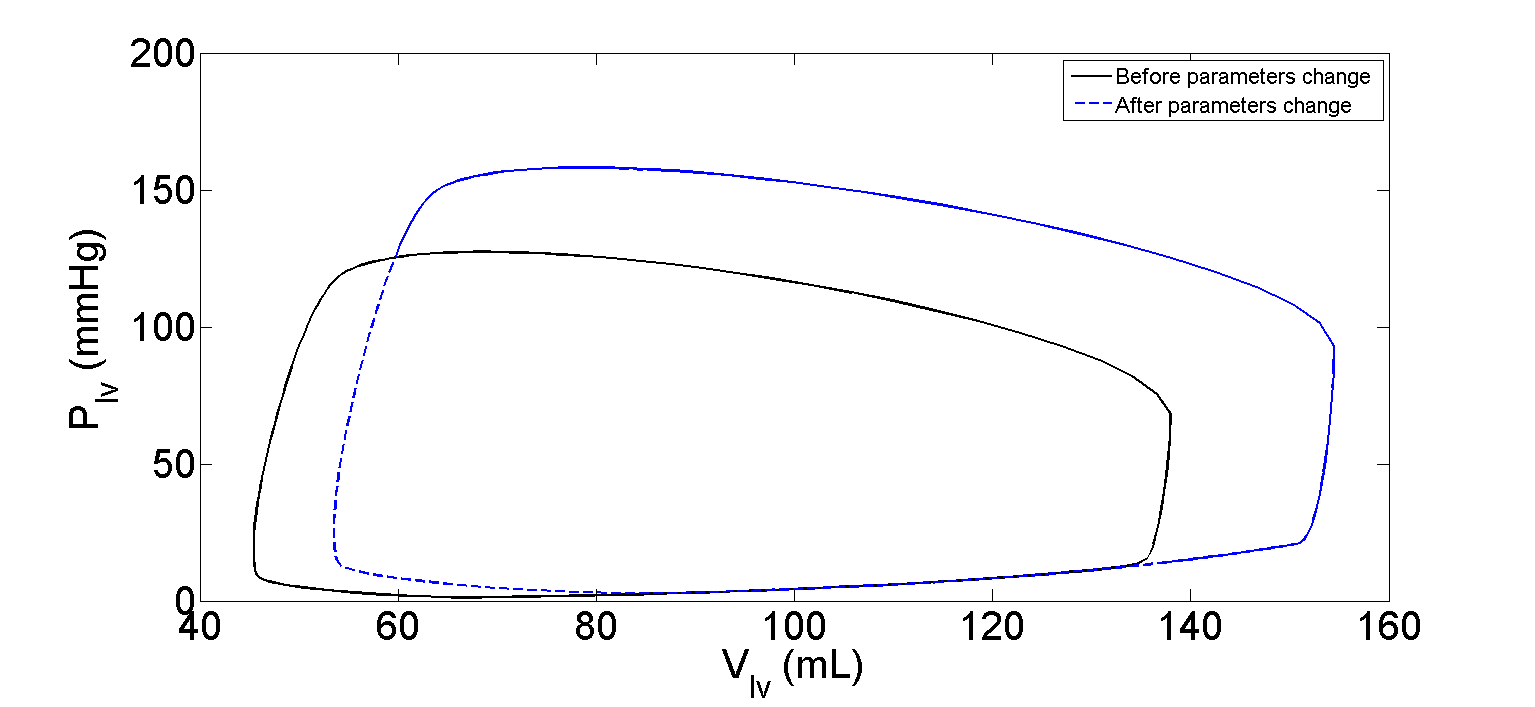}
   \label{56ea}
 }
\subfigure[RV volume versus RV pressure before and after Parameter Change.]{
   \includegraphics[scale =0.16752] {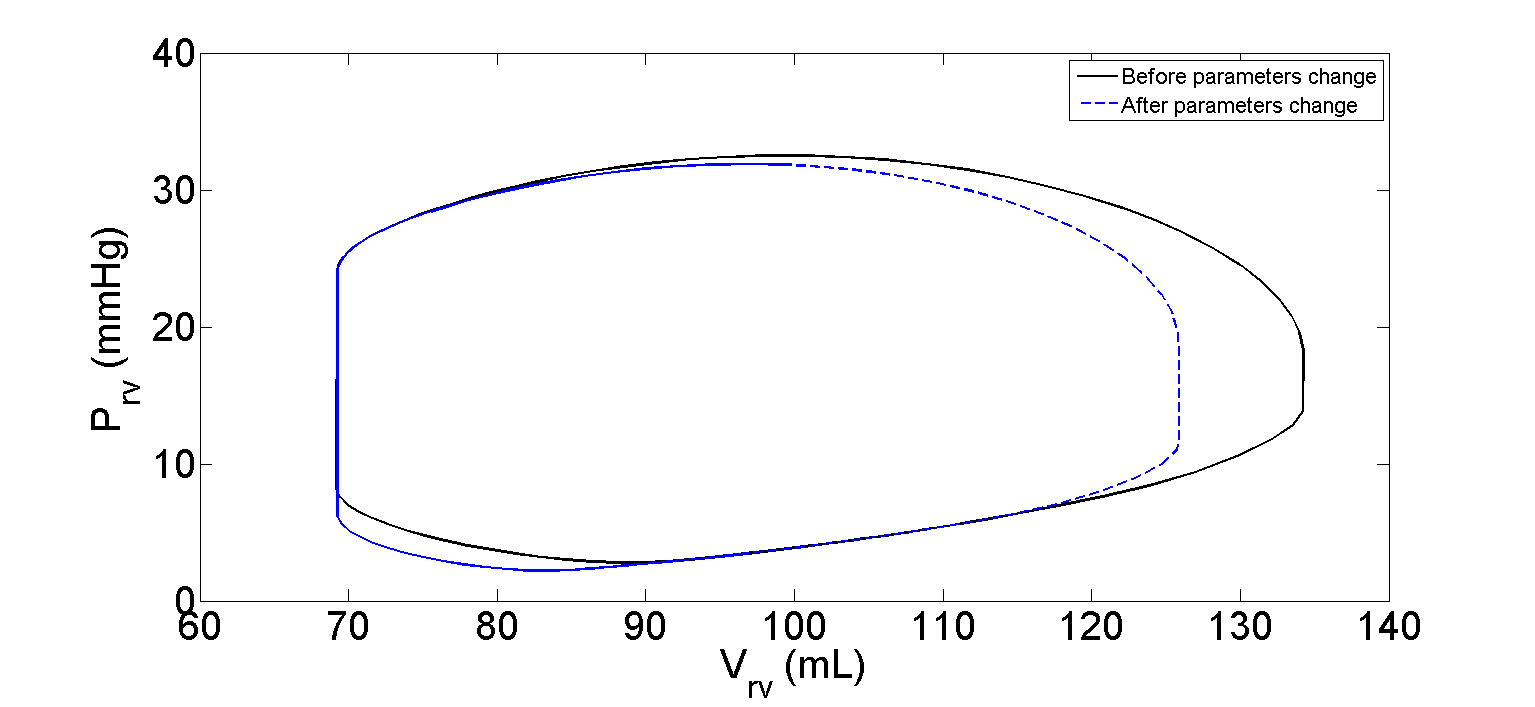}
   \label{56eb}
 }

 \subfigure[Aortic pressure.]{
   \includegraphics[scale =0.16752] {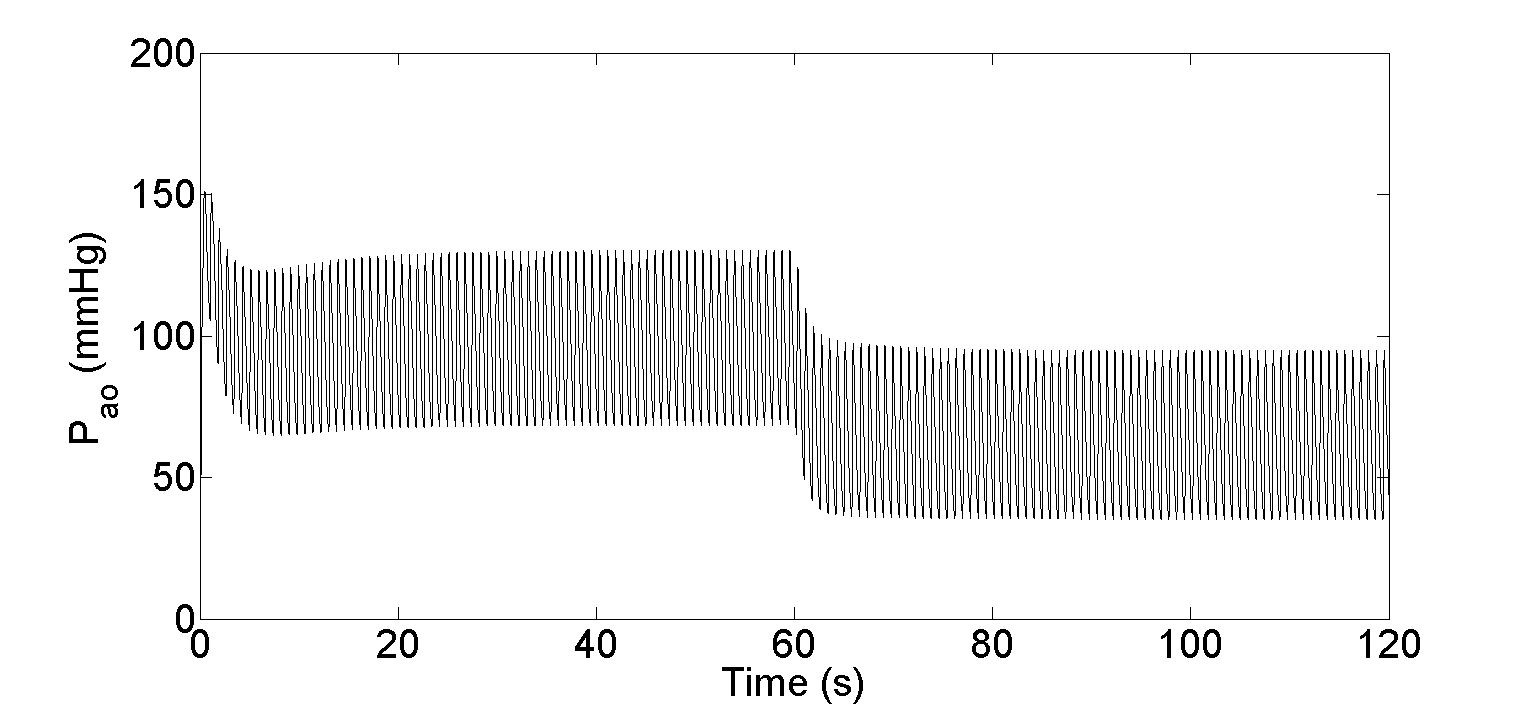}
   \label{56ec}
 }
  \subfigure[Left atrial pressure.]{
   \includegraphics[scale =0.16752] {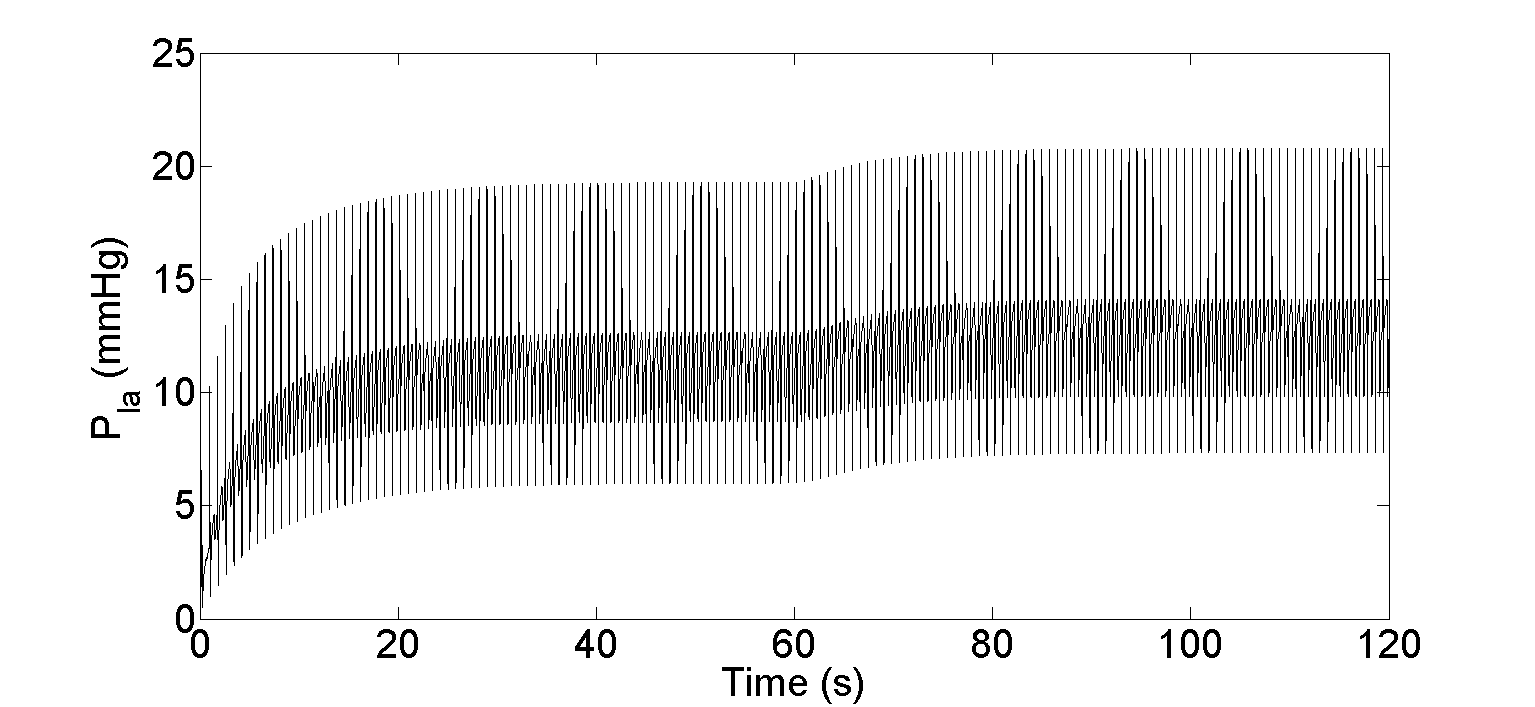}
   \label{56ed}
 }

\subfigure[Right atrial pressure.]{
   \includegraphics[scale =0.16752] {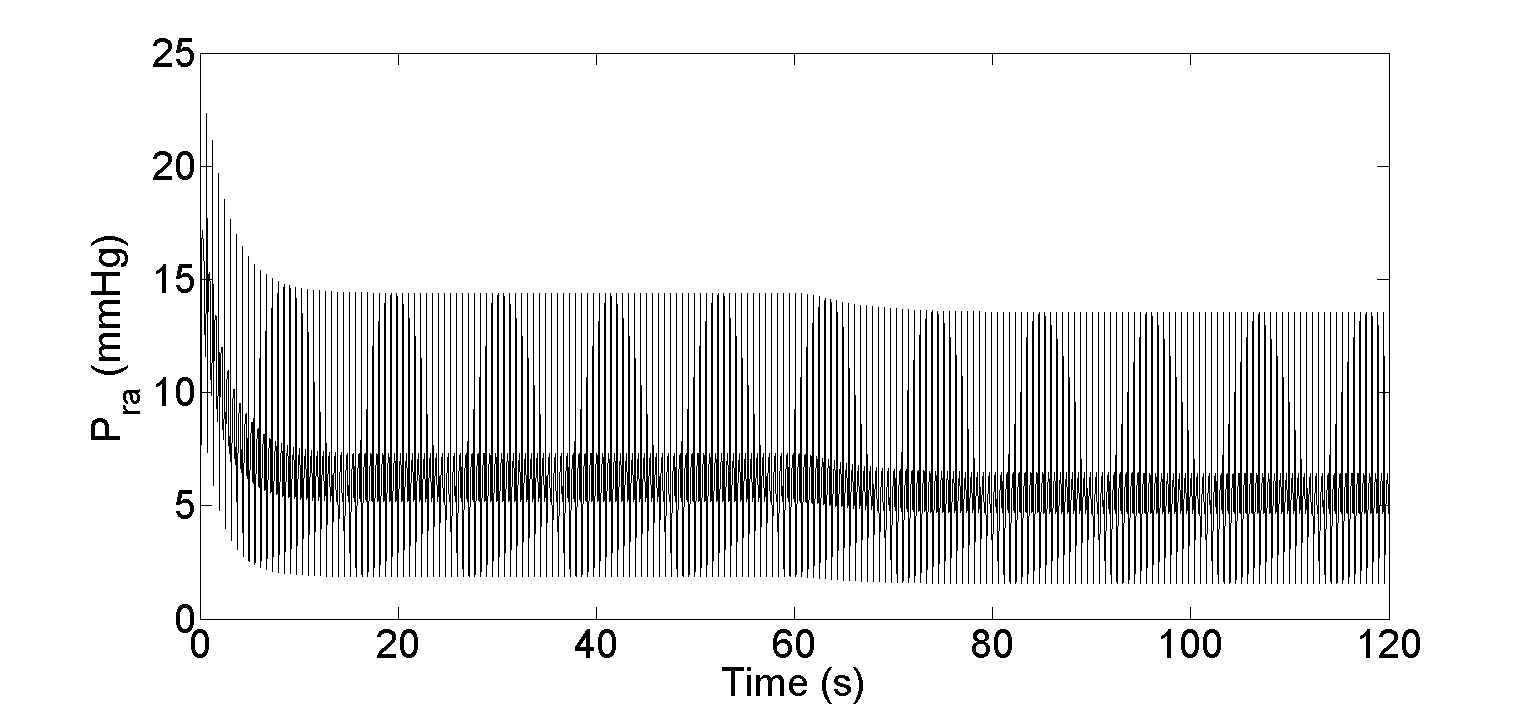}
   \label{56ee}
 }
\caption{Hemodynamic variables results in rest condition when the system induced at 60s.}
\label{5:60ea}
\end{figure*}

\begin{figure}[htbp]
\centering
\subfigure[Average pump speed.]{
   \includegraphics[scale =0.16752] {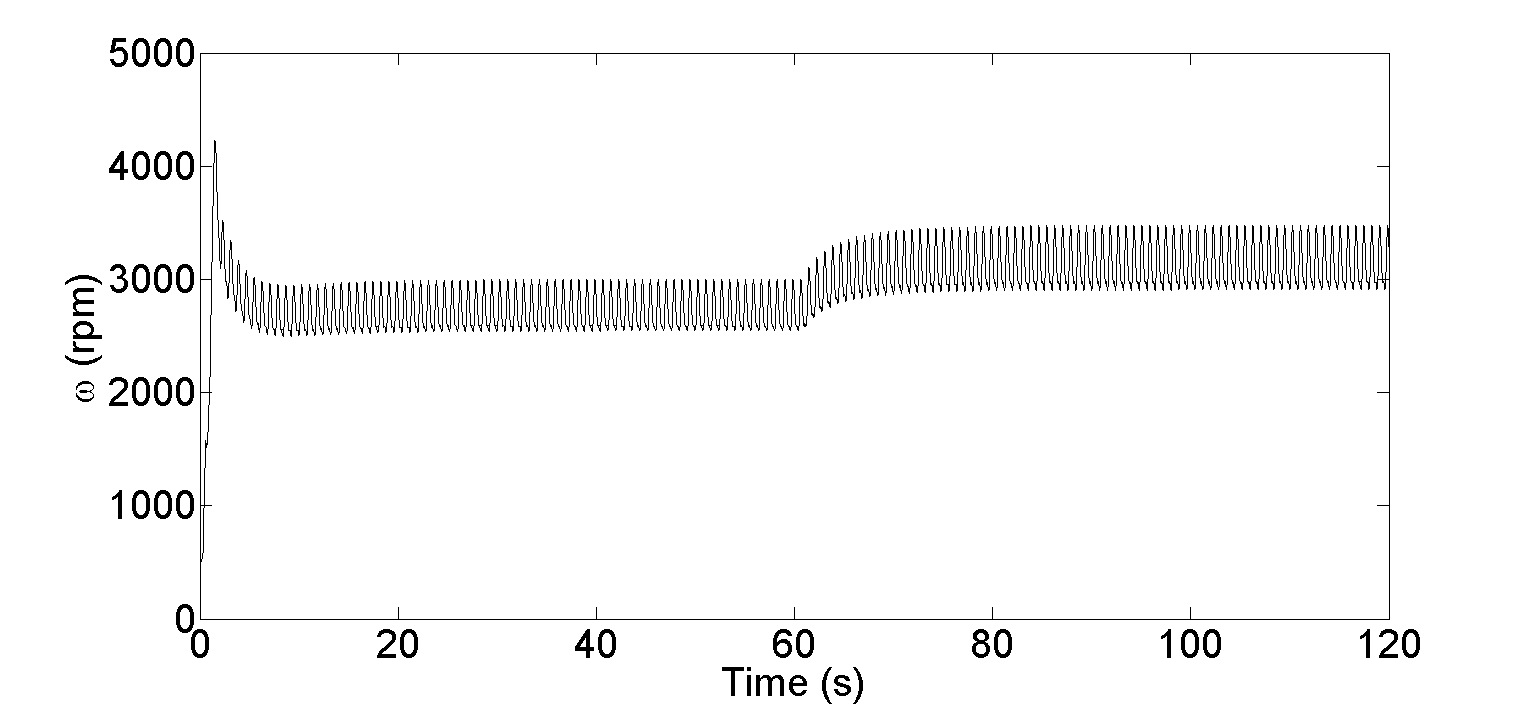}
   \label{56ef}
 }

  \subfigure[Pump flow compared with reference signal.]{
   \includegraphics[scale =0.16752] {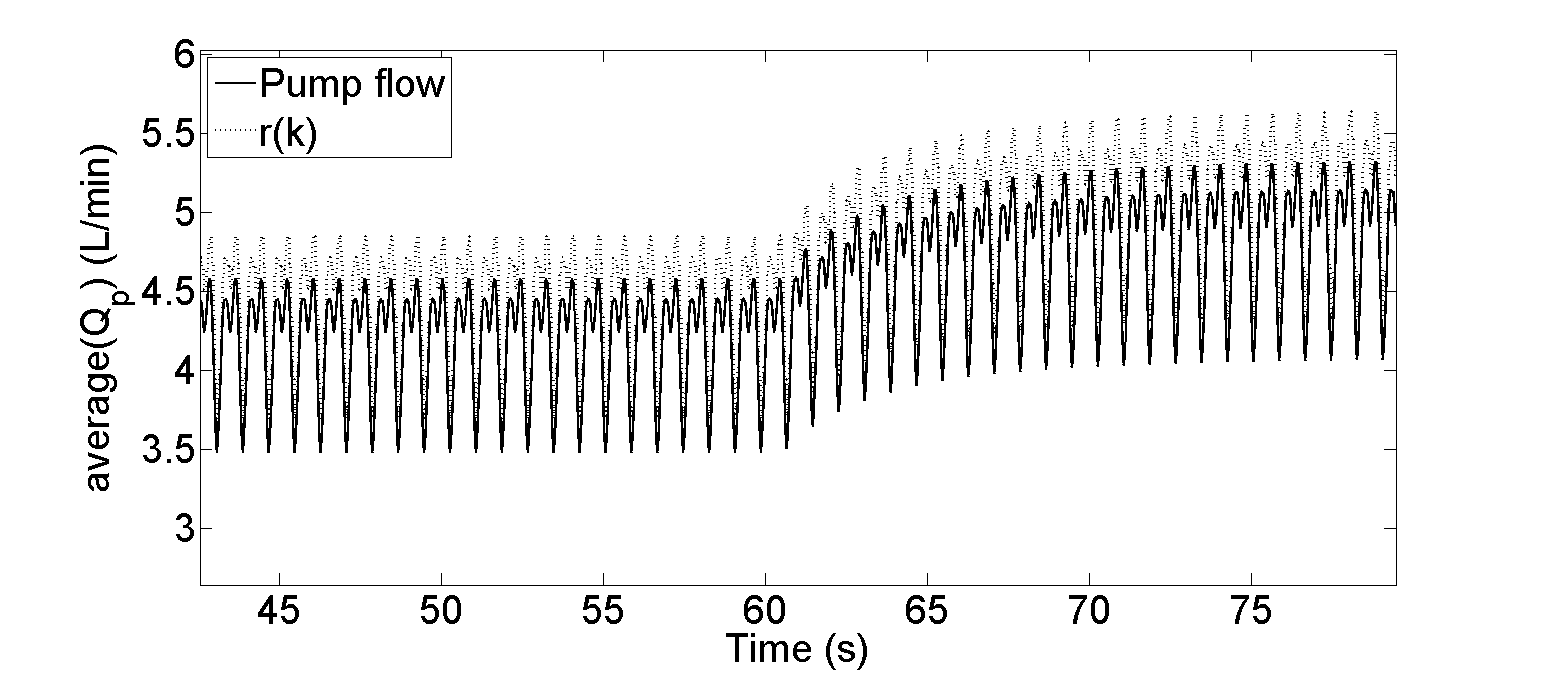}
   \label{56eh}
 }

\subfigure[Measured steady state pump flow against estimated pump flow.]{
   \includegraphics[scale =0.16752] {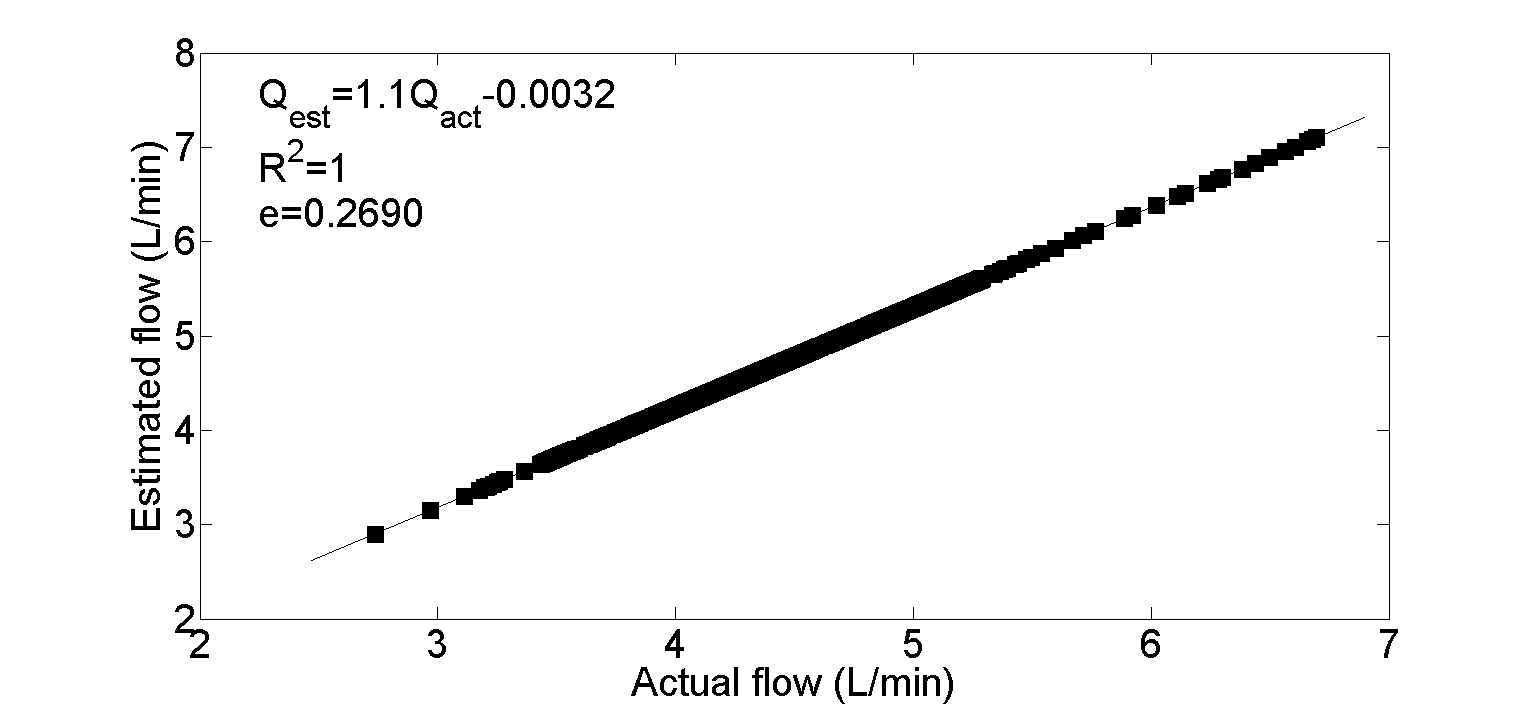}
   \label{56ei}
 }

\caption{Pump variable results in exercise condition when the system induced at 60s.}
\label{5:60eb}
\end{figure}

\begin{figure*}[htbp]
\centering
\subfigure[LV volume versus LV pressure before and after Parameter Change.]{
   \includegraphics[scale =0.16752] {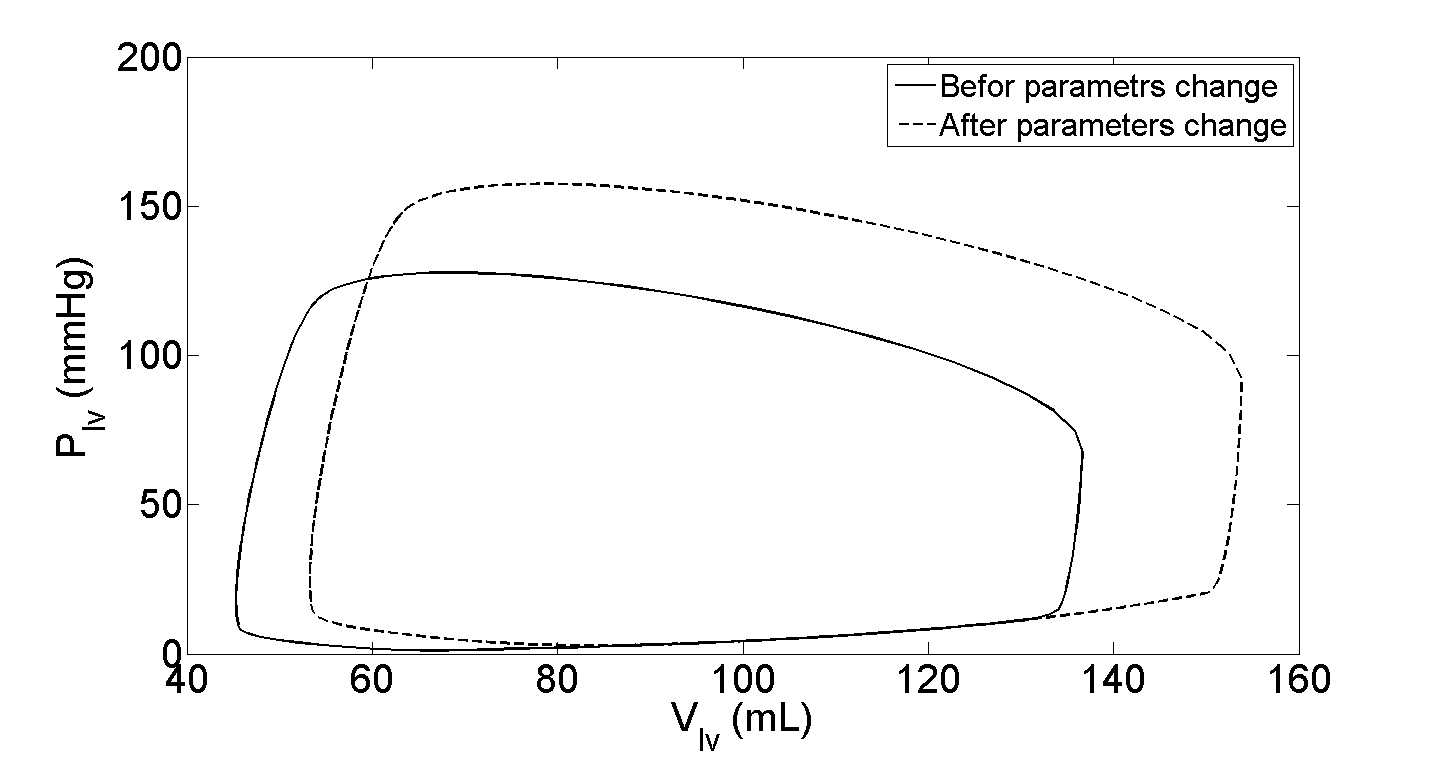}
   \label{59ea}
 }
\subfigure[RV volume versus RV pressure before and after Parameter Change.]{
   \includegraphics[scale =0.16752] {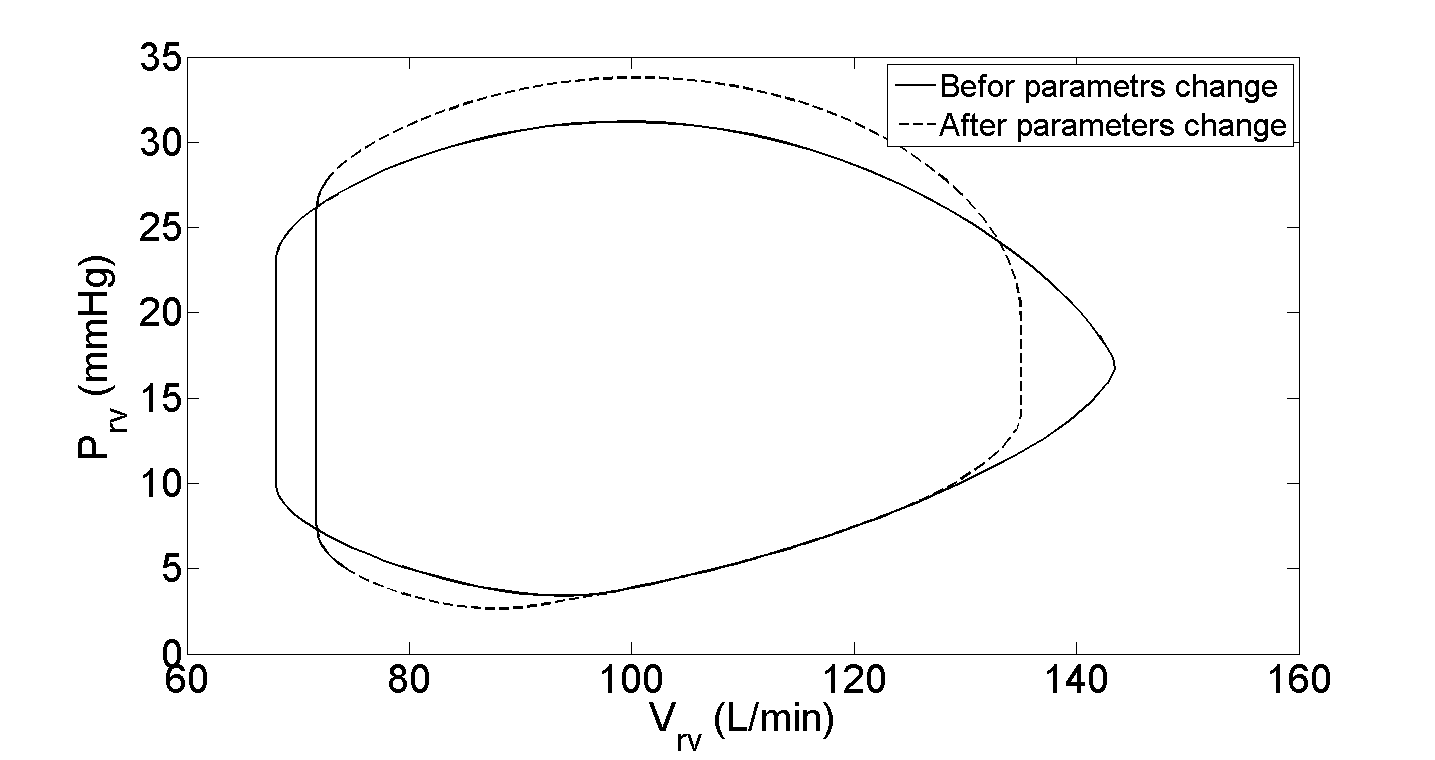}
   \label{59eb}
 }

 \subfigure[Aortic pressure.]{
   \includegraphics[scale =0.16752] {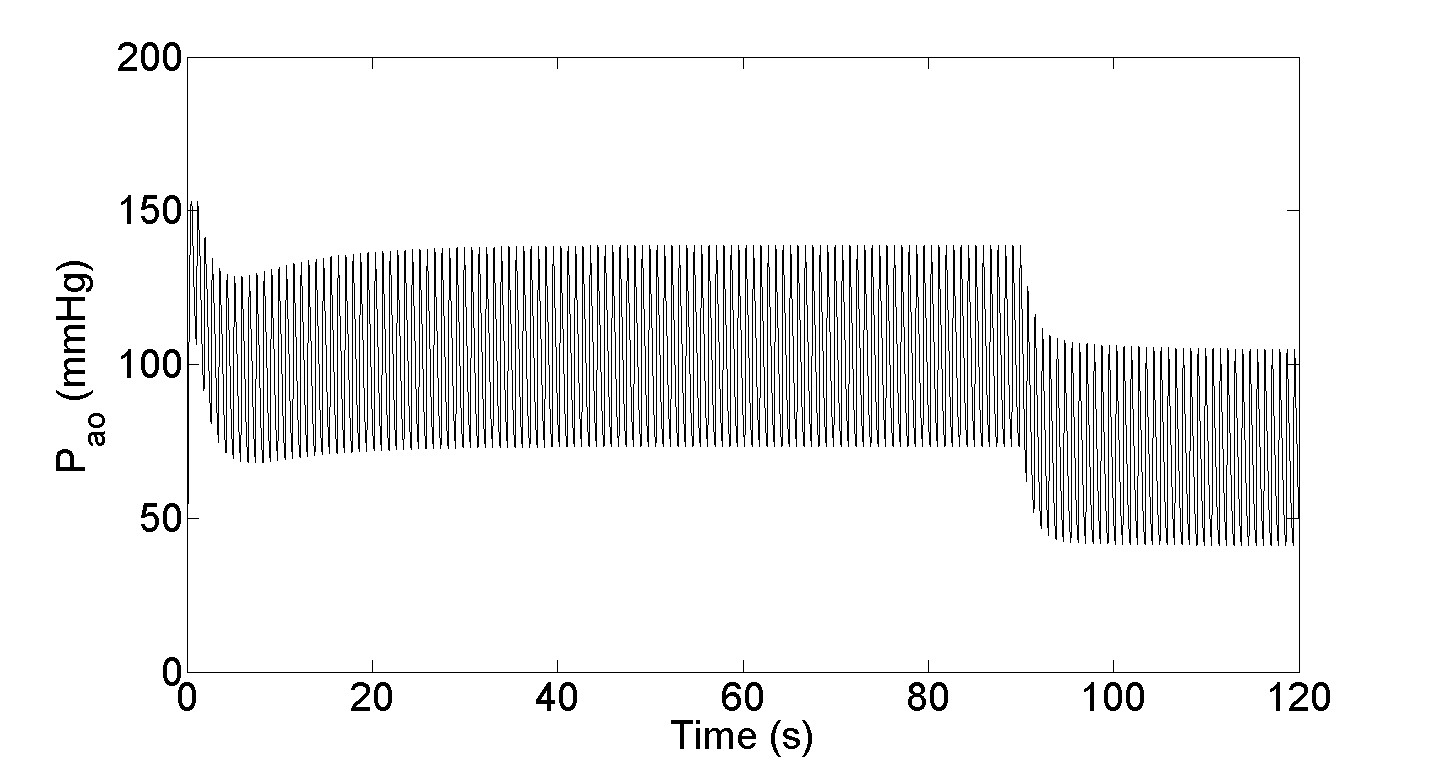}
   \label{59ec}
 }
  \subfigure[Left atrial pressure.]{
   \includegraphics[scale =0.16752] {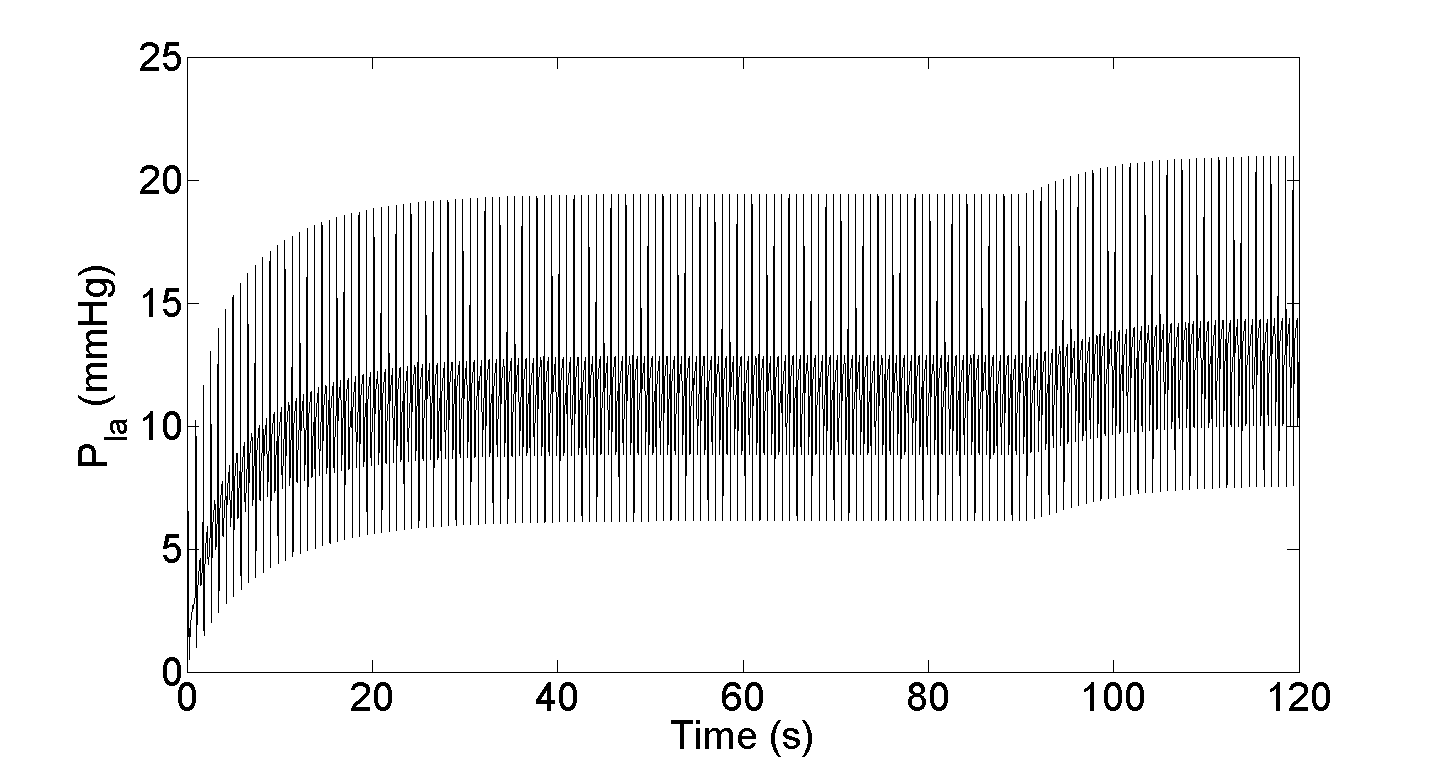}
   \label{59ed}
 }

\subfigure[Right atrial pressure.]{
   \includegraphics[scale =0.16752] {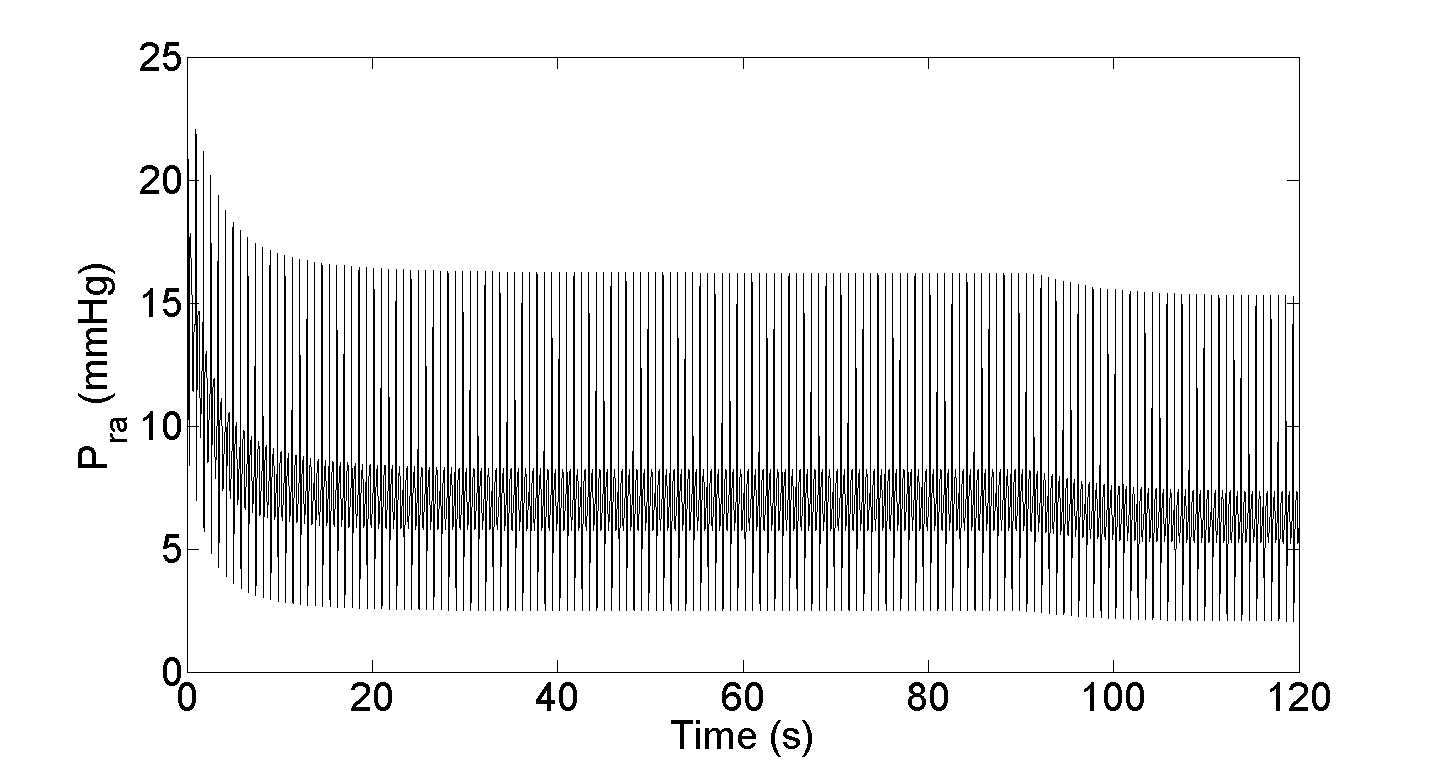}
   \label{59ee}
 }
\caption{Hemodynamic variables results in exercise condition when the system induced at 90s.}
\label{5:90ea}
\end{figure*}

\begin{figure}[htbp]
\centering
\subfigure[Average pump speed.]{
   \includegraphics[scale =0.1752] {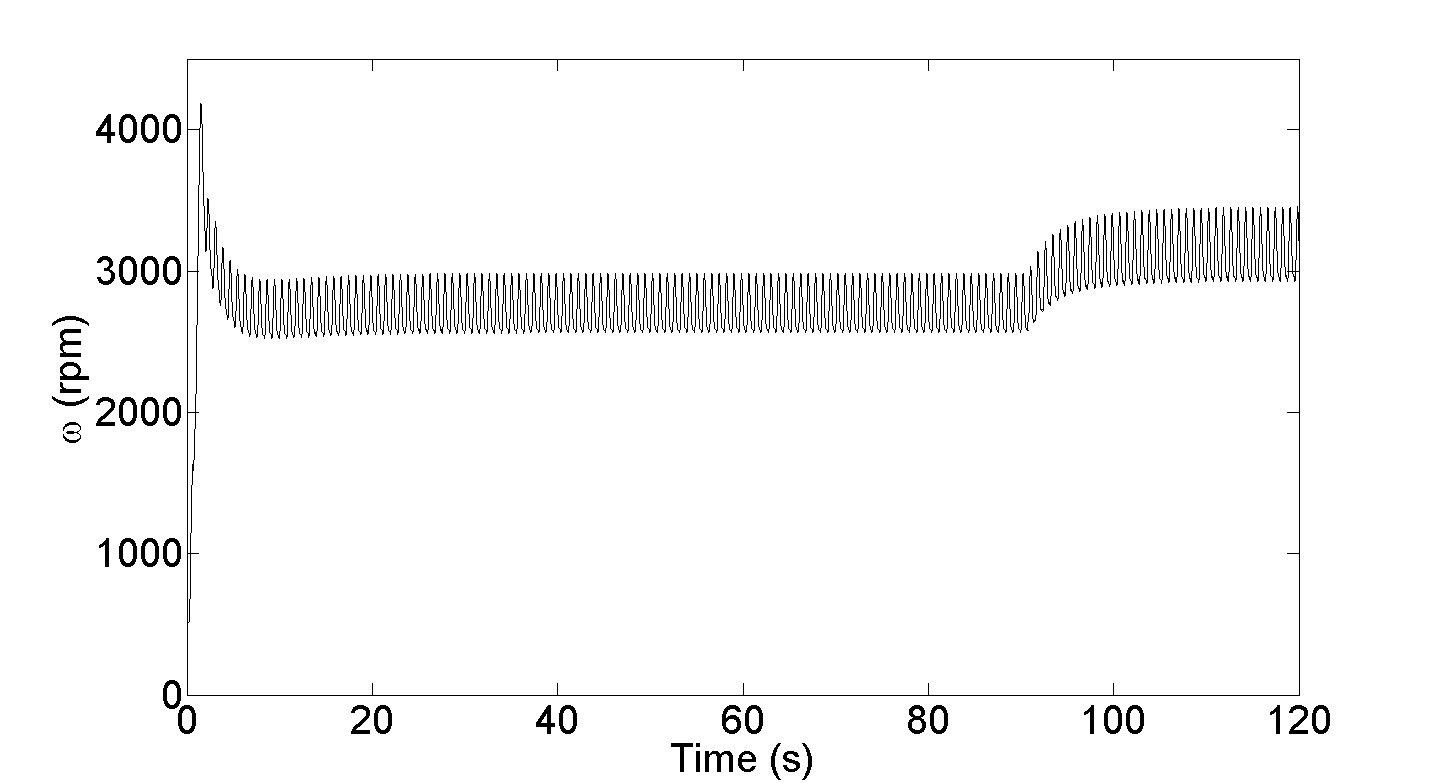}
   \label{59ef}
 }

   \subfigure[Pump flow compared with reference signal.]{
   \includegraphics[scale =0.1752] {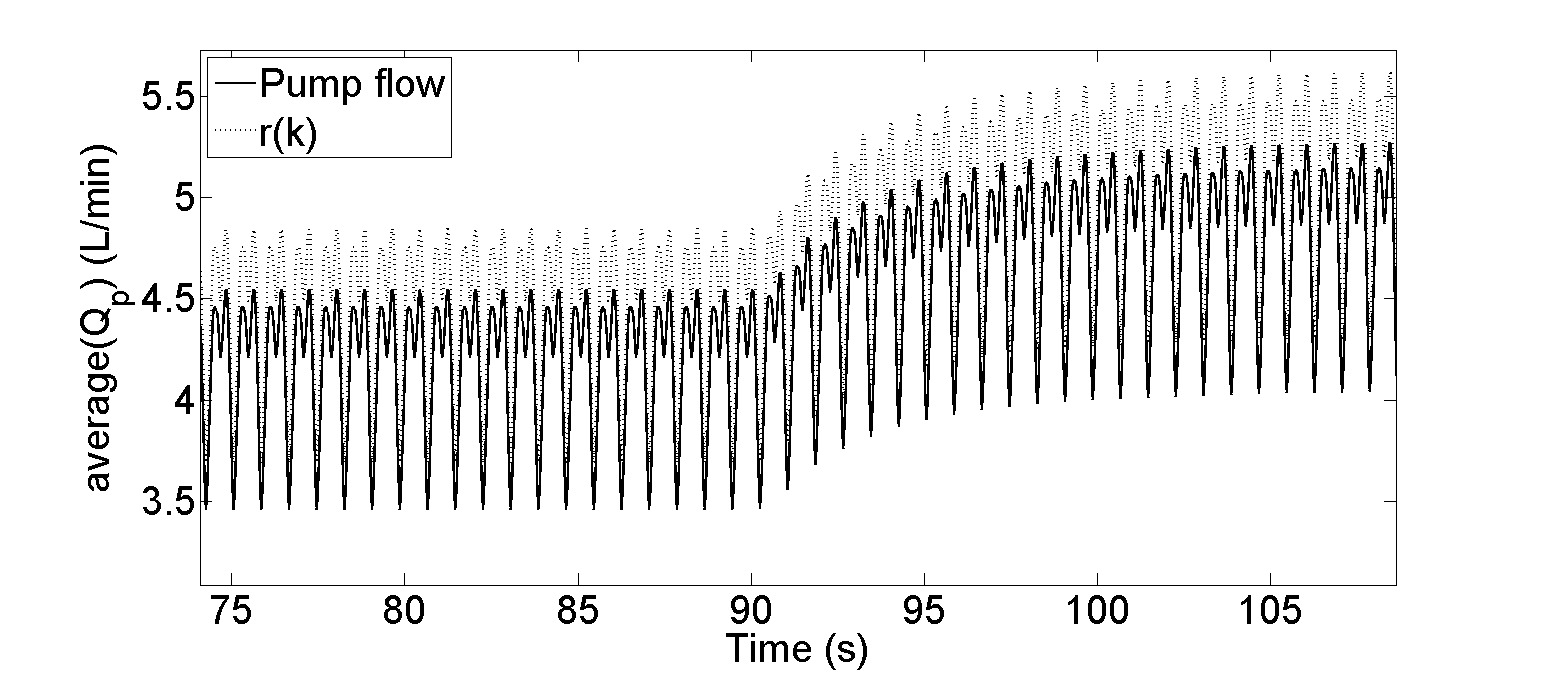}
   \label{59eh}
 }

\subfigure[Measured steady state pump flow against estimated pump flow.]{
   \includegraphics[scale =0.1752] {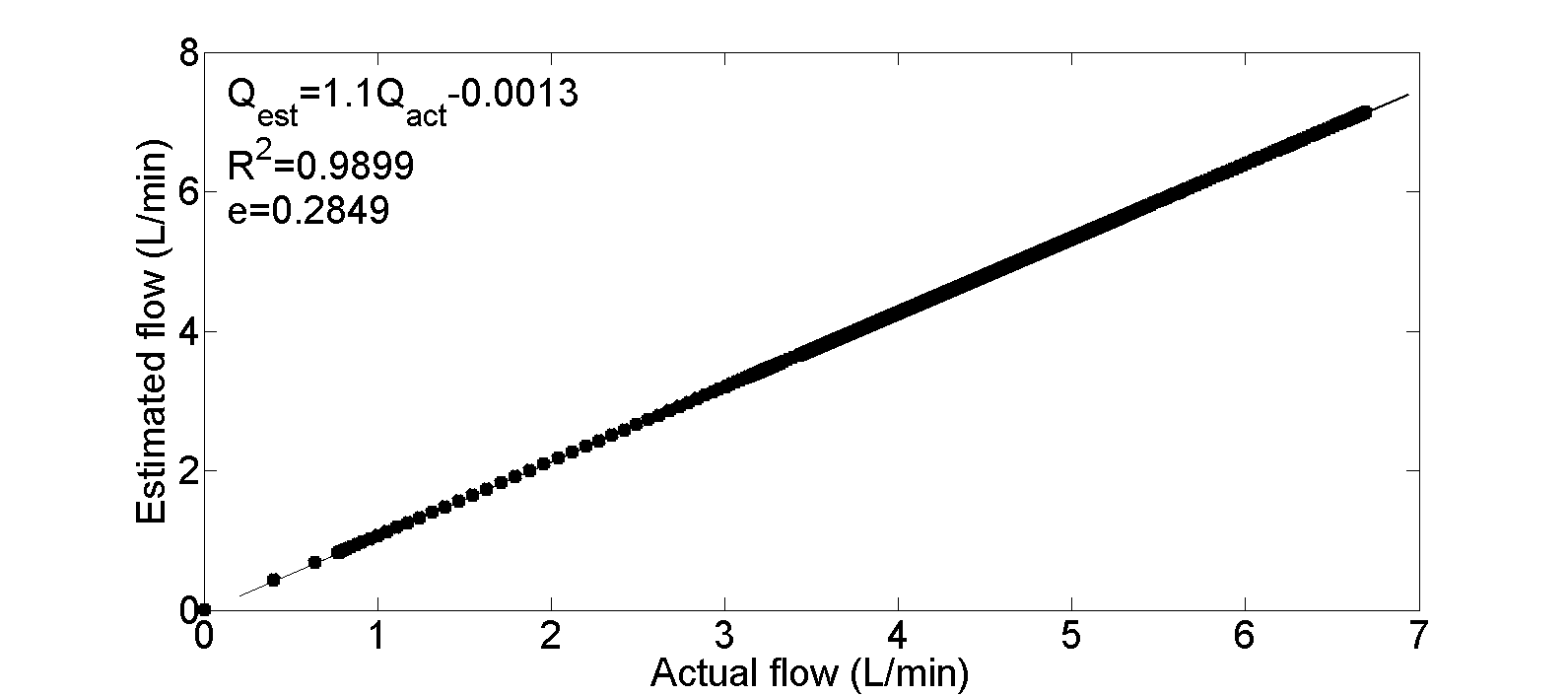}
   \label{59ei}
 }

\caption{Pump variable results in exercise condition when the system induced at 90s.}
\label{5:90eb}
\end{figure}

\begin{figure*}[htbp]
\centering
\subfigure[LV volume versus LV pressure before and after Parameter Change.]{
   \includegraphics[scale =0.16752] {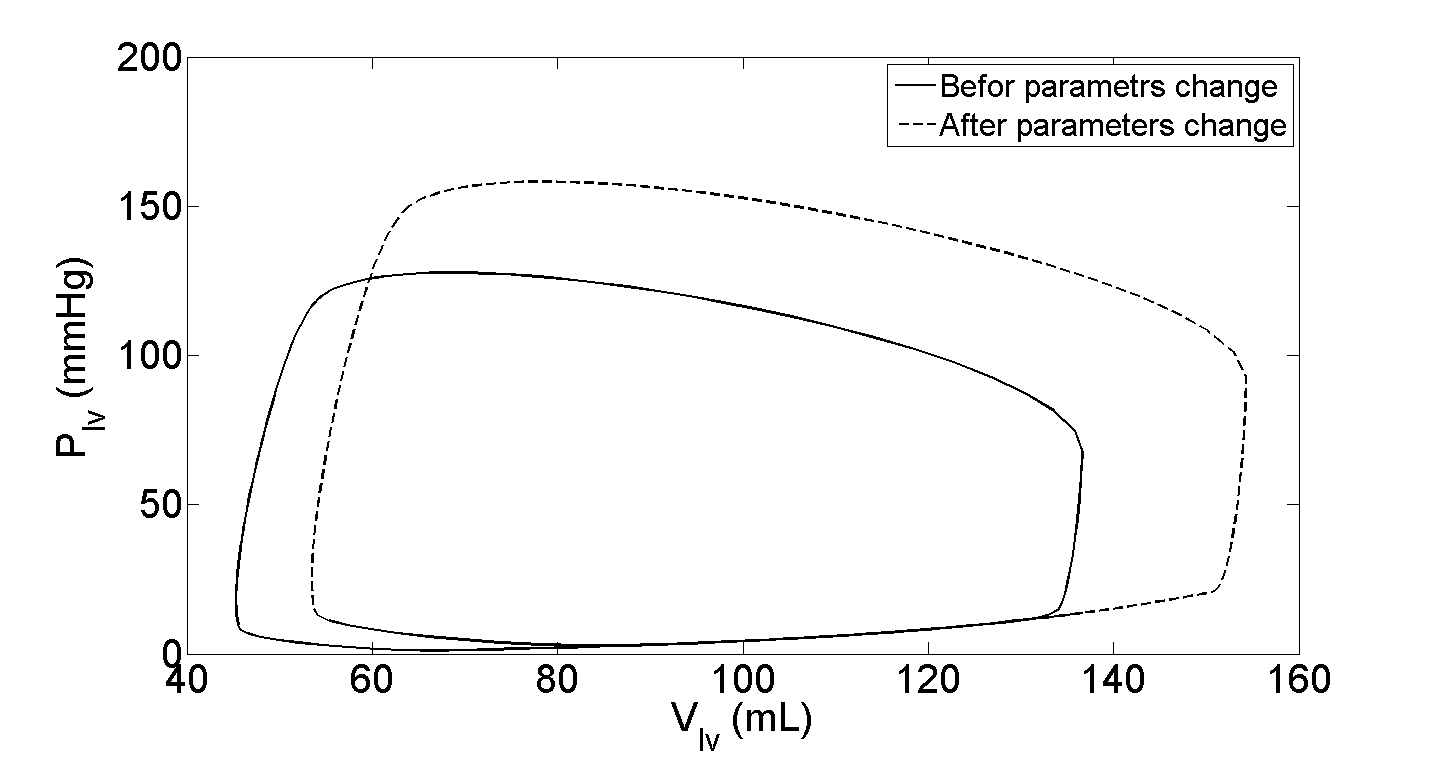}
   \label{52ea}
 }
\subfigure[RV volume versus RV pressure before and after Parameter Change.]{
   \includegraphics[scale =0.16752] {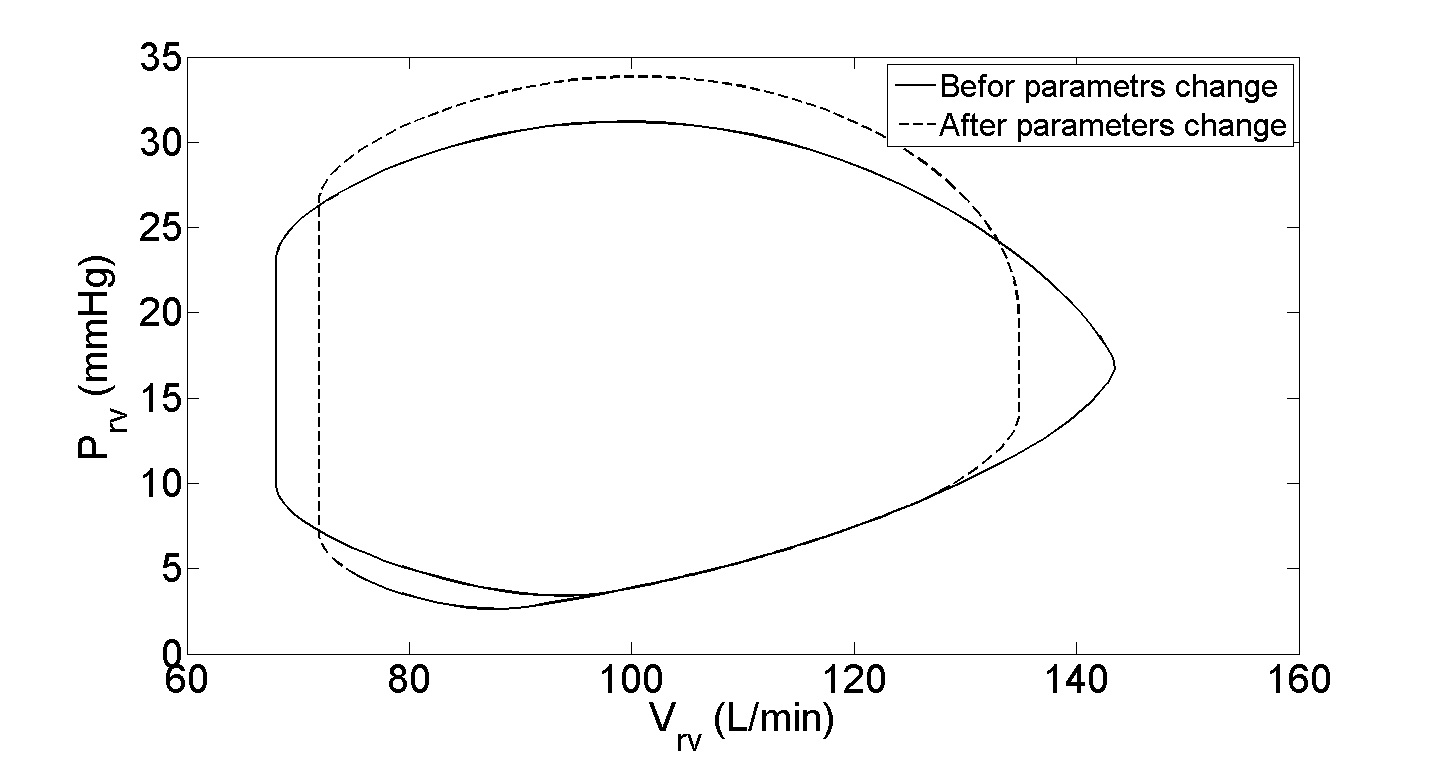}
   \label{52eb}
 }

 \subfigure[Aortic pressure.]{
   \includegraphics[scale =0.16752] {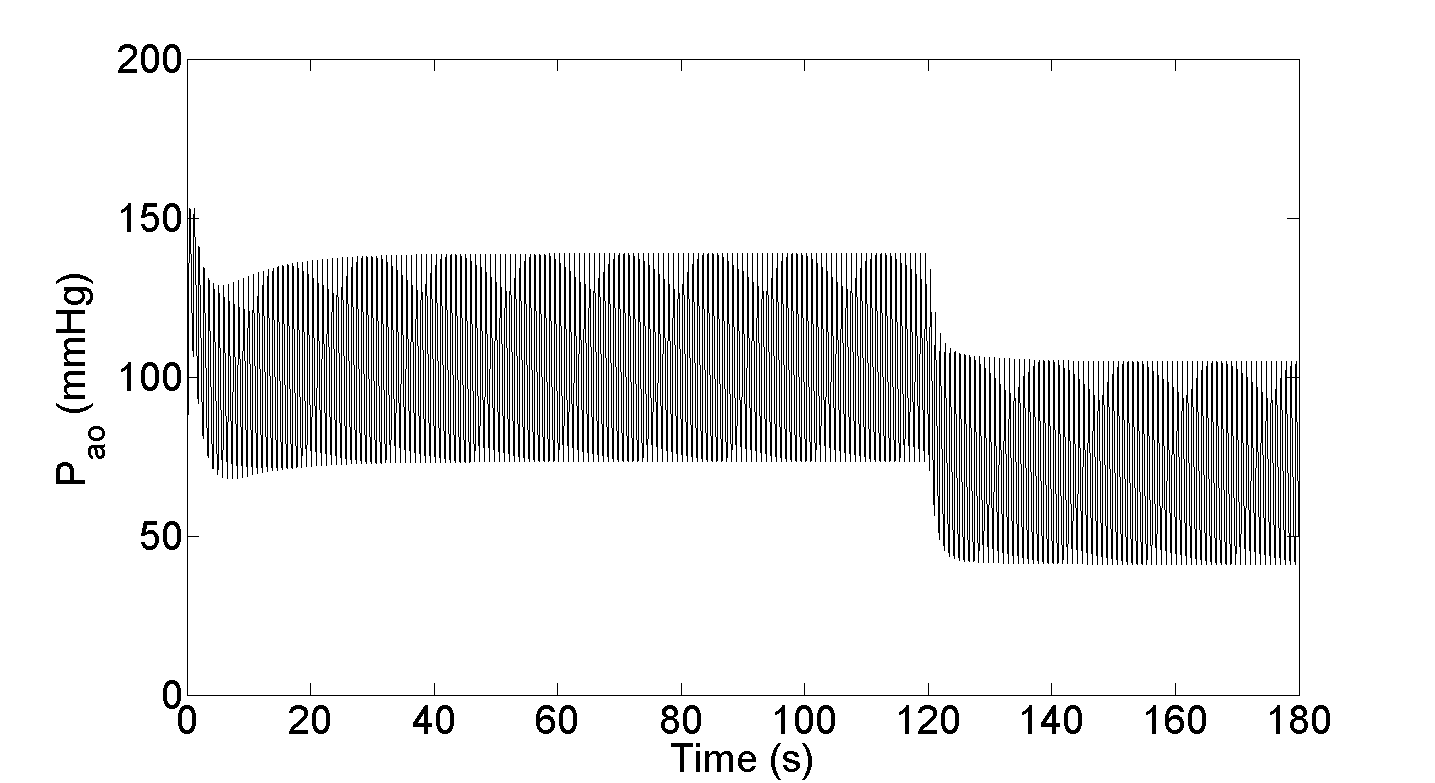}
   \label{52ec}
 }
  \subfigure[Left atrial pressure.]{
   \includegraphics[scale =0.16752] {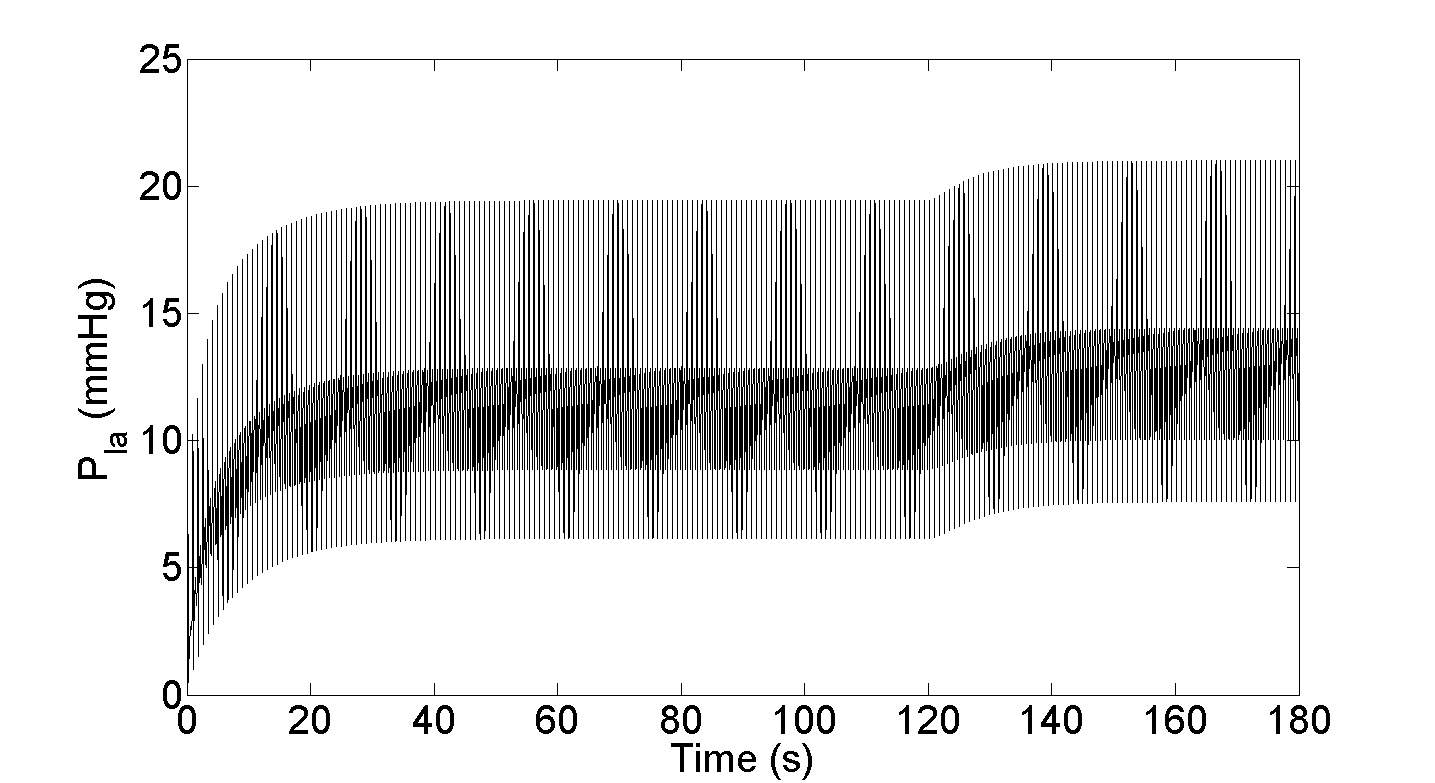}
   \label{52ed}
 }

\subfigure[Right atrial pressure.]{
   \includegraphics[scale =0.16752] {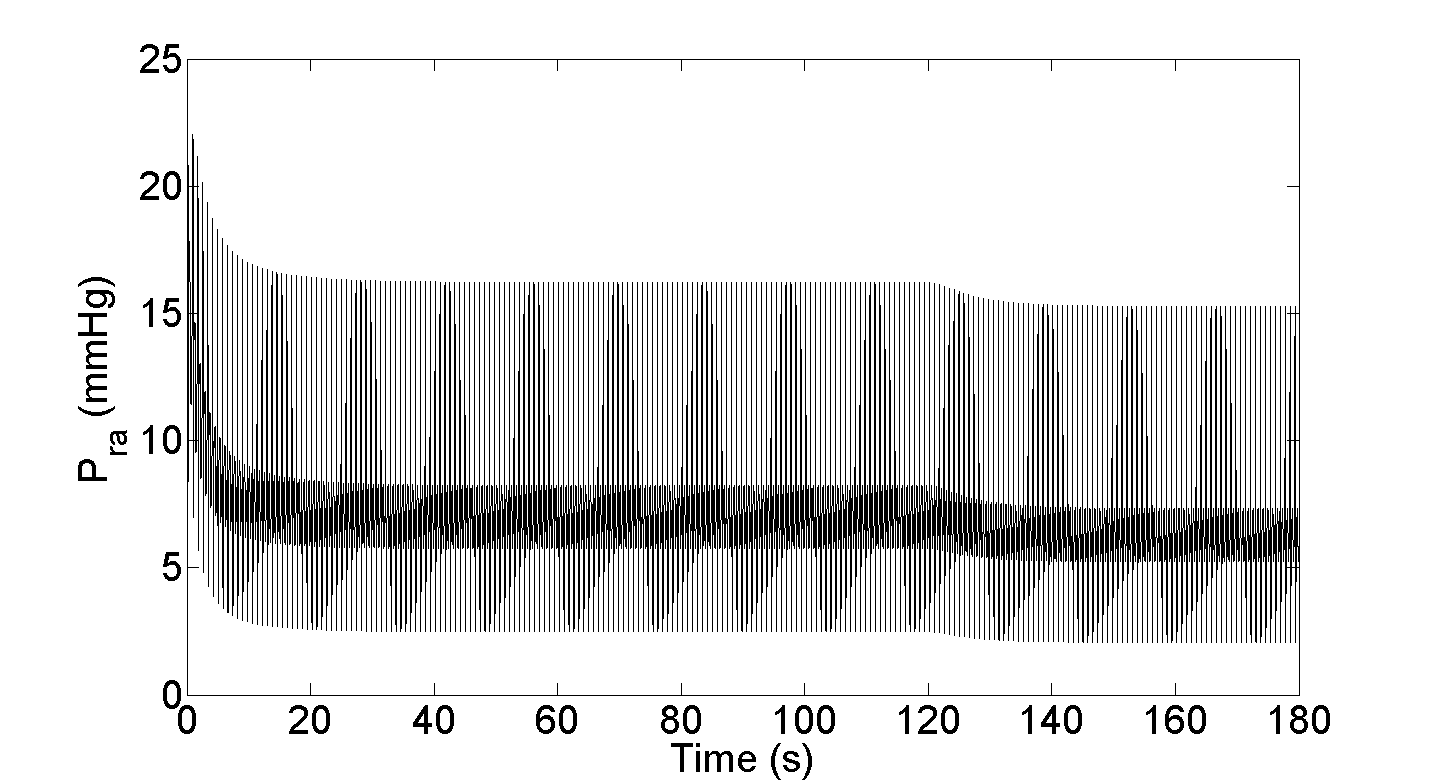}
   \label{52ee}
 }
\caption{Hemodynamic variables results in exercise condition when the system induced at 120s.}
\label{5:20ea}
\end{figure*}

\begin{figure}[htbp]
\centering
\subfigure[Average pump speed.]{
   \includegraphics[scale =0.16752] {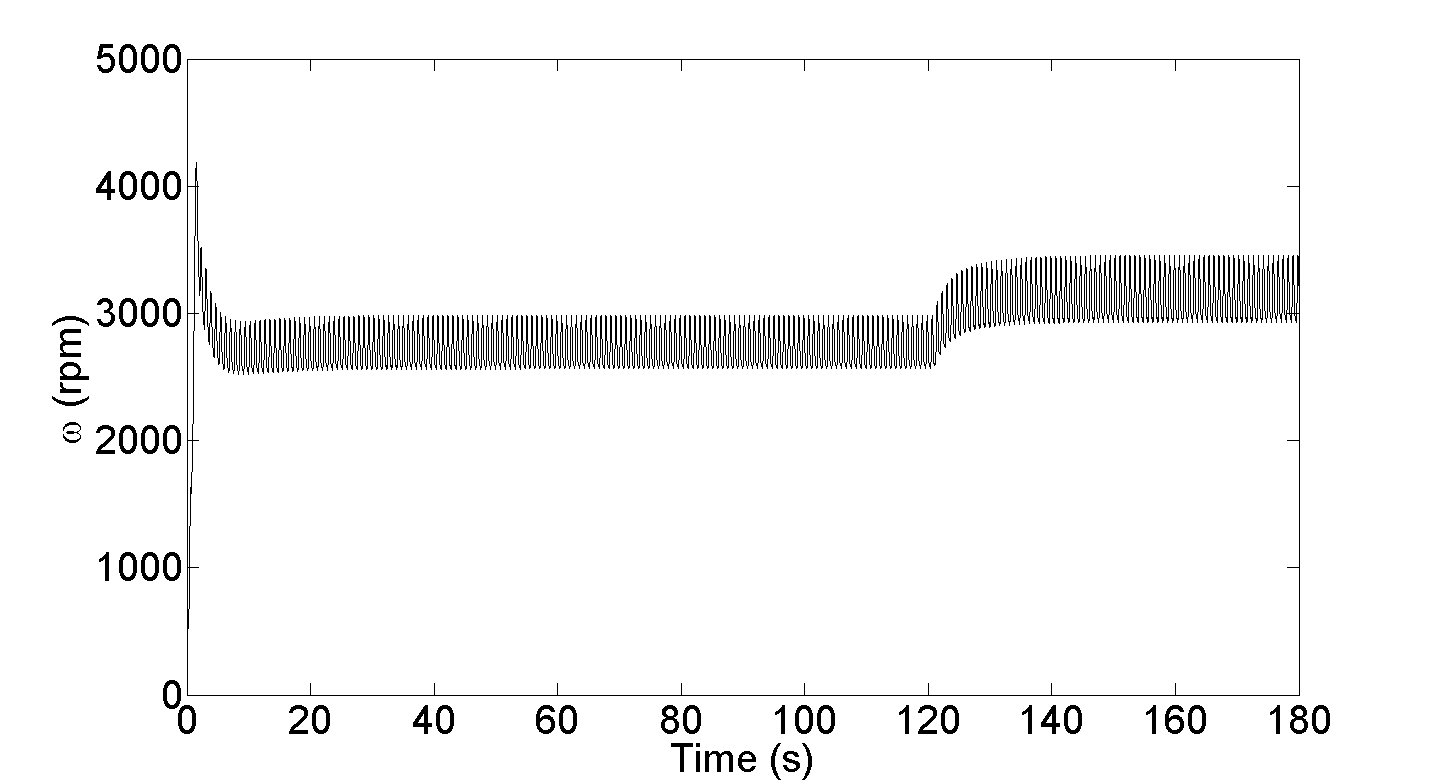}
   \label{52ef}
 }

   \subfigure[Pump flow compared with reference signal.]{
   \includegraphics[scale =0.16752] {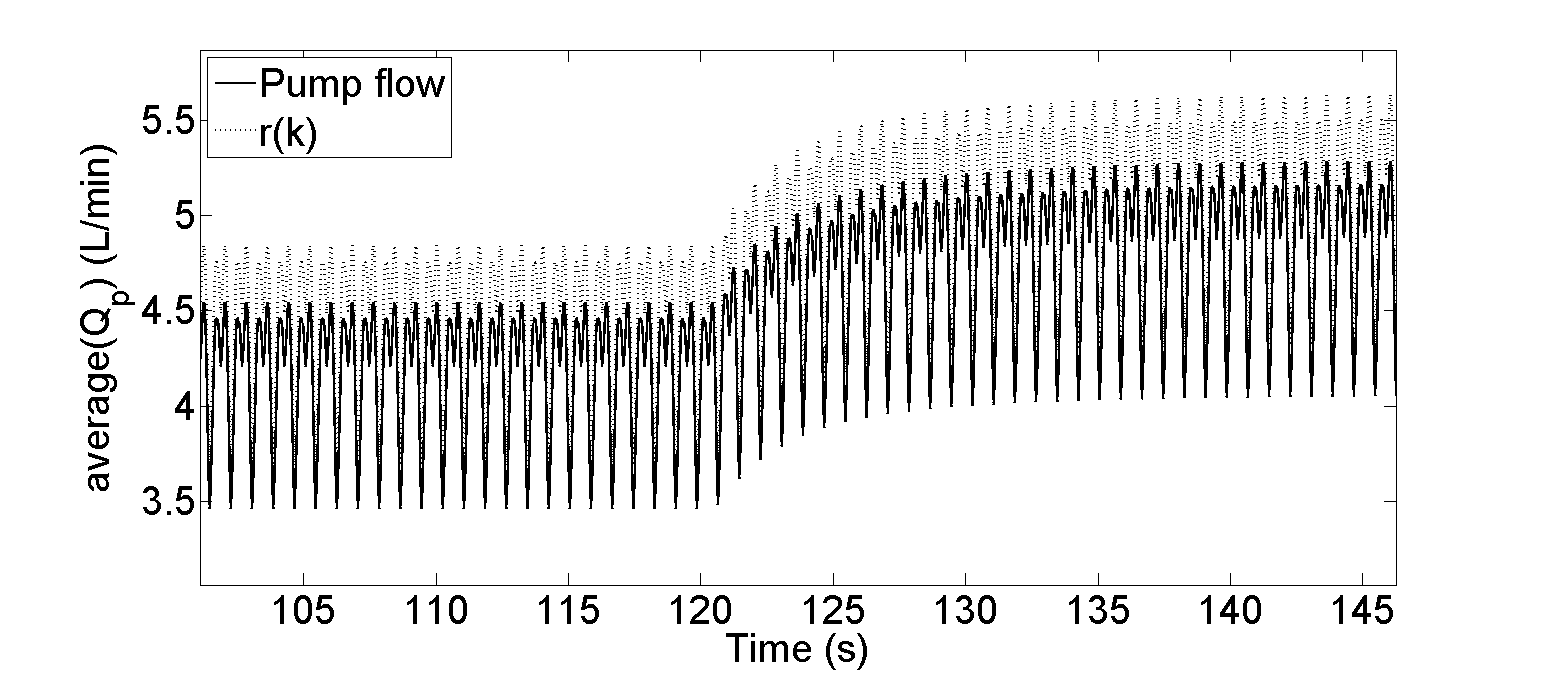}
   \label{52eh}
 }

\subfigure[Measured steady state pump flow against estimated pump flow.]{
   \includegraphics[scale =0.16752] {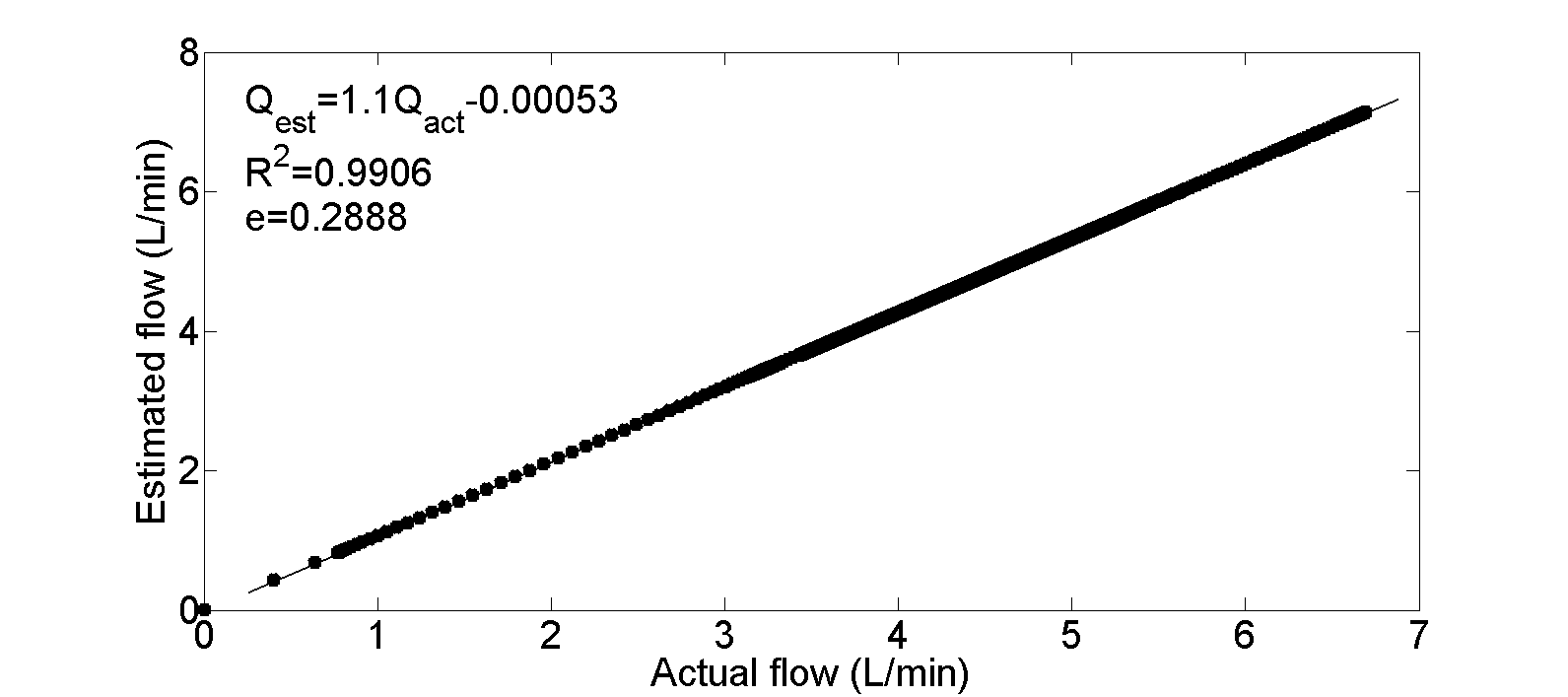}
   \label{52ei}
 }

\caption{Pump variable results in exercise condition when the system induced at 120s.}
\label{5:20eb}
\end{figure}

In this scenario, simulations have been performed by changing the model parameters so as to simulate the transition from rest to exercise. Figures \ref{5:30ea} - \ref{5:20eb} illustrate the immediate response of the controller to the parameter changes at the middle of each periodic of time 30s, 60s, 90s and 120s respectively. As the result of these changes, a rightward shift of LV pressure volume loops combined with a major increase in LV stroke volume and similar increase in LV end-systolic pressure have been produced. This is associated with a shift to the left of the RV pressure-volume loop, causing a reduction in LV end-diastolic and end-systolic volumes and pressure. The LVAD is successful in decreasing the aortic pressure $P_{ao}$ and increasing the left atrial pressure $P_{la}$ and is capable of keeping the right atrial pressure $P_{ra}$ within safe operating modes as shown in Figures (\ref{5:30ea}, \ref{5:60ea}, \ref{5:90ea} and \ref{5:20ea}).

The results of pump variable are depicted in the Figures \ref{5:30eb}, \ref{5:60eb}, \ref{5:90eb} and \ref{5:20eb}.  The controller responds to the parameter change in such a way that it  increases  average pump rotational speed from 2900 rpm to 3400 rpm and average pulsatile flow from 4.6 L/min to 5.4 L/min.   Consequently, estimated average pulsatile flow increases from 4.8 L/min to 5.6 L/min. These changes have been substantially completed within five heartbeats. Also, it can be observed that the simulated pump flow accurately tracks the reference signal within an error of $\pm$ 0.32 L/min. Furthermore, there is significantly high correlation between actual and estimated pump flows and the slope is close to unity.

Table \ref{5tab:t2}  summarises the salient hemodynamic variables  specifically for the heart failure condition before and after perturbations of blood loss and exercise. While, Table \ref{5t2} presents a brief comparison between the values of the model correlation $R^{2}$, slope $S$ and mean absolute error $e$ for each period of time.


\begin{table}[htbp]
 \caption{Associated hemodynamic variables at rest and exercise conditions.}
 \small\addtolength{\tabcolsep}{5pt}
  \label{5tab:t2}
	\begin{center}
		\begin{tabular}{ c  c  c  c  c }
			\hline
	     \multirow{2}{*}{Variables} & \multirow{2}{*}{Unit}       &  \multicolumn{3}{c }{Heart failure  plus LVAD} \\\cline{3-5} 	
	                            &       &   Normal     &  Rest (Blood loss)   &   Exercise  \\  \hline
			 $P_{lved}$         & mmHg      &  9.50   &  8.50      &  21.79    \\
	     $P_{rved}$         & mmHg      &   7.00  &  6.76      &  7.52   \\
	     $V_{lves}$         & L/min     &  40.50  &  39.00     &  62.00     \\
			 $V_{lved}$	        & L/min     &  140.70 &  141.90    &  152.4    \\
			 $SV$               & mL        &  102.00 &  100.0     &  100.0    \\
			 $\overline P_{ao}$ & mmHg      &   85.50 &   181.00   &  104.00   \\
			 $\overline P_{la}$ & mmHg      &   14.50 &   12.19    &  14.27    \\
			 $\overline P_{ra}$ & mmHg      &   6.75  &   8.20     &  7.18    \\
			 $\overline Q_{act}$& L/min     &   4.50  &   3.40     &   5.05        \\
			 $\overline Q_{est}$& L/min     &   4.95  &   3.65     &   5.52        \\
			\hline
		\end{tabular}
	\end{center}
\end{table}

\begin{table}[htbp]
 \caption{Values of the model correlation (R), slope (S) and mean absolute error ($e$) in different period of times}
 \small\addtolength{\tabcolsep}{5pt}
  \label{5t2}
	\begin{center}
		\begin{tabular}{ c  c  c  c c c c}
			\hline
			\multirow {3} {*} {Time (s)} & & &\multicolumn{2}{c}{Heart failure  + LVAD} \\ \cline{3-6}	
	                  &   \multicolumn{3}{c }{Blood loss}    &    \multicolumn{3}{c }{Exercise}  \\ \cline{2-7}
										&    $R^{2}$   &  S        &   $e$ (L/min)         &   $R^{2}$ & S    & $e$ (L/min)       \\   \hline
	     30           &    0.9815         &  1.1      &  0.2245      &    0.9890      &  1.1    &  0.2952  \\
	     60           &    1.0000         &  1.1      &  0.2141      &    1.0000      &  1.1    &  0.2690   \\
	     90           &    0.9800         &  1.1      &  0.2650      &    0.9899      &  1.1    &  0.2849   \\
			 120	        &    0.9902         &  1.1      &  0.2585      &    0.9906      &  1.1    &  0.2888    \\
			 \hline
		\end{tabular}
	\end{center}
\end{table}

\section{Discussions}

In order to  maximise the quality of life of the implant recipients so that they could regain a normal lifestyle in a long term unsupervised environment, it is believed that a physiologically responsive pump control strategy with automatic adjustment of the pump rotational speed to cater for changes in metabolic demand is needed. If pump control is not properly implemented, under-pumping or over-pumping may occur which may lead to some unacceptable risks such as collapse of the left ventricle or pulmonary edema \cite{vollkron2005development}. It  becomes further complicated by the insensitivity of IRBPs to preload \cite{mason2008reliable} and the remaining intrinsic ventricular function.

In general, the main goal required to improve the clinical application of an LVAD technology includes the development of a control strategy that automatically adjusts the pump speed to cater for cardiovascular system perturbations and the changing metabolic demand. In a healthy individual, the frank-starling mechanism ensures that the stroke volume of  LV is adjusted to compensate for changes in LV end-diastolic pressure such that the LV ejects whatever volume of blood it receives from the right ventricle \cite{guytontextbook}.   The design of  the physiological controller is based on the assumption that while the aortic valve is closed, total cardiac output can be approximated as average pump flow. Therefore, the output of an LVAD is used as a part to design a non-linear function which is used to update the reference signal according to the physiological requirements of body. The designed controller is used to regulate estimated average flow and reference signal considered as pump flow. In addition, the controller regulates average pulsatile flow instead of total cardiac output due to the difficulty in measuring total cardiac output.

It is important to highlight that this approach requires the availability of pump inlet pressure and flow signals. Direct measurements of pressure and flow by implanted sensors provide immediate and accurate information for the controller. However, the implantation of sensors is not desirable due to some limitations such that they can result in thrombus formation, reduce system reliability, increase cost and limited availability to date because of the need for regular in-vivo calibration to correct measurement drifts \cite{lim2010parameter, alomari2009non}. As an alternative, pulsatility in pump parameters across the heart beat (will discussed in next chapter) such as flow and speed can form surrogates for LV stroke work and by inference LV preload \cite{salamonsen2011response}. While less specific than LV end-diastolic pressure the variables can be estimated non-invasively \cite{alomari2011non}.

Generally, SMC is subjected to chattering in the output signal with continuous systems and this is considered to be a drawback. However, in our case there is no problem of this nature as our system is a discrete SMC system. Also, there are techniques to minimise and eliminate chattering such as using the 'sat' function instead of the 'sign' function. This approach will smooth out the control discontinuity in a thin boundary layer neighbouring the sliding surface with a linear interpolation of the 'sign' function within the boundary layer \cite{hung1993variable}.

\section{Conclusion}
A novel physiological controller based on SMC has been developed in this chapter to drive LVADs. The performance  of the proposed controller has been assessed in the presence of model  uncertainties with varying degrees of heart failure using a previously developed lumped parameter model of the CVS. The simulation results have been shown for two physiological conditions ranging from rest to exercise using sensorless measurements. It has been observed that the controller tracks the reference signal with minimum mean absolute error. Furthermore, the controller ensures the smooth and prompt transient system response with good disturbance rejection ability and consequently restores the abnormal hemodynamic variables of LVADs back to normal.

\chapter{Pole Placement Sliding Mode Control Approach Based Starling-Like Controller for IRBPs: Control of Pulsatility Index}

\section{Overview}

In a recent review of physiological control mechanisms by Alomari et al. \cite{alomari2013developments}, it is indicated that no fewer than 30 novel approaches to this problem have been proposed over the past few years. However none of them have achieved widespread acceptance by the medical profession. In this research, we attempt to fill this hiatus by proposing  a new approach combining a Starling-like control strategy with a sliding mode controller \cite{utkin1992sliding} to determine the physiologically optimum pump flow and calculate the appropriate speed to achieve it. The essence of the proposed approach is that it is intuitively understood by medical staff \cite{salamonsen2012theoretical}. The Starling mechanism emulates the natural processes of the heart to synchronise the outputs of left and right ventricles and to match the pump output to the fluctuating metabolic requirements of the body. The technique avoids over-pumping which consequently leads to  ventricular collapse and also under pumping which can cause left ventricular insufficiency \cite{hall2008physiologic}. The proposed sliding mode control (SMC) combines a rapid response of pump rotational speed without significant overshoot and with high tolerance to noisy reference signals.

In this chapter we exploit the linear relation between estimated average pulsatile flow ($\overline{Q}_{est}$) and pump flow pulsatility ($PI_{Q_{p}}$) in a tracking control algorithm based on pole placement sliding mode control \cite{bakouri2013sliding, bakouri2013physiological}. The immediate response of the controller has been assessed using a lumped parameter model of the CVS and pump  where  both $\overline{Q}_{est}$ and $PI_{Q_{p}}$ could be easily extracted. Two different perturbations from the resting state in the presence of left ventricular failure have been tested. The first is the blood loss requiring a reduction in pump flow to match the reduced output from the right ventricle and to avoid the complication of ventricular suction. The second is the exercise requiring an increase in pump flow. The sliding mode controller induces the required changes in $Q_{p}$ within approximately 5 heart beats in the blood loss simulation and 8 heart beats in the exercise simulation without significant clinical transients or steady-state errors.

\section{Control Strategy}

According to the Frank-Starling mechanism, an increased volume of blood progressively stretches the ventricular wall causing the cardiac muscle to contract more forcefully. This effect underlies the Starling characteristic relating  LV preload to LV output \cite{salamonsen2012theoretical}. When this relation is expressed in terms of pump pulsatility rather than preload and applied to an IRBP, the Starling characteristics define the relation between $\overline Q_{p}$ and $PI_{Q_{p}}$.  Although this relation is basically linear, there is a point of inflexion which is associated with the opening of the aortic valve \cite{salamonsen2012theoretical}. Before this inflexion point, the flow of blood through the pump constitutes the cardiac output and the linear relation between estimated pump output and flow pulsatility in this zone defines the pump control line as used in this study.

\begin{figure}[htbp]
\centering
\includegraphics[width=2.8in]{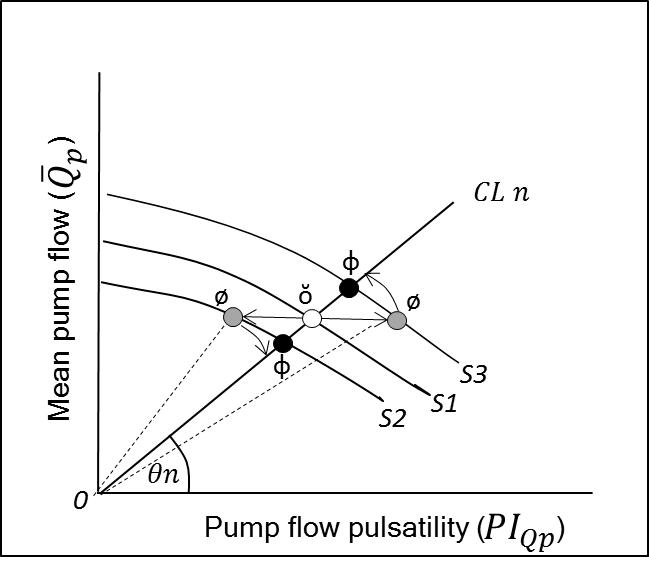}
\caption{Immediate response of the Starling like pump flow controller to change in system states.}
\label{fig:starling}
\end{figure}

The previous statement enables to implement the Starling-like controller working as a physiological controller to maintain the body perfusion. This approach is based on an assumption that while the aortic valve is closed, the total cardiac output can be approximated as pump flow. This leads to the linear relation between average pulsatile flow and flow pulsatility which forms the target flow for our Starling-like controller. Fig.  \ref{fig:starling} illustrates the concept of immediate response of the Starling-like controller to change in system states. When a change in system state from $S1$ to $S2$ or $S3$ causes a deviation in the operating point from the intersection of $S1$ and $CLn$ ($\circ$) to ($\bullet$), on $S2$ or $S3$, the controller responds by returning the operating points along a radial path with center of rotation at origin of axes, to the control line $CLn$, settling to positions ($\bullet$) at the intersection of $CLn$ and state lines $S2$ or $S3$. This operation is controlled by the following final equation; 

\begin{equation}
\overline{Q}_{pr,t}=\left(\sqrt{(\overline{Q}_{p,t-1})^{2}+(PI_{Qp,t-1})^{2}}\right)\cdot \sin \theta _{n} \label{6eq:1}
\end{equation}
where $\overline{Q}_{r,t}$ is the desired pump  flow, $\overline{Q}_{p,t-1}$  is the estimated average pulsatile flow, $PI_{Qp,t-1}$ is pulsatility of pump flow and $\theta _{n}$ is the angle which defines the gradient of the pump control line. Upper and lower limits for both average pulsatile flow and pump flow pulsatility are applied to modify the control line gradient when necessary.

There is a host of body conditions (causing migration of the operating point to S2 or S3 as in Fig. \ref{fig:starling}), which calls for more or less outputs from the heart. It includes exercise, posture, anxiety states, sleep/wake cycles, and other changes in medical treatment.

\section{Methods and Approaches}

\subsection{Software Model}

A software model incorporating a lumped-parameter model of the CVS in combination with a stable dynamic model of an LVAD has been used to evaluate the control strategy. The CVS model has been developed based on experimental measurements obtained from healthy pigs implemented with an IRBP over a wide range of operating conditions including variations in total circulatory volume, systemic vascular resistance and cardiac contractility. A detailed description of the model as well as parameter values can be found in  \cite{lim2010parameter}. 

\begin{figure}[htbp]
\centering
\includegraphics[scale =0.6]{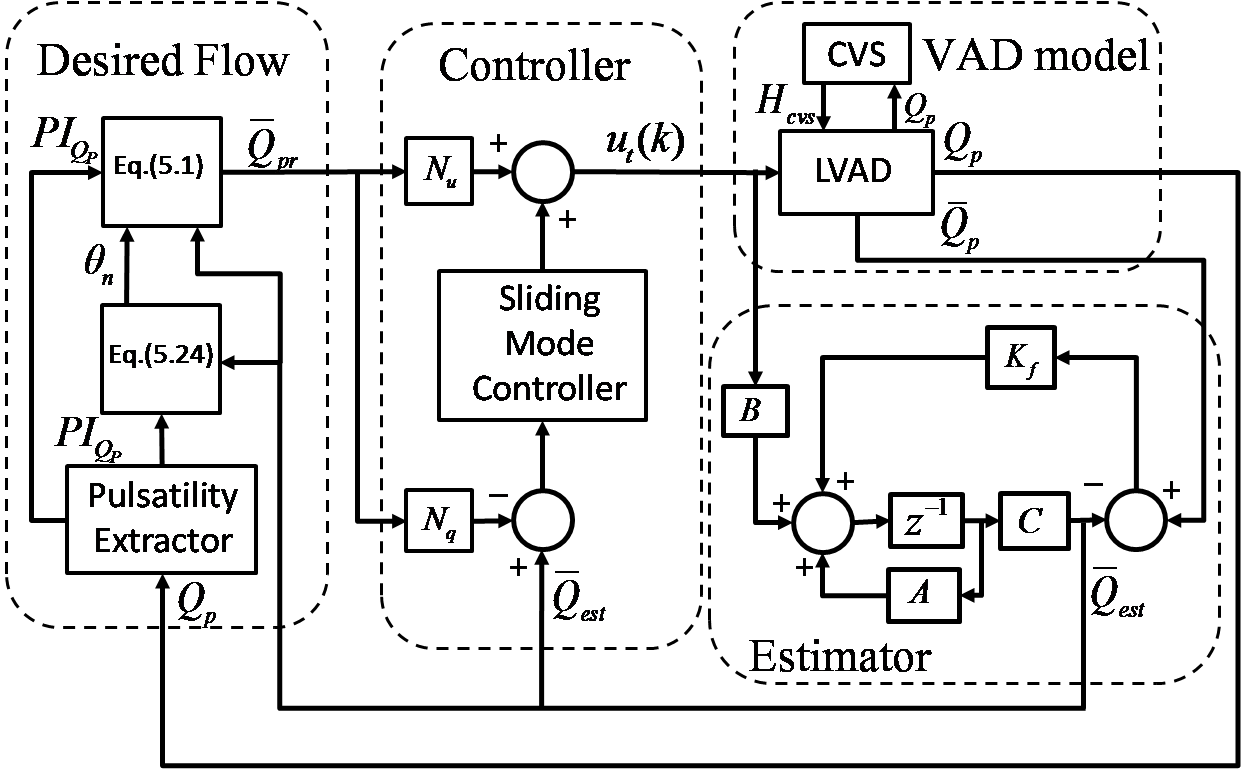}
\caption{Block diagram of control system.}
\label{6fig:block}
\end{figure}

A schematic diagram  of the proposed control strategy is shown in Fig.  \ref{6fig:block}. In this scheme, the block entitled �VAD model� represents a model of the CVS and LVAD. The blocks entitled �Estimator, �Controller� and �Desired flow� consist of the physiological control system. The input to the controller is the desired pump flow (desired $\overline{Q}_{pr}$) and the estimated average pulsatile flow $\overline{Q}_{p}$ for a cardiac cycle, while the output of the controller is the PWM voltage signal to the rotary pump, represented as $u_{pwm}$.

\subsection{Software Simulation Protocols}

In order to evaluate the immediate response of the controller to different physiological states, defined by system parameter variations, the LVAD with the designed controller has been tested under two physiological conditions. Firstly, total circulatory volume ($V_{total}$) is linearly decreased by 500 mL at the middle of each period of time. This periodicity has been chosen to be 30s, 60s, 90s and 120s respectively to verify the system stability. The simulation continues up to the end of each period to allow the system to reach to a steady state corresponding to new parameters values. The purpose of this test is to determine whether the LVAD could provide essential support to the HF patient under minimal stress.

Secondly, the simulation has been carried out to validate the controller response during the transition from rest to exercise. In this test, the system parameters have been changed linearly at the middle of each period (30s, 60s, 90s and 120s). These changes include left and right ventricular contractility (increased linearly by 15\%), systemic peripheral resistance (decreased linearly by 50\%) and systemic veins unstressed volume (decreased linearly by 50\%). The purpose of this test is to determine whether the controller is capable enough in combination with the CVS to provide the haemodynamic support required during normal daily activities. The other parameters of the model have been kept constant during the simulations.

\subsection{Implementation of the Control Algorithm}

Consider the model of an LVAD identified by ARX model as in  (\ref{5eq:3}) without disturbance and uncertainties as:

\begin{equation} \begin{array}{rcl}
q(k+1) & = & Aq(k)+Bu(k) \\
    y(k) &=& Cq(k),     \label{6eq:2}
\end{array} \end{equation}
where $q\epsilon \mathbb{R}^{n}$ is the state vectors of the system, $u\epsilon \mathbb{R}^{m}$ is the input vector, $y\epsilon \mathbb{R}^{r}$ is the assumed model output. The system is controllable and observable.

This design  is based on discrete-time sliding mode control approach which performs measurements and control signal applications at regular intervals of time by keeping the control signal constant between the time intervals. Discrete-time sliding mode offers invariance to uncertain parameters, compensating for  uncertainties that exist in real dynamic applications, thus making it a good choice for the error trajectory tracking problems. The proposed SMC controller is designed based on pole placement method as described on the following steps:
\begin{itemize}
\item Switching function:
Robust pole placement method is used to design this function. The aim of this method is to make the non-zero sliding mode eigen-values insensitive to perturbation using robust eigen-structure assignment . This will minimise the effect of parameter variations outside the range space \cite{edwards1998sliding}. Consider that the switching function in discrete time can be defined as:
\begin{equation}
\boldsymbol{\eta}(k)=\boldsymbol{\Gamma}q(k),  \label{6eq:7} \\
\end{equation}
where  $\boldsymbol{\Gamma}$ is a constant vector, stated with $[1*2]$ matrix chosen to ensure that  $q(k)$  is asymptotically stable on  $\boldsymbol{\eta}(k)=0$  \cite{spurgeon1992hyperplane}. Fig. \ref{fig:ss} illustrates the convergence of the state variables $q_{1}(k)$ and $q_{2}(k)$ as the move on the switching plane.

\begin{figure}[htbp]
\centering
\includegraphics[width=4in]{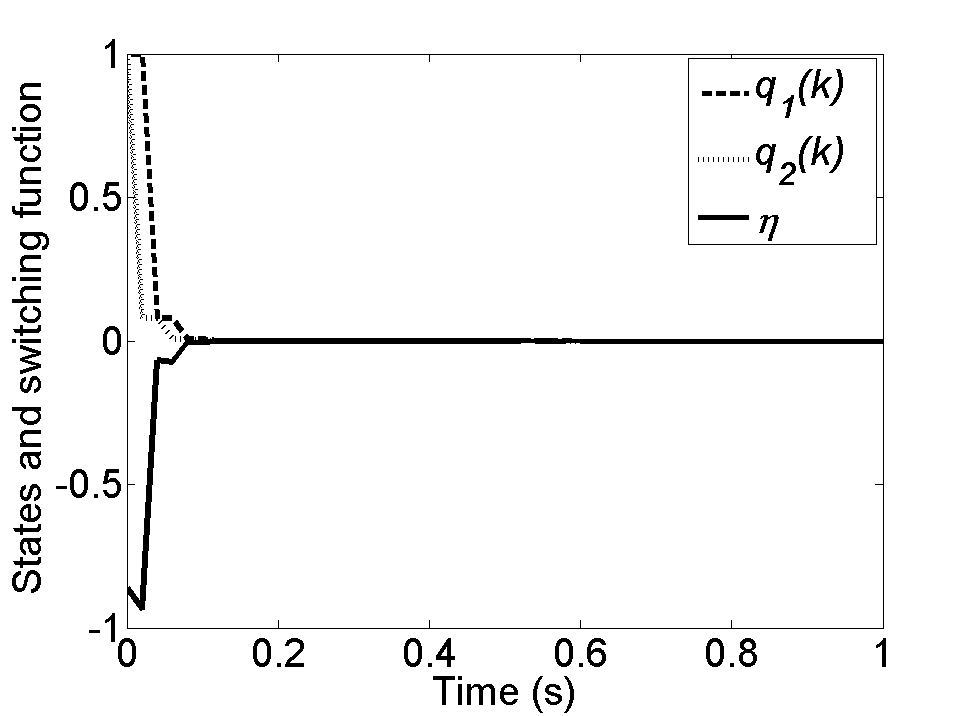}
\caption{System response.}
\label{fig:ss}
\end{figure}

\item Reaching law:
In \cite{gao1995discrete}, Gao et al proposed quasi-sliding mode (QSM) approach that describes the reaching law conditions to guarantee the ideal sliding motion. The equivalent discrete-time reaching law that satisfies our model and design with strong reachability is:
\begin{equation}
\boldsymbol{\eta}(k+1)=(1-\tau T)\boldsymbol{\eta}(k)-\sigma T \text{sign}(\boldsymbol{\eta}(k)), \label{6eq:8} \\
\end{equation}
where , $T>0$ is the sampling period,
$\sigma>0$
$\tau \geq0$
such that $0<(1-\tau T)<1$ and $0<\sigma T<1$.

\item Control law:
the control law is synthesised from the reaching law in conjunction with a known model of the plant and known perturbations.
To satisfy the reaching law in   (\ref{6eq:8}), we obtain:
\begin{equation}
\boldsymbol{\eta}(k+1)=\boldsymbol{\Gamma}q(k+1),   \label{6eq:9}  \\
\end{equation}
from   (\ref{6eq:2}) and   (\ref{6eq:9}) we get:
\begin{equation}\begin{array}{rcl}
\boldsymbol{\eta}(k+1)&=&\boldsymbol{\Gamma}q(k+1)\\
                      &=&\boldsymbol{\Gamma}(A-BK)q(k)+\boldsymbol{\Gamma}Bu(k), \label{6eq:10}  \\
\end{array}\end{equation}
where $K\in \mathbb{R}^{n}$ is a gain matrix obtained by assigning $n$ desired eigenvalues

$(\lambda_{1},..\lambda_{n-m})$ of $A-BK$,
the equalisation of   (\ref{6eq:8}) and   (\ref{6eq:10}) gives:
\begin{equation} \begin{array}{rcl}
\boldsymbol{\eta}(k+1)&=&\boldsymbol{\Gamma}(A-BK)q(k)+\boldsymbol{\Gamma}Bu(k)\\
                      &=&(1-\tau T)\boldsymbol{\eta}(k)-\sigma T \text{sign}(\boldsymbol{\eta}(k)),  \label{6eq:11}  \\
\end{array} \end{equation}
solving the above equation for the command signal $u(k)$ yields:
\begin{equation} \begin{array}{rl}
u(k) = & -(\boldsymbol{\Gamma}B)^{-1}(\boldsymbol{\Gamma}(A-BK)q(k) \\
       & +(\tau T-1)\boldsymbol{\Gamma}q(k)+\sigma T \text{sign}(\boldsymbol{\eta}(k))), \\
        \label{6eq:12}  \\
\end{array} \end{equation}
where $\boldsymbol{\Gamma}$ is chosen such that $\boldsymbol{\Gamma}B$ is non-singular \cite{spurgeon1992hyperplane, edwards1998sliding}. 

So far, the design of control law is based  on nominal plant model that has been defined by (\ref{6eq:2}). However, in presence of disturbances or system parameter uncertainties, the control algorithm  follows the following procedure. In a similar fashion, if we invoke the LVAD model in   (\ref{3eq:14}) as:
\begin{equation} \begin{array}{rcl}
q(k+1) & = & Aq(k)+\delta Aq(k)+Bu(k)+\zeta(k) \\
    y(k) &=& Cq(k)+\psi(k),     \label{6eq:112}
\end{array} \end{equation}
where $\delta A$ is system parameter variation, $\zeta (k)\in \mathbb{R}^{n}$ and $\psi (k)\in \mathbb{R}^{r}$ are system disturbance and measurement noise respectively.

The design of control algorithm with the reaching law (\ref{6eq:8}) can be constructed as follows:

From   (\ref{6eq:9}) we obtain:

\begin{equation}\begin{array}{rcl}
\boldsymbol{\eta}(k+1)&=&\boldsymbol{\Gamma} q(k+1)\\
                      &=&\boldsymbol{\Gamma} ((A-BK)q(k)+\delta Aq(k)+Bu(k)+\zeta(k)), \label{6eq:113}  \\
\end{array}\end{equation}
re-arranging the above equation we obtain:

\begin{equation}
\boldsymbol{\eta}(k+1)=\boldsymbol{\Gamma} (A-BK)q(k)+\boldsymbol{\Gamma} \delta Aq(k)+\boldsymbol{\Gamma} Bu(k)+\boldsymbol{\Gamma} \zeta(k), \label{6eq:114}  \\
\end{equation}
by the equalisation of reaching law in   (\ref{6eq:8}) with   (\ref{6eq:114}) we   get:

\begin{equation}\begin{array}{cr}
\boldsymbol{\Gamma} (A-BK)q(k)+\boldsymbol{\Gamma} \delta Aq(k)+\boldsymbol{\Gamma} Bu(k)+\boldsymbol{\Gamma} \zeta(k) \\
         = (1-\tau T)\boldsymbol{\eta}(k)-\sigma T \text{sign}(\boldsymbol{\eta}(k)), \label{6eq:115}  \\
\end{array}\end{equation}
by solving the above equation for the command signal $u(k)$ yields:

\begin{equation} \begin{array}{rl}
u(k) = & -(\boldsymbol{\Gamma} B)^{-1}(\boldsymbol{\Gamma} (A-BK)q(k)-(\tau T-1)\boldsymbol{\Gamma} q(k) \\
       &+\sigma T\text{sign}(\boldsymbol{\eta}(k)))-(\boldsymbol{\Gamma} B)^{-1}(\boldsymbol{\Gamma} \delta Aq(k)+\boldsymbol{\Gamma} \zeta(k)). \\
        \label{6eq:116}  \\
\end{array} \end{equation}

As $\delta A$ and $\zeta(k)$ are unknown, the control algorithm cannot be implemented unless we assume that the upper and lower boundaries of the value $(\boldsymbol{\Gamma} \delta Aq(k)+\boldsymbol{\Gamma} \zeta(k))$ is known as:
\begin{equation}
-\beta<(\boldsymbol{\Gamma} \delta Aq(k)+\boldsymbol{\Gamma} \zeta(k))<\beta, \label{6eq:117}  \\
\end{equation}

then the control algorithm can be re-written as:
\begin{equation} \begin{array}{rl}
u(k) = & -(\boldsymbol{\Gamma} B)^{-1}(\boldsymbol{\Gamma} (A-BK)q(k)-(\tau T-1)\boldsymbol{\Gamma} q(k) \\
       &+(\sigma T+\beta)\text{sign}(\boldsymbol{\Gamma} q(k))). \\
        \label{6eq:118}  \\
\end{array} \end{equation}

In order to achieve output tracking control, a reference input $\overline Q_{pr}$ is introduced into the system by modifying the state feedback control law $u_{h}(k)=-K q(k)$ with pole-placement design method  \cite{franklin1997digital}:

\begin{equation} \begin{array}{lcr}
 u_{h}(k) & = -K[q(k)-N_{q}\overline Q_{pr}]+N_{u}\overline Q_{pr} \\
          & = -K q(k)+(KN_{q}+N_{u})\overline Q_{pr} \\
          & = -K q(k)+\overline{N}* \overline Q_{pr},  \label{6eq:13}
\end{array} \end{equation}
where
\begin{equation}
\overline {N}=K N_{q}+N_{u},
\label{6eq:14}
\end{equation}
and
\begin{equation}
 \begin{bmatrix}
 N_{q}\\
 N_{u}\\
 \end{bmatrix}=
 \begin{bmatrix}
 A-I  &  B \\
C & 0 \\
\end{bmatrix}^{-1}
\begin{bmatrix}
 0\\
 I\\
 \end{bmatrix}. \label{6eq:15}
\end{equation}

The proposed SMC input based on   \ref{6eq:14} is assumed to be:

\begin{equation}
u_{t}(k)=u_{h}(k)+u(k)=(-K q(k)+\overline{N}*\overline Q_{pr})+u(k),
\label{6eq:16}
\end{equation}
substituting   (\ref{6eq:12}) into   (\ref{6eq:16}) gives the proposed SMC input as:

\begin{equation} \begin{array}{cl}
 u_{t}(k) & = (-Kq(k)+\overline{N}*\overline Q_{pr}) \\
          & -(\boldsymbol{\Gamma} B)^{-1}(\boldsymbol{\Gamma} Aq(k)-(\tau T-1)\boldsymbol{\Gamma} q(k) \\
          &+(\sigma T+\beta)\text{sign}(\boldsymbol{\Gamma} q(k))), \\  \label{6eq:17}
\end{array} \end{equation}
The pole-placement SMC design method utilises the feedback of all the state variables to form the desired vector.
\end{itemize}

In our problem, the whole state $q(k)$ is not available to our controller, so we need to estimate $q(k)$ based on the measured output $y(k)$. For this purpose, we use the steady-state Kalman estimator \cite{welch1995introduction, simon2006optimal} given as:

\begin{equation} \begin{array}{rcl}
\hat {q}(k+1) & = & A\hat {q}(k)+Bu(k)+K_{f}(y(k)-C\hat{q}(k)) \\
    \hat{y}(k) &=& C\hat{q}(k),     \label{6eq:20}
\end{array} \end{equation}
where $\hat{q}(k)$ is the estimated of the state $q(k)$ and $K_{f}$ is the  "optimal Kalman" gain given as:

\begin{equation}
K_{f}=PC^{T}R^{-1}
\label{6eq:211}
\end{equation}
and $P$ is the solution of the following algebraic Riccati equation which is given as:

 \begin{equation} \begin{array}{rcl}
AP+PA^{T}-PC^{T}R^{-1}CP+Q=0, \\
       \label{6eq:21}
\end{array} \end{equation}

As shown in Fig.  \ref{6fig:block}, the reference pump flow $\overline Q_{pr}$ has been  calculated using gradient angle ($\theta$) of the control line as:
\begin{equation}
\theta = K_{p,\theta}(e_{\overline{Q}_{p}}+e_{PI_{Q_{p}}})+K_{i,\theta}\int(e_{\overline{Q}_{p}}+e_{PI_{Q_{p}}}),        \label{6eq:22}
\end{equation}
where $K_{p}$ and $K_{i}$ are the proportional and integral gains respectively. The gradient angle $\theta$ is automatically adjusted the $\overline Q_{p}$ and $PI_{Q_{p}}$ to their corresponding upper or lower limits for each cycle of the model using a proportional integral controller.

\section{Simulation Results}

The design parameters of the switching function in   (\ref{6eq:17})  are  $K$ = [0.9500 -0.0800] and those of the control law in the same equation are $\tau T = 0.015$ and $\sigma T= 0.025$. The resulting value of $\boldsymbol{\Gamma}$ is [0.9413 -0.0805]. In addition, the lower and upper limits for pump flow pulsatility were set to 1.5 L/min and 4 L/min, while lower and upper limits for average pulsatile flow were set to 3 L/min and 6 L/min. Study Protocol Model parameters are adjusted from the �Healthy� to �Heart failure� condition as a precondition for simulations. The most relevant parameters include systemic peripheral resistance, blood volume and contractilities of the left and right ventricles expressed as maximum end-diastolic elastances are given in Table \ref{6tab:t1}.

\begin{table}[htbp]
 \caption{Changes in important model parameters to simulate the HF condition.}
  \small\addtolength{\tabcolsep}{5pt}
  \label{6tab:t1}
\begin{center}
\scalebox{0.85}{
		\begin{tabular}{ c c  c  c c c}
			\hline
No.&	Variable & Symbol & Unit & Healthy & Heart failure\\ \hline	
1&	Left ventricular contractility  & $E_{lv}$ & $ mmHg.mL^{-1}$      &    3.54    & 0.71   \\
2&	Right ventricular contractility & $E_{rv}$ & $ mmHg.mL^{-1}$      &    1.75    & 0.53   \\
3&	Systemic peripheral resistance  & $R_{sa}$ &  $mmHg.s.mL^{-1}$    &    0.74    & 1.11   \\
4&	Total blood volume              & $V_{total}$ & $(mL)$            &    5300    & 5800    \\
			\hline
		\end{tabular}}
		\end{center}
\end{table}

\subsection{Results in Rest Condition (Blood Loss)}

\begin{figure*}[htbp]
\centering
\subfigure[LV volume versus LV pressure before and after Parameter Change.]{
   \includegraphics[scale =0.16752] {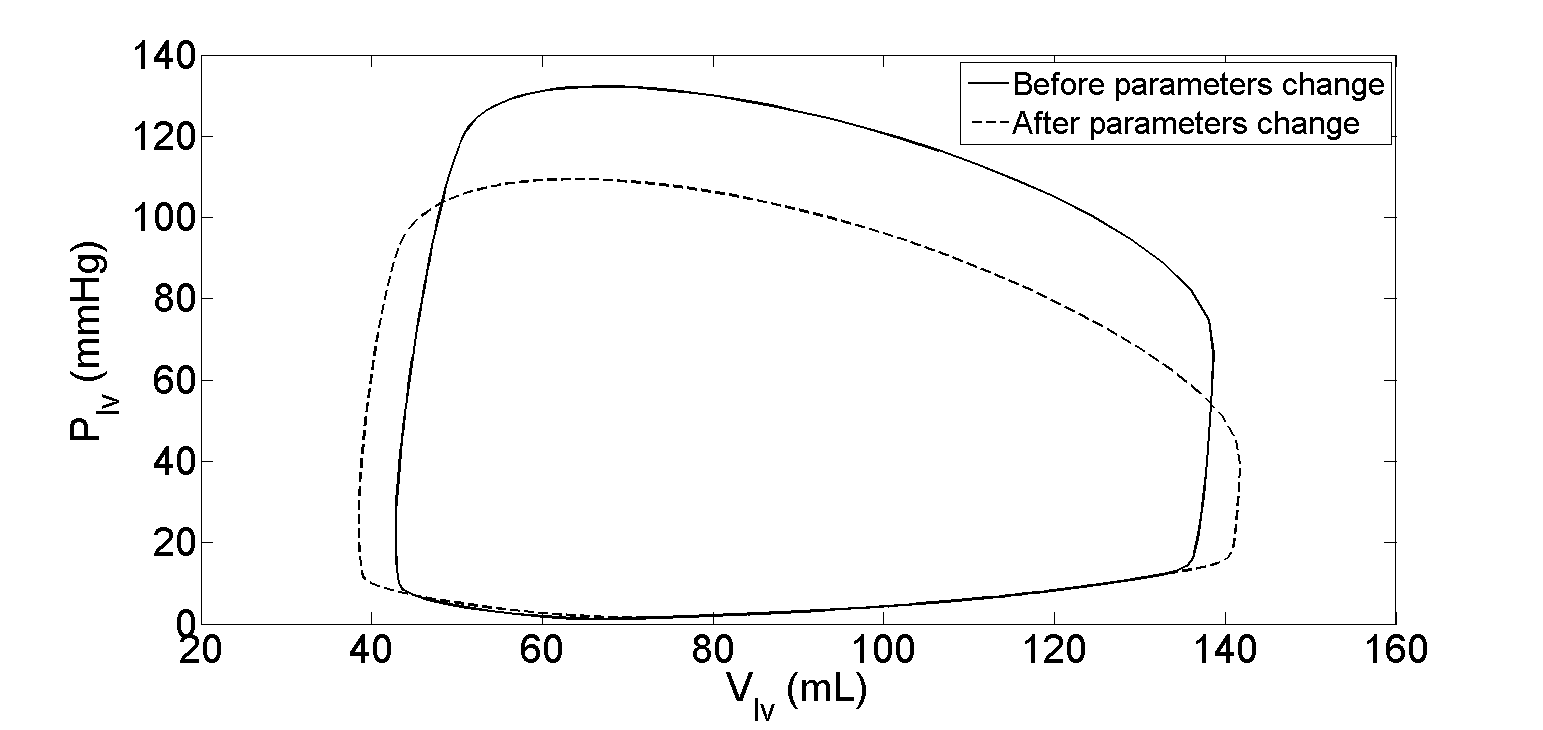}
   \label{63a}
 }
\subfigure[RV volume versus RV pressure before and after Parameter Change.]{
   \includegraphics[scale =0.16752] {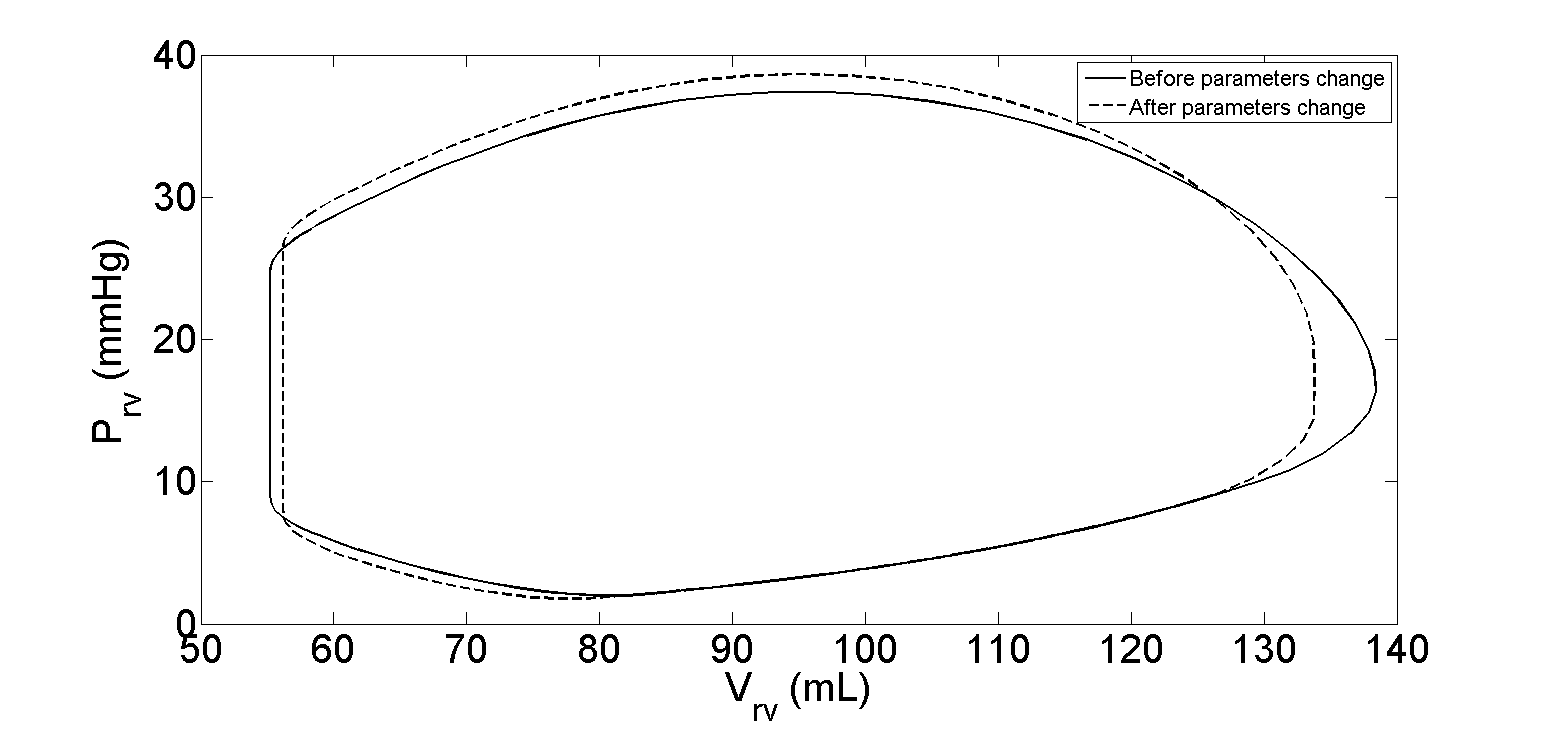}
   \label{63b}
 }

 \subfigure[Aortic pressure.]{
   \includegraphics[scale =0.16752] {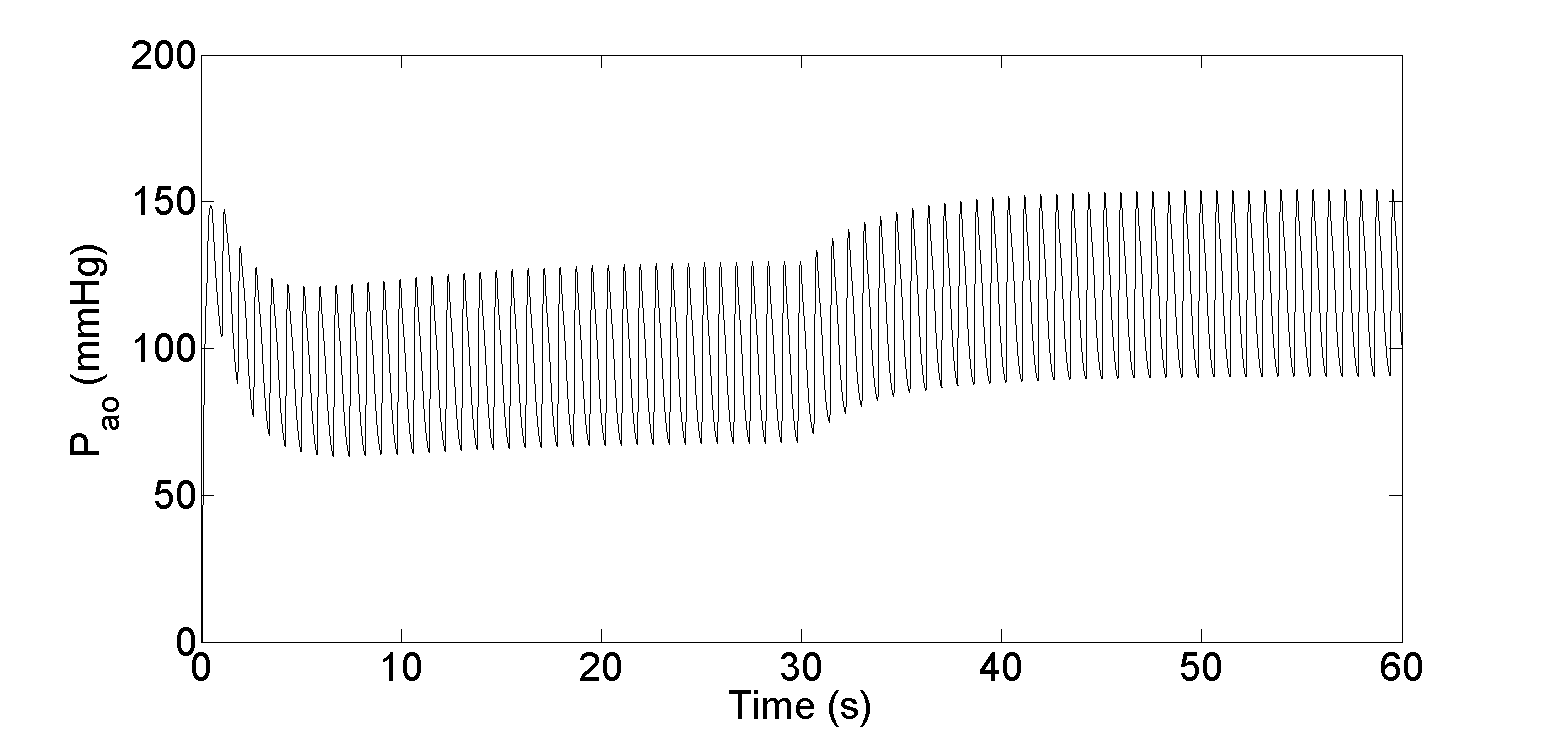}
   \label{63c}
 }
  \subfigure[Left atrial pressure.]{
   \includegraphics[scale =0.16752] {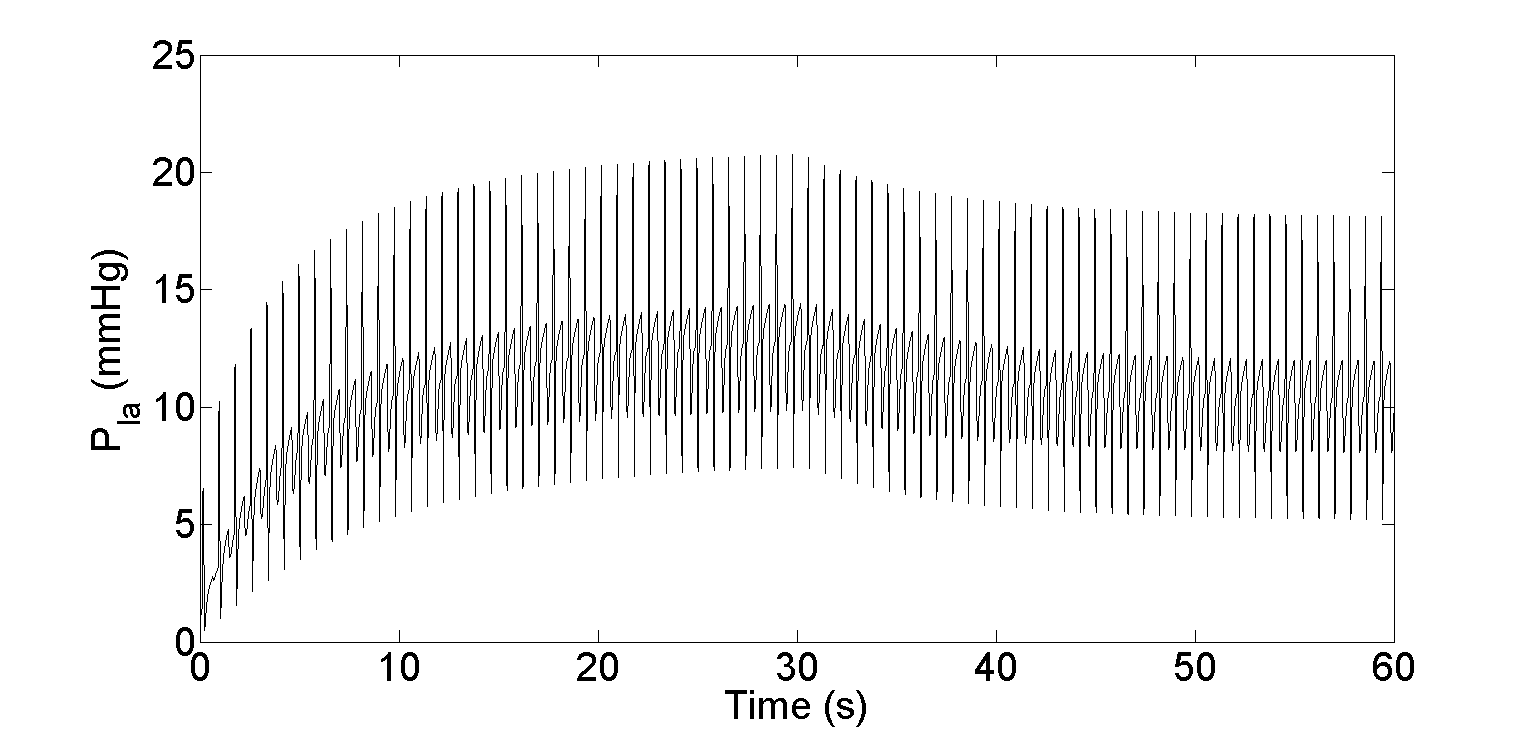}
   \label{63d}
 }

\subfigure[Right atrial pressure.]{
   \includegraphics[scale =0.16752] {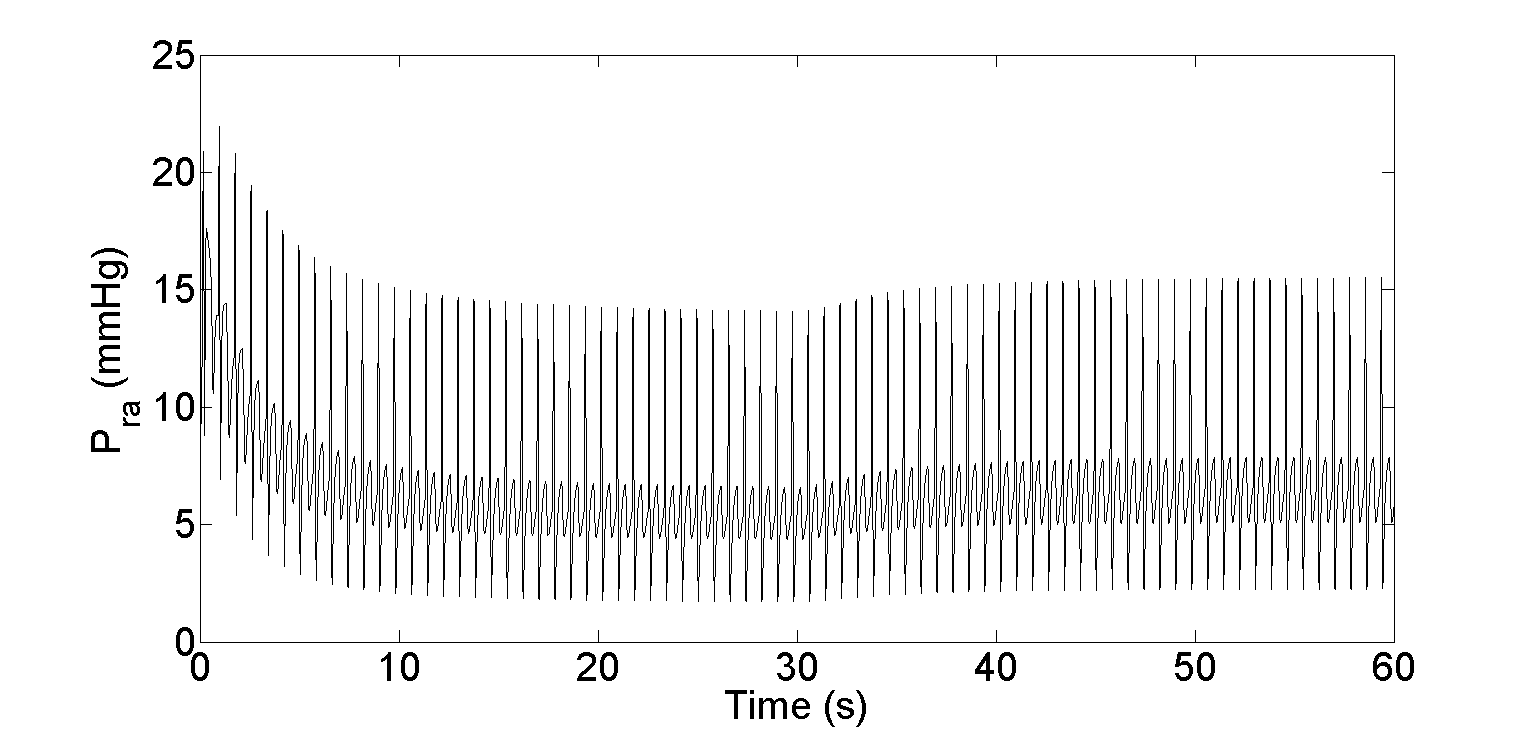}
   \label{63e}
 }
\caption{Hemodynamic variables results in rest condition when the system induced at 30s.}
\label{6:30a}
\end{figure*}

\begin{figure*}[htbp]
\centering
\subfigure[Average pump speed.]{
   \includegraphics[scale =0.16752] {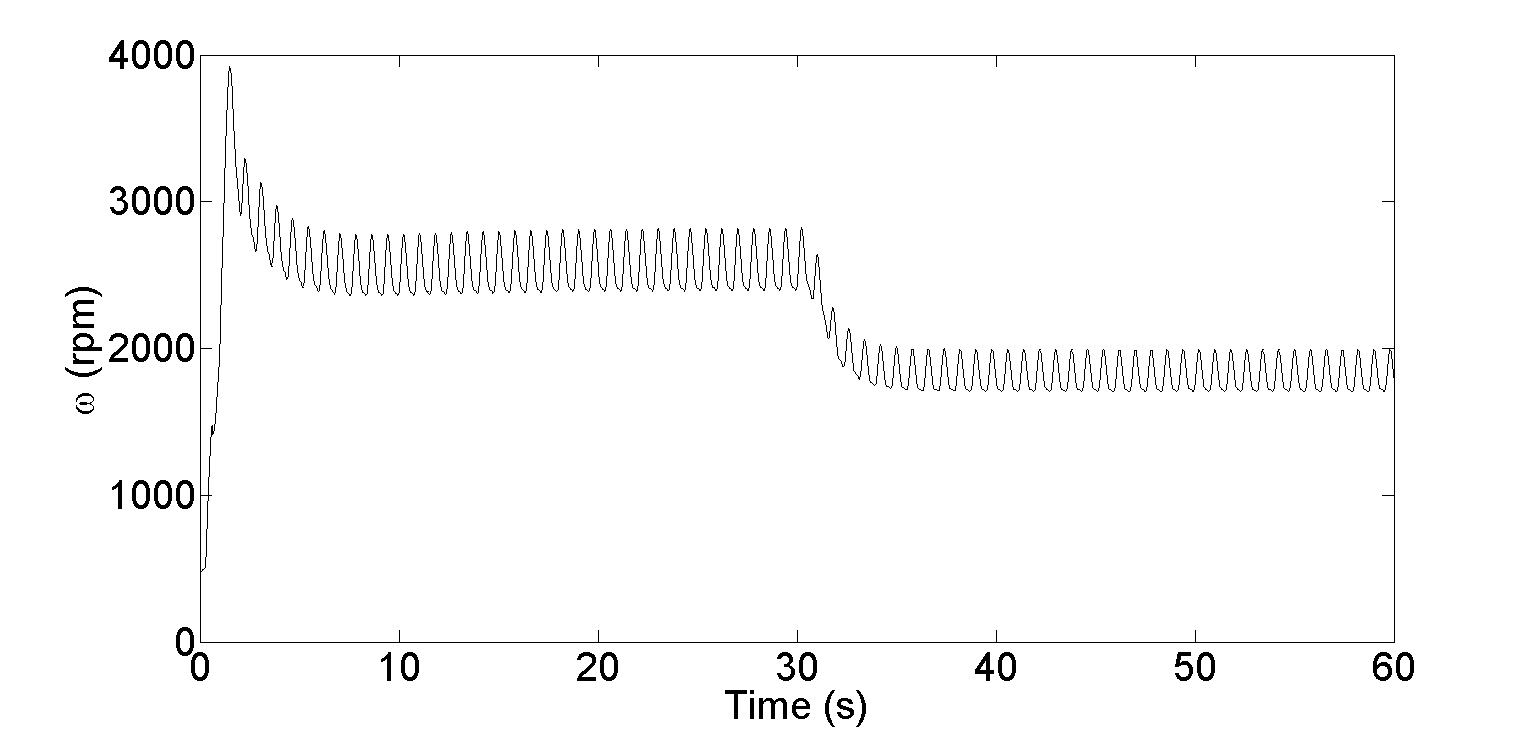}
   \label{63f}
 }
 \subfigure[Pump flow pulsatility versus average pulsatile flow.]{
   \includegraphics[scale =0.16752] {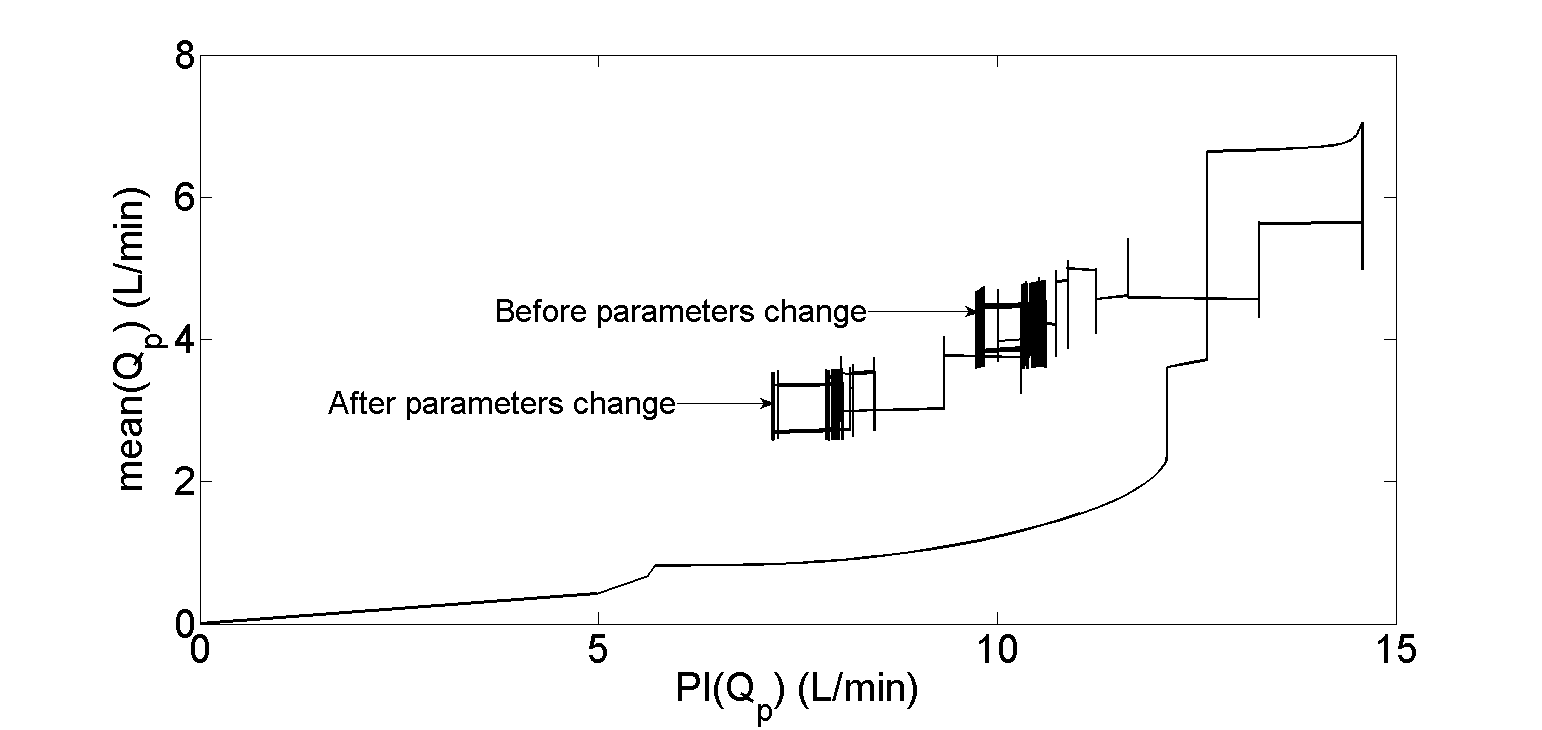}
  \label{63g}
 }

  \subfigure[Pump flow compared with desired reference flow at initial time.]{
   \includegraphics[scale =0.16752]{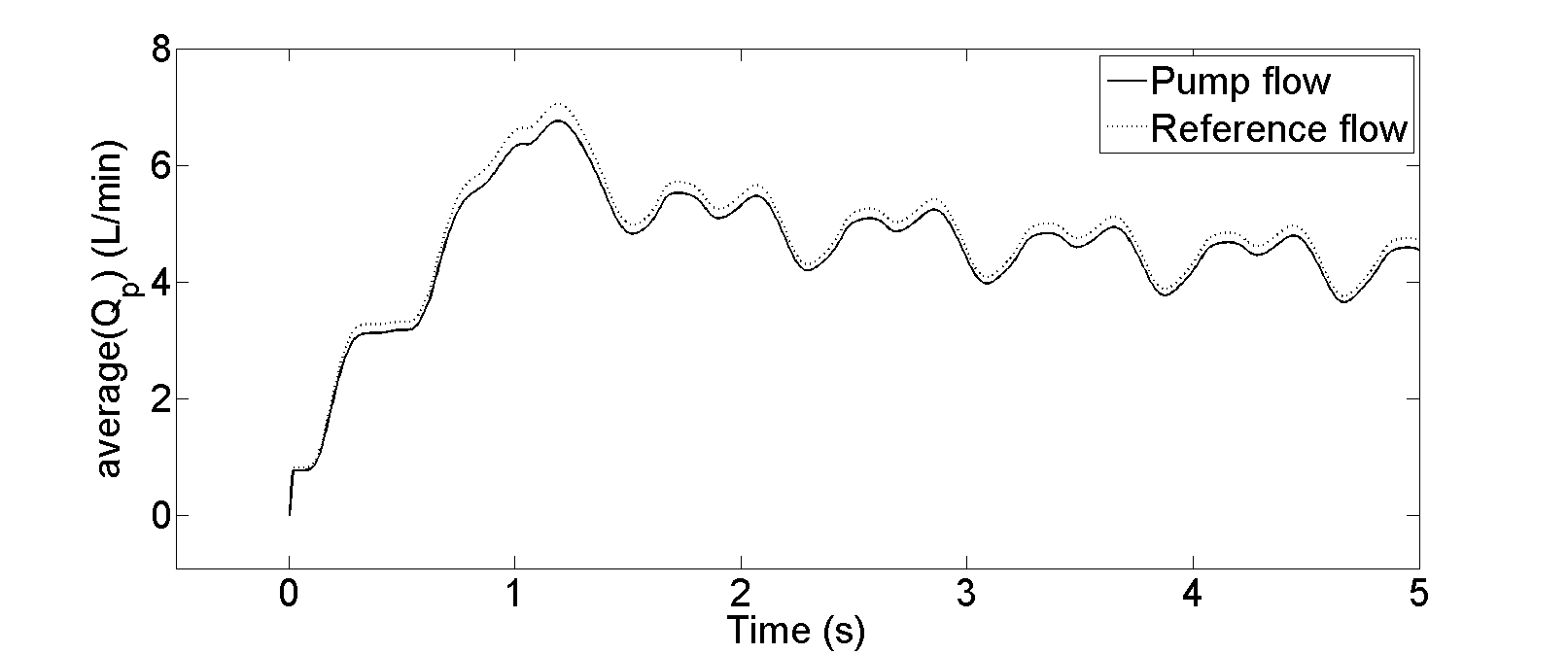}
   \label{i63h}
 }
\subfigure[Pump flow compared with desired reference flow at induced time.]{
   \includegraphics[scale =0.16752]{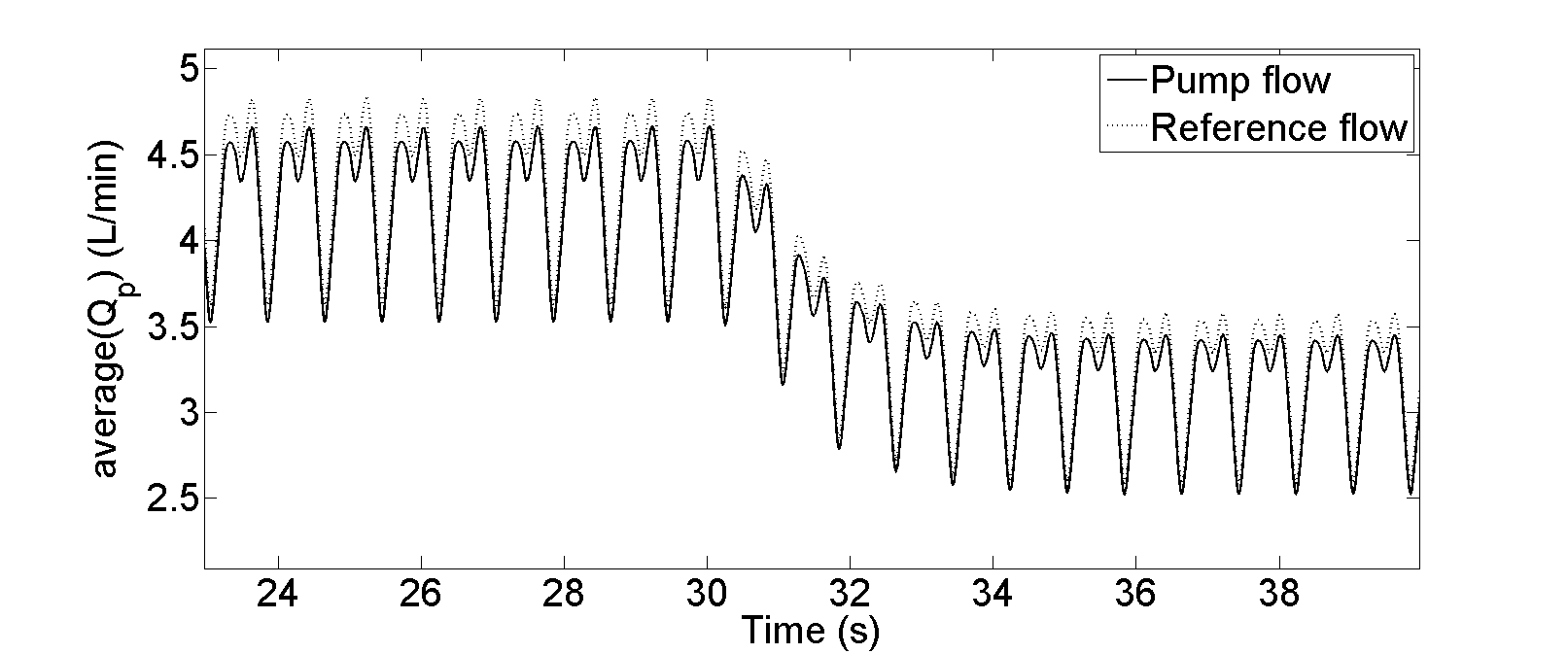}
   \label{63h}
 }

\subfigure[Measured steady state pump flow against estimated pump flow.]{
   \includegraphics[scale =0.16752] {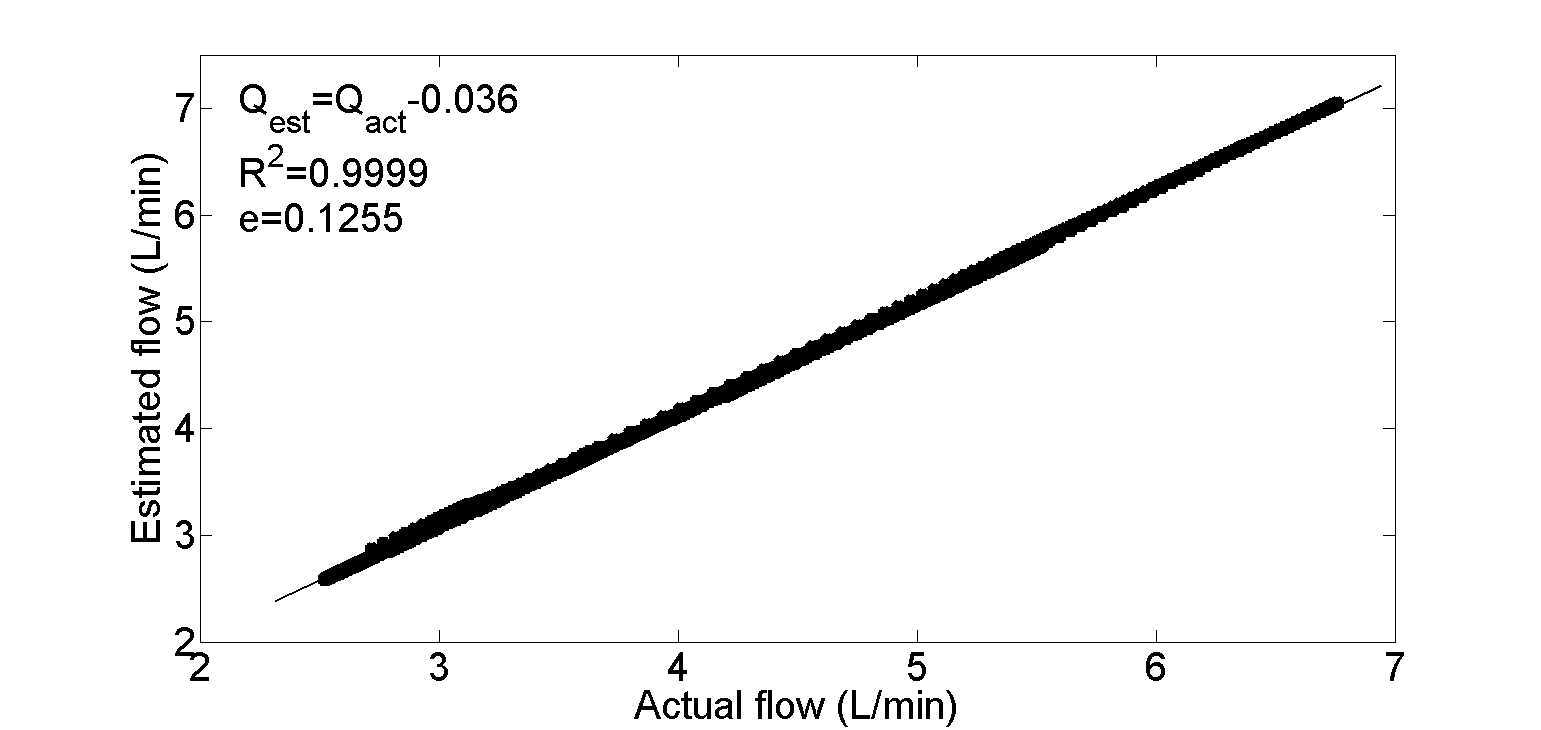}
   \label{63i}
 }

\caption{Pump variable results in rest condition when the system induced at 30s.}
\label{6:30b}
\end{figure*}

\begin{figure*}[htbp]
\centering
\subfigure[LV volume versus LV pressure before and after Parameter Change.]{
   \includegraphics[scale =0.16752] {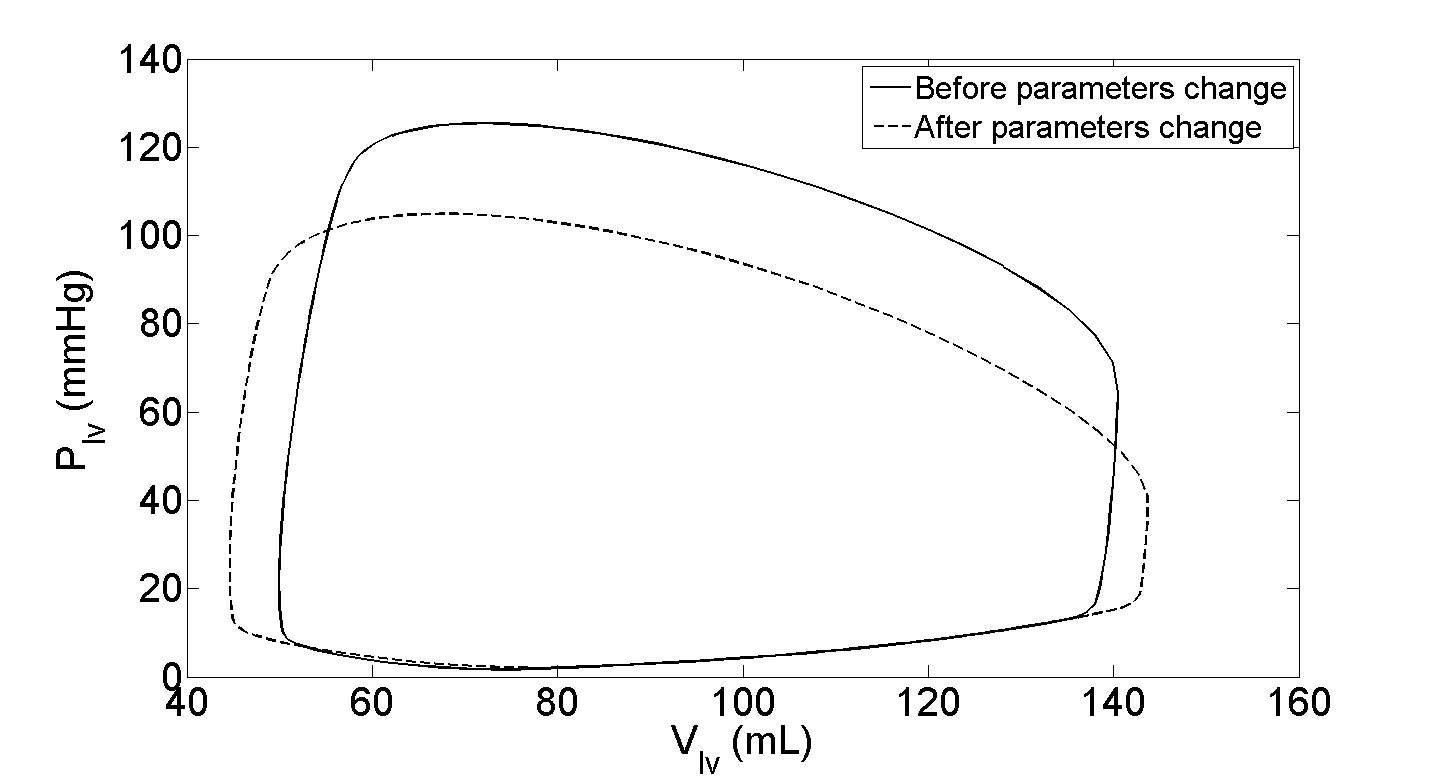}
   \label{66a}
 }
\subfigure[RV volume versus RV pressure before and after Parameter Change.]{
   \includegraphics[scale =0.16752] {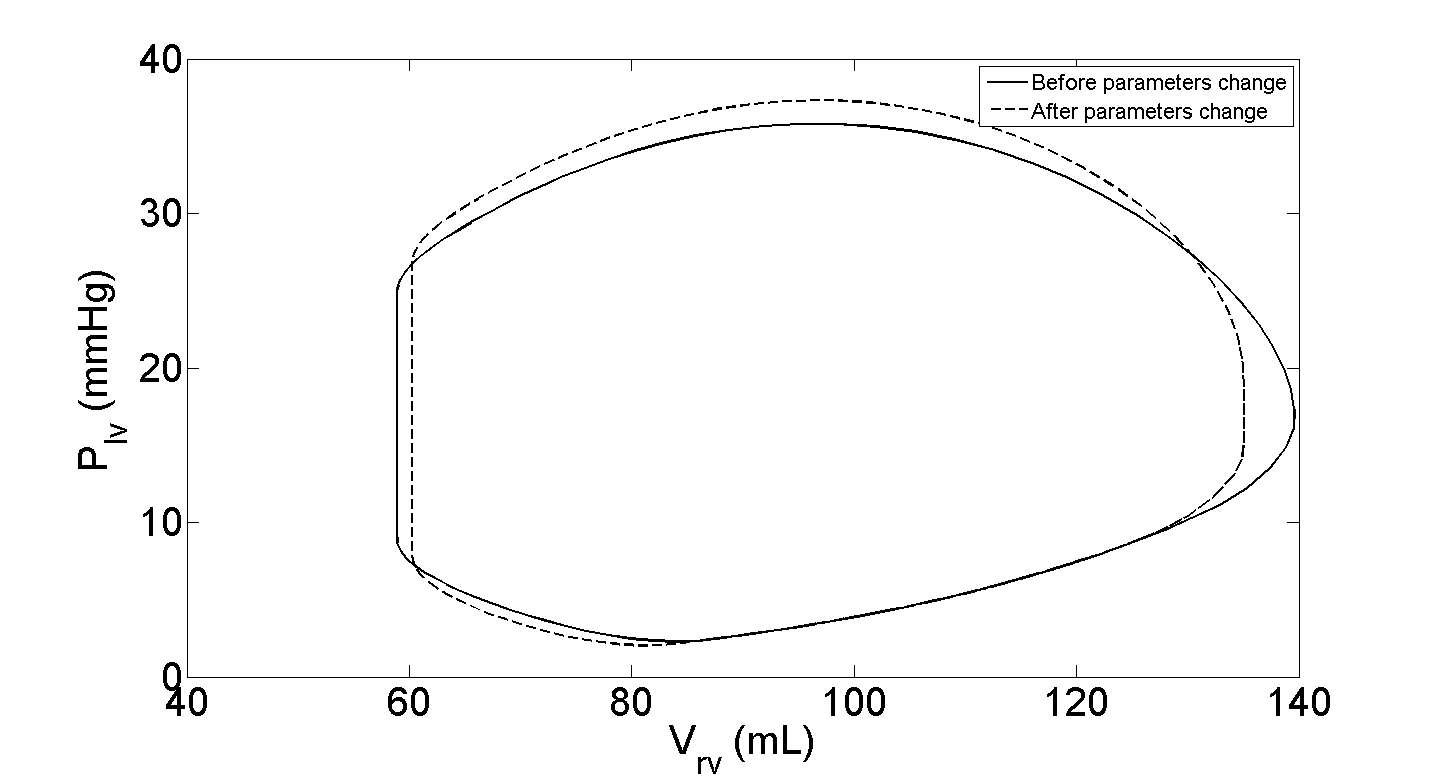}
   \label{66b}
 }

 \subfigure[Aortic pressure.]{
   \includegraphics[scale =0.16752] {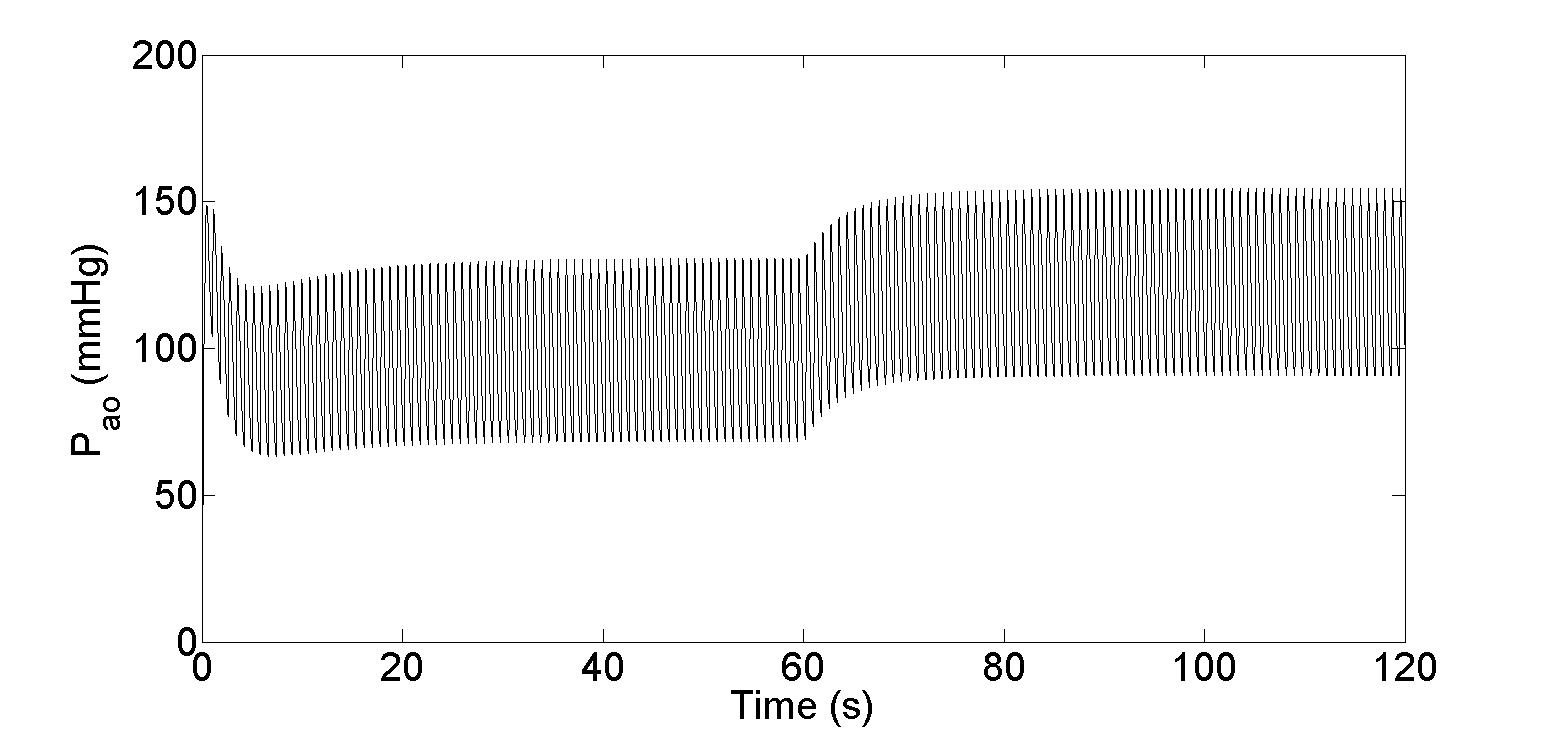}
   \label{66c}
 }
  \subfigure[Left atrial pressure.]{
   \includegraphics[scale =0.16752] {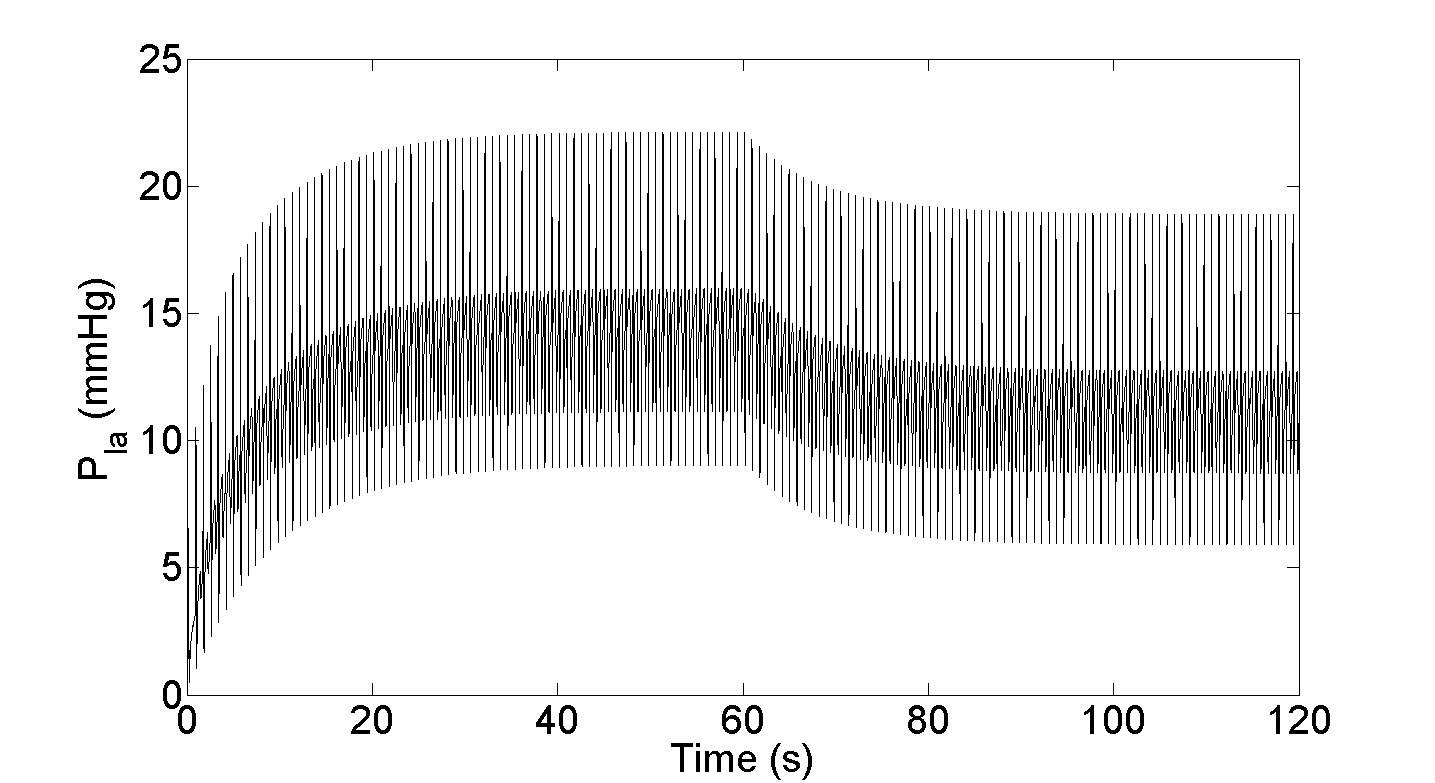}
   \label{66d}
 }

\subfigure[Right atrial pressure.]{
   \includegraphics[scale =0.16752] {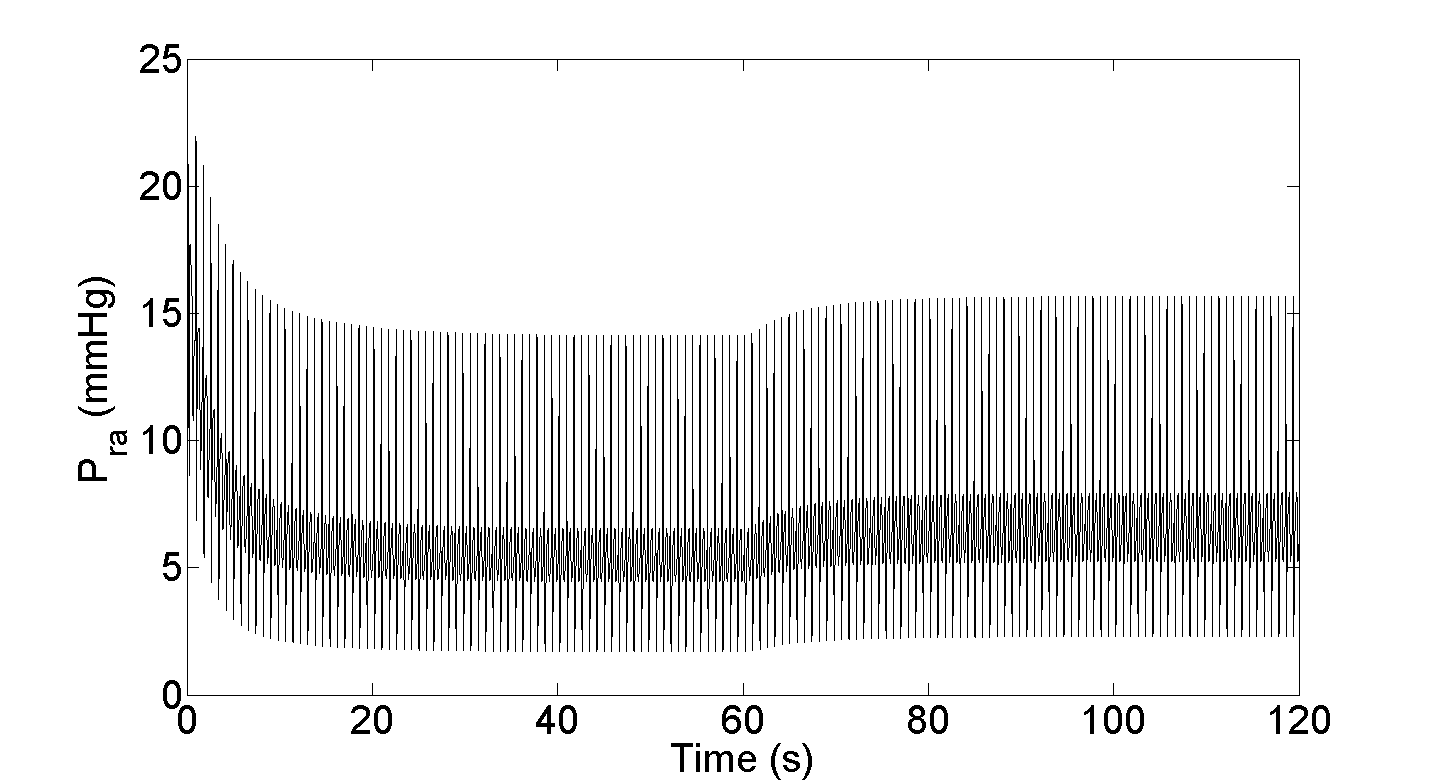}
   \label{66e}
 }
\caption{Hemodynamic variables results in rest condition when the system induced at 60s.}
\label{6:60a}
\end{figure*}

\begin{figure*}[htbp]
\centering
\subfigure[Average pump speed.]{
   \includegraphics[scale =0.16752] {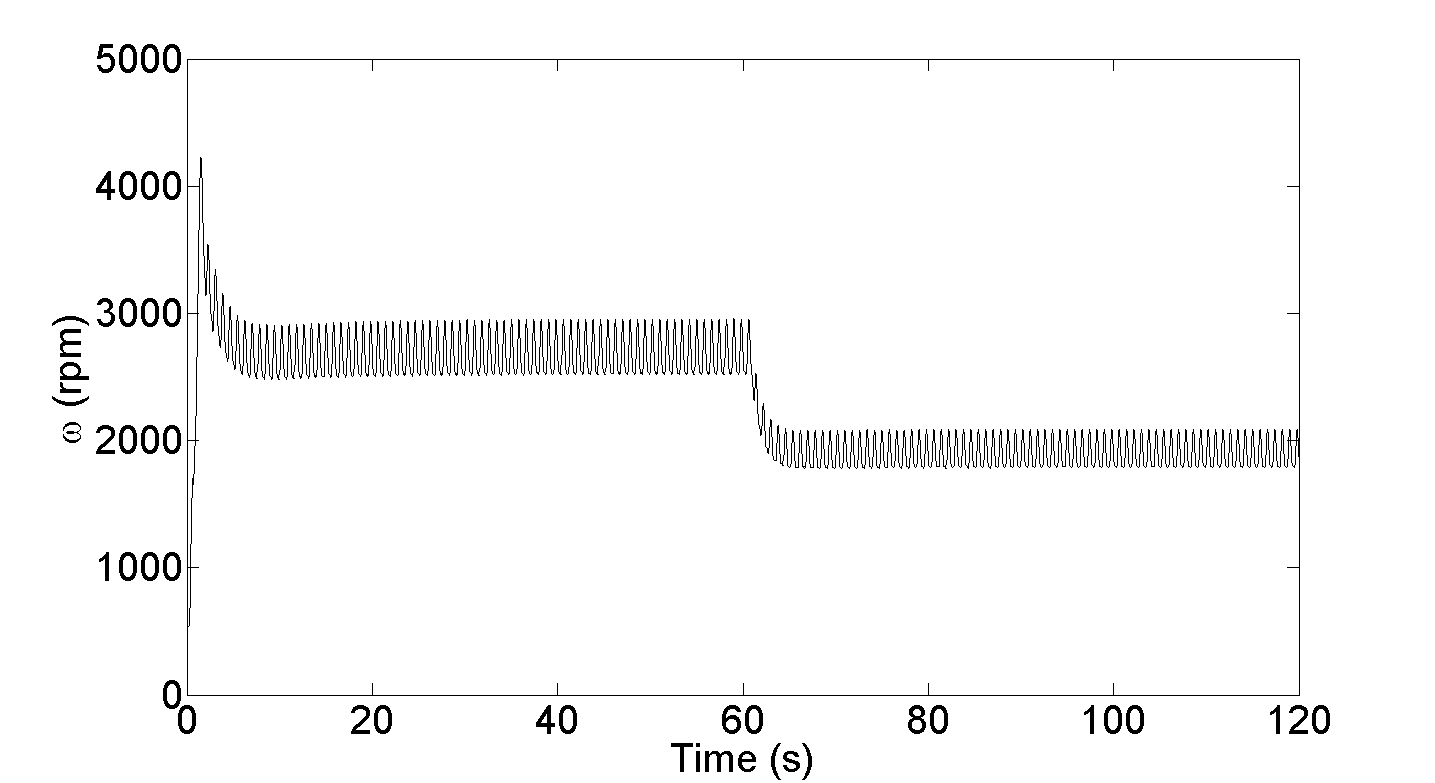}
   \label{66f}
 }
 \subfigure[Pump flow pulsatility versus average pulsatile flow.]{
   \includegraphics[scale =0.16752] {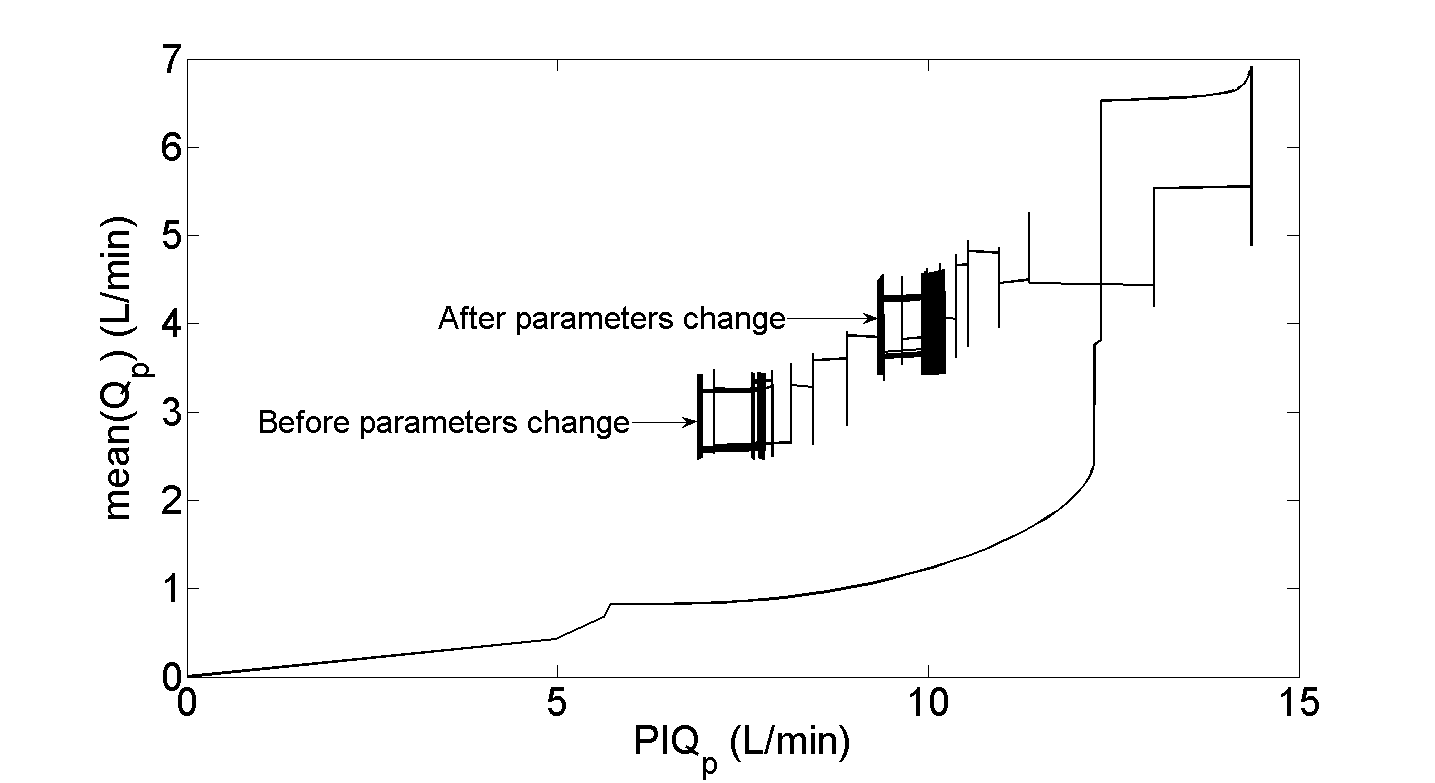}
  \label{66g}
 }

\subfigure[Pump flow compared with desired reference flow at initial time.]{
   \includegraphics[scale =0.16752] {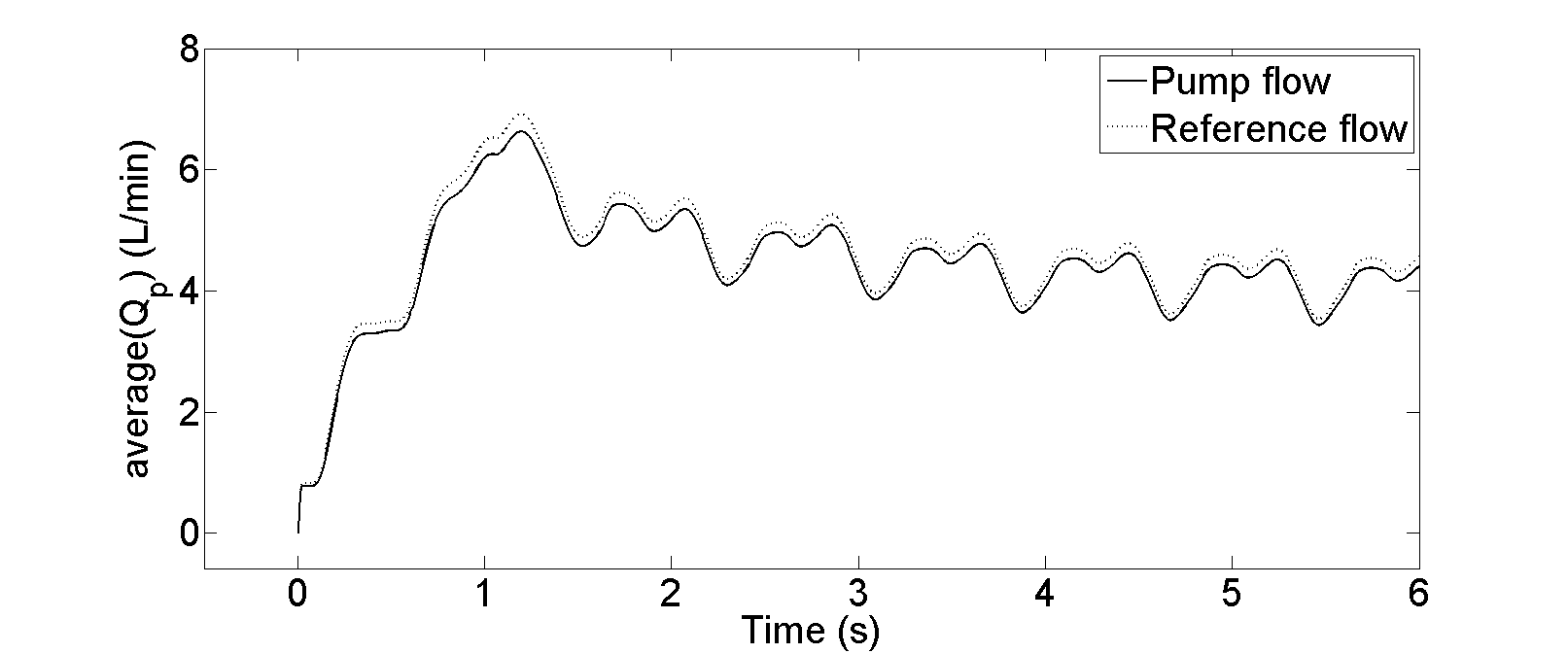}
   \label{i66h}
 }
  \subfigure[Pump flow compared with desired reference flow at induced time.]{
   \includegraphics[scale =0.16752] {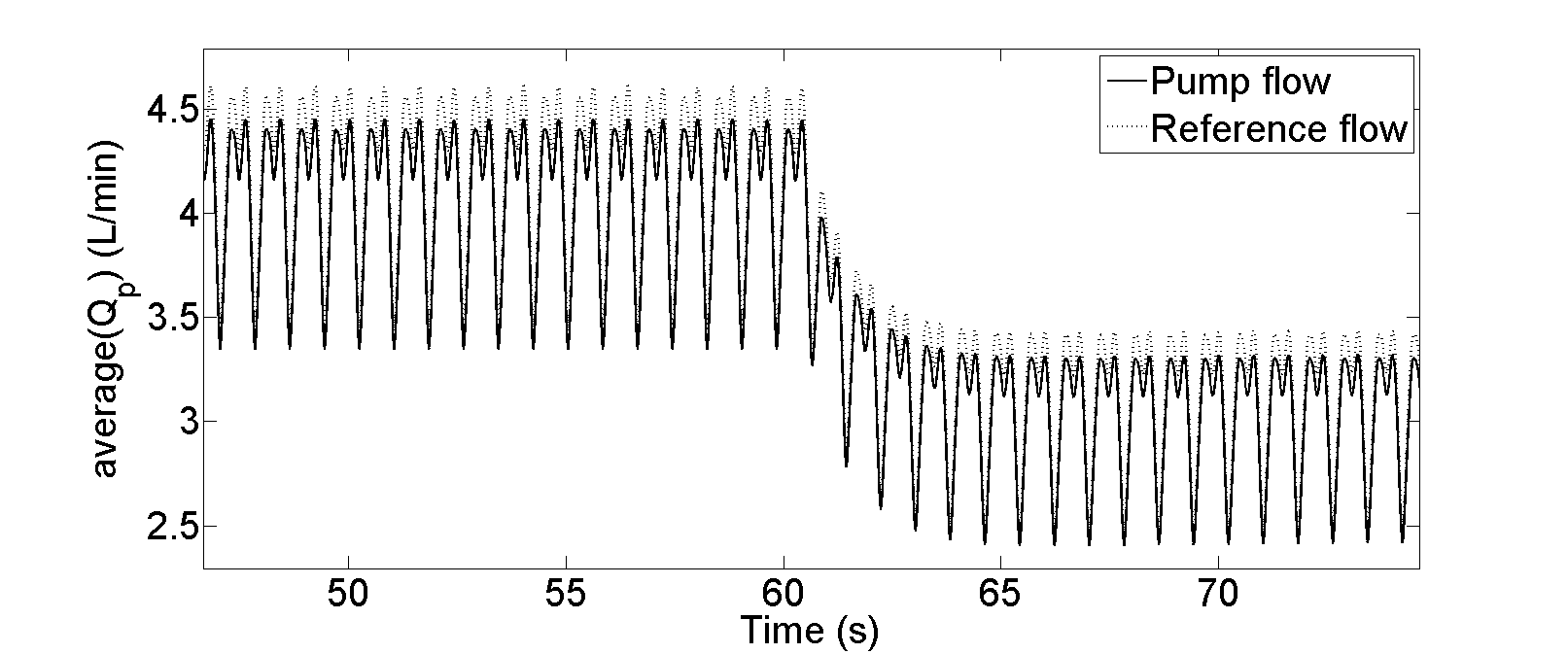}
   \label{66h}
 }

\subfigure[Measured steady state pump flow against estimated pump flow.]{
   \includegraphics[scale =0.1752] {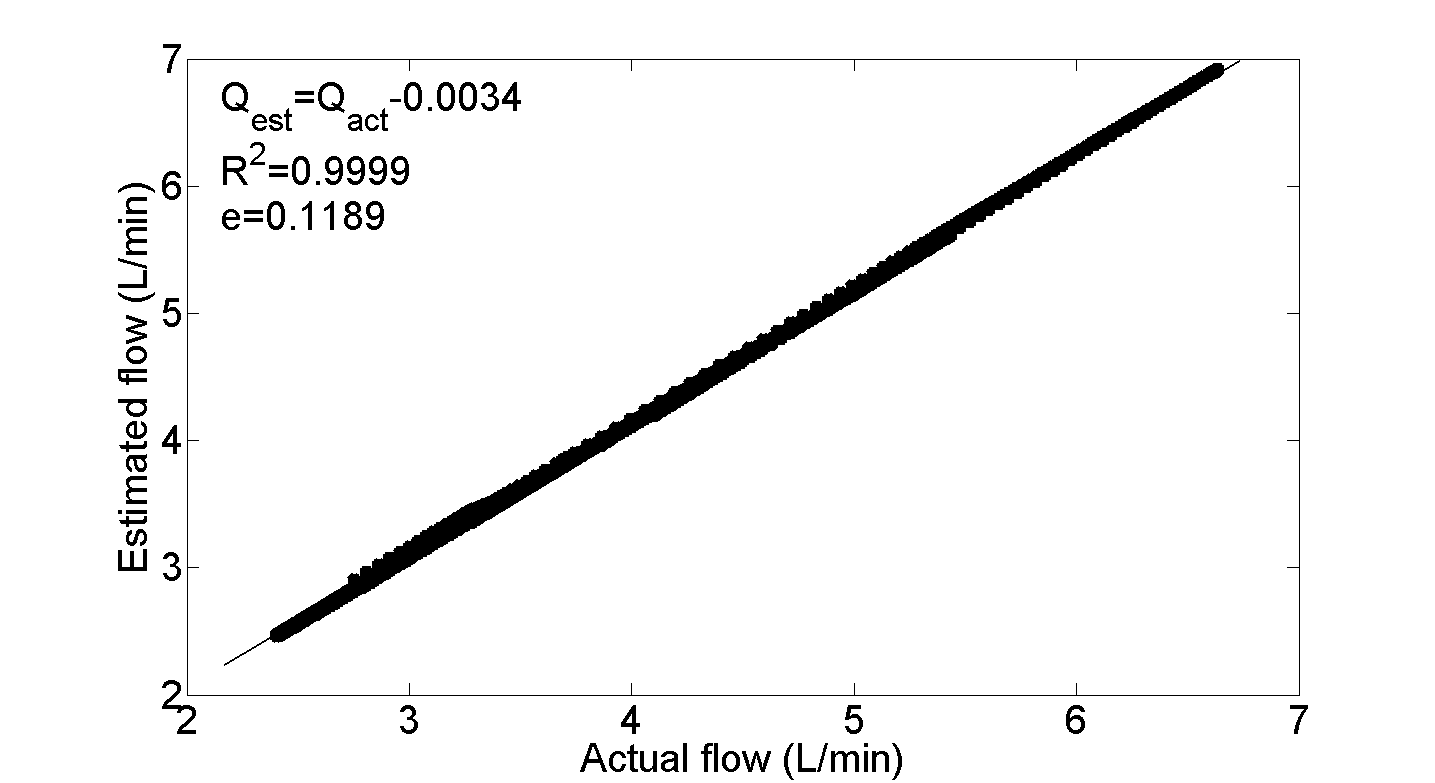}
   \label{66i}
 }

\caption{Pump variable results in rest condition when the system induced at 60s.}
\label{6:60b}
\end{figure*}

\begin{figure*}[htbp]
\centering
\subfigure[LV volume versus LV pressure before and after Parameter Change.]{
   \includegraphics[scale =0.16752] {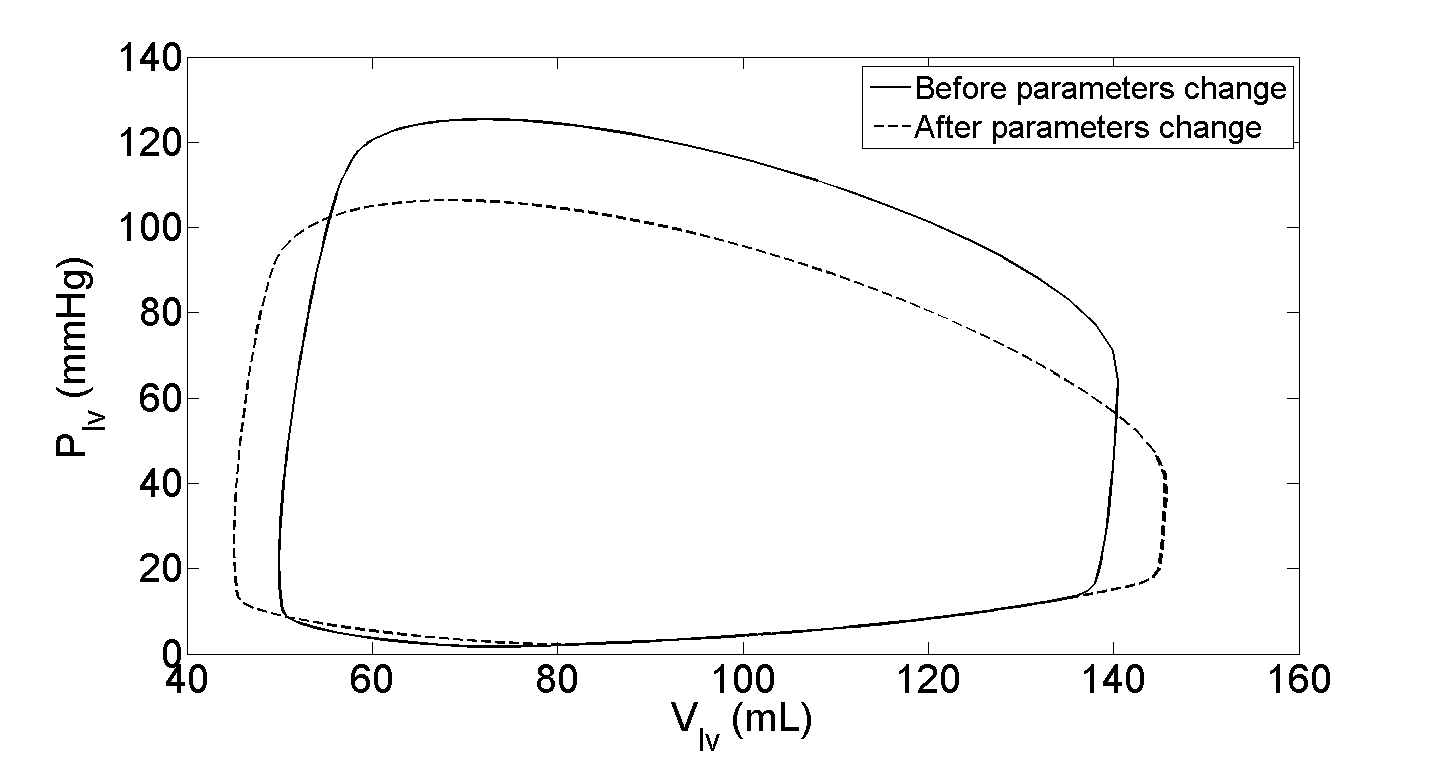}
   \label{69a}
 }
\subfigure[RV volume versus RV pressure before and after Parameter Change.]{
   \includegraphics[scale =0.16752] {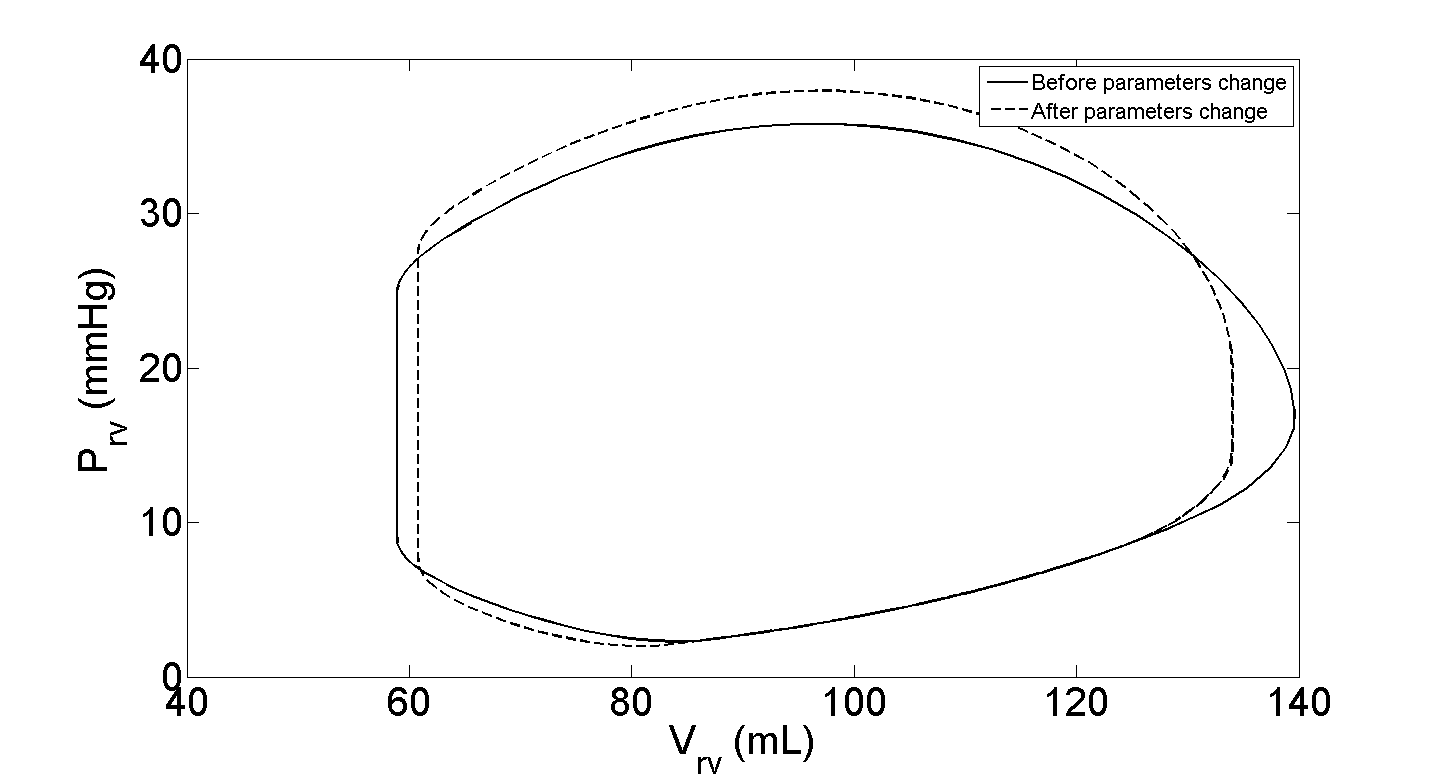}
   \label{69b}
 }

 \subfigure[Aortic pressure.]{
   \includegraphics[scale =0.16752] {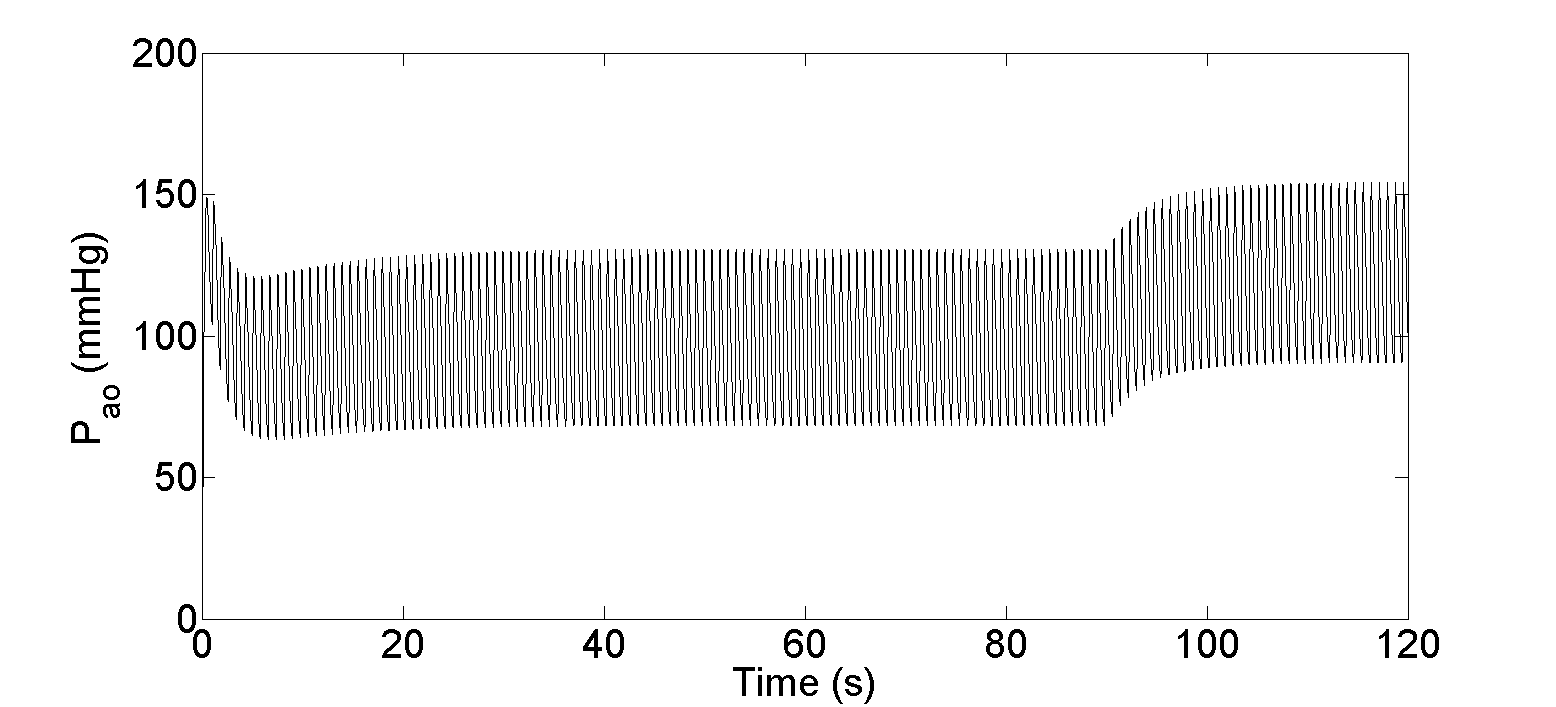}
   \label{69c}
 }
  \subfigure[Left atrial pressure.]{
   \includegraphics[scale =0.16752] {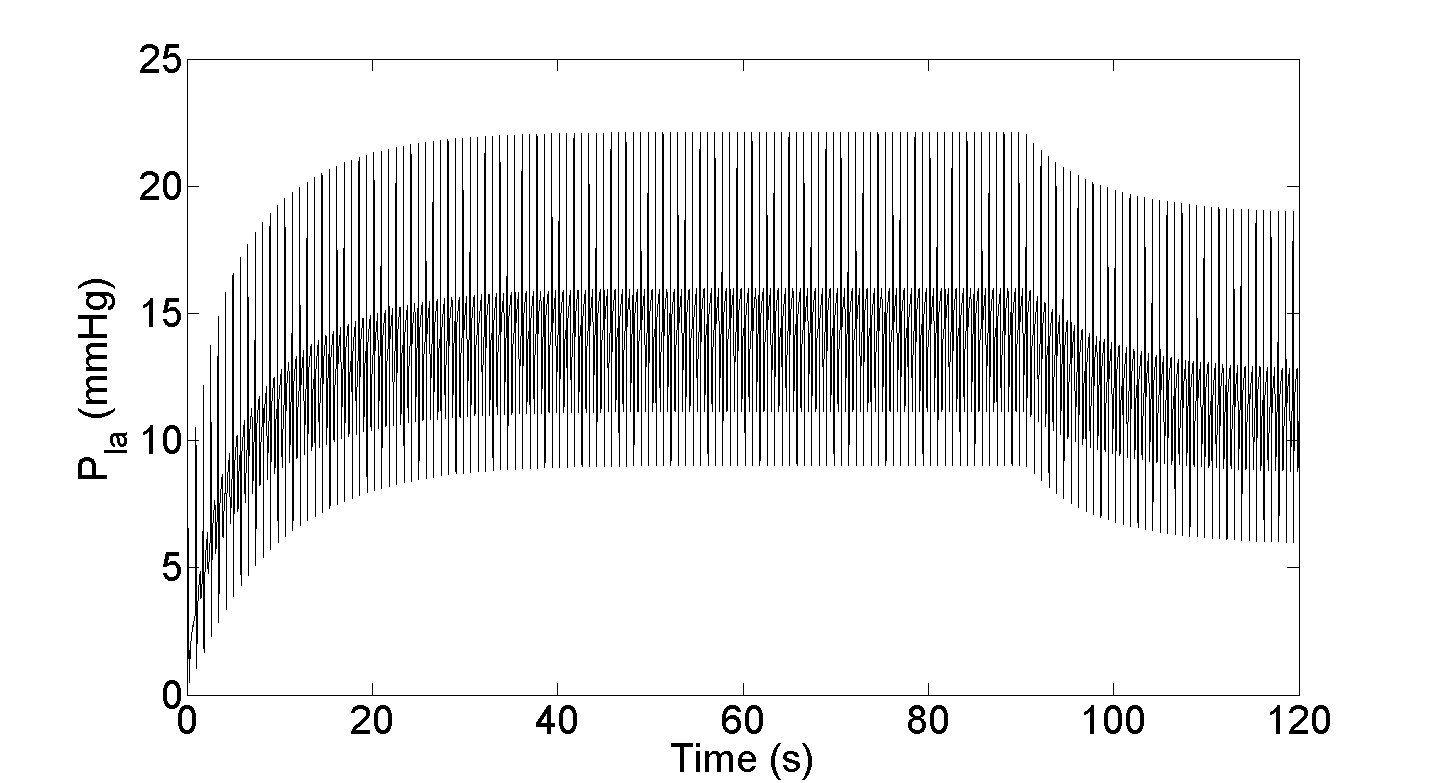}
   \label{69d}
 }

\subfigure[Right atrial pressure.]{
   \includegraphics[scale =0.16752] {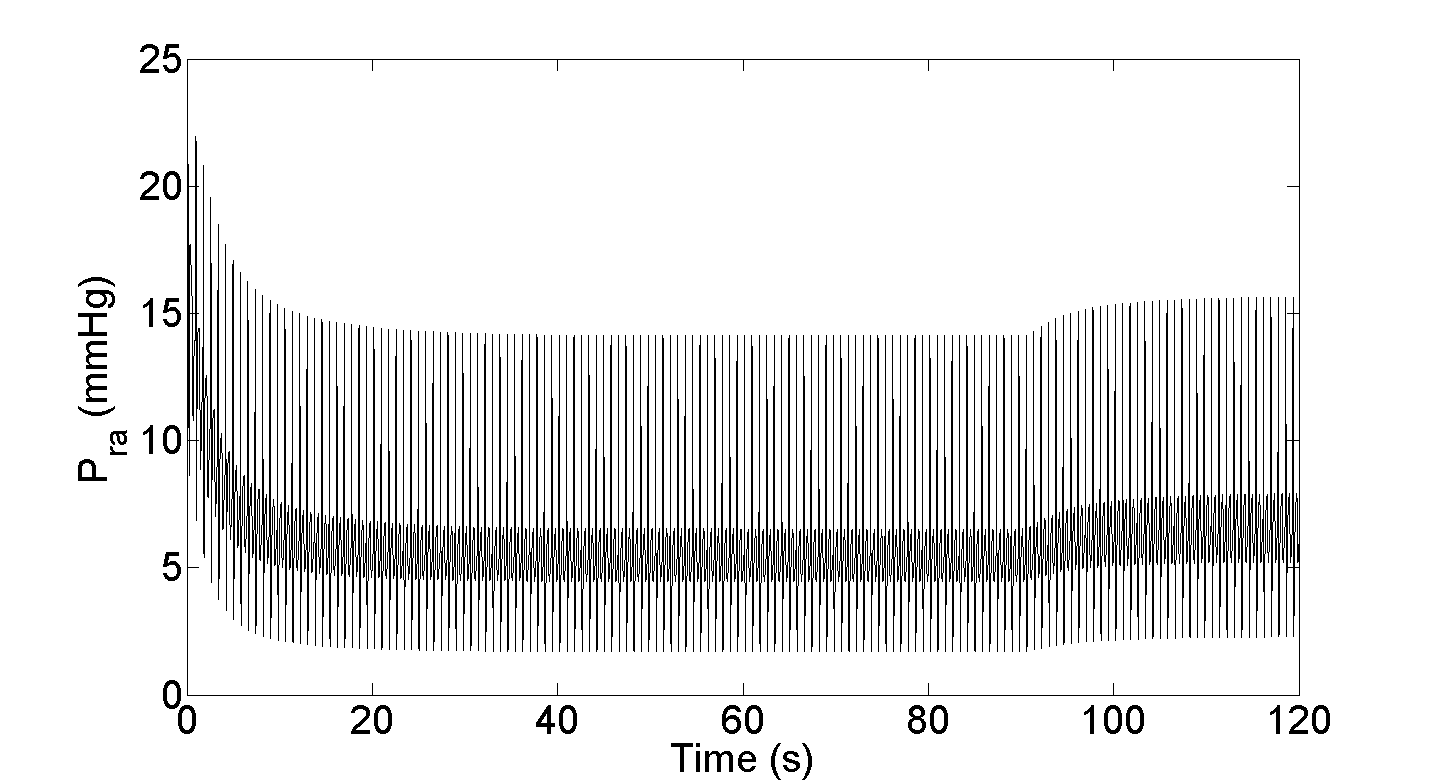}
   \label{69e}
 }
\caption{Hemodynamic variables results in rest condition when the system induced at 90s.}
\label{6:90a}
\end{figure*}

\begin{figure*}[htbp]
\centering
\subfigure[Average pump speed.]{
   \includegraphics[scale =0.16752] {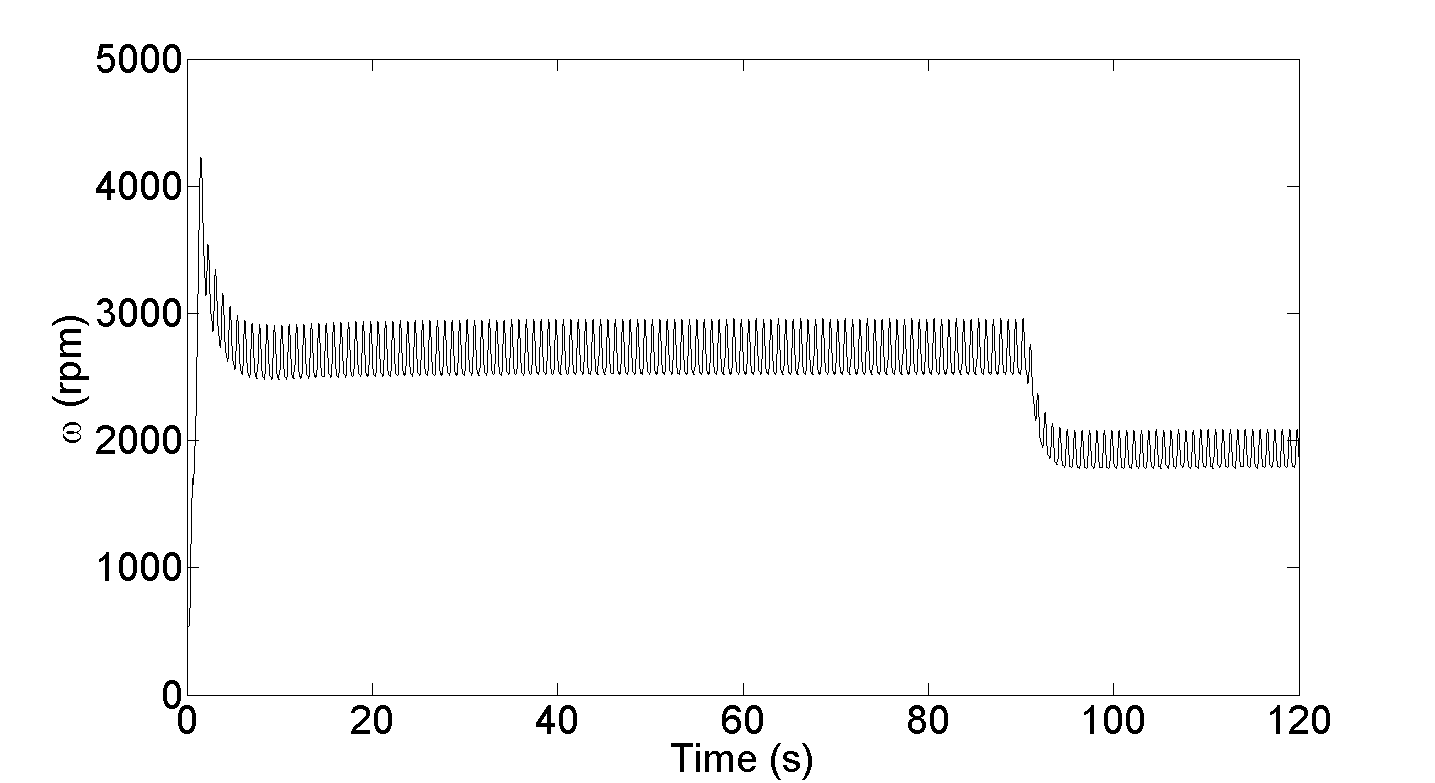}
   \label{69f}
 }
 \subfigure[Pump flow pulsatility versus average pulsatile flow.]{
   \includegraphics[scale =0.16752] {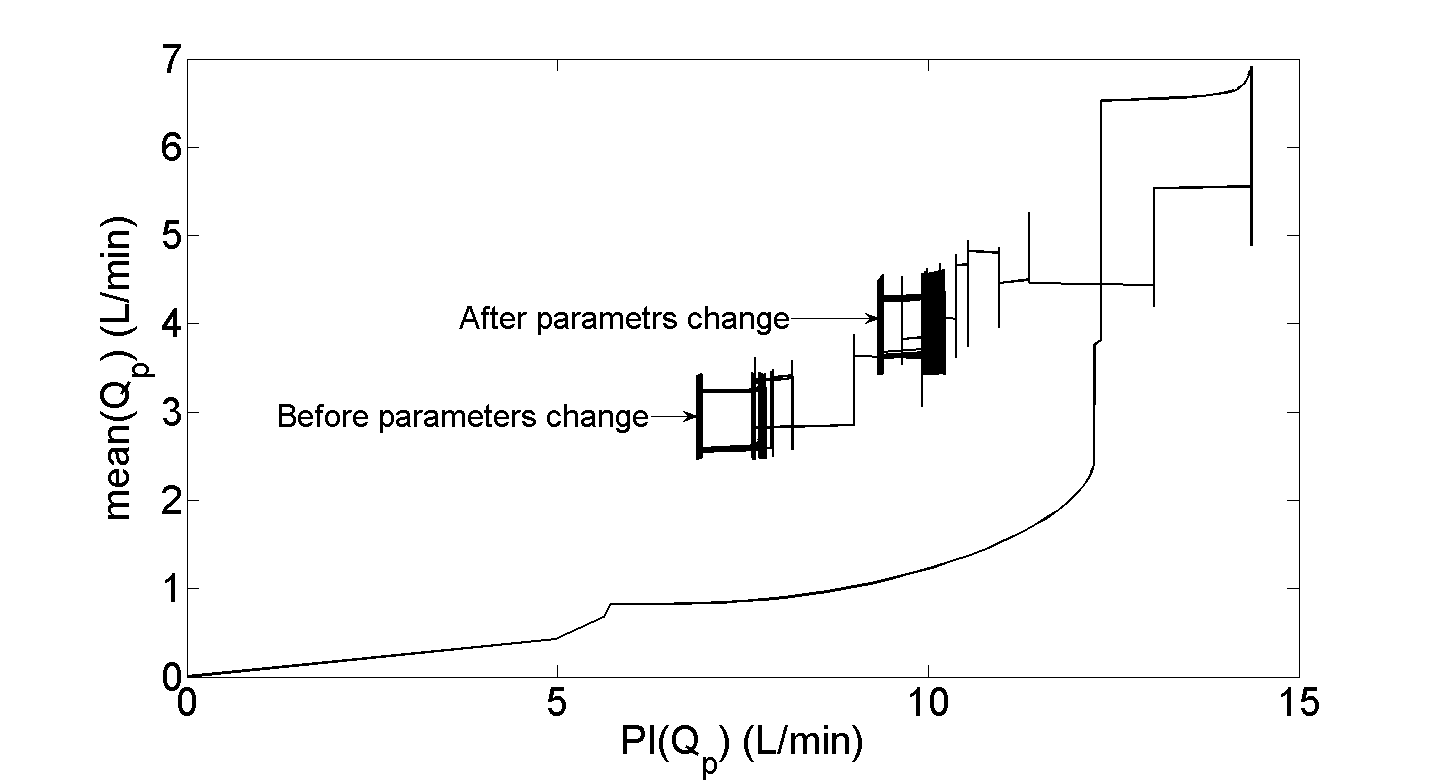}
  \label{69g}
 }

\subfigure[Pump flow compared with desired reference flow at initial time.]{
   \includegraphics[scale =0.16752] {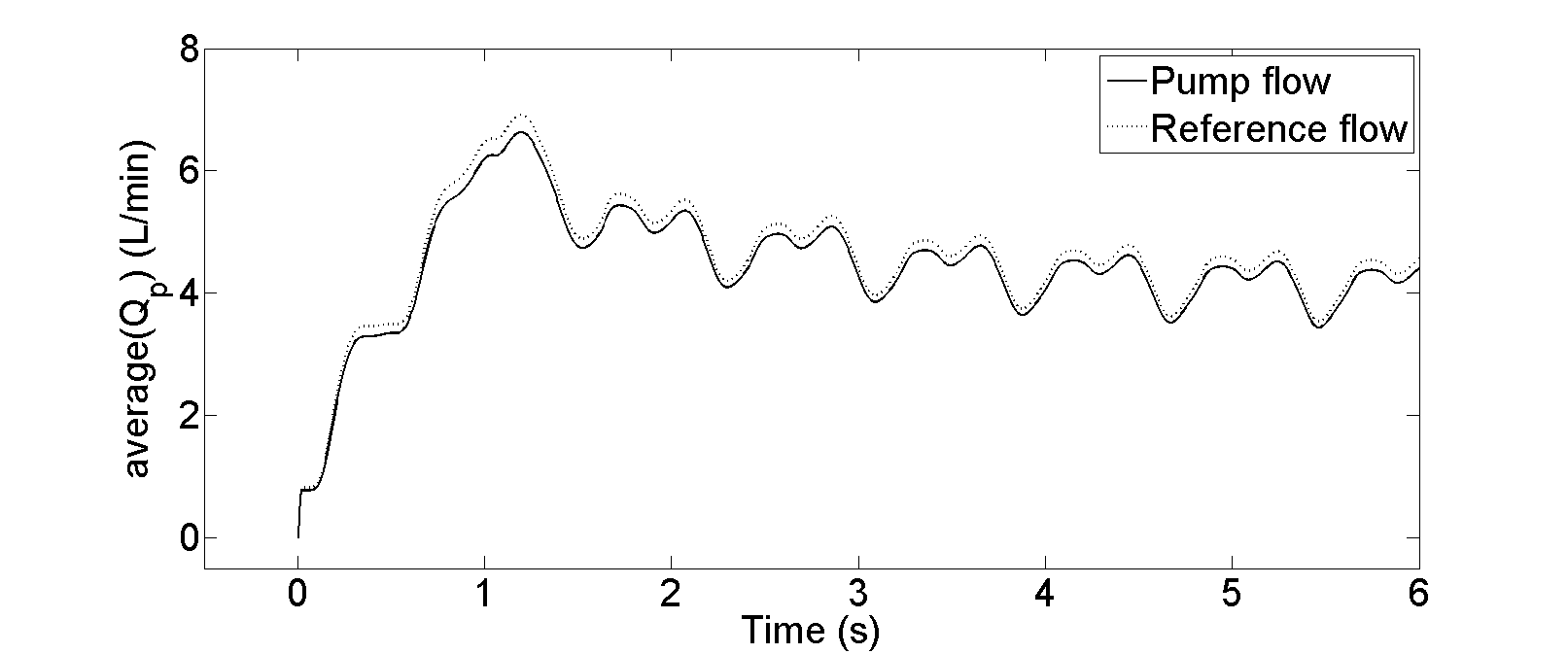}
   \label{i69h}
 }
  \subfigure[Pump flow compared with desired reference flow at induced time.]{
   \includegraphics[scale =0.16752] {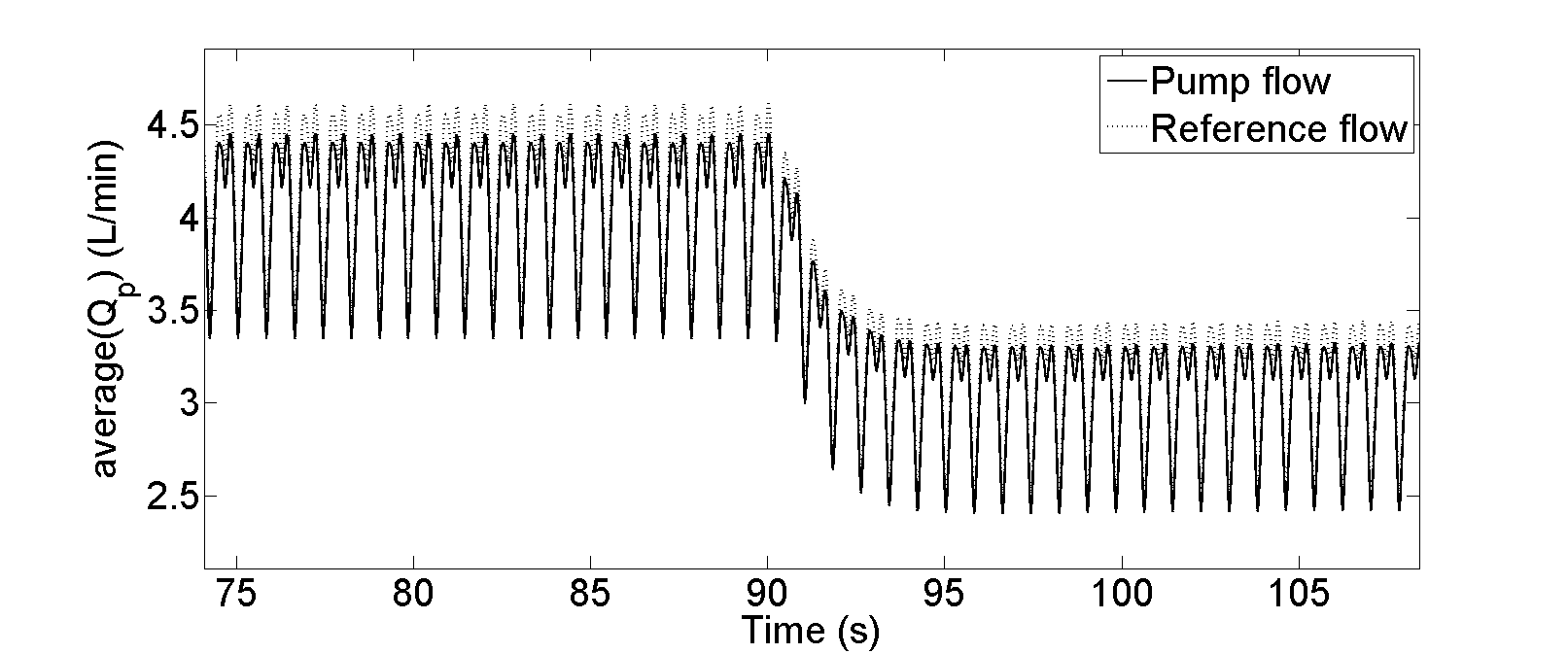}
   \label{69h}
 }

\subfigure[Measured steady state pump flow against estimated pump flow.]{
   \includegraphics[scale =0.1752] {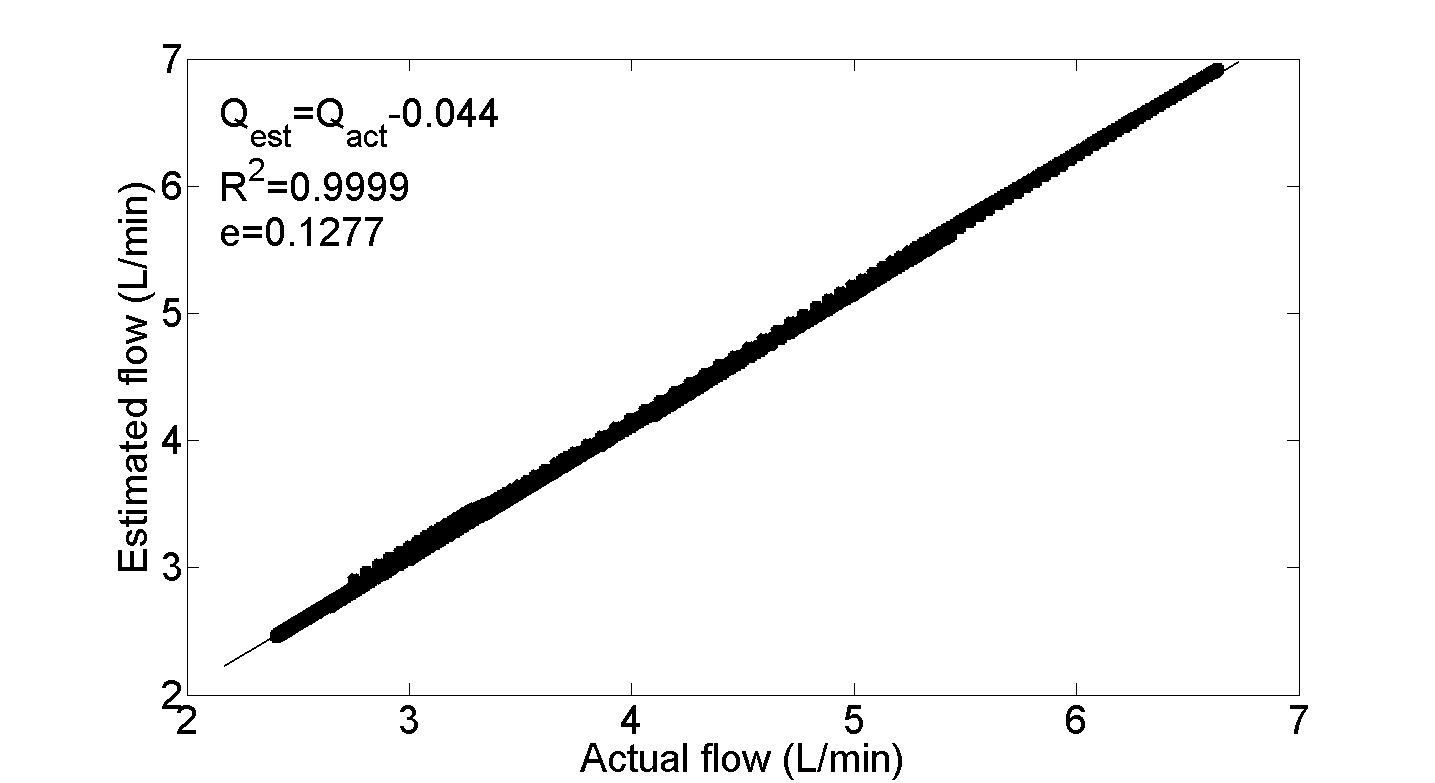}
   \label{69i}
 }

\caption{Pump variable results in rest condition when the system induced at 90s.}
\label{6:90b}
\end{figure*}

\begin{figure*}[htbp]
\centering
\subfigure[LV volume versus LV pressure before and after Parameter Change.]{
   \includegraphics[scale =0.16752] {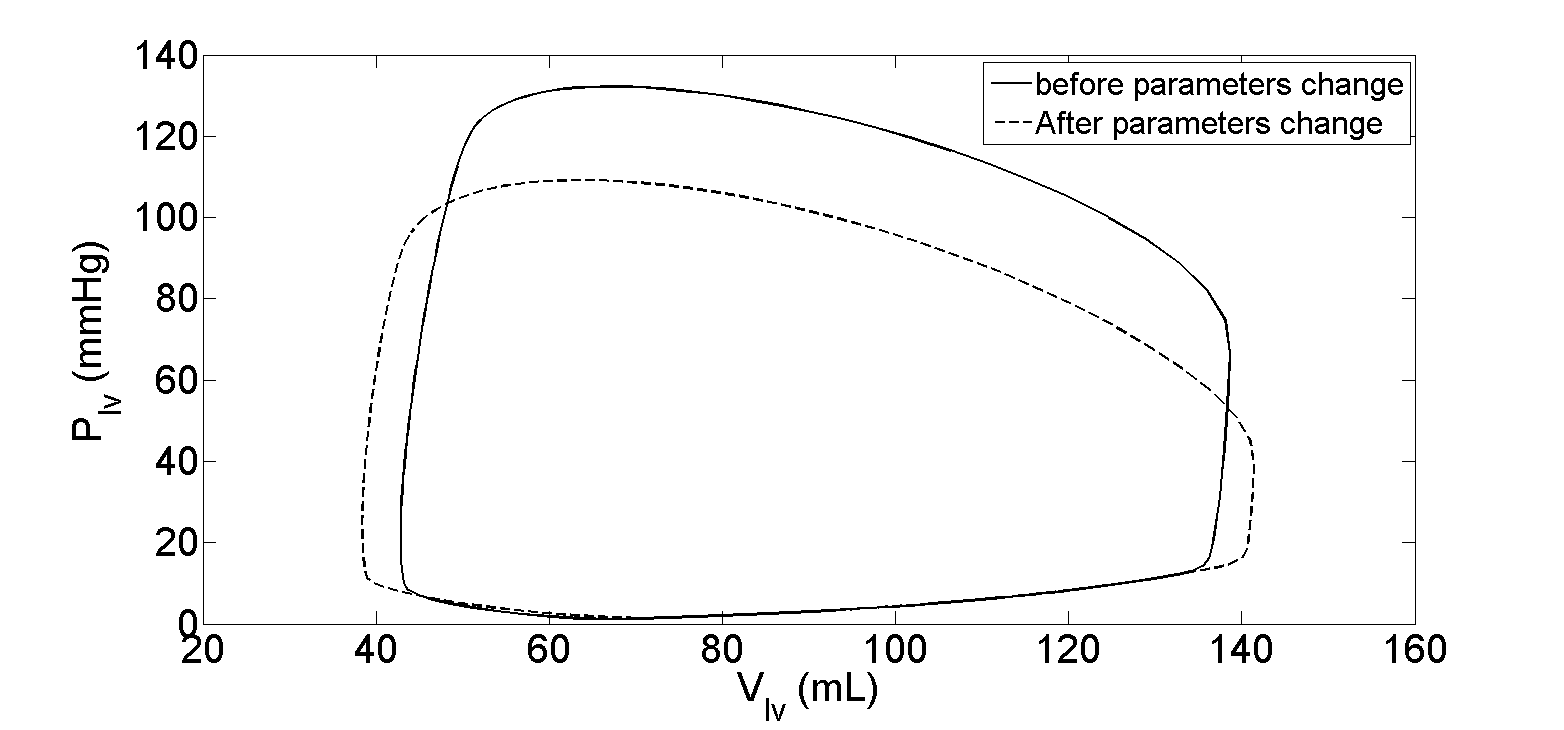}
   \label{62a}
 }
\subfigure[RV volume versus RV pressure before and after Parameter Change.]{
   \includegraphics[scale =0.16752] {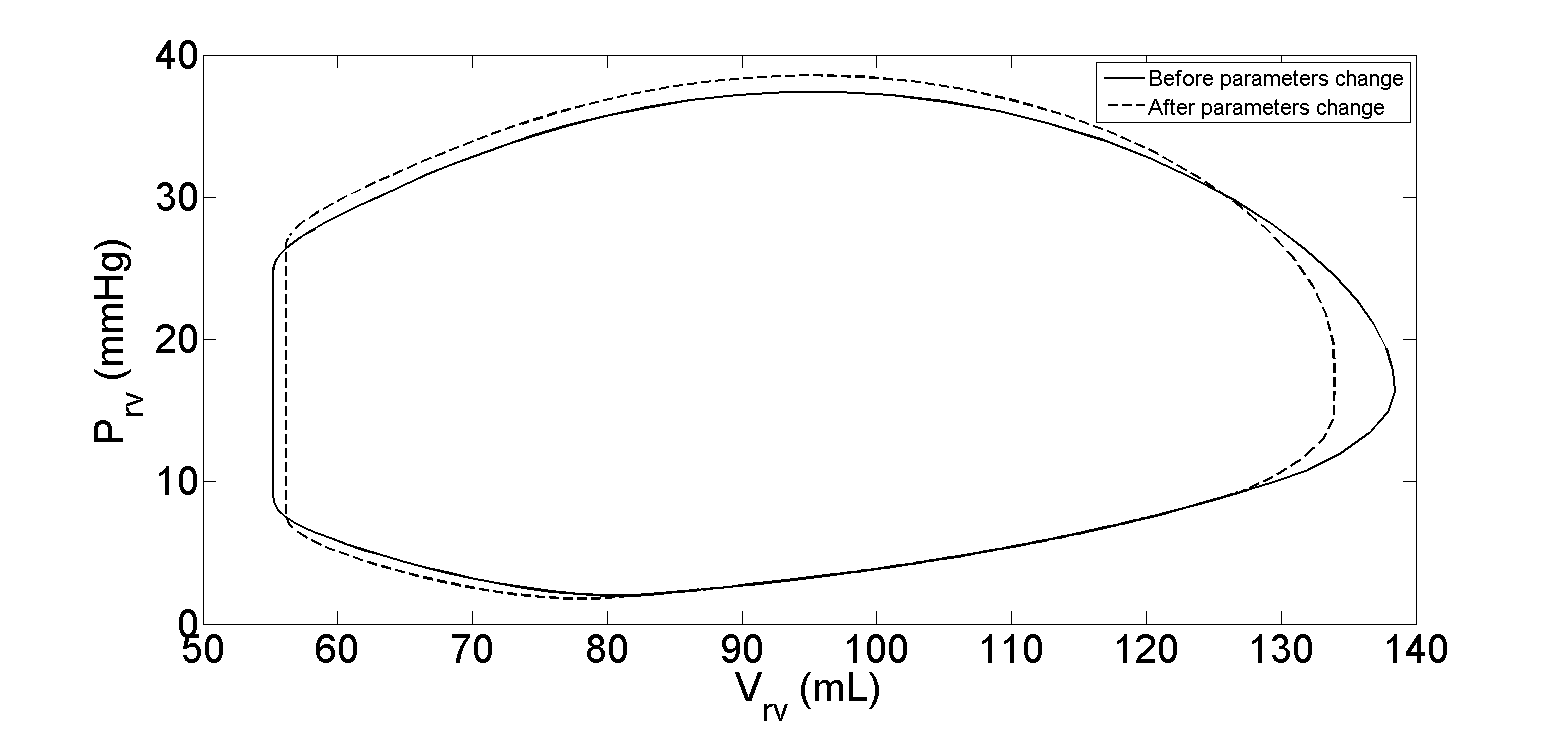}
   \label{62b}
 }

 \subfigure[Aortic pressure.]{
   \includegraphics[scale =0.16752] {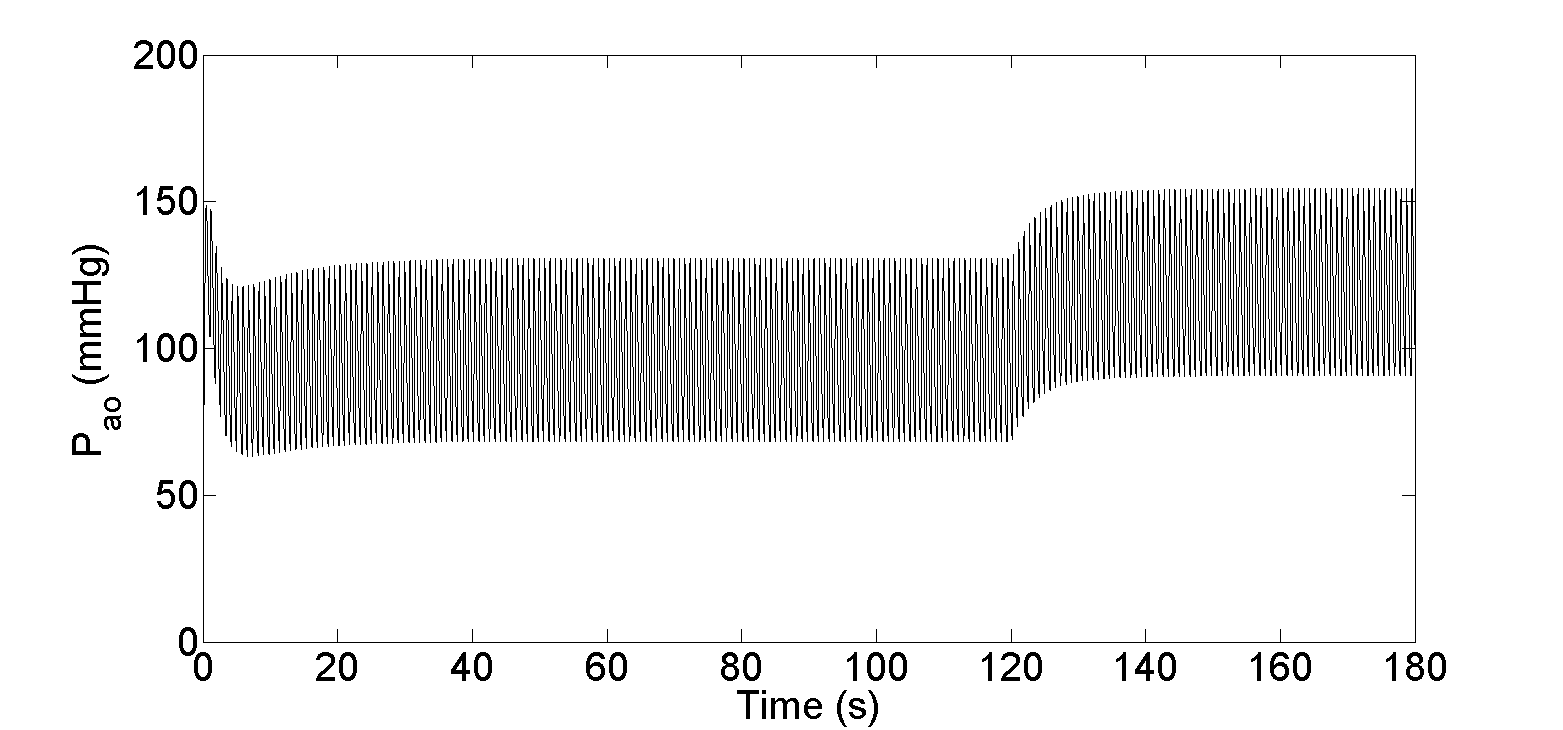}
   \label{62c}
 }
  \subfigure[Left atrial pressure.]{
   \includegraphics[scale =0.16752] {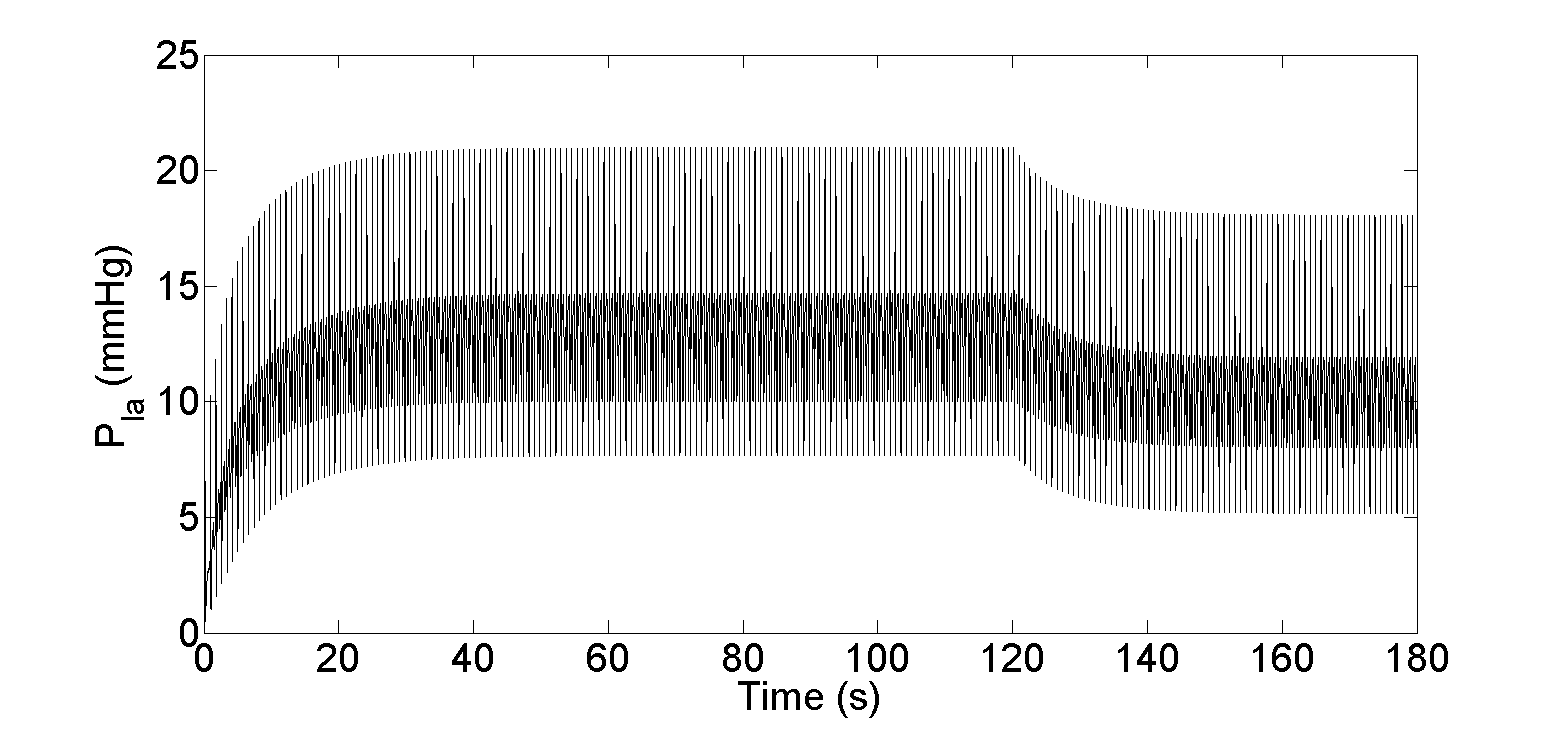}
   \label{62d}
 }

\subfigure[Right atrial pressure.]{
   \includegraphics[scale =0.16752] {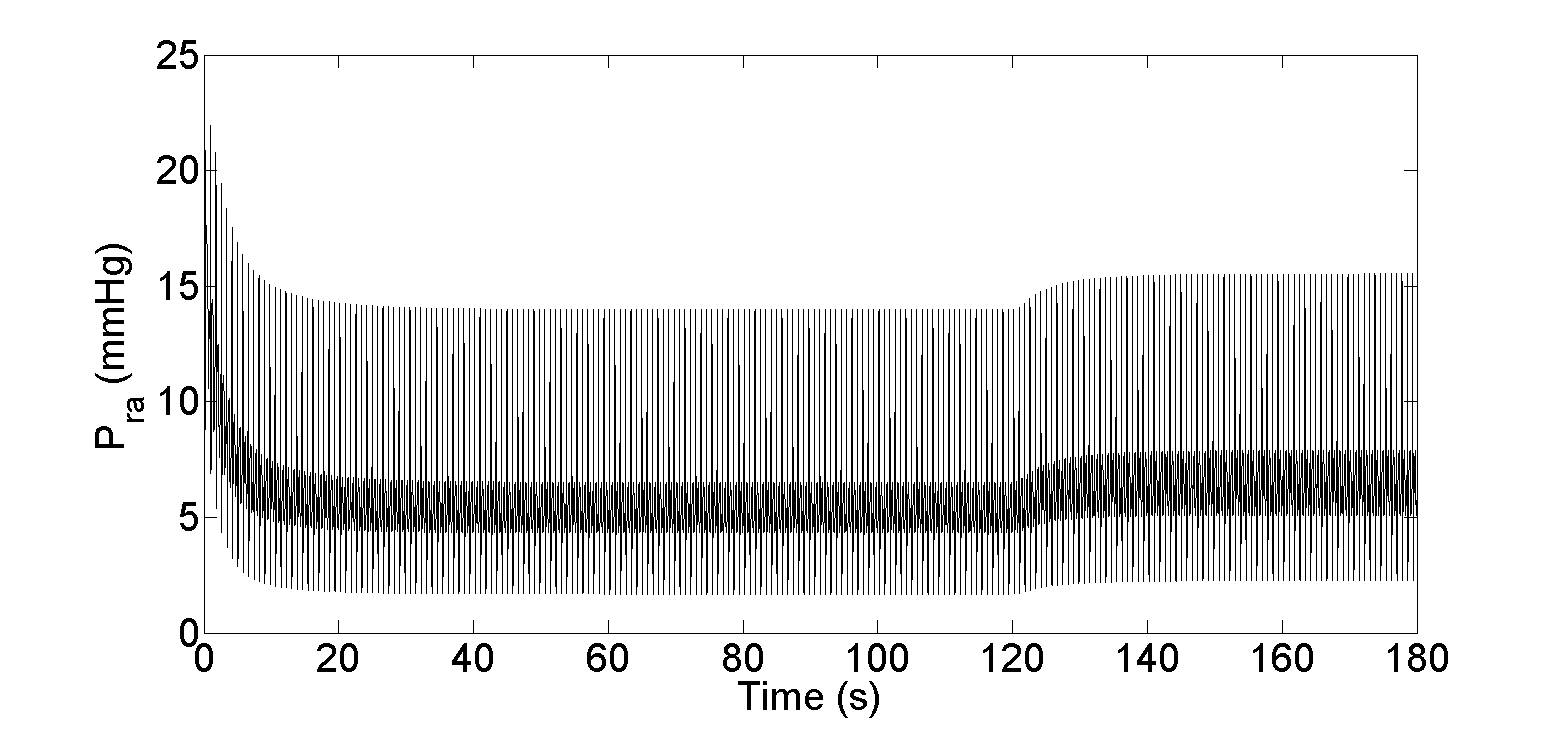}
   \label{62e}
 }
\caption{Hemodynamic variables results in rest condition when the system induced at 120s.}
\label{6:20a}
\end{figure*}

\begin{figure*}[htbp]
\centering
\subfigure[Average pump speed.]{
   \includegraphics[scale =0.16752] {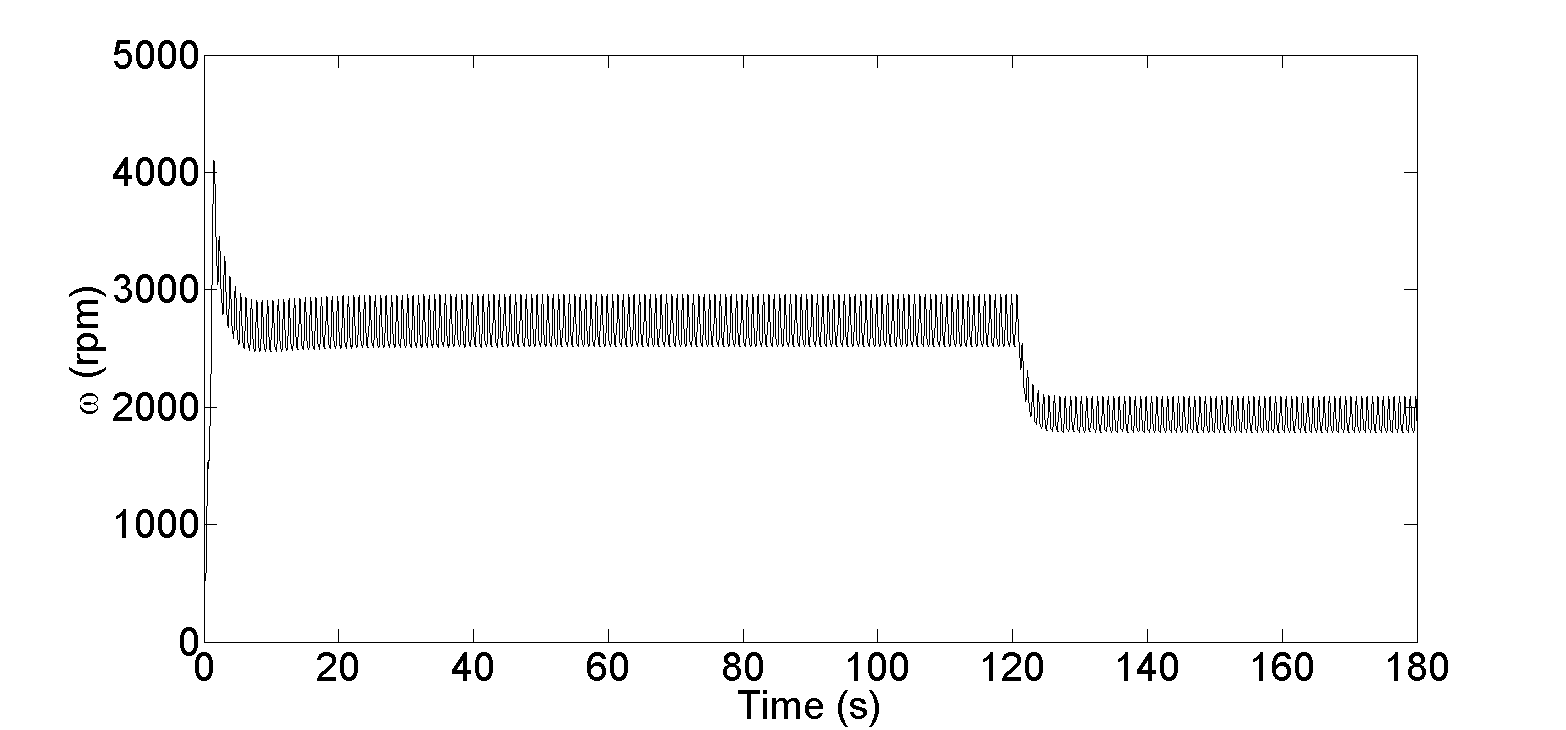}
   \label{62f}
 }
 \subfigure[Pump flow pulsatility versus average pulsatile flow.]{
   \includegraphics[scale =0.16752] {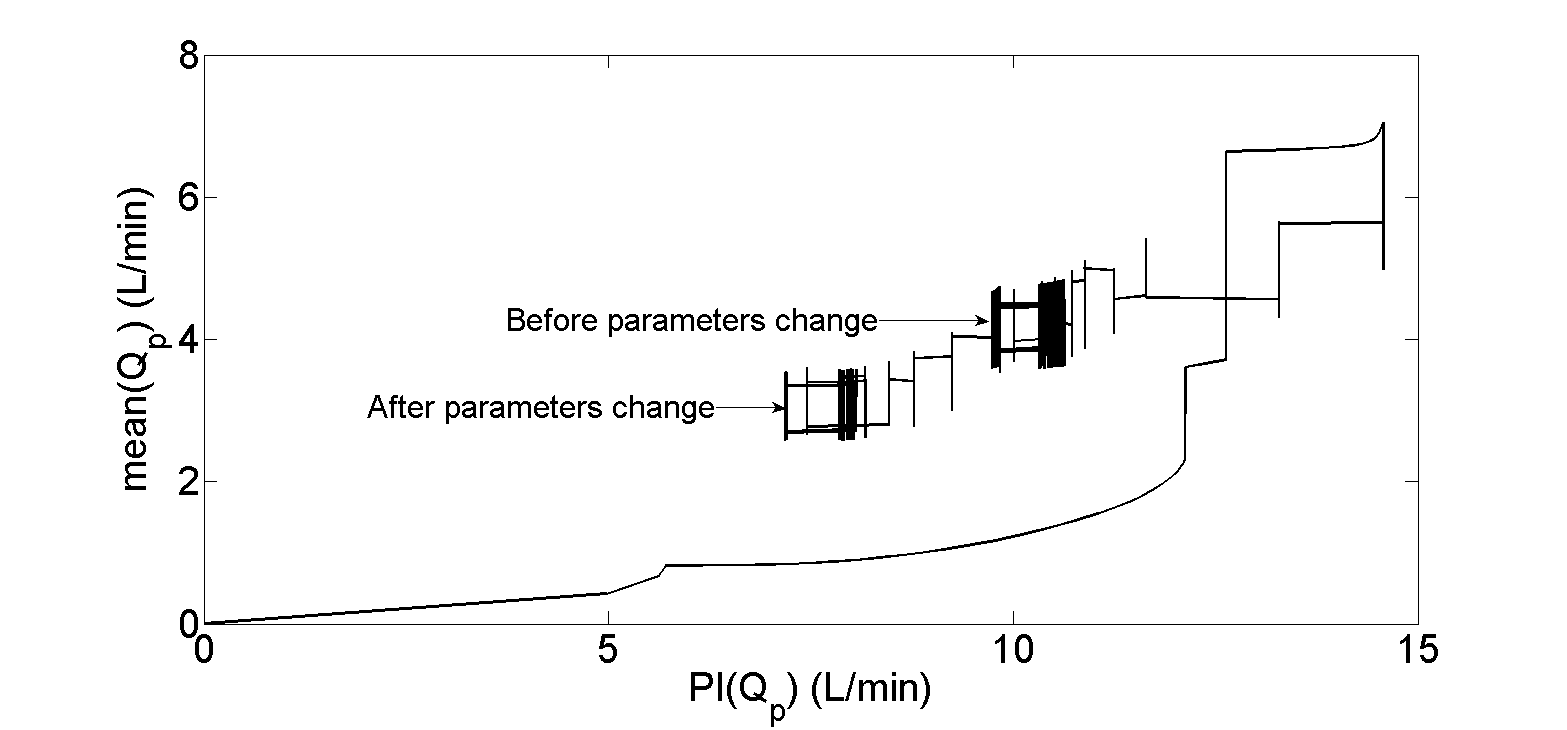}
  \label{62g}
 }

\subfigure[Pump flow compared with desired reference flow at initial time.]{
   \includegraphics[scale =0.16752] {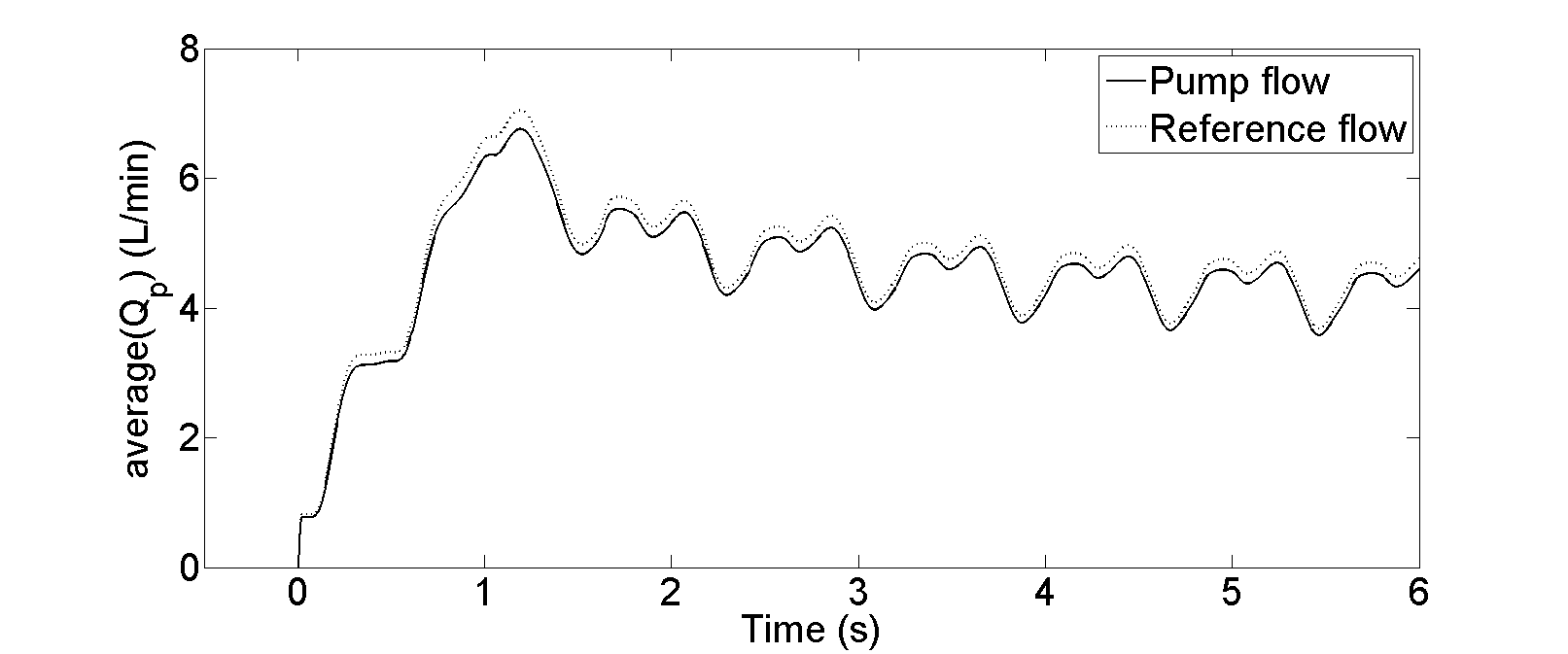}
   \label{i62h}
 }
  \subfigure[Pump flow compared with desired reference flow at induced time.]{
   \includegraphics[scale =0.16752] {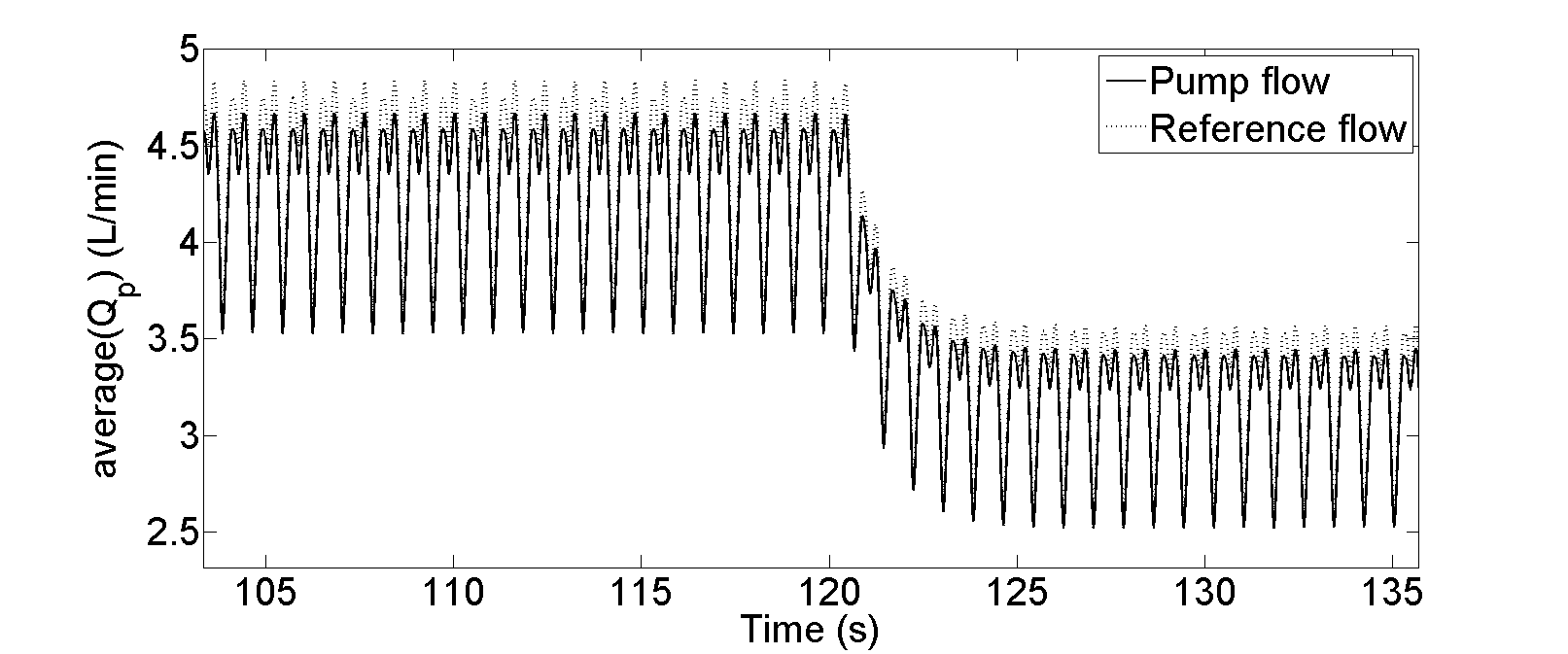}
   \label{62h}
 }

\subfigure[Measured steady state pump flow against estimated pump flow.]{
   \includegraphics[scale =0.16752] {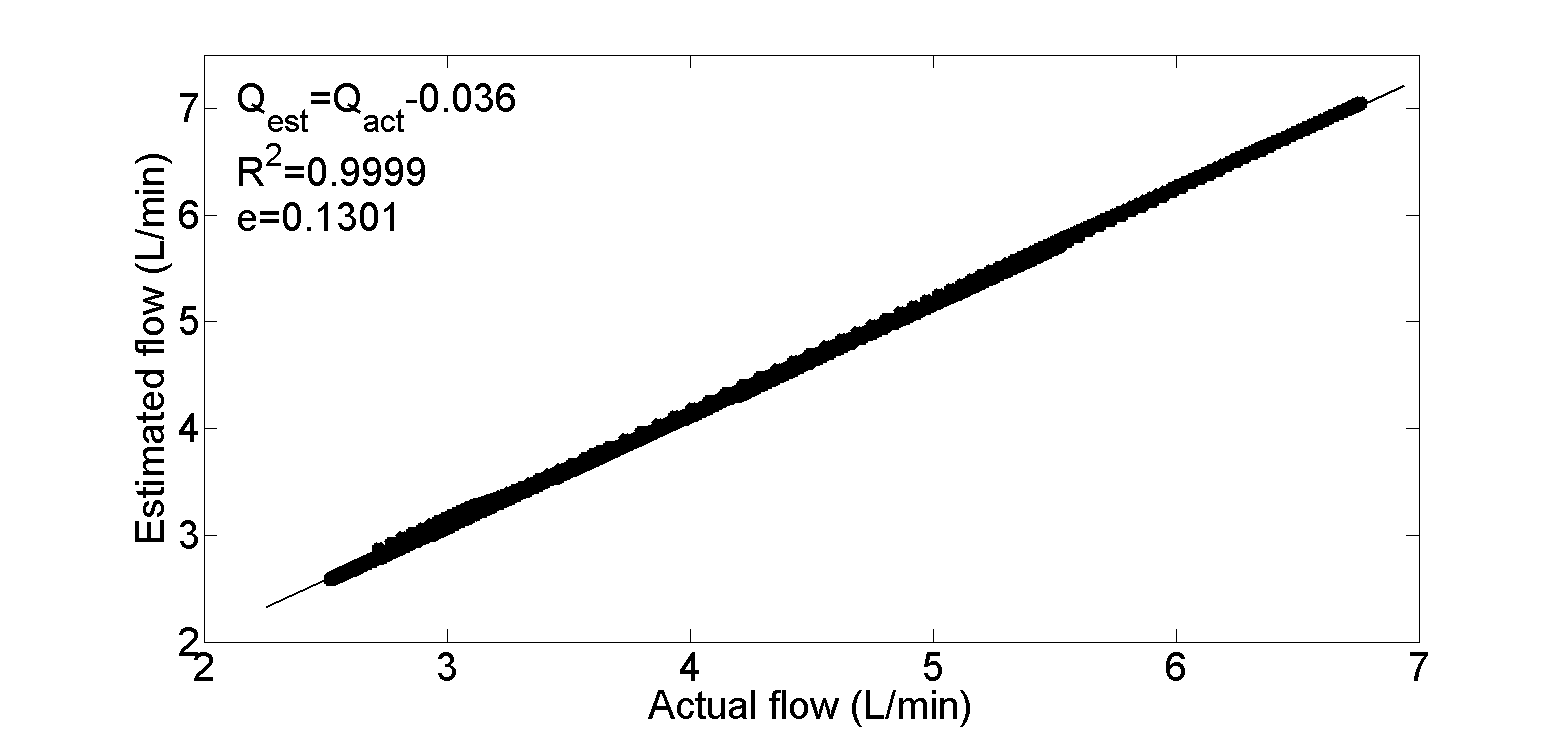}
   \label{62i}
 }

\caption{Pump variable results in rest condition when the system induced at 120s.}
\label{6:20b}
\end{figure*}

Figures \ref{6:30a} - \ref{6:20b} show the immediate response of the controller corresponding to the blood loss at different period of times (30s, 60s, 90s and 120s). The system is induced at the middle of each period by reduction of the blood volume. The reduction in blood volume caused a reduction in stroke volume of the right ventricle. This was associated with a shift to the left of the LV pressure-volume loop and slightly shifts to the right of the RV pressure-volume loops causing a reduction in LV end-diastolic and end-systolic volumes and pressure. In addition, the LVAD successfully increases the aortic pressure $P_{ao}$ and decrease left atrial pressure $P_{la}$ and keep the right atrial pressure $P_{ra}$ within the safe operating mode as shown in Figures (\ref{6:30a}, \ref{6:60a}, \ref{6:90a} and \ref{6:20a}).

The results of pump variables are illustrated in Figures (\ref{6:30b}, \ref{6:60b}, \ref{6:90b} and \ref{6:20b}). During parameters change, the controller responds to the decrease in LV preload and subsequently pump flow pulsatility by decreasing average pump rotational speed from around 3000 rpm to 2100 rpm, actual average pulsatile flow from around 4.50 L/min to 3.40 L/min and estimated average pulsatile flow from around 4.80 L/min to 3.60 L/min. These changes are substantially completed within four heartbeats. Also, it can be observed that the simulated pump flow accurately tracks the desired reference flow within an error of $\pm$ 0.2 L/min. In addition, a strong correlation can be observed between both actual and estimated average pulsatile flows from the Figures \ref{63h}, \ref{66h}, \ref{69h} and \ref{62h}..


\subsection{Results in Exercise Condition}

\begin{figure*}[htbp]
\centering
\subfigure[LV volume versus LV pressure before and after Parameter Change.]{
   \includegraphics[scale =0.16752] {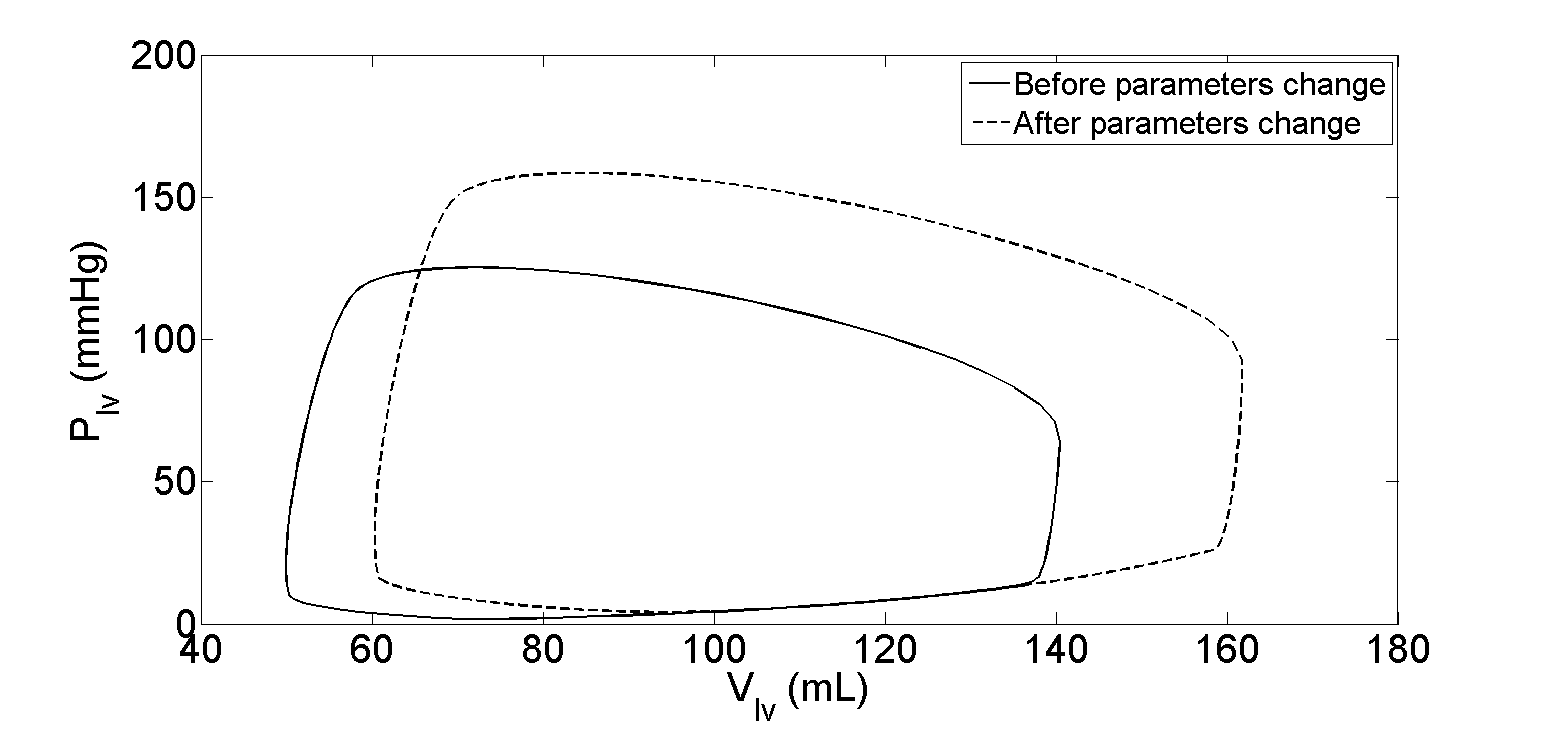}
   \label{63ea}
 }
\subfigure[RV volume versus RV pressure before and after Parameter Change.]{
   \includegraphics[scale =0.16752] {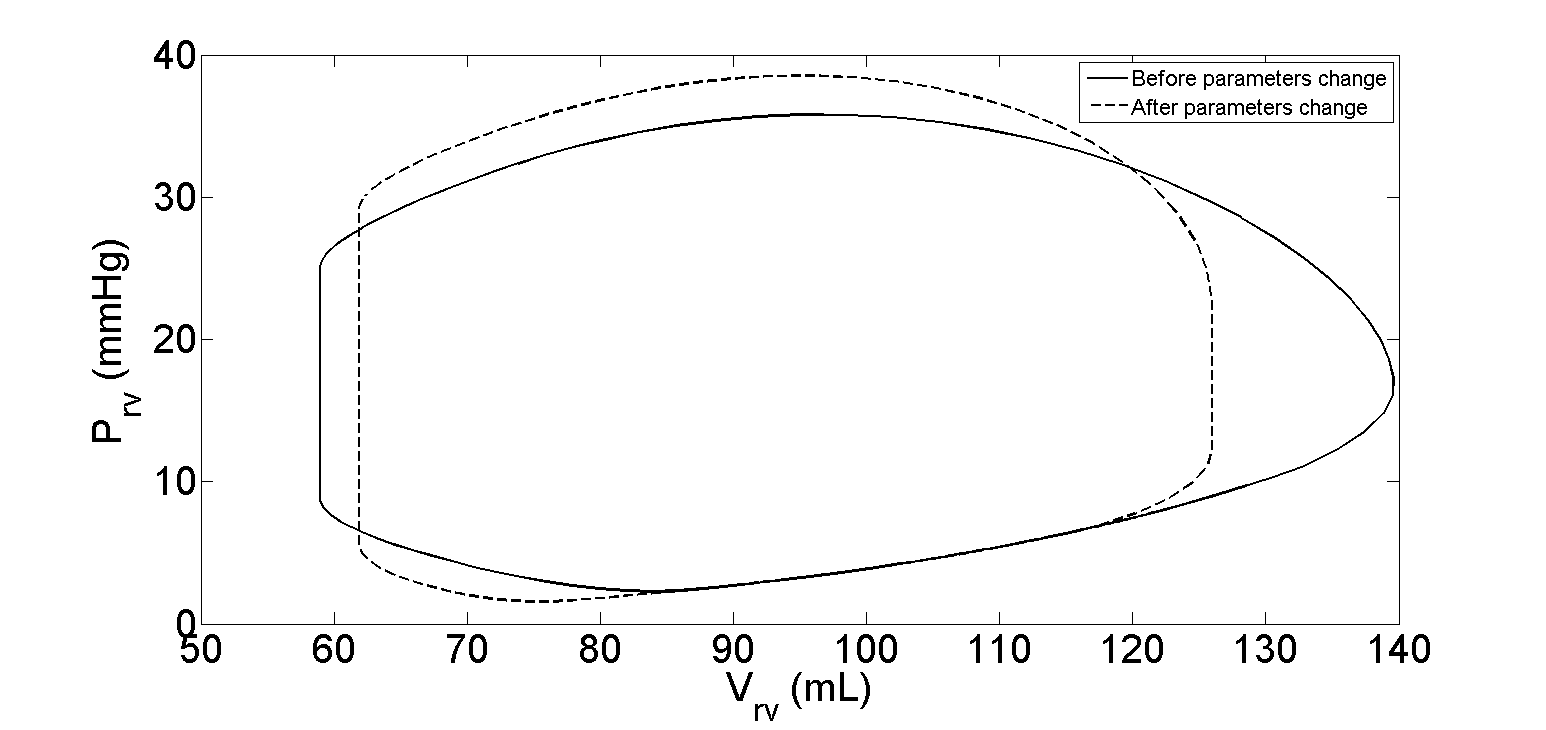}
   \label{63eb}
 }

 \subfigure[Aortic pressure.]{
   \includegraphics[scale =0.16752] {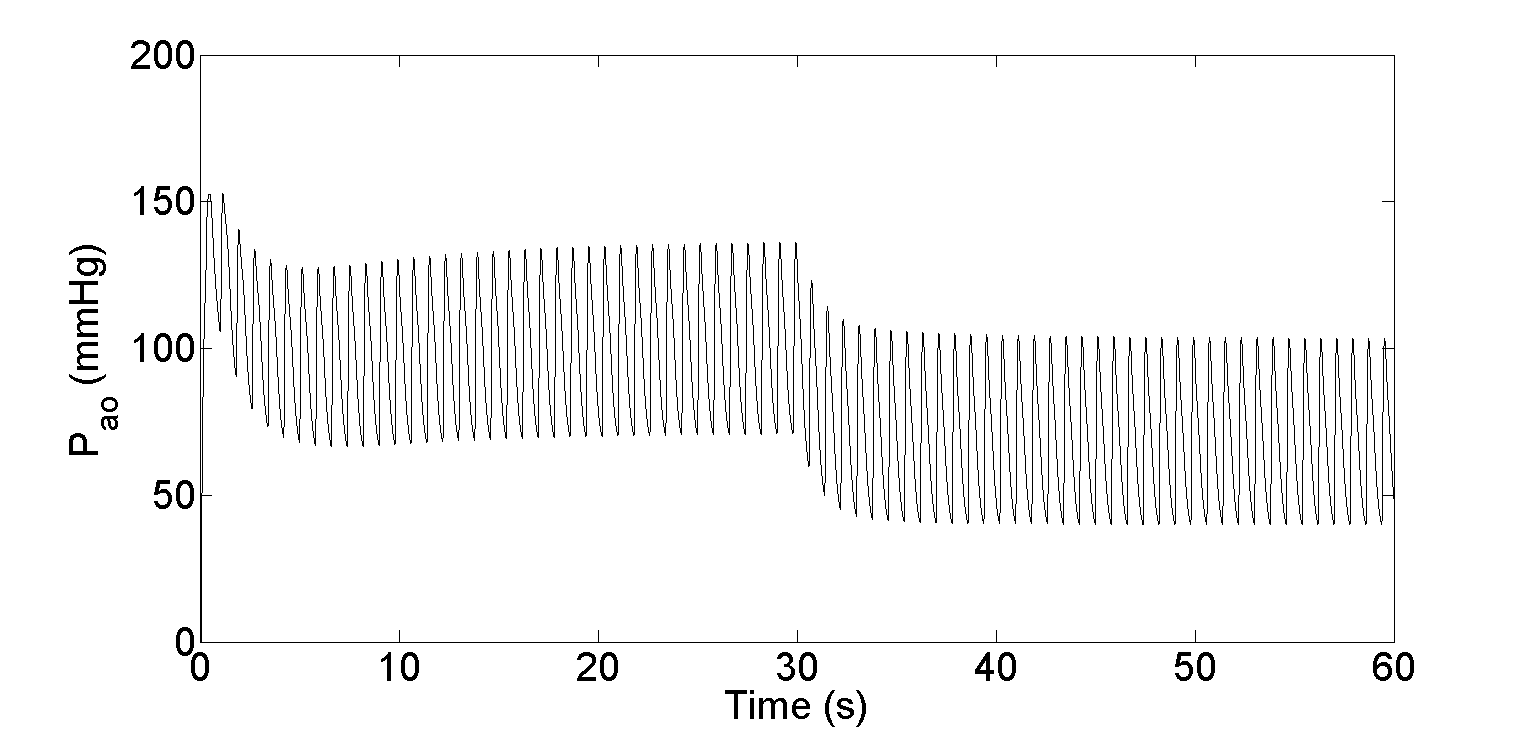}
   \label{63ec}
 }
  \subfigure[Left atrial pressure.]{
   \includegraphics[scale =0.16752] {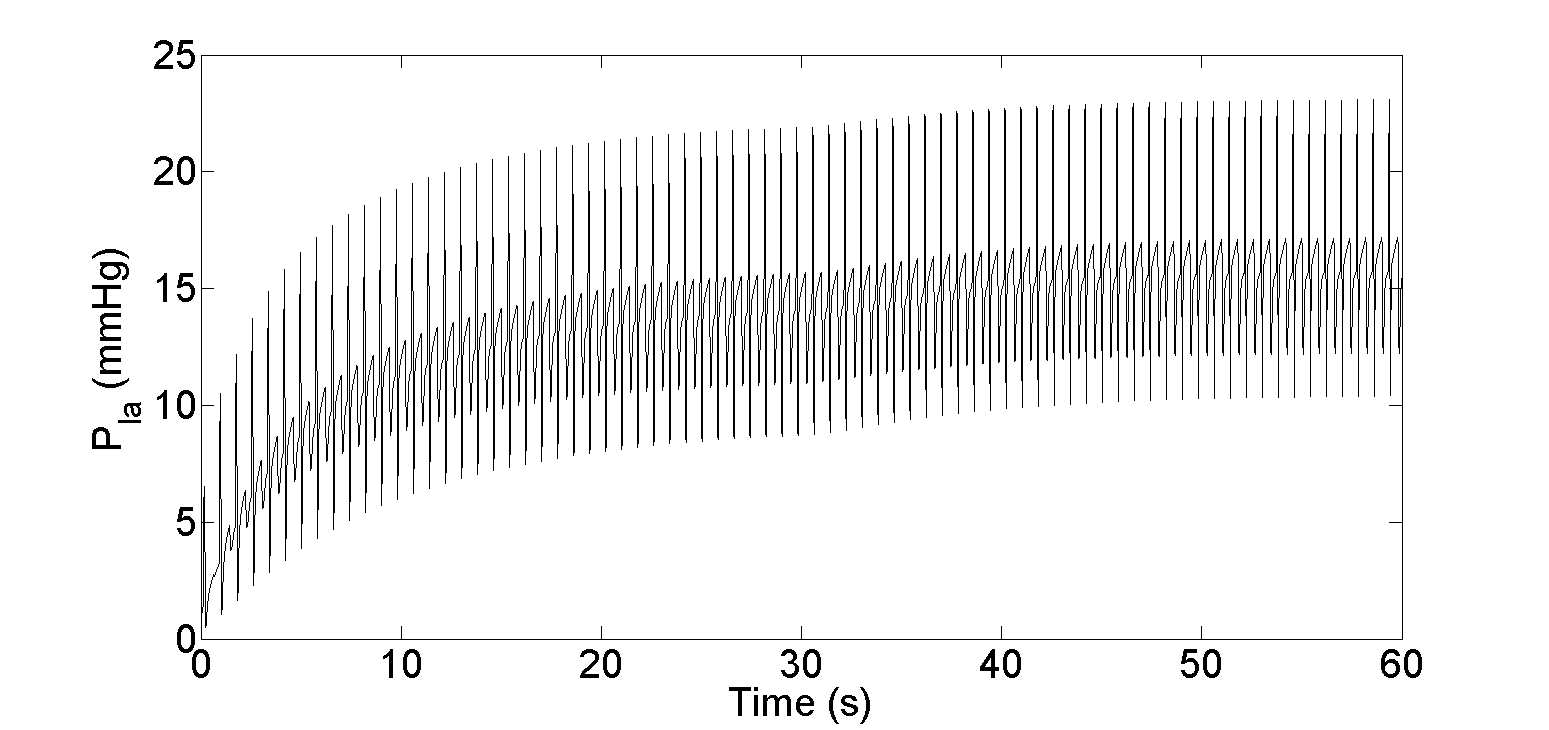}
   \label{63ed}
 }

\subfigure[Right atrial pressure.]{
   \includegraphics[scale =0.16752] {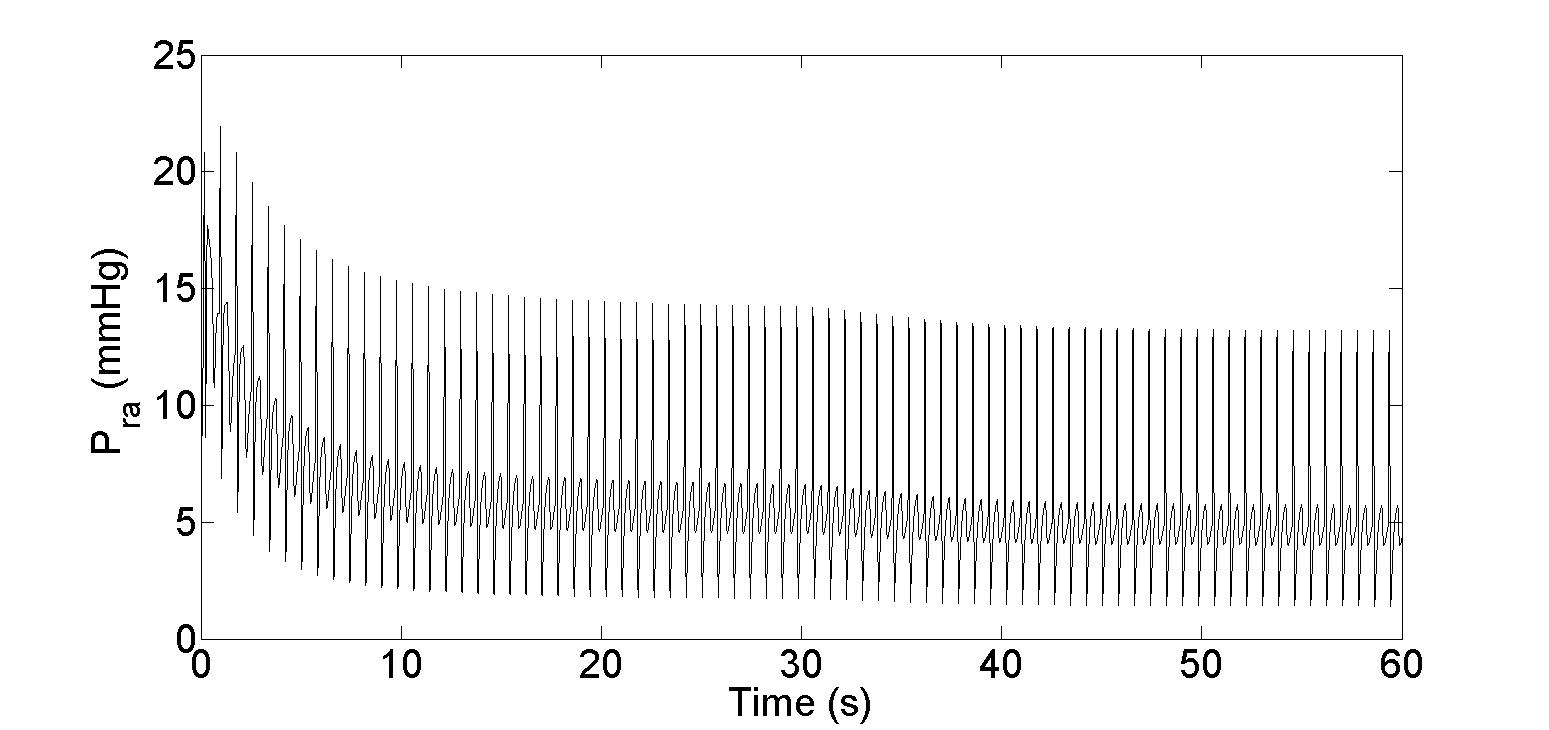}
   \label{63ee}
 }
\caption{Hemodynamic variables results in exercise condition when the system induced at 30s.}
\label{6:30ea}
\end{figure*}

\begin{figure*}[htbp]
\centering
\subfigure[Average pump speed.]{
   \includegraphics[scale =0.16752] {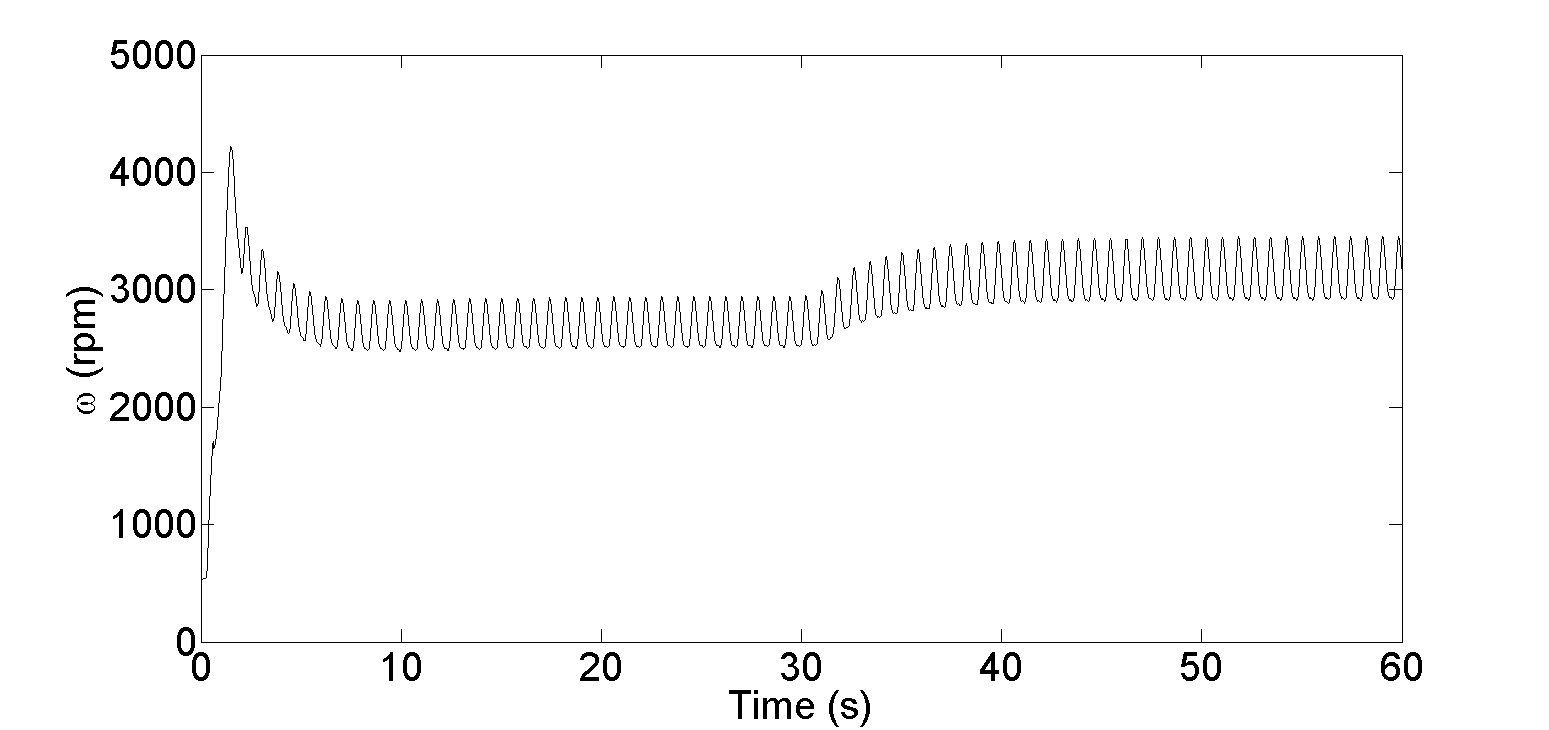}
   \label{63ef}
 }
 \subfigure[Pump flow pulsatility versus average pulsatile flow.]{
   \includegraphics[scale =0.16752] {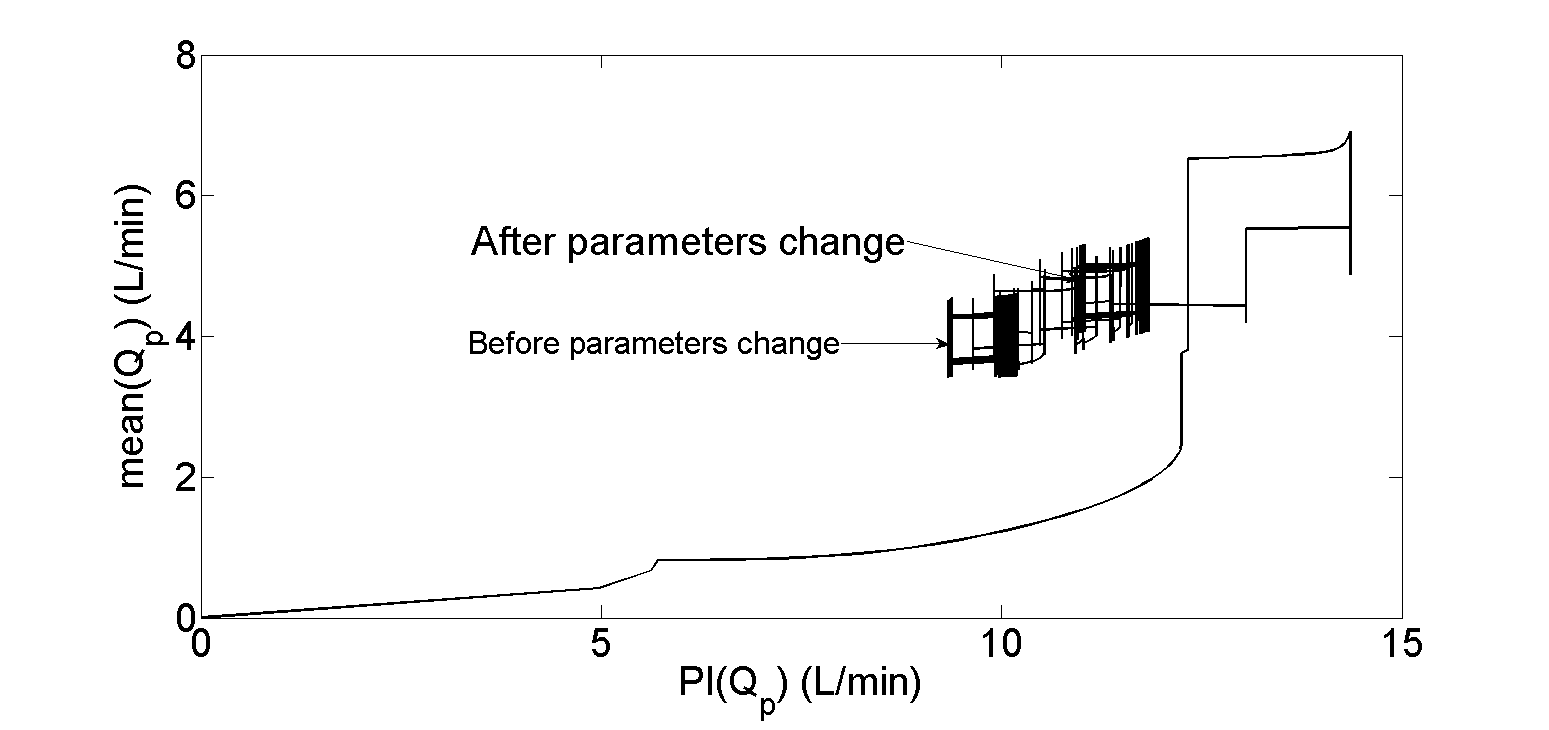}
  \label{63eg}
 }

\subfigure[Pump flow compared with desired reference flow at initial time.]{
   \includegraphics[scale =0.16752] {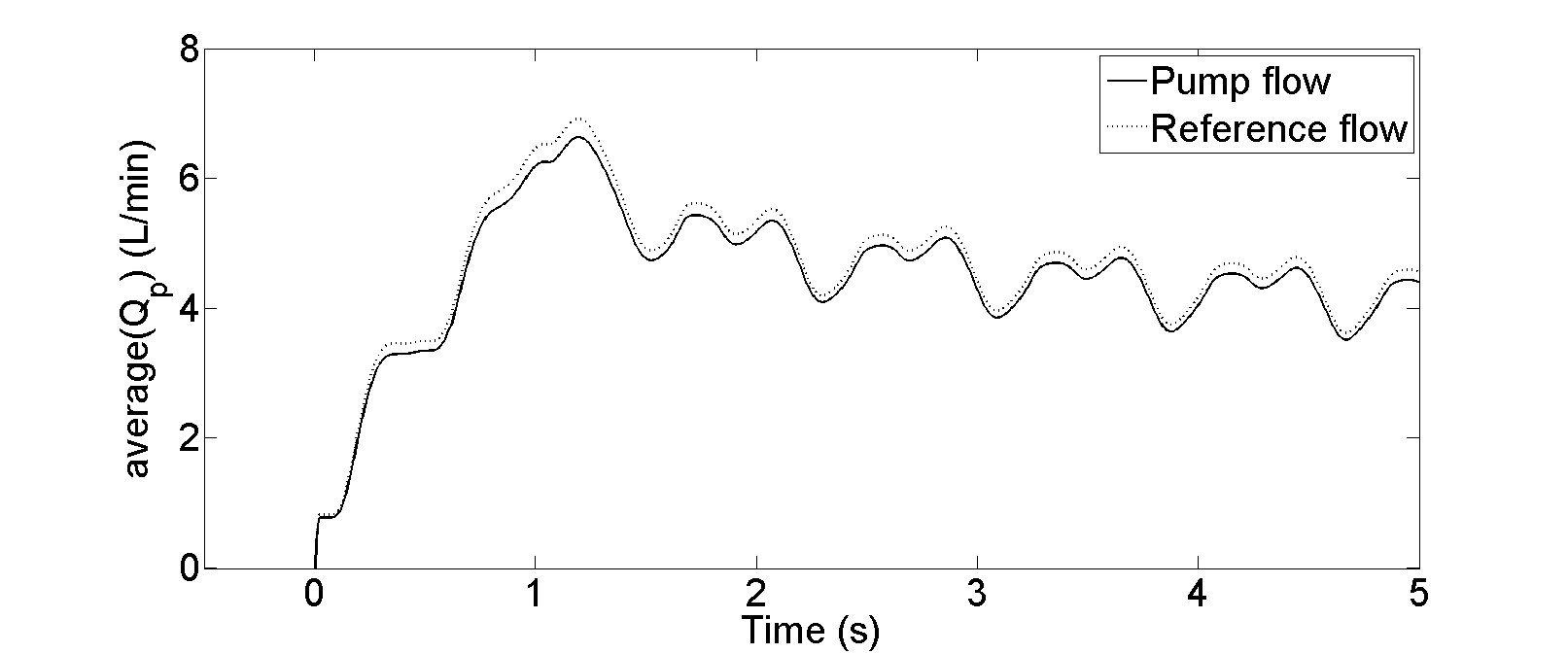}
   \label{i63eh}
 }
  \subfigure[Pump flow compared with desired reference flow at induced time.]{
   \includegraphics[scale =0.16752] {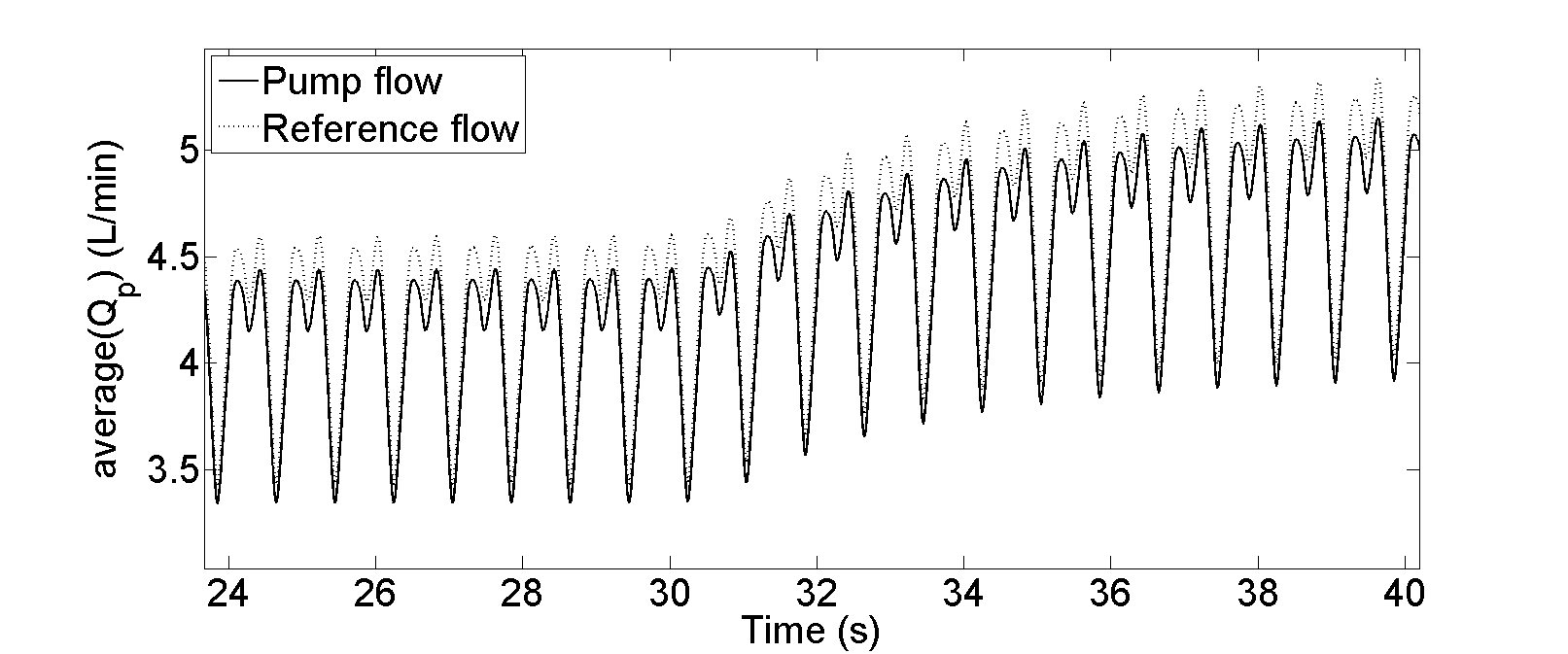}
   \label{63eh}
 }

\subfigure[Measured steady state pump flow against estimated pump flow.]{
   \includegraphics[scale =0.16752] {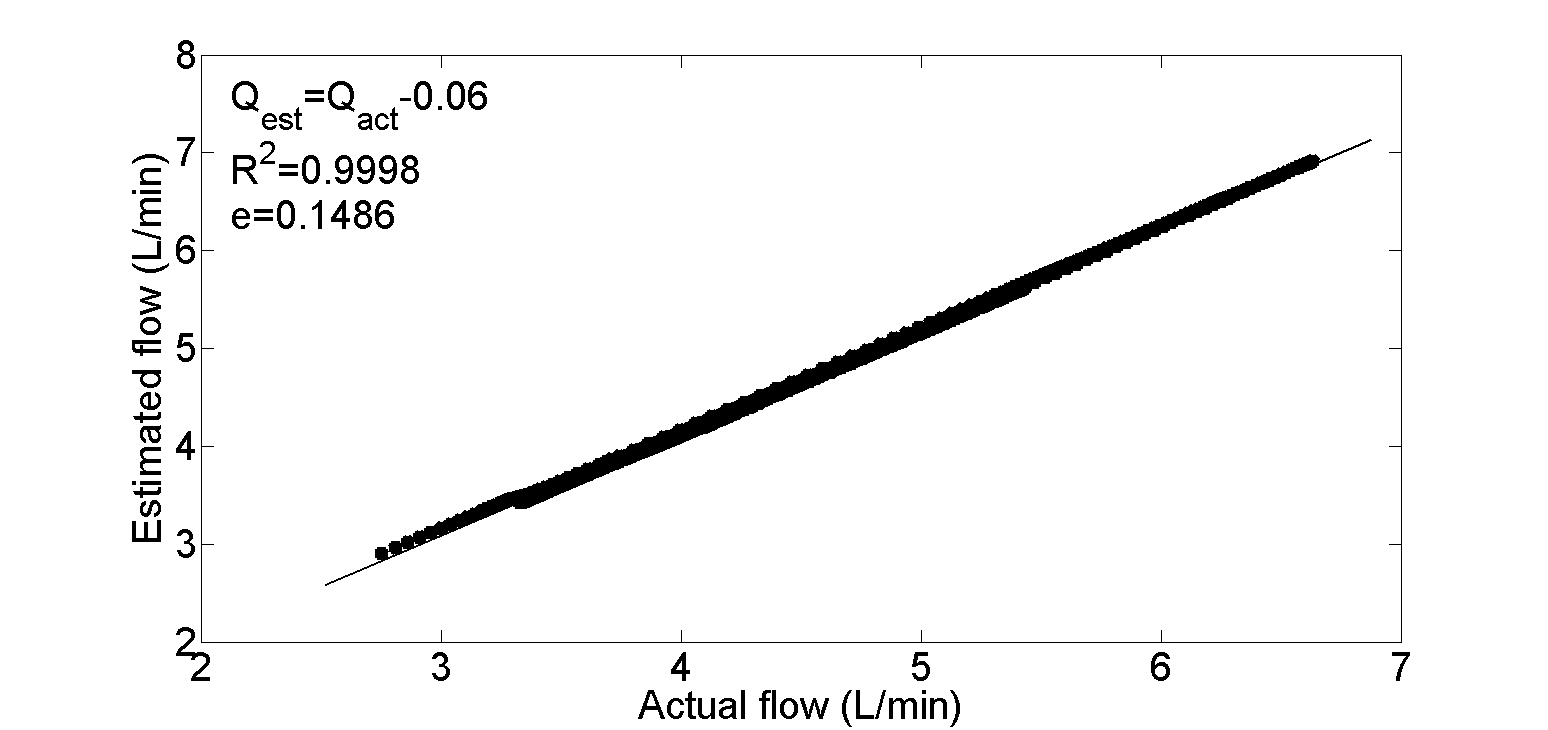}
   \label{63ei}
 }

\caption{Pump variable results in exercise condition when the system induced at 30s.}
\label{6:30eb}
\end{figure*}

\begin{figure*}[htbp]
\centering
\subfigure[LV volume versus LV pressure before and after Parameter Change.]{
   \includegraphics[scale =0.16752] {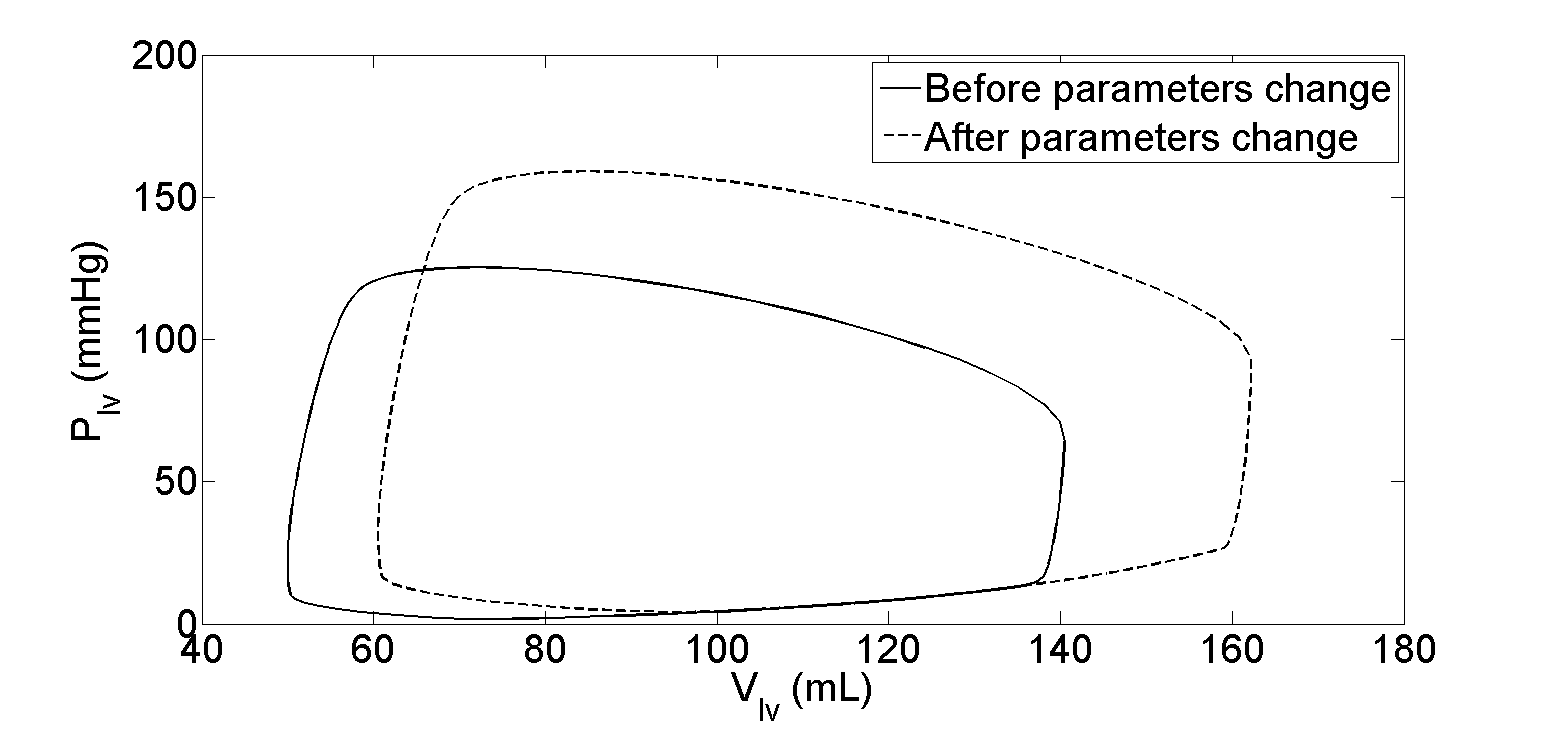}
   \label{66ea}
 }
\subfigure[RV volume versus RV pressure before and after Parameter Change.]{
   \includegraphics[scale =0.16752] {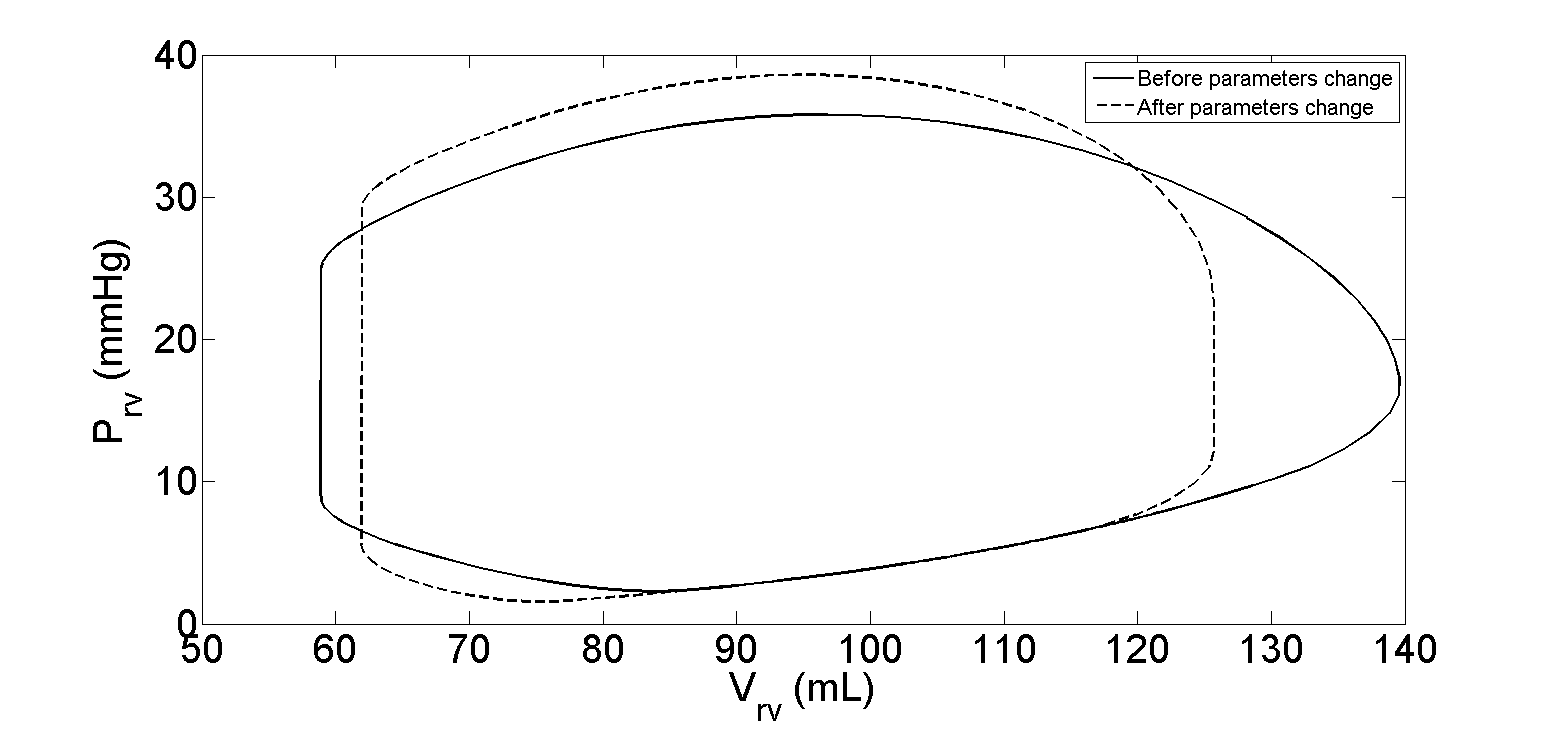}
   \label{66eb}
 }

 \subfigure[Aortic pressure.]{
   \includegraphics[scale =0.16752] {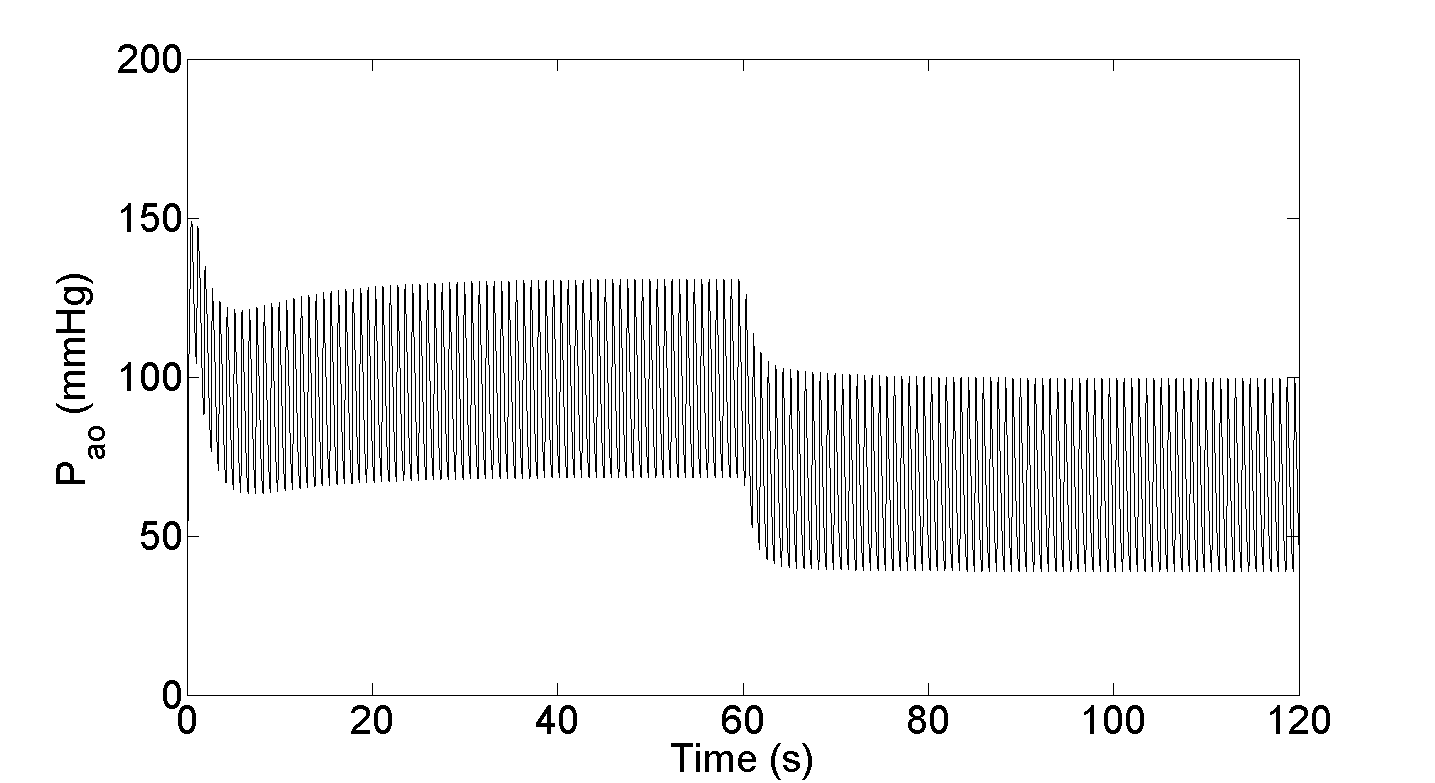}
   \label{66ec}
 }
  \subfigure[Left atrial pressure.]{
   \includegraphics[scale =0.16752] {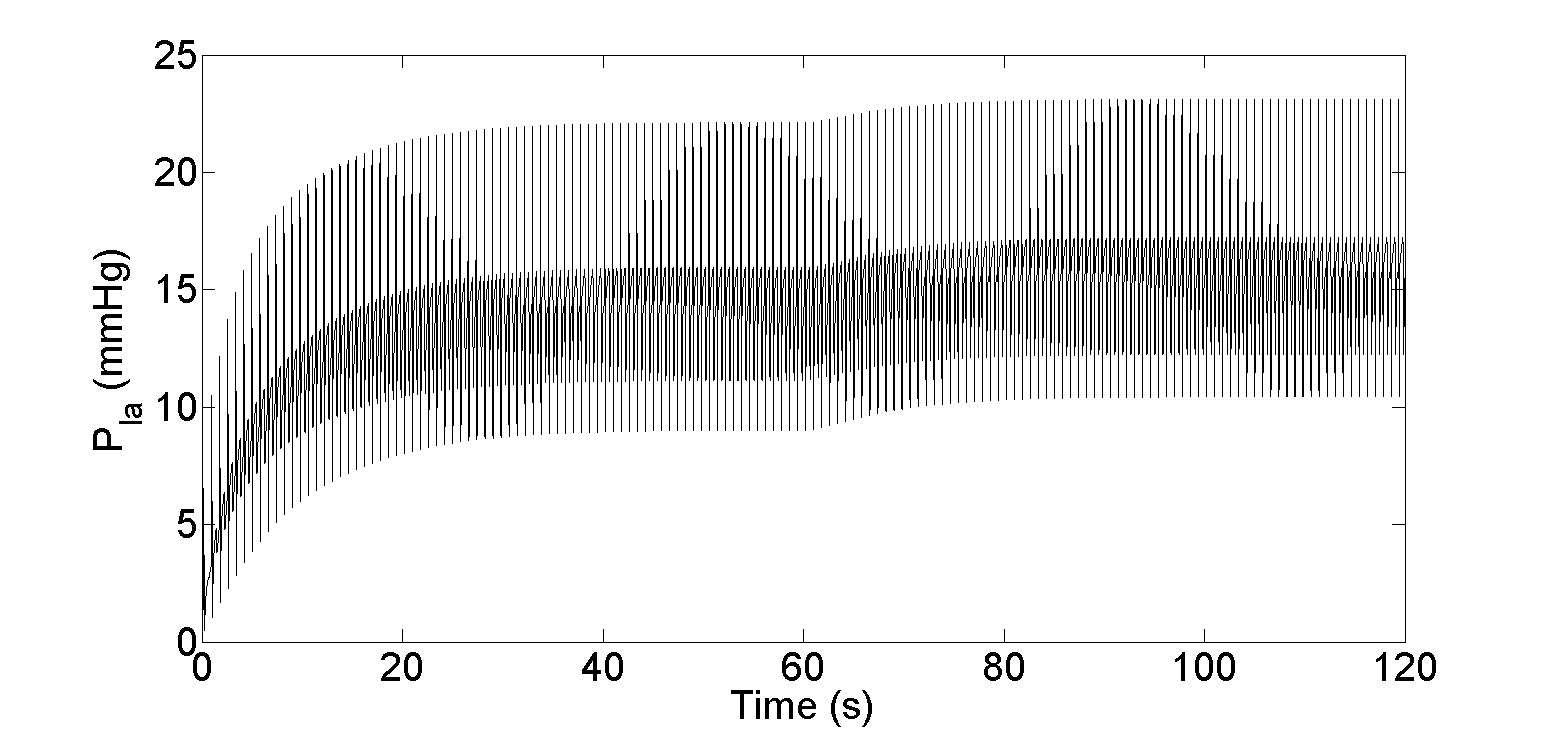}
   \label{66ed}
 }

\subfigure[Right atrial pressure.]{
   \includegraphics[scale =0.16752] {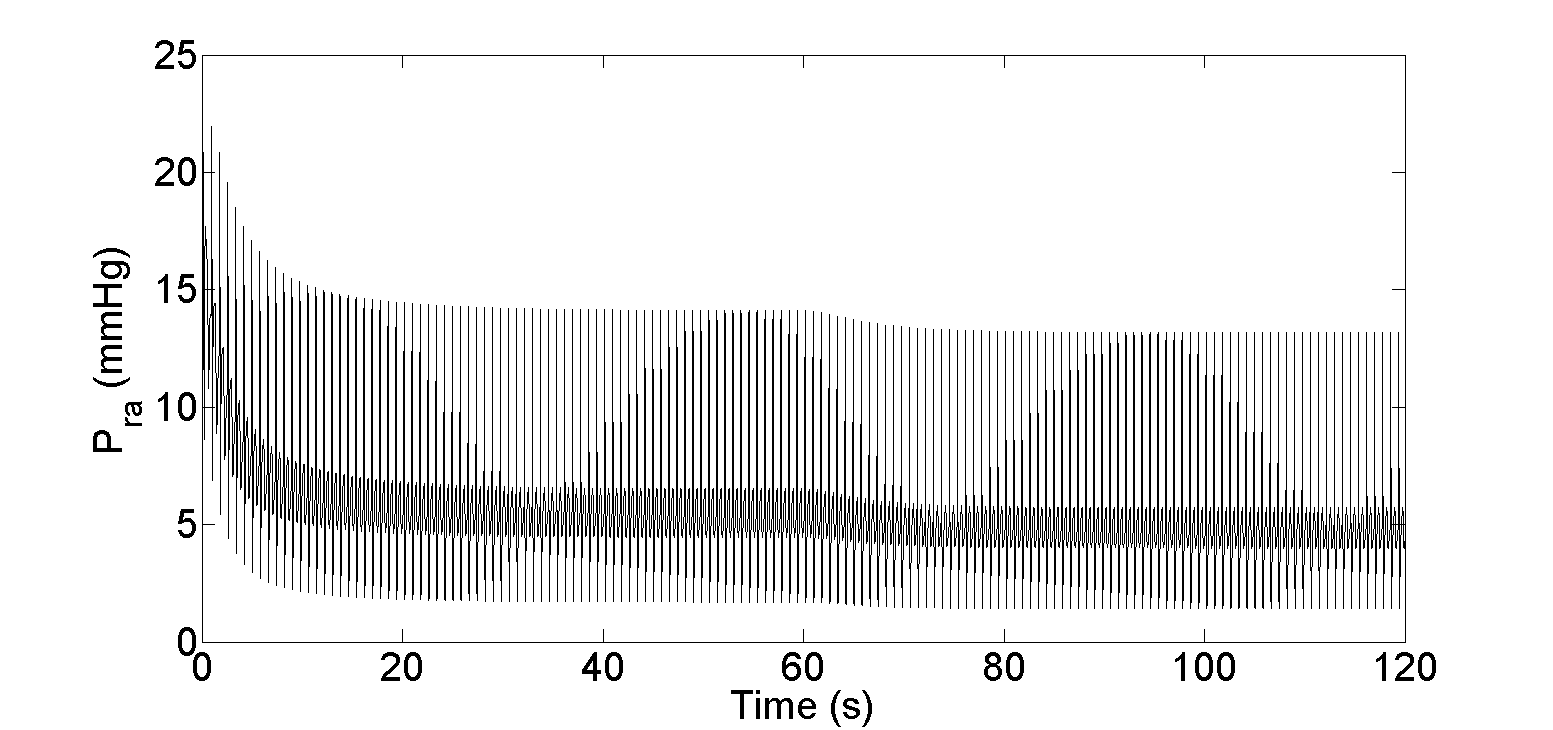}
   \label{66ee}
 }
\caption{Hemodynamic variables results in exercise condition when the system induced at 60s.}
\label{6:60ea}
\end{figure*}

\begin{figure}[htbp]
\centering
\subfigure[Average pump speed.]{
   \includegraphics[scale =0.16752] {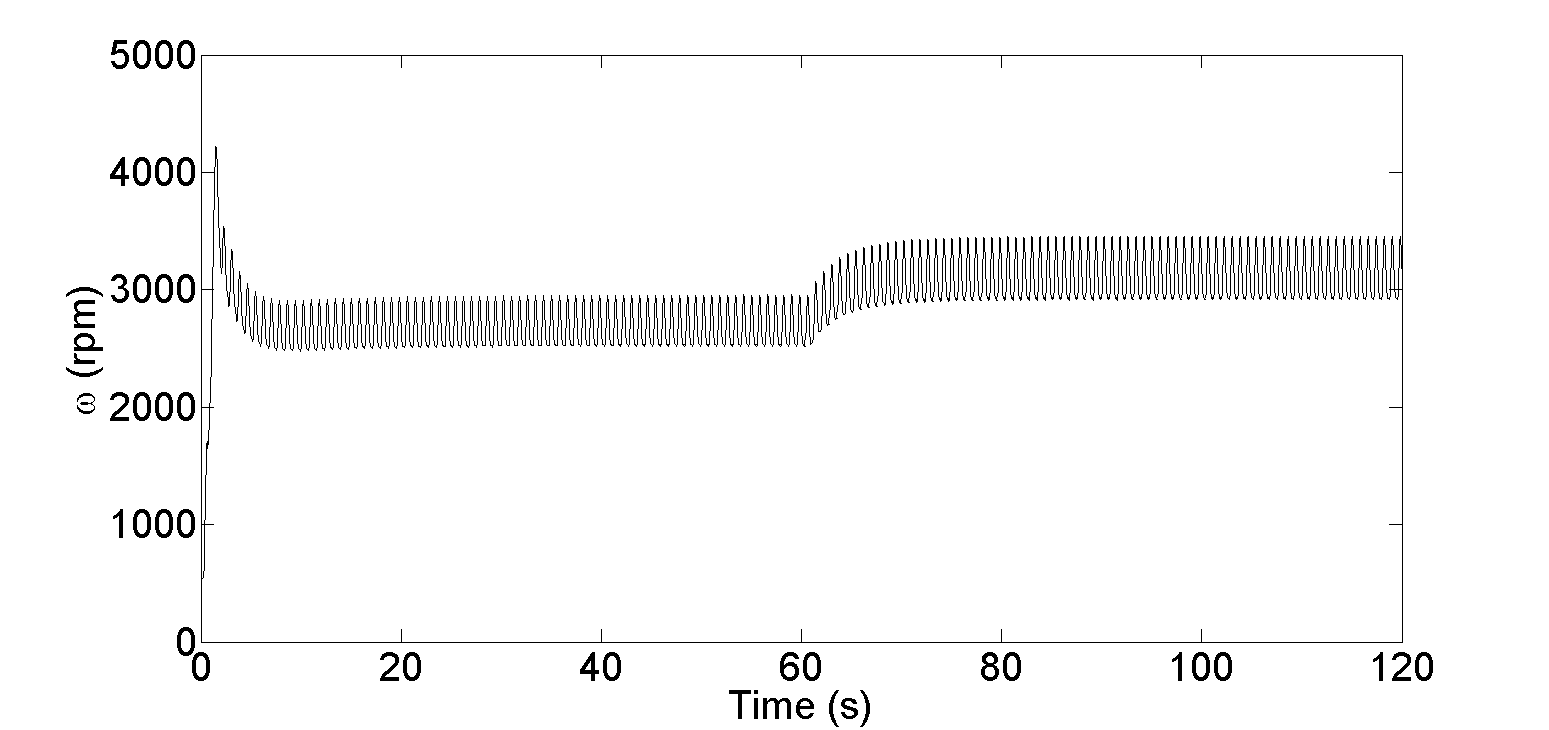}
   \label{66ef}
 }
 \subfigure[Pump flow pulsatility versus average pulsatile flow.]{
   \includegraphics[scale =0.16752] {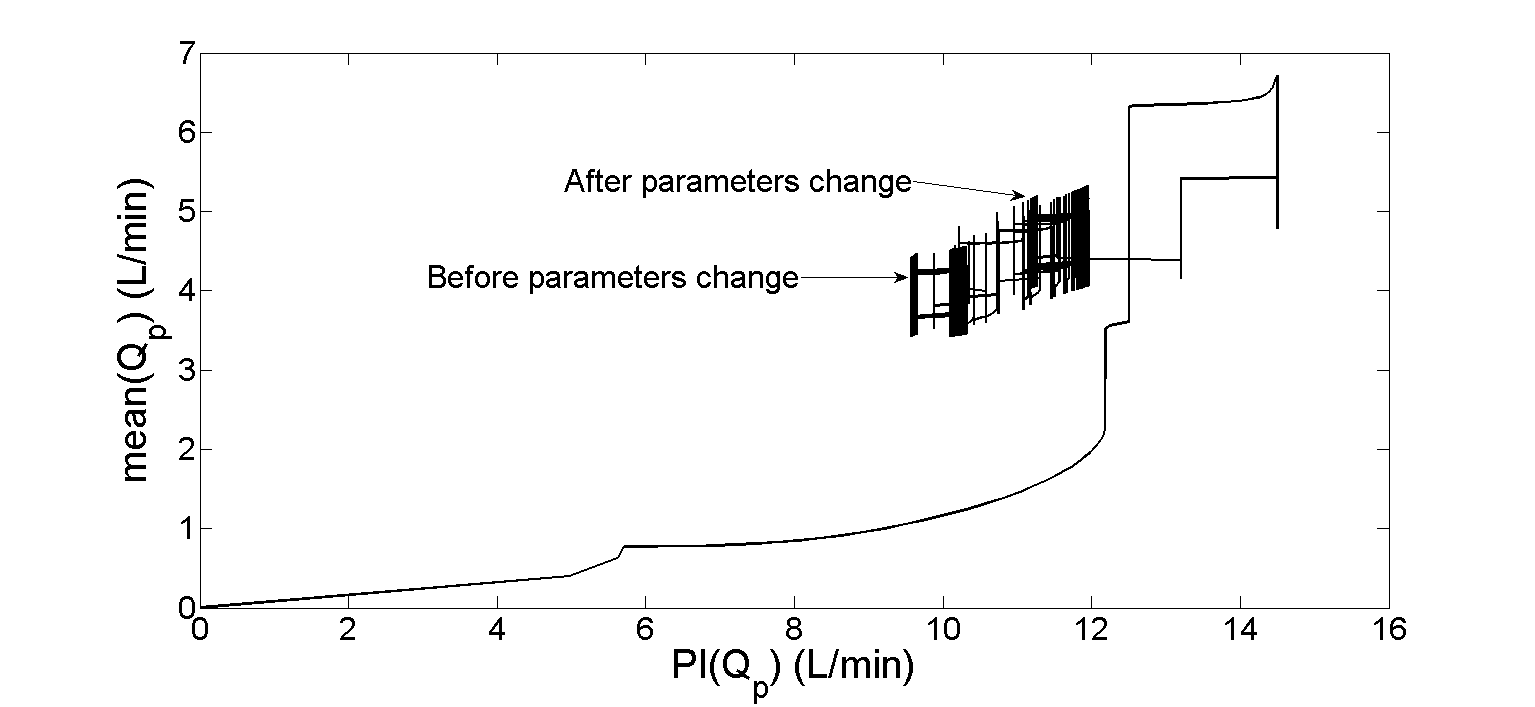}
  \label{66eg}
 }

\subfigure[Pump flow compared with desired reference flow at initial time.]{
   \includegraphics[scale =0.16752] {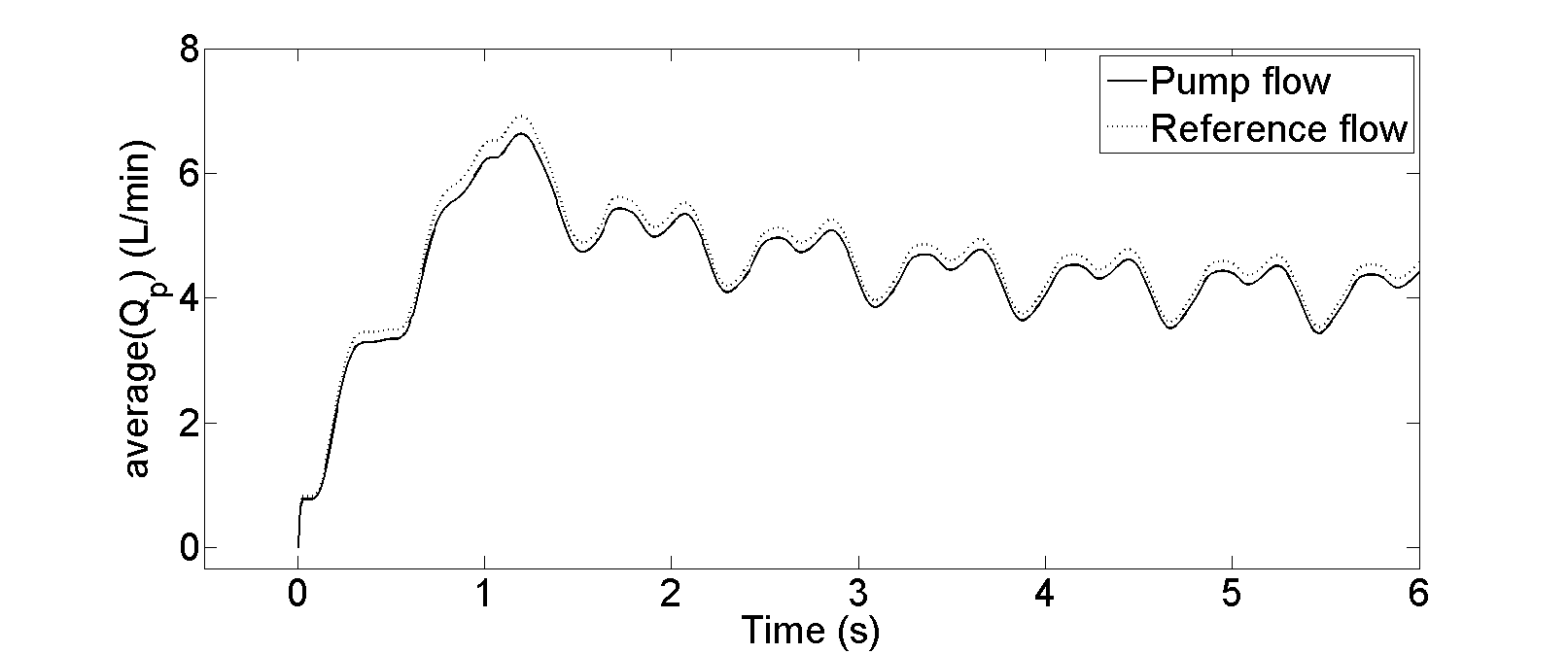}
   \label{i66eh}
 }
  \subfigure[Pump flow compared with desired reference flow at induced time.]{
   \includegraphics[scale =0.16752] {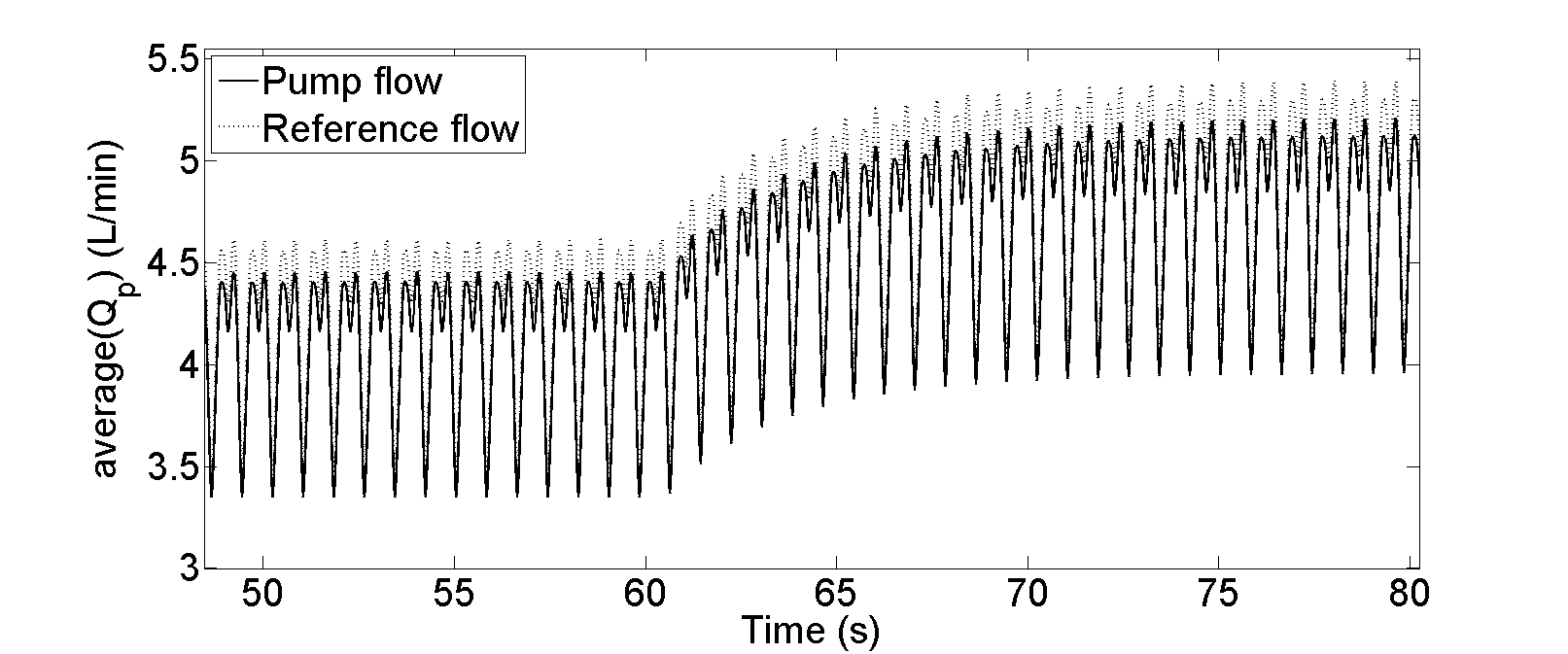}
   \label{66eh}
 }

\subfigure[Measured steady state pump flow against estimated pump flow.]{
   \includegraphics[scale =0.16752] {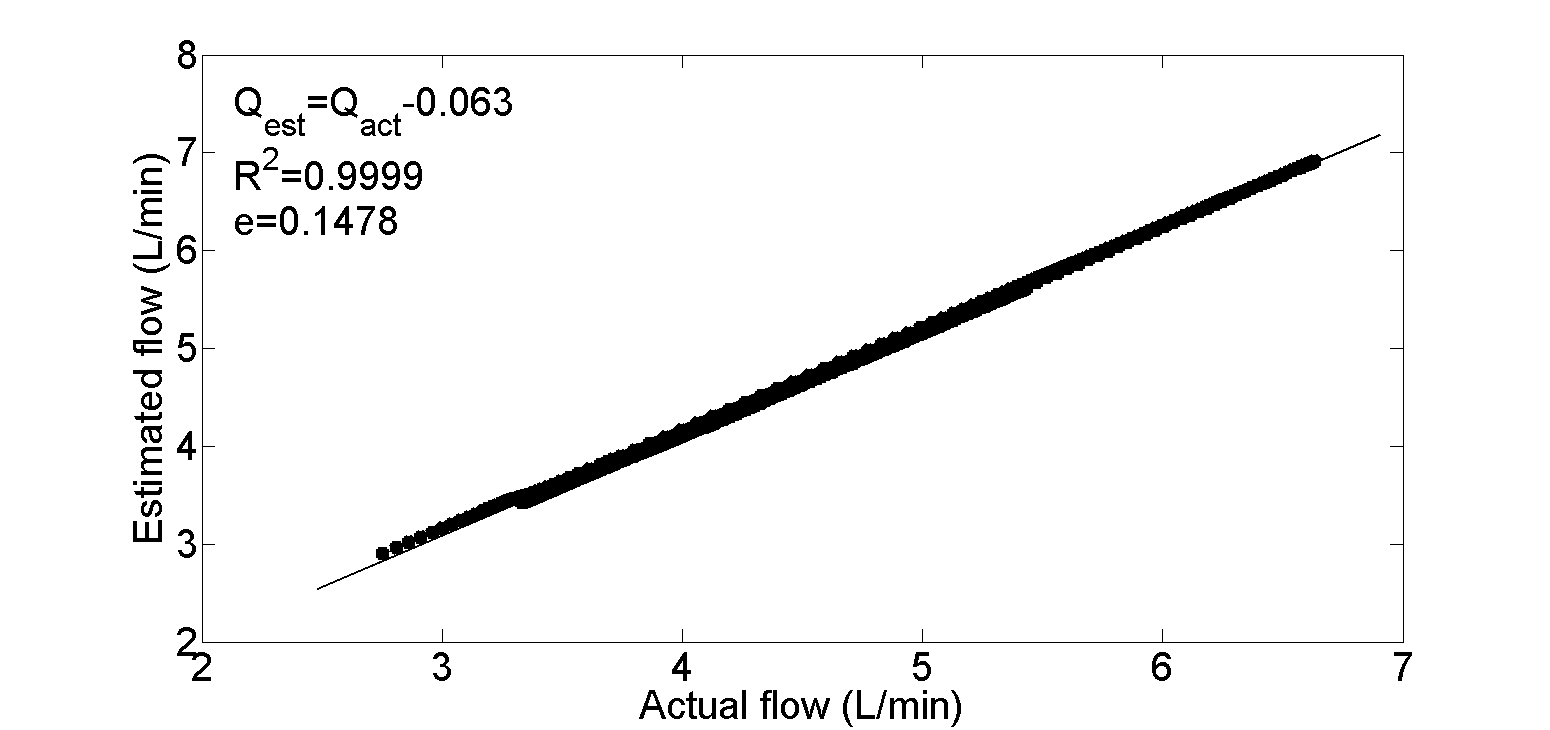}
   \label{66ei}
 }

\caption{Pump variable results in exercise condition when the system induced at 60s.}
\label{6:60eb}
\end{figure}

\begin{figure*}[htbp]
\centering
\subfigure[LV volume versus LV pressure before and after Parameter Change.]{
   \includegraphics[scale =0.16752] {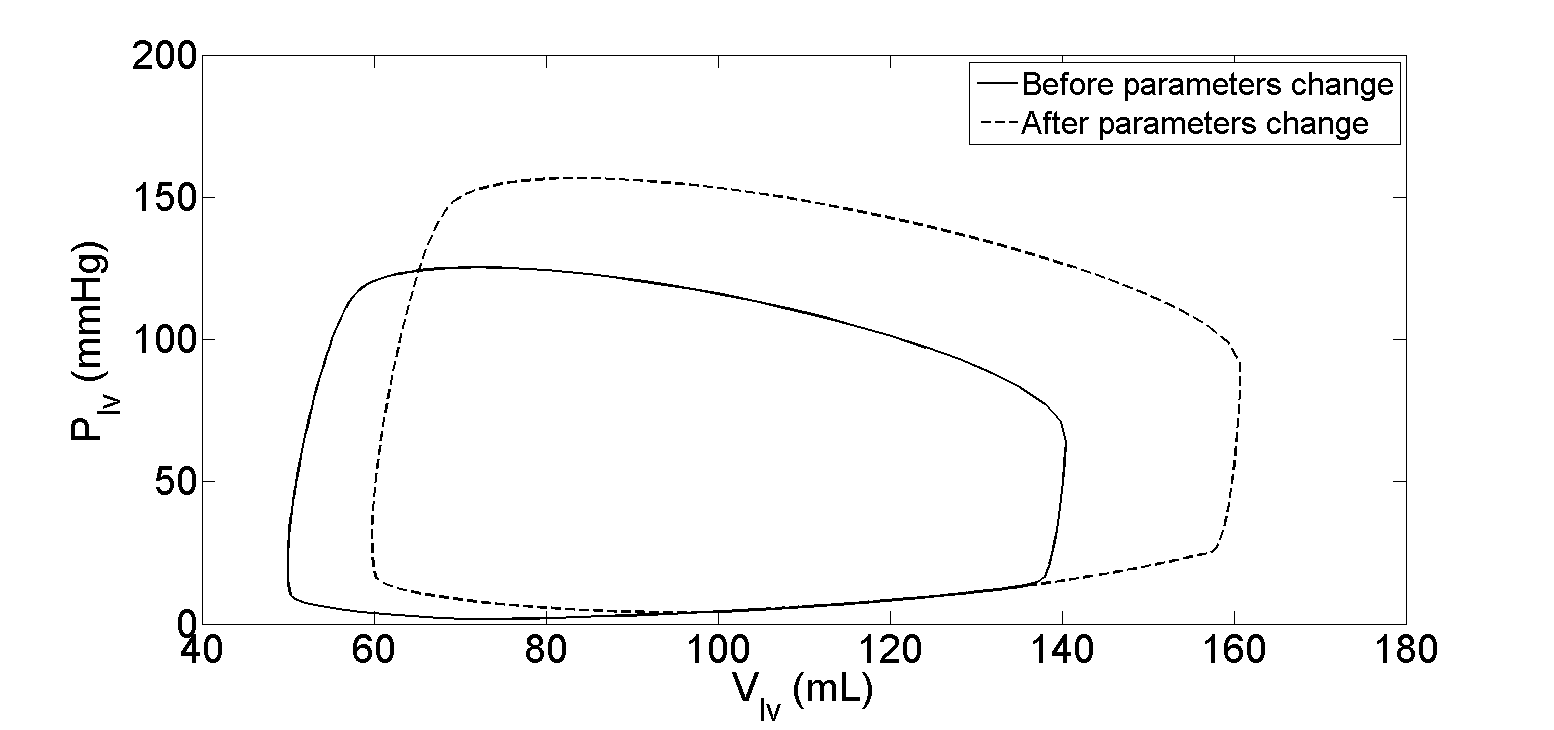}
   \label{69ea}
 }
\subfigure[RV volume versus RV pressure before and after Parameter Change.]{
   \includegraphics[scale =0.16752] {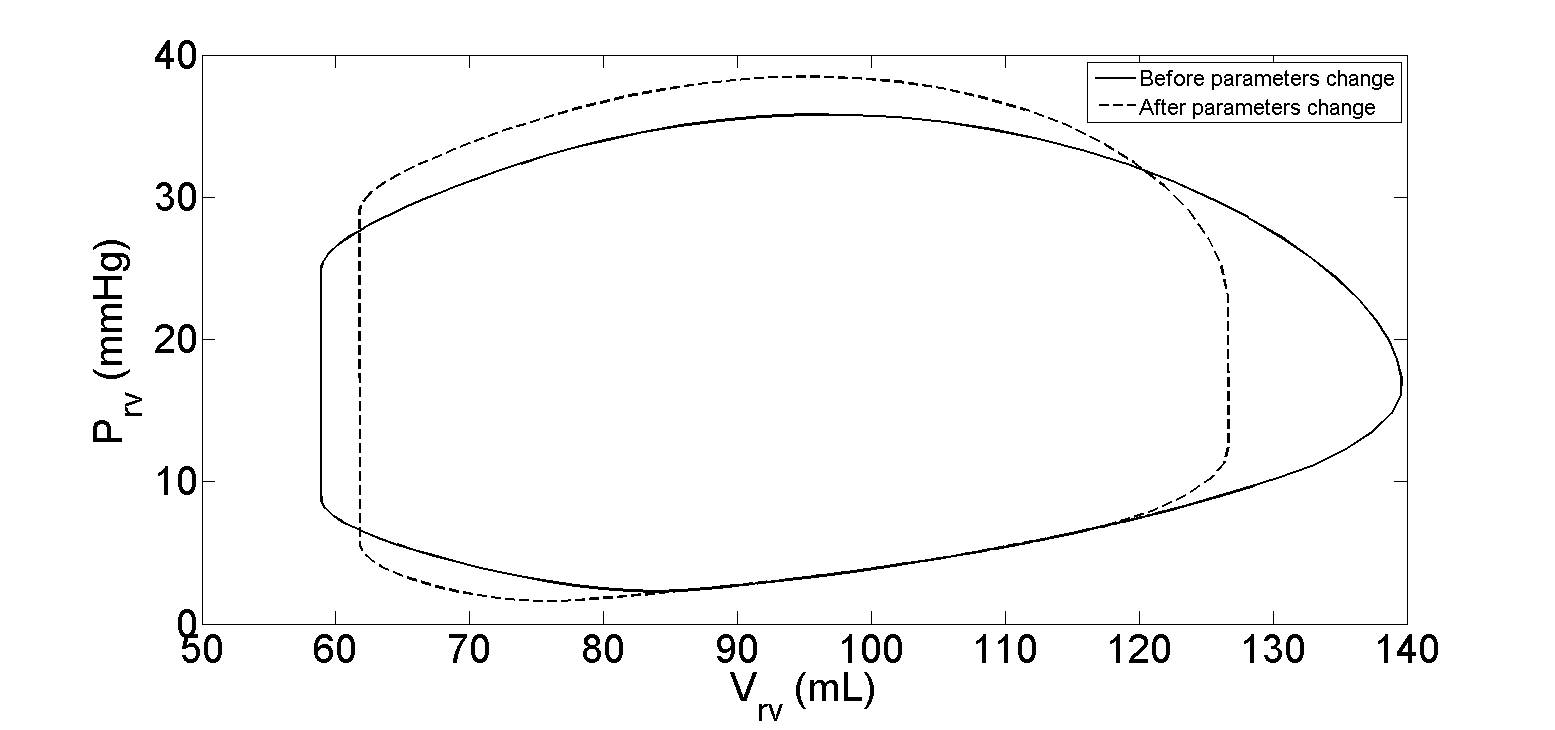}
   \label{69eb}
 }

 \subfigure[Aortic pressure.]{
   \includegraphics[scale =0.16752] {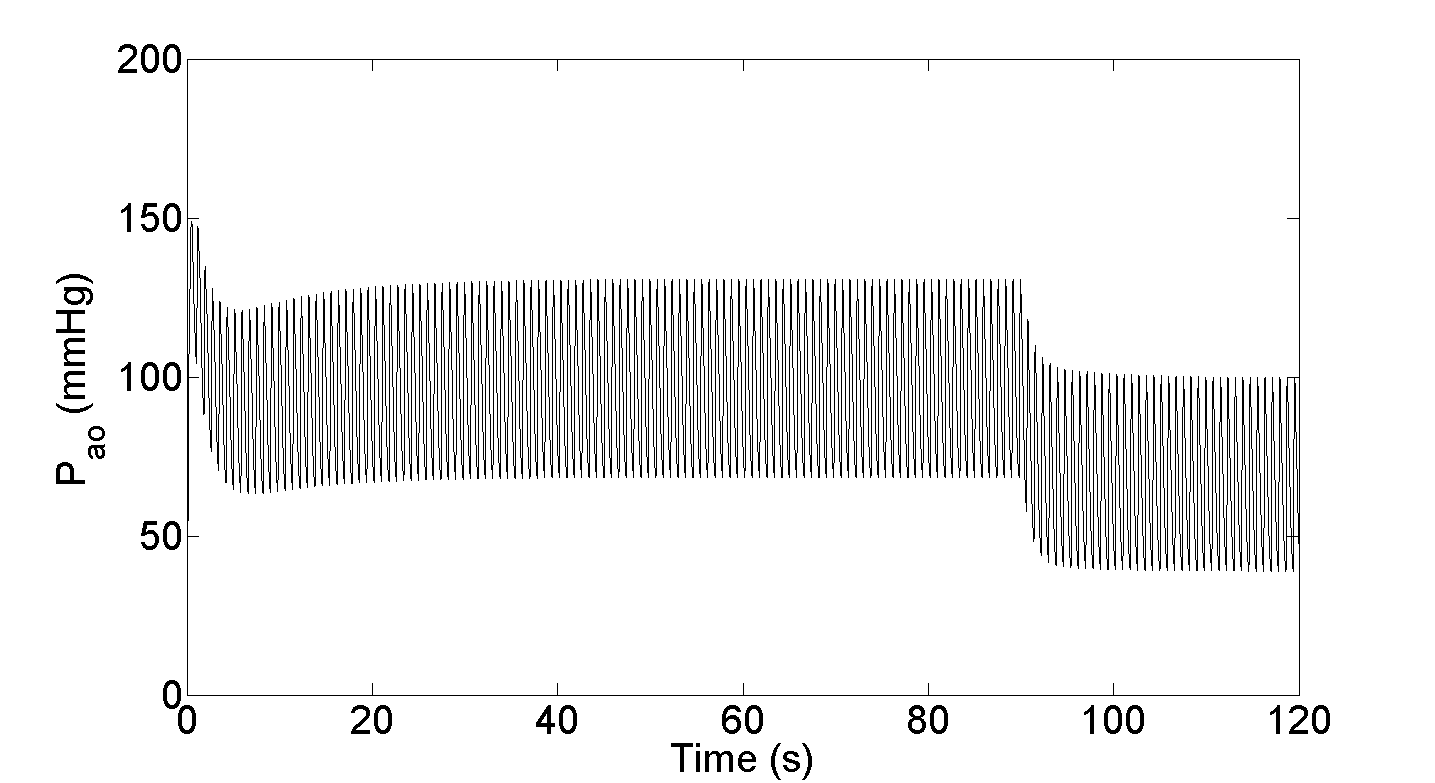}
   \label{69ec}
 }
  \subfigure[Left atrial pressure.]{
   \includegraphics[scale =0.16752] {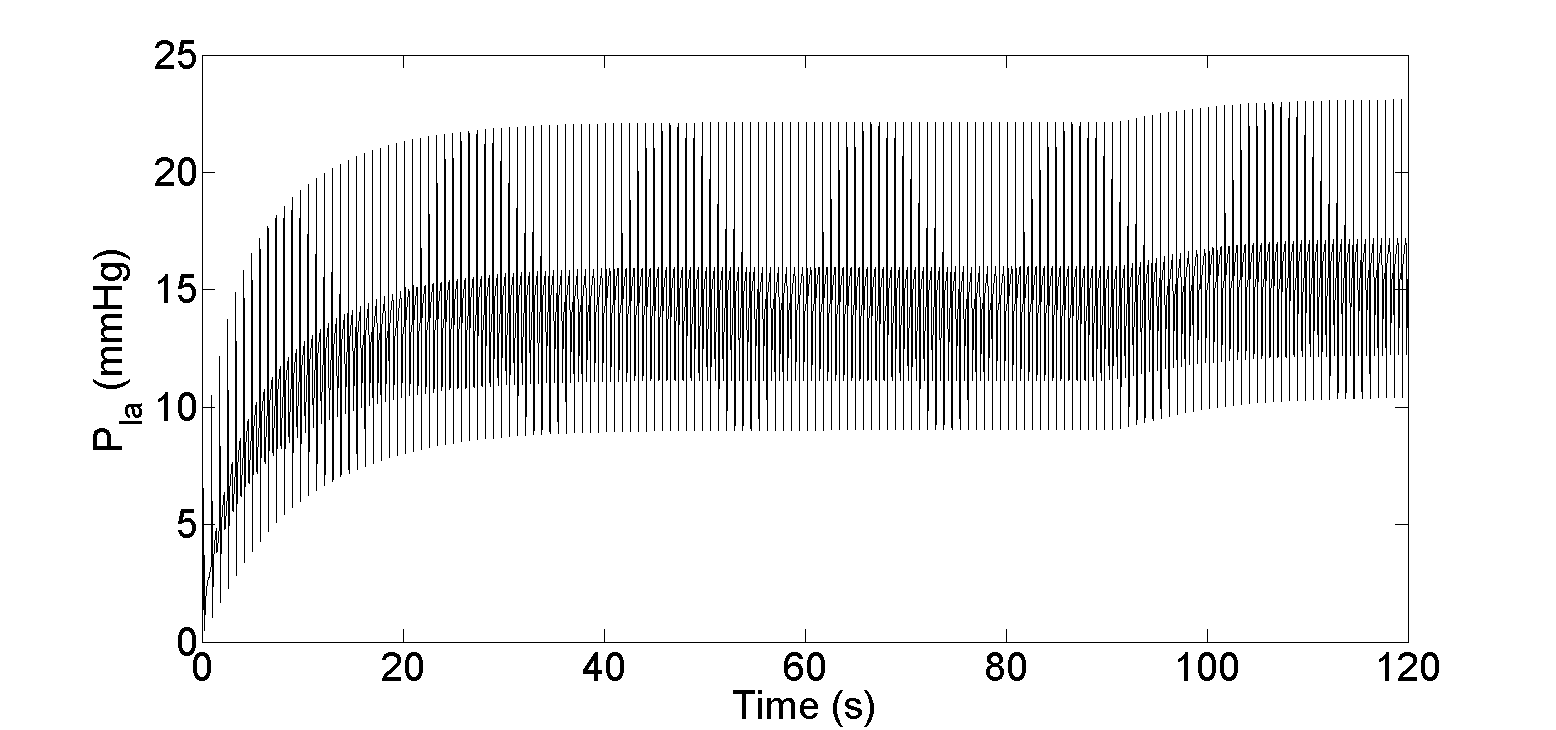}
   \label{69ed}
 }

\subfigure[Right atrial pressure.]{
   \includegraphics[scale =0.16752] {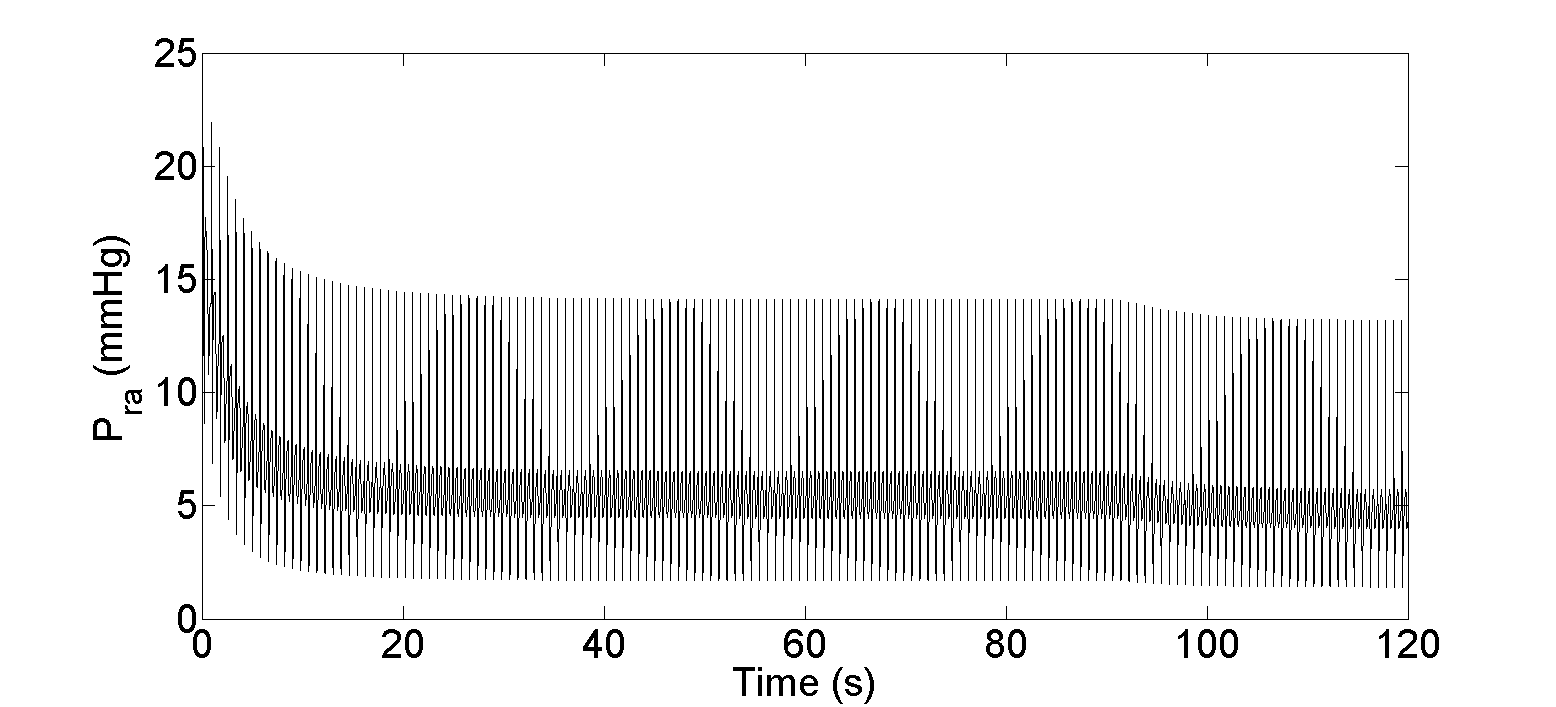}
   \label{69ee}
 }
\caption{Hemodynamic variables results in exercise condition when the system induced at 90s.}
\label{6:90ea}
\end{figure*}

\begin{figure}[htbp]
\centering
\subfigure[Average pump speed.]{
   \includegraphics[scale =0.16752] {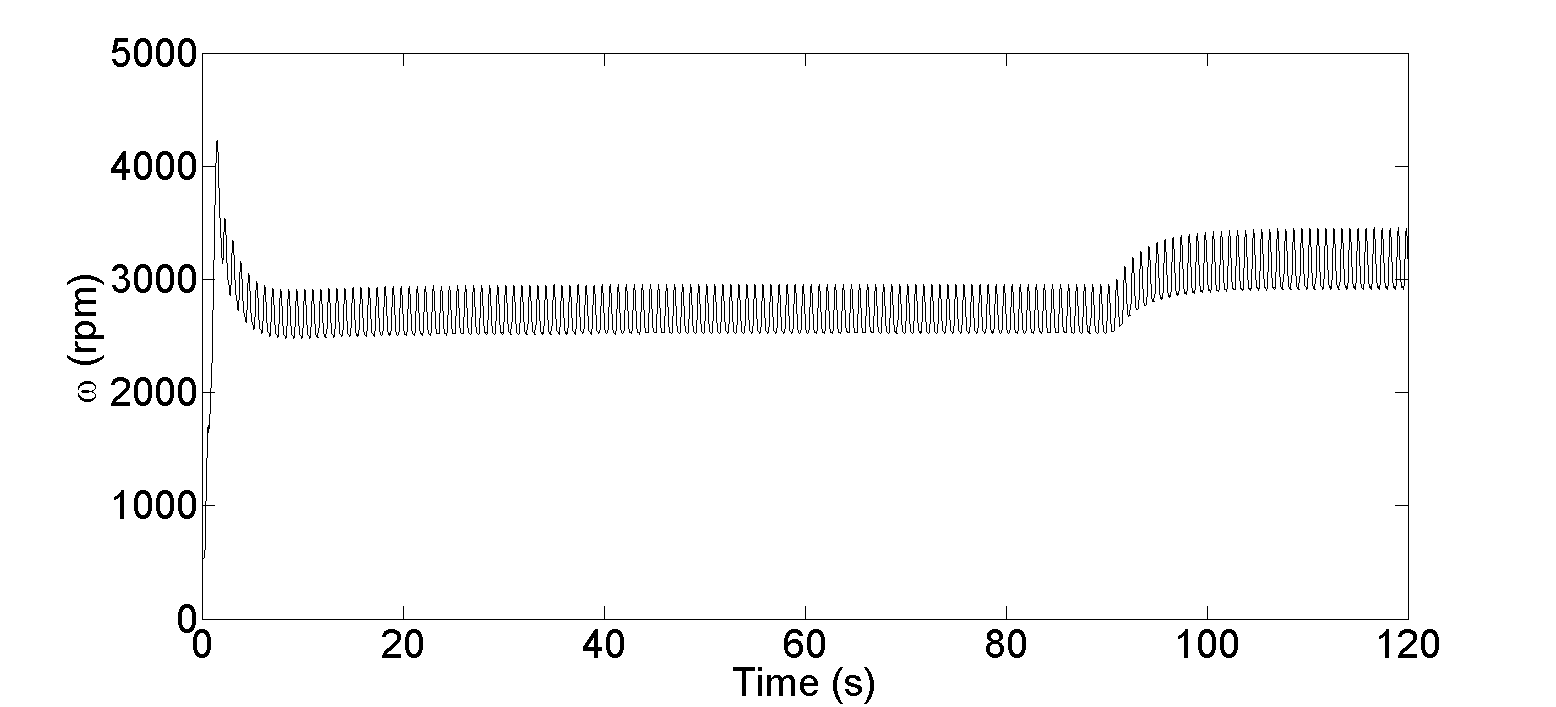}
   \label{69ef}
 }
 \subfigure[Pump flow pulsatility versus average pulsatile flow.]{
   \includegraphics[scale =0.16752] {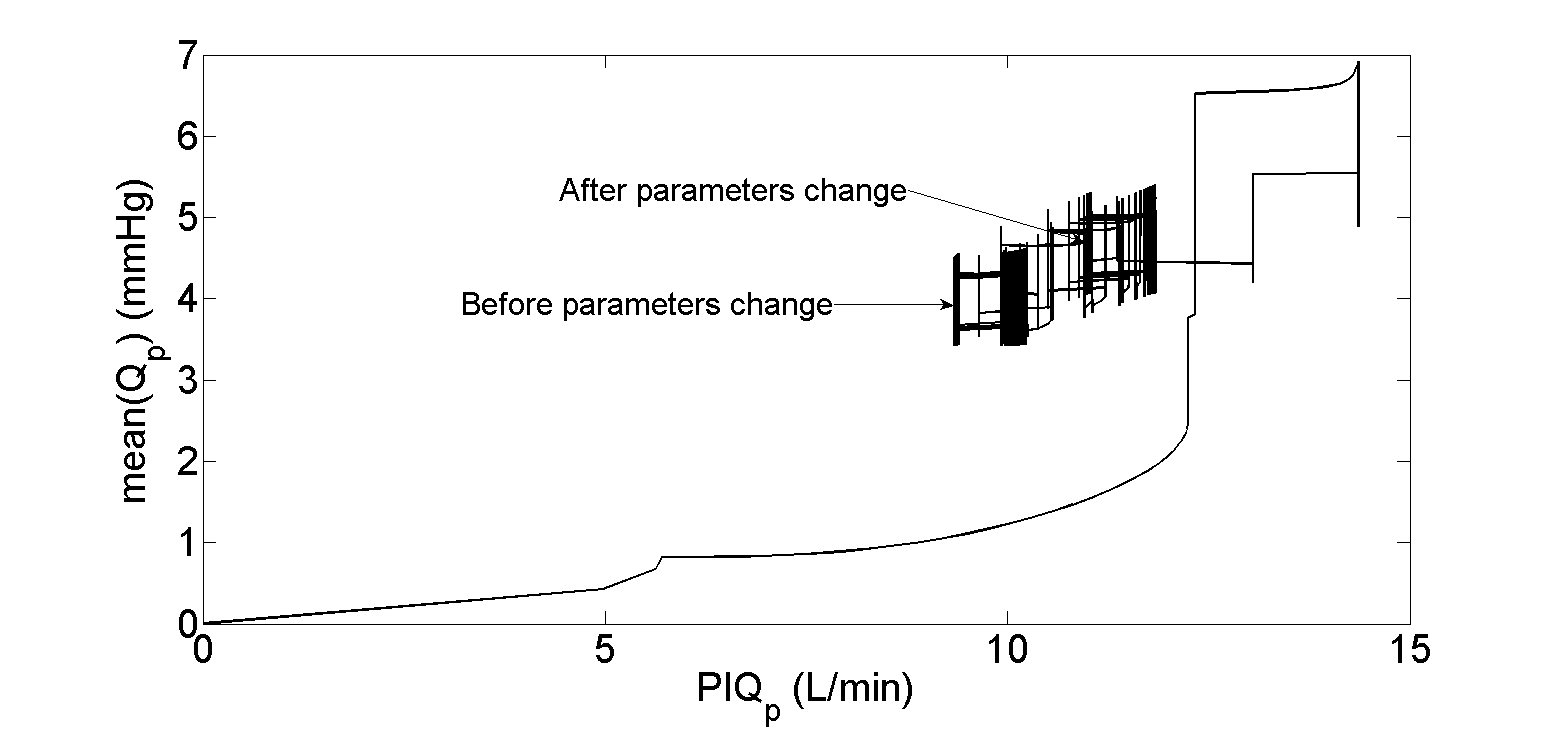}
  \label{69eg}
 }

\subfigure[Pump flow compared with desired reference flow at initial time.]{
   \includegraphics[scale =0.16752] {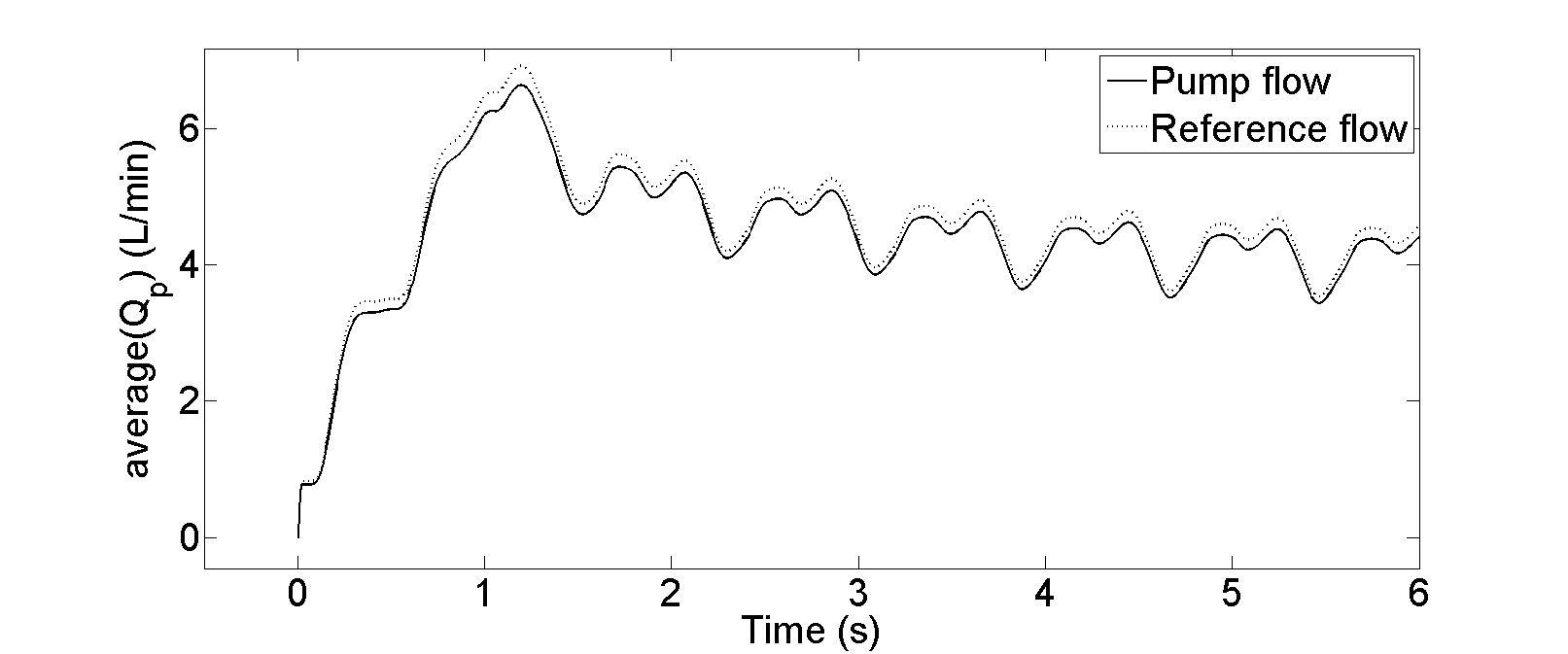}
   \label{i69eh}
 }
  \subfigure[Pump flow compared with desired reference flow at induced time.]{
   \includegraphics[scale =0.16752] {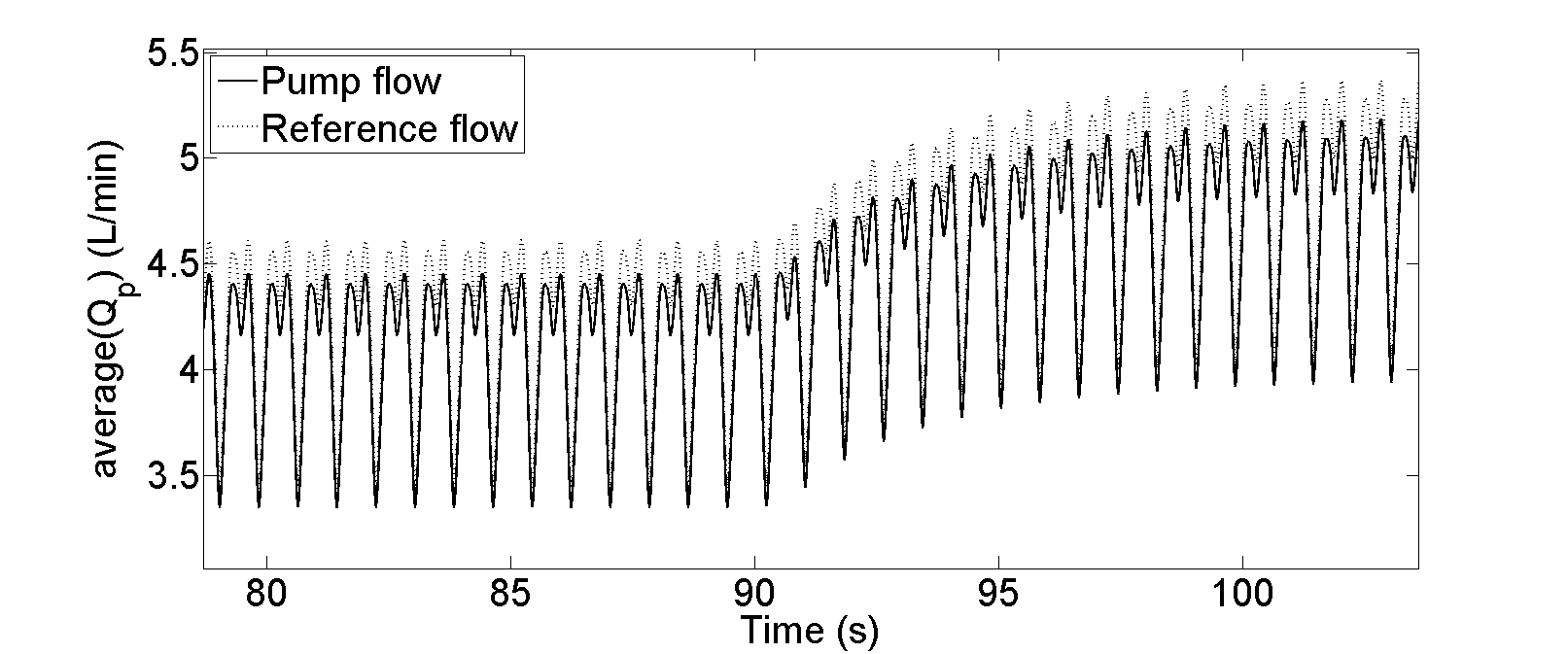}
   \label{69eh}
 }

\subfigure[Measured steady state pump flow against estimated pump flow.]{
   \includegraphics[scale =0.16752] {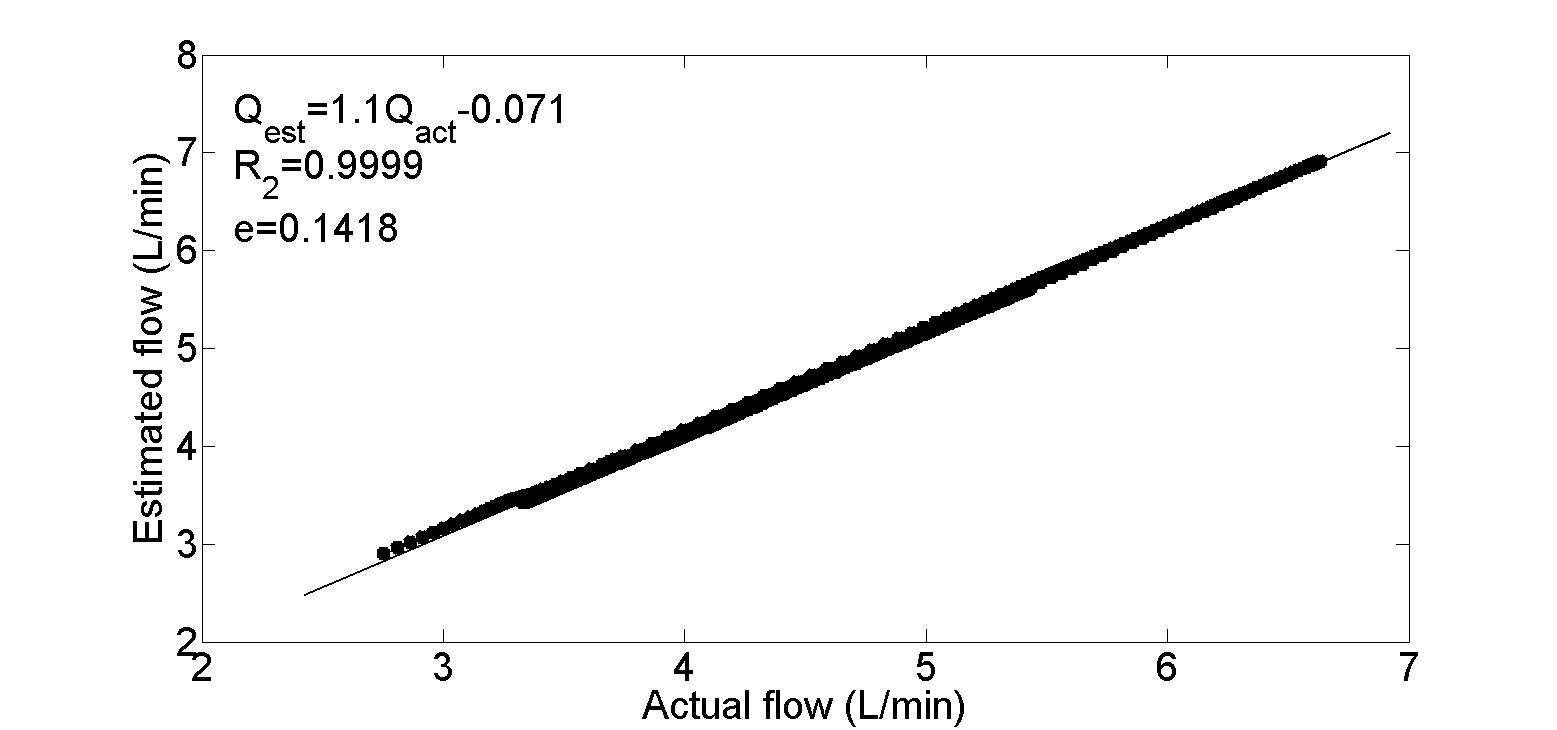}
   \label{69ei}
 }

\caption{Pump variable results in exercise condition when the system induced at 90s.}
\label{6:90eb}
\end{figure}

\begin{figure*}[htbp]
\centering
\subfigure[LV volume versus LV pressure before and after Parameter Change.]{
   \includegraphics[scale =0.16752] {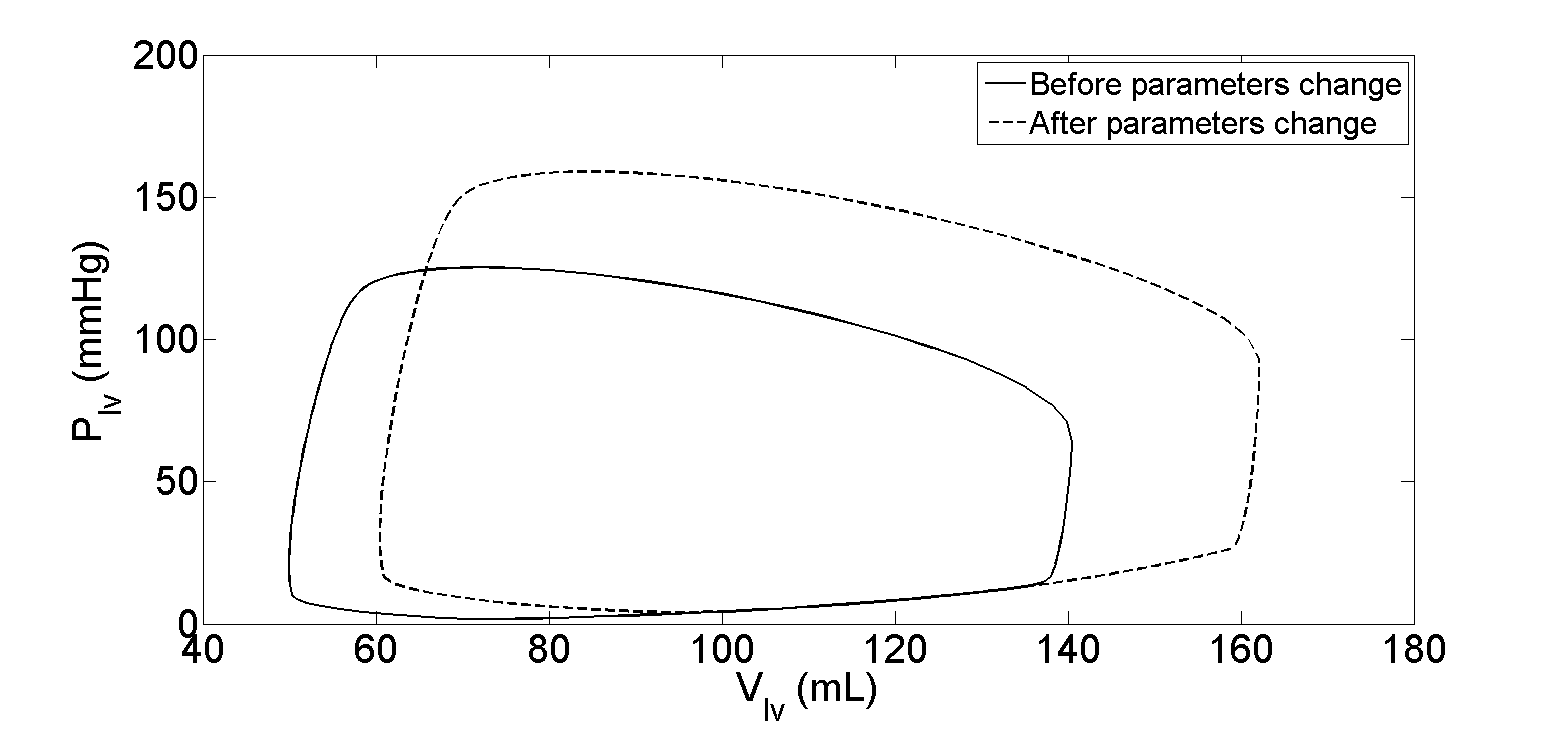}
   \label{62ea}
 }
\subfigure[RV volume versus RV pressure before and after Parameter Change.]{
   \includegraphics[scale =0.156752] {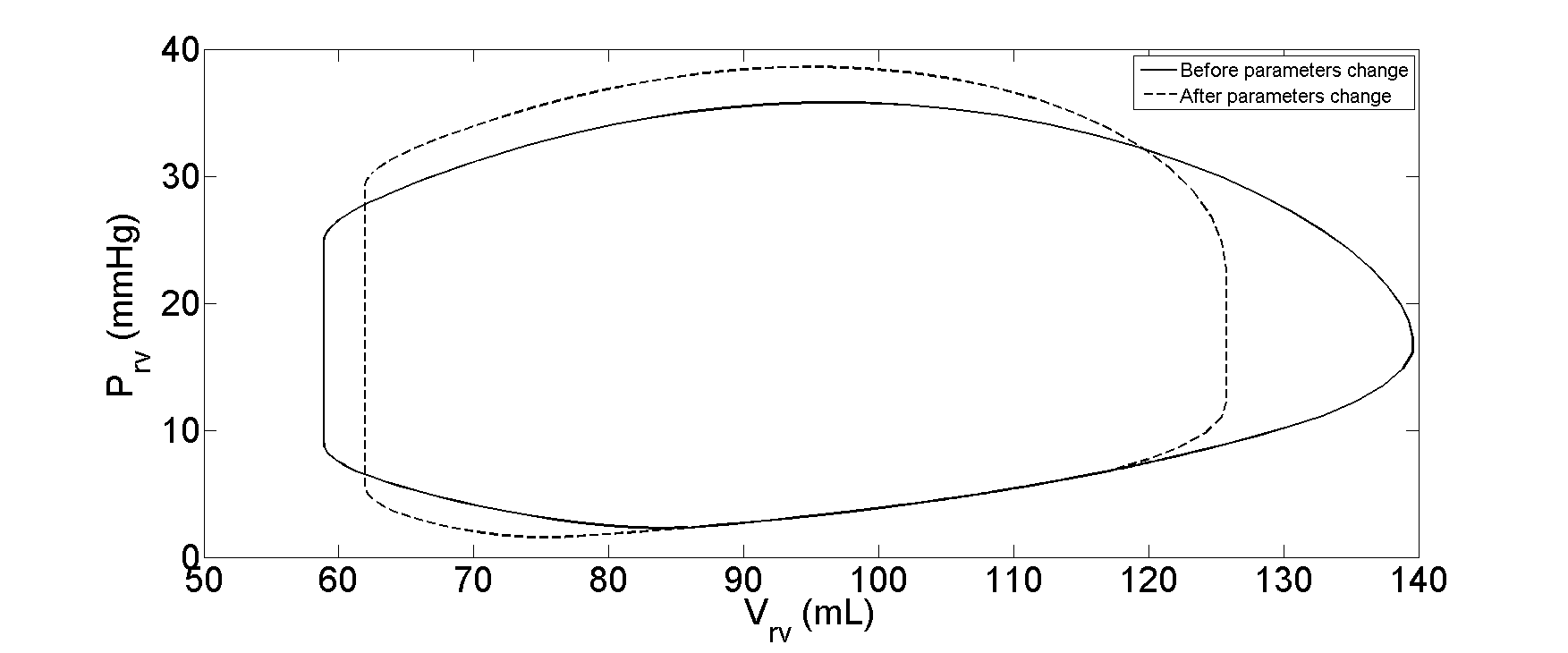}
   \label{62eb}
 }

 \subfigure[Aortic pressure.]{
   \includegraphics[scale =0.16752] {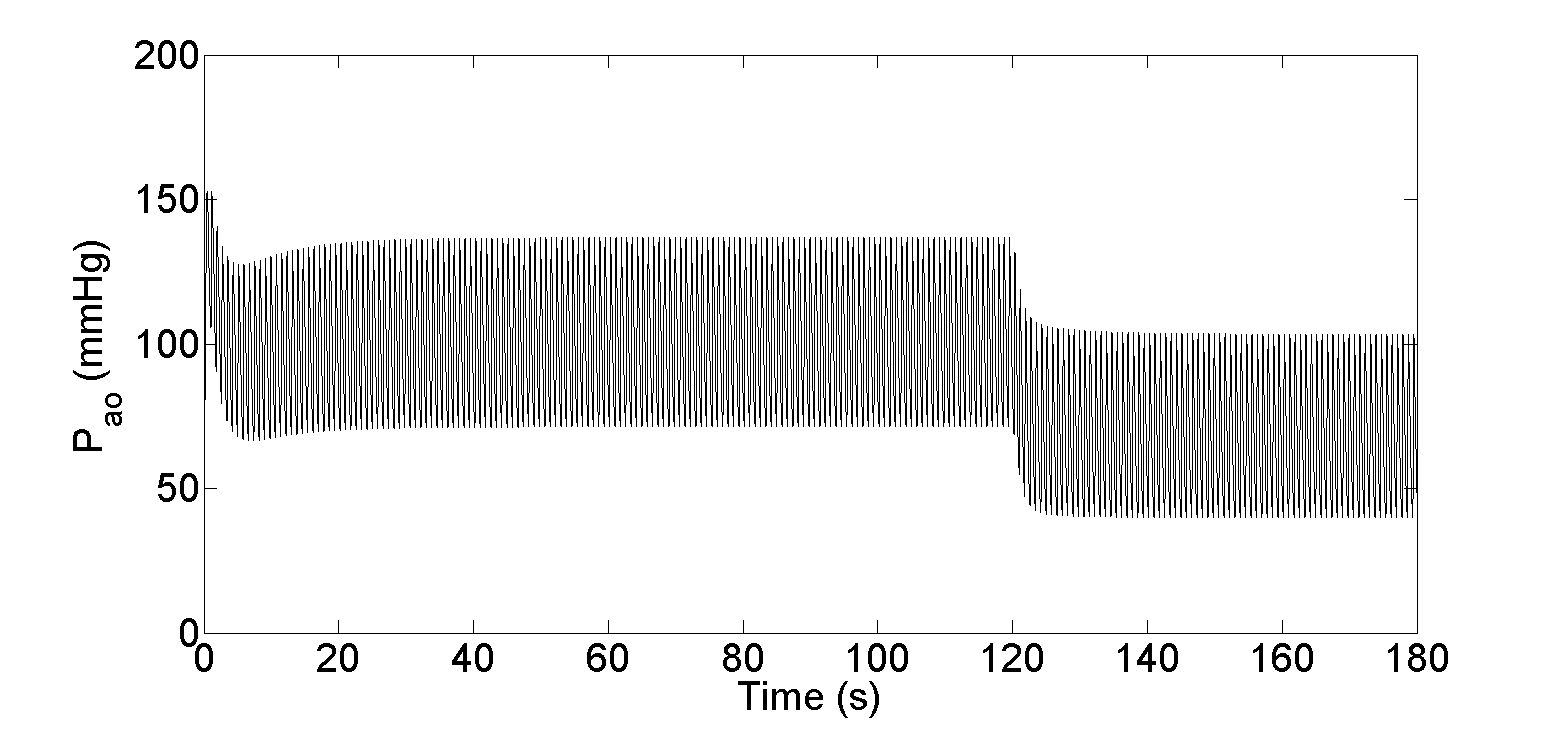}
   \label{62ec}
 }
  \subfigure[Left atrial pressure.]{
   \includegraphics[scale =0.16752] {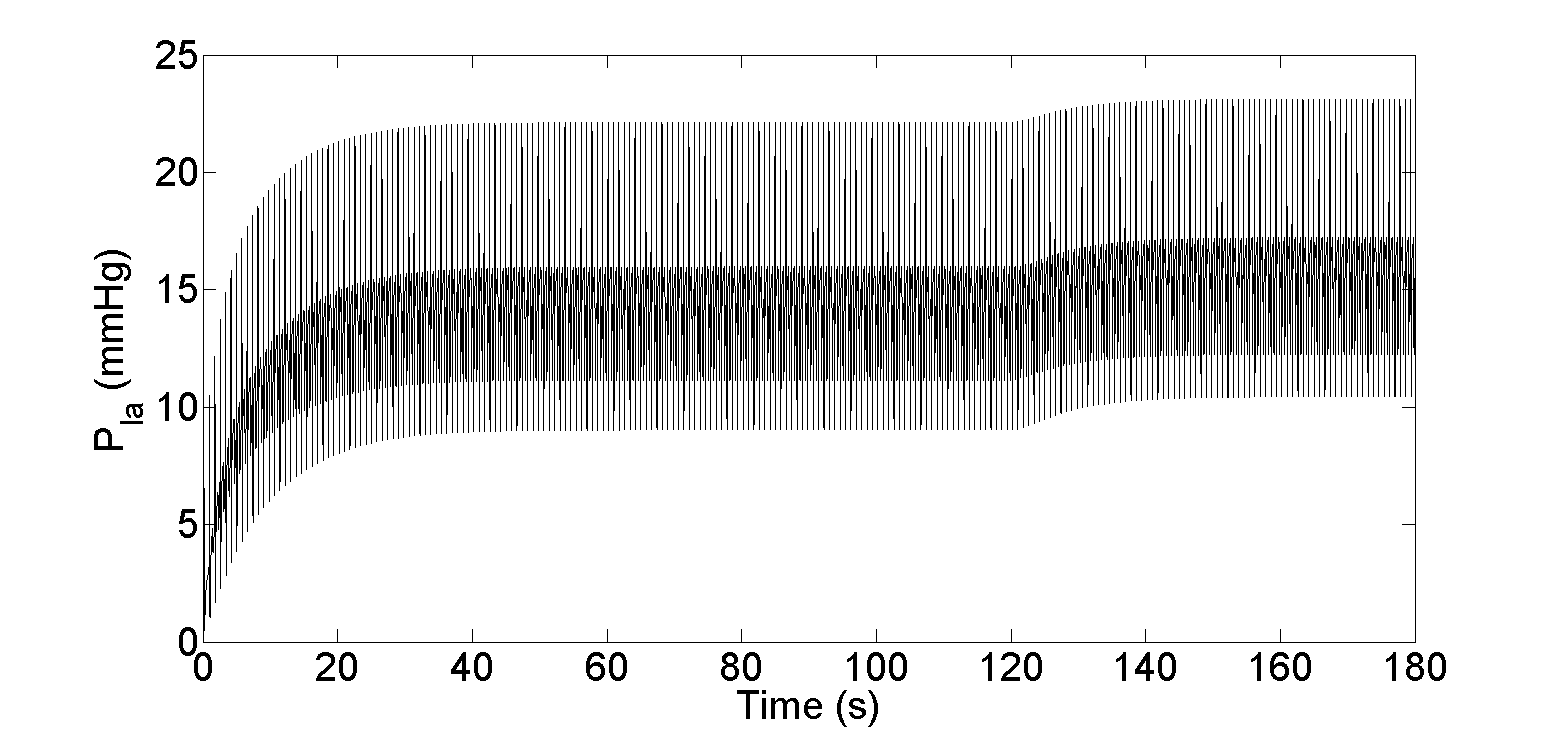}
   \label{62ed}
 }

\subfigure[Right atrial pressure.]{
   \includegraphics[scale =0.16752] {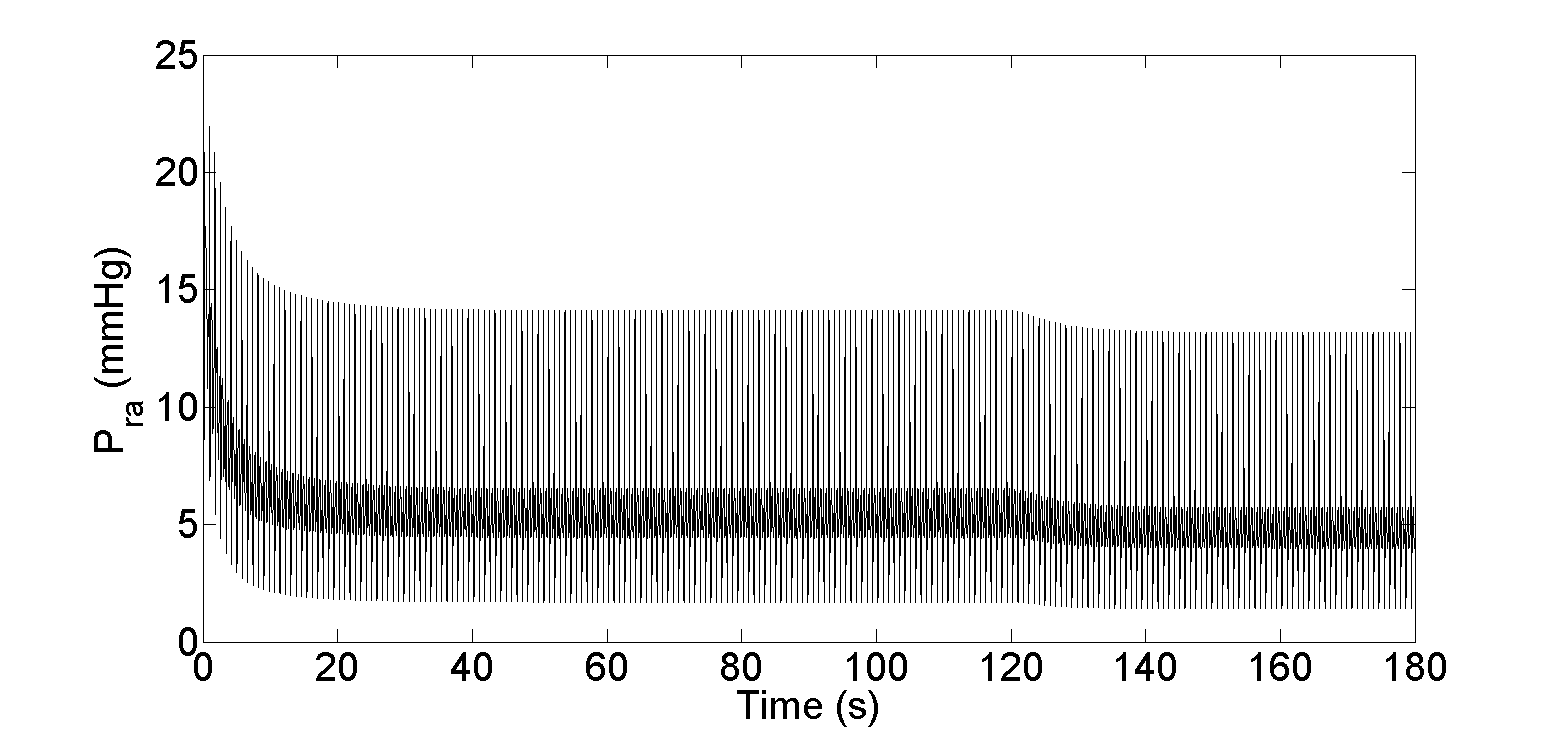}
   \label{62ee}
 }

\caption{Hemodynamic variables results in exercise condition when the system induced at 120s.}
\label{6:20ea}
\end{figure*}

\begin{figure}[htbp]
\centering
\subfigure[Average pump speed.]{
   \includegraphics[scale =0.16752] {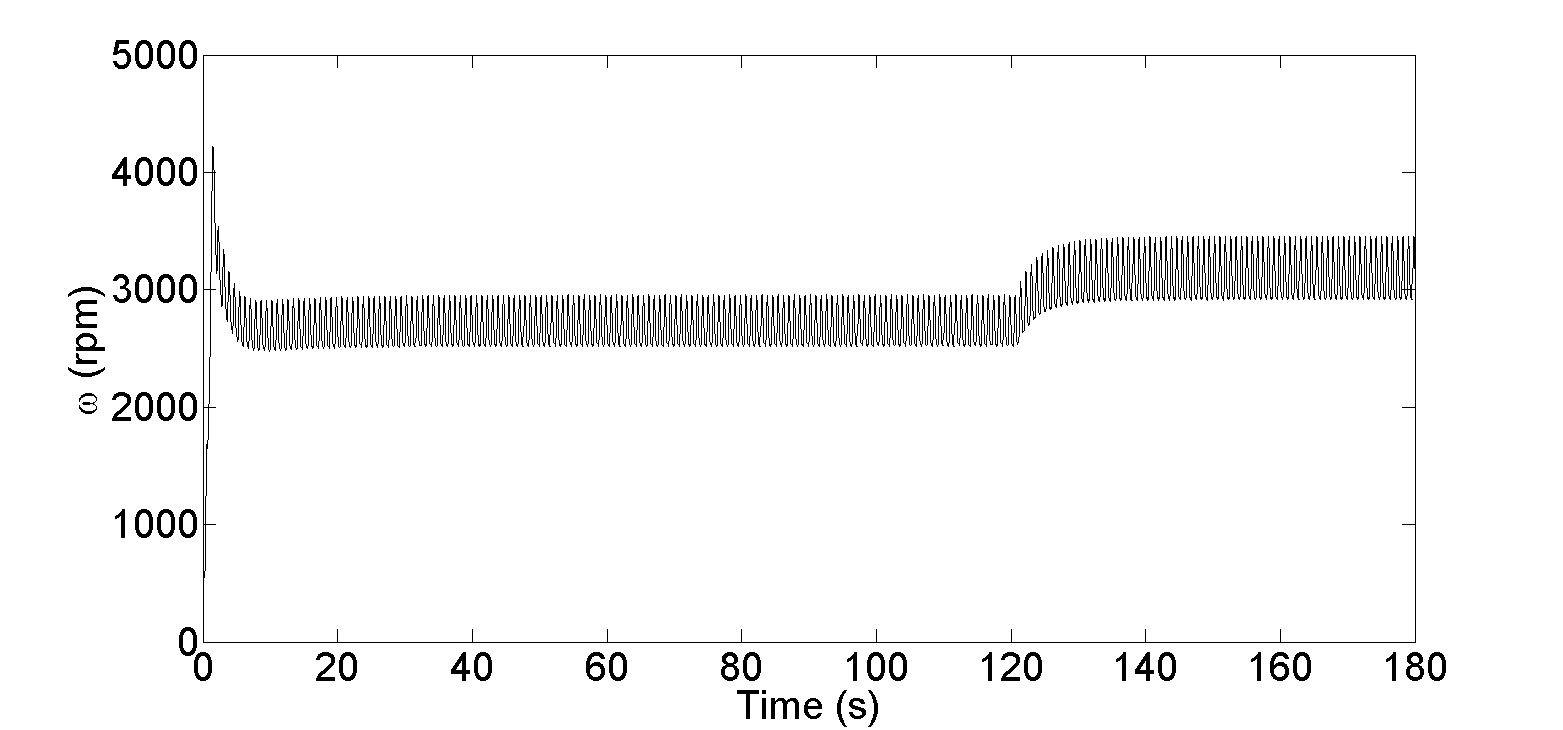}
   \label{62ef}
 }
 \subfigure[Pump flow pulsatility versus average pulsatile flow.]{
   \includegraphics[scale =0.16752] {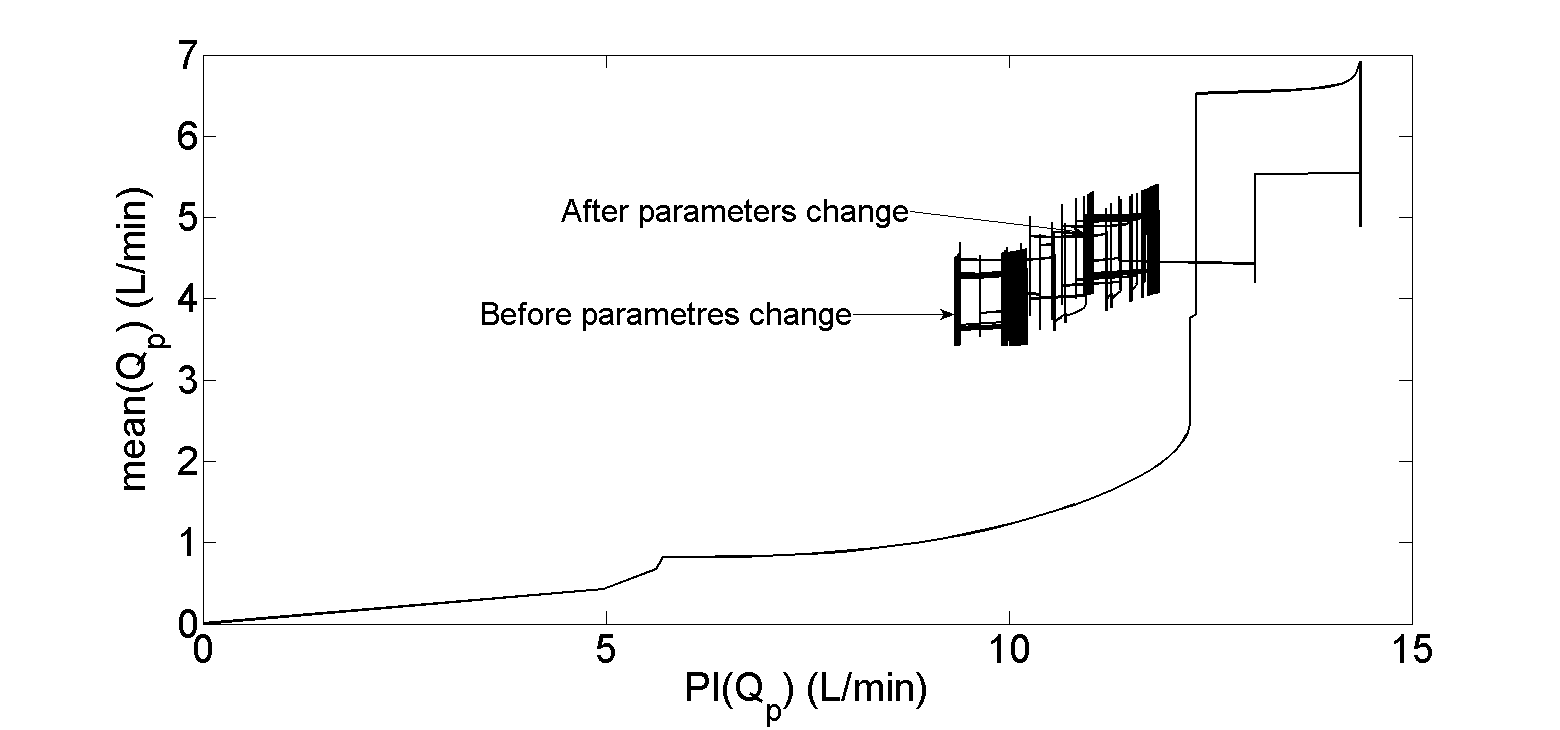}
  \label{62eg}
 }

\subfigure[Pump flow compared with desired reference flow at initial time.]{
   \includegraphics[scale =0.16752] {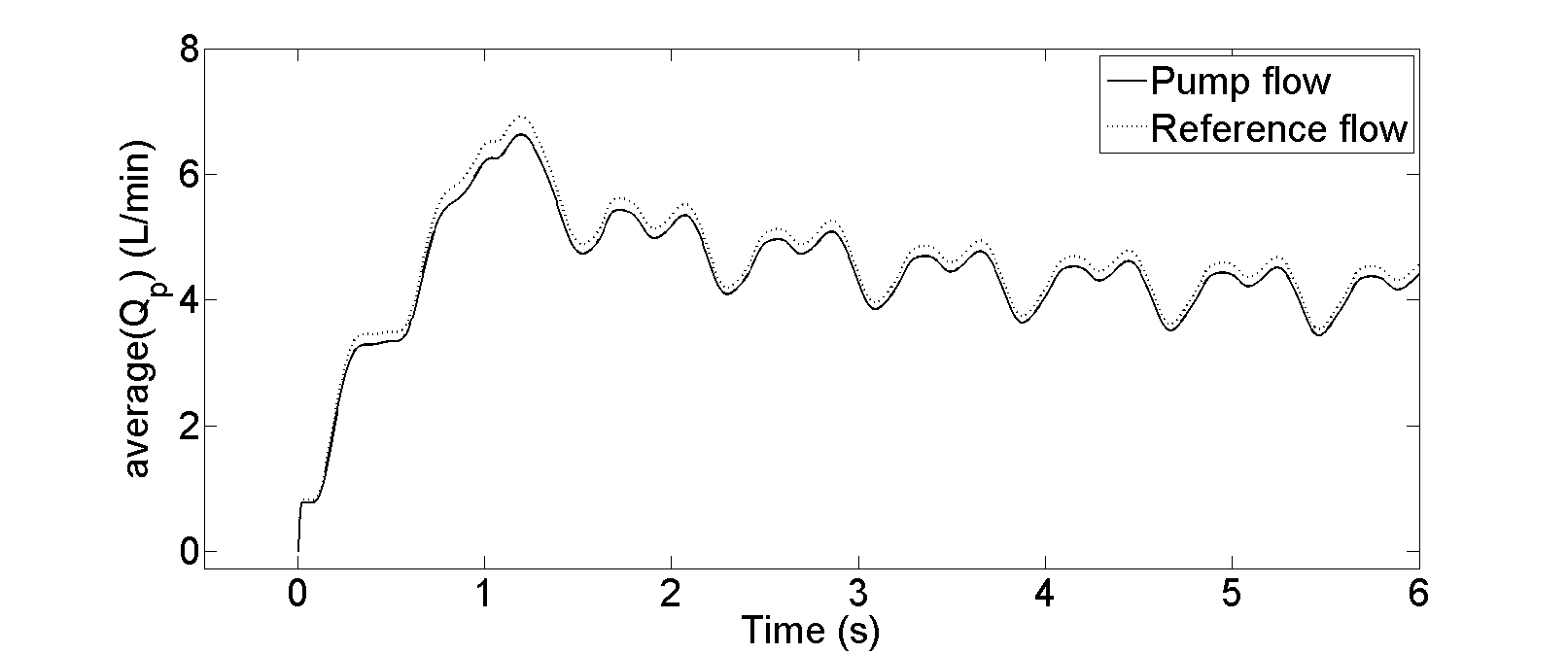}
   \label{i62eh}
 }
  \subfigure[Pump flow compared with desired reference flow at induced time.]{
   \includegraphics[scale =0.16752] {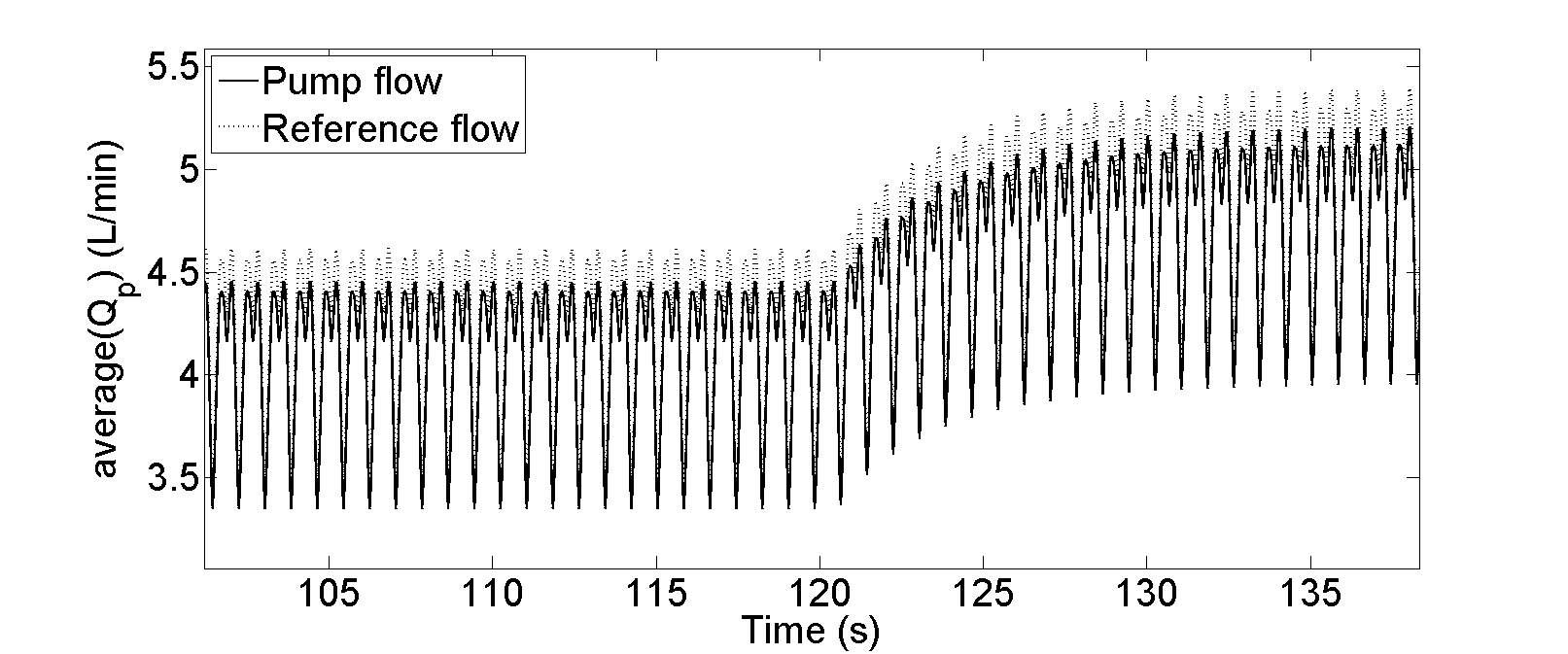}
   \label{62eh}
 }

\subfigure[Measured steady state pump flow against estimated pump flow.]{
   \includegraphics[scale =0.16752] {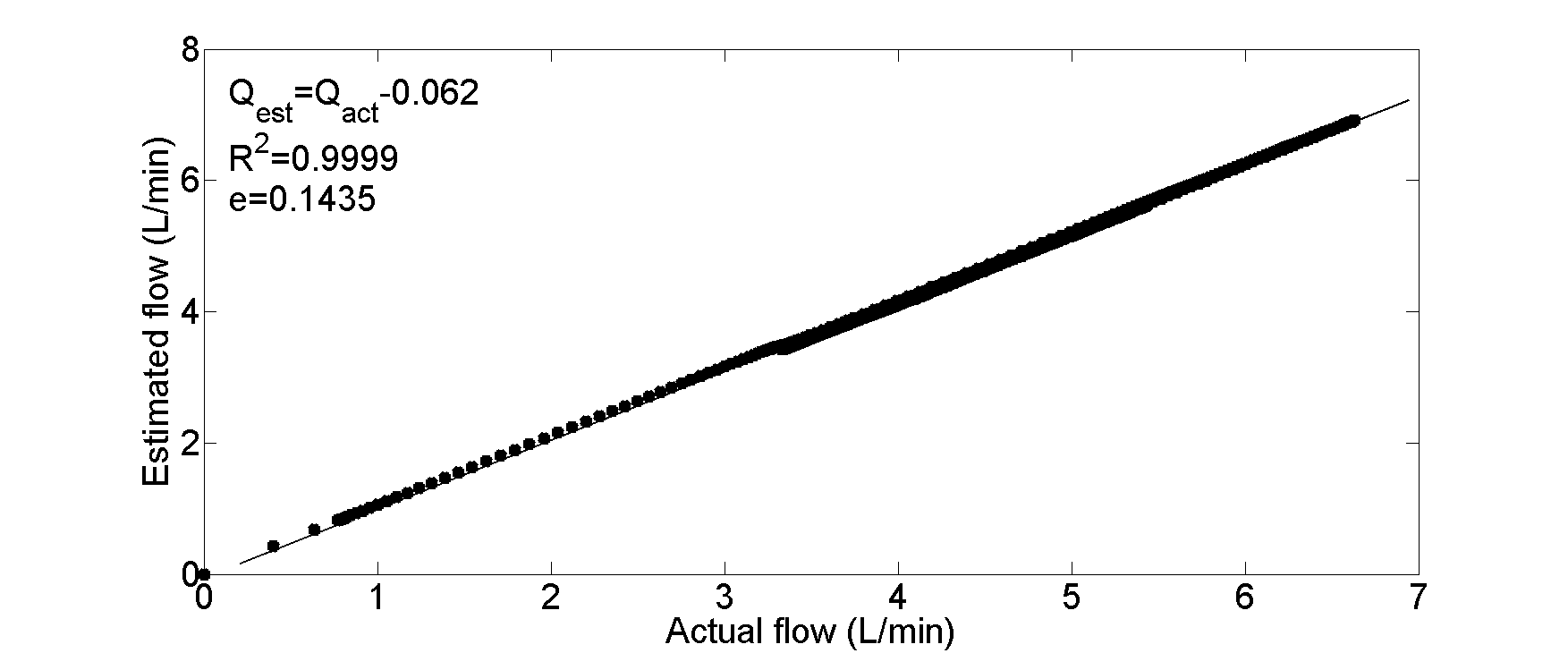}
   \label{62ei}
 }

\caption{Pump variable results in exercise condition when the system induced at 120s.}
\label{6:20eb}
\end{figure}

In the second scenario, the model parameters have been varied to simulate the transition from rest to exercise. Figures \ref{6:30ea} - \ref{6:20eb} illustrate the immediate response of the controller to the parameter changes at the middle of each period of 30s, 60s, 90s, and 120s respectively. These changes produce a rightward shift of LV pressure volume loops combined with a major increase in LV stroke volume, and similar increase in LV end-systolic pressure. In addition, the LVAD successfully decreases the aortic pressure $P_{ao}$ and increases the left atrial pressure $P_{la}$ and keeps the right atrial pressure $P_{ra}$ within safe operating mode (see Figures \ref{6:30ea}, \ref{6:60ea}, \ref{6:90ea} and \ref{6:20ea}).

In regards to pump variables, the controller responds to an increase in LV preload and subsequent pump flow pulsatility by increasing average pump rotational speed from 2900 rpm to 3400 rpm, actual average pulsatile flow from around 4.50 L/min to 5.05 L/min and estimated average pulsatile flow from around 4.8 L/min to 5.25 L/min. These changes are substantially completed within five heartbeats. Also, it can be observed that the simulated pump flow accurately tracks the desired reference flow within an error of $\pm$ 0.22 L/min. Figures \ref{63eh}, \ref{66eh}, \ref{69eh} and \ref{62eh} show an extremely close correlation between actual and estimated pump flows. Figures \ref{63ei}, \ref{66ei}, \ref{69ei} and \ref{62ei} show that the correlation between actual and estimated flows is highly significant, and the slope is close to unity for the linear regression.

Table \ref{6tab:t2} presents a summary of the salient hemodynamic variables, specific for the heart failure condition, before and after perturbations of blood loss and exercise. While Table \ref{6t2} shows the comparison between the values of the model correlation $R^{2}$, slope $S$ and mean absolute error $e$ for each period of time.


\begin{table}[htbp]
 \caption{Associated hemodynamic variables at rest and exercise conditions.}
 \small\addtolength{\tabcolsep}{5pt}
  \label{6tab:t2}
	\begin{center}
		\begin{tabular}{ c  c  c  c  c }
			\hline
	     \multirow{2}{*}{Variables} & \multirow{2}{*}{Unit}       &  \multicolumn{3}{c }{Heart failure  plus LVAD} \\\cline{3-5} 	
	                            &       &   Normal     &  Rest (Blood loss)   &   Exercise  \\  \hline
			 $P_{lved}$         & mmHg      &  10.40    &   6.50      &  21.50    \\
	     $P_{rved}$         & mmHg      &  7.80    &   5.20      &  13.90   \\
	     $V_{lves}$         & L/min     &  38.80    &   62.70     &  54.10     \\
			 $V_{lved}$	        & L/min     &  142.70   &   132.30    &  152.40    \\
			 $SV$               & mL        &  100.00   &   96.70     &  100.0    \\
			 $\overline P_{ao}$ & mmHg      &   83.50   &   125.10     &  97.80    \\
			 $\overline P_{la}$ & mmHg      &   14.20  &   11.80    &  17.15    \\
			 $\overline P_{ra}$ & mmHg      &   6.50   &   7.80     &  5.65    \\
			 $\overline Q_{act}$& L/min     &   4.50   &   3.40     &  5.05        \\
			 $\overline Q_{est}$& L/min     &   4.80   &   3.60     &  5.25        \\
			\hline
		\end{tabular}
	\end{center}
\end{table}

\begin{table}[htbp]
 \caption{Values of the model correlation (R), slope (S) and mean absolute error (e) in different period of times.}
 \small\addtolength{\tabcolsep}{5pt}
  \label{6t2}
	\begin{center}
		\begin{tabular}{ c  c  c  c c c c}
			\hline
			\multirow {3} {*} {Time (s)} & & &\multicolumn{2}{c}{Heart failure  + LVAD} \\ \cline{3-6}	
	                  &   \multicolumn{3}{c }{Blood loss}    &    \multicolumn{3}{c }{Exercise}  \\ \cline{2-7}
										&    $R^{2}$        &  S        &   $e$ (L/min)          &   $R^{2}$ & S       &  $e$ (L/min)       \\   \hline
	     30           &    0.9999         &  1      &  0.1255      &    0.9999      &  1    &  0.1486  \\
	     60           &    0.9999         &  1      &  0.1189      &    0.9999      &  1    &  0.1478   \\
	     90           &    0.9999         &  1      &  0.1277      &    0.9999      &  1    &  0.1418   \\
			 120	        &    0.9999         &  1      &  0.1301      &    0.9999      &  1    &  0.1435    \\
			 \hline
		\end{tabular}
	\end{center}
\end{table}

\section{Discussions}

This study demonstrates the application of the SMC technique to the recently described Starling-like controller for physiological control of pump rotational speed in IRBPs. To author's best knowledge, it is an innovative approach and implemented for the first time for such type of problems. It is important to note that the presented  simulations covering both perturbations in blood volume and exercise resulting in decreases and increases in pump rotational speed respectively, appropriate changes in pump rotational speed and flow are induced extremely rapidly without significant transient or steady state errors. We believe that this represents significantly good progress on two fronts.

As described previously in the literature review, there are two main limitations of previously designed control algorithms as compared with our work: the first is the need to estimate differential pressure and flow using steady-state models without data relating to the transient response of the pump. The second is that other systems measure pressure and/or flow and thus require the implantation of additional sensors to provide measurements of parameters used as inputs to their control algorithm. In this work, we have tried to minimise/eliminate those limitations.

In our opinion, the Starling-like controller, which emulates the inherent control mechanism in the natural heart to synchronise the outputs of left and right ventricles, is superior to other reported physiological strategies like the  control of differential pressure across the pump \cite{bullister2002physiologic,giridharan2002nonlinear}; afterload control \cite{wu2003advanced}; flow control \cite{lim2011method,fu2000computer}; preload control \cite{choi2001sensorless}; control of pulsatility ratio \cite{choi2007hemodynamic} and control of pulsatility gradient \cite{arndt2008physiological}. In the healthy cardiovascular system the cardiac output is normally generated ultimately not by the  LV but by the metabolic requirements of the tissues as transferred to the left heart by the right heart via the pulmonary circulation \cite{menicanti2007surgical}. Hence the ultimate control goal of the LVAD should therefore be to maintain a stable appropriate level of work for the failing LV and at the same time faithfully deliver the output from the pulmonary circulation to the arterial compartment. Achieving the goal means that a average arterial pressure in excess of 60 mmHg is ensured, which is a fundamental pre-requirement for auto-regulation of flows by the tissues \cite{mcintyre1971hemodynamic}.

The second important prerequisite for any physiological control system particularly at low preload, where there is an imminent risk of ventricular suction, is a rapid speed of response. Some physiological events like changes in posture can cause ventricular suction within ten heart beats. The proposed sliding mode controller meets this requirement very competently. Our simulations indicate that the change in pump rotational speed is completed efficiently within four to five heartbeats. Therefore this performance is superior to standard proportional-integral-derivative (PID) controllers where the high gain for the proportional coefficient, necessary for a rapid response, leads to under damped transient responses causing not desired overshoots and oscillations in pump rotational speed \cite{goodwin2001control}. In addition, the operation of the derivative componentis compromised by the noisy nature of estimated flow and flow pulsatility signals.

In addition to inherent advantages, SMC has some drawbacks like chattering in the output signal in case of continuous SMC systems. However, this is not valid for our discrete SMC system. Furthermore, there are some specific techniques to eliminate the chattering in case of continuous plants as well such as using the �sat� function instead of the; �sign� function can be applied \cite{utkin1992sliding}.

The Starling-like controller requires accurate data on pump rotational speed and speed pulsatility which are actually estimated in this application. While estimation avoids the inherent problems in flow and pressure sensors, it is subjected to the problems of its own particularly inaccuracy and slow speeds \cite{slaughter2009intraoperative}. Secondly estimation is mostly applicable to steady states only \cite{alomari2011non}.  Also, the accuracy of estimation methods can easily be severely compromised \cite{slaughter2009intraoperative}. So, implanted sensors give a better  immediate response comparatively.

In the existing literature, many other alternative methods have been proposed for error trajectory tracking but mostly all of them contain some inherent drawbacks. For example, Fuzzy logic control suffers from a slow response time due to heavy computational loads \cite{passino1998fuzzy}; a feedback linearisation approach is limited by the convergence conditions while  it is difficult to find a suitable Lyapunov candidate function in the case of Lyapunov function based control \cite{khalil1996nonlinear}.

Although the proposed control algorithm was shown to be able to respond quickly enough to sudden perturbations in the cardiovascular system accordingly without inducing suction, but still another issue introduced in this method. This issue is the transient overshoot which can be observed at first two seconds in each scenario on Figs.\ref{i63h}, \ref{i66h}, \ref{i69h} and \ref{i62h} at blood loss scenario and Figs. \ref{i63eh}, \ref{i66eh}, \ref{i69eh} and \ref{i62eh} at exercise scenario. Despite, the overshoot is increases the average pulsatile flow up to 7.2 L/min, overperfusion is not occurred with this level. Further studies to eliminate these limitations using H-infinity control \cite{petersen2000robust, savkin1996robust} is in progress.

\section{Conclusion}

In this chapter we have  presented an  innovative physiological controller that mimics the Frank-Starling law of the heart  coupled with a novel robust  SMC. The controller is capable of adjusting the average pulsatile flow efficiently using pump flow pulsatility as the feedback parameter. The immediate response of the controller subjected to varying conditions such as at rest and from rest to exercise has been evaluated using a lumped parameter model of the cardiovascular system. Simulation results depict that the abnormal hemodynamic variables of heart failure patients are restored back to a normal physiological range. The future work includes further in-depth validation of the proposed strategy using a circulation mock loop and in-vivo animal experiments.

\chapter{Conclusions and Future  Work}

\section{Conclusions} \label{7sec:con}
This report aims to design, develop and validate extensively physiological control algorithms to drive LVADs. The LVAD is a newly third generation device designed as a permanent alternative to the heart transplants. The feedback signals to the physiological control system include the motor signals and the pump flow obtained from sensorless estimator. The main objective of the physiological control system is to restore the abnormal hemodynamic variables of heart failure patients back to a normal physiological range,  and to maintain average pulsatile flow properly in order to prevent ventricular suction or over perfusion using appropriate advanced estimation and control techniques. We summarise the research work of this dissertation as follows:

In $\boldsymbol{Chapter}$ $\boldsymbol{2}$, we have proposed two auto-regressive (ARX) models using linear time variant (LTV) system to estimate the average pulsatile flow. The first ARX model uses pulse-width modulation (PWM)  signal that is acquired sensorless from the pump controller to estimate the pulsatility index of rotational speed. The second ARX model is used to model the pulsatility index of rotational speed to  estimate average pulsatile flow. This model has been developed by incorporating the cardiovascular system (CVS) model along with the model of an LVAD. These models are based on actual dogs' data that is collected in the real-time experiments. The final outcome of the proposed system is significantly high correlation between estimated and measured flows coupled with as minimum absolute error as $(e=0.22)$ L/min as compared to existing techniques with a minimum $e$ as 0.45 L/min. 

In $\boldsymbol{Chapter}$ $\boldsymbol{3}$, an innovative robust control algorithm has been proposed for a left ventricular assist device (LVAD) using model reference sliding mode control (MRSMC) approach in order to properly adjust average pulsatile flow. The controller is developed with the main objective of the tracking the desired reference input without inducing any suction in the ventricle. The average pulsatile flow is estimated non-invasively using a stable dynamical model developed with actual data  obtained from animal experiments ($\boldsymbol{Chapter}$ $\boldsymbol{3}$). The validation of the proposed controller is based on a lumped parameter model of the cardiovascular system (CVS) that has been developed previously using healthy pigs data over a wide range of operating conditions. The simulation  is carried out using both constant and sinusoidal reference pump flow input of varying mean, amplitude and phase shift to study the hemodynamic response of the CVS under varying conditions. The immediate response of the proposed controller is evaluated using two physiological conditions ranging from rest to exercise. It has been observed from simulations that the proposed controller tracks the reference input signal very well with minimum mean absolute error of 0.46 L/min and is fairly robust against model uncertainties and external disturbances.

In $\boldsymbol{Chapter}$ $\boldsymbol{4}$, a physiological control algorithm based on feedforward - sliding mode control  has been proposed to regulate the rotational speed of LVADs according to the metabolic demand of the body. We proposed the reference signal that is updated based on a non-linear dynamical function representing the metabolic demand of the body and designed a tracking controller for it. The main purpose of proposing the non-linear reference function is to prevent suction (under pumping) and over perfusion (over pumping). The proposed controller is validated using a lumped CVS - RBP model obtained from a variety of actual pigs' data experiments. The tracking  performance of the controller is evaluated subjected to different circulatory changes from rest to exercise.  Simulations demonstrate that the proposed controller is capable of tracking the complex desired reference function very accurately even in presence of external disturbances and uncertainties with a minimal mean absolute error.

In $\boldsymbol{Chapter}$ $\boldsymbol{5}$, we have designed and developed a physiological control algorithm emulating the Frank Starling-like mechanism to control the operation of LVADs. The proposed approach is based on pole placement sliding mode control to construct the Frank Starling-like law of the heart which demonstrates that the left ventricular (LV) outflow increases with an increase in the LV filling pressure. The developed controller adjusts the average pulsatile flow automatically using pump flow pulsatility and estimated average pulsatile flow. Migration to different controller gradients is allowed to compensate for longer term and larger variations in LV contractility and the metabolic requirements of the body and restricted within the upper and lower limits for both average pulsatile flow and pump flow pulsatility. The controller is assessed under a variety of varying operating conditions such as the changes at rest and changes from rest to using a lumped parameter model of the cardiovascular system. Simulation results prove the effectiveness of the proposed control methodology and it has been shown that the proposed controller is capable of restoring any abnormal hemodynamic variables of HF patients back to normal.

\section{Directions for Future Research}

Although the merits of proposed control algorithms have been studied carefully and validated using parameter optimized model of the CVS - RBP that was previously developed and validated in our laboratory, however, we propose the following future research directions to further enhance the quality of the work done and based on our experience we suggest a few important areas where improvements can be made more effectively and efficiently:

\subsection{Control Algorithms Development in Pulsatile Mock-loop Studies}

Further evaluation and validation of the developed control algorithms using pulsatile circulatory mock loop system are needed. This will entail the development of a LabVIEW/MATLAB environment to implement control strategies for implementation on the physical LVAD controller. In this validation, the postural changes and valsalva manoeuvres can be simulated in realistic rates of change in the physical mock loop resistance and compliance.

\subsection{Algorithms Validation in Acute Animal Studies}

Validation of the developed control algorithms in acute animal studies using thiopentone infusions to induce controllable levels of heart failure is needed. Where, the control algorithms under range of different conditions including variable ventricular function and systemic vascular resistance via pharmacological investigation should be evaluated. In addition, online testing of pump state detection algorithms including suction detection and regurgitant flow in animals can be evaluated.

\subsection{Advanced Modelling and State Estimation Techniques}

One of directions for the future research is the use of modern advanced modelling and state estimation techniques. Applications of such methods have the potential to significantly improve the accuracy and the robustness of the mathematical models for an LVAD interacting with human cardiovascular system. The advanced estimation and validation techniques include robust Kalman state estimation \cite{pathirana2005node, savkin1995recursive, moheimani1998robust}, model validation \cite{savkin1996modell, petersen1999robust, savkin1998robust} and Kalman state estimation with with communication constraints \cite{matveev2003problem, savkin2003set, matveev2001optimal}.

\subsection{Advanced Modern  Control Techniques}
Although the proposed control algorithms have been shown to be able to respond quickly enough to sudden perturbations in the cardiovascular system accordingly without inducing suction; future studies may include other advanced modern control strategies such as robust linear quadratic and LQG control \cite{savkin1995minimax, matveev2004problem}, H-infinity control based robust stabilization \cite{savkin1995nonlinear, ugrinovskii2000decentralized}, robust controller switching \cite{matveev2000qualitative, savkin2002hybrid, savkin1999robust, skafidas1999stability}, communication limited control \cite{matveev2009estimation, savkin2006analysis, matveev2005multirate, savkin2007detectability, matveev2007analogue}, back stepping control \cite{li2004robust} and geometric control \cite{jurdjevic1997geometric}.

The outcomes of the proposed research will be the improved and optimised potential of generally poorly managed biomedical devices, and consequently enabling the improved life style and mortality for patients with end stage heart failures.



\end{document}